\documentclass [PhD] {uclathes}

\usepackage{graphicx}% Include figure files
\usepackage{dcolumn}% Align table columns on decimal point
\usepackage{bm}% bold math
\usepackage{bbold}
\usepackage{amsmath}
\usepackage{amssymb}
\usepackage{amsfonts}
\usepackage[usenames, dvipsnames]{color}
\usepackage [cp1250]{inputenc}

\usepackage{lipsum}    
\usepackage{textcomp}
\usepackage[T1]{fontenc}
\usepackage{multirow}

\usepackage{url}
\usepackage[colorlinks=true,
pagebackref=true,
linkcolor=blue,anchorcolor=blue,citecolor=blue,filecolor=blue,menucolor=blue,runcolor=blue,
urlcolor=VioletRed
]{hyperref}

\usepackage{titlesec}
\titleformat{\section}{\bfseries\Large}{\thesection.}{0.5em}{}
\titleformat{\subsubsection}{}{\thesubsubsection}{1em}{\itshape}

\def\hfp{\hspace{5pt}}
\def\hffp{\hspace{15pt}}

\def\distand{\hspace{15pt} {\textrm{and}} \hspace{15pt}}

\newcommand{\bhat}[1]{\bm{\hat{#1}}}

\newcommand{\infrac}[2]{{#1}/{#2}}
\newcommand{\inbfrac}[2]{\left({#1}/{#2}\right)}

\newcommand{\derr}[2]{\frac{d{#1}}{d{#2}}}
\newcommand{\inderr}[2]{d{#1}/d{#2}}
\newcommand{\inbderr}[2]{\left(d{#1}/d{#2}\right)}

\newcommand{\parr}[2]{\frac{\partial{#1}}{\partial{#2}}}
\newcommand{\inparr}[2]{\partial{#1}/\partial{#2}}

\newcommand{\parrtwo}[3]{\frac{\partial^2{#1}}{\partial{#2}\partial{#3}}}

\newcommand{\txt}[1]{\textrm{#1}}

\newcommand{\cb}[1]{{\color{blue}{#1}}}
\newcommand{\cred}[1]{{\color{red}{#1}}}
\newcommand{\cg}[1]{{\color{PineGreen}{#1}}}
\newcommand{\co}[1]{{\color{Orange}{#1}}}

\makeatletter
\newsavebox\myboxA
\newsavebox\myboxB
\newlength\mylenA
\newcommand*\barr[2][0.75]{%
	\sbox{\myboxA}{$\m@th#2$}%
	\setbox\myboxB\null% Phantom box
	\ht\myboxB=\ht\myboxA%
	\dp\myboxB=\dp\myboxA%
	\wd\myboxB=#1\wd\myboxA% Scale phantom
	\sbox\myboxB{$\m@th\overline{\copy\myboxB}$}%  Overlined phantom
	\setlength\mylenA{\the\wd\myboxA}%   calc width diff
	\addtolength\mylenA{-\the\wd\myboxB}%
	\ifdim\wd\myboxB<\wd\myboxA%
	\rlap{\hskip 0.5\mylenA\usebox\myboxB}{\usebox\myboxA}%
	\else
	\hskip -0.5\mylenA\rlap{\usebox\myboxA}{\hskip 0.5\mylenA\usebox\myboxB}%
	\fi}
\makeatother

\def\unity{\mathbb{1}}

\def\bnabla{\bm{\nabla}}

\def\llangle{\left\langle}
\def\rrangle{\right\rangle}
\newcommand{\non}{\nonumber}

\def\pdens{\frac{d^3p}{(2\pi)^3}~}

\def\ptilde{d^3\tilde{p}~}

\newcommand{\eps}{\varepsilon}

\usepackage{tikz}
\usetikzlibrary{shapes.geometric, arrows}

\tikzstyle{io} = [rectangle, rounded corners, minimum width=3cm, minimum height=1cm, text centered, text width = 5.0 cm, draw=black, fill=cyan!20]

\tikzstyle{startingpoint} = [rectangle, rounded corners, minimum width=3cm, minimum height=1cm,text centered, text width=5.0cm, draw=black, fill=green!30]

\tikzstyle{operationsmall} = [rectangle, minimum width=3cm, minimum height=1cm, text centered, text width = 5.5 cm,  draw=black, fill=yellow!30]

\tikzstyle{operation} = [rectangle, minimum width=3cm, minimum height=1cm, text centered, text width = 9.0 cm,  draw=black, fill=yellow!30]

\tikzstyle{question} = [rectangle, minimum width=3cm, minimum height=1cm, text centered, draw=black, fill=red!25]

\tikzstyle{arrow} = [thick,->,>=stealth]

\usepackage{tikz-feynman}

\title          {Density Functional Equation of State \\ %of Nuclear Matter \\
                 and Its Application to the Phenomenology of Heavy-Ion Collisions}

\author         {Agnieszka Ma{\l}gorzata Sorensen}
\department     {Physics}
\degreeyear     {2021}

\chair          {Huan Z.\ Huang}        
\member         {Volker Koch}
\member         {Michail Bachtis}
\member         {Zhongbo Kang}
\member         {Giovanni Zocchi}

\abstract       {
	
A prominent goal within the field of modern heavy-ion collisions is to uncover the phase diagram of QCD. Studies of the properties of systems created in heavy-ion collisions strongly suggest that a new state of matter described by quark and gluon degrees of freedom, the quark-gluon plasma, is created when nuclei are collided at very high-energies. Consequently, the QCD phase diagram may contain a rich structure in regions currently accessible to heavy-ion experiments, including a possible critical point where the transformation between hadronic and partonic matter changes from a smooth crossover to a first-order phase transition. Whether this is the case will have to be born out through a combination of experimental analyses and state-of-the-art simulations of heavy-ion collisions.

We present a mean-field model of the dense nuclear matter equation of state designed for use in computationally demanding hadronic transport simulations. Our approach, based on the relativistic Landau Fermi-liquid theory, allows us to construct a family of equations of state spanning a wide range of possible bulk properties of dense QCD matter. For the application to simulations of heavy-ion collisions at intermediate beam energies, and in particular having in mind studies centered on probing the regions of the QCD phase diagram most relevant to the search for the QCD critical point, we further present and discuss parametrizations of the developed equation of state describing dense nuclear matter with two phase transitions: the known nuclear-liquid gas phase transition in ordinary nuclear matter, with its experimentally observed properties, and a postulated phase transition at high temperatures and high baryon number densities, meant to model the QCD phase transition from hadronic to quark and gluon degrees of freedom.

We implement the developed model in the hadronic transport code \texttt{SMASH}, and show that the resulting dynamic behavior reproduces theoretical expectations for the thermodynamic properties of the system based on the underlying equation of state. In particular, we discuss simulations of systems initialized in regions of the phase diagram affected by the conjectured  QCD critical point, and we demonstrate that they reproduce effects due to critical behavior. Specifically, we show that pair distribution functions calculated from hadronic transport simulation data are consistent with theoretical expectations based on the second-order cumulant ratio, and can be used as a signature of crossing the phase diagram in the vicinity of a critical point. Through this, we validate the use of hadronic transport codes as a tool to study signals of a phase transition in dense nuclear matter.

We additionally present a novel method that may enable a measurement of the speed of sound and its derivative with respect to the baryon number density in heavy-ion collisions. The devised approach is based on a connection between the speed of sound and the cumulants of the net baryon number, which in the context of the search for the QCD critical point are given considerable attention due to their potential to signal critical fluctuations. We confirm the applicability of the proposed method in two models of dense nuclear matter, including the parametrization of the equation of state developed in this work. Application of our approach to available experimental data implies that the derivative of the speed of sound is non-monotonic in systems created in collisions at intermediate to low energies, which in turn may be connected to non-trivial features in the underlying equation of state.

}

\acknowledgments { 
	
This work would not have been possible without the guidance and support I have received at numerous stages of my PhD.  

I could not have had better advisors than Huan Huang and Volker Koch. 

Huan graciously opened his group to me at a point when I was failing by usual standards of graduate school performance. Not only did he put his support behind me at that difficult time, but he then sought opportunities for my research at Lawrence Berkeley National Laboratory (LBNL), where I could follow my strongest interests in nuclear physics, and arranged for my two-year stay there as a visiting student. Throughout my time in his group, Huan always encouraged me to continue pursuing my next goal, and his strong belief that it can all be done often gave me a well-needed boost of optimism.

Volker took on the role of my LBNL mentor, and he persisted in his commitment to this role even though advising me in my PhD research was not always free from difficulties. His collaborative style and openness to discussing various facets of our project at length, including answering my unceasing questions, enriched my understanding of the field beyond what could be learned from the literature or at scientific meetings. Volker's candid involvement in all steps of my work was truly extraordinary, and his daily advice and support allowed me to produce research results that I am confident in and that I am proud of. I cannot imagine completing this thesis without his help and patience.

Dima Oliinychenko, who was a postdoc at LBNL throughout my time there, first helped me make my early steps in contributing to a hadronic transport code, and then, through daily discussions and advice, became one of the strongest influences on my PhD experience. His kindness, enthusiasm, and willingness to delve into minor details of heavy-ion phenomenology, programming, and physics subjects in general were invaluable. Moreover, when starting his next postdoc at the Institute for Nuclear Theory, Dima invited Volker and me to work on a research project he was pursuing with Larry McLerran, which turned into a great collaboration and was the highlight of my final year as a PhD student.

The intellectual and financial support of the Beam Energy Scan Theory (BEST) Topical Collaboration allowed me to create connections beyond my closest scientific circle. In particular, the weekly virtual meetings of the BEST Interfaces group -- comprised of Dima Oliinychenko, Scott Pratt, Chun Shen, Karina Martirosova, and Sangwook Ryu -- led to many fruitful discussions on subjects related to the Beam Energy Scan and beyond. The contribution of these meetings to my knowledge and to my sense of community was priceless. 

Hannah Elfner, who heads the group developing the hadronic transport code \texttt{SMASH} at the Frankfurt Institute for Advanced Studies, generously gave me access to the \texttt{SMASH} development branch, which allowed me to not only stay on top of the latest code developments, but also to observe best coding practices as they were implemented. 

All of the above could not have taken place without crucial institutional support. The Nuclear Theory group at LBNL, led by Feng Yuan, not only provided a vigorous scientific environment, but also helped fund my stay at LBNL. The UCLA Department of Physics and Astronomy gave me key moral and financial aid. In particular, David Saltzberg, both as the Faculty Graduate Advisor and then as the Chair of the Department, was always willing to discuss best strategies to move forward and offered help when I needed it. Stephanie Krilov, as the Graduate Student Affairs officer, was invaluable not only in managing forms and procedures, but also in doing so while projecting a sense of calm and optimism.

To all of them I would like to say: Thank you. You have helped me in ways that amount to more than the sum of your actions.

\hspace{20mm}

Many results contributing to this thesis can also be found in Refs.\ [\cb{177}, \cb{212}]. These references are the primary sources of that material, and should be given preference in citations. The writing of this thesis was supported by the UCLA Dissertation Year Fellowship.

}

\vitaitem   {2012}
                {M.S.~Theoretical Physics \\
                 University of Wroc{\l}aw, Poland}
\vitaitem   {2012-2013}
                {Researcher\\
                 University of Wroc{\l}aw, Poland}
\vitaitem   {2013--2018}
                {Teaching Assistant \\
                 Physics and Astronomy Department, UCLA}
\vitaitem   {2016--2018}
                {Teaching Assistant Consultant \\ 
                 Office of Instructional Development, UCLA}
\vitaitem   {2018--2021}
                {Graduate Student Researcher \\
                 Physics and Astronomy Department, UCLA}
\vitaitem   {2018--2021}
                {Visiting student at Lawrence Berkeley National Laboratory	}

\publication {A.\ Sorensen, V.\ Koch \\
	\textit{Phase transitions and critical behavior in hadronic transport with a relativistic density functional equation of state}\\
	Phys.\ Rev.\ C \textbf{104}, no.\ 3, 034904 (2021)
}

\publication {X.\ An \textit{et al.}\\
	\textit{The BEST framework for the search for the QCD critical point and the chiral magnetic effect}\\
	e-print: arXiv:2108.13867
}

\publication {M.\ Colonna \textit{et al.} \\
                     \textit{Comparison of Heavy-Ion Transport Simulations: Mean-field Dynamics in a Box} \\
                     Phys.\ Rev.\ C \textbf{104}, no.\ 2, 024603 (2021)
                 }
                   
\publication {A.\ Sorensen, D.\ Oliinychenko, V.\ Koch, L.\ McLerran \\
                   	 \textit{Speed of Sound and Baryon Cumulants in Heavy-Ion Collisions}\\
                   	 Phys.\ Rev.\ Lett.\ \textbf{127}, no.\ 4, 042303 (2021)
                    }

\publication {D.\ Blaschke, A.\ Dubinin, A.\ Radzhabov and A.\ Wergieluk \\
					\textit{Mott dissociation of pions and kaons in hot, dense quark matter} \\
					Phys.\ Rev.\ D \textbf{96}, no.\ 9, 094008 (2017)
					}
				
\publication {A.\ Wergieluk, D.\ Blaschke, Y.\ L.\ Kalinovsky and A.\ Friesen \\
					\textit{Pion dissociation and Levinson`s theorem in hot PNJL quark matter} \\
				Phys.\ Part.\ Nucl.\ Lett.\ \textbf{10}, 660-668 (2013)
					}

\graphicspath{{Figures/}}

\begin {document}
\makeintropages

%%%%%%%%%%%%%%%%%%%%%%%%%%%%%%%%%%%%%%%%%%%%%%%%%%%%%%%%%%%%%%%%%%%%%%

\chapter{Introduction to studies of the QCD phase diagram}
\label{introduction}

The goal of heavy-ion collisions is to study the properties of matter composed of quarks and gluons: the fundamental particles and force carriers associated with the strong interaction. Although the theory of strong interactions, quantum chromodynamics (QCD) \cite{Fritzsch:1973pi}, has been tested and confirmed in an overwhelming number of experiments, its intrinsic computational complexity poses a challenge to answering many outstanding questions in nuclear physics. In particular, while first-principle approaches together with perturbation theory can be used in studies of processes involving large momentum transfers, where the QCD coupling constant is small and the theory approaches asymptotic freedom \cite{Gross:1973id,Politzer:1973fx}, the large value of the coupling at energies of interest to nuclear physics means that the underlying processes cannot presently be understood through QCD calculations. Even more importantly, the description of studied problems predominantly involves many-body physics, where using first-principle methods is notoriously difficult. Therefore heavy-ion experiments and theoretical models, devoted to investigating the evolution of systems created in collisions of nuclei moving at relativistic velocities, are presently the only methods of studying the dynamic behavior of nuclear matter at high temperatures and high densities.

Current understanding of QCD matter at extreme conditions is facilitated by numerous experimental and theoretical advancements to date. The evidence gathered over 30 years of experiments carried out at the Super Proton Synchrotron (SPS), the Relativistic Heavy Ion Collider (RHIC), and the Large Hadron Collider (LHC) strongly suggests that a new state of matter is produced in high-energy heavy-ion collisions \cite{Heinz:2000bk,BRAHMS:2004adc,PHENIX:2004vcz,PHOBOS:2004zne,STAR:2005gfr,Muller:2012zq}. This new state of matter, characterized by deconfined and, at the same time, strongly-interacting color charges, is called the quark-gluon plasma (QGP), and if it is indeed achieved in high-energy heavy-ion collisions, then these experiments probe two states of strongly-interacting matter: the state in which the dominant degrees of freedom are hadrons, and the state in which the dominant degrees of freedom are quarks and gluons. This in turn means that it is possible to study the nature of the QGP-hadron transition in the laboratory. The characteristics of this phase transition are expected to vary at different temperatures and densities, and rich structures are predicted to be present in the QCD phase diagram. Studies devoted to this subject will not only lead to a better understanding of the properties of dense QCD matter, but also have the potential for revealing insights about fundamental aspects of the underlying theory, such as confinement. Moreover, among the several predicted phase transitions involving fundamental degrees of freedom of the Standard Model, the QGP-hadron phase transition is the only one that can be feasibly studied experimentally.

Below, we present a broad overview of heavy-ion collision studies devoted to uncovering the QCD phase diagram. In particular, we briefly recall what is known about the states of QCD matter (Section \ref{the_QCD_phase_diagram}) and sketch the physics behind probing different regions of the QCD phase diagram through heavy-ion collisions (Section \ref{probing_the_QCD_phase_diagram}). Next, after introducing the experimental programs devoted to the exploration of the phases of QCD (Section \ref{BES-I_and_BES-II}), we focus on prominent experimental results to date (Sections \ref{softening_of_the_equation_of_state}, \ref{turning_off_the_QGP}, and \ref{non-statistical_event-by-event_fluctuations_of_conserved_charges}), followed by a summary of challenges to finding the QCD critical point (Section \ref{Challenges_for_finding_the_QCD_critical_point}) and an overview of simulations of heavy-ion collisions (Section \ref{simulations_of_heavy-ion_collisions}). 

A reader familiar with the field can comfortably omit these introductory parts of this chapter and proceed directly to Section \ref{The_need_for_generalized_mean-field_potentials_in_hadronic_transport}, where we introduce the problem addressed in this thesis, and Section \ref{overview_of_the_thesis}, where we give a short description of the work presented in each of the following chapters.

We note here that throughout this work, we adopt the natural units (see Appendix \ref{units_and_notation} for more details on units and notation).

\section{The QCD phase diagram}
\label{the_QCD_phase_diagram}

\begin{figure}[t]
	\centering\mbox{
	\includegraphics[width=0.63\textwidth]{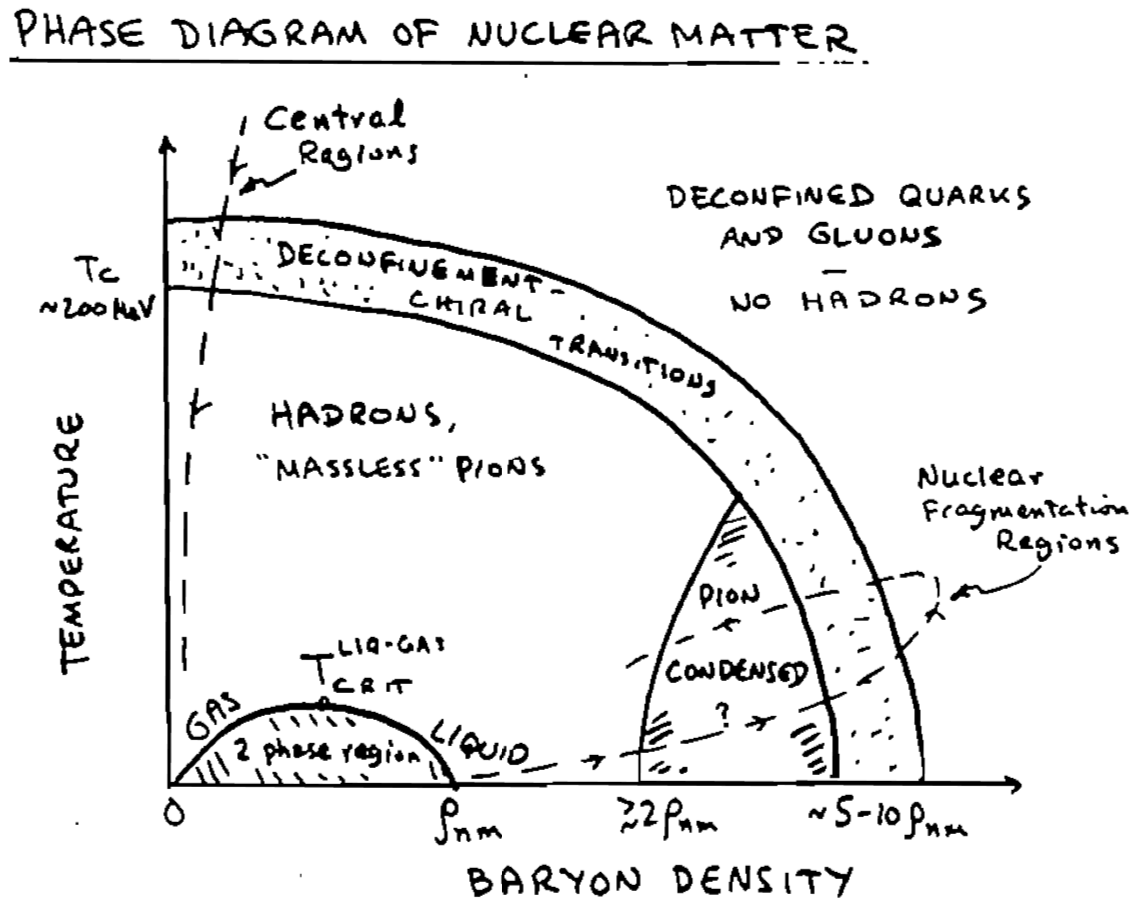}
	}
	\caption[The conjectured phase diagram of QCD matter (1983 Long Range Plan)]{A sketch of the conjectured phase diagram of QCD matter in the $(T,n_B)$ plane, included in the 1983 DOE/NSF Long Range Plan for Nuclear Science \cite{LongRangePlan1983}. The saturation density of nuclear matter is here denoted as $\rho_{nm}$. The original caption states: ``\textit{Expected phases of nuclear matter at various temperatures and baryon (or nucleon) densities, showing the ``hadronic phase'' including a gas-liquid phase transition region, and the transition region to deconfined quarks and gluons. The dashed lines illustrate trajectories in this phase diagram that can be explored in ultra-relativistic heavy-ion collisions.}'' The sketch was made by Gordon Baym.}
	\label{QCD_phase_diagram_Long_Range_Plan_1983}
\end{figure}

\begin{figure}[t]
	\centering\mbox{
	\includegraphics[width=0.6\textwidth]{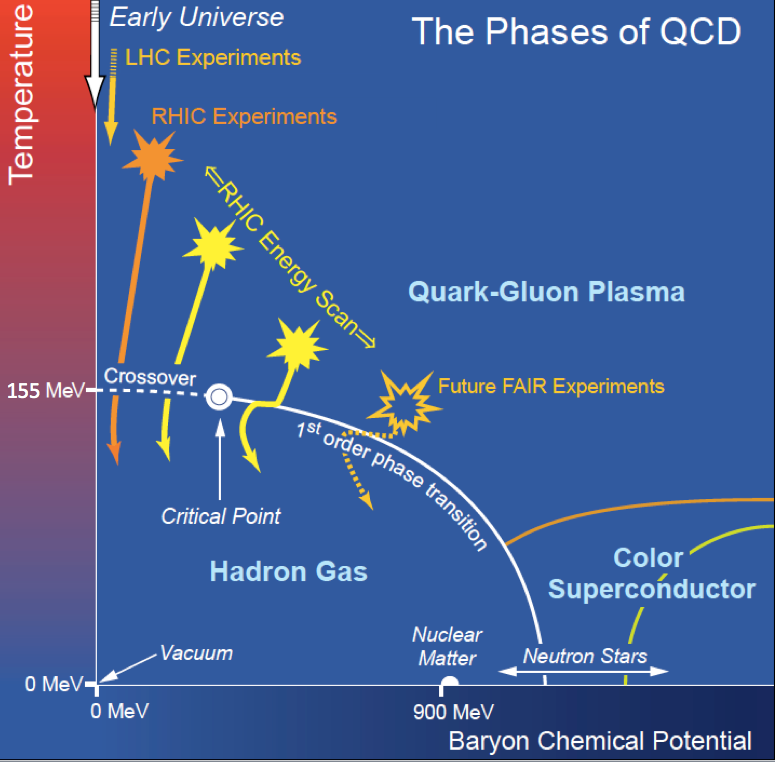}
    }	
	\caption[The conjectured phase diagram of QCD matter (2015 Hot QCD White Paper)]{A sketch of the conjectured phase diagram of QCD matter in the $(T,\mu_B)$ plane \cite{Akiba:2015jwa}. Current knowledge on the states of matter at near-zero temperatures and moderate baryon chemical potentials as well as at high temperatures and near-zero baryon chemical potentials is relatively well established, while regions of the phase diagram characterized by finite temperatures and high baryon chemical potentials are subject of ongoing experimental and theoretical research.}
	\label{QCD_phase_diagram_hot_QCD_white_paper}
\end{figure}

Studying the QGP phase transition involves mapping out the QCD phase diagram, that is identifying the phases of QCD matter that exist at given values of temperature $T$ and baryon number density $n_B$ or, alternatively, baryon chemical potential $\mu_B$. Fig.\ \ref{QCD_phase_diagram_Long_Range_Plan_1983} shows the conjectured QCD phase diagram in the $(T,n_B)$ plane as included in the 1983 DOE/NSF Long Range Plan for Nuclear Science \cite{LongRangePlan1983}, while Fig.\ \ref{QCD_phase_diagram_hot_QCD_white_paper} shows the conjectured QCD phase diagram in the $(T,\mu_B)$ plane as included in the 2015 Hot QCD White Paper \cite{Akiba:2015jwa}. Although significant portions of both of these diagrams are speculative, a few regions can be described with a reasonable certainty, and we briefly introduce them below.

The most experimentally well-known is the region of low temperatures, $T \lesssim 30\ \txt{MeV}$, and moderate baryon chemical potentials, $\mu_B \approx 800$--$1000\ \txt{MeV}$, or alternatively moderate baryon densities, $n_B \approx 0.1$--$1.5 n_0$, where $n_0 \approx 0.160\ \txt{fm}^{-3}$ is the nuclear saturation density, that is the average central density of nuclei. Here, QCD matter is encountered in the form of ordinary nuclear matter, that is systems composed of nucleons, with mesons acting as effective carriers for the strong force (see Section \ref{nuclear_matter_quasiparticles_and_the_Landau_Fermi-liquid_theory} for more details). Nuclear matter appears in one of two states: as a dilute gas of nucleons or as a dense concentration of nucleons, known as the nuclear liquid. The phase transition between these two states is of the first order, which means that there exists a range of densities and temperatures at which the two phases coexist with each other (see Appendix \ref{phase_transitions} for a brief description of properties of first-order phase transitions). The critical point of nuclear matter is located at values of the temperature and the baryon number density at which the densities of the gaseous and the liquid phases have the same value, that is at which it is no longer possible to distinguish between the gaseous and the liquid phase. For all temperatures higher than the critical temperature, this distinction is likewise impossible. Based on extrapolations from experimental data, the nuclear matter critical point has been identified at the critical temperature $T_c^{(N)} \approx 18 \ \txt{MeV}$ and the critical baryon number density $n_c^{(N)} \approx 0.4 n_0$ \cite{Elliott:2013pna}.

QCD matter at even higher chemical potentials is encountered in neutron stars, which are composed of highly isospin-asymmetric nuclear matter at $T \approx 0$ and whose central densities can reach about $n_B \approx 5$--$10 n_0$ \cite{Ozel:2016oaf}. Despite an astounding difference in scales (the diameter and mass of a heavy nucleus are about $10^{-14}\ \txt{m}$ and $10^{-25}\ \txt{kg}$, respectively, while the corresponding values for a neutron star are on the order of $10^4 \ \txt{m}$ and $10^{30}\ \txt{kg}$), the equation of state (EOS) of asymmetric nuclear matter is central to neutron star research. This is because the behavior of the pressure of asymmetric nuclear matter as a function of the energy density determines the relationship between the masses and radii of neutron stars \cite{Silbar:2003wm}. In consequence, currently known values of masses and radii of neutron stars put strong constraints on the EOS of nuclear matter at large densities and small temperatures \cite{Ozel:2016oaf,Fujimoto:2019hxv,Raaijmakers:2021uju}. Nevertheless, it is currently not established whether the very dense cores of neutron stars could be described in terms of quark and gluon degrees of freedom. In addition to this possibility, a few exotic phenomena are predicted for nuclear matter at very high densities, including systems described by a mixture of nucleons and meson condensates \cite{Pethick:2015jma}, or systems with a color superconducting phase \cite{Alford:2007xm}.

In the opposite regime, at relatively low temperatures and near-zero densities (vanishing chemical potentials), QCD matter is well-described by chiral effective field theories \cite{Kapusta:2006pm} and can be shown to be well-approximated by an interacting gas of pions \cite{Gasser:1983yg}. As the temperature increases, other hadronic species as well as their excited states become relevant, and the description in terms of the hadron resonance gas (HRG) model \cite{Hagedorn:1965st,Andronic:2005yp} is appropriate.

At moderate to high temperatures and negligible baryon chemical potentials, the behavior of QCD matter is well-understood theoretically from the first-principle calculations in lattice QCD (LQCD). These calculations confirm that for temperatures satisfying $100\ \txt{MeV} \lesssim T \lesssim 140\ \txt{MeV}$ and $\mu_B \approx 0$, the HRG model gives a very good description of hot QCD matter. For higher temperatures, however, LQCD shows that QCD matter undergoes a crossover transition \cite{Aoki:2006we} from a hadron gas to a strongly-interacting QGP. (We note that the strongly-interacting nature of QGP is in opposition to the early expectation that above the phase transition, QCD matter would be composed of free quarks and gluons.) This result has been further supported with a Bayesian inference approach applied to heavy-ion collisions probing this region of the phase diagram \cite{Pratt:2015zsa}, where the range of EOSs most consistent with experimental data has been identified and shown to include the LQCD EOS. Based on LQCD, the pseudocritical temperature of the crossover QGP-hadron transition at $\mu_B \approx 0$ is equal $T_{pc} = 156.6 \pm 1.5\ \txt{MeV}$ \cite{HotQCD:2018pds} (see also Refs.\ \cite{Aoki:2006br,Borsanyi:2020fev}), with the restoration of the approximate chiral symmetry of QCD occurring at high temperatures. 

The region of the QCD phase diagram characterized by both moderate-to-high temperatures and moderate-to-high baryon chemical potentials is not known well due to the lack of first-principle approaches available in this regime: at finite chemical potentials, $\mu_B \neq 0$, LQCD suffers from a calculational difficulty known as the fermion sign problem \cite{Karsch:2001cy}. However, a number of theoretical considerations lead to the conclusion that the phase diagram of QCD at moderate ranges of temperature and baryon chemical potential may contain interesting structures. Starting from a more accessible region of $\mu_B  = 0$, theoretical calculations on the chiral phase transition in QCD suggest that the QGP-hadron phase transition is a first-order phase transition in the limit of massless quarks \cite{Pisarski:1983ms}, known as the ``chiral limit'' due to the chiral symmetry displayed by the QCD Lagrangian for zero quark masses. If only the two lightest quarks, up and down, are considered massless, while the strange quark remains sufficiently heavy, the transition at $\mu_B = 0$ is instead of second-order \cite{Stephanov:2004wx}. Finally, if the up and down quarks are given small masses, corresponding to the situation found in nature, the transition becomes a crossover \cite{Stephanov:2004wx}, just as obtained in LQCD. (Considerations of a similar type are also studied in LQCD, and recent results can be found, e.g., in Ref.\ \cite{Cuteri:2021ikv}.) In the last case numerous chiral effective field theory models predict that the first-order QGP-hadron phase transition line must begin at a critical point located at some finite value of the baryon chemical potential \cite{Stephanov:2004wx}, see Fig.\ \ref{QCD_phase_diagram_hot_QCD_white_paper}. If this is true, there are two critical points related to the strong interaction in the phase diagram of QCD: one corresponding to the ordinary nuclear liquid-gas phase transition, and one corresponding to the QGP-hadron phase transition. Studies devoted to this possibility, as well as to understanding the boundary between the ordinary nuclear matter and QGP in general, are at the forefront of heavy-ion collision research.

\section{Probing the QCD phase diagram}
\label{probing_the_QCD_phase_diagram}

Heavy-ion collision experiments probe different regions of the QCD phase diagram primarily by changing the energy of the colliding beams. Additionally, experiments can also probe different baryon densities by choosing particular rapidity acceptance windows in data analysis. We briefly describe the physics behind these two possibilities below. For a rudimentary introduction to the kinematic variables employed in the description of heavy-ion collisions, as well as to the heavy-ion collision geometry and baryon transport, see Appendix \ref{kinematic_variables_and_geometry_in_heavy-ion_collisions}. 

\begin{figure}[t]
	\centering\mbox{
	\includegraphics[width=0.95\textwidth]{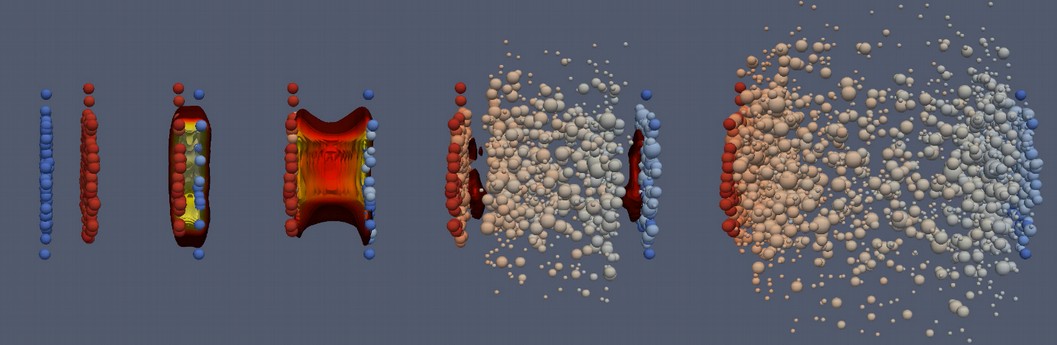} 
	}
	\caption[Visualization of a Au+Au collision at $\sqrt{s_{NN}} = 200\ \txt{GeV}$]{Visualization of a gold-gold (Au+Au) collision at the center-of-mass energy $\sqrt{s}_{NN} = 200 \ \txt{GeV}$, based on a hybrid UrQMD simulation \cite{Petersen:2008dd} utilizing the Ultra-relativistic Quantum Molecular Dynamics (\texttt{UrQMD}) hadronic transport code \cite{Bass:1998ca,Bleicher:1999xi} with an intermediate hydrodynamic stage based on the ideal hydrodynamics simulation code \texttt{SHASTA} \cite{Rischke:1995ir,Rischke:1995mt}. Highly Lorentz-contracted nuclei interpenetrate each other, depositing their energy in the collision region; the created system, most often assumed to be the QGP, quickly thermalizes and expands hydrodynamically; when the matter expands enough to cool down to the transition temperature, the QGP hadronizes into discrete particles that continue moving outward, scattering among each other, until the system is too dilute for any collisions to occur and particles travel in straight lines to the detector. See text for more details. Different heavy-ion collision simulations will be discussed in Section \ref{simulations_of_heavy-ion_collisions}. Figure from Hannah Elfner (Petersen), MADAI.us \cite{MADAI_collaboration}.}
	\label{contracted_pancakes}
\end{figure}

\begin{figure}[t]
	\centering\mbox{
	\includegraphics[width=0.95\textwidth]{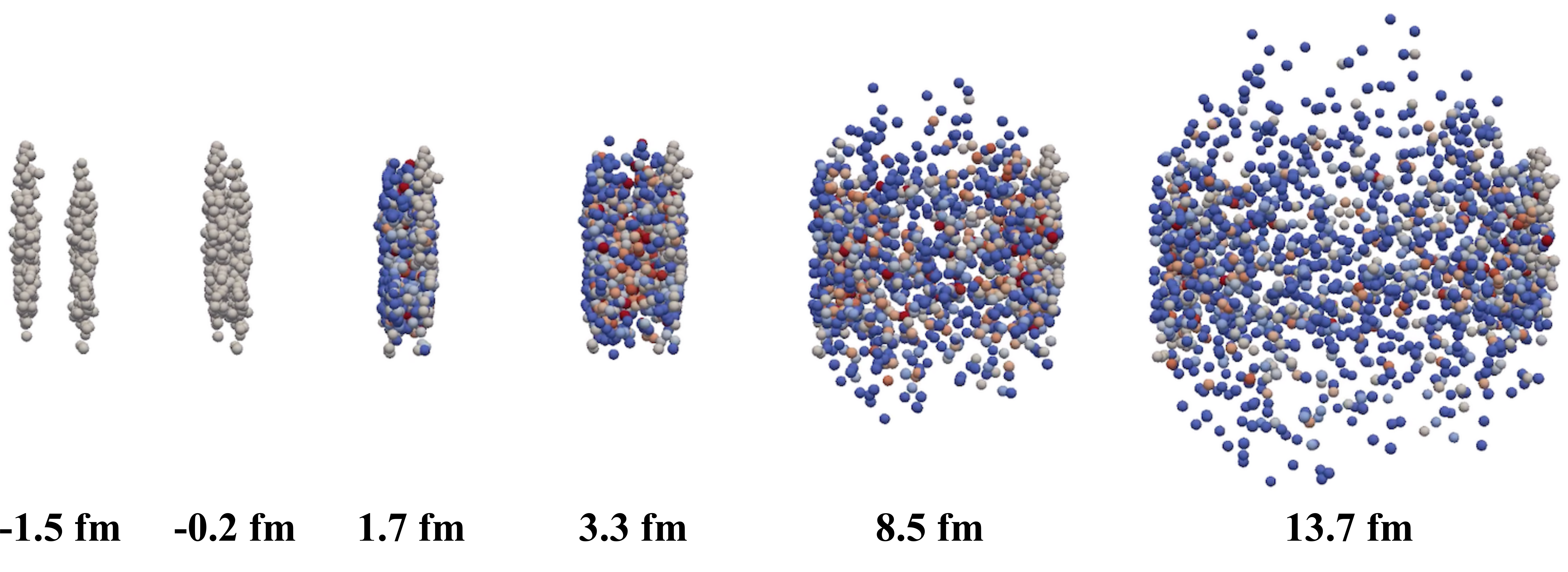} 
	}
	\caption[Visualization of a Pb+Pb collision at $\sqrt{s_{NN}} = 17.3\ \txt{GeV}$]{Visualization of a lead-lead (Pb+Pb) collision at the center-of-mass energy $\sqrt{s_{NN}} = 17.3\ \txt{GeV}$, based on a \texttt{SMASH} hadronic transport simulation \cite{Weil:2016zrk}. Significantly Lorentz-contracted nuclei collide, depositing their energy in the collision region; the collision energy is transformed into produced particles and the system naturally expands, with scatterings occurring between all participating particles; finally, the system is too dilute for any collisions to occur and particles travel in straight lines to the detector. Different heavy-ion collision simulations will be discussed in Section \ref{simulations_of_heavy-ion_collisions}. Figure from Justin Mohs \cite{SMASH_website}.}
	\label{contracted_pancakes_17,3}
\end{figure}

Varying the energy of beams of colliding nuclei changes the fraction of the initial baryon number ($N_{B,\txt{initial}} \approx 400$, originating from the projectile and the target, which are often gold or lead nuclei) transported in the course of the collision to the central rapidity region in the center-of-mass frame. In highly energetic collisions, where the colliding nuclei are traveling with velocities approximately equal 99.9\% of the speed of light, their Lorentz contraction in the laboratory frame is significant (as seen in Figs.\ \ref{contracted_pancakes} and \ref{contracted_pancakes_17,3}). The two contracted nuclei can be roughly thought of as very dense ``mixtures'' of valence quarks and the strong force interaction bosons: the gluons. However, while to a good approximation one can assume that for each nucleon there are three valence quarks (sharing between each other about half of the energy of the nucleon), bound in the nucleus speeding towards the collision, the density of gluons in the nucleon increases with the increasing colliding energy, with a very strong prevalence of low-momentum gluon states (this had been found by the H1 and ZEUS collaborations from experiments in deep inelastic $ep^{\pm}$ scattering \cite{H1:2009pze}). Importantly, the strong interaction coupling constant diminishes with the energy of a particle, and so do the associated cross sections. Consequently, within a zeroth-order description, when a heavy-ion collision takes place, the valence quarks belonging to different nuclei ``fly through'' each other, while the majority of the gluons, composed of the low-momentum gluon states, collide and create the highly energetic medium that becomes a QGP.

Beyond the zeroth-order description, even though at very high energies the valence quarks have very small cross sections for interaction with quarks belonging to the other nucleus, they still interact with the low-momentum gluons as well as with quarks within the same nucleus, where the latter interaction is mediated by gluon fields. Since many of the gluons are stopped in the collision region while the valence quarks continue to move apart, the fields between the quarks are ``stretched'' and form a ``string'' of gluons. As this process continues, the potential energy in the gluon strings increases (at the cost of the valence quarks' kinetic energy) and it becomes high enough to enable particle production. This proceeds by breaking individual strings, with a quark-antiquark pair produced at the two new ends of each broken string. This string-breaking process often further diminishes the energy of the ``original'' valence quarks as well as causes them to develop some transverse momentum. Once the string breaks, however, the further propagation of quarks can be thought of as largely unimpeded. 

Overall, string creation and breaking significantly decreases the initial energy of the valence quarks (by about a half) and slightly changes their transverse momenta (on the order of $0.3\ \txt{GeV}$), which corresponds to a rapidity change of about one. When the system hadronizes (or ``freezes out''), the valence quarks form baryons that are detected at high absolute values of rapidity; this is a direct consequence of the early evolution of the collision system described above: after the collision takes place, the valence quarks continue to move with a momentum $p \approx p_{\txt{beam}}$. On the other hand, the quark-antiquark pairs which appear through string-breaking are produced more isotropically, so the final state distributions of mesons and baryons into which they hadronize (with an overwhelming dominance of mesons, as their production is favored energetically) are peaked at midrapidity. Note that the produced baryons satisfy $N_{B,\txt{prod}} - N_{\bar{B},\txt{prod}} = 0$, so that their contribution to the net baryon density is zero. Altogether, at very high-energy collisions the net baryon distribution displays a minimum at midrapidity and rises with increasing $|y|$. (For a brief review of properties of rapidity and pseudorapidity distributions, see Appendix \ref{kinematic_variables_and_geometry_in_heavy-ion_collisions}.)

The situation in low-energy collisions differs from the description sketched above in two ways. First, in nuclei moving at a smaller speed the cross sections for quark-quark and quark-gluon interactions between the two colliding nuclei increase and a significant fraction of the initial baryon number can be ``stopped'' in the collision region by scattering, leading to a subsequent detection at midrapidity. Moreover, the initial rapidity of the participating valence quarks is smaller, so that if string creation and breaking processes occur (which result in a reduction of the rapidity of valence quarks by about one unit, similarly as in the high-energy collisions), the final rapidity of detected baryons is smaller as well. Altogether, this means that in low-energy collisions relatively large values of the net baryon number are detected at midrapidity, which is in opposition to the behavior of matter in high-energy collisions. The behavior of the baryon number in collisions at intermediate energies should interpolate between these two scenarios.

\begin{figure}[t]
	\centering\mbox{
		\includegraphics[width=0.69\textwidth]{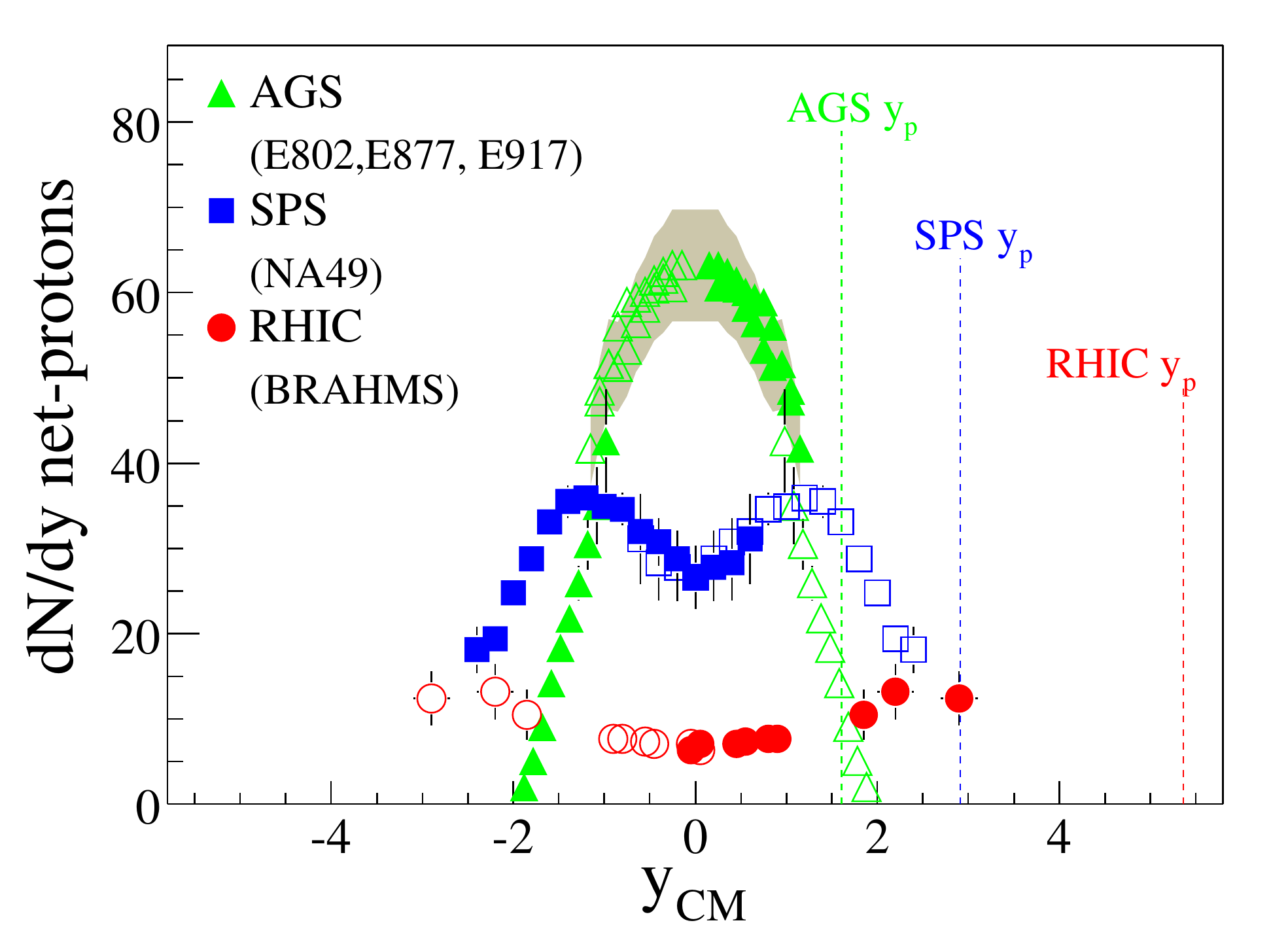} 
	}
	\caption[Rapidity distributions of net protons in 0--5\% Au+Au collisions]{Rapidity distributions of net protons \cite{BRAHMS:2003wwg} measured in collisions with 0--5\% centrality at the AGS (Au+Au collisions at $\sqrt{s_{NN}} = 5\ \txt{GeV}$, $y_{\txt{beam}} = 1.64$) \cite{E802:1999hit}, SPS (Pb+Pb collisions at $\sqrt{s_{NN}} = 17.2\ \txt{GeV}$, $y_{\txt{beam}} = 2.91$) \cite{NA49:1998gaz}, and RHIC (Au+Au collisions at $\sqrt{s_{NN}} = 200\ \txt{GeV}$, $y_{\txt{beam}} = 5.36$) \cite{BRAHMS:2003wwg}. For collisions at $\sqrt{s_{NN}} = 5\ \txt{GeV}$, the net proton distribution is peaked around $y \approx 0$, while for $\sqrt{s_{NN}} = 17.2\ \txt{GeV}$ the distribution develops two separate peaks at relatively large values of rapidity $y \approx \pm 1.3$. For collisions at $\sqrt{s_{NN}} = 200\ \txt{GeV}$ the peaks of the distribution are beyond the reach of the detector, but fits to data establish them at $y \approx \pm 4.3$.}
	\label{BRAHMS_net-proton_dNdy}
\end{figure}

\begin{figure}[t]
	\centering\mbox{
		\includegraphics[width=0.69\textwidth]{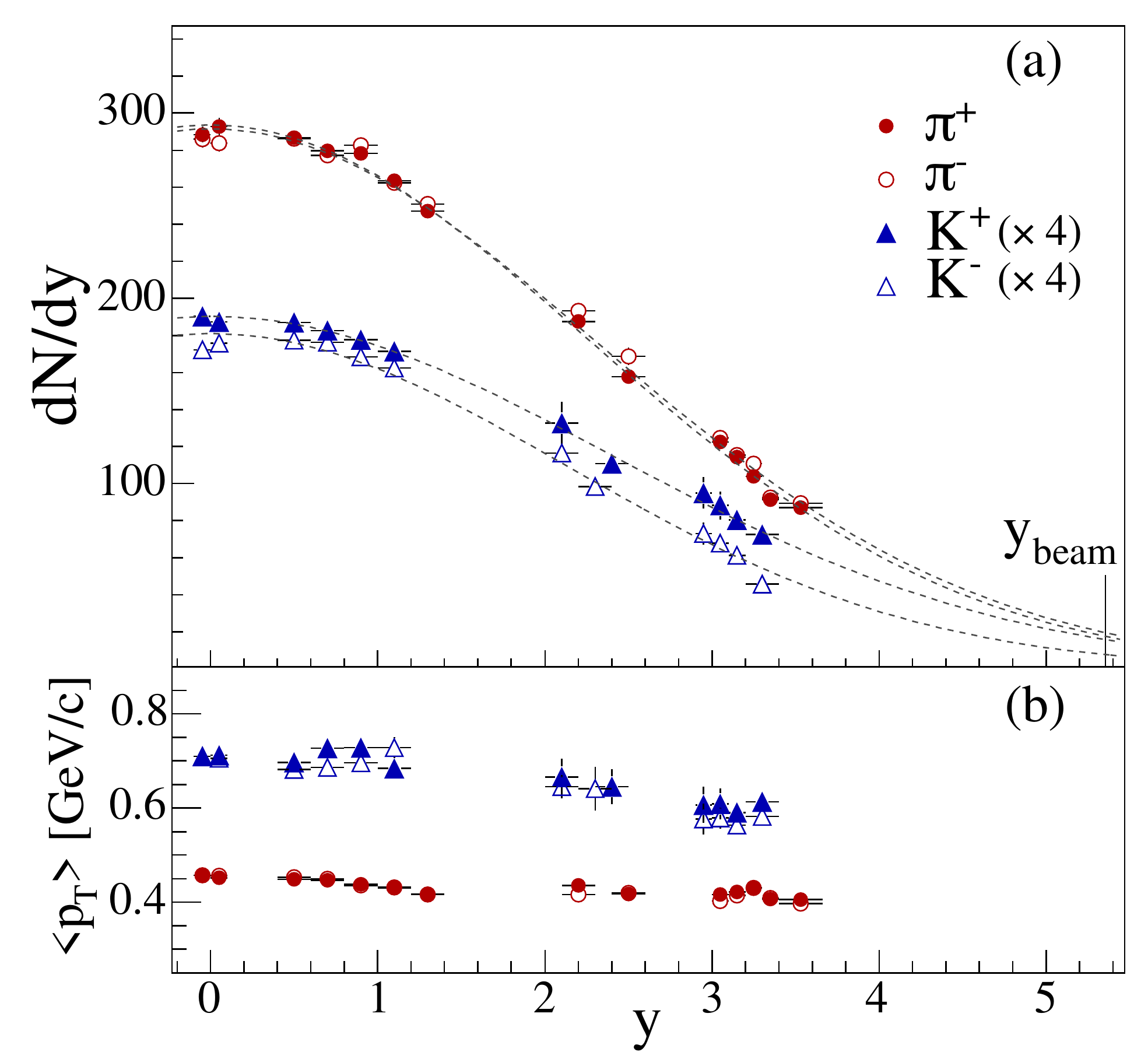} 
	}
	\caption[Rapidity distributions and mean transverse momenta of pions and kaons in 0--5\% Au+Au collisions at $\sqrt{s_{NN}} = 200\ \txt{GeV}$]{Rapidity distributions (a) and mean transverse momenta (b) of pions and kaons \cite{BRAHMS:2004dwr} measured in collisions with 0--5\% centrality at RHIC (Au+Au collisions at $\sqrt{s_{NN}} = 200\ \txt{GeV}$). The kaon yields are multiplied by a factor of 4 for better visibility, and the dashed lines are Gaussian fits to the distributions.}
	\label{BRAHMS_mesons_dNdy}
\end{figure}

\begin{figure}[t]
	\centering\mbox{
		\includegraphics[width=0.6\textwidth]{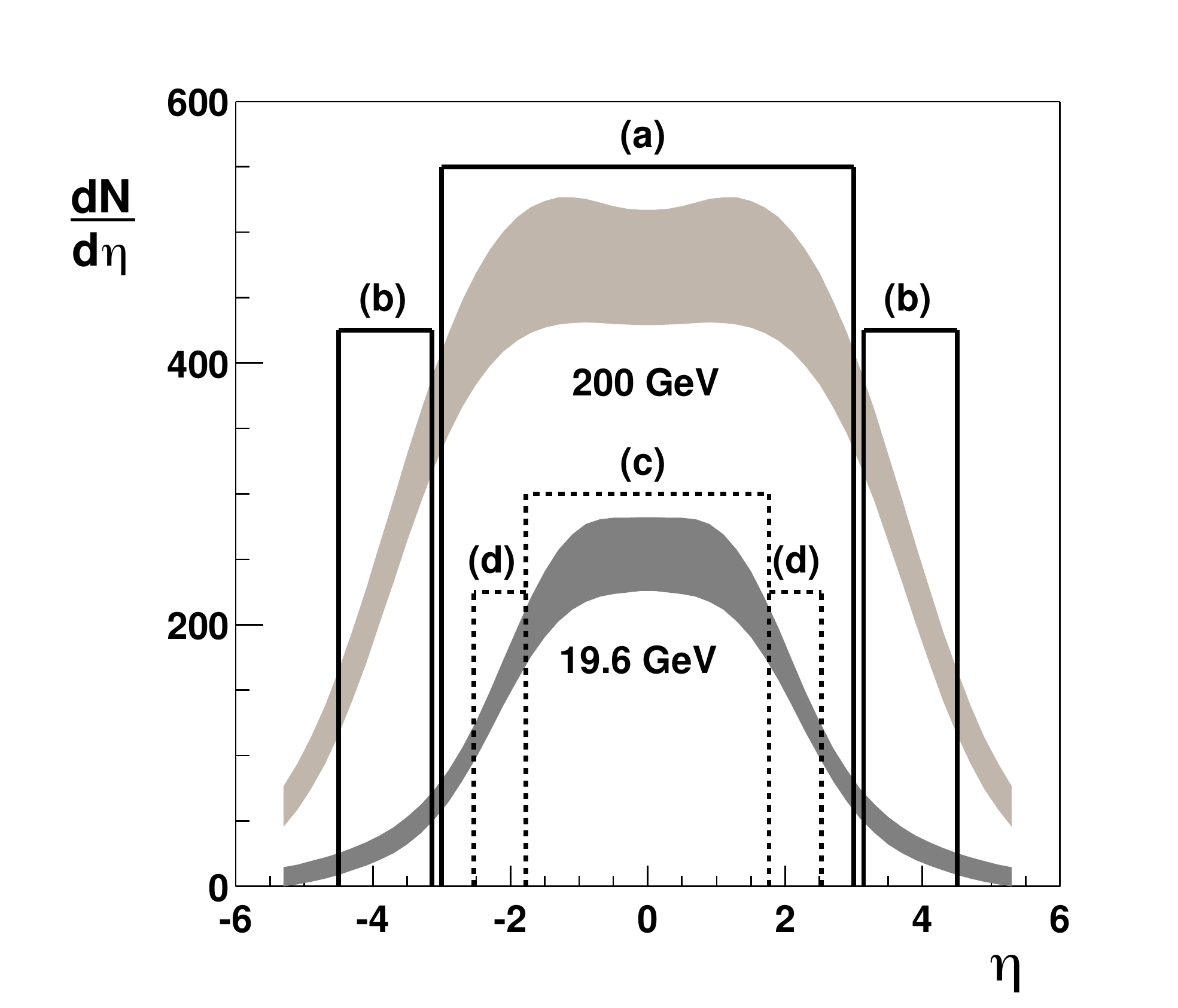} 
	}
	\caption[Rapidity distributions of charged particles in 0--25\% Au+Au collisions]{Rapidity distributions of charged particles \cite{PHOBOS:2004hlv} measured in collisions with 0--25\% centrality at RHIC (Au+Au collisions at $\sqrt{s_{NN}} = 19.6\ \txt{GeV}$ and $\sqrt{s_{NN}} = 200\ \txt{GeV}$). The boxed areas (a)-(d) indicate separate regions in pseudorapidity used for the centrality determination at each energy.}
	\label{PHOBOS_charged_particles_dNdy}
\end{figure}

Indeed, the data supports this picture. Fig.\ \ref{BRAHMS_net-proton_dNdy} shows the rapidity spectra of net protons $dN_{p - \bar{p}}/dy$ \cite{BRAHMS:2003wwg} as measured in collisions with 0--5\% centrality at the AGS (Au+Au collisions at $\sqrt{s_{NN}} = 5\ \txt{GeV}$, $y_{\txt{beam}} = 1.64$) \cite{E802:1999hit}, SPS (Pb+Pb collisions at $\sqrt{s_{NN}} = 17.2\ \txt{GeV}$, $y_{\txt{beam}} = 2.91$) \cite{NA49:1998gaz}, and RHIC (Au+Au collisions at $\sqrt{s_{NN}} = 200\ \txt{GeV}$, $y_{\txt{beam}} = 5.36$) \cite{BRAHMS:2003wwg}. For collisions at $\sqrt{s_{NN}} = 5\ \txt{GeV}$, the net proton distribution is peaked around $y \approx 0$, while for $\sqrt{s_{NN}} = 17.2\ \txt{GeV}$ the distribution develops two separate peaks at relatively large values of rapidity $y \approx \pm 1.3$. For collisions at $\sqrt{s_{NN}} = 200\ \txt{GeV}$ the peaks of the distribution are beyond the reach of the detector, but fits to data establish them at $y \approx \pm 4.3$. On the other hand, Fig.\ \ref{BRAHMS_mesons_dNdy} shows rapidity distributions of charged pions and mesons as well as their mean transverse momenta as measured in collisions with 0--5\% centrality at RHIC (Au+Au collisions at $\sqrt{s_{NN}} = 200\ \txt{GeV}$) \cite{BRAHMS:2004dwr}. The meson distributions are peaked at midrapidity, and their transverse momentum is approximately constant as a function of rapidity. Similarly, Fig.\ \ref{PHOBOS_charged_particles_dNdy} shows pseudorapidity distributions of charged hadrons as measured in 0--25\% most central collisions at RHIC for very high and low beam energies (Au+Au collisions at $\sqrt{s_{NN}} = 200\ \txt{GeV}$ and $\sqrt{s_{NN}} = 19.6\ \txt{GeV}$, respectively) \cite{PHOBOS:2004hlv}. Rapidity and pseudorapidity distributions of charged particles are dominated by mesons, and so the distributions presented in Fig.\ \ref{PHOBOS_charged_particles_dNdy} further confirm those shown in Fig.\ \ref{BRAHMS_mesons_dNdy}. Overall, the behavior of both the net proton and charged meson rapidity distributions reflects the energy dependence of the evolution of a heavy-ion collision sketched above.

We stress that while this simplified picture of the net baryon number evolution is useful for developing intuition, the exact mechanism of baryon transport in heavy-ion collisions, often referred to as ``baryon stopping'', is not known. Therefore baryon stopping is a subject of active research, both within the phenomenological approaches \cite{Capella:1999cz,Mehtar-Tani:2009wji,McLerran:2018avb}, as well as within simulations of heavy-ion collisions \cite{Denicol:2018wdp, Li:2018fow}.

Since the net baryon number measured at given values of rapidity changes with the collision energy, it follows that varying the beam energy allows one to probe systems characterized by different values of the net baryon number. Additionally, collisions at varying $\sqrt{s_{NN}}$ are also characterized by different initial temperatures, corresponding to different amounts of initial energy deposition in the collision region. It is possible to get an intuition about which regions of the phase diagram are probed in a given class of collisions by fitting the energy spectra of the final state particles to the HRG model, in this context also often referred to as the statistical hadronization model (SHM) \cite{Cleymans:2005xv}. By doing so one arrives at a good estimate of the temperature and baryon chemical potential of the system at the moment of the evolution, occurring some time after the hadronization and known as the chemical freeze-out, when the particle-changing processes cease and particle yields of the collision are established. Fits to particle yield ratios in 0--5\% central collisions using SHM lead to the freeze-out temperatures $T_{\txt{fo}}$ and baryon chemical potentials $\mu_{\txt{fo}}$ as listed in Table \ref{STAR_freeze-out} for collisions from $\sqrt{s_{NN}} = 7.7 \ \txt{GeV}$ to $\sqrt{s_{NN}} = 200 \ \txt{GeV}$ \cite{STAR:2017sal}. These values reflect the conclusion, made above based on rapidity distributions, that particles detected at midrapidity in high-energy collisions carry close to zero net baryon number (corresponding to very low values of the baryon chemical potential), while in low-energy collisions the midrapidity region probes systems with significantly higher net baryon number (corresponding to moderate values of the baryon chemical potential). 

\begin{table}[t]
	\caption[Freeze-out parameters in 0--5\% central Au+Au collisions]{Freeze-out parameters $T_{\txt{fo}}$ and $\mu_{\txt{fo}}$ \cite{STAR:2017sal} (in MeV) for Au+Au collisions at different center-of-mass energies $\sqrt{s_{NN}}$ (in GeV), at 0--5\% centrality and a rapidity window $|y| \leq 0.5$ (both values and errors have been rounded up to nearest integers here).}
	\begin{center}
		\def\arraystretch{1.4}
		\small\addtolength{\tabcolsep}{1pt}
		\begin{tabular}{|c|r|r|r|r|r|r|r|r|}
			\hline
			\hline
			$\sqrt{s_{NN}}$ &  \hspace{-2mm}7.7 &  \hspace{0mm} 11.5  & \hspace{0mm}14.5 & \hspace{0mm}19.6 & \hspace{0mm}27 & \hspace{0mm}39 & \hspace{0mm}62.4 & \hspace{0mm} 200  \\ 
			\hline
			$\mu_{\txt{fo}}$ & $398 \pm 13$  & $287\pm13$   & $264\pm 10$ & $188\pm8$ & $144\pm8$ & $103\pm10$ & $70 \pm 11$  & $28 \pm 8$  \\  
			\hline
			$T_{\txt{fo}}$ &  $144\pm 2$  & $149\pm3$   & $152\pm3$ & $154 \pm 3$  & $155 \pm 3$  & $156\pm 3$  & $160 \pm 4$  & $164 \pm 4$  \\  
			\hline
		\end{tabular}
		\label{STAR_freeze-out}
	\end{center}
\end{table}

As a result, varying the beam energy as well as analyzing data from specific rapidity windows allows one to probe different points on the phase diagram of dense nuclear matter. By systematically exploring different beam energies, the experiment is effectively performing a scan across the QCD phase diagram. 

While the means to search for signatures of the QCD phase transition are clear, the success of this endeavor is premised on the ability to experimentally uncover a number of effects born out in systems of immense complexity. Some of these predicted signatures involve Hanbury-Brown-Twiss (HBT) interferometry measurements (discussed in Section \ref{HBT_correlations}), quark number scaling of the elliptic flow (discussed in Section \ref{quark_number_scaling_of_the_elliptic_flow}), or enhanced multiplicity fluctuations of produced hadrons (discussed in Section \ref{non-statistical_event-by-event_fluctuations_of_conserved_charges}), and their dependence on the beam energy. Often, the magnitudes of these effects and their interaction with other experimental signals, as well as the influence of the finite lifetime and size of the collision remain elusive to purely theoretical predictions. A clear interpretation of the experimental data will have to be supported by comparisons with results of dynamical simulations of heavy-ion collisions, developed to account for the complex evolution of relevant observables.

\section{BES-I and BES-II}
\label{BES-I_and_BES-II}

Probing the phase diagram of QCD matter was one of the main goals behind the Beam Energy Scan I (BES-I) program and is the driving motivation behind the ongoing Beam Energy Scan II (BES-II) program at RHIC, pursued by the Solenoidal Tracker at RHIC (STAR) experiment \cite{Odyniec:2019kfh}. The names of the programs refer to heavy-ion collisions performed at a series of different beam energies, allowing for a systematic study of the beam-energy--dependence of observables. Due to the dynamics of heavy-ion collisions (see Section \ref{probing_the_QCD_phase_diagram}), this in fact allows for studying the dependence of observables on the temperature and the baryon chemical potential.

Running during the years 2010-2017, the BES-I collided two beams of gold nuclei ($^{197}$Au) at a series of center-of-mass energies, $\sqrt{s_{NN}} = 62.4,\ 54.4, \ 39.0, \ 27.0, \ 19.6, \ 14.5, \ 11.5, \ 7.7 \ \txt{GeV}$, and had three major objectives \cite{STAR:2010vob}:  

i) to search for the ``onset'' of specific features of collective behavior, such as constituent quark scaling of the elliptic flow associated with the formation of the QGP, measured in systems created in the high-energy collisions at $\sqrt{s_{NN}} = 200 \ \txt{GeV}$ \cite{STAR:2003wqp},

ii) to search for the evidence of the softening of the equation of state, which would aid in locating the region of the phase diagram where the phase transition occurs,

iii) to search for fluctuations of conserved charges, expected to be enhanced in the vicinity of the critical point.

The results from BES-I, which was an exploratory run, were encouraging (we provide a brief overview in Sections \ref{softening_of_the_equation_of_state}, \ref{turning_off_the_QGP}, and \ref{non-statistical_event-by-event_fluctuations_of_conserved_charges}), but at the same time underscored the need for better experimental statistics at particular collision energies. In response to this need, the collider mode of the BES-II, starting in 2018 and continuing through 2021, ran at the center-of-mass energies of $\sqrt{s_{NN}} = 27.0, \ 19.6, \ 14.5, \ 11.5, \ 9.2, \ 7.7 \ \txt{GeV}$. Additionally, BES-II also ran in the fixed target mode, colliding a beam of gold nuclei on a thin gold foil, covering low-energy collisions at the center-of-mass energies of $\sqrt{s_{NN}} = 7.7,\ 6.2, \ 5.2, \ 4.5, \ 3.9, \ 3.5, \ 3.2, \ 3.0 \ \txt{GeV}$. Through the fixed target mode, freeze-out baryon chemical potentials on the order of $\mu_B \approx 800 \ \txt{MeV}$ can be reached, thus significantly extending the phase diagram coverage of the program. The BES-II data-taking campaign concluded in 2021, and the program will be followed by several years of analyzing the produced data.

We note that BES-II is not the only ongoing experimental effort probing QCD matter at high values of the baryon chemical potential. Other experiments include the High Acceptance Di-Electron Spectrometer (HADES) experiment \cite{Galatyuk:2014vha} at the GSI Helmholtz Center for Heavy Ion Research, Germany, colliding various nuclei in the fixed target mode at center-of-mass energies $\sqrt{s}_{NN} \approx 1$--$4\ \txt{GeV}$, and the NA61/SHINE \cite{NA49-future:2006qne} experiment (where SHINE stands for ``SPS Heavy Ion and Neutrino Experiment'') at CERN, colliding various nuclei at different energies in a fixed target mode (for lead-lead collisions, the energy range is $\sqrt{s}_{NN} \approx 5$--$17\ \txt{GeV}$). Additionally, several experiments are expected to begin in the near future, including the Compressed Baryonic Matter (CBM) experiment at the Facility for Antiproton and Ion Research in Europe (FAIR) in Darmstadt, Germany, experiments at the Nuclotron-based Ion Collider fAcility (NICA) at the Joint Institute for Nuclear Research (JINR) in Dubna, Russia, or the Cooling-storage-ring External-target Experiment (CEE) at the Heavy Ion Research Facility in Lanzhou (HIRFL), China.

\section{Tantalizing results from BES-I: Softening of the equation of state}
\label{softening_of_the_equation_of_state}

Some of the observables studied in BES-I showed behavior that could be interpreted as consistent with systems evolving in the vicinity of the QGP-hadron phase transition, where softening of the EOS should lead to smaller pressure gradients driving the evolution of the fireball. These include HBT correlations and the slope of the directed flow at midrapidity, collectively identifying collision energies in the range $\sqrt{s_{NN}} \approx 10$--$40 \ \txt{GeV}$ as possibly probing the boundary of the QGP-hadron phase transition.

\subsection{HBT correlations}
\label{HBT_correlations}

\begin{figure}[t]
	\centering\mbox{
		\includegraphics[width=0.5\textwidth]{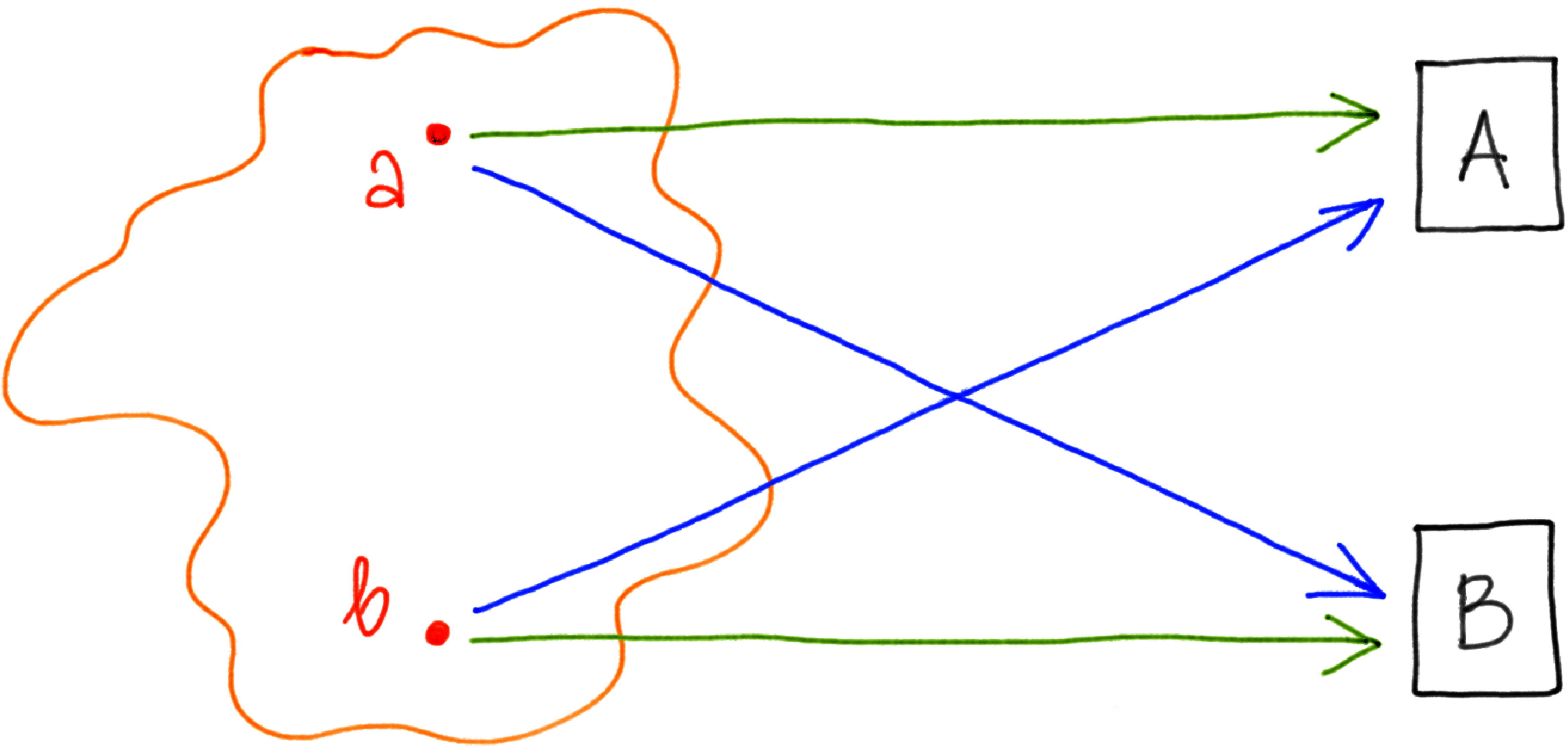} 
	}
	\caption[A sketch illustrating the origin of HBT correlations]{A sketch illustrating the origin of the HBT correlations due to the two ways in which a particle emitted from one of the sources, $a$ or $b$, can be registered by one of the detectors, $A$ or $B$; see text for more details.}
	\label{HBT_geometry}
\end{figure}

The spatial and temporal size of the collision system can be established using a technique known as femtoscopy or Hanbury-Brown--Twiss (HBT) interferometry, referring to Robert Hanbury Brown's and Richard Twiss' 1956 works on photon interferometry \cite{HanburyBrown:1954amm,HanburyBrown:1956bqd}. The HBT analysis has been used in studying nuclear matter since the early years of heavy-ion collisions, including experiments performed at the Bevalac \cite{Zajc:1984vb,Fung:1978eq}, and is based on a simple premise. If a source emits two particles at two points $a$ and $b$, and these particles are then detected in two detectors $A$ and $B$, there are two ways in which this can happen: either particle emitted at $a$ is detected by detector $A$ and particle emitted at $b$ is detected by detector $B$, or particle emitted at $a$ is detected by detector $B$ and particle emitted at $b$ is detected by detector $A$ (see Fig.\ \ref{HBT_geometry}). Quantum-mechanically, these possibilities correspond to two probability amplitudes, which may be symbolically denoted as $\langle A| a\rangle \langle B | b \rangle$ and $\langle B| a\rangle \langle A | b \rangle$. The probability of detecting the particles is then described by the sum of these amplitudes, which can constructively or destructively interfere depending on the details of the problem such as the distance between the points $a$ and $b$ and the distance between the detectors. Measuring the  enhancement or suppression in the signal (HBT correlations) for different values of the spacing between the detectors allows one to determine the distance between the point sources or, more generally, the size of a continuous source. Naturally, the details become more complicated for more complex systems such as heavy-ion collisions (see Refs.\ \cite{Heinz:1999rw} or \cite{Lisa:2005dd} for a review). In particular, because in heavy-ion collisions the measured emitted particles are hadrons, HBT correlations reveal the geometry of the system at hadronization (or ``freeze-out''); nevertheless, this geometry is naturally affected by the evolution of the system up to that point, and so the HBT interferometry can be used to, e.g., constrain the dynamics of the early stages of the fireball evolution.

\begin{figure}[t]
	\centering\mbox{
		\includegraphics[width=0.75\textwidth]{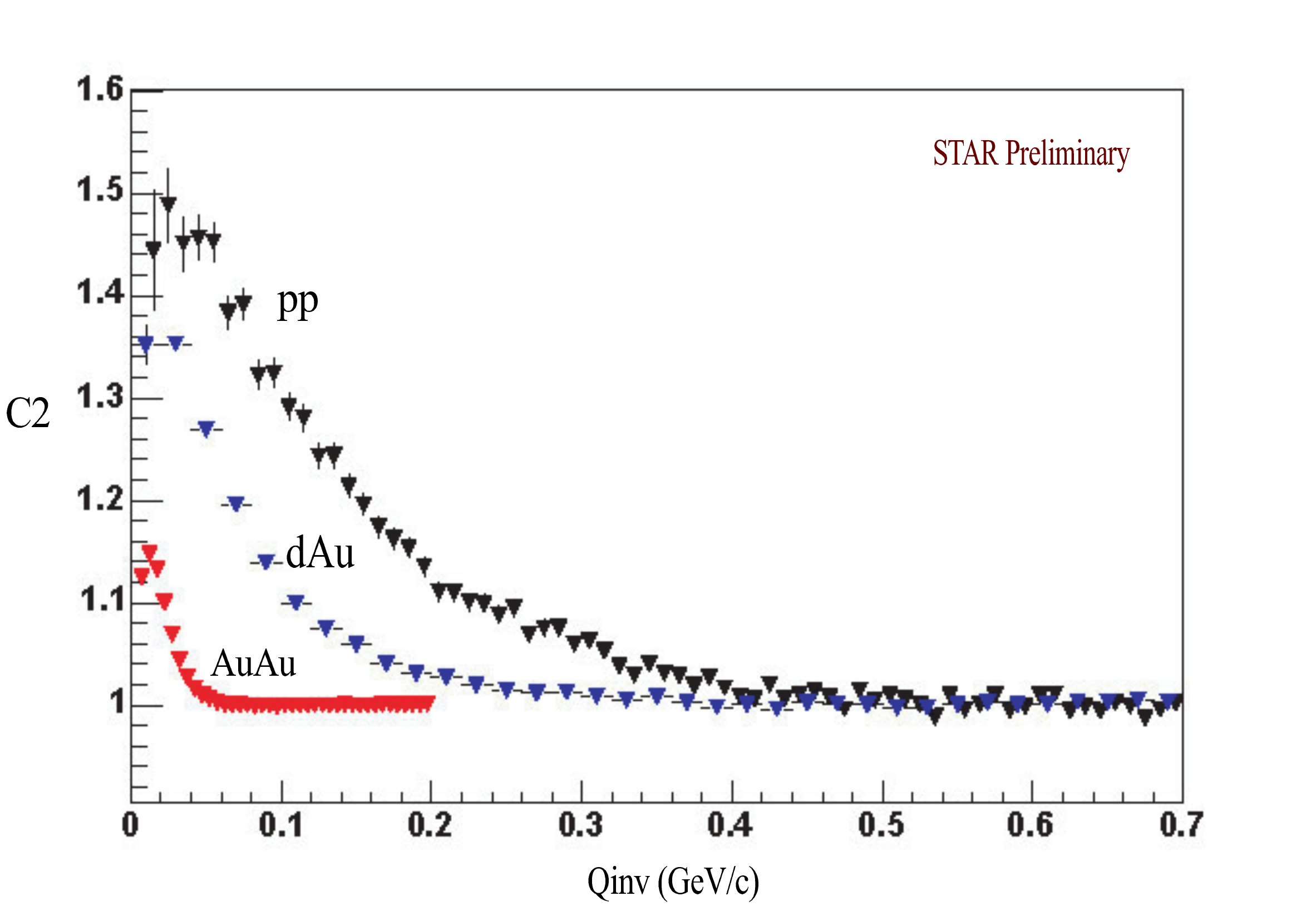} 
	}
	\caption[System size dependence of the width of the pion-pion correlation]{One-dimensional pion-pion correlation functions for p+p (upper), d+Au (middle), and Au+Au (lower) collisions at $\sqrt{s_{NN}}$ = 200\ \txt{GeV} \cite{Gutierrez:2004rm}; the p+p and d+Au results are based on minimum bias data, while the correlation function for Au+Au is shown for central collisions.}
	\label{HBT_system_size}
\end{figure}

If one focuses on a one-dimensional extent of the fireball $R$ (which, for example, in the case of a spherical system would directly correspond to its radius $r$), then the HBT 2-particle correlation function can be parametrized with a Gaussian
\begin{eqnarray}
C_{2,\text{HBT}} (Q_{\txt{inv}}) = 1 \pm \lambda \exp \Big( - R^2 Q_{\txt{inv}}^2   \Big) ~,
\label{HBT_one_dimensional}
\end{eqnarray}
where $\lambda$ represents the strength of the correlation at zero momentum difference and $Q_{\txt{inv}} \equiv \sqrt{ (p_1 - p_2)_{\mu} (p_1 -p_2)^{\mu}   }$, where $p_1$, $p_2$ are four-momenta of the two particles, is a Lorentz invariant related to the relative momentum of the particles. From Eq.\ \eqref{HBT_one_dimensional} it is clear that a source characterized by a large extent $R$ will lead to a small value of the correlation, while the opposite is true for sources characterized by a small $R$. Indeed, measurements of one-dimensional pion-pion correlations for proton-proton (p+p), deuteron-gold (d+Au), and gold-gold (Au+Au) collisions \cite{Gutierrez:2004rm} show that as the system size increases, the width of the correlation function decreases, see Fig.\ \ref{HBT_system_size}.

\begin{figure}[t]
	\centering\mbox{
		\includegraphics[width=0.75\textwidth]{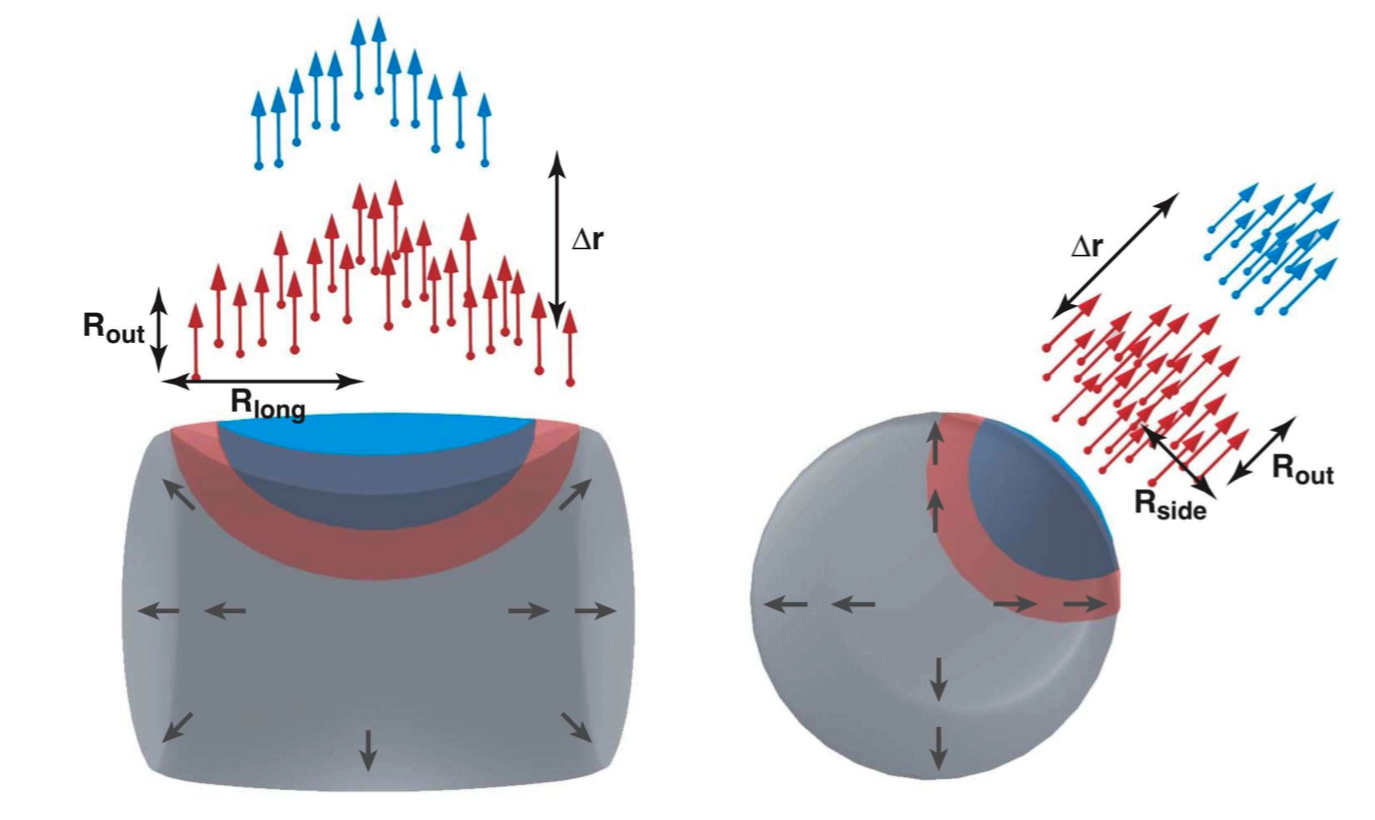} 
	}
	\caption[Orientation of the $R_\txt{long}$, $R_\txt{out}$, and $R_\txt{side}$ directions with respect to the collision geometry]{Orientation of the $R_\txt{long}$, $R_\txt{out}$, and $R_\txt{side}$ directions with respect to the collision geometry at high collision energies \cite{Lisa:2005dd}. Particles emitted at different correlated values of the transverse momentum $p_T$ (indicated with red and blue arrows) correspond to different source regions in the spacetime volume of the collision.}
	\label{HBT_R_long,R_side,R_out}
\end{figure}

In analyses devoted to studying the 3-dimensional geometry of the collisions, the size of the collision system is usually encoded in three variables: $R_{\txt{long}}$, $R_{\txt{out}}$, and $R_{\txt{side}}$. In Fig.\ \ref{HBT_R_long,R_side,R_out}, these three directions are shown in two views: the view in the left panel is along the rapidity $y = 0$ axis (that is the beam axis goes left to right, while the $y=0$ axis goes into the page), while the view in the right panel is along the center of the transverse plane axis (that is the beam axis goes into the page, while the $y=0$ axis goes left to right). The ``long'' axis is defined to point along the beam, that is in the $z$-direction, the ``out'' axis points along the average transverse momentum of the contributing correlated pair, and the ``side'' axis is perpendicular to the ``out'' and ``long'' axis. The figure shows particle emission corresponding to two different values of the transverse momentum $p_T$, indicated with blue (larger values) and red (smaller values) arrows. On average, particles emitted with larger momenta correspond to earlier times in the collision, where the system has a smaller size. One can see that $R_{\txt{long}}$ reflects the longitudinal extent of the source, $R_{\txt{out}}$ reflects the ``transverse depth'' of the source, and $R_{\txt{side}}$ reflects the ``transverse width'' of the source when looking along the $R_{\txt{out}}$ axis.

The HBT 2-particle correlation function can then be decomposed according to
\begin{eqnarray}
C_{2,\text{HBT}} (\bm{q}) = 1 \pm \lambda \exp \Big( - R_{\txt{long}}^2 q_{\txt{long}}^2    - R_{\txt{out}}^2 q_{\txt{out}}^2 - R_{\txt{side}}^2 q_{\txt{side}}^2 \Big) ~,
\end{eqnarray}
where $q_{\txt{long}}$, $q_{\txt{out}}$, and $q_{\txt{side}}$ are the relative momenta of the particle pair in the long, out, and side directions. It can be further shown that $R_{\txt{long}}$, $R_{\txt{out}}$, and $R_{\txt{side}}$ are given by the following averages involving the position differences in the long, out, and side directions ($\Delta x_{\txt{long}}$, $\Delta x_{\txt{out}}$, $\Delta x_{\txt{side}}$), the average pair velocity in the long and transverse directions ($v_L$, $v_T$), and the difference between the emission times of the particles ($\Delta t$):
\begin{eqnarray}
&& R_{\txt{long}}^2= \langle \big( \Delta x_{\txt{long}}  - v_{L} \Delta t \big)^2 \rangle ~, \\
&& R_{\txt{out}}^2 = \langle \big( \Delta x_{\txt{out}}  - v_{T} \Delta t \big)^2 \rangle ~, \\
&& R_{\txt{side}}^2  = \langle \big( \Delta x_{\txt{side}}^2   \rangle 
\end{eqnarray}
(for a detailed derivation, see Ref.\ \cite{Heinz:1999rw}). Combinations of $R_{\txt{long}}$, $R_{\txt{out}}$, and $R_{\txt{side}}$ can then reveal the characteristics of the spacetime geometry of the system. In particular, it can be shown that $R_{\txt{out}}^2 - R_{\txt{side}}^2$ is proportional to the duration of the emission of detected particles \cite{Bertsch:1989vn,Pratt:1990zq}.

\begin{figure}[t]
	\centering\mbox{
		\includegraphics[width=0.53\textwidth]{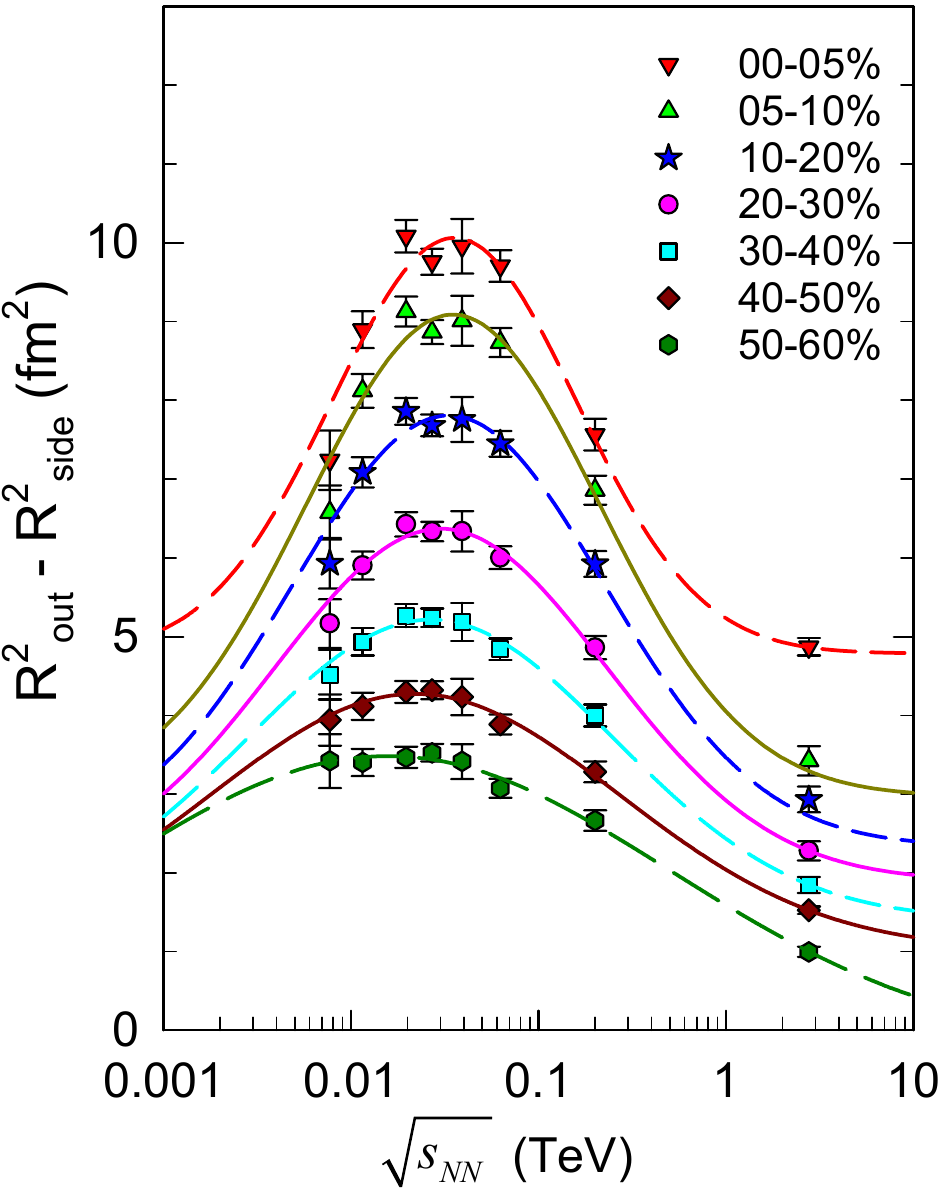}
	}
	\caption[The beam energy dependence of $R_{\txt{out}}^2 - R_{\txt{side}}^2$ extracted from two-pion correlation functions]{The beam energy dependence of $R_{\txt{out}}^2 - R_{\txt{side}}^2$ \cite{Lacey:2014wqa} extracted from two-pion correlation functions as measured by the STAR Collaboration \cite{STAR:2014shf} and by the ALICE Collaboration \cite{ALICE:2011dyt}. The results imply that the lifetime of the system as a function of the collision energy has a maximum around $\sqrt{s_{NN}} \approx 20$--$40 \ \txt{GeV}$.}
	\label{HBT_RoyLacey}
\end{figure}

Fig.\ \ref{HBT_RoyLacey} shows the beam energy dependence of interferometry measurements of two-pion correlation functions \cite{Lacey:2014wqa} as obtained by the STAR Collaboration during the BES-I program \cite{STAR:2014shf} as well as by the ALICE Collaboration \cite{ALICE:2011dyt}. The shown behavior of $R_{\txt{out}}^2 - R_{\txt{side}}^2$ implies that the lifetime of the system as a function of the collision energy has a maximum around $\sqrt{s_{NN}} \approx 20$--$40 \ \txt{GeV}$; a slight dependence on centrality can also be observed. As suggested in Refs.\ \cite{Pratt:1988vn,Bertsch:1988db, Hung:1994eq}, these results can be interpreted to mean that around $\sqrt{s_{NN}} \approx 20$--$40\ \txt{GeV}$, systems created in heavy-ion collisions evolve through regions of the phase diagram where the EOS is soft, and the corresponding relatively small pressure gradients result in an expansion that takes place at a slower rate and, consequently, for a longer time. Because softening of the EOS is expected in the vicinity of a phase transition, the data implies that collisions at $\sqrt{s_{NN}} \approx 20$--$40\ \txt{GeV}$ may be accessing a region of the phase diagram containing the QCD critical point. 

While these results are encouraging, to date they have not been reproduced in simulations. Existing studies utilizing models including an EOS with a first-order phase transition reproduce the qualitative, but the quantitative behavior of the HBT correlations; for example, Ref.\ \cite{Rischke:1996em} predicts significantly larger values of $R_{\txt{out}}/R_{\txt{side}}$ (which is another measure of the lifetime of the system expected to be less sensitive to effects due to flow) than obtained in experiment \cite{STAR:2020dav}.

\subsection{Slope of the directed flow at midrapidity}
\label{slope_of_directed_flow_at_midrapidity}

The flow coefficients (or flow harmonics) $v_n$ describe the asymmetry of the particle azimuthal distribution $dN/d\phi$, where $\phi$ is the azimuthal angle, that is the angle measured with respect to rotations around the beam axis (with $\phi =0$ usually coincident with the positive $x$-axis of the transverse plane; see Appendix \ref{collision_geometry_appendix} for more details). Formally, given the invariant particle distribution~,
\begin{eqnarray}
\frac{E~d^3N}{d^3p} = \frac{d^3N}{p_T dp_T~ dy ~d\phi} ~,
\label{particle_distribution}
\end{eqnarray} 
$v_n$ are defined as Fourier decomposition coefficients of $dN/d\phi$,
\begin{eqnarray}
\frac{d^3N}{p_T dp_T~ dy ~d\phi}  =\frac{d^2N}{p_T dp_T ~dy}  ~\frac{1}{2\pi} \left[ 1 + \sum_{n=1}^{\infty} 2 v_n(p_T, y) \cos \big( n \phi \big)  \right]~.
\label{vn_Fourier_decomposition}
\end{eqnarray} 
From this definition it is straightforward to obtain the expression for $v_n$, 
\begin{eqnarray}
v_n(p_T, y)   = \frac{ \int_{0}^{2\pi} d\phi ~ \cos \big( n\phi \big) ~  \frac{d^3N}{p_T dp_T~ dy ~d\phi} }{\int_{0}^{2\pi} d\phi ~   \frac{d^3N}{p_T dp_T~ dy ~d\phi}} \equiv \llangle \cos \big(n\phi_i \big)  \rrangle ~.
\label{vn_definition}
\end{eqnarray}
In practice, given experimental data, $v_n$ is simply given by
\begin{eqnarray}
v_n (p_T, y)  =\frac{1}{N}\sum_{i=1}^N  \cos \big(n\phi_i \big)  ~,
\label{flow_in_terms_of_phi}
\end{eqnarray}
where $N$ is the number of detected particles characterized by a transverse momentum $p_T$ and rapidity $y$, and $\phi_i$ is the azimuthal angle of the $i$-th particle. One can also calculate integrated $v_n$, that is $v_n$ calculated from the particle distribution, Eq.\ \eqref{particle_distribution}, integrated over, e.g., the transverse momentum $p_T$,
\begin{eqnarray}
v_n(y)   = \frac{\int dp_T \int_{0}^{2\pi} d\phi ~ \cos \big( n\phi \big) ~  \frac{d^3N}{p_T dp_T~ dy ~d\phi} }{\int dp_T \int_{0}^{2\pi} d\phi ~   \frac{d^3N}{p_T dp_T~ dy ~d\phi}} ~.
\label{integrated_v_n}
\end{eqnarray}
Often, $v_n (y)$ is calculated for particles in a given range of rapidity, for example $v_n(|y| < 0.5)$. 

We note here that while the concept of measuring the angle $\phi$ with respect to the reaction plane is simple, its realization in experiment is far from trivial; in practice, it is approximately done either by utilizing the transverse distribution of the spectators or particle-particle correlations. For an in-depth review of flow observables and relevant calculation methods, see Ref.\ \cite{Voloshin:2008dg}.

The directed flow (used to be known as the ``sideways flow'') is obtained by taking $n=1$ in Eq.\ \eqref{vn_definition}, yielding
\begin{eqnarray}
v_1 = \llangle \frac{p_x}{p_T} \rrangle~,
\end{eqnarray}
where $p_x$ is the component of the transverse momentum along the $x$-axis of the transverse plane. The directed flow is often calculated for particles characterized by different values of rapidity $y$, $v_1 (y)$, and averaged over events within the same centrality class. The behavior of the directed flow as a function of rapidity is affected by both the collision geometry and the collective expansion of the system, which we explain below.

\begin{figure}[t]
	\centering\mbox{
	\includegraphics[width=0.7\textwidth]{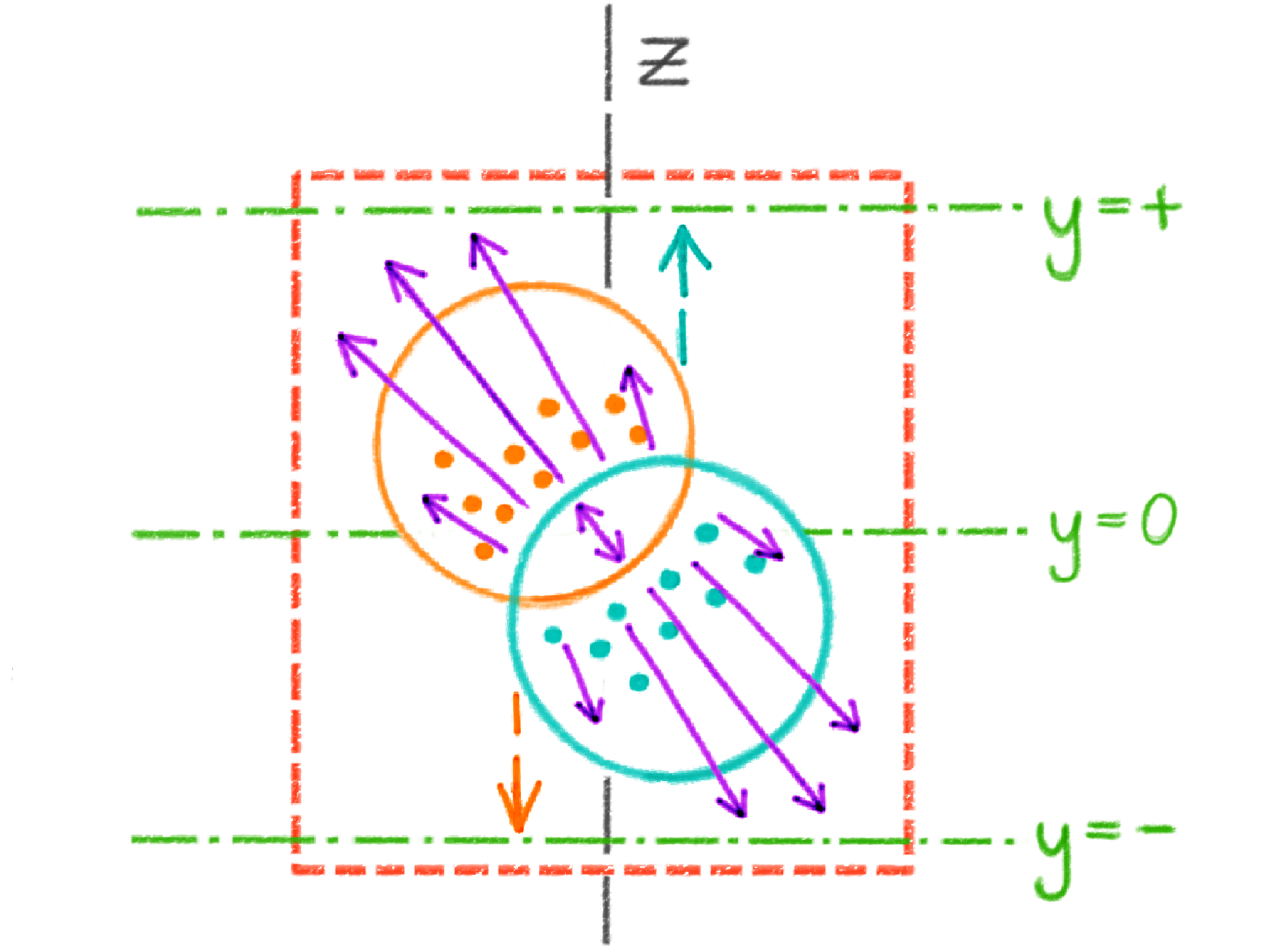} 
    }
	\caption[A sketch illustrating the geometric and thermodynamic origin of directed flow]{A sketch illustrating the geometric and thermodynamic origin of the directed flow $v_1$, with the beam axis indicated by the long-dashed gray line, the reaction plane indicated by the short-dashed red line, and the planes of constant values of rapidity indicated by the green dash-dotted lines. The geometry of a non-central collision between two nuclei, depicted with orange and turquoise circles, creates a pressure gradient along the short axis of the elliptical overlap region in the reaction plane, leading to asymmetrical particle trajectories in the forward and backward directions. See text for more details.}
	\label{v1_intro}
\end{figure}

Before the collision, the total transverse momentum of the system is zero, and by conservation of momentum it is also zero after the collision has taken place. This means, in particular, that the directed flow at midrapidity is by construction equal zero, $v_1 (y=0) = 0$. This does not have to be the case, however, for $v_1$ measured at finite rapidity (naturally, one still has $\int dy ~  v_1 (y)  =0$). Let us consider a mid-central collision as depicted in Fig.\ \ref{v1_intro}. Using a perspective from above the reaction plane, the figure shows that the collision will lead to a formation of an almond-shaped overlap region (also referred to as the collision or participants zone) at an angle to the beam axis. Matter in this region (manifesting itself in the detector mostly through produced particles such as pions) will naturally experience large pressure gradients; importantly, because of the shape of the overlap region, the largest pressure gradient will occur in the direction of the short axis of the region, denoted in Fig.\ \ref{v1_intro} with a purple double-headed arrow. As a result, trajectories of the particles in the forward and backward directions will not be symmetric with respect to the $y=0$ axis, leading to non-zero values of $v_1$ at $|y| > 0$. Note that, for example, the value of $v_1$ at $y = +1$ will be approximately opposite to the value at $y = -1$.

\begin{figure}[t]
	\centering\mbox{
	\includegraphics[width=0.99\textwidth]{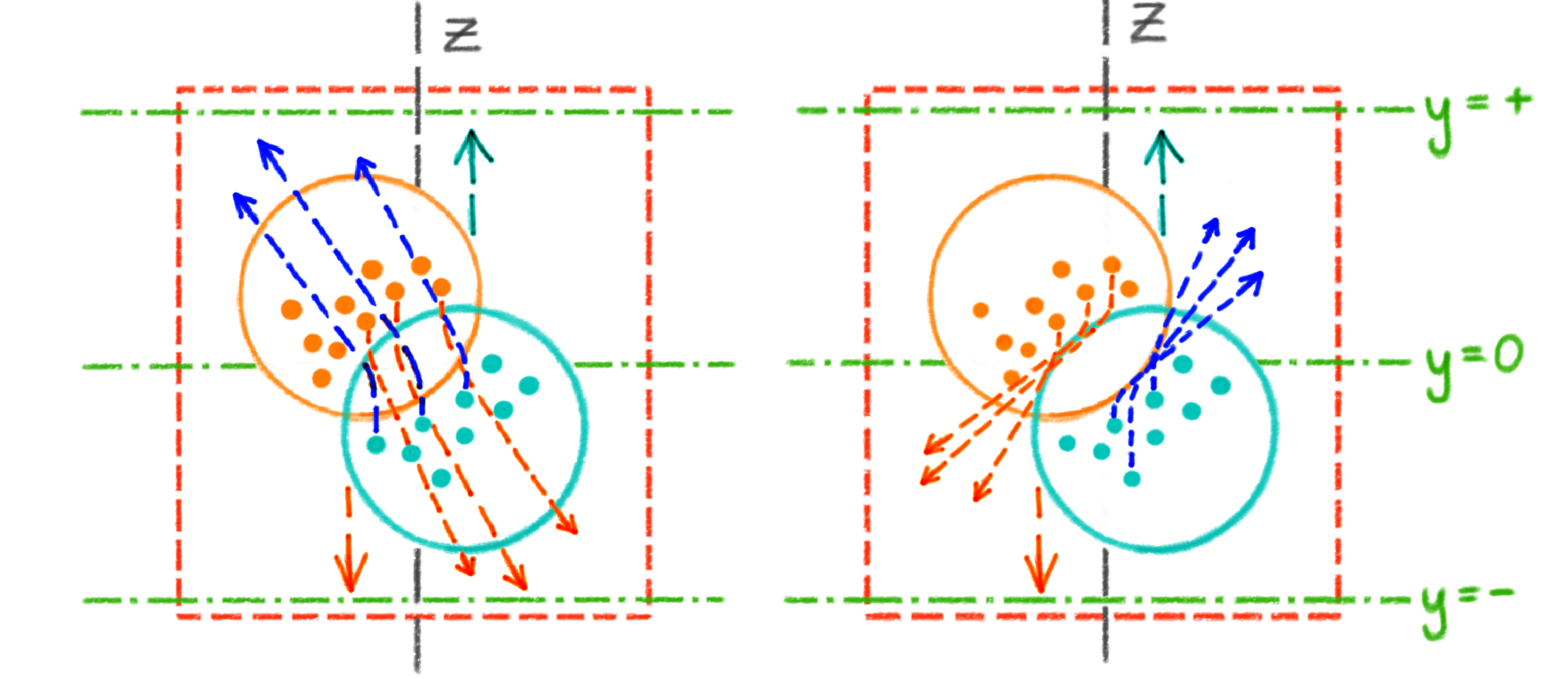} 
    }
	\caption[Sketches illustrating the relation between the directed flow of nucleons and the equation of state]{Sketches illustrating the relation between the directed flow of nucleons and the EOS, with the beam axis indicated by the long-dashed gray line, the reaction plane indicated by the short-dashed red line, and the planes of constant values of rapidity indicated by the green dash-dotted lines. Nucleons incoming into the collision region, depicted with orange and turquoise dots, will likely penetrate it if the EOS is soft (corresponding to small pressure gradients in the collision zone), or be pushed away if the EOS if hard (corresponding to large pressure gradients in the collision zone), resulting in two different directed flow patterns as shown on the left and right panel, respectively. See text for more details.}
	\label{v1_cont}
\end{figure}

The behavior of $v_1(y)$ of net protons can be connected to the EOS of dense nuclear matter \cite{Rischke:1995pe, Stoecker:2004qu} in the following way, sketched in Fig.\ \ref{v1_cont}: At early stages of the evolution, the initial collision zone is formed from nucleons that were the first to participate in the collision. At that time, the remaining nucleons in the two nuclei, depicted with orange and turquoise dots, are still coming into the collision zone, and their further trajectories depend on the magnitude of the pressure gradients present in the overlap region. If the EOS is soft (corresponding to smaller pressure gradients), the incoming nucleons will be able to penetrate the initial overlap region; once this happens, the pressure gradients present in the overlap region will deflect these nucleons such that they will contribute negatively to the directed flow at forward rapidity, $v_1\big(y>0\big)< 0$, and positively to the directed flow at backward rapidity, $v_1\big(y<0\big) > 0$ (left panel). On the other hand, if the EOS is hard (which corresponds to large pressure gradients), the incoming nucleons are going to be ``pushed'' out of the initial overlap region, and will contribute positively to the directed flow at forward rapidity, $v_1\big(y > 0\big)>0$, and negatively to the directed flow at backward rapidity, $v_1\big(y <0\big) <0$ (right panel). As a result, values of directed flow for different signs of rapidity reveal the stiffness of the EOS. In practice, a more convenient measure of the EOS is not the value of $v_1$ itself, but its slope at midrapidity, $dv_1/dy~ (y=0)$: this slope will be negative for a soft EOS and positive for a hard EOS.

In reality, the behavior of $dv_1/dy$ is a result of a rather complicated interplay of the collision geometry, pressure gradients, and scattering off of the collision zone; nevertheless, it is perceived as a promising measure of the EOS of nuclear matter. Indeed, if some of the collision energies studied in the BES create systems evolving in the proximity of the critical point, then the pressure gradients characterizing the overlap region created in these collisions should be much smaller than in collisions evolving far from the critical point, and one expects the following behavior of the slope of directed flow: $dv_1/dy ~(y=0) > 0$ for regions away from the critical point, then $dv_1/dy ~(y=0)  < 0$ for regions in the vicinity of the critical point, and again $dv_1/dy ~(y=0) > 0$ for points away from the critical point. 

\begin{figure}[t]
	\centering\mbox{
	\includegraphics[width=0.50\textwidth]{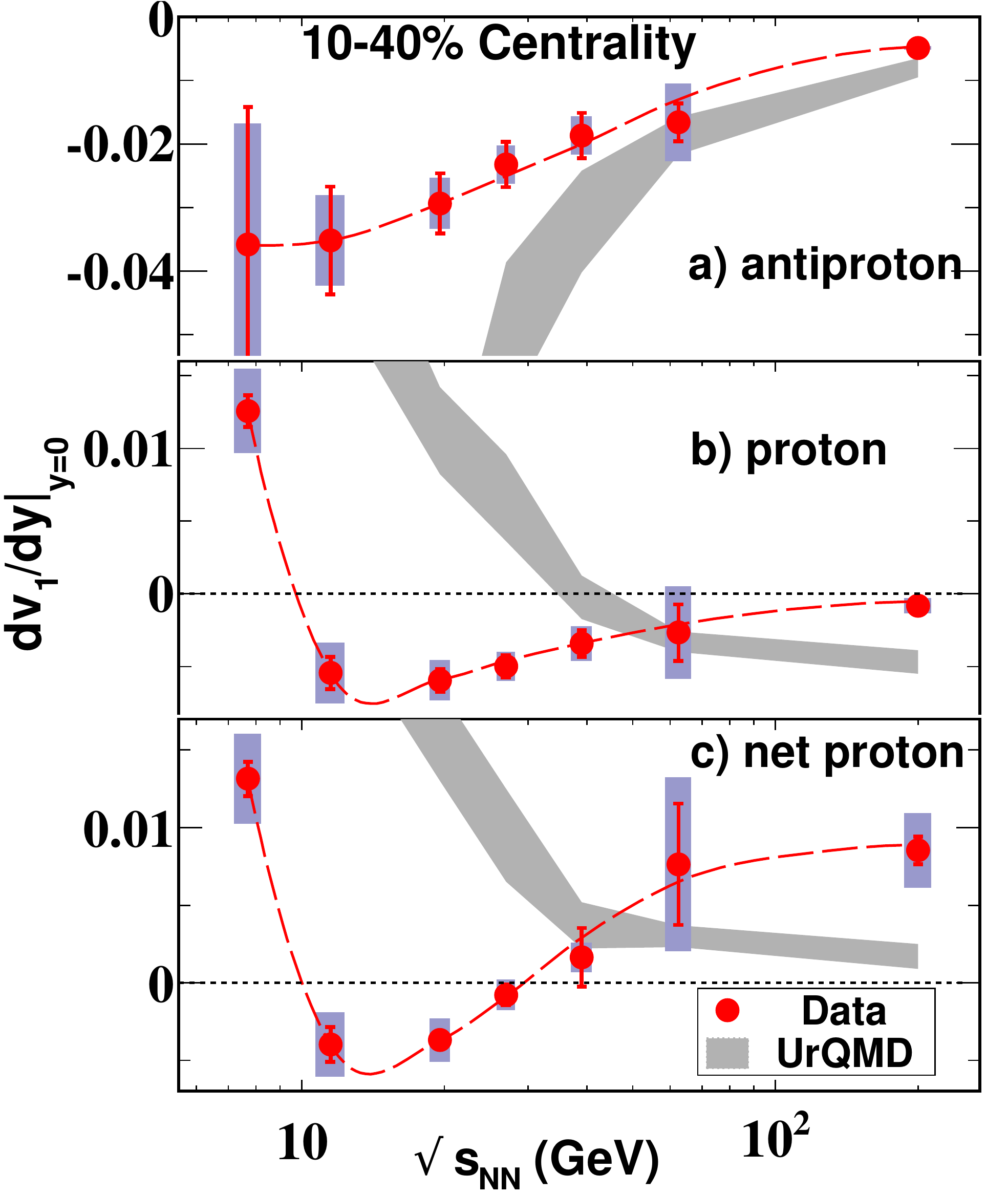} 
	}
	\caption[The slope of the directed flow of protons, anti-protons, and net protons in 10--40\% Au+Au collisions]{The slope of the directed flow of protons, anti-protons, and net protons as measured in 10--40\% Au+Au collisions, with \texttt{UrQMD} simulation results for comparison \cite{STAR:2014clz}. For antiprotons, which are produced baryons created at the hadronization stage of the collision, $dv_1/dy$ is a monotonic function of the beam energy; in contrast, for protons and net protons (the latter being a good proxy for the transported baryon number) $dv_1/dy$ displays a pronounced minimum around $\sqrt{s_{NN}} \approx 10$--$20\ \txt{GeV}$. See text for more details.}
	\label{STAR_dv1_dy}
\end{figure}

Such non-monotonic behavior has indeed been observed \cite{STAR:2014clz}: Fig.\ \ref{STAR_dv1_dy} shows the slope of the directed flow of antiprotons, protons, and net protons as a function of the beam energy in 10--40\% central Au+Au collisions. The figure also shows the behavior of $dv_1/dy$ as obtained from Ultra-relativistic Quantum Molecular Dynamics (\texttt{UrQMD}) simulations \cite{Bass:1998ca,Bleicher:1999xi}, which are simulations of heavy-ion collision evolution that do not include effects related to the QCD EOS and therefore are often used as a baseline expectation, deviations from which highlight the influence of the EOS on the dynamics of the collisions (various simulations of heavy-ion collisions will be further discussed in Section \ref{simulations_of_heavy-ion_collisions}). The upper panel of Fig.\ \ref{STAR_dv1_dy} shows that the slope of the directed flow of antiprotons does not display any non-monotonic behavior as a function of the beam energy, while the $dv_1/dy$ of protons, shown in the middle panel, is indeed non-monotonic and displays a pronounced negative minimum at beam energies around $\sqrt{s_{NN}} \approx 10$--$20\ \txt{GeV}$. This difference in behavior can be explained by the fact that antiprotons are produced baryons, created at the hadronization stage of the collision, while protons embody both the produced and the transported baryon number, the latter of which is affected by the stages of the evolution leading to the formation of the $v_1$ signal. The slope of the directed flow of net protons, which are considered to be the best indicator of the behavior of the transported baryon number, is shown in the bottom panel of Fig.\ \ref{STAR_dv1_dy}; it is, like the $dv_1/dy$ of protons, characterized by a non-monotonic behavior with a minimum that suggests the softest point of the EOS at low energies.

Unfortunately, to date efforts to reproduce the directed flow in dynamical models have not been successful \cite{Steinheimer:2014pfa,Nara:2016hbg,Hillmann:2018nmd}. This must mean that there are effects, contributing significantly to the $v_1$, that are as of now missing from simulations of heavy-ion collisions. Uncovering these effects would greatly increase our understanding of the microscopic origin of the $v_1$ signal.

\section{Tantalizing results from BES-I: Turning off the quark-gluon plasma}
\label{turning_off_the_QGP}

Mapping the QCD phase diagram consists not only of searching for signals of the QCD phase transition, but also includes identifying collision energies at which QGP ceases to be produced. The occurrence of QGP in high-energy collisions is argued based on several signatures which are expected to vanish in collisions where a QGP state is not created. Many of these signatures rely on the collective expansion of the QGP phase of the collisions, and one can ask: when does the collective behavior of the systems, and with it the evidence for QGP, turn off? A number of measurements addressed this question, and even though the answer remains elusive, these studies showed inconsistencies in the behavior of the observables that could lead to identifying collision energies, and through that regions of the QCD phase diagram, in which QGP is not produced.

\subsection{Quark number scaling of the elliptic flow}
\label{quark_number_scaling_of_the_elliptic_flow}

\begin{figure}[t]
	\centering\mbox{
		\includegraphics[width=0.70\textwidth]{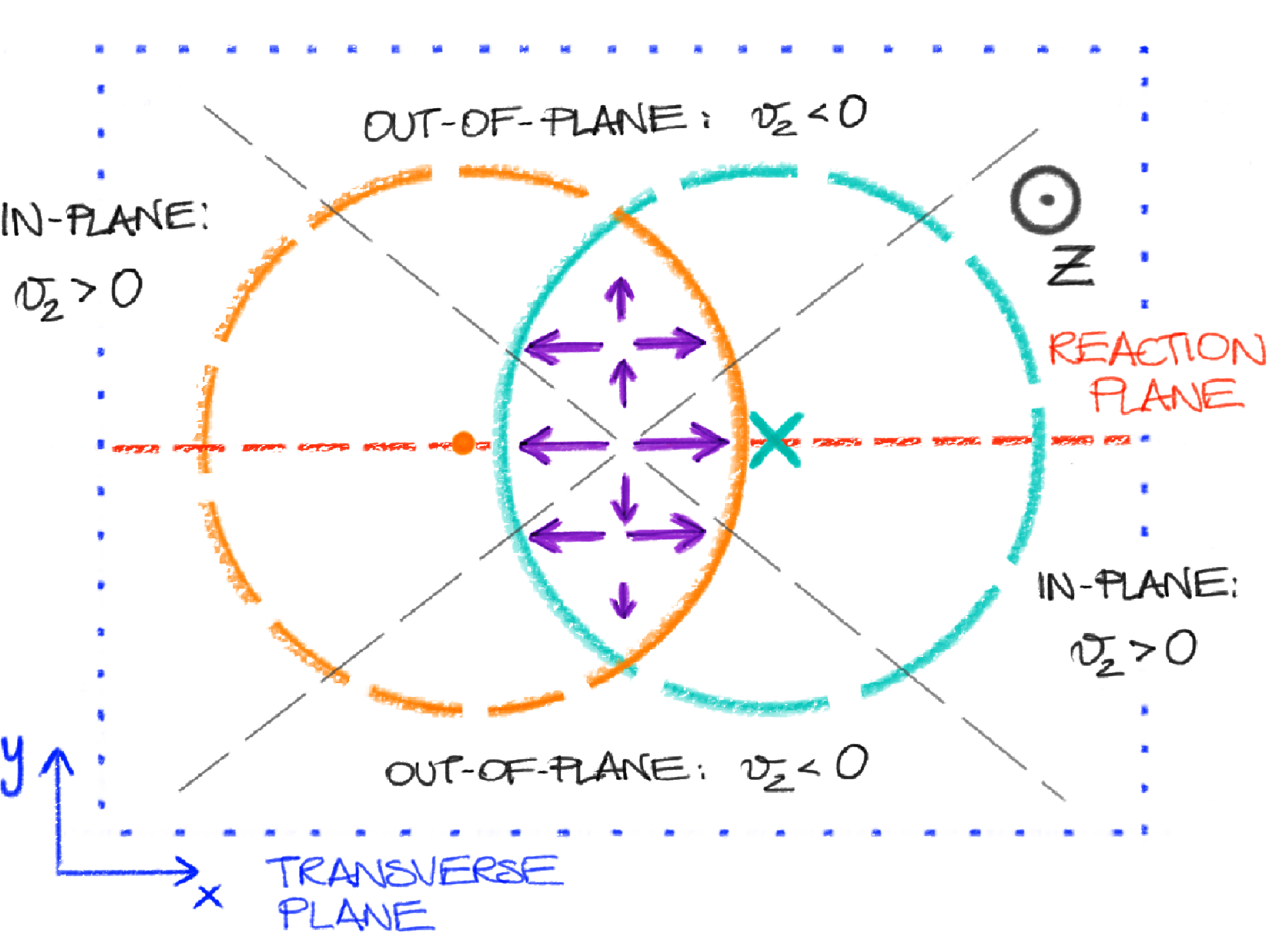}
	}
	\caption[A sketch illustrating the geometric and thermodynamic origin of elliptic flow]{A sketch of a mid-central heavy-ion collision in a view along the beam axis, with the reaction plane indicated by the short-dashed red line and the transverse plane indicated by the blue dotted line. The sketch illustrates the geometric and thermodynamic origin of the elliptic flow $v_2$: a non-central collision between two nuclei, depicted with orange and turquoise circles, creates strong pressure gradients along the short axis of the collision zone and results in larger transverse momenta of particles emitted in the ``in-plane'' direction, which contribute positively to the elliptic flow. See text for more details.}
	\label{v2_intro}
\end{figure}

The integrated elliptic flow $v_2( y)$ is the second Fourier decomposition coefficient of the particle azimuthal distribution integrated over transverse momenta of the particles, given by Eq.\ \eqref{integrated_v_n} with $n=2$, and experimentally it is calculated from
\begin{eqnarray}
v_2(y) = \frac{1}{N} \sum_{i=1}^N \cos \big( 2\phi_i \big) = \langle  \cos \big( 2\phi_i \big) \rangle ~,
\label{v2_equation}
\end{eqnarray}
where the sum is performed over all particles characterized by a given rapidity $y$. The elliptic flow vanishes in systems with an azimuthal symmetry, while non-zero values of $v_2$ are a consequence of an asymmetric initial geometry of the collision and the subsequent thermodynamics-driven expansion. To explain how an asymmetry in the momentum space can arise from the initial geometry, it is helpful to consider a mid-central collision, depicted in Fig.\ \ref{v2_intro} in a view along the beam axis. Parts of the nuclei that do not overlap during the collision (indicated by the dashed orange and turquoise lines) are ``sheared off'' of the overlap region and continue moving along the $z$-axis in their respective directions, leaving behind the collision zone. The collision zone (indicated by the solid orange and turquoise lines) displays a largely elliptic azimuthal asymmetry: its size along the principal axis coincident with the $y$-direction of the transverse plane is larger than its size along the principal axis coincident with the reaction plane. The pressure is the largest in the center of the overlap region while it is equal zero outside of it, and as a consequence of the geometry described above, the pressure gradient is larger along the short axis of the ellipse than along it's long axis (indicated in the figure by long and short purple arrows, respectively). This means that the expansion rate in directions coincident with the reaction plane (``in-plane'') will be larger than the expansion rate in directions perpendicular to the reaction plane (``out-of-plane''), so that particles from the collision zone will overall gain more transverse momentum in the ``in-plane'' direction than in  the ``out-of-plane'' direction. In this way, the initial azimuthal asymmetry in the coordinate space will be transformed into an azimuthal asymmetry in the transverse momenta of detected particles. Since particles moving in the ``in-plane'' direction will contribute positively to Eq.\ \eqref{v2_equation}, while particles moving ``out-of-plane'' will contribute negatively, in general one obtains a positive elliptic flow, $v_2>0$, for overlap regions of asymmetry characterized by a long axis that is perpendicular to the reaction plane, as in Fig.\ \ref{v2_intro}.

This simple picture becomes more complicated at very low beam energies where the spectators cannot be neglected, as the combination of smaller velocities of the nuclei and the resulting smaller Lorentz contraction means that the spectators largely remain in the vicinity of the collision when the collision region is being formed. As a result, for beam energies in the range $\sqrt{s}_{NN} \approx 2$--$5\ \txt{GeV}$, the spectators intercept particles emitted from the collision region in the ``in-plane'' direction, while paths of particles emitted ``out-of-plane'' are unobstructed. This effect, known as ``squeeze-out'', leads to negative values of the elliptic flow, $v_2<0$, and it has been both reported experimentally \cite{Voloshin:2008dg} as well as reproduced in simulations \cite{Danielewicz:2002pu}. 

\begin{figure}[t]
	\centering\mbox{
		\includegraphics[width=0.6\textwidth]{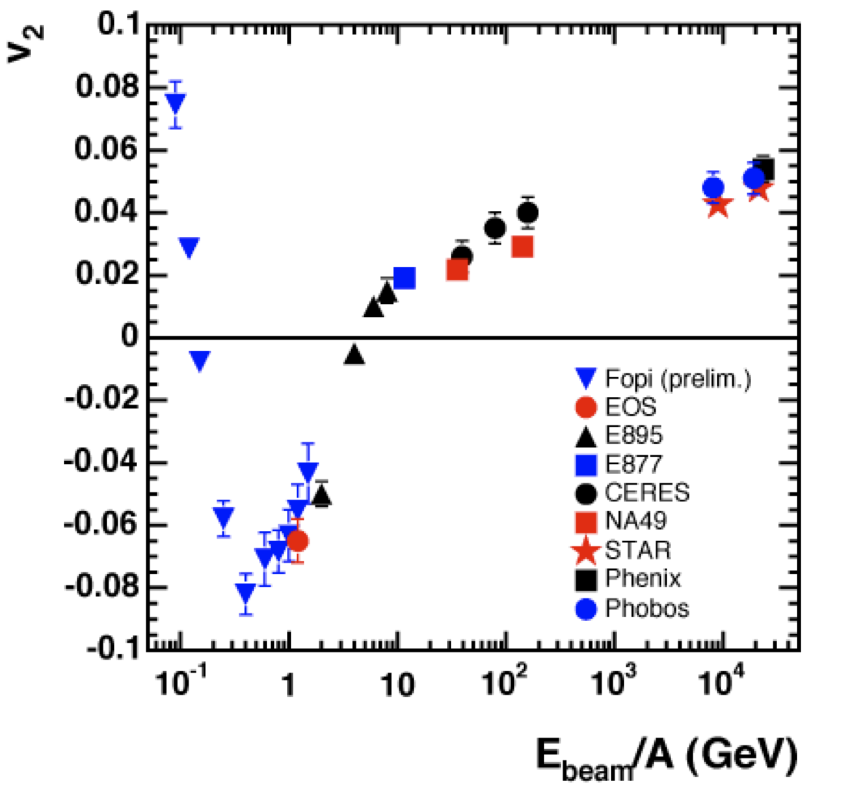} 
	}
	\caption[Elliptic flow at midrapidity for 0--25\% collisions as a function of projectile energies $E_{\txt{kin}} \approx 0.2$--$10^4\ A\txt{GeV}$]{Elliptic flow at midrapidity for 0--25\% most central collisions, integrated over particle species and transverse momenta, as a function of projectile energies $E_{\txt{kin}} \approx 0.2$--$10^4\ A\txt{GeV}$, corresponding to $\sqrt{s_{NN}} \approx 2$--$200\ \txt{GeV}$ \cite{Voloshin:2008dg}. The leading mechanism behind the elliptic flow signal changes as the collision energy is varied, resulting in different behavior in the respective energy regimes; see text for more details.}
	\label{v2_squeeze-out}
\end{figure}

At even lower energies the very concepts of participants and spectators cease to be applicable, as the colliding nuclei may fuse to form a rotating compound nucleus. For these systems, emitting particles ``in-plane'' is more favorable due to the system's non-zero angular momentum, which again yields a positive elliptic flow, $v_2 >0$. 

Altogether, the physics from which $v_2$ originates changes as the beam energy is varied, and the corresponding behavior can be seen in Fig.\ \ref{v2_squeeze-out}, showing elliptic flow of charged particles at midrapidity, integrated over particle transverse momenta, for 0--25\% most central collisions as a function of the beam energy. Nevertheless, for $\sqrt{s_{NN}} \gtrsim 5\ \txt{GeV}$ (corresponding to $E_{\txt{kin}} \approx 12 \ A\txt{GeV}$) the physics of the elliptic flow is driven by the collision geometry and the thermodynamics-driven expansion as sketched in Fig.\ \ref{v2_intro}. The beam energy dependence of $v_2$ observed in this region may convey information about the EOS, viscosity, or the number of degrees of freedom characterizing the system. In particular, changes in any of these characteristics could lead to a change in the magnitude or slope of the elliptic flow.

One of the most striking arguments supporting the creation of QGP in high-energy heavy-ion collisions utilizes the concept of quark number scaling of the elliptic flow \cite{STAR:2003wqp}, based on the following simple picture: 

If the collision region contains the QGP, the degrees of freedom undergoing the transverse expansion due to the pressure gradients are those of quarks and gluons, and in the course of the expansion any given quark will gain some transverse momentum $\Delta p_T$. On average, due to the elliptic flow, the magnitude of the ``in-plane'' component of $\Delta p_T$ will be larger than the ``out-of-plane'' component, so that more quarks will be traveling in the approximately ``in-plane'' direction. Then, following the expansion of the system, the quarks eventually hadronize into various mesons and baryons. This is thought to occur largely through two processes: fragmentation and coalescence. During fragmentation, the energy of the interaction between a given system of partons increases as the system expands, eventually leading to a production of a quark-antiquark pairs, which then contribute to the formation of hadrons; notably, the emerging hadrons carry a fraction of the initial quark momentum. On the other hand, in a quark coalescence mechanism \cite{Molnar:2003ff} quarks that are close enough to each other both in the position and in the momentum space (that is, which are close to each other in the phase space) will form a hadron; in this case the produced hadrons carry momenta which are sums of the momenta of the initial quarks. Importantly, this implies that hadrons characterized by intermediate to high values of $p_T$ are predominantly produced \textit{via} quark coalescence. (On a side note, at very high values of $p_T$ the probability of production through quark coalescence becomes negligibly small, and these hadrons are thought to be produced by fragmentation of high-energy jets.) 

It follows that within the coalescence picture the invariant spectrum of produced particles is proportional to the product of invariant spectra of their constituents \cite{Schwarzschild:1963zz,Sato:1981ez}, and in particular, assuming that quarks and antiquarks are described by the same invariant distribution $dN_q/d^2 p_{T}$, independent of flavor, the meson and baryon distributions are given by
\begin{eqnarray}
&& \frac{dN_M}{d^2 p_T}(\bm{p}_T) = C_M(p_T) \left[ \frac{dN_q}{d^2p_T}\left( \frac{\bm{p}_T}{2} \right)   \right]^2 ~,
\label{coalescence_mesons} \\
&&\frac{dN_B}{d^2 p_T}(\bm{p}_T) = C_B(p_T) \left[ \frac{dN_q}{d^2p_T}\left( \frac{\bm{p}_T}{3} \right)   \right]^3   ~, 
\label{coalescence_baryons}
\end{eqnarray}
where coefficients $C_M$ and $C_B$ are the probabilities for the meson and baryon coalescence, respectively. Using these relations allows one to relate the elliptic flow of mesons and baryons to the flow of partons: for example, if partons are characterized by a purely elliptical anisotropy,
\begin{eqnarray}
\frac{dN_q}{p_T dp_T ~ d\phi} = \frac{1}{2\pi } \frac{dN_q}{p_T dp_T} \Big[1 + 2 v_{2,q} \cos(2\phi) \Big]~,
\end{eqnarray}
then utilizing Eqs.\ (\ref{coalescence_baryons}-\ref{coalescence_mesons}) in Eq.\ \eqref{vn_definition} immediately leads to
\begin{eqnarray}
v_{2,M}(p_T) = \frac{2v_{2,q}\left( \frac{p_T}{2} \right) }{1 + 2 v_{2,q}^2\left( \frac{p_T}{2} \right) } 
\end{eqnarray}
and
\begin{eqnarray}
v_{2,B}(p_T) = \frac{3v_{2,q}\left( \frac{p_T}{3} \right) + 3v^2_{2,q}\left( \frac{p_T}{3} \right)}{1 + 6 v_{2,q}^2\left( \frac{p_T}{3} \right) }  ~.   
\end{eqnarray}
Provided that $v_{2,q} \ll 1$, we obtain the following relation
\begin{eqnarray}
\frac{v_2^{(B)}}{3} \approx \frac{v_2^{(M)} }{2} ~,
\label{q_number_scaling}
\end{eqnarray} 
which states that, for particles produced via coalescence of deconfined quarks, the elliptic flow calculated for a specific particle species scales with the number of constituent quarks of that species $n = \{2, 3\}$. Eq.\ \eqref{q_number_scaling} can be understood intuitively: since the probability for any three quarks to be found in the same region of the phase space is smaller than finding any two quarks in the same region, in the context of the elliptic flow hadronization occurring through coalescence means that producing baryons in the more dilute ``out-of-plane'' regions is more suppressed, relative to the dense ``in-plane'' regions, than the corresponding meson production; therefore, relative to mesons, baryons are more likely to be found in the ``in-plane'' regions than in the ``out-of-plane'' regions.

\begin{figure}[t]
	\centering\mbox{
		\includegraphics[width=0.8\textwidth]{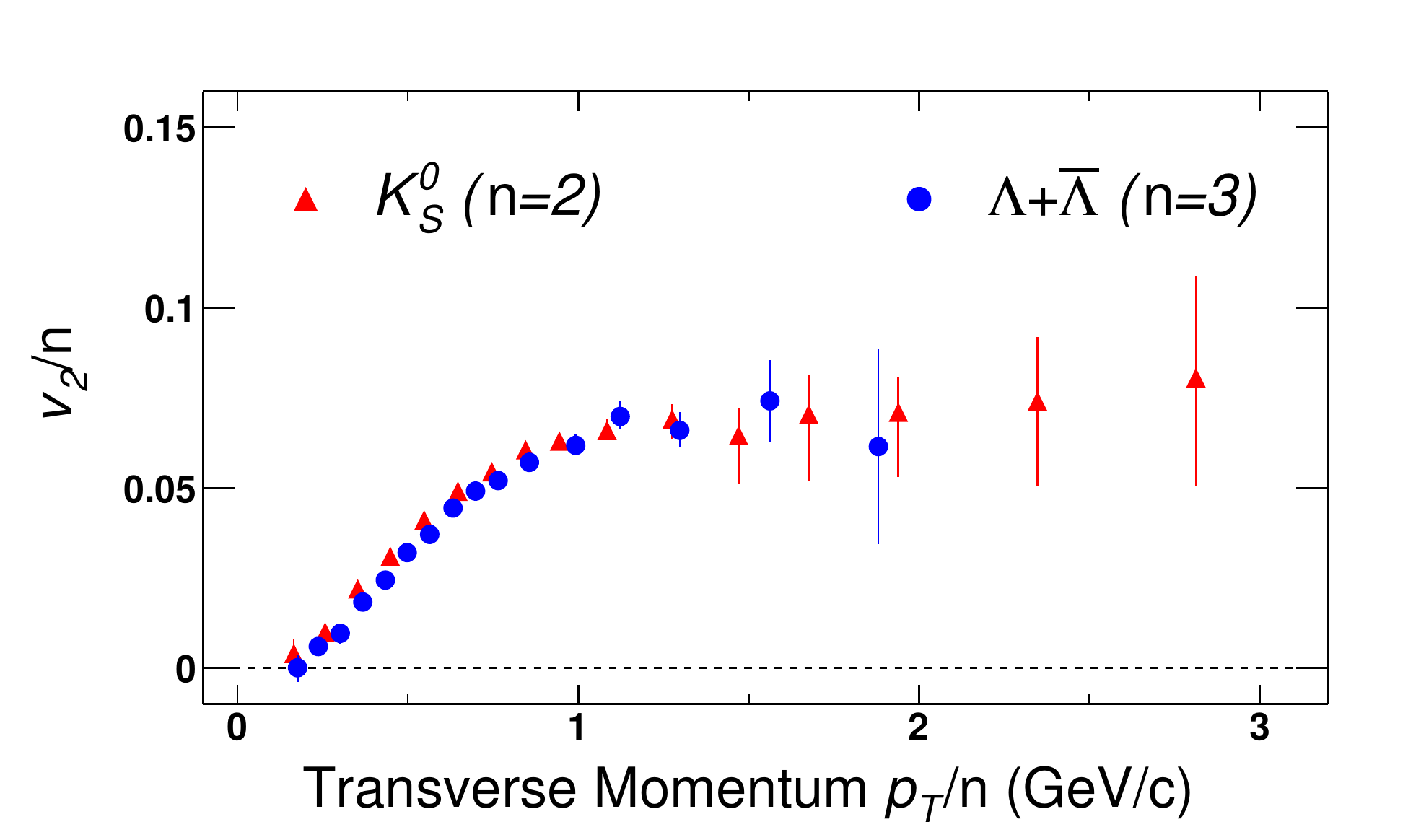} 
	}
	\caption[Quark number scaling of the elliptic flow of neutral kaons and Lambda baryons in minimum bias Au+Au collisions at $\sqrt{s_{NN}} = 200 \ \txt{GeV}$]{Quark number scaling of the elliptic flow of neutral kaons and the sum of Lambda baryons and antibaryons in minimum bias Au+Au collisions at $\sqrt{s_{NN}} = 200 \ \txt{GeV}$ \cite{STAR:2003wqp}. The elliptic flow and transverse momenta of kaons are divided by a factor of 2, while the elliptic flow and transverse momenta of Lambda baryons and antibaryons are divided by a factor of 3. See text for more details.}
	\label{v2_quark_number_scaling}
\end{figure}

Fig.\ \ref{v2_quark_number_scaling} shows scaled elliptic flows of neutral kaons and the sum of Lambda baryons and antibaryons, denoted by $v_2/n$, against transverse momentum, also scaled by the number of constituent quarks, $p_T/n$ (the latter scaling is introduced as the momentum of hadrons emerging from the fireball through coalescence is a sum of the momentum carried by the quarks from which they are formed, and through that the momentum of baryons is on average trivially larger than the momentum of mesons). The scaling of the elliptic flow with the number of constituent quarks is evident. 

\begin{figure}[t]
	\centering\mbox{
	\includegraphics[width=0.99\textwidth]{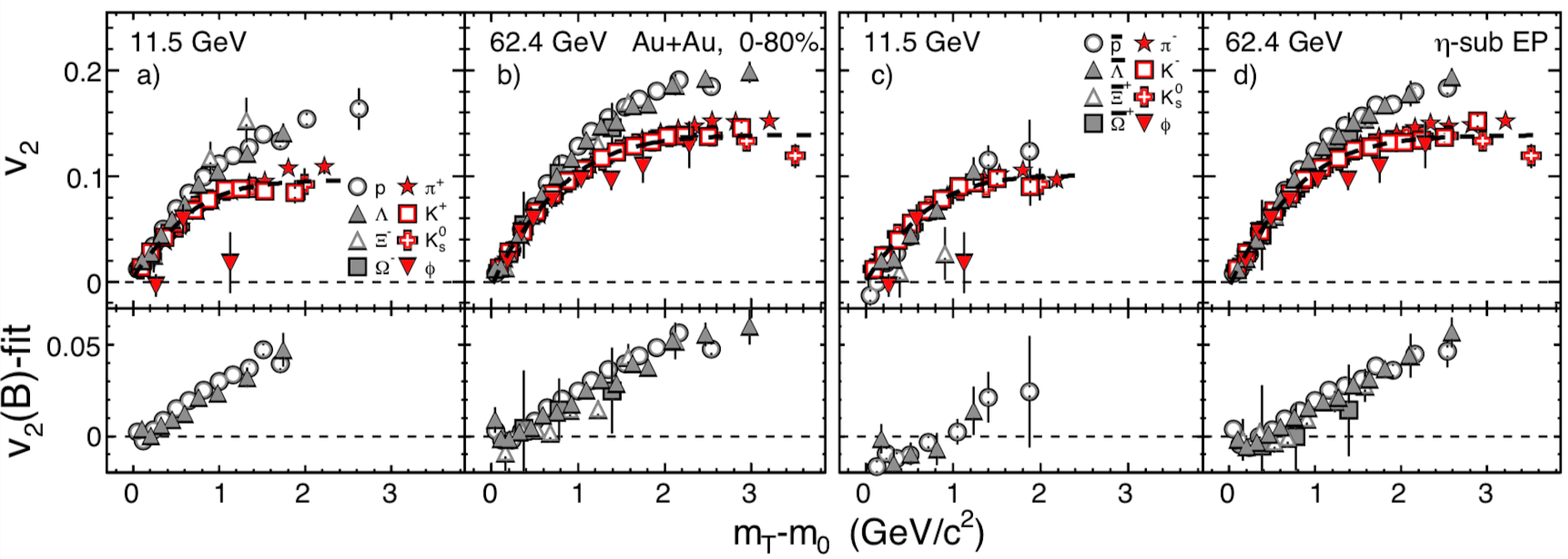} 
	}
	\caption[Elliptic flow of identified particles in minimum bias Au+Au collisions at $\sqrt{s_{NN}} = 11.5 \ \txt{GeV}$ and $62.4\ \txt{GeV}$]{Elliptic flow of identified particles in minimum bias Au+Au collisions at $\sqrt{s_{NN}} = 11.5 \ \txt{GeV}$ and $62.4\ \txt{GeV}$ as a function of the transverse part of the kinetic energy $m_T - m_0$ \cite{STAR:2013cow}. Panels b) and d): at $\sqrt{s_{NN}} = 64\ \txt{GeV}$, both baryons and antibaryons show quark number scaling of the elliptic flow. Panels a) and c): at $\sqrt{s_{NN}} = 11.5\ \txt{GeV}$, only baryons behave consistently with quark number scaling. See text for more details.}
	\label{v2_quark_number_scaling_low_energies}
\end{figure}

Because quark number scaling is based on an assumption that the evolution of a heavy-ion collision involves a deconfined phase followed by quark coalescence, deviation from the scaling shown in Fig.\ \ref{v2_quark_number_scaling} could signal that this assumption is broken. Fig.\ \ref{v2_quark_number_scaling_low_energies} shows the elliptic flow of identified particles in minimum bias Au+Au collisions at beam energies $\sqrt{s_{NN}} = 11.5 \ \txt{GeV}$ and $62.4\ \txt{GeV}$ as a function of the transverse part of the kinetic energy $m_T - m_0$ \cite{STAR:2013cow}. Panels a) and b) show $v_2$ of baryons as well as of neutral and positively charged mesons, while panels c) and d) show $v_2$ of antibaryons as well as of neutral and negatively charged mesons (the division of mesons between the two sets of panels is arbitrary and serves to provide largely equal baselines for the elliptic flows of baryons and antibaryons). In panels b) and d), corresponding to collisions at $\sqrt{s_{NN}} = 62.4\ \txt{GeV}$, the elliptic flow of both baryons and antibaryons is enhanced relative to the elliptic flow of mesons at high values of the transverse kinetic energy (where the coalescence mechanism should dominate), supporting the notion that quark number scaling applies in  this case. In panels a) and c), corresponding to collisions at $\sqrt{s_{NN}} = 11.5\ \txt{GeV}$, baryons exhibit enhancement expected from coalescence, however, antibaryons do not. While one could argue, based on the behavior of produced baryons (for which antibaryons are a good proxy), that this may be a signal for the regime in which the coalescence mechanism stops being applicable, the lack of consistency between the behavior displayed on panels a) and c) makes the interpretation less clear. Additionally, the deviations in the scaling of the elliptic flow could also be caused by final state interactions such as baryon-antibaryon annihilation or large decay contributions.

\subsection{Disappearance of the triangular flow}

\begin{figure}[t]
	\centering\mbox{
		\includegraphics[width=0.95\textwidth]{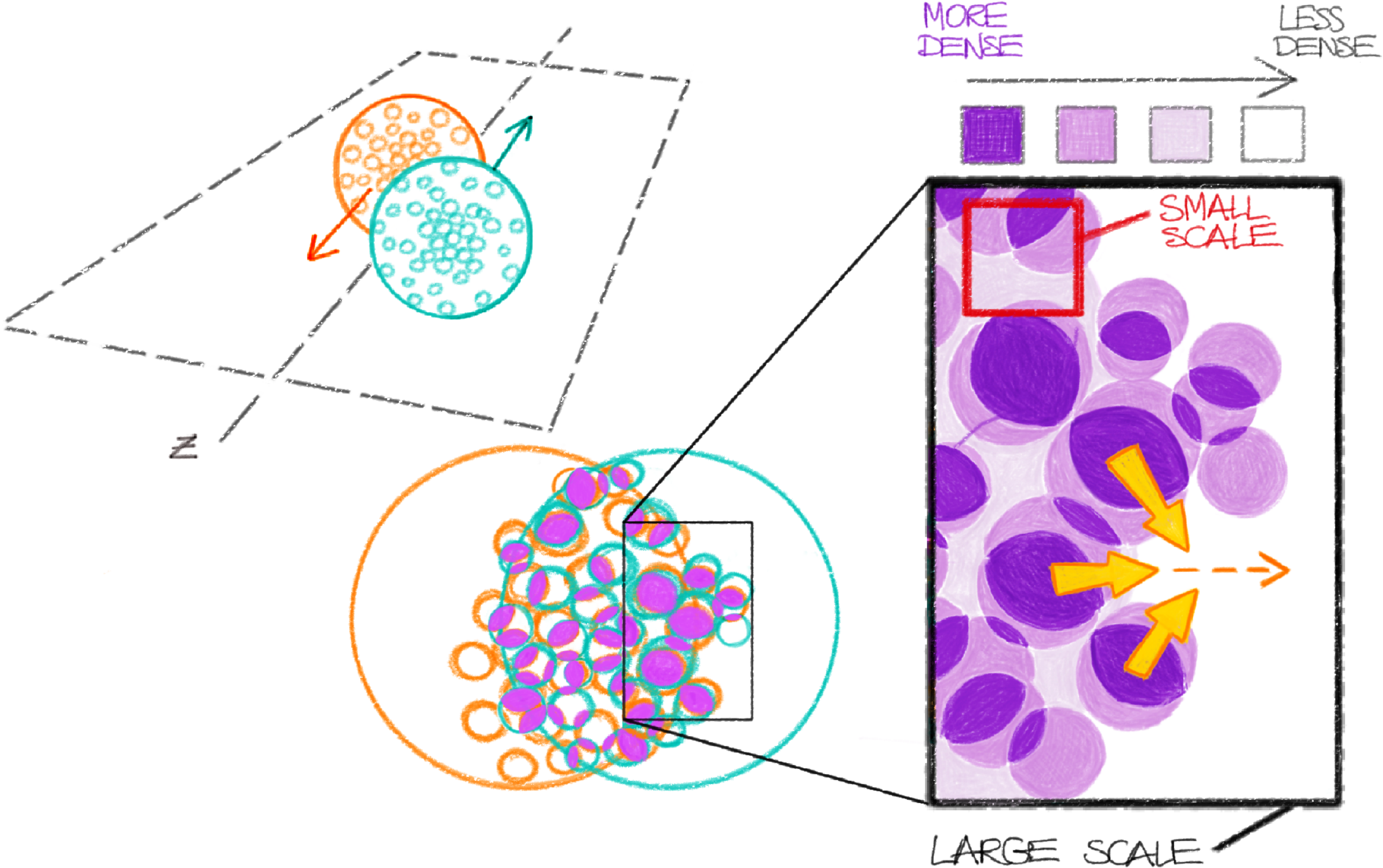} 
	}
	\caption[A sketch illustrating the geometric and thermodynamic origin of triangular flow]{A sketch illustrating the geometric and thermodynamic origin of the triangular flow $v_3$. The granularity of the density distribution in the collision region, originating from the ``overlapping'' of individual nucleons in the colliding nuclei, leads to small scale fluctuations in pressure gradients. Locally, matter expands preferentially in the direction of the steepest pressure gradient, resulting in a non-zero contribution to the triangular flow $v_3$. See text for more details.}
	\label{v3_intro}
\end{figure}

The integrated triangular flow $v_3(y)$ is the third Fourier decomposition coefficient of the particle azimuthal distribution integrated over transverse momenta of the particles, given by Eq.\ \eqref{integrated_v_n} with $n=3$, and in practice it is calculated from
\begin{eqnarray}
v_3(y) = \frac{1}{N} \sum_{i=1}^N \cos \big( 3\phi \big) = \langle  \cos \big( 3\phi_i \big) \rangle ~,
\label{v3_equation}
\end{eqnarray}
where the sum is performed over all particles characterized by a given rapidity $y$. Until fairly recently \cite{Mishra:2007tw,Sorensen:2010zq}, it had been thought that all odd harmonics (that is $v_1$, $v_3$, $v_5$, etc.) must average to zero at midrapidity due to the symmetry of the overlap region. Indeed, it is easy to convince oneself, looking at Fig.\ \ref{v2_intro}, that sums over particles emerging from the collision weighted with $\cos \big( \phi \big)$, $\cos \big(3 \phi \big)$, etc., will add up to zero. However, this conclusion is only true for an overlap region with a density profile that has an approximately elliptical symmetry (see Fig.\ \ref{collision_geometry} in Appendix \ref{collision_geometry_appendix}). In any given collision this is not, in fact, the case, as the density profile of the colliding nuclei is not a continuous and spherically symmetric function, $\rho = \rho(r)$, but instead it is a sum of discrete contributions, $\rho = \sum_{i=1}^A \rho_i (\bm{r})$, which only yields $\rho(r)$ on average (see the left-hand side part of Fig.\ \ref{v3_intro}). When two nuclei collide, these intrinsic fluctuations in the densities of the nuclei seed fluctuations in the density profile of the overlap region. Because of that, the density profile of the collision zone is not in fact elliptical, and the odd flow harmonics can take finite values at $y=0$ (see the lower center part of Fig.\ \ref{v3_intro}). (We note here that intrinsic fluctuations also contribute to the even flow harmonics at $y=0$, however, in this case it's a second-order effect as compared to the contributions stemming from the collision geometry.)

Although the length scale of the intrinsic density fluctuations is small compared to the size of the overlap region, generating $v_3$ proceeds through a similar mechanism as in the case of $v_2$, and both depend on pressure gradients. On the right-hand side part of Fig.\ \ref{v3_intro}, a close-up sketch of a part of the overlap zone is shown where a more dense and a less dense region are formed in the initial stage of the collision due to the underlying anisotropies in the positions of the nucleons. The pressure gradient will drive the matter from the more dense to the less dense region; this process diminishes the spacial anisotropy and at the same time creates a momentum anisotropy (in the example sketched in the Figure, particles will on average gain some momentum to the right, in the $\phi \approx 0$ direction). This is the azimuthal momentum anisotropy that drives the $v_3$ signal \cite{Mishra:2007tw,Sorensen:2010zq}. Notably, the fact that the scale of the initial anisotropies is small means that they only survive for a short time, and so it is likely that $v_3$ is a snapshot of pressure anisotropies in the earliest stages of the collision \cite{Petersen:2012qc}.

Importantly, matter whose evolution is sensitive to small scale structures (such as the density fluctuations in the initial overlap region) must be characterized by a small viscosity. Indeed, the damping rate in a hydrodynamic medium is proportional to $\eta/ \lambda^2$ \cite{Landau_Fluid_Mechanics}, where $\eta$ is the viscosity and $\lambda$ is the wavelength of the propagated fluctuation, and therefore small-scale fluctuations can be expected to dissipate more quickly as compared with larger-scale fluctuations. Consequently, as an example, effects due to the density profile arising from the shape of the overlap region, leading to the formation of the elliptic flow, should be more robust against dissipation than fluctuations due to the positions of individual nucleons, leading to the formation of the triangular flow. The fact that the magnitudes of $v_2$ and $v_3$ are often of the same order indicates that damping does not play a big role in the evolution of the system, which can only be the case for a small value of $\eta$. Moreover, the hadronic state of nuclear matter is found to be characterized by a relatively large viscosity, and so it can be argued that the measurement of a significant $v_3$ in high-energy collisions is a signal for the creation of a new state of matter, most likely the QGP, characterized by a very low viscosity.

\begin{figure}[t]
	\centering\mbox{
		\includegraphics[width=0.82\textwidth]{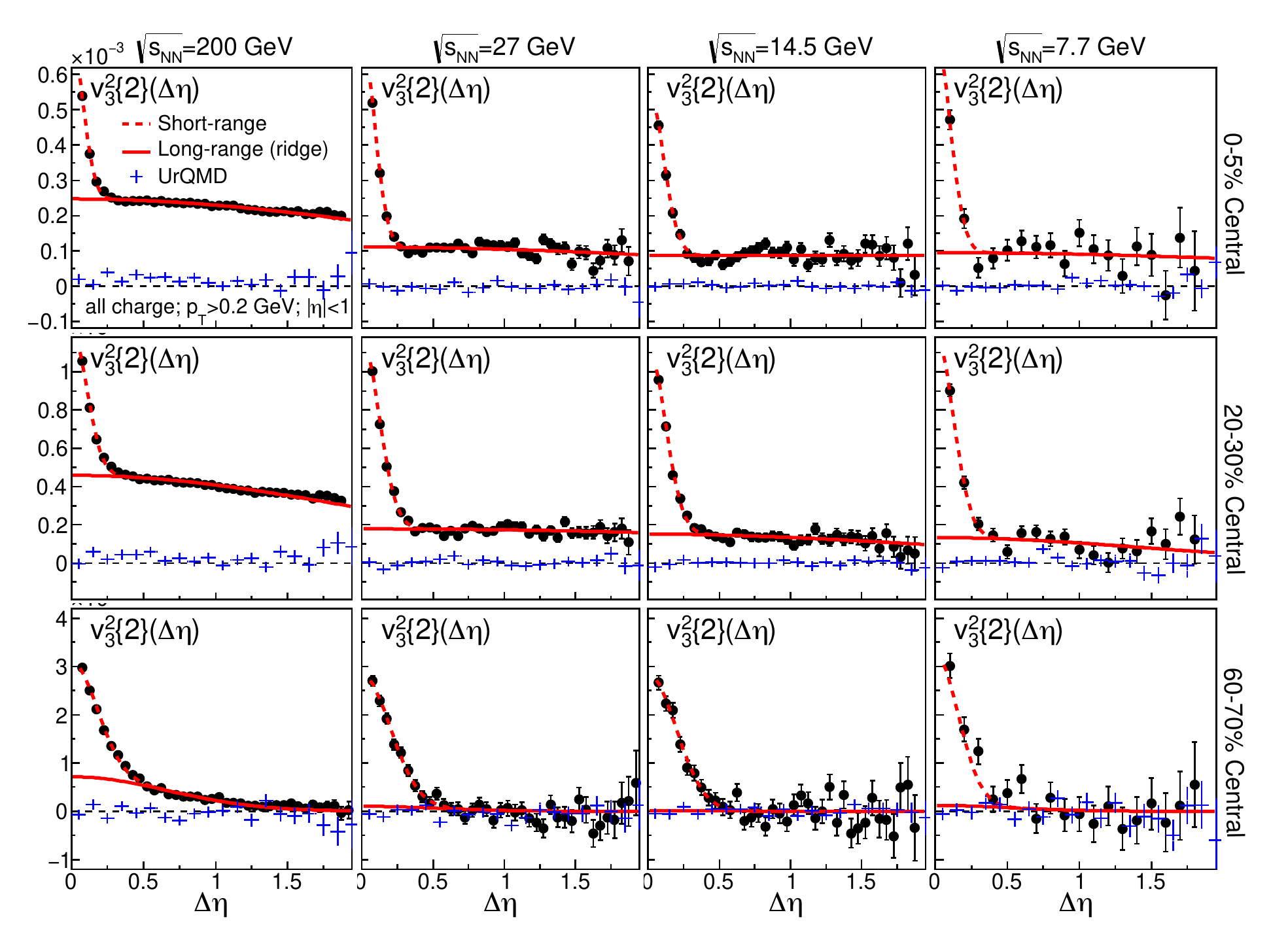} 
	}
	\caption[Triangular flow of charged hadrons as a function of $\Delta \eta$ for $\sqrt{s_{NN}} = 200,\ 27,\ 14.5$, $7.7 \ \txt{GeV}$ in 0--5\%, 20--30\%, and 60--70\% Au+Au collisions]{Triangular flow of charged hadrons as a function of $\Delta \eta$ for beam energies $\sqrt{s_{NN}} = 200,\ 27,\ 14.5, \ 7.7 \ \txt{GeV}$ in 0--5\%, 20--30\%, and 60--70\% central Au+Au collisions, with \texttt{UrQMD} simulation results for comparison \cite{STAR:2016vqt}. For the most central collisions, the triangular flow displays a strong signal at most energetic collisions and diminishes with the collision energy, but remains finite. For peripheral collisions, the triangular flow is consistent with zero for all but the most energetic collisions. See text for more details.}
	\label{STAR_triangular_flow}
\end{figure}

Consequently, vanishing of the triangular flow could mean that the collisions probe regions of the phase diagram in which QGP is not produced. It has been found that $v_3$ indeed disappears in low-energy peripheral collisions. Fig.\ \ref{STAR_triangular_flow} shows plots of the triangular flow of charged hadrons as a function of particle-particle pseudorapidity difference $\Delta \eta$ for beam energies $\sqrt{s_{NN}} = 200,\ 27,\ 14.5$, and $7.7 \ \txt{GeV}$ in 0--5\%, 20--30\%, and 60--70\% central Au+Au collisions \cite{STAR:2016vqt}. The figure also shows the behavior of the triangular flow as calculated in \texttt{UrQMD} simulations \cite{Bass:1998ca,Bleicher:1999xi}, used as a baseline expectation for systems in which QGP is not created (various simulations of heavy-ion collisions will be further discussed in Section \ref{simulations_of_heavy-ion_collisions}). The top row displays results for central collisions, where $v_3$ is measured to be non-zero for all collision energies analyzed. In the bottom row, corresponding  to peripheral collisions, $v_3$ is only non-zero at the highest considered energy (we note here that the peak present in all plots at small values of $\Delta \eta$ is found to correspond to nonflow correlations like the HBT correlations, resonance decays, and effects due to Coulomb interactions, and can be considered as background to the $v_3$ measurement \cite{STAR:2013qio,STAR:2016vqt}). This suggests that a stage of the collision necessary for the production of the $v_3$ signal is not reached in small systems created in low-energy peripheral collisions. Such systems are characterized by a relatively small energy density, and it is therefore possible that QGP ceases to be produced in these collisions. On the other hand, the fact that the $v_3$ signal persists in the most central collisions down to the lowest of the studied energies, where QGP was not expected to occur, needs to be understood better in order to draw firm conclusions.

\section{Tantalizing results from BES-I: Non-statistical event-by-event fluctuations of conserved charges}
\label{non-statistical_event-by-event_fluctuations_of_conserved_charges}

A set of observables that gained significant attention in the context of the search for the QCD critical point are fluctuations of the net baryon number distribution, which are related to derivatives of the pressure with respect to the order parameter and which can be shown to diverge in the critical region. If such critical fluctuations can be measured in experiment, they would constitute a signal for systems evolving in the vicinity of the critical point.

The behavior of pressure derivatives with respect to the order parameter can be encoded in the susceptibilities of the conserved charge, which in the context of heavy-ion collisions means susceptibilities of the net baryon number, defined as 
\begin{eqnarray}
\chi_j \equiv \left( \frac{d^j P}{d\mu_B^j} \right)_T~,
\label{susceptibilities_definition}
\end{eqnarray}
where $P$ is the pressure and $\mu_B$ is the baryon chemical potential. In particular, we have $\chi_1 = \left( dP/d\mu_B\right)_T \equiv n_B$, the net baryon number density of the system. Notably, susceptibilities of the net baryon number are directly related to cumulants of the net baryon number, where the latter can be defined as
\begin{eqnarray}
\kappa_j \equiv T^j \frac{d^j}{d\mu_B^j}\bigg|_{T} \ln \mathcal{Z} (T,V, \mu_B) ~.
\label{cumulants_definition_2}
\end{eqnarray}
Indeed, with the grand canonical partition function given by
\begin{eqnarray}
\mathcal{Z} (T,V, \mu_B) = \sum_{N=0}^{\infty} e^{\beta \mu_B N} Z_N (T,V,N)~,
\end{eqnarray}
where $\beta = 1/T$ and $Z_N (T,V,N)$ is the canonical partition function, it is straightforward to show, based on Eq.\ (\ref{cumulants_definition_2}), that the following relations hold
\begin{eqnarray}
&& \hspace{-5mm}\kappa_1 = \frac{1}{\mathcal{Z}(T,V,\mu_B)} \sum_{N=0}^{\infty} N e^{\beta \mu_B N} Z_N (T,V,N)  \equiv \langle N \rangle   ~, 
\label{cumulant_1_in_terms_of_moments} \\
&& \hspace{-5mm} \kappa_2 =  \Bigg(  - \bigg[\frac{1}{\mathcal{Z}(T,V,\mu_B)}  \sum_{N=0}^{\infty} N e^{\beta \mu_B N} Z_N (T,V,N)\bigg]^2  +  \non\\
&& \hspace{10mm} + ~  \frac{1}{\mathcal{Z}(T,V,\mu_B)} \sum_{N=0}^{\infty} N^2 e^{\beta \mu_B N} Z_N (T,V,N)  \Bigg) = \mu_2 \equiv \Big\langle \Big( N - \big\langle N \big\rangle \Big)^2 \Big\rangle~, 
\label{cumulant_2_in_terms_of_moments} \\
&& \hspace{-5mm} \kappa_3 =  \mu_3 = \Big\langle \Big( N - \big\langle N \big\rangle \Big)^3 \Big\rangle ~, 
\label{cumulant_3_in_terms_of_moments} \\
&& \hspace{-5mm} \kappa_4 =   \mu_4 - 3 \mu_2^2 = \Big\langle \Big( N - \big\langle N \big\rangle \Big)^4 \Big\rangle - 3 \kappa_2^2 ~,
\label{cumulant_4_in_terms_of_moments}
\end{eqnarray}
where we only showed explicit expressions for the first two cumulants, and where $\mu_j$ are central moments of the net baryon distribution. At the same time, because the pressure is related to the grand canonical partition function through $P =  T \ln \mathcal{Z} (T,V,\mu_B)/V$, cumulants of the net baryon number are related to the thermodynamic fluctuations in the system by
\begin{eqnarray}
\kappa_j = VT^{j-1} \chi_j 
\label{cumulants_from_susceptibilities}
\end{eqnarray}
where $V$ is the volume and $T$ is the temperature. From this representation it is clear that the susceptibilities of the net baryon number reflect the fluctuations in the conserved charge that occur in the system at hand. Using Eq.\ \eqref{cumulants_from_susceptibilities}, one can obtain explicit expressions for the first four cumulants expressed in terms of derivatives of the pressure with respect to the net baryon number density (see Appendix \ref{cumulants_of_the_net_baryon_number} for the explicit calculation),
\begin{eqnarray}
&&\hspace{-15mm} \kappa_1 = Vn_B = N_B ~, 
\label{cumulant_1} \\
&&\hspace{-15mm}  \kappa_2 = \frac{VTn_B}{\left( \frac{dP}{dn_B} \right)_T} ~, 
\label{cumulant_2}  \\
&&\hspace{-15mm}  \kappa_3 = \frac{VT^2 n_B}{\left( \frac{dP}{dn_B} \right)_T^2} \left[ 1 - \frac{n_B}{\left( \frac{dP}{dn_B} \right)_T}  \left( \frac{d^2P}{dn_B^2} \right)_T  \right] ~, 
\label{cumulant_3} \\
&&\hspace{-15mm}  \kappa_4 = \frac{VT^3 n_B}{\left(\frac{dP}{dn_B}  \right)_T^3}  \left[  1    -   \frac{4n_B}{\left(\frac{dP}{dn_B}  \right)_T} \left(\frac{d^2P}{dn_B^2}\right)_T +       \frac{3n_B^2}{\left(\frac{dP}{dn_B}  \right)_T^2}   \left(\frac{d^2P}{dn_B^2}\right)_T^2    -   \frac{n_B^2}{\left(\frac{dP}{dn_B}  \right)_T} \left(\frac{d^3P}{dn_B^3}  \right)_T \right] ~.
\label{cumulant_4} 
\end{eqnarray}

\begin{figure}[t]
	\centering\mbox{
	\includegraphics[width=0.45\textwidth]{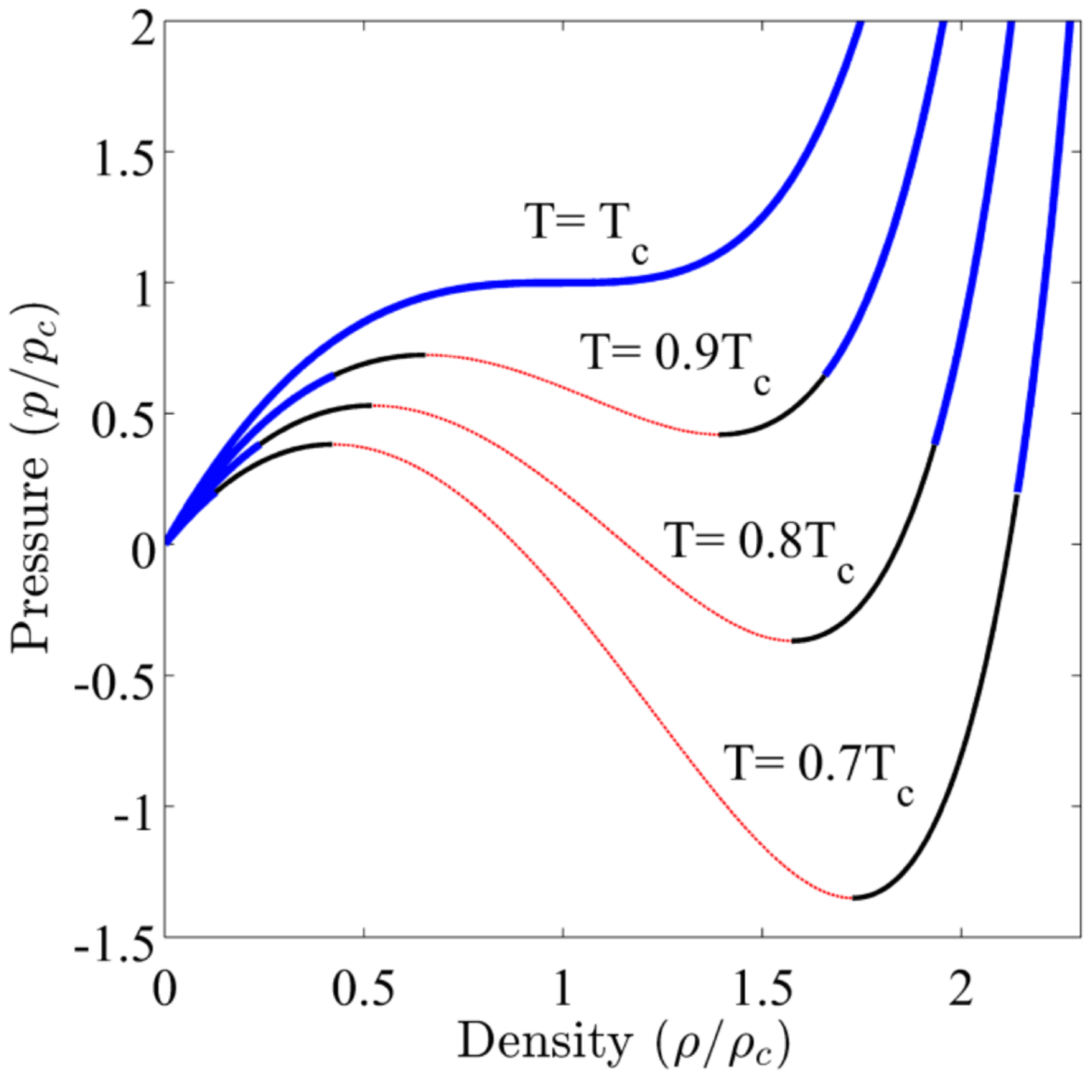} 
	}
	\caption[The isotherms of pressure as a function of density for the Van der Waals EOS]{The isotherms of pressure as a function of density for the Van der Waals EOS \cite{Cueto-Felgueroso_VdW_EOS}; the density is shown in units of the critical density and the pressure is shown in units of the critical pressure. Blue lines mark paths along the isotherms which are thermodynamically stable, and for temperatures lower than the critical temperature they terminate at coexistence densities for a given temperature. Black lines correspond to regions which are thermodynamically unstable. Red lines correspond to parts of the isotherms for which the derivative of the pressure with respect to density is negative, that is to the mechanically unstable spinodal regions of the transition. At the critical point both the first and the second derivative of the pressure with respect to the density vanish.}
	\label{VDF_EOS}
\end{figure}

As an example of pressure in a model with a first-order phase transition, Fig.\ \ref{VDF_EOS} shows the pressure as a function of density at a series of temperatures for the Van der Waals EOS (a brief overview of first-order phase transitions is given in Appendix \ref{phase_transitions}). By analyzing the pressure curves in this figure, one can notice that the behavior of the cumulants at a given temperature identifies the position on a pressure isotherm with respect to the critical region. For example, because the derivative of the pressure with respect to the net baryon density approaches zero in the vicinity of the spinodal lines in general and at the critical point in particular, the second-order cumulant $\kappa_2$, Eq.\ (\ref{cumulant_2}), diverges in the corresponding regions. Similarly, one sees that because the curvature of the pressure tends to be negative for densities smaller than the critical density, the third-order cumulant $\kappa_3$, Eq.\ (\ref{cumulant_3}), is positive, with increasing magnitude when approaching the critical region due to the diverging factors of $\left( dP/dn_B \right)_T$ in the denominator. Conversely, for densities larger than the critical density, $\kappa_3$ is negative in the vicinity of the critical point due to the positive curvature of the pressure. Finally, as one approaches the critical region from below, the curvature of the pressure first becomes more negative, then increases across the critical region until it reaches a positive maximum, and then decreases again. Thus as one goes from densities below the critical point to densities above it, the fourth-order cumulant $\kappa_4$, driven in the critical region by the derivative of the pressure curvature entering in the last term in Eq.\ (\ref{cumulant_4}), will be first positive, then negative, and then again positive. This behavior is magnified in the vicinity of the critical point, where $\kappa_4$ diverges. In general, the divergence of the higher order cumulants in the vicinity of a phase transition is directly connected to small values of the first-order pressure derivative, $\left(dP / dn_B\right)_T \approx 0$, and signals the softness of the EOS (pressure) near the critical region.

\begin{figure}[t]
	\centering\mbox{
		\includegraphics[width=0.95\textwidth]{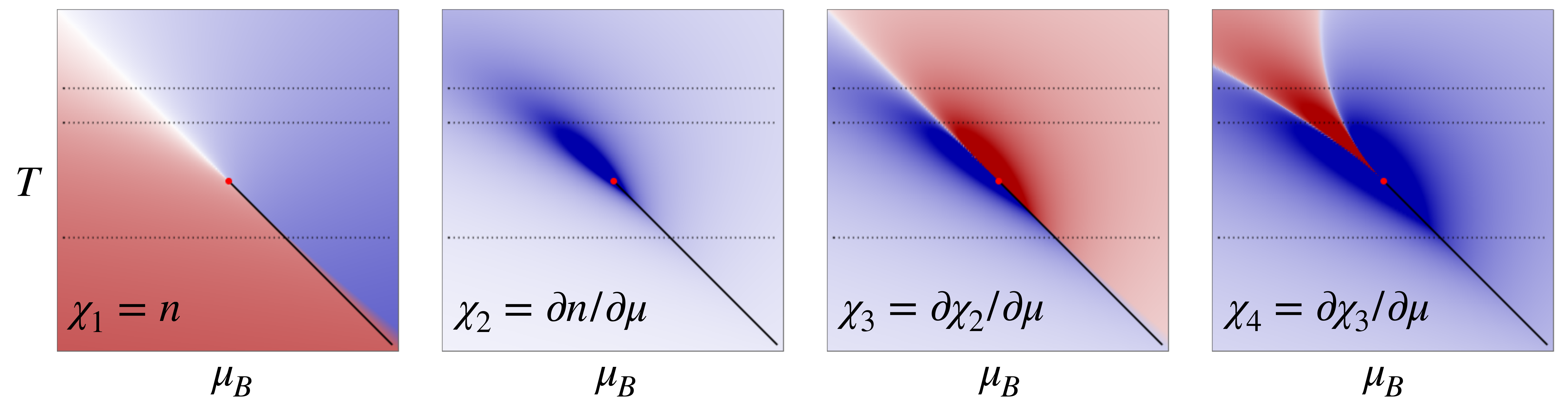} 
	}
	\caption[Density plots of the first four susceptibilities of pressure from a 3D Ising EOS]{Density plots of the first four susceptibilities of pressure, obtained using a universal 3D Ising EOS exhibiting a critical point, mapped onto coordinates of the QCD phase diagram, $T$ and $\mu_B$  \cite{Bzdak:2019pkr}. Red and blue denote regions of positive and negative susceptibilities, respectively, and the intensity of color corresponds to the magnitude of plotted values. The behavior of the susceptibilities in the vicinity of the critical point, along lines of constant temperature, follows a universal pattern. See text for more details.}
	\label{cumulants_bzdak}
\end{figure}

Such behavior of cumulants as a function of the order parameter (directly connected to the behavior of the corresponding susceptibilities) will be qualitatively the same for every theory with a critical point, leading to detailed expectations for their behavior \cite{Stephanov:1998dy,Stephanov:1999zu,Koch:2008ia}, see Fig.\ \ref{cumulants_bzdak}. In particular, one again sees that in the vicinity of the critical point $\kappa_3$ is expected to change its sign once \cite{Asakawa:2009aj}, while $\kappa_4$ is expected to change its sign twice \cite{Stephanov:2011pb}. The dependence of the magnitude of the cumulants on the location in the phase diagram, and in particular their divergent behavior near the critical point, can be expressed through their dependence on the correlation length, $\xi$. To the leading order in $\xi$ \cite{Stephanov:2008qz}, 
\begin{eqnarray}
\kappa_2 \propto \xi^2 ~, \hspace{5mm} \kappa_3 \propto \xi^{9/2} ~, \hspace{5mm} \kappa_4 \propto \xi^7 ~.
\label{cumulants_in_terms_of_the_correlation_length}
\end{eqnarray}
As the correlation length diverges at the critical point \cite{Zinn-Justin:2002ecy}, so do the values of the cumulants.

Crucially, cumulants of the net proton distribution can be measured in experiment. If either the net proton distribution measured in experiment can be considered as a good proxy for the net baryon distribution of systems of infinite extent considered in the theory, or the connection between the two distributions can be made (\cite{Kitazawa:2011wh} makes such connection in the case of non-interacting systems, while \cite{Vovchenko:2020gne} takes into account systems of finite size), then there exists a direct link between the thermodynamics driving the evolution of heavy-ion collisions and the experimental observables. In order to compare the values of the cumulants across different energies and centralities (characterized by different volumes of the created systems), one excludes the volume dependence by considering ratios of the cumulants,
\begin{eqnarray}
\frac{\kappa_2}{\kappa_1} ~, \hspace{5mm} \frac{\kappa_3}{\kappa_2}~, \hspace{5mm} \frac{\kappa_4}{\kappa_2} ~. 
\end{eqnarray}

\begin{figure}[t]
	\centering\mbox{
	\includegraphics[width=0.45\textwidth]{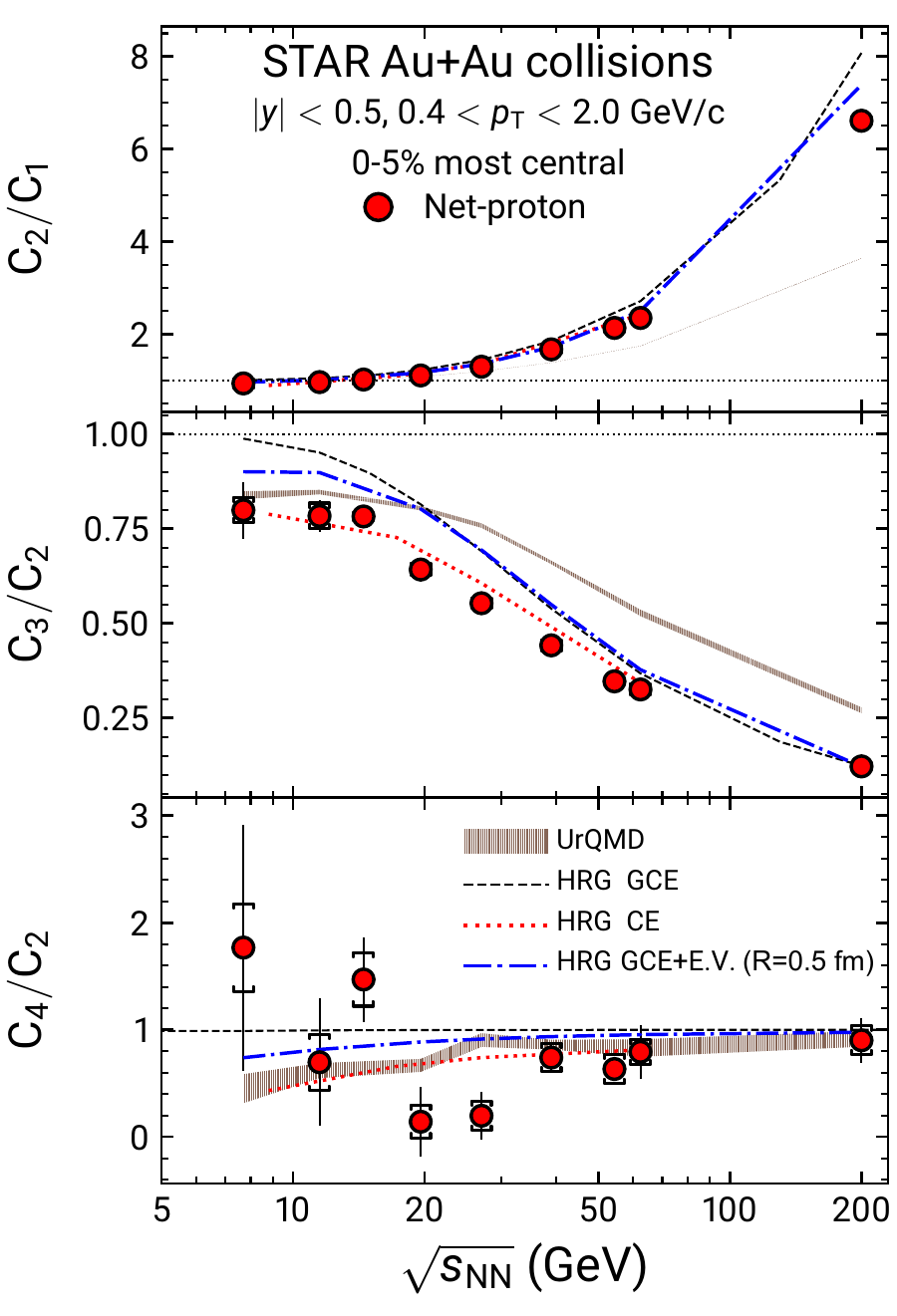} 
	}
	\caption[Ratios of cumulants of the net proton number for $|y| < 0.5$ in 0--5\% Au+Au collisions]{Ratios of cumulant of the net proton number for $|y| < 0.5$ in 0--5\% central Au+Au collisions \cite{STAR:2021iop}; also shown are values of the cumulant ratios obtained from \texttt{UrQMD} simulations and from HRG models within experimental acceptances. Note the different notation used for the cumulant ratios. See text for more details.}
	\label{STAR_cumulant_ratios}
\end{figure}

The results for measured cumulants of the net proton number \cite{STAR:2021iop} are shown in Fig.\ \ref{STAR_cumulant_ratios}. In the bottom panel, the fourth-order cumulant ratio $\infrac{\kappa_4}{\kappa_2}$ (denoted in the Figure as $C_4/C_2$) is seen to show hints of a non-monotonic behavior in the collision energy range $\sqrt{s_{NN}} \approx 5$--$20 \ \txt{GeV}$, where it also differs significantly from results obtained in \texttt{UrQMD} simulations \cite{Bass:1998ca,Bleicher:1999xi}, which do not include effects related to the QCD EOS and therefore can be used as a baseline expectation (various simulations of heavy-ion collisions will be further discussed in Section \ref{simulations_of_heavy-ion_collisions}). This suggests that collisions in this region may be probing the QCD phase diagram near the critical point. However, the limited statistical precision precludes making a definitive statement. Moreover, the second- and third-order cumulants (denoted in the Figure as $C_3/C_2$ and $C_2/C_1$, respectively) fail to simultaneously show behavior that would also correspond to probing the vicinity of a critical region; in fact, they are shown to behave fairly consistently with \texttt{UrQMD}.

Arguments can be made \cite{Stephanov:2008qz} that the fourth-order cumulant should be the most sensitive to critical fluctuations due to its higher order dependence on the correlation length, Eq.\ (\ref{cumulants_in_terms_of_the_correlation_length}), and that therefore it is more robust against dissipation. However, it can be also shown that the finite size and, more importantly, finite lifetime of the system \cite{Berdnikov:1999ph} affect the magnitudes of the cumulants, and furthermore, simulations show that the magnitudes of all cumulants should be affected in a similar way \cite{Nahrgang:2018afz}. In view of this, the fact that the experimental results for the cumulants do not provide a firm conclusion calls for more research with improved statistics.

\section{Challenges for finding the QCD critical point}
\label{Challenges_for_finding_the_QCD_critical_point}

In Sections \ref{softening_of_the_equation_of_state}, \ref{turning_off_the_QGP}, and \ref{non-statistical_event-by-event_fluctuations_of_conserved_charges} we described several experimental observables expected to behave in specific ways for systems evolving in the vicinity of the critical point, and thus to help uncover features of the QCD phase diagram at moderate-to-high temperatures and moderate-to-high baryon chemical potentials. Each of these signatures, however, carries with it significant theoretical uncertainties. Additionally, the dynamics of heavy-ion collisions, relatively simple at ultra-relativistic energies, becomes significantly more complicated at lower energies which are central to the BES program. Below, we list a few experimental and theoretical challenges for locating the critical point on the QCD phase diagram.

First, at lower energies the nuclei move at smaller velocities, resulting in a smaller Lorentz contraction. Consequently, each of the nuclei is characterized by an appreciable depth in the longitudinal (beam) direction, and the assumption that the colliding nuclei cross each other instantaneously (well-justified at high energies) doesn't apply. This means that the evolution of the initial stage of the collision is significantly different from that at high energies; in particular, the extended duration of energy deposition in the collision region likely significantly affects final-state observables.

Next, at low energies some of the baryon number characterizing the colliding nuclei is trapped in the collision zone and transported into the midrapidity region (a process known as ``baryon stopping'', already described in Section \ref{probing_the_QCD_phase_diagram} and Appendix \ref{baryon_transport}). Naturally, this is a necessary component of exploring the behavior of QCD matter at finite baryon densities, however, understanding the influence of baryon transport dynamics on the experimental observables, in particular baryon number fluctuations, will be required for their interpretation.

Further, matter created in a heavy-ion collision samples large ranges of both temperatures and baryon number densities, or alternatively baryon chemical potentials, during its evolution. This means that any single collision, rather than following a narrow path in the phase diagram, is affected by a significant area of the phase diagram (see, e.g., Refs.\ \cite{Shen:2020gef,Batyuk:2016qmb}). In particular, probing exclusively regions close to the critical point is in practice impossible, meaning that any signal coming from a possible QCD critical point will be contaminated by signals from adjacent regions of the phase diagram. 

Moreover, it is likely that a significant portion (if not all) of the collision evolution takes place out of equilibrium. The extent to which observables reflect an equilibrated state depends on the ratio of their equilibration time $\tau_{\txt{eq}}$ to the expansion time $\tau_{\txt{exp}}$. Crucially, while the correlation length diverges in the vicinity of the critical point, the time needed for critical fluctuations to equilibrate and reflect that fact, $\tau_{\txt{eq}}$, diverges as well, which is a phenomenon referred to as ``critical slowing down'' \cite{Hohenberg:1977ym,Berdnikov:1999ph}. Consequently, the magnitude of critical fluctuations in the vicinity of the critical point, instead of following the equilibrium dependence on the correlation length given by Eq.\ (\ref{cumulants_in_terms_of_the_correlation_length}), will also depend on the interplay of $\tau_{\txt{eq}}$ and $\tau_{\txt{exp}}$ \cite{Akamatsu:2018vjr,Asakawa:2019kek}. (Additionally, it should also be kept in mind that in a system of a finite size characterized by a length $L$, the correlation length can at most become comparable to $L$, but not larger.) On the other hand, while finite $\tau_{\txt{eq}}$ will likely suppress the magnitude of critical fluctuations (which won't have enough time to equilibrate to their diverging values), it will also work towards ``preserving'' the developed correlations as the system evolves away from the critical point towards hadronization and hadron rescattering. 

Most notably, while initial correlations in the colliding system are correlations in the position space, measurable observables are related to momentum space correlations. The details of transforming the former into the latter are not trivial at low energies (in contrast to high energies, where an approximate boost-invariance along the beam-direction near midrapidity can be assumed, and a mapping between the two types of correlations exists \cite{Bjorken:1982qr} and is shown to be valid \cite{Teaney:2000cw, Teaney:2001av}), and must be fully understood to interpret the data. This is especially important in view of the fact that long-distance correlations characterized by low relative momentum would be subject to a significant background from the HBT correlations.

Finally, the hadronic phase of the collision, occurring after the QGP hadronizes and lasting over an ever increasing fraction of the total collision time for decreasing beam energies, is likely to influence many of the observables. Among others, hadronic scatterings and decays will change the number of detected baryons and modify the magnitude of measured baryon number fluctuations \cite{Asakawa:2019kek,Steinheimer:2016cir}. These effects should be quantitatively understood in order to interpret the experimental data and draw conclusions about fluctuations in a hot and expanding QGP.

\section{Simulations of heavy-ion collisions}
\label{simulations_of_heavy-ion_collisions}

Experimental observables in heavy-ion collisions are likely influenced by an array of thermodynamic and dynamic factors, as outlined in the previous section, and therefore a definite statement on the sensitivity of given observables to probing the QCD phase diagram as well as on their expected quantitative behavior at different beam energies can only be made by utilizing reliable models and state-of-the-art simulations. In particular, such simulations can test the influence of different possible EOSs on the observables and identify EOSs most consistent with nature through comparisons against data. In this way, the search for the properties of the QCD phase transition is a natural extension of the search for the QGP, in which simulations were essential for interpreting the experimental results.

The multi-stage evolution of the nuclear matter fireball created in a heavy-ion collision is reflected in the many modules comprising modern heavy-ion collision simulations. Below, we give a brief overview of most sophisticated setups, often referred to as hybrid models or hybrid simulations.

A hybrid model of heavy-ion collisions begins with a simulation of the initial state, which primarily describes the initial energy density distribution in the collision zone, and may also model the initial evolution of the system towards thermalization. Popular approaches to the primary task of the initial state modeling include those based on the Glauber model \cite{Miller:2007ri}, the IP-Glasma (impact-parameter--dependent saturation model with glasma fields) model \cite{Schenke:2012wb, Schenke:2012hg}, or \texttt{T\raisebox{-.5ex}{R}ENTo} (Reduced Thickness Event-by-event Nuclear Topology) model \cite{Moreland:2014oya} (whose parameters can be adjusted to reproduce either the Glauber model or the IP-Glasma model), while simulations of the approach to equilibrium can be achieved by using a hadronic transport simulation of the collision as the initial state \cite{Karpenko:2015xea,Du:2018mpf}, through a 3D Monte-Carlo--Glauber model with string deceleration \cite{Shen:2017bsr}, or with a non-equilibrium linear response model \texttt{K{\o}MP{\o}ST} (named for the authors of the code) \cite{Kurkela:2018vqr}.

Such obtained distributions of initial density, energy density, and four-velocities present in the system are then used as the starting point for a relativistic hydrodynamics simulation. The applicability of hydrodynamics in simulations of heavy-ion collisions has been confirmed \cite{Teaney:2000cw, Teaney:2001av} through reproducing measurements of collective behavior of matter created in heavy-ion collisions \cite{STAR:2000ekf}. This in turn indicates that for a considerable fraction of a heavy-ion collision's evolution, it can be thought of as a locally thermal system described by an EOS. Hydrodynamic simulations explicitly depend on the EOS used to describe the system, and using different EOSs leads to different behavior of the simulated QGP, although these differences are by no means trivial (especially taking into the account other model parameters such as the value of viscosity) and the produced effects are rather subtle (for constraining the EOS in high-energy collisions through Bayesian analysis of hydrodynamic simulations, see Ref.\ \cite{Pratt:2015zsa}). State-of-the-art 3+1D relativistic viscous hydrodynamics simulations evolve the matter according to equations of non-ideal hydrodynamics; we note that at very high energies it is possible to use the boost invariance of the system and employ hydrodynamics with reduced dimensionality, for example 1+1D or 2+1D, however, a full spacetime evolution is necessary for realistic simulations at BES energies. Examples of hydrodynamic simulation codes include \texttt{MUSIC} \cite{Schenke:2010nt}, \texttt{v-USPhydro} (viscous Ultrarelativistic Smoothed Particle hydrodynamics) \cite{Noronha-Hostler:2013gga,Noronha-Hostler:2014dqa}, and \texttt{BEShydro} \cite{Du:2019obx}; notably, only \texttt{MUSIC} and \texttt{BEShydro} include effects due to diffusion of the net baryon current, important for the collision energy range studied in BES. (We also note that another 3+1D viscous hydrodynamics code with baryon diffusion has been recently developed \cite{Wu:2021fjf}.)

As the system expands, it becomes more dilute, and eventually it becomes too dilute for a description in terms of a hydrodynamically evolved bulk to be applicable; this can be thought of as the time of the collision when hadronization occurs. For different points in the system this takes place at a different time, and a suitable criterion for ending the hydrodynamic simulation must be used. Studies suggest \cite{Ahmad:2016ods} that evolving the system hydrodynamically until the energy density decreases to a chosen fixed value, usually in the range
\begin{eqnarray}  
\mathcal{E} \in  (0.2, 0.5)\ \txt{GeV}/\txt{fm}^3~,
\label{particllization_energy_density}
\end{eqnarray} 
yields results most consistent with data. Using Eq.\ (\ref{particllization_energy_density}) as a condition to stop the hydrodynamic evolution yields a hypersurface in the 4-dimensional spacetime characterized by varying values of the parameters of the propagated bulk. Because hydrodynamic simulations evolve systems discretized  on a 3-dimensional lattice, this hypersurface is not a continuous field, but rather a collection of cells, each characterized by the chosen fixed value of $\mathcal{E}$ and individual values of position $x^{\mu}$, temperature $T$, baryon chemical potential $\mu_B$, etc. (We stress that by construction, hydrodynamics evolves energy density $\mathcal{E}$ and baryon number density $n_B$; values of $(\mathcal{E},n_B)_i$ in the $i$-th cell of the hypersurface are used together with the EOS to obtain the corresponding values of $(T,\mu_B)_i$.) 

To obtain a description in terms of discrete particles, one performs particlization, in which the values of $T$ and $\mu_B$ in a given cell of the hypersurface are used to sample hadrons. Sampling requires choosing a probability distribution describing the expected yields of hadrons for a given $T$ and $\mu_B$ provided by the hydrodynamic simulation. Most often the distribution is an ideal relativistic Boltzmann distribution, but one can also use, for example, a relativistic Bose distribution to sample pions, whose quantum-mechanical character can become important at low temperatures (in this case the correction is on the order of 5--10\% already at $T \simeq 150\ \txt{MeV}$). Some particle samplers in addition take into account viscous corrections (see, e.g., the \texttt{iS3D} code \cite{McNelis:2019auj}), finite decay widths of resonances obtained through spectral functions \cite{Ryu:2019atv}, or local event-by-event conservation of energy, momentum, and charges \cite{Oliinychenko:2019zfk,Oliinychenko:2020cmr}.

The final stage of a heavy-ion collision simulation is devoted to propagating and scattering hadrons that emerge from the fireball until the system is so dilute that all interactions cease to take place. This is done through a hadronic transport code such as, among others, \texttt{UrQMD} (Ultra-relativistic Quantum Molecular Dynamics) \cite{Bass:1998ca,Bleicher:1999xi} or \texttt{SMASH} (Simulating Many Accelerated Strongly-interacting Hadrons) \cite{Weil:2016zrk}. (We note here that since version 3.4, \texttt{UrQMD} is a hybrid simulation with an intermediate hydrodynamic stage, activated at $\sqrt{s_{NN}} \geq 17.3\ \txt{GeV}$, based on the ideal hydrodynamics simulation code \texttt{SHASTA} \cite{Rischke:1995ir,Rischke:1995mt}.) Hadronic transport evolves the final stage hadrons from particlization onward by numerically solving the Boltzmann equation. In practice, this means propagating hadrons according to equations of motion consistent with the Boltzmann equation and allowing the hadrons to scatter and decay. The extensive list of evolved particle species, including their scattering cross sections and decay widths, is compiled based on the best currently known experimental results aided by phenomenological models when needed, with resonances accounted for through spectral functions. 

Notably, while hadronic transport is often used as part of a hybrid heavy-ion simulation framework, where it propagates hadrons from particlization to the final state as described above, transport codes are also capable of simulating the entire span of a heavy-ion collision, starting from the mutual approach of the two colliding nuclei and ending when the evolved particles are too far apart to interact. Because most hadronic transport codes do not include any effects due to the possible phase transition between hadrons and a QGP, these simulations are often used as a baseline for comparisons with experimental data, deviations from which could signal the presence of effects related to the creation of the QGP; see, e.g., Figs.\ \ref{STAR_dv1_dy}, \ref{STAR_triangular_flow}, and \ref{STAR_cumulant_ratios}.

Altogether, while there has been a significant progress in modern heavy-ion collision simulations over the recent years, there is still a lot to be done. At low energies, initial state models should include a description of all conserved charges relevant to heavy-ion collisions, namely the baryon, strangeness, and electric charge; this feature is already present in a recently developed model \texttt{ICCING} (Initial Conserved Charges in Nuclear Geometry) \cite{Martinez:2019rlp} which can be applied at high energies. Hydrodynamic evolution codes need to not only include transport of all conserved charges listed above (while this is a highly non-trivial task, initial theoretical work on transport coefficients in a fluid with multiple charges \cite{Rougemont:2017tlu,Fotakis:2021diq} as well as a study in a 1+1D kinetic theory for baryon and strangeness diffusion \cite{Fotakis:2019nbq} have been done), but should also include dynamics of correlations (notably, a framework including both the bulk hydrodynamic evolution and the non-equilibrium evolution of critical fluctuations has been recently presented in the form of \texttt{Hydro+} \cite{Stephanov:2017ghc,Rajagopal:2019xwg}, however, this approach hasn't been used yet to make a connection with experimental data). Particlization, so far predominantly using ideal grand-canonical probability distributions which do not conserve charge or energy on an event-by-event basis, should in fact conserve baryon number and energy exactly and locally (an approach that enables this, as well as the conservation of charge, strangeness, and momentum, has been recently developed in Refs.\ \cite{Oliinychenko:2019zfk,Oliinychenko:2020cmr}, but requires further work on, among others, implementing viscous corrections), as well as take fluctuations and interactions between the hadrons into account (a recent work has addressed the problem of particlization with critical fluctuations through an effective coupling between the particles and the critical field \cite{Pradeep2021}). Finally, hadronic transport codes need to include a more sophisticated description of hadronic interactions. Below, we describe this last problem in more detail, while Section \ref{overview_of_the_thesis} introduces the work presented in this thesis which aims at providing a solution.

\section{The need for generalized mean-field interactions in hadronic transport simulations}
\label{The_need_for_generalized_mean-field_potentials_in_hadronic_transport}

With a few exceptions (see, e.g., Ref.\ \cite{Nara:2016hbg}), state-of-the-art hadronic transport simulations, also known as ``afterburners'', typically neglect bulk hadronic interactions (in which case a transport simulation is often called a ``cascade''). This is motivated mainly by the fact that simulations including hadronic potentials are computationally expensive. Additionally, while nuclear potentials are well-known at low temperatures and moderate densities characterizing ordinary nuclear matter, they are not well-constrained in other parts of the QCD phase diagram. In consequence, the role of many-body interactions in the hadronic stage of a heavy-ion collision evolution and their influence on final state observables are largely unexplored. While this may be approximately correct for studies at high-energies, where the hadronic phase is relatively short, for low-energy collisions both the baryon densities (which are related to the strength of the mean-field interactions) and the fraction of the collision time that the system spends in a hadronic state are substantial. Because of this, hadronic interactions described by mean fields may have a significant influence on the system's evolution, including the diffusion dynamics affecting, among others, the propagation of fluctuations which may signal the existence of the critical point \cite{Asakawa:2019kek} (a study addressing this question utilizing \texttt{UrQMD} in the cascade mode was performed in Ref.\ \cite{Steinheimer:2016cir}). Therefore it is possible that in simulations of low-energy collisions, afterburners without potentials are missing important effects.

Moreover, understanding the behavior of QCD matter at finite temperature and baryon chemical potential, where it cannot at this time be known from first principles, depends on inferring the QCD EOS from systematic model comparisons with experimental data. In order to reach this goal, elements of hybrid heavy-ion collision simulations should consistently treat the interactions occurring in the evolved systems. In the case of afterburners, this means employing hadronic interactions that reproduce the properties (such as, e.g., the conjectured position of the QCD critical point) of a given EOS used in the hydrodynamic stage of the evolution. However, while continued theoretical efforts to model different variants of the QCD EOS with criticality, intended for use in hydrodynamic simulations, are being undertaken (see, e.g., Refs.\ \cite{Parotto:2020fwu,Stafford:2021wik}), hadronic afterburners are often only equipped with mean-field potentials modeling the behavior of the ordinary nuclear matter without the possible QGP phase transition. To address this issue, one needs a hadronic EOS that can be easily parametrized to reproduce given properties of the conjectured QCD phase transition, and that at the same time provides the corresponding relativistic single-particle dynamics in a form that is feasible to implement in a hadronic transport code.

\section{Overview of the thesis}
\label{overview_of_the_thesis}

The work presented in this thesis addresses the need for including generalized mean-field potentials in hadronic transport simulations, described in the previous section. 

In Chapter \ref{a_flexible_model_of_dense_nuclear_matter}, after a pedagogical introduction to modeling the EOS of nuclear matter and to the Landau Fermi-liquid theory, we develop a model in which the EOS of hadronic matter and the corresponding single-particle dynamics are both obtained from a relativistic density functional with fully parametrizable vector- and scalar-current interactions. The model we construct is Lorentz covariant and thermodynamically consistent, and is shown to obey conservation laws.

In Chapter \ref{parametrization_and_model_results}, with application to simulations of heavy-ion collisions in mind, we constrain the developed model to describe two phase transitions: the known nuclear liquid-gas phase transition with its experimentally observed properties as well as a postulated phase transition at high temperature and high baryon density, meant to model the QGP-hadron phase transition. We discuss the thermodynamic properties of the obtained family of EOSs based on several representative examples.

In Chapter \ref{implementation}, we discuss the implementation our model of mean-field hadronic interactions in the hadronic transport code \texttt{SMASH} \cite{Weil:2016zrk}, where we pay particular attention to details of the simulation such as baryon density and mean-field calculation algorithms.

In Chapter \ref{results}, we present results of hadronic transport simulations utilizing the developed model of the nuclear matter EOS. First, we confirm that \texttt{SMASH} simulations using the obtained single-particle equations of motion reproduce the thermodynamic behavior described by the underlying EOS. Next, we study systems initialized at various points of the nuclear matter phase diagram, including inside the spinodal region of the phase transition and in the vicinity of the critical point, and we investigate the collective behavior of simulated systems. We also note the effects of finite number statistics on obtained observables.

In Chapter \ref{the_speed_of_sound_in_heavy-ion_collisions}, we apply the developed EOS in a study addressing the possibility of using data from heavy-ion collision experiments to infer the magnitude and the behavior of the speed of sound in dense nuclear matter. To this end, we show a connection between the cumulants of the baryon number distribution and the speed of sound, and we test the applicability of our approach within two models of nuclear matter. We then apply our method to experimental data and use model calculations to interpret the results.

Finally, in Chapter \ref{thesis_summary}, we summarize the presented work.

\newpage
\chapter{Model of dense nuclear matter based on a relativistic density functional equation of state}
\label{a_flexible_model_of_dense_nuclear_matter}

The  goal of this chapter is to introduce a framework within which one can obtain a flexible equation of state (EOS) of dense nuclear matter with the corresponding single-particle equations of motion that can be feasibly implemented in a hadronic transport simulation. 

In Section \ref{nuclear_matter_quasiparticles_and_the_Landau_Fermi-liquid_theory}, we give a short overview of two popular methods of describing strongly interacting dense matter, the Walecka model and the Landau Fermi-liquid theory, with some emphasis on the underlying fundamental assumptions; a reader who is familiar with these two approaches can safely skip this part. In Section \ref{Relativistic_Landau_Fermi-liquid_theory_for_hadronic_transport}, we give a short motivation for the particular approach to describing nuclear matter developed in this thesis. In Section \ref{flexible_equation_of_state_for_nuclear_matter}, we present the theoretical framework that allows one to obtain a flexible, relativistically covariant, and thermodynamically consistent EOS, and we summarize in Section \ref{summary_of_VSDF_formulas}.

\section{Nuclear matter, quasiparticles, and the Landau Fermi-liquid theory}
\label{nuclear_matter_quasiparticles_and_the_Landau_Fermi-liquid_theory}

This section is based on works discussing various theoretical approaches to describing nuclear matter \cite{Day:1967zza, Walecka1986_lectures, Blaettel:1993uz} (see also Ref.\ \cite{Kapusta:2006pm}), as well as sources devoted to the Landau Fermi-liquid theory \cite{Abrikosov_Methods_of_QFT_in_SP, BaymPethickLandauFermiLiquidTheory, Brown:1971zza}.

\subsection{Microscopic models of nuclear matter}
\label{models_of_nuclear_matter}

Nuclear physics aims at understanding the structure of nuclei in terms of their elementary constituents (nucleons) and their mutual interactions. Due to difficulties in describing the nuclei from first principles (arising not only from the intrinsic non-perturbative nature of QCD at the energies of interest, but also predominantly due to the complexity of the many-body problem), one often uses some simplifying limit of the fundamental theory. A natural question that arises in this situation is what characteristics of nuclei should emerge within such a simplified approach. Here some guidance can be obtained from one of the most noteworthy descriptions of nuclei provided by the nuclear shell model (first developed, independently, by Maria Goeppert Mayer and Johannes Jensen in 1949 \cite{Mayer:1949pd,Haxel:1949fjd}), in which each nucleon moves in a central potential created by the remaining nucleons. Based on the success of this model, one can postulate that an approach satisfactorily reproducing the known properties of nuclei should be reducible to a description utilizing a set of single-particle states and effective interactions among nucleons described by those states.

Because calculations involving finite nuclei are a formidable task even when using effective approaches, one often considers a simplified scenario: nuclear matter, that is a uniform, infinite, isospin-symmetric (composed of an equal number of protons and neutrons) system in which the Coulomb interaction is neglected. In this limit, nuclear matter is well-described by an interacting Fermi gas at zero temperature. By definition, the system is translationally invariant and therefore must be composed out of the eigenstates of the momentum operator, which means that the single-particle states are plane waves. Having in this way sidestepped the need for calculating the wave-functions describing the states of the system (a task central to problems involving finite nuclei), the core quantity of interest in this approach is the energy per particle $E/A$ (where $E$ denotes the total energy and $A$ denotes the total number of nucleons) as a function of baryon (nucleon) density $n_B$.

\begin{figure}[t]
	\centering\mbox{
	\includegraphics[width=0.42\textwidth]{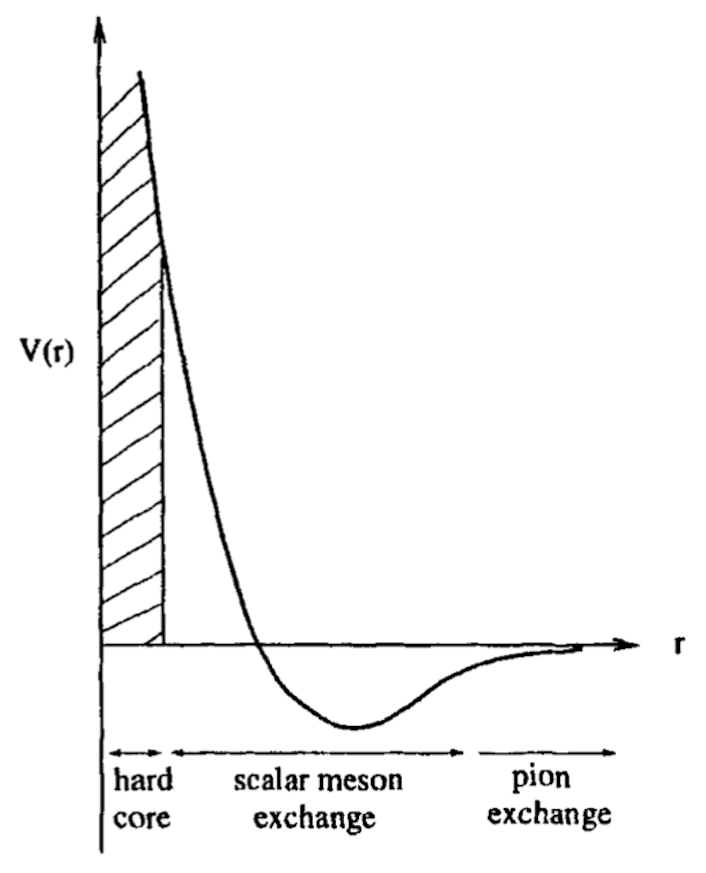} 
	}
	\caption[A sketch of the nucleon-nucleon potential]{A sketch of the nucleon-nucleon potential as a function of distance $r$ between two nucleons \cite{WongIntroductoryNuclearPhysics}. The hard core radius is around $0.4\ \txt{fm}$, the attraction at intermediate ranges, at radius $\approx 1\ \txt{fm}$, is believed to be dominated by the exchange of scalar mesons, and the long-range attraction, starting at around $2\ \txt{fm}$, is due to the single-pion exchange. The mediation of the interaction by mesons is discussed further in the text.}
	\label{nucleon-nucleon_potential}
\end{figure}

An expectation for the behavior of $E/A$ as a function of $n_B$ can be formed based on the known properties of nuclear matter. Scattering experiments and phenomenology \cite{WongIntroductoryNuclearPhysics} show that the volume of a nucleus is a linear function of $A$. This means that the compression of nucleons in a nucleus is approximately the same for all nuclei, and in particular it follows that the central densities of all nuclei are similar \cite{WongIntroductoryNuclearPhysics} (note that this is different than in the case of long range forces, such as for example gravity, which causes more massive stars to have denser cores). This implies that the range of the strong interaction is small, so that while adjacent nucleons are very strongly bound to each other, their effect on nucleons in a more distant part of the nucleus is much smaller. Additionally, it means that the nuclear forces become repulsive at very small separations, and as a consequence the nuclear binding forces are the largest (saturate) at some close distance $r_0$. Indeed, experiments show that the nucleon-nucleon interaction exhibits a strong attraction for an internucleonic distance of about $1.4\ \txt{fm}$, while for smaller distances the potential sharply rises and becomes strongly repulsive at about $0.5 \ \txt{fm}$, see Fig.\ \ref{nucleon-nucleon_potential}. 

This property of the nuclear force is often referred to as saturation of nuclear matter, and it means that there is a certain density of nucleons $n_0$ (corresponding  to some average distance $r_0$ between the nucleons) at which the energy of the system is minimized. If one then considers a uniform many-body system composed of nucleons, the energy per particle must exhibit a minimum occurring at the equilibrium density $n_0$ and characterized by some value $B_0$, $B = \frac{E(n_0)}{A} - m_N = B_0$ (here, we subtract the trivial contribution to the energy per particle due to the nucleon mass $m_N$; in result, $B$ is nothing else than the binding energy). The values of parameters $B_0$ and $n_0$ can be established based on the experimentally observed properties of nuclei, for example as encoded in the semi-empirical Bethe-Weizs\"{a}cker mass formula extrapolated to the case of infinite nuclear matter, and are approximately equal $B_0 \approx -16\ \txt{MeV}$ and $n_0 \approx 0.160\ \txt{fm}^{-3}$ \cite{Bethe:1971xm}.

In view of these universal properties of nuclei, any model claiming to describe nuclear matter must reproduce its saturation properties encoded in the values of $B_0$ and $n_0$. Interestingly, two different potentials that lead to the same scattering phase shifts may at the same time yield disparate results for the properties of nuclear matter; naturally, a potential implying unphysical properties of nuclear matter should be abandoned. For example, it is known that nuclear potentials based on even very sophisticated two-body forces, fit to scattering experiments, do not satisfactorily reproduce the saturation properties of nuclear matter, while at the same time, inclusion of somewhat simplistic three-body forces results in a much better agreement with data \cite{Bethe:1971xm}.

On the other hand, one can also reverse the problem and fit the parameters of a given potential such that the saturation properties of nuclear matter are reproduced. This is the approach employed in a model developed by John Walecka in 1974 \cite{Walecka:1974qa, Chin:1974sa}, known as the Walecka model, which is a relativistic model of nuclear matter with nucleons interacting through the exchange of scalar and vector mesons. (In other words, the Walecka model utilizes an effective field theory picture of nuclear interactions, first proposed by Hideki Yukawa in 1935 \cite{Yukawa:1935xg}, in which mesons are treated as interaction bosons. This approach is also used in numerous other models, such as, e.g., the Bonn potential \cite{Machleidt:1987hj} or the Argonne potential \cite{Wiringa:1994wb}. Naturally, neither baryons nor mesons are elementary particles and ultimately these interactions should be described by QCD; however, such an effective approach invoking QCD bound states lying close to the ground state (vacuum), in this case nucleons and low-mass mesons, should be correct at distances $\gtrsim 1\ \txt{fm}$  which do not probe the internal structure of baryons or mesons.) The choice of the omega meson, $\omega$, and the sigma meson, $\sigma$, as the sole carriers of the interaction can be argued based on the fact that the most significant meson exchanges in uniform isospin-symmetric nuclear matter are indeed those coming from $\omega$'s and $\sigma$'s, as measured by experiments on intermediate-energy nucleon-nucleon scattering which report a large Lorentz scalar and a large four-vector interaction \cite{Walecka:1985my,Clark:1982sa,McNeil:1983yh,Clark:1983wa}. (In uniform symmetric nuclear matter, the nuclear forces coming from a single-pion exchange average to zero unless parity is broken, and contributions from the $\rho$ meson vanish due to the isospin symmetry. Note that this is not true in more general applications, in particular because the attractive interaction between the nucleons is effectively realized through a single-pion and a correlated-pion exchange, and for example the Bonn potential has explicit contributions from both pions and $\rho$ mesons. Nevertheless, it is possible to show that the correlated-pion exchange, responsible for the short-range attraction, can be re-expressed as a sigma meson exchange \cite{Machleidt:1987hj,Durso:1980vn}, which is why for applications to isospin-symmetric nuclear matter it is possible to neglect explicit contributions from both $\pi$'s and $\rho$'s.)

The Walecka Lagrangian with interactions proceeding through the $\sigma$ and $\omega$ mesons can be written down as
\begin{eqnarray}
\mathcal{L}_W &=&  \bar{\psi}\big( i  \gamma^{\mu} \partial_{\mu} - m_N  \big) \psi + \frac{1}{2} \Big( \partial_{\mu} \sigma \partial^{\mu} \sigma - m_{\sigma}^2 \sigma^2   \Big) +  g_{\sigma} \bar{\psi} \psi \sigma \non \\
&& \hspace{10mm} -~  g_\omega \bar{\psi} \gamma_{\lambda}  \omega^{\lambda}  \psi   - \frac{1}{4} F_{\lambda \nu}F^{\lambda \nu} + \frac{1}{2} m_{\omega}^2 \omega_{\lambda} \omega^{\lambda} ~.
\label{Walecka_lagrangian}
\end{eqnarray}
Here, $\psi$ is a nucleon field of rest mass $m_N$, $\sigma$ is a neutral scalar meson field of rest mass $m_{\sigma}$, $\omega^{\lambda}$ is a neutral vector meson of rest mass $m_{\omega}$, and the field tensor $F_{\lambda \nu}$, capturing the dynamics of the vector meson, is given by
\begin{eqnarray}
F_{\lambda \nu} = \partial_{\lambda} \omega_{\nu} - \partial_{\nu} \omega_{\lambda}~.
\end{eqnarray}
It can be shown that in the static limit with heavy nucleons, the above Lagrangian corresponds to the following effective nucleon-nucleon potential (see Appendix \ref{static_limit_of_the_Walecka_model} for more details),
\begin{eqnarray}
V_{\txt{eff}} = \frac{g_{\omega}^2}{4\pi} \frac{e^{- m_{\omega}r}}{r} - \frac{g_{\sigma}^2}{4\pi} \frac{e^{- m_{\sigma} r}}{r} ~,
\end{eqnarray}
which mimics the long-range attraction and short-range repulsion characteristic of the nucleon-nucleon interaction.

It is clear from the Lagrangian in Eq.\ (\ref{Walecka_lagrangian}) that the neutral vector meson field couples to the baryon current $j_{\lambda} \equiv \bar{\psi} \gamma_{\lambda} \psi$ with a strength $g_{\omega}$, while the scalar meson field couples to the baryon scalar density $n_s \equiv \bar{\psi} \psi$ with a strength $g_{\sigma}$. Using Lagrange equations, one arrives at the field equations for nucleons and mesons,
\begin{eqnarray}
&& \Big[ i \gamma^{\mu} \partial_{\mu} -  m_N   \Big] \psi (x)  + \Big[ g_{\sigma} \sigma(x)  - g_{\omega} \gamma_{\lambda} \omega^{\lambda} (x)  \Big] \psi (x) = 0 ~,
\label{nucleon_eq} \\
&& \Big[ \partial_{\mu} \partial^{\mu} + m_{\sigma}^2  \Big] \sigma(x) = g_{\sigma} \bar{\psi} (x) \psi(x) ~, 
\label{sigma_eq} \\
&& \partial_{\nu} F^{\nu \mu} + m_{\omega}^2 \omega^{\mu} (x)= g_{\omega} \bar{\psi}(x) \gamma^{\mu} \psi (x)~,
\label{omega_eq} 
\end{eqnarray}
where we explicitly wrote the $x$-dependence of the fields. Importantly, the nucleon-meson couplings lead to the appearance of nucleons as sources in the meson field equations, and this means that a finite net baryon density leads to a non-zero expectation values of the meson fields.

To calculate properties of the system described by the Walecka Lagrangian, Eq.\ (\ref{Walecka_lagrangian}), the wave-function $\psi(x)$ must be found. Since the fields are strongly-interacting, the couplings $g_{\omega}$ and $g_{\sigma}$ are large, which precludes obtaining solutions using trivial perturbation theory. Several simplifying assumptions can be made to proceed. First, one can take the local density approximation, in which one neglects the derivative terms present in Eqs.\ (\ref{sigma_eq}) and (\ref{omega_eq}), leading to
\begin{eqnarray}
&& m_{\sigma}^2  \sigma = g_{\sigma} \bar{\psi} \psi ~, 
\label{heavy_sigma_eq} \\
&&  m_{\omega}^2 \omega^{\mu} = g_{\omega} \bar{\psi} \gamma^{\mu} \psi ~.
\label{heavy_omega_eq} 
\end{eqnarray}
Formally, the local density approximation is equivalent to the limit of $m_{\sigma} \to \infty, ~m_{\omega}\to \infty$, and physically it implies that the meson masses are significantly larger than any gradients present in the system; importantly, while this implies that the system is somewhat smooth, the density is still allowed to vary. Next, the non-zero expectation values of the meson fields, already mentioned above, suggest that the fields can be rewritten as sums of expectation values and fluctuations around those values,
\begin{eqnarray}
&& \sigma = \bar{\sigma} + \sigma ' ~,
\label{heavy_sigma_mean_plus_fluct} \\
&& \omega_{\mu} = \bar{\omega}_\mu + \omega'_{\mu} ~, 
\label{heavy_omega_mean_plus_fluct}
\end{eqnarray} 
where $\bar{\sigma}$ and $\bar{\omega}_\mu$ are field averages. Further, if the rest frame baryon density $n_B$ is large, then the source terms on the right-hand side of Eqs.\ (\ref{heavy_sigma_eq}) and (\ref{heavy_omega_eq}) are also large. In this case, it is a valid approximation to consider the meson fields as well-described by classical fields and replace their source terms by expectation values of the scalar baryon density and the baryon 4-current, $\langle \bar{\psi} \psi \rangle \equiv n_s$ and $\langle \bar{\psi} \gamma^{\mu} \psi \rangle \equiv j^{\mu}$ (note that this is exactly the way one treats classical electromagnetic fields). This approach, known as the mean-field approximation, is equivalent to neglecting the meson field fluctuations in Eqs.\ (\ref{heavy_sigma_mean_plus_fluct}) and (\ref{heavy_omega_mean_plus_fluct}), and leads to replacing Eqs.\ (\ref{heavy_sigma_eq}) and (\ref{heavy_omega_eq}) by their expectation values,
\begin{eqnarray}
&& m_{\sigma}^2   \bar{\sigma} = g_{\sigma} n_s ~, 
\label{heavy_sigma_final} \\
&& m_{\omega}^2 \bar{\omega}^{\mu} = g_{\omega} j^\mu ~.
\label{heavy_omega_final}
\end{eqnarray}
Physically, the mean-field approximation corresponds to a situation in which the propagation of the nucleon field, rather than being affected by the particular values of the meson fields at a given point in space, can be treated as a practically independent propagation in an average constant background of the meson fields. Computationally, this means that the meson-field operators are reduced to complex-valued fields.

Eqs.\ (\ref{heavy_sigma_final}) and (\ref{heavy_omega_final}) can be used to replace the meson fields in the nucleon field equation, Eq.\ (\ref{nucleon_eq}), by the baryon vector and scalar currents, 
\begin{eqnarray}
\Bigg[ \gamma^{\mu}\bigg( i \partial_{\mu} - \frac{g_{\omega}^2}{m_{\omega}^2}  j_{\mu} (x) \bigg) -  \bigg(m_N  - \frac{g_{\sigma}^2}{m_{\sigma}^2} n_s(x) \bigg)  \Bigg] \psi (x)   = 0 ~;
\label{nucleon_field_eq_2}
\end{eqnarray}
here, the terms have been rearranged in a way which makes it apparent that the coupling to the vector field is a ``minimal coupling'', appearing through a covariant derivative term (similarly as in electrodynamics), while the coupling to the scalar field results in an effective mass,
\begin{eqnarray}
m^* \equiv m_N  - \frac{ g_{\sigma}^2}{m_{\sigma}^2} n_s~.
\label{effective_mass}
\end{eqnarray}

In the case of equilibrated, infinite nuclear matter the system is uniform (translationally and rotationally invariant), so that $\bar{\omega}_{\mu} = \delta^{0}_{\mu} \bar{\omega}_0$ (as the rotational invariance of the system warrants that $\bar{\omega}_i =0$), which means that all currents in the system vanish, $\bm{j} = 0$. In that case, Eq.\ (\ref{nucleon_field_eq_2}) reduces to 
\begin{eqnarray}
\Bigg[ \gamma^{\mu}\bigg( i \partial_{\mu} - \delta_{\mu}^0 \frac{g_{\omega}^2}{m_{\omega}^2}  j_{0} \bigg) - m^* \Bigg] \psi (x)   = 0 ~.
\label{nucleon_field_eq_3}
\end{eqnarray}
Note that now the nucleon field equation, Eq.\ (\ref{nucleon_field_eq_3}), can be rewritten as $\gamma^{\mu}\partial_{\mu} \psi(x) = C \psi(x)$, from which it follows that $\psi(x)$ is a plane wave; as mentioned at the beginning of this section, one can expect this to be a reasonable result based on the wide applicability of the nuclear shell model. Looking for normal-mode solutions, $\psi(x) = U(p) e^{-i p^{\nu} x_{\nu}}$, further leads to
\begin{eqnarray}
\Big[  \bm{\gamma} \cdot \bm{ p}   + m^*  \Big] U(p)  = \gamma^{0}\bigg( p_{0} - \frac{g_{\omega}^2}{m_{\omega}^2}  j_{0} (x) \bigg) U(p)~,
\end{eqnarray}
and multiplying both sides by $\gamma^0$ and squaring yields the eigenvalue relation
\begin{eqnarray}
p^0 = \pm \sqrt{\bm{ p} ^2 + {m^*}^2} + \frac{g_{\omega}^2}{m_{\omega}^2}  j^{0} ~,
\label{Walecka_dispersion_relation_uniform}
\end{eqnarray}
in which the influence of the vector and scalar interactions on the nucleon energy are clear. Finally, the general plane wave solution for the nucleon field operator $\psi(x)$ can always be expanded in the basis of solutions to the Dirac equation,
\begin{eqnarray}
\psi (x) = \frac{1}{\sqrt{V}} \sum_{\bm{p}} \sum_{s} \bigg[ U(\bm{p}, s) a_{\bm{p}, s} e^{i \bm{p} \cdot \bm{x} } + V (-\bm{p}, s) b^{\dagger}_{\bm{p},s} e^{ - i \bm{p} \cdot \bm{x}}     \bigg]~,
\label{nucleon_field_expansion}
\end{eqnarray}
where the amplitudes for the normal modes, $a_{\bm{p}, s}$ and $b^{\dagger}_{\bm{p},s}$, are destruction and creation operators satisfying the anticommutation relations
\begin{eqnarray}
\big\{ c_{\bm{p},s} , c^{\dagger}_{\bm{p}', s'}  \big\} = \delta_{\bm{p}\bm{p}'} \delta_{ss'} ~, \hspace{10mm} c = \{a, b \} ~.
\end{eqnarray}

Importantly, it is possible to solve Eq.\ (\ref{nucleon_field_eq_2}) without making an assumption of uniform nuclear matter, that is for $\bm{j} \neq 0$. This is done using the relativistic Wigner function formalism \cite{Blaettel:1993uz}, and leads to the eigenvalues
\begin{eqnarray}
p^0 = \pm \sqrt{ \bigg(\bm{ p} - \frac{g_{\omega^2}}{m_{\omega}^2} \bm{j}(x) \bigg)^2 + {m^*}^2} + \frac{g_{\omega}^2}{m_{\omega}^2}  j^{0} ~.
\label{Walecka_dispersion_relation}
\end{eqnarray}
Here, it is apparent that the mean-field vector interaction results in the emergence of an effective nucleon momentum, $\Pi^{\mu} \equiv p^{\mu} - \frac{g_{\omega}^2}{m_{\omega}^2}  j^{\mu} (x) $, often called a kinetic momentum. (It's worth noting that while the canonical momentum $p^{\mu}$ is the momentum subject to momentum conservation, the kinetic momentum determines the motion of the nucleon.) The effects of vector and meson fields can be then summarized in terms of ``shifts'' of the energy (including the effective mass) and momentum of the nucleon.

From Eq.\ \eqref{Walecka_dispersion_relation} it is also evident that the effects due to the mean-field vector potential ``mimic'' the influence of the electromagnetic four-potential $(\phi, \bm{A})$. This is to be expected based on the ``minimal coupling'' of the mean-field vector field to the nucleon field, and we further elucidate this fact in Appendix \ref{Dirac_field_with_vector_and_scalar_interactions}, where in particular in Section \ref{the_Pauli_equation} we show that the mean-field Dirac equation, Eq.\ \eqref{nucleon_field_eq_2}, leads to a Pauli-like equation in the non-relativistic limit.

To obtain the energy-momentum tensor of the theory, one uses the Noether theorem \cite{Noether:1918zz},  
\begin{eqnarray}
T^{\mu\nu} \equiv \sum_a  \parr{\mathcal{L}}{ \left( \parr{\phi_a}{x^{\mu}} \right)  } \parr{\phi_a}{x_{\nu}}  - g^{\mu \nu} \mathcal{L} ~,
\end{eqnarray}
where the sum goes over the fields present in the Lagrangian; in particular, $T_W^{00} \equiv \mathcal{E}_W$, the energy density. Using the nucleon field expanded in terms of the mean-field Dirac equation solutions, Eq.\ (\ref{nucleon_field_expansion}), leads to summations over contributions from given momentum states and the corresponding probabilities of these states being occupied. In the language of thermodynamics, this is nothing else than sums weighted with a distribution function of the system, and thus one arrives at
\begin{eqnarray}
\mathcal{E}_W = g \int \pdens \sqrt{\Pi^2 + {m^*}^2} ~  f_{\bm{p}} + \frac{g_{\omega^2}}{m_{\omega}^2} (j^0)^2  + \frac{1}{2}\left( \frac{g_{\sigma}^2}{m_{\sigma}^2} \right) n_s^2 - \frac{1}{2}\left( \frac{g_{\omega}^2}{m_{\omega}^2} \right) j_\mu j^\mu~,
\label{Walecka_energy_density}
\end{eqnarray}
where $g$ is the degeneracy of the system (for isospin-symmetric nuclear matter, $g = g_{\txt{spin}} g_{\txt{iso}} = 4$) and $f_{\bm{p}}$ is the distribution function, and where the vector and scalar currents are given by
\begin{eqnarray}
j^\mu \equiv \langle  \bar{\psi} \gamma^{\mu} \psi \rangle = g \int \pdens \frac{\Pi^{\mu}}{\Pi^0}~ f_{\bm{p}}
\end{eqnarray}
and
\begin{eqnarray}
n_s  \equiv \langle  \bar{\psi} \psi \rangle = g \int \pdens \frac{m^*}{\sqrt{ \bm{\Pi}^2 + {m^*}^2}} ~ f_{\bm{p}} ~. 
\label{self-consistent}
\end{eqnarray}
Note that in view of Eq.\ (\ref{effective_mass}), Eq.\ (\ref{self-consistent}) is in fact a self-consistent equation for the scalar density, which has to be solved numerically for a given temperature $T$ and baryon density $n_B$ (or equivalently baryon chemical potential $\mu_B$).

In the case of uniform nuclear matter, the Walecka energy density becomes
\begin{eqnarray}
\mathcal{E}_W = g \int \pdens \sqrt{p^2 + {m^*}^2} ~  f_{\bm{p}} + \frac{1}{2}\left( \frac{g_{\omega}^2}{m_{\omega}^2} \right) n_B^2 + \frac{1}{2}\left( \frac{g_{\sigma}^2}{m_{\sigma}^2} \right) n_s^2 ~.
\end{eqnarray}
Importantly, both ratios of the coupling constants to the meson masses, $C_{\omega} = \left( \frac{g_{\omega}^2}{m_{\omega}^2} \right)$ and $C_{\sigma} = \left( \frac{g_{\sigma}^2}{m_{\sigma}^2} \right) $, are unknown constants (the meson masses are in fact known relatively well from the experiment, however, the coupling constants are not, making the ratios unconstrained). These constants are established by demanding that the binding energy per particle, $B = \frac{E_W}{A} - m_N = \frac{\mathcal{E}_W}{n_B} - m_N$, has a minimum value of $B_0$ at $n_B = n_0$. In this way, the mean-field Yukawa couplings are fitted to be used as effective couplings that capture, at least up to the first order, many-body effects not included in the original Lagrangian. 

Having fitted the coupling constants $C_{\omega}$ and $C_{\sigma}$, one can proceed to calculate other properties of nuclear matter. In particular, one of the early successes of the Walecka model was not only that it allowed for a correct description of the saturation properties of ordinary nuclear matter (a feature that was not attained in nonrelativistic models of nucleon-nucleon interactions), but also that it accurately identified the critical point of the nuclear liquid-gas phase transition (see Section \ref{the_QCD_phase_diagram} for more details).

\subsection{Quasiparticles}

Within the approach used in the Walecka model, outlined in the previous section, the starting point is a microscopic model. Employing the mean-field approximation is equivalent to assuming that the states of the system are single-particle wave functions, with dispersion relations ``shifted'' due to uniform background meson fields, Eq.\ (\ref{Walecka_dispersion_relation}). Based on this picture, one can obtain macroscopic observables, such as the energy density, which are then used to fit the coupling constants such that they reproduce macroscopic properties of nuclear matter. The use of plane waves (corresponding to freely propagating particles) with some of the free-particle properties ``shifted'' due to the interactions is equivalent to using ``quasiparticles'', a concept that we introduce in this section.

The notion of quasiparticles originates from the many-body theory, which aims at describing phenomena occurring in systems with many degrees of freedom. This includes, for example, the behavior of metals or the behavior of liquid helium. In such systems there are macroscopic numbers of elementary constituents, such as ions and electrons of a crystal lattice, or liquid $^3$He atoms. These constituents interact with each other in some very complex ways, which are not only challenging to capture on the level of particle-particle interactions, but would be absolutely impossible to handle when considering a macroscopic sample of the system. However, it can be argued that such detailed description of all interactions occurring in a given system is simply unnecessary for capturing the macroscopic properties of that system, which largely reflect average properties of the interactions in the system as a whole.

This average behavior of the system can be described by means of collective excitations. One of the most illustrative examples of collective excitations is provided by vibrations of a lattice of $N$ atoms. If the oscillations are assumed to be small, such lattice can be represented by a set of coupled harmonic oscillators. Through introducing normal coordinates one can arrive at a description in which the system is represented by $3N$ linear oscillators, each described by a given eigenfrequency $\omega_i$. From quantum mechanics, the energy of such a system is then given by the formula 
\begin{eqnarray}
E = \sum_{i=1}^{3N} \Big(n_i + \frac{1}{2} \Big) \omega_i~,
\end{eqnarray}
and taking different sets of values of $n_i$ leads to constructing the energy spectrum of the system. In the case where the potential energy of the lattice takes into the account anharmonic terms (which become more important with the amplitude of the vibrations, that is with increasing temperature), the system will start to exhibit non-zero probability of transitioning between states with different values of the $n_i.$

On the other hand, lattice vibrations can be always decomposed into a sum of monochromatic waves propagating in the crystal. Based on the correspondence principle of quantum mechanics, a plane wave can be associated with a moving ``particle'': the wave vector $\bm{k}$ is directly related to the momentum of such a particle, and the frequency $\omega(\bm{k})$ is related to its energy. Any excited state of the lattice can be described as a set of such ``particles'' moving throughout the volume of the considered solid, and so the energy levels of the system will be very similar to energy states of an ideal gas of these ``particles'', called ``phonons'' (a name first introduced by Igor Tamm in 1932, in an analogy to ``photon'' and based on a Greek word meaning ``sound''). In consequence, one can reinterpret the number $n_i$ as the number of phonons in the state $i$. Because $n_i$ can take any values, evidently phonons are subject to Bose-Einstein statistics (this holds also in the case when the lattice atoms have half-odd-integer spin). The number of phonons increases with temperature, which leads to an increased role of phonon-phonon interactions (in analogy to an increasing importance of anharmonic terms). In particular, one can reinterpret the finite probability of the system going from one energy state to another, characterized by different values of $\{n_i\}$, in terms of different interactions between the phonons, such as scattering, creation, or decay.

Based on the above, one can see that the excitations of a lattice, energy levels of which are a result of an incomprehensible number of interactions between a macroscopic number of particles, can to a good approximation be constructed out of energy levels of free ``particles'', which are entirely independent for sufficiently low temperatures and are weakly-interacting for higher temperatures. These ``particles'' are often called ``collective excitations'' or ``quasiparticles'' (the former is often used when the ``particles'' are governed by the Bose-Einstein statistics, while the latter when they are governed by the Fermi-Dirac statistics, although this is by no means a strict rule, and we will not always follow it). Intuitively, quasiparticles can be understood as emergent phenomena occurring when a microscopically complex system of ``real'' particles can be described as if it was made of different, weakly interacting ``quasiparticles'' in free space, which are characterized by a particular value of the momentum $\bm{p}$ and the corresponding dispersion relation $\eps(\bm{p})$. A widely known example of a system that can be described using this concept is a semiconductor, in which the behavior of an electron interacting with the ion lattice can be described as a motion of a free electron with a different, ``effective'' mass.

One needs to remember that the collective excitations are born out of the collective behavior of the studied system, and as such do not exist ``on their own''; rather, their properties are determined by the state of the system. Importantly, the number of quasiparticles can, but need not be equal to the total number of ``real'' particles considered. Furthermore, because of a finite probability of a transition from one state of the system to another, each described by a different set of quasiparticles, the quasiparticles are characterized by a finite decay width. Finally, it should be stressed that the concept of collective excitations only makes sense for length scales larger then some fundamental scale of the problem such as the lattice spacing; for example, if one resolves the individual microscopic motions of the lattice ions, the description in terms of phonons becomes meaningless.

In view of the above, it is clear that the nucleon fields described in the mean-field Walecka model are in fact quasiparticles. We can think of them as nucleons ``dressed'' by their interactions with mesons, which results in a ``quasi-nucleon'' field characterized by an effective, density- and temperature-dependent mass, and an effective (kinetic) momentum, both reflecting the fact that each nucleon is interacting with baryon currents created by all other nucleons in the system. The next section is devoted to introducing a quasiparticle theory for spin-$\frac{1}{2}$ particles which indeed can be used to reproduce Walecka model results.

\subsection{Landau Fermi-liquid theory}
\label{Landau_Fermi-liquid_theory}

A system of non-interacting particles of spin-$\frac{1}{2}$ (which are therefore governed by the Fermi-Dirac statistics) is called a Fermi gas. A system of interacting spin-$\frac{1}{2}$ particles at low temperatures is called a Fermi liquid. Examples of such systems include liquid $^3$He, electrons in metals, or nuclear matter. One of the most fruitful phenomenological theories of Fermi liquids was developed by Lev Landau between 1956 and 1959 \cite{Landau:1956_1,Landau:1957_2,Landau:1959_3}, and is nowadays known as the Landau Fermi-liquid theory. 

Landau's theory is based on an assumption that the excitation spectrum of a Fermi liquid has a structure similar to that of an ideal Fermi gas. This is a strong assumption on the dominating type of interactions occurring between the particles of the liquid which must preserve their quantum statistics (an example of a system in which the opposite is the case is deuterium atoms, in which nucleons, governed by the Fermi-Dirac statistics, interact in a way leading to the formation of molecules governed by the Bose-Einstein statistics). Furthermore, the theory assumes that the number of particles in the liquid and the number of quasiparticles are equal, and the Fermi momentum of the system is related to the particle density of the liquid, $n$, in a way completely mirroring the ideal gas case,
\begin{eqnarray}
p_F = \left( \frac{6 \pi^2 n}{g} \right)^{1/3}~,
\label{Fermi_momentum}
\end{eqnarray}
where $g$ is the degeneracy (the above equation follows from the fact that the particle density $n$ is given by the momentum integral of the distribution function $f_{\bm{p}}$, $n = g \int \pdens f_{\bm{p}}$, which for the ideal Fermi gas at $T=0$, where the system occupies all momentum states from 0 to $p_F$ and no momentum states above $p_F$, reduces to $n_B = \frac{g}{2\pi^2} \int_0^{p_F} dp ~  p^2$).

The bulk behavior of a system of quasiparticles is assumed to be described by a quasiparticle distribution function, constructed based on the one-to-one correspondence between quasiparticles and ``real'' particles. This correspondence can be introduced formally as follows: One begins by considering an ideal (nonrelativistic) Fermi gas, in which the dispersion relation (neglecting spin) is
\begin{eqnarray}
\eps_{\bm{p}}^{\txt{free}} = \frac{\bm{p}^2 }{2m}~.
\label{dispersion_relation_free_gas}
\end{eqnarray}
The state of the system as whole can be specified by giving the number of particles $N_{\bm{p}} = \{0, 1\}$ in each of the single-particle states defined by a specific value of the momentum $\bm{p}$. Thus for example in the ground state, each of the states with momenta less than the Fermi momentum, $p_F$, is occupied ($N_{\bm{p}} = 1$), and all other states are empty ($N_{\bm{p}} = 0$). One can then imagine that interactions in the system are slowly turned on in such a way that the process is adiabatic. Quantum mechanics shows that while such an adiabatic change will lead to a distortion of the energy levels, it will preserve their number. This means that the distribution function $N_{\bm{p}}$, while also smoothly distorted, preserves its functional form. Now, however, the dispersion relation $\eps_{\bm{p}}^{\txt{int}} $ takes interactions into the account, and so it differs from that of a free particle, Eq.\ (\ref{dispersion_relation_free_gas}). (Note that the dependence of the construction of the quasiparticle distribution function on the one-to-one correspondence between quasiparticles and ``real'' particles means that the formalism is not appropriate for describing phenomena in which the number of particles in the system changes as a result of the interactions, such as formation or dissolution of bound states.) For describing macroscopic properties of a Fermi liquid, it is sufficient to use a mean or smoothed quasiparticle distribution function, often denoted by $f_{\bm{p}}$, which is an average of $N_{\bm{p}}$ over a group of neighboring single-particle states. While $N_{\bm{p}}$ is a discrete function of $\bm{p}$, $f_{\bm{p}}$ and is a continuous function of $\bm{p}$.  

It follows from the above that the quasiparticle distribution function $f_{\bm{p}}$ has the same functional form as the Fermi-Dirac distribution function; a more formal derivation of this fact can be found in Appendix \ref{form_of_the_quasiparticle_distribution_function}. Importantly, the Fermi distribution function satisfies the constraint that the quasiparticles are subject to the Pauli exclusion principle, which is an essential ingredient of the theory. At low temperatures ($T \ll p_F$), all states deep in the Fermi sphere are occupied, so that available states only appear in the vicinity of the Fermi surface. Excitation from one of the states from deep within the Fermi sphere to one of the available states is then strongly suppressed by a factor of $1/\Delta E$, where $\Delta E$ is the energy difference between the two considered states. In consequence, only the states in the vicinity of the Fermi surface can actively contribute to the dynamics of the system, while the rest of the states constitute a necessary (from the point of view of Pauli blocking) but kinematically suppressed ``background''.

As the next step in developing the theory, Landau assumed that the interaction between the quasiparticles takes the form of a self-consistent field, whose effect on any given quasiparticle is considered to be produced by the remaining quasiparticles. He then proceeded to construct a theory of quasiparticles in which energy and momentum were conserved. To this end, he utilized the Boltzmann equation for quasiparticles,
\begin{eqnarray}
\frac{df_{\bm{p}}}{dt} = \parr{f_{\bm{p}}}{t} + \frac{d \bm{x}}{dt} \parr{f_{\bm{p}}}{\bm{x}} + \frac{d \bm{p}}{dt} \parr{f_{\bm{p}}}{\bm{p}} = \left(\derr{f_{\bm{p}}}{t}\right)_{\txt{coll}}~,
\label{Boltzmann_eq_LFLT}
\end{eqnarray}
where $f_{\bm{p}}$ denotes the quasiparticle distribution function and $\big(\inderr{f_{\bm{p}}}{t}\big)_{\txt{coll}}$ is the change in the distribution function due to collisions, often referred to as the ``collision term'' or the ``collision integral'' (a more thorough discussion of the Boltzmann equation can be found in Chapter \ref{implementation} and Appendix \ref{basics_of_the_kinetic_theory}). From Hamilton's equations we know that $d \bm{x}/dt \equiv \partial H_{(1)}/\partial \bm{p}$ and $d \bm{p}/dt = - \partial H_{(1)}/\partial \bm{x}$, where $H_{(1)}$ is the single-particle Hamiltonian. Equating $H_{(1)}$ with the energy of a quasiparticle $\eps_{\bm{p}}$ yields
\begin{eqnarray}
\parr{f_{\bm{p}}}{t} - \parr{ \eps_{\bm{p}}}{p_i} \parr{f_{\bm{p}}}{x^i} + \parr{ \eps_{\bm{p}}}{ x_i } \parr{f_{\bm{p}}}{p^i} =  \left(\derr{f_{\bm{p}}}{t}\right)_{\txt{coll}} ~,
\label{Boltzmann_quasiparticle_energy}
\end{eqnarray}
where we utilize the Einstein summation convention as well as co- and contravariant vector notation (as summarized in Appendix \ref{units_and_notation}), the latter of which leads to a change in signs as compared with Eq.\ \eqref{Boltzmann_eq_LFLT}. 

To ensure momentum conservation, Landau wanted to obtain an expression for the continuity of the momentum density (leading to the conservation of the quasiparticle momentum by assuming that the momentum density flux is zero on the surface of the integrated volume), which formally can be written as
\begin{eqnarray}
\parr{}{t} ~g \int \pdens p^i ~ f_{\bm{p}} + \parr{}{x^k} ~\Pi^{ik}  = 0~,
\label{Landau_momentum_flux_0}
\end{eqnarray}
where $\Pi^{ik}$ is the momentum density flux tensor of an as of yet unspecified form. To arrive at the momentum conservation law through the Boltzmann equation, Eq.\ (\ref{Boltzmann_quasiparticle_energy}), one multiplies both sides of the equation by $p^k$ and integrates over all possible momenta, 
\begin{eqnarray}
\hspace{-5mm}g \int \pdens p^k  ~\parr{f_{\bm{p}}}{t} - g  \int \pdens p^k \parr{ \eps_{\bm{p}}}{ p_i} ~\parr{f_{\bm{p}}}{x^i} +  g \int \pdens p^k \parr{ \eps_{\bm{p}}}{ x_i } ~\parr{f_{\bm{p}}}{p^i}  = 0~;
\label{Landau_momentum_flux_1}
\end{eqnarray}
here, the right hand side vanishes as the integral $g \int \pdens p^k ~ \big(\derr{f_{\bm{p}}}{t}\big)_{\txt{coll}}$ is equal zero due to the conservation of momentum in collisions (see Appendix \ref{vanishing_of_the_collision_integral} for more details). Using the product rule for derivatives and the fact that the distribution function $f_{\bm{p}}$ vanishes at the boundary, $f_{\bm{p}}\big(|\bm{p}|=+\infty\big) = 0$, it is possible to rewrite Eq.\ (\ref{Landau_momentum_flux_1}) as (a similar calculation is performed in detail in Appendix \ref{the_energy-momentum_tensor})
\begin{eqnarray}
&& g \int \pdens p^k ~ \parr{f_{\bm{p}}}{t} - \parr{}{x^i} ~ g \int \pdens p^k \parr{\eps_{\bm{p}}}{p_i}  ~ f_{\bm{p}}   \non \\ 
&& \hspace{10mm} - ~  \parr{}{x_i} g^{k}_{~i} ~ g \int \pdens  \eps_{\bm{p}} ~f_{\bm{p}} + g^{k}_{~i} ~ g \int \pdens \eps_{\bm{p}} ~ \parr{f_{\bm{p}}}{x_i} = 0 ~,
\label{Landau_momentum_flux_2}
\end{eqnarray}
where we have used $\parr{p^k }{p^i} = \delta^k_i = g^{kj} g_{ji} =  g^{k}_{~i} $. The first three terms in the above equation are already of the from required by Eq.\ (\ref{Landau_momentum_flux_0}), however, the same cannot be achieved trivially for the last term. The only way to write that term as a divergence of some quantity is to assume that the integrand is a functional differential of some functional $\mathcal{E}$, that is to assume that
\begin{eqnarray}
\delta \mathcal{E}[f_{\bm{p}}] = g \int \pdens  \eps_{\bm{p}} ~ \delta f_{\bm{p}} ~. 
\label{quasiparticle_energy_definition_LFLT}
\end{eqnarray}
If this is the case, then one achieves the goal of bringing Eq.\ (\ref{Landau_momentum_flux_2}) to the form of Eq.\ (\ref{Landau_momentum_flux_0}) (which is equivalent to establishing that momentum density is conserved), with
\begin{eqnarray}
\Pi^{ik} = -g\int \pdens p^i \frac{d \eps_{\bm{p}}}{d p_k} ~f_{\bm{p}}  +  g^{ik}~\left[ \mathcal{E} - g  \int \pdens \eps_{\bm{p}}~ f_{\bm{p}}   \right] ~.
\end{eqnarray}

Looking at Eq.\ (\ref{quasiparticle_energy_definition_LFLT}), it is quite natural to assume that $\mathcal{E}$ is the energy density of the system. Then the quasiparticle energy $\eps_{\bm{p}}$, which based on the above derivation is a functional derivative of the energy density,
\begin{eqnarray}
\eps_{\bm{p}} \equiv \frac{\delta \mathcal{E}}{\delta f_{\bm{p}}} ~, 
\label{Landau_quasiparticle_energy_definition}
\end{eqnarray}
can be interpreted as the change in the energy density of the system due to an addition of a quasiparticle with a momentum $\bm{p}$. So defined, the quasiparticle energy naturally contains information about the interactions in the liquid. Consequently, the energy of a quasiparticle includes potential energy terms arising due to the interactions with all other quasiparticles, and from this it follows that the energy of the system as a whole is not equal to the sum of the energies of individual quasiparticles. (Note that this is similar to the case of a classical electrostatic system, where in the calculation of the total electrostatic potential energy $U$ one needs to avoid summing over the same particle pair twice, $U = \sum_{i=0}^N \sum_{j >i} U_{ij} = \frac{1}{2} \sum_{i=0}^N \sum_{j \neq i} U_{ij} $, where $U_{ij} = (4\pi \eps_0)^{-1} q_i q_j /r_{ij} $ is the potential energy of the particle $i$ due to the presence of the particle $j$, with $q_i$ and  $q_j$ denoting charges of the $i$-th and the $j$-th particle and $r_{ij}$ denoting the distance between these particles. Defining the overall potential energy of the $i$-th particle as $U_i = \sum_{j \neq i} U_{ij} $, the potential energy of the system can be rewritten as $U = \frac{1}{2} \sum_i U_i $. Then the total energy of the system can be expressed as $E = \sum_{i=0}^N \big( K_i + U_i \big) - \frac{1}{2} \sum_i U_{i}$, where $K_i$ is the kinetic energy contribution from the $i$-th particle and $K_i  + U_i = \eps_i$, the total energy of that particle. The term $-\frac{1}{2} \sum_i U_i$ is known as the ``counterterm'', and its purpose is to cancel out the double-counting of potential energy terms that occurs in the sum over the energies of all particles. In a complete analogy, the energy density of a system of quasiparticles always has the form $\mathcal{E} = g \int \pdens \eps_{\bm{p}} + \mathcal{E}_{\txt{counter}}$, where $\mathcal{E}_{\txt{counter}}$ is the counterterm.)

With Eq.\ (\ref{Landau_quasiparticle_energy_definition}) established, it is straightforward to show that the developed Landau Fermi-liquid theory conserves not only the quasiparticle momentum, but also the number of quasiparticles and the quasiparticle energy. The conservation of the number of quasiparticles is obtained by integrating both sides of Eq.\ \eqref{Boltzmann_quasiparticle_energy} over all momenta; here, the right-hand side of the obtained expression vanishes due to the conservation of the number of particles in collisions (see Appendix \ref{vanishing_of_the_collision_integral} for more details), and after some algebra we get
\begin{eqnarray}
\parr{}{t} \int \pdens f - \parr{}{x^i} \int \pdens \parr{\eps_{\bm{p}}}{p_i} ~ f = 0 ~.
\end{eqnarray}
Using $-\inparr{\eps_{\bm{p}}}{p_i} = \inparr{\eps_{\bm{p}}}{p^i} \equiv v^i$, where $v^i$ is the velocity of a quasiparticle, we immediately obtain
\begin{eqnarray}
\parr{}{t} \int \pdens f + \parr{}{x^i} \int \pdens v^i ~ f = 0 ~,
\label{conservation_of_number}
\end{eqnarray}
where identifying $\int \pdens f $ and $\int \pdens v^i ~ f $ with the number density $n$ and the number current $j^i$, respectively, reveals Eq.\ \eqref{conservation_of_number} as the continuity equation for the quasiparticle number density. Similarly, the law of the conservation of the quasiparticle energy is obtained by multiplying both sides of Eq.\ \eqref{Boltzmann_quasiparticle_energy} by $\eps_{\bm{p}}$ and integrating over all momenta. As before, the right-hand side of the obtained expression vanishes (see Appendix \ref{vanishing_of_the_collision_integral} for more details), leading to
\begin{eqnarray}
\int \pdens \eps_{\bm{p}} ~ \parr{f}{t} - \parr{}{x^i} \int \pdens \eps_{\bm{p}} \parr{\eps_{\bm{p}}}{p_i} ~ f = 0 ~. 
\end{eqnarray}
Using Eq.\ \eqref{quasiparticle_energy_definition_LFLT} to rewrite the first term and the definition of the quasiparticle velocity $v^i$, the above equation becomes
\begin{eqnarray}
\parr{\mathcal{E}}{t}  + \parr{}{x^i} \int \pdens \eps_{\bm{p}} v^i ~ f = 0 ~,
\end{eqnarray}
where identifying $\int \pdens \eps_{\bm{p}} v^i ~ f $ as the quasiparticle energy density flux yields the continuity equation for the quasiparticle energy density.

Let us further discuss two features of the Landau theory. 

First, because the system is interacting, the concept of quasiparticles only has a well-defined meaning for energies close to the Fermi energy $\eps_{F}$, or equivalently momenta close to the Fermi momentum $p_F$. This is because for excitations far away from the Fermi surface, processes such as scattering into other states become kinematically allowed. In the ``real'' system, these correspond to a finite probability for the system to transition from one state to another, leading to damping (finite decay width) of the excitation that the quasiparticles are to represent. This means that the description in terms of quasiparticles is strictly valid only for temperatures which are low in comparison with the Fermi momentum. (Note, however, a recent work which shows that the expectation of broadening and disappearance of quasiparticle states at high energies, correct for weak interactions, is invalid when strong interactions are considered \cite{Verresen_2019}.)

Second, one notes that the above derivation assumes a uniform distribution of quasiparticles in space. Such an assumption can be considered valid as long as any spatial non-uniformity occurs at distances not probed by quasiparticles, that is distances that exceed the quasiparticle wavelength. At the Fermi surface, the momenta are close to the Fermi momentum $p_F = \big[ \inbfrac{6 \pi^2 N}{g V} \big]^{1/3}$, and consequently the quasiparticle wavelengths are of order $\left( \infrac{V}{N}\right)^{1/3}$, which is the average inter-particle spacing. It follows that the system only needs to be approximately uniform on a length scale of the average distance between the particles, which in practice doesn't impose any restrictions for systems at low temperatures.

How does one use the Landau-Fermi liquid theory in practice? First, one postulates the energy density of the system, and in particular its dependence on the distribution function, $\mathcal{E} = \mathcal{E}[f_{\bm{p}}]$. (Note that this is very much different than the strategy used in microscopic approaches, which start from microscopic interactions and then endeavor to calculate macroscopic observables such as the energy density or pressure.) Next, one calculates the quasiparticle energy using Eq.\ (\ref{Landau_quasiparticle_energy_definition}). Then one can also calculate the second functional derivative of the energy density,
\begin{eqnarray}
\phi \equiv \frac{\delta \mathcal{E}}{\delta f_{\bm{p}'} \delta f_{\bm{p}}}  ~,
\end{eqnarray}
which can be interpreted as a quasiparticle interaction energy and generally depends on the two quasiparticles' momenta, $\bm{p}$ and $\bm{p'}$, as well as their spins, $s$ and $s'$. By neglecting the spin dependence and assuming that both quasiparticles are close to the Fermi surface, $|\bm{p}| \approx |\bm{p}'| \approx p_F$, one has $\phi = \phi (p_F, \theta)$, where $\theta = \measuredangle(\bm{p}, \bm{p}')$. It is then possible to expand $\phi$ in spherical harmonics, and coefficients of this expansion are known as the Landau coefficients; often, it is enough to know the two first coefficients of the expansion, $F_0$ and $F_1$. Many properties of matter, such as the quasiparticle velocity at the Fermi surface or the zero sound, can then be calculated and expressed in terms of the Landau coefficients. One can utilize experimental measurements of some of these quantities to obtain phenomenological values of $F_0$ and $F_1$, and then use these to predict the values of other experimental observables. In general, this has been a very successful approach in describing phenomena occurring in Fermi liquids, including the behavior of liquid $^3$He \cite{Wheatley1970}, electrons in metals \cite{Platzman1973}, and nuclear matter \cite{Migdal_theory_of_finite_fermi_systems}.

In a 1975 paper, Gordon Baym and Siu Chin \cite{Baym:1975va} generalized the Landau Fermi-liquid theory to include relativistic effects. Soon after, the formalism was used by Tetsuo Matsui \cite{Matsui:1981ag} as an alternative way of obtaining some of the results predicted by the Walecka model, such as the incompressibility or the speed of sound of nuclear matter. This further underscored the equivalency between the Landau Fermi-liquid theory and the Walecka model field theory in the mean-field approximation. In particular, both of these approaches are appropriate in sufficiently dense regimes, where excitations are low-energetic in comparison with the Fermi surface and the collective effects due to a self-consistent background are more important than contributions from fluctuations in the fields. For nuclear matter, the Fermi momentum at saturation density is $p_F(n_0) \approx 260\ \txt{MeV}$, resulting in a relatively substantial range of temperatures in which the description in terms of quasiparticles is appropriate.

\section{Relativistic Landau Fermi-liquid theory for hadronic transport}
\label{Relativistic_Landau_Fermi-liquid_theory_for_hadronic_transport}

In the previous section, we reviewed two phenomenological methods of describing nuclear matter: First, we introduced the Walecka model as an example of a Lagrangian-based, self-consistent approach at the mean-field level. Then we introduced the Landau Fermi-liquid theory, in which the relevant physics is entirely encoded in the postulated energy density and its variations with respect to the distribution function governing the system. Both methods utilize the concept of quasiparticles, which can be thought of as ``dressed'' real particles whose properties, defined through self-consistent equations, incorporate the effects of interactions. Based on that, we concluded that the two approaches describe the same physics. The advantage of the Landau Fermi-liquid theory is that one can freely postulate any energy density functional (as long as it transforms like the $00$-component of the energy-momentum tensor under a Lorentz boost and includes correct counterterms) and rely on the formalism to yield a description of the studied system in terms of quasiparticles whose dynamics conserve particle number, energy, and momentum. At the same time, the advantage of the Walecka model is that it provides a clear example of what such an energy density functional could look like based on a Lagrangian incorporating nucleon-meson interactions.

In the following, we will draw on both approaches to form a flexible equation of state for nuclear matter. We will use the energy density stemming from the Walecka model as an inspiration for postulating a generalized energy density in a model with an arbitrary number of scalar- and vector-type interactions. We will then obtain the single-particle energies of the quasiparticles using the Landau Fermi-liquid theory, from which all other properties of the system will be derived. Unlike in the original Landau Fermi-liquid approach, which uses the properties of the interactions between the quasiparticles (as encoded in the Landau parameters) to constrain the free parameters of the model, we will instead fit these parameters to reproduce a set of chosen bulk properties of nuclear matter (this will take place in Chapter \ref{parametrization_and_model_results}). In this regard, the formalism we develop is similar to a large class of effective approaches to describing nuclear matter using self-consistent models based on the density functional theory (DFT) \cite{Hohenberg:1964zz,Kohn:1965zzb}. Such models are a starting point for numerous Skyrme-like potentials of varying degrees of complexity which are successfully applied in low-energy nuclear physics \cite{Bender:2003jk}.

\section{Relativistic density functional equation of state for nuclear matter}
\label{flexible_equation_of_state_for_nuclear_matter}

The contents of this section are a generalization of the model presented in Ref.\ \cite{Sorensen:2020ygf}.

\subsection{Formalism}

We adopt the relativistic Landau Fermi-liquid theory \cite{Baym:1975va} with vector- and scalar-density--dependent interactions as the basis for constructing a vector and scalar density functional (VSDF) model of dense nuclear matter EOS. To simplify the notation, we will introduce a VSDF model with a single vector-current- and a single scalar-current--dependent interaction term, however, it is straightforward to generalize to a model with multiple interaction terms of both kinds, which we do in the next subsection. Some of the details of the derivation will be referred to Appendix \ref{model_derivations}.

We introduce the energy density of a system composed of one species of fermions of rest mass $m_0$, interacting through a single mean-field vector interaction term and a single mean-field scalar interaction term, 
\begin{eqnarray}
\mathcal{E}_{\txt{v1s1}} &=& g \int \pdens  \epsilon_{\txt{kin}}^* ~ f_{\bm{p}} + C_1 \big(j_{\mu}j^{\mu}\big)^{\frac{b_1}{2} - 1} \big(j^0\big)^2 \non \\
&& \hspace{5mm } - ~ g^{00} C_1 \left(\frac{b_1 - 1}{b_1}\right) \big(j_{\mu} j^{\mu} \big)^{\frac{b_1}{2}}  + G_1 \left(\frac{d_1 - 1}{d_1}\right) n_s^{d_1}~.
\label{energy_density_one_scalar_one_vector_term}
\end{eqnarray}
Here, the subscript ``$\txt{v1s1}$'' underscores the fact that we have one vector and one scalar interaction term, $g$ is the degeneracy, $\epsilon_{\txt{kin}}^*$ is the kinetic energy of a single quasiparticle,
\begin{eqnarray}
\epsilon_{\txt{kin}}^* = \sqrt{ \bigg( \bm{p} - C_1 \big(j_{\mu} j^{\mu} \big)^{\frac{b_1}{2} - 1} \bm{j} \bigg)^2 + {m^*}^2  } ~,
\label{kinetic_energy_definition}
\end{eqnarray}
$\bm{j}$ and $j^0$ are the spatial and temporal components of the number current $j^{\mu}$, given by
\begin{eqnarray}
\bm{j} = g \int \pdens \frac{\bm{p} - C_1 \big(j_{\mu} j^{\mu}\big)^{\frac{b_1}{2} - 1} \bm{j}}{\epsilon_{\txt{kin}}^*} ~  f_{\bm{p}} 
\label{current_definition}
\end{eqnarray}
and 
\begin{eqnarray}
j^0 = g \int \pdens f_{\bm{p}}~,
\label{density_definition}
\end{eqnarray}
respectively, $f_{\bm{p}}$ is the quasiparticle distribution function, $m^* = m^*(x)$ is an effective mass given by a self-consistent equation,
\begin{eqnarray}
m^* = m_0 - G_1 n_s^{d_1 - 1} ~,
\label{effective_mass_definition}
\end{eqnarray}
where the number density $n_s$ is given by 
\begin{eqnarray}
n_s = g \int \pdens \frac{m^*}{\epsilon_{\txt{kin}}^*} ~ f_{\bm{p}} ~,
\label{scalar_density_definition}
\end{eqnarray}
$g^{00}$ is the $00$-component of the metric tensor (see Appendix \ref{units_and_notation} for the metric convention), and finally $C_1$, $G_1$, $b_1$, $d_1$ are constants specifying the interaction, as of yet undetermined. The coefficients and powers of the interaction terms have number indeces in anticipation of adding more interaction terms. We note that Eq.\ (\ref{energy_density_one_scalar_one_vector_term}) reduces to the Walecka model energy density, Eq.\ (\ref{Walecka_energy_density}), for $b_1 = 2$ and $d_1 = 2$ (and an appropriate identification of the coefficients of the interaction terms).

The quasiparticle energy $\eps_{\bm{p}}$ is obtained from the definition (details of the calculation can be found in Appendix \ref{the_quasiparticle_energy}),
\begin{eqnarray}
\eps_{\bm{p}}^* \equiv  \frac{\delta \mathcal{E}_{\txt{v1s1}} }{\delta f_{\bm{p}}} = \epsilon_{\txt{kinetic}}^*  +  C_1\big( j_{\mu} j^{\mu} \big)^{\frac{b_1}{2} - 1} j_0  ~. 
\label{quasiparticle_energy_00}
\end{eqnarray}
Also here setting $b_1 = 2$ and $d_1 = 2$ reduces the quasiparticle energy, Eq.\ (\ref{quasiparticle_energy_00}), to the expression known from the Walecka model, Eq.\ (\ref{Walecka_dispersion_relation}).

To simplify the notation, we introduce a vector field,
\begin{eqnarray}
A^{\lambda}_{(1)}  = A^{\lambda}(x; C_1, b_1) = C_1 \big(j_{\mu} j^{\mu} \big)^{\frac{b_1}{2} - 1} j^{\lambda} ~,
\label{Agnieszka_variable}
\end{eqnarray}
which allows us to concisely write
\begin{eqnarray}
\eps_{\bm{p}}^* = \sqrt{ \Big(\bm{p} - \bm{A}_{(1)}  \Big)^2  + {m^*}^2 } + A_0^{(1)} 
\label{quasiparticle_energy_generalized}
\end{eqnarray}
and
\begin{eqnarray}
\mathcal{E}_{v1s1} = g \int \pdens \eps_{\bm{p}}^* ~  f_{\bm{p}} - g^{00} \left(\frac{b_1 - 1}{b_1} \right) A_{\lambda}^{(1)} j^{\lambda} + G_1 \left(\frac{d_1 - 1}{d_1} \right)n_s^{d_1}~.
\label{energy_density_generalized}
\end{eqnarray}
For clarity, in the following derivation we will temporarily suppress the interaction term index $(1)$ and refer to this variable simply as $A^{\lambda}(x)$.

Quite generally, the quasiparticle energy is equivalent to the single-particle Hamiltonian, $\eps_{\bm{p}}^* = H_{(1)}$. Given the single-particle Hamiltonian, the equations of motion governing the evolution of quasiparticles follow immediately from Hamilton's equations, 
\begin{eqnarray}
\frac{dx^i}{dt} &\equiv& - \parr{H_{(1)}}{p_i} =  - \parr{\eps_{\bm{p}}^*}{p_i}  ~,
\label{x_equation_of_motion_general} \\
\frac{dp^i}{dt} &\equiv& \parr{H_{(1)}}{x_i} =  \parr{\eps_{\bm{p}}^*}{x_i}   ~.
\label{p_equation_of_motion_general}
\end{eqnarray}
where we utilize the co- and contravariant vector notation as summarized in Appendix \ref{units_and_notation}. Inserting Eqs.\ (\ref{x_equation_of_motion_general}) and (\ref{p_equation_of_motion_general}) into the Boltzmann equation (where we use the Einstein summation convention) gives
\begin{eqnarray}
\parr{f_{\bm{p}}}{t} - \parr{\eps_{\bm{p}}^*}{p_i} \parr{f_{\bm{p}}}{x^i} + \parr{\eps_{\bm{p}}^*}{x_i} \parr{f_{\bm{p}}}{p^i} = \left(\derr{f_{\bm{p}}}{t} \right)_{\txt{coll}}~, 
\label{Boltzmann_eq}
\end{eqnarray}
where $\big(\inderr{f_{\bm{p}}}{t} \big)_{\txt{coll}}$ is the change in the distribution function due to collisions occurring in the system (a thorough discussion of the Boltzmann equation can be found in Chapter \ref{implementation} and Appendix \ref{basics_of_the_kinetic_theory}). Multiplying both sides of Eq.\ (\ref{Boltzmann_eq}) by $X = \{  1,  \eps_{\bm{p}}^*,  p^j \}$ and integrating over all momenta, $g \int \frac{d^3p}{(2\pi)^3}$, yields the conservation laws for quasiparticle number ($X = 1$), energy ($X = \eps_{\bm{p}}^*$), and momentum ($X = p^j$). In particular, terms entering the energy and momentum conservation laws can be identified with the components of the energy-momentum tensor (see Appendix \ref{the_energy-momentum_tensor} for the derivation), given by
\begin{eqnarray}
&& T^{00} = \mathcal{E} ~, 
\label{T00_general} \\
&& T^{0i} = - g\int \pdens \eps_{\bm{p}}^* \parr{\eps_{\bm{p}}^*}{p_i} ~ f_{\bm{p}} ~,
\label{T0i_general} \\
&& T^{i0} = g \int \pdens p^i ~  f_{\bm{p}}~, 
\label{Ti0_general} \\
&& T^{ij} = - g \int \pdens  p^i \parr{\eps_{\bm{p}}^*}{p_j} ~  f_{\bm{p}} + g^{ij} \left(\mathcal{E} -  g\int \pdens \eps_{\bm{p}}^* ~  f_{\bm{p}}  \right) ~.   
\label{Tij_general}              
\end{eqnarray}
One can show \cite{Baym:1975va} that such obtained $T^{\mu\nu}$ has the correct transformation properties under a Lorentz boost. Additionally, the conservation of energy and momentum, $\partial_{\nu} T^{\mu\nu} =0$, is ensured by construction.

Inserting the expression for the quasiparticle energy, Eq.\ (\ref{quasiparticle_energy_generalized}), into Eqs.\ (\ref{x_equation_of_motion_general}) and (\ref{p_equation_of_motion_general}) yields
\begin{eqnarray}
\frac{dx^i}{dt} &=& - \parr{\eps_{\bm{p}}^*}{p_i} = \frac{p^i - A^i }{\epsilon_{\txt{kin}}^*}  ~,
\label{x_equation_of_motion} \\
\frac{dp^i}{dt} &=&   \parr{\eps_{\bm{p}}^*}{x_i} = \frac{  \big(p^k - A^k \big)\parr{A_k}{x_i}  +  m^* \parr{m^*}{x_i} }{ \epsilon_{\txt{kin}}^* } + \parr{A_0}{x_i}   ~,
\label{p_equation_of_motion}
\end{eqnarray}
while the energy-momentum tensor becomes
\begin{eqnarray}
&& T^{00} = \mathcal{E}_{v1s1} ~, 
\label{T00} \\
&& T^{0i} =  g\int \pdens \eps_{\bm{p}}^* ~ \frac{p^i - A^i }{\epsilon_{\txt{kin}}^*}  ~ f_{\bm{p}} ~,
\label{T0i} \\
&& T^{i0} = g \int \pdens p^i ~  f_{\bm{p}}~, 
\label{Ti0} \\
&& T^{ij} = g \int \pdens  p^i ~ \frac{p^j - A^j }{\epsilon_{\txt{kin}}^*}  ~  f_{\bm{p}} + g^{ij} \left(\mathcal{E} -  g\int \pdens \eps_{\bm{p}}^* ~  f_{\bm{p}}  \right) ~.   
\label{Tij}              
\end{eqnarray}
In particular, we can immediately check that the energy current $T^{0i}$ is equal to the momentum density $T^{i0}$, as required by hydrodynamics,
\begin{eqnarray}
T^{0i} &=& g \int \pdens \Big[\epsilon_{\txt{kinetic}}^* + A_0 \Big] \left[\frac{p^i - A^i }{\epsilon_{\txt{kinetic}}^*} \right] ~ f_{\bm{p}} \non  \\ 
&=& g \int \pdens  p^i  ~ f_{\bm{p}} -  A^i ~ g \int \pdens    ~ f_{\bm{p}} + A_0 ~ g \int \pdens  \frac{p^i - A^i }{\epsilon_{\txt{kinetic}}^*}  ~ f_{\bm{p}} \non \\
&=& g \int \pdens  p^i  ~ f_{\bm{p}} -  A^i j^0 + A_0 j^i \non \\
&=& g  \int \pdens  p^i  ~ f_{\bm{p}} = T^{i0} ~,
\end{eqnarray}
where we have used the definition of vector current, Eq.\ (\ref{current_definition}), the definition of vector density, Eq.\ (\ref{density_definition}), and the fact that from the definition of $A^{\lambda}$, Eq.\ (\ref{Agnieszka_variable}), we have $-  A^i j^0 + A_0 j^i = 0$. 

For completeness, we note that the quasiparticle number conservation law as obtained by taking the zeroth moment ($X = 1$) of the Boltzmann equation, Eq.\ (\ref{Boltzmann_eq}), is
\begin{eqnarray}
\parr{ }{t}~ g \int \pdens f_{\bm{p}} - \parr{ }{x^i }~ g \int \pdens \parr{\eps_{\bm{p}}^*}{p_i} ~ f_{\bm{p}} = 0 ~.
\end{eqnarray}
Using Eq.\ (\ref{x_equation_of_motion}), the above equation becomes 
\begin{eqnarray}
\parr{ }{t} g \int \pdens f_{\bm{p}} + \parr{ }{x^i } g \int \pdens \frac{p^i - A^i}{\epsilon_{\txt{kin}}^*} ~ f_{\bm{p}} = \partial_t j^0 + \partial_i j^i = 0 ~, 
\label{Boltzmann_eq_particle_number_conservation}
\end{eqnarray}
where we used the defining equations for the number current $j^i$ and density $j^0$, Eqs.\ (\ref{current_definition}) and (\ref{density_definition}). The above result can be taken as a confirmation that $j^i$ and $j^0$ have been correctly defined.

\subsection{Multiple terms, kinetic momentum, and manifestly covariant equations of motion}
\label{multiple_terms}

Having derived the properties of the VSDF model with one vector and one scalar interaction term, we can easily extend the formalism to an arbitrary number of interaction terms of each kind. Here, we are dealing with multiple vector fields labeled by an index $k$ (where one should remember that $k$ is a numbering subscript and not a space-time index),
\begin{eqnarray}
A_k^{\lambda}(x; C_k, b_k) = C_k \big(j_{\nu}j^{\nu} \big)^{\frac{b_k}{2} - 1} j^{\lambda} ~,
\end{eqnarray}
in terms of which a general energy density with $K$ vector-density--dependent terms and $M$ scalar-density--dependent terms can be written as
\begin{eqnarray}
\mathcal{E}_{(K,M)} &=& \int \ptilde \epsilon^{*}_{\txt{kin}}~  f_{\bm{p}} + \sum_{k=1}^K A^{0}_k j_{0} - g^{00} \sum_{k=1}^{K} \left(\frac{b_k - 1}{b_k} \right) A_k^{\mu} j_{\mu} + \sum_{m=1}^M G_m \left(\frac{d_m - 1}{d_m}\right) n_s^{d_m} \non \\
&=& \int \ptilde \eps^{*}_{\bm{p}}~  f_{\bm{p}} - g^{00} \sum_{k=1}^{K} \left(\frac{b_k - 1}{b_k} \right) A_k^{\mu} j_{\mu} + \sum_{m=1}^M G_m \left(\frac{d_m - 1}{d_m} \right) n_s^{d_m}~.
\label{energy_density_generalized_multiple}
\end{eqnarray}
We note that taking $K=2$, $M=0$ and evaluating $\mathcal{E}_{2,0}$ in the rest frame results in the interaction of the same form as in commonly used parametrizations of the Skyrme model (see, e.g., Ref.\ \cite{Kruse:1985hy}); in particular, using $b_1 = 2$ and $b_2 = 3$ ($b_2 = \frac{13}{6}$) leads to the stiff (soft) parametrization. In our approach, however, we leave the interaction parameters, including the powers of vector and scalar number densities in the interaction terms, unspecified until a later time, when we will fit them to match chosen properties of nuclear matter. 

The generalization of the remaining parts of the VSDF model is straightforward, and in particular we arrive at the quasiparticle energy,
\begin{eqnarray}
\eps_{\bm{p}}^{*} =  \sqrt{ \bigg( \bm{p} - \sum_{k=1}^K \bm{A}_k \bigg)^2 + {m^*}^2} + \sum_{k=1}^K A^0_k ~, 
\label{quasiparticle_energy_multiple_terms}
\end{eqnarray}
with the conserved current and the effective mass given by
\begin{eqnarray}
\bm{j} = \int \ptilde \frac{\bm{p} - \sum_{k=1}^K \bm{A}_k }{\epsilon^{*}_{\txt{kin}}}~  f_{\bm{p}}
\end{eqnarray}
and 
\begin{eqnarray}
m^* = m_0 -  \sum_{m=1}^M G_m n_s^{d_m - 1}~,
\label{effective_mass_definition_generalized}
\end{eqnarray}
respectively, while the equations of motion become 
\begin{eqnarray}
\frac{dx^i}{dt} &=& \frac{p^i - \sum_{k=1}^K (A_k)^i }{\epsilon_{\txt{kin}}^*}  ~,
\label{x_equation_of_motion_generalized} \\
\frac{dp^i}{dt} &=&  \frac{  \bigg(p^j - \sum_{k=1}^K (A_k)^j \bigg) \left( \sum_{k=1}^K\parr{(A_k)_j}{x_i}\right)  +  m^* \parr{m^*}{x_i} }{ \epsilon_{\txt{kin}}^* } + \sum_{k=1}^K \parr{A_k^0}{x_i}   ~.
\label{p_equation_of_motion_generalized}
\end{eqnarray}
We stress that the generalization to $(K,M)$ interaction terms preserves the conservation laws and the relativistic covariance of the $T^{\mu\nu}$ tensor. 

Finally, the equations of motion, Eqs.\ (\ref{x_equation_of_motion_generalized}) and (\ref{p_equation_of_motion_generalized}), can be rewritten in a manifestly covariant way. First, we rewrite Eq.\ (\ref{quasiparticle_energy_multiple_terms}) as
\begin{eqnarray}
\eps_{\bm{p}}^{*} - \sum_{k=1}^K A^0_k =  \sqrt{ \bigg( \bm{p} - \sum_{k=1}^K \bm{A}_k \bigg)^2 + {m^*}^2} ~.
\end{eqnarray}
It is then natural to define a quantity known as the kinetic momentum $\Pi^{\mu}$ \cite{Blaettel:1993uz},
\begin{eqnarray}
\Pi^{\mu} \equiv p^{\mu} - \sum_{k=1}^K (A_k)^{\mu} ~,
\label{kinetic_momentum}
\end{eqnarray}
which by construction satisfies
\begin{eqnarray}
\Pi^0 = \sqrt{\bm{\Pi}^2 + m^2} ~.
\end{eqnarray}
Note that in terms of the kinetic momentum $\Pi^{\mu}$, the baryon 4-current can be naturally defined as
\begin{eqnarray}
j^{\mu} = g \int \pdens \frac{\Pi^{\mu}}{\Pi_0} ~ f_{\bm{p}} ~.
\label{baryon_4current}
\end{eqnarray}
Using the kinetic momentum, one can rewrite the equations of motion as (see Appendix \ref{EOMs_covariant} for details)
\begin{eqnarray}
&& 
\frac{dx^{\mu}}{dt}  = \frac{\Pi^{\mu}}{\Pi_0}   ~, 
\label{EOM_covariant_formulation_x}\\
&&
\frac{d \Pi^{\mu}}{dt} = \sum_{\nu} \frac{\Pi_{\nu}}{\Pi_0} \sum_{k=1}^K \Big[ \partial^{\mu} (A_k)^{\nu} - \partial^{\nu} (A_k)^{\mu}  \Big]  + \frac{m^*}{\Pi_0} \parr{m^*}{x_i}~.
\label{EOM_covariant_formulation_p}
\end{eqnarray}
It is apparent that the first force term on the right-hand side of Eq.\ (\ref{EOM_covariant_formulation_p}), stemming from the vector interaction terms, is similar to the electromagnetic force known from covariantly formulated electrodynamics \cite{Landau_Classical_Theory_of_Fields}, except that in our case there is an arbitrary number $K$ of vector fields. In particular, this means that the equation of motion in Eq.\ (\ref{EOM_covariant_formulation_p}) naturally includes a Lorentz force (see Appendix \ref{Lorentz_force} for an explicit demonstration).

\subsection{Thermodynamics and thermodynamic consistency}

The thermodynamic properties of the VSDF model are obtained in the following way. In equilibrium, the distribution function must have the form of the Fermi-Dirac distribution (for details, see Appendix \ref{form_of_the_quasiparticle_distribution_function}),
\begin{eqnarray}
f_{\bm{p}} = \frac{1}{e^{\beta ( \eps_{\bm{p}}^* - \mu  )} + 1 }~,
\label{distribution_function}
\end{eqnarray}
where $\beta$ is the inverse temperature, $\beta = 1/T$, and $\mu$ is the chemical potential. We consider the energy-momentum tensor in the rest frame, where it has the form
\begin{eqnarray}
T^{\mu\nu} = \txt{diag} \big(\mathcal{E}, P, P, P \big) 
\end{eqnarray}
and where the spatial components of the current vanish, $j^i = 0$, while $j_{\mu}j^{\mu} = n^2$, with $n$ being the rest frame density. Then the pressure is given by
\begin{eqnarray}
P_{(K,M)} &=& \frac{1}{3} \sum_k T^{kk} \bigg|_{\substack{\text{rest} \\ \text{frame}}} = \non \\
&=&  g \int \pdens T ~ \ln \bigg[ 1 + e^{- \beta \big(\eps_{\bm{p}}^* - \mu_B \big)}  \bigg]_{\substack{\text{rest} \\ \text{frame}}}  \non \\
&& \hspace{5mm} + ~   \sum_{k=1}^K C_k \left(\frac{b_k - 1}{b_k} \right) n^{b_k } - \sum_{m=1}^M G_m \left(\frac{d_m - 1}{d_m}\right) n_s^{d_m}   ~.
\label{VSDF_pressure}
\end{eqnarray}
We note that in an equilibrated system, vector-density--dependent interactions can be described in terms of a shift of the chemical potential $\mu$. Using the expression for the quasiparticle energy, Eq.\ (\ref{quasiparticle_energy_multiple_terms}), in the exponent of the distribution function, Eq.\ (\ref{distribution_function}), we see that it is convenient to define an effective chemical potential $\mu^*$ through
\begin{eqnarray}
\mu^* \equiv \mu - \sum_{k=1}^K A^0_k ~,
\label{effective_chemical_potential}
\end{eqnarray}
so that 
\begin{eqnarray}
\eps_{\bm{p}}^{*} - \mu = \epsilon_{\txt{kin}}^* - \mu^*  = \Pi^0 - \mu^*~. 
\end{eqnarray}
Consequently, the dependence of the thermal part of the pressure, Eq.\ (\ref{VSDF_pressure}), on temperature $T$ and effective chemical potential $\mu^*$ is just like that of an ideal Fermi gas, with the effects due to scalar interactions ``folded into'' the effective mass $m^*$ and the effects due to vector interactions ``folded into'' the effective chemical potential $\mu^*$.

The grand canonical potential is related to the pressure through $\Omega(T,\mu, V) = - PV$, and we can immediately calculate the entropy density,
\begin{eqnarray}
s &\equiv& - \frac{1}{V}  \left( \frac{d \Omega}{dT} \right)_{V,\mu}  = g \int \pdens \bigg( \ln \Big[ 1 + e^{-\beta (\eps_{\bm{p}}^* - \mu)}  \Big] +    \frac{\eps_{\bm{p}}^* - \mu }{T} ~  f_{\bm{p}} \bigg) ~, 
\label{VSDF_entropy}
\end{eqnarray}
and the number density,
\begin{eqnarray}
n \equiv - \frac{1}{V} \left( \frac{d \Omega}{d\mu} \right)_{V,T} = g\int  \pdens f_{\bm{p}}~,
\end{eqnarray}
where the latter equation proves the correct normalization of the distribution function $f_{\bm{p}}$. We note that the entropy density, Eq.\ (\ref{VSDF_entropy}), can be rewritten as
\begin{eqnarray}
s  =\frac{1}{T}\Big[ \big(P_{FG}^* + \mathcal{E}_{FG}^* \big) -    \mu^*  n \Big] = s_{FG}^*~,
\end{eqnarray}
where we marked the dependence of the ideal Fermi gas pressure, energy density, and entropy density on $m^*$ and $\mu^*$ by a star superscript. Thus we see that the entropy density in the VSDF model is just like that of an ideal Fermi gas with the effective mass $m^*$ and effective chemical potential $\mu^*$.

Minimizing the thermodynamic potential (which in the grand canonical ensemble is equivalent to finding the extremum of the pressure) with respect to the auxiliary fields, $n$ and $n_s$, confirms the defining equations for the effective mass and the baryon density, Eqs.\ (\ref{effective_mass_definition_generalized}) and (\ref{baryon_4current}) (see Appendix \ref{minimization_of_the_auxiliary_fields} for more details). Finally, calculating the energy density using the fundamental equation of thermodynamics, $\mathcal{E} \equiv sT - P + \mu n$, yields Eq.\ (\ref{energy_density_generalized_multiple}) evaluated in the rest frame, thus proving that the VSDF model is thermodynamically consistent.

\section{Summary}
\label{summary_of_VSDF_formulas}

In this chapter, we have given a short review of the mean-field approach to describing nuclear matter and of the Landau Fermi-liquid theory, followed by a derivation of the flexible vector and scalar density functional (VSDF) model. The VSDF model, which obeys Lorentz covariance, preserves conservation laws, and is shown to be thermodynamically consistent, allows one to construct a parametrized dense nuclear matter EOS. In particular, the energy density and pressure as obtained in the VSDF model can be used to fix the thermodynamic properties of the system, while the corresponding quasiparticle energies and quasiparticle equations of motion can be implemented in a hadronic transport simulation.

For convenience, we recall the most important equations from the derivation above. Starting from the energy density of the form
\begin{eqnarray}
\hspace{-10mm}\mathcal{E}_{(K,M)} &=& g \int \pdens \epsilon^{*}_{\txt{kin}}~  f_{\bm{p}} + \sum_{k=1}^K A^{0}_k j_{0} - g^{00} \sum_{k=1}^{K} \left(\frac{b_k - 1}{b_k} \right) A_k^{\mu} j_{\mu} \non \\
&& \hspace{10mm} + ~  \sum_{m=1}^M G_m \left(\frac{d_m - 1}{d_m}\right) n_s^{d_m} ~,
\label{summary_energy_density}
\end{eqnarray}
where the vector $A_k^{\lambda}$ field is defined as
\begin{eqnarray}
A_k^{\lambda}(x; C_k, b_k) = C_k \big(j_{\nu}j^{\nu} \big)^{\frac{b_k}{2} - 1} j^{\lambda} ~,
\label{summary_vector_field}
\end{eqnarray}
the kinetic energy is given by
\begin{eqnarray}
\epsilon_{\txt{kin}}^* =  \sqrt{ \bigg( \bm{p} - \sum_{k=1}^K \bm{A}_k \bigg)^2 + {m^*}^2}  ~, 
\label{summary_kinetic_energy}
\end{eqnarray}
and the conserved current and the scalar density are given by
\begin{eqnarray}
j^{\mu} = g \int \pdens \frac{p^{\mu} - \sum_{k=1}^K A_k^{\mu} }{\epsilon^{*}_{\txt{kin}}}~  f_{\bm{p}}
\label{summary_current_definition}
\end{eqnarray}
and 
\begin{eqnarray}
n_s = g \int \pdens \frac{m^*}{\epsilon_{\txt{kin}}^*} ~ f_{\bm{p}} ~.
\label{summary_scalar_density_definition}
\end{eqnarray}
respectively, with the effective mass satisfying
\begin{eqnarray}
m^* = m_0 -  \sum_{m=1}^M G_m n_s^{d_m - 1}~,
\label{summary_effective_mass_definition}
\end{eqnarray}
we arrive at the quasiparticle energy
\begin{eqnarray}
\eps_{\bm{p}}^{*} =  \sqrt{ \bigg( \bm{p} - \sum_{k=1}^K \bm{A}_k \bigg)^2 + {m^*}^2} + \sum_{k=1}^K A^0_k ~, 
\label{summary_quasiparticle_energy}
\end{eqnarray}
the quasiparticle equations of motion,
\begin{eqnarray}
\frac{dx^i}{dt} &=& \frac{p^i - \sum_{k=1}^K (A_k)^i }{\epsilon_{\txt{kin}}^*}  ~,
\label{summary_EOM_x} \\
\frac{dp^i}{dt} &=&  \frac{  \bigg(p^j - \sum_{k=1}^K (A_k)^j \bigg) \left( \sum_{k=1}^K\parr{(A_k)_j}{x_i}\right)  +  m^* \parr{m^*}{x_i} }{ \epsilon_{\txt{kin}}^* } + \sum_{k=1}^K \parr{A_k^0}{x_i}   ~,
\label{summary_EOM_p}
\end{eqnarray}
and the pressure of the system, 
\begin{eqnarray}
P_{(K,M)} &=&  g \int \pdens T ~ \ln \bigg[ 1 + e^{- \beta \big(\eps_{\bm{p}}^* - \mu_B \big)}  \bigg]_{\substack{\text{rest} \\ \text{frame}}}  \non \\
&& \hspace{5mm} + ~   \sum_{k=1}^K C_k \left(\frac{b_k - 1}{b_k} \right) n^{b_k } - \sum_{m=1}^M G_m \left(\frac{d_m - 1}{d_m}\right) n_s^{d_m}   ~.
\label{summary_pressure}
\end{eqnarray}
We again note that all of the presented expressions reduce to the Walecka model for $K = M = 1$ and $b_1 = d_1 = 2$.  

Let us once more stress that within the Landau Fermi-liquid theory, the form of the energy density $\mathcal{E}$, Eq.\ (\ref{summary_energy_density}), determines all other quantities of interest. A natural question that arises here is: Does the particular choice of $\mathcal{E}$ presented in Eq.\ \eqref{summary_energy_density} provide maximal flexibility? If one considers a mean-field model with self-consistently defined baryon current and effective mass, and one demands that the coefficients of the interaction are constant, then it can be shown (see Appendix \ref{form_of_the_energy_density_functional}) that the interaction terms must have the form of polynomials of vector and scalar number densities, $n$ and $n_s$. While it is not entirely inconceivable to construct a model within which the interaction coefficients are functions of $n$ and $n_s$, such model would present significant computational challenges. Therefore for the purposes of this thesis, which centers on applications in hadronic transport models, an energy density of the form displayed in Eq.\ (\ref{summary_energy_density}) provides a maximal flexibility for a mean-field model of vector and scalar interactions.

\newpage
\chapter{Parametrization and model results}
\label{parametrization_and_model_results}

The contents of this chapter are largely based on Ref.\ \cite{Sorensen:2020ygf}. We note, however, that unlike in Ref.\ \cite{Sorensen:2020ygf}, we include antiparticles in our description of nuclear matter.

The VSDF model (see Section \ref{summary_of_VSDF_formulas} for the model summary) has been constructed with applications to hadronic transport simulations in mind. Most importantly, in opposition to theories in which the chemical potential $\mu$ is the independent variable, we chose the number density $n$, naturally accessible in hadronic transport, to be the dynamical variable of the theory (the scalar number density is a function of $n$ and temperature $T$). The constructed EOS is easily parametrizable through fitting the interaction parameters $(C_k, b_k, G_m, d_m)$ to reproduce a chosen set of nuclear matter properties. Here, the number of vector and scalar interaction terms, $K$ and $M$ in Eq.\ (\ref{summary_energy_density}), is entirely determined by the number of constraints that one want to impose on the EOS. In particular, it is straightforward to construct a family of EOSs that on the one hand reproduces the known properties of ordinary nuclear matter, and on the other allows one to postulate and explore critical behavior in dense nuclear matter over vast regions of the phase diagram. The former will ensure that the model takes into the account the behavior of nuclear matter known from the experiment (see Section \ref{models_of_nuclear_matter} for more details), while the latter will allow for meaningful comparisons of the influence of different possible properties of the QGP phase transition on observables. Such comparisons can be made, among others, through Bayesian analysis \cite{Novak:2013bqa,Bernhard:2016tnd}.

Apart from the flexibility of the EOS allowing for a systematic exploration of possible phase diagrams, another factor to consider in simulations is numerical efficiency. This is especially important for afterburner studies, which often require simulating thousands or even millions of events, depending on the studied observables, to reach statistical significance comparable with the experiment. Here, vector-type interactions are more convenient for numerical evaluation of mean-field potentials than scalar-type interactions, which require solving a self-consistent equation at each point where mean-fields are calculated (usually, this means at every node of a discrete 3-dimensional lattice representing the fields). Therefore, as an initial step in studies of critical behavior in hadronic transport, in this thesis we develop the parametrization of the flexible VSDF EOS that includes only interactions of the vector type, that is we set $M=0$ in Eqs.\ (\ref{summary_energy_density}-\ref{summary_pressure}). In this case, the energy density reduces to a vector density functional (VDF) model of the EOS. While using an energy density that is a VDF instead of a VSDF leads to a smaller flexibility of the obtained family of EOSs (which is connected with the fact that vector interactions depend only on the baryon density, which in our model, along with the temperature $T$, is one of the two independent variables), it nevertheless allows for creating a large family of EOSs that can be easily utilized in simulations and whose behavior can be easily understood. (We note that in spite of the significantly increased numerical cost of simulations utilizing scalar mean-field interactions, this cost is not entirely prohibitive, especially for observables requiring lower statistics such as particle yields or elliptic flow, and future studies will be devoted to this area; among the existing approaches, a prominent example of simulations with scalar-type interactions are those using equations of motion stemming from the Walecka model, see Refs.\ \cite{Ko:1987gp,Blaettel:1993uz,Buss:2011mx}.)

To apply our approach to studies of heavy-ion collisions and the search for the location of the conjectured QCD critical point, we want to parametrize the VDF model so that it describes hadronic matter whose phase diagram contains two first-order phase transitions. The first of these is the experimentally observed low-temperature, low-density phase transition in nuclear matter, sometimes known as the nuclear liquid-gas transition. The second is a postulated high-temperature, high-density phase transition that is intended to correspond to the QCD phase transition. 

We want to stress that while the latter may, in principle, coincide with the location of the phase transition in the real QCD phase diagram, its nature is fundamentally different. This is because within Landau Fermi-liquid theory, unlike in QCD, the degrees of freedom do not change across the phase transition. Interestingly, this is also the case in some other approaches to the QCD EOS, for example in models based on quarkyonic matter \cite{McLerran:2018hbz}, where the active degrees of freedom at the Fermi surface remain hadronic even after quark degrees of freedom appear (although the extent to which such dynamics may be captured in the VDF or VSDF model remains to be seen). The nature of the phase transition that we can simulate in the VDF (or VSDF) model is that of going from a less organized to a more organized state. This is easily visualized in the case of the transition from gas to liquid (nucleon gas to nuclear drop). In the case of the high-temperature, high-density phase transition, we may think of it as a transition from a fluid to an even more dense, and more organized, fluid (nuclear matter to quark matter). This interpretation is supported by the functional dependence of entropy per particle on the order parameter (number density), which decreases across the phase transition from a less dense to a more dense state (for an extended discussion, see Ref.\ \cite{Hempel:2013tfa}). For brevity, in the following we will refer to the high-temperature, high-density phase transition within the VDF model as the ``QGP-like'' or ``quark-hadron'' phase transition, with the expectation that it is understood as a useful moniker rather than a statement on the nature of the described transformation. 

We also emphasize that the degrees of freedom used in the VDF model agree with those expected after hadronization by design. Since the VDF model is ultimately intended to be used in the hadronic afterburner stage of a hybrid heavy-ion collision simulation, the issue of hadronic degrees of freedom present above the QGP-like phase transition will never arise in realistic calculations. At the same time, in parts of the phase diagram close to the critical region, the hadronic systems studied will display effects typical for approaching a phase transition, such as for example those expected due to the softening of the EOS.

\section{Parametrization}
\label{parametrization}

We present a rather simplified version of the VDF model, in which we choose the degrees of freedom to be those of isospin symmetric nuclear matter, that is nucleons with nucleon mass $m_N = 938$ MeV and degeneracy factor $g_N=4$. Additionally, we also consider the case including thermally induced $\Delta$ resonances, whose mass and degeneracy factor are taken to be $m_{\Delta} = 1232$ MeV and $g_{\Delta} = 16$, respectively. We note that the model can be easily extended to include arbitrarily many baryon resonances (which is done through a substitution $g f_{\bm{p}} \to \sum_i g_i  f^{(i)}_{\bm{p}}$), however, a thorough study of the corresponding effects is beyond the scope of this thesis. Finally, we modify the defining expressions of our model, Eqs.\ (\ref{summary_energy_density}-\ref{summary_pressure}), to include contributions from antiparticles. For example, in this case $j^0$ represents the net baryon density given by
\begin{eqnarray}
j^0 \equiv j^0_{\txt{baryons} } - j^0_{\txt{antibaryons}} = g \int \pdens \Big[ f_{\bm{p}} - \bar{f}_{\bm{p}} \Big] ~,
\end{eqnarray}
where we introduce the distribution function for antibaryons,
\begin{eqnarray}
\bar{f}_{\bm{p}}(T, \mu_B) \equiv  \frac{1}{e^{\beta\big( \eps_{\bm{p}} + \mu_B\big)} + 1 } = f_{\bm{p}} (T, - \mu_B)~,
\end{eqnarray}
and the remaining VDF model equations are likewise modified, in particular the equations for the energy density
\begin{eqnarray}
\hspace{-10mm}\mathcal{E}_{(K,M)} &=& g \int \ptilde \epsilon^{*}_{\txt{kin}}~ \Big[ f_{\bm{p}} +  \bar{f}_{\bm{p}} \Big] + \sum_{k=1}^K A^{0}_k j_{0} - g^{00} \sum_{k=1}^{K} \left(\frac{b_k - 1}{b_k} \right) A_k^{\mu} j_{\mu}  
\label{summary_energy_density_antiparticles}
\end{eqnarray}
and the pressure, 
\begin{eqnarray}
P_{(K,M)} &=&  g \int \pdens T ~ \ln \bigg[ 1 + e^{- \beta \big(\eps_{\bm{p}}^* - \mu_B \big)}  \bigg] + g \int \pdens T ~ \ln \bigg[ 1 + e^{- \beta \big(\eps_{\bm{p}}^* + \mu_B \big)}  \bigg] \non \\
&& \hspace{5mm} + ~   \sum_{k=1}^K C_k \left(\frac{b_k - 1}{b_k} \right) n_B^{b_k }    ~,
\label{summary_pressure_antiparticles}
\end{eqnarray}
where  $n_B \equiv \sqrt{j_{\mu} j^{\mu}}$ is the rest frame net baryon number density.

In a system that undergoes two first-order phase transitions, the pressure exhibits two mechanically unstable regions, known as the spinodal regions (see Appendix \ref{phase_transitions} for more details), defined by the condition that the first derivative of the pressure with respect to the order parameter is negative \cite{Landau_Stat, Chomaz:2003dz}. In a minimal VDF model realizing such behavior, the pressure needs to be a four-term polynomial in the order parameter, and thus we adopt a version of the VDF model in which we utilize four interaction terms. (We note that to describe only one of the phase transitions mentioned above, it is enough to adopt a model with two interaction terms. In the case of the nuclear liquid-gas phase transition, the resulting model will be similar to many Skyrme-based parametrizations of the EOS \cite{Bender:2003jk}). The energy density, Eq.\ (\ref{summary_energy_density_antiparticles}), is easily adapted to include $K=4$ interaction terms, and in the rest frame we have
\begin{eqnarray}
\mathcal{E}\big|_{\substack{\text{rest} \\ \text{frame}}} = g \int \pdens \epsilon_{\txt{kin}} ~  \Big[f_{\bm{p}} + \bar{f}_{\bm{p}} \Big]+ \sum_{i=1}^4 \frac{C_i}{b_i} n_B^{b_i}~.
\end{eqnarray}
Similarly, we take $K=4$ in the expression for the pressure, Eq.\ (\ref{summary_pressure_antiparticles}).

To construct an EOS with a general QGP-like phase transition properties while ensuring that the known properties of ordinary nuclear matter are well reproduced, we choose the following constraints to fix the eight free parameters $\{b_1, b_2, b_3, b_4, C_1, C_2, C_3, C_4\}$ in the VDF model:\\
1) the position of the minimum of energy per particle of nuclear matter at the saturation density $n_B = n_0$,
\begin{eqnarray}
\frac{d \left( \frac{\mathcal{E}_{(K=4)} }{n_B} \right)}{dn_B} \bigg|_{\substack{T =0 \\ n_B=n_0 }} = 0~,
\label{minimum}
\end{eqnarray}
2) the value of the binding energy at the minimum,
\begin{eqnarray}
\frac{\mathcal{E}_{(K=4)} }{n_B}\bigg|_{\substack{T =0 \\n_B=n_0  }} - m_N = B_0~,
\label{binding_energy}
\end{eqnarray}
3, 4) the position of the critical point $\big(T^{(N)}_c, n^{(N)}_c\big)$ for the nuclear liquid-gas phase transition,
\begin{eqnarray}
&& \frac{dP}{dn_B} \Big(T =T^{(N)}_c,n_B= n^{(N)}_c\Big) = 0~, 
\label{nuclear_CP_1} \\
&&\frac{d^2P}{dn_B^2} \Big(T = T^{(N)}_c, n_B = n^{(N)}_c\Big) = 0~,
\label{nuclear_CP_2}
\end{eqnarray}
5, 6) the position of the critical point $ \big(T^{(Q)}_c, n^{(Q)}_c\big)$ for the quark-hadron phase transition,
\begin{eqnarray}
&& \frac{dP}{dn_B} \Big(T=T^{(Q)}_c, n_B=n^{(Q)}_c\Big) = 0~,
\label{QGP_like_CP_1} \\ 
&& \frac{d^2P}{dn_B^2} \Big(T=T^{(Q)}_c, n_B=n^{(Q)}_c\Big) = 0~,
\label{QGP_like_CP_2} 
\end{eqnarray}
7, 8) the position of the lower (left) and upper (right) boundaries of the spinodal region, $\eta_L$ and $\eta_R$, for the quark-hadron phase transition at $T=0$,
\begin{eqnarray}
&& \frac{dP}{dn_B} \Big(T=0, n_B=\eta_L\Big) = 0~,
\label{spinodal_1} \\
&& \frac{dP}{dn_B} \Big(T=0, n_B=\eta_R\Big) = 0~.
\label{spinodal_2}
\end{eqnarray}
(Explicit forms of the above equations are shown in Appendix \ref{parametrization_equations}.) The set of quantities $(n_0 , B_0, T_c^{(N)}, n_c^{(N)}, T_c^{(Q)}, n_c^{(Q)} , \eta_L, \eta_R)$ is referred to as the characteristics of an EOS.

\begin{table}[t]
	\caption[Example characteristics $\big( T_c^{(Q)}, n_c^{(Q)}, \eta_L, \eta_R\big)$ of the QGP-like phase transition in the VDF model]{Example characteristics $\big( T_c^{(Q)}, n_c^{(Q)}, \eta_L, \eta_R\big)$ of the QGP-like phase transition: critical temperature $T_c^{(Q)}$, critical baryon number density $n_c^{(Q)}$, and the boundaries of the spinodal region at $T =0$, $\eta_L$ and $\eta_R$. The corresponding parameter sets can be found in Appendix \ref{parameter_sets}. Characteristics in sets I-V are obtained based on systems composed only of nucleons, while in set VI we consider a system composed of nucleons and thermally produced $\Delta$-resonances. We also show the incompressibility at saturation density and zero temperature $K_0$ calculated for the parametrized EOSs.}
	\label{example_characteristics}
	\begin{center}
		\bgroup
		\def\arraystretch{1.4}
		\begin{tabular}{c c c c c c c}
			\hline
			\hline
			set & \hspace{3mm}$T_c^{(Q)}[\txt{MeV}]$\hspace{3mm} & \hspace{3mm}$n_c^{(Q)}[n_0]$\hspace{3mm} & 
			\hspace{3mm}$\eta_L[n_0]$\hspace{3mm} &  \hspace{3mm}$\eta_R[n_0]\hspace{5mm}$ \hspace{3mm}& particle species & \hspace{3mm}$K_0[\txt{MeV}]$\hspace{3mm}\\ 
			\hline
			I &50  & 3.0 & 2.70 & 3.22 & N &260 \\  
			II & 50  & 3.0 & 2.85 & 3.12 & N & 279\\  
			III & 50  & 4.0 & 3.90 & 4.08 & N & 280\\  
			IV & 100& 3.0 & 2.50 & 3.32 & N & 261\\  
			V & 125 & 4.0 & 3.60 & 4.28 & N & 278\\  
			VI & 125 & 4.0 & 3.60 & 4.28 & N + $\Delta$& 277\\
			\hline
			\hline
		\end{tabular}
		\egroup
	\end{center}
\end{table}

We choose the properties of the ordinary nuclear matter, encoded in conditions (\ref{minimum}-\ref{nuclear_CP_2}), based on experimentally determined values \cite{Elliott:2013pna,Bethe:1971xm}:
\begin{eqnarray}
&& n_0 = 0.160~ \txt{fm}^{-3}~, \hspace{5mm} B_0 = - 16.3 ~\txt{MeV}~, \label{enforced_saturation_density_and_binding_energy} \\
&& T^{(N)}_c = 18~ \txt{MeV}~, \hspace{5mm} n^{(N)}_c = 0.06 ~ \txt{fm}^{-3} ~. 
\label{enforced_nuclear_critical_point}
\end{eqnarray}
On the other hand, the properties of dense nuclear matter, $n_B \gg n_0$, are only weakly constrained by experiment at this time. We are then in a position to create a family of possible EOSs based on a number of different postulated characteristics (\ref{QGP_like_CP_1}-\ref{spinodal_2}), while ensuring that nuclear matter properties are preserved. The resulting family of EOSs encompasses QGP-like phase transition characteristics spanning vast regions of the dense nuclear matter phase diagram. This allows for a systematic comparison with experimental data, with the goal of constraining the properties of allowed EOSs to a small subfamily with qualitatively similar properties.

In the remainder of this chapter, we illustrate properties of the VDF model by discussing key results for a few representative EOSs which exhibit chosen sets of the QGP-like phase transition characteristics $\big( T_c^{(Q)}, n_c^{(Q)}, \eta_L, \eta_R\big)$ listed in Table \ref{example_characteristics}. The corresponding parameter sets can be found in Appendix \ref{parameter_sets}.

\section{VDF model results: Pressure, the speed of sound, and energy per particle}
\label{pressure,_speed_of_sound,_energy_per_particle}

\begin{figure}[t]
	\centering\mbox{
	\includegraphics[width=0.65\textwidth]{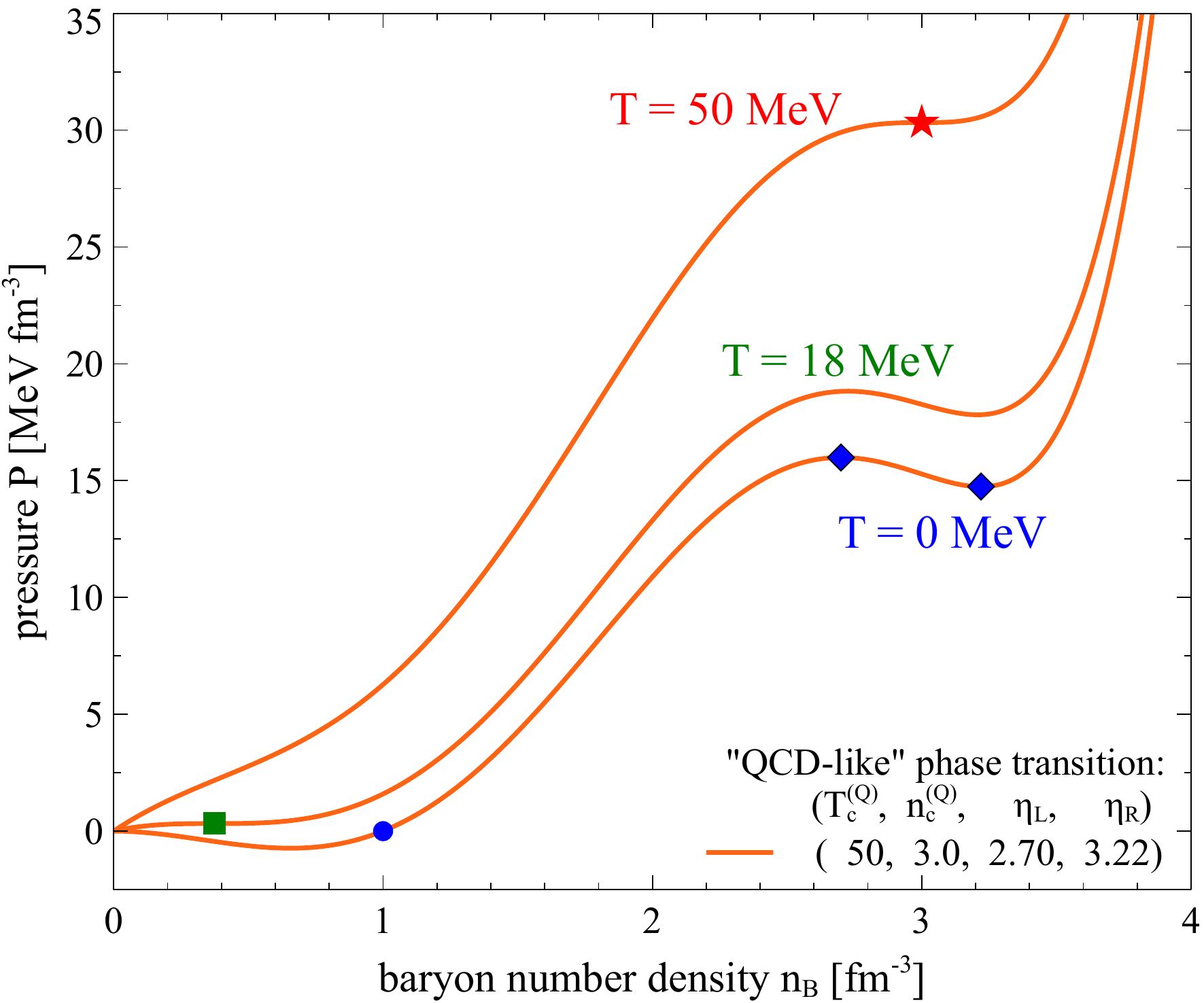} 
	} 
	\caption[An example of the fitting procedure used for the VDF model]{An example of the fitting procedure used for the VDF model. Pressure is plotted as a function of baryon number density at three significant temperatures ($T=0$, nuclear critical temperature $T_c^{(N)}$, and quark-hadron critical temperature $T_c^{(Q)}$) for an EOS with characteristics from set I, see Table \ref{example_characteristics} or the legend (where $T_c^{(Q)}$ is given in MeV, while the critical density $n_c^{(Q)}$ and the boundaries of the spinodal region at $T=0$, $\eta_{L}$ and $\eta_R$, are given in units of saturation density, $n_0 = 0.160 \ \txt{fm}^{-3}$). Specific points at which the parameters of the EOS are fixed are indicated on the plot; see text for more details. Figure from \cite{Sorensen:2020ygf}.}
	\label{pressures_one}
\end{figure}

We illustrate the fitting procedure in Fig.\ \ref{pressures_one}, where we show pressure as a function of baryon number density at three significant temperatures ($T = 0$, nuclear critical temperature $T_c^{(N)}$, and quark-hadron critical temperature $T_c^{(Q)}$) for an EOS with characteristics from set I (see Table \ref{example_characteristics}) and where we also indicate the location of key features that determine the fit parameters. At temperature $T = 0$, conditions (\ref{minimum}) and (\ref{binding_energy}) are applied at the saturation density of nuclear matter, marked with a blue circle; we note here that because at $T=0$ the pressure is given by
\begin{eqnarray}
P \equiv n_B^2 \frac{d}{dn_B} \left(\frac{\mathcal{E}}{n_B} \right) ~,
\label{pressure_Tzero}
\end{eqnarray}
condition (\ref{minimum}) is equivalent to demanding that $P=0$. Also at $T =0$, conditions (\ref{spinodal_1}) and (\ref{spinodal_2}) fix the positions of the lower (left) and upper (right) boundary of the high density spinodal region, $\eta_L$ and $\eta_R$; these are denoted with blue diamonds. At the critical point of nuclear matter, $T = T_c^{(N)}$ and $n_B = n_c^{(N)}$, denoted with a green square, conditions (\ref{nuclear_CP_1}) and (\ref{nuclear_CP_2}) are enforced. Finally, conditions (\ref{QGP_like_CP_1}) and (\ref{QGP_like_CP_2}) are applied to set the position of the QGP-like critical point $\big(T_{c}^{(Q)}, n_c^{(Q)} \big)$, denoted with a red star. Note that in general, the values of pressure at baryon densities corresponding to ordinary nuclear matter are significantly smaller than the values at large baryon densities, including the region where the QGP-like phase transition and the corresponding softening of the EOS occur.

\begin{figure}[t]
	\centering\mbox{
	\includegraphics[width=0.65\textwidth]{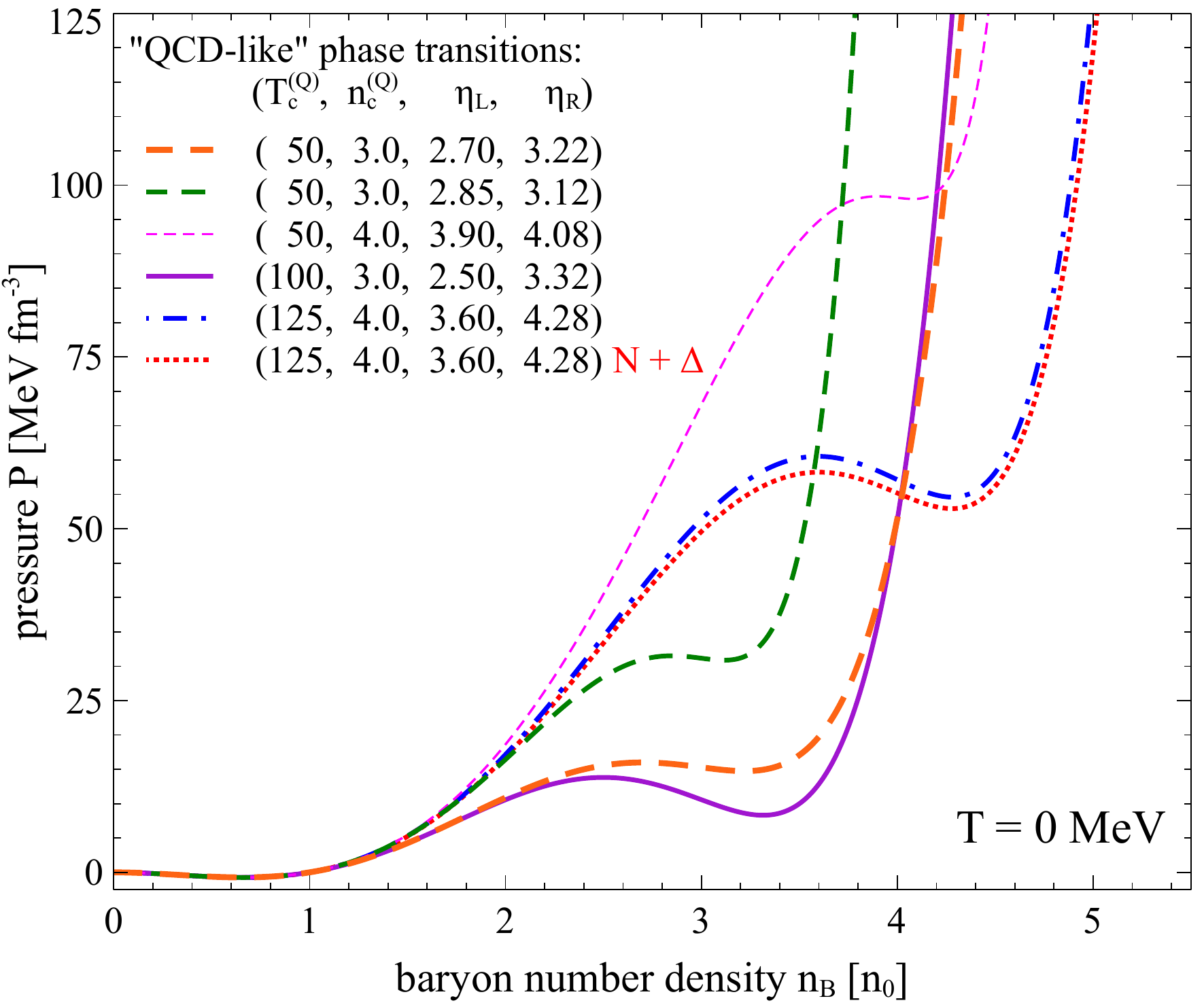} 
	}
	\caption[Pressure at $T=0$ as a function of baryon number density for several VDF EOSs]{Pressure is plotted as a function of baryon number density at temperature $T=0$ for all sets of characteristics listed in Table \ref{example_characteristics}. In the legend, the critical temperature of the QGP-like phase transition $T_c^{(Q)}$ is given in MeV, while the critical density $n_c^{(Q)}$ and the boundaries of the spinodal region at $T=0$, $\eta_{L}$ and $\eta_R$, are given in units of saturation density, $n_0 = 0.160 \ \txt{fm}^{-3}$. The hardness of the EOSs is noticeable for densities above the quark-hadron transition regions, and is a consequence of employing interaction terms with high powers ($b_i > 2$) of baryon number density $n_B$ (see text for details).}
	\label{pressures_all}
\end{figure}

Fig.\ \ref{pressures_all} shows pressure as a function of baryon number density at zero temperature, where the curves correspond to all EOSs defined by the sets of characteristics listed in Table \ref{example_characteristics}. While most of the results are calculated in the presence of nucleons only, the thin dotted red line shows pressure for a system with both nucleons (protons and neutrons) and thermally excited $\Delta$ resonances. By construction, all of the EOSs display the same behavior for baryon number densities corresponding to ordinary nuclear matter and only start differing from each other in regions currently not constrained by experimental data, $n_B \gtrsim 1.5 n_{0}$.

A few regularities are apparent in the behavior of the pressure curves at zero temperature in regions corresponding to the QGP-like phase transition. Let us focus on the value of the pressure at the lower boundary of the spinonal region $P(\eta_L)$ (which is directly related to the average value of the pressure across the transition region), and compare its values for sets of characteristics between which only one property of the QGP-like phase transition changes substantially. First, $P(\eta_L)$ increases with the critical baryon number density $n_c^{(Q)}$, which can be seen by comparing the pressure curves for the second and third sets of characteristics (delineated with medium dashed green and thin dashed magenta lines, respectively). Second, $P(\eta_L)$ decreases with the critical temperature $T_c^{(Q)}$, as evidenced by pressure curves for the first and fourth sets of characteristics (delineated with thick dashed orange and solid purple lines, respectively). Third, $P(\eta_L)$ decreases with the width of the spinodal region, $\Delta \eta = \eta_R - \eta_L$, which can be seen by comparing pressure curves for the first and second sets of characteristics (thick dashed orange and medium dashed green lines, respectively). Furthermore, the magnitude of the drop in the pressure across the spinodal region $\Delta P = P(\eta_R) - P(\eta_L)$, increases with the critical temperature, as seen by comparing curves for the first and fourth sets of characteristics (thick dashed orange and solid purple lines, respectively). Importantly, these features create a physical bound on which QGP-like transitions are allowed in the VDF model. A transition with a wide spinodal region, with a critical point at a relatively low baryon number density and at the same time a relatively high critical temperature can often be excluded, as it leads to such a significant drop in the pressure across the spinodal region that the pressure becomes negative in some parts of the quark-hadron coexistence region, which would correspond to an unphysical ``QGP bound state''. This is because at $T=0$ the pressure is given by Eq.\ \eqref{pressure_Tzero}, and locally negative pressure implies that there exists a baryon density for which $d\inbfrac{\mathcal{E}}{n_B}/dn_B = 0$ and $d^2 \inbfrac{\mathcal{E}}{n_B}/ dn_B^2> 0$, corresponding to a local minimum in energy per particle, $\infrac{\mathcal{E}}{n_B}$. While such a minimum is in fact expected in the region of the phase diagram corresponding to ordinary nuclear matter, where $d\inbfrac{\mathcal{E}}{n_B}/dn_B = 0$ at the nuclear saturation density $n_0$, it is forbidden for large baryon number densities, where it would correspond to a metastable or even stable state of QGP. As an example, most obtained phase transitions with $n_c^{(Q)} = 2.5n_0$ and $T_c^{(Q)} \geq 125 \  \txt{MeV}$ are rejected based on this argument.

\begin{figure}[t]
	\centering\mbox{
	\includegraphics[width=0.65\columnwidth]{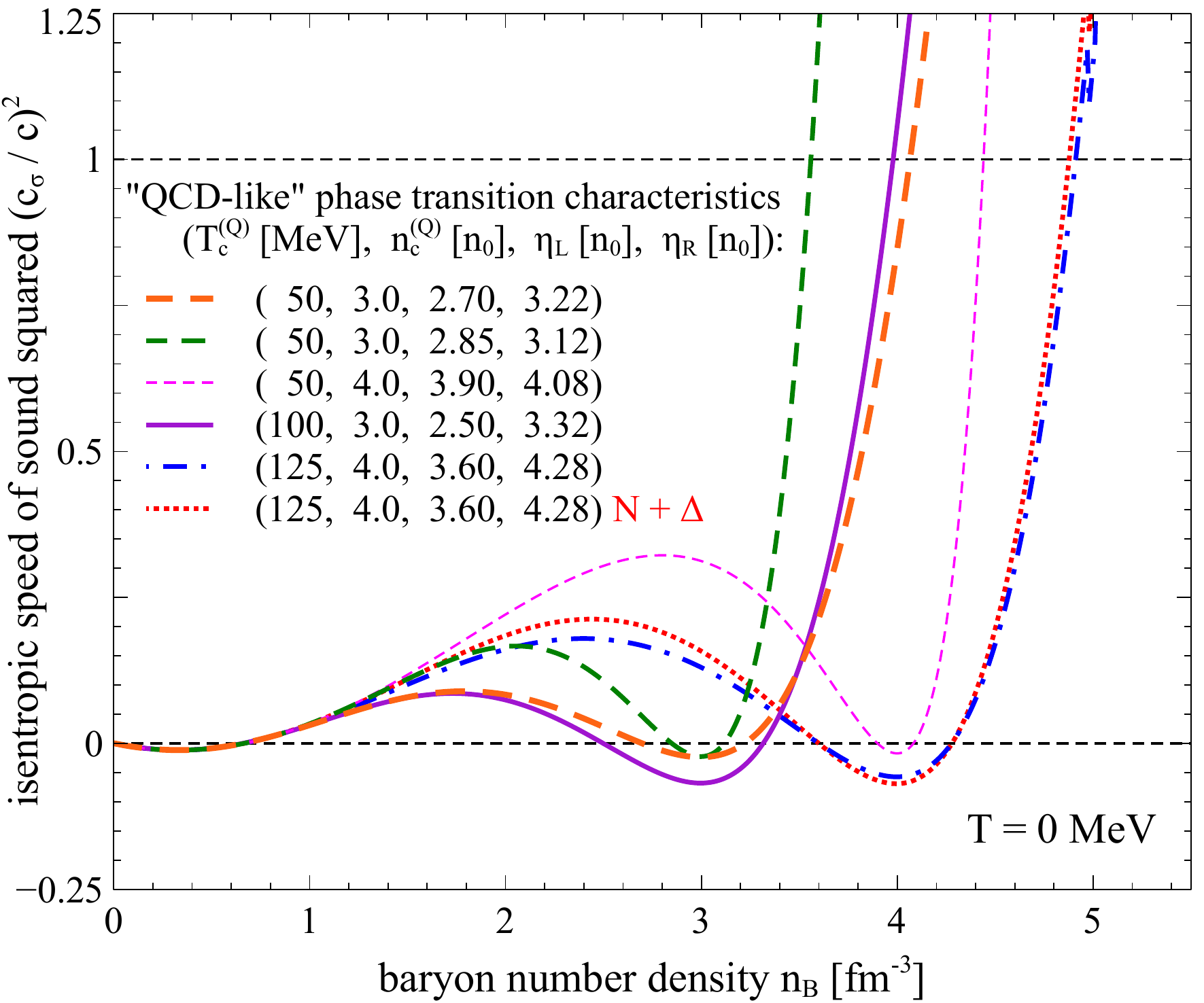} 
	}
	\caption[The isentropic speed of sound at $T=0$ as a function of baryon number density for several VDF EOSs]{The isentropic speed of sound $c_{\sigma}^2$ at $T=0$ as a function of baryon number density, plotted for all sets of characteristics listed in Table \ref{example_characteristics}. In the legend, the critical temperature of the QGP-like phase transition $T_c^{(Q)}$ is given in MeV, while the critical density $n_c^{(Q)}$ and the boundaries of the spinodal region at $T=0$, $\eta_{L}$ and $\eta_R$, are given in units of saturation density, $n_0 = 0.160 \ \txt{fm}^{-3}$. The speed of sound becomes acausal for relatively large baryon number densities above the quark-hadron transition region, which is a consequence of the hardness of the equation of state in the same region (see Fig.\ \ref{pressures_all}). See text for details.}
	\label{speed_of_sound} 
\end{figure}
Next, it is easy to notice that the pressure rises rapidly after leaving the quark-hadron transition region. This hardness of the EOS is a general feature of models based on high powers of baryon number density (specifically, with exponents higher than 2), and is ubiquitous among various Skyrme-type models (see, e.g., Ref.\ \cite{Dutra:2012mb}). In fact, it can be shown that any relativistic Lagrangian with vector-type interactions leading, in the mean-field approximation, to terms of the form $n_B^\alpha$, where $\alpha>2$, results in acausal phenomena at high baryon number densities \cite{Zeldovich:1962emp}. Indeed, Fig.\ \ref{speed_of_sound} shows the isentropic speed of sound squared $\left(c_{\sigma}/c\right)^2$ at $T=0$ for the chosen sets of phase transition characteristics listed in Table \ref{example_characteristics} (we note that at $T = 0$, the isothermal and isentropic speeds of sound are identical; for the derivation of the formula used to calculate the speed of sound, see Appendix \ref{the_speed_of_sound}). The speed of sound squared is negative within the spinodal region, as expected for a first-order phase transition \cite{Chomaz:2003dz}, while for large baryon number densities above the quark-hadron phase transition it eventually becomes acausal. Although the latter feature of the speed of sound is not ideal, such pathological behavior of the EOS can be, in fact, expected outside of the region in which its parameters are fitted. Moreover, it is not going to pose challenges to using the VDF model in a hadronic afterburner, where nuclear matter is simulated at densities below the coexistence region which in general are not affected by this problem. Out of the abundance of caution, in creating parameter sets we make sure that the speed of sound preserves causality for all baryon number densities below the upper boundary of the quark-hadron coexistence region. We note that in some of the studied phase transitions this still allows for a speed of sound above the conformal bound of $\left(\infrac{c_\sigma}{c}\right)^2 \leq \infrac{1}{3}$; it is presently unclear whether this bound is satisfied in dense nuclear matter, see, e.g., Refs.\ \cite{Fujimoto:2019hxv, Bedaque:2014sqa, McLerran:2018hbz, Annala:2019puf}.

\begin{figure}[t]
	\centering\mbox{
	\includegraphics[width=0.65\textwidth]{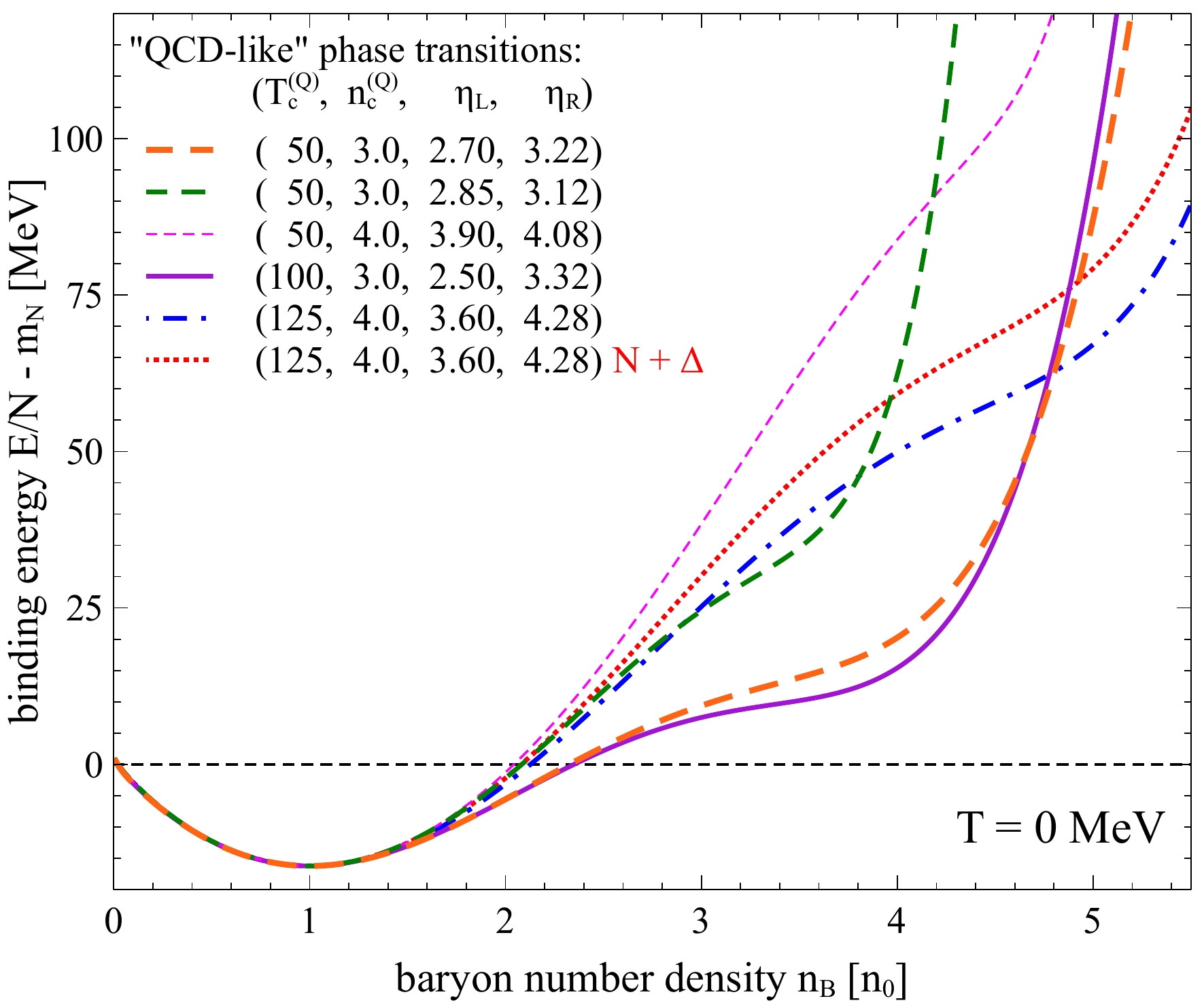} 
	}
	\caption[Binding energy at $T=0$ as a function of baryon number density for several VDF EOSs]{The binding energy at $T=0$ as a function of baryon number density, plotted for all sets of characteristics listed in Table \ref{example_characteristics}. In the legend, the critical temperature of the QGP-like phase transition $T_c^{(Q)}$ is given in MeV, while the critical density $n_c^{(Q)}$ and the boundaries of the spinodal region at $T=0$, $\eta_{L}$ and $\eta_R$, are given in units of saturation density, $n_0 = 0.160 \ \txt{fm}^{-3}$. The degree of the softening in energy per particle at high baryon number density is directly related to the width of the spinodal region of a given EOS (see text for more details).}
	\label{binding_energy_plots} 
\end{figure}

Finally, in Fig.\ \ref{binding_energy_plots} we show the binding energy $\mathcal{E}_{(4)}/n_B - m_N$ at $T=0$ as a function of baryon number density for EOSs corresponding to all sets of characteristics listed in Table \ref{example_characteristics}. All obtained EOSs describe the same physics in the region $n_B \lesssim 1.5 n_0$, where the behavior of nuclear matter is relatively well known; in particular, all curves reproduce the value of the chosen binding energy at nuclear matter saturation as well as the location of the saturation density, Eq.\ \eqref{enforced_saturation_density_and_binding_energy}. On the other hand, at high densities the binding energy displays a softening related to the postulated QGP-like phase transition, which is different for each of the considered EOS. We note that the extent of this softening is directly related to the width of the spinodal region of a given EOS. This can again be seen from the fact that at zero temperature the pressure is given by Eq.\ \eqref{pressure_Tzero}, from which it immediately follows that the curvature of the energy density, $\inbfrac{d^2 \mathcal{E}}{dn_B^2}$, must be negative in the spinodal region; consequently, the region over which $\inbfrac{d^2 \mathcal{E}}{dn_B^2} <0$ is related to $(\eta_L, \eta_R)$.

\begin{figure}[t]
	\centering\mbox{
		\includegraphics[width=0.89\textwidth]{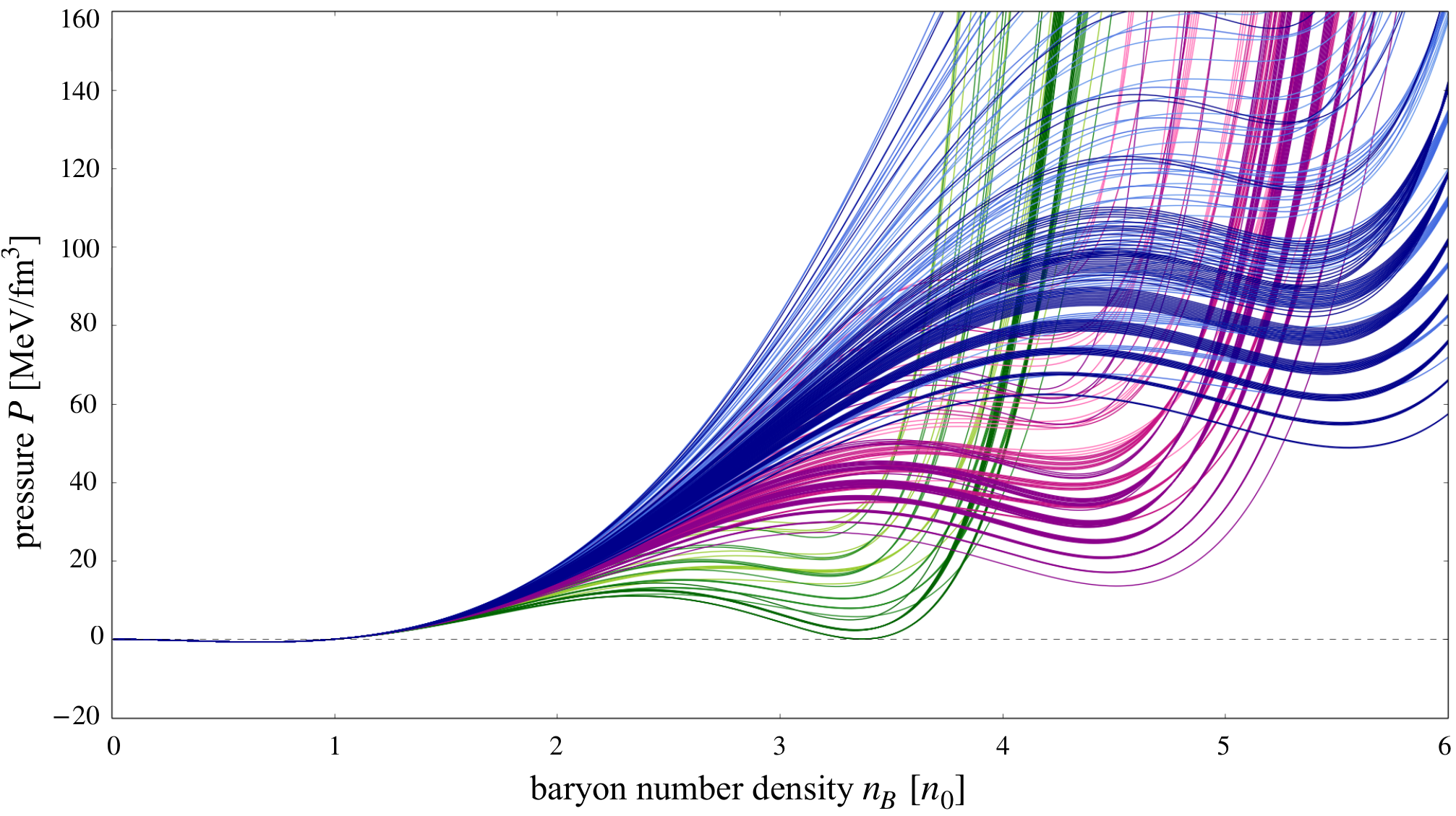} 
	}
	\caption[Pressure at $T=0$ as a function of baryon number density for 283 VDF EOSs]{Pressure as a function of baryon number density at $T=0$ for 283 VDF EOSs. Green, magenta, and blue curves mark EOSs with the critical density of $n_c = 3.0 n_0$, $4.0 n_0$, and $5.0 n_0$, respectively, while light, medium dark, and dark curves mark EOSs with the critical temperature $T_c = 50$, $100$, and $150\ \txt{MeV}$. For each $(n_c, T_c)$, the left boundary of the spinodal region at $T=0$ was set a series of values given by $\eta_L = n_c - 0.1 - i\times0.05n_0$, with $i = \{0, 1, 2, \dots\}$. Small variations in  the right boundary of the spinodal region at zero temperature $\eta_R$, explained further in Section \ref{phase_diagrams}, result in ``bundles'' of EOSs that correspond to slightly differing $\eta_R$'s for a given set of the EOS characteristics $(n_c, T_c, \eta_L)$.}
	\label{all_pressures_favorite_plot_2}
\end{figure}

Although we have only shown results corresponding to a few possible QGP-like phase transitions, it should be understood that arbitrarily many versions of the dense nuclear matter EOS can be obtained in the VDF model, as can be seen in Fig.\ (\ref{all_pressures_favorite_plot_2}), which shows pressure curves at $T=0$ for all EOSs obtained by demanding the critical density of $n_c^{(Q)} \in \{ 3.0, 4.0, 5.0\}\  [n_0]$ (green, magenta, and blue curves, respectively), the critical temperature of $T_c^{(Q)} \in \{ 50, 100, 150 \} \  [\txt{MeV}]$ (light, medium, and dark curves, respectively), and the lower boundary of the spinodal region varied according to $\eta_L = n_c - 0.1 - i \times 0.05n_0$, where $i = \{0, 1, 2, \dots\}$. While the obtained EOSs vary widely in the high baryon density region, by construction they all reproduce the same physics in the range of baryon number densities corresponding to ordinary nuclear matter, $n_B \lesssim 1.5n_0$. In fact, fitting the VDF model to reproduce the experimental values of the saturation density, the binding energy, and the nuclear critical point gives a remarkably good prediction for the value of pressure at the nuclear critical point $P_c$ and the value of incompressibility at the saturation point $K_0$ as compared with experiment and against other models (summarized in Table \ref{table_comparison_vdf_vs_other_models}). This is partially expected, as the value of the incompressibility $K_0$ depends strongly on the critical temperature \cite{Kapusta:1984ij}, and the latter is used in the VDF model as an input parameter fixed based on the known experimental results. Nevertheless, it is noteworthy that the minimal VDF model, based on a few characteristics taken at their experimentally established values (here $n_0$, $B_0$, $T_c^{(N)}$, $n_c^{(N)}$), leads to predictions for other properties of nuclear matter agreeing remarkably well with experimental data. Apparently, constraining four properties of the EOS is enough to reproduce the thermodynamic behavior of nuclear matter in the fitted region. The same could be true in the case of nuclear matter at high baryon number density. We may be hopeful that postulating QGP-like phase transition characteristics that happen to lay close to their true QCD values will lead to a VDF model parametrization correctly describing other properties of dense nuclear matter in the transition region. We expect that this correct description would manifest itself through agreement of simulation results with experimental data.

\begin{table}[t]
	\caption[Comparison of values of $T_c^{(N)}$, $n_c^{(N)}$, $P_c$, and $K_0$ as obtained in experiment and in models]{Comparison of the nuclear phase transition critical temperature $T_c^{(N)} \ [\txt{MeV}]$, the critical baryon number density $n_c^{(N)} \ [\txt{fm}^{-3}]$, pressure at the critical point $P_c \ [\txt{MeV fm}^{-3}]$, and incompressibility $K_0 \ [\txt{MeV}]$ obtained in experiment \cite{Elliott:2013pna} and in various models: the Walecka model \cite{Walecka:1974qa, Chin:1974sa}, the quantum Van der Waals model \cite{Poberezhnyuk:2017yhx} (denoted with ``Quantum VdW''), the VDF model with nuclear phase transition only (using two interaction terms and denoted by ``V2DF (N)''), and the VDF model with both nuclear and quark-hadron phase transitions (using four interaction terms and denoted by ``V4DF (N+Q)''). For the latter, values of $P_c$ and $K_0$ are given as averages calculated across all obtained EOSs for quark-hadron critical temperatures $T_c^{(Q)} \in \{ 50, 100, 150 \} \  [\txt{MeV}]$ and critical baryon number densities $n_c^{(Q)} \in \{ 3.0, 4.0, 5.0\}\  [n_0]$, see Fig.\ \ref{all_pressures_favorite_plot_2}. Asterisks mark input values of the models.}  
	\label{table_comparison_vdf_vs_other_models}
	\begin{center}
		\bgroup
		\def\arraystretch{1.4}
		\begin{tabular}{ l c c c c c}
			\hline
			\hline
			& \hspace{1mm}Experiment\hspace{1mm} & \hspace{1mm}Walecka\hspace{1mm} & 
			\hspace{0mm}Quantum VdW\hspace{0mm} & 
			\hspace{3mm}V2DF (N)\hspace{3mm}& \hspace{3mm}V4DF (N+Q)\hspace{3mm} \\ 
			\hline
			$T_c^{(N)}$ & $17.9 \pm 0.4$& 18.9 & 19.7 & 18*& 18* \\  
			$n_c^{(N)}$ & $0.06 \pm 0.01$ & 0.070 & 0.072 & 0.06*& 0.06* \\  
			$P_c $ & $0.31 \pm 0.07$  & 0.48 & 0.52 & 0.311 & $\langle P_c \rangle = 0.3067 \pm 0.0014$ \\  
			$K_0 $ & 230-315 & 553 & 763 & 282 & $\langle K_0 \rangle = 273.1 \pm 5.1$\\
			\hline
			\hline
		\end{tabular}
		\egroup
	\end{center}
\end{table}

\section{VDF model results: Phase diagrams}
\label{phase_diagrams}

\begin{figure}[t]
	\centering\mbox{
	\includegraphics[width=0.65\textwidth]{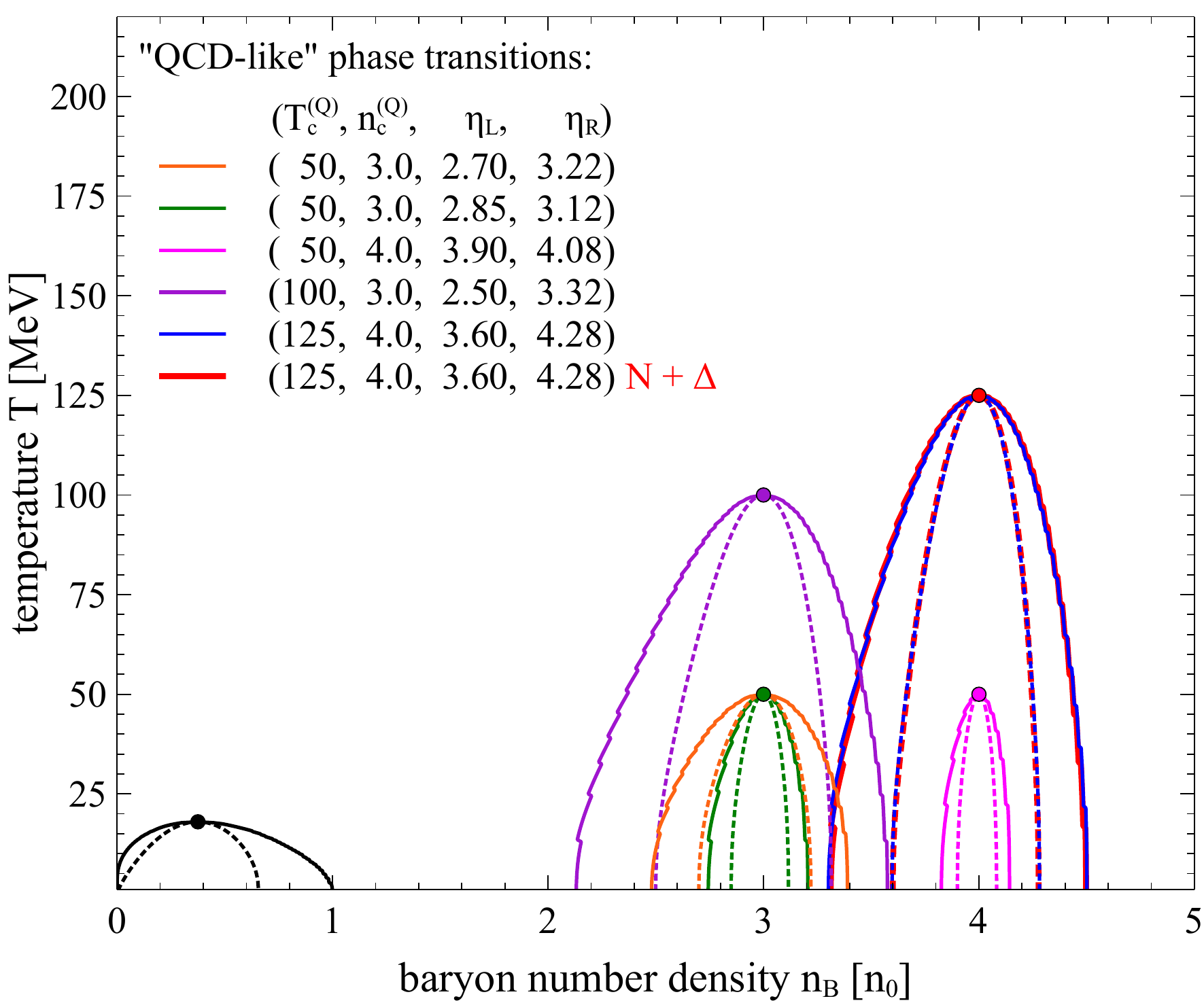}}
	\caption[Phase diagrams in the $(T,n_B)$ plane for several VDF EOSs]{Phase diagrams in the $(T,n_B)$ plane for sets of characteristics listed in Table \ref{example_characteristics}. Solid (dashed) lines represent the boundaries of the coexistence (spinodal) regions. In the legend, the critical temperature of the QGP-like phase transition $T_c^{(Q)}$ is given in MeV, while the critical baryon number density $n_c^{(Q)}$ and the boundaries of the spinodal region, $\eta_{L}$ and $\eta_R$, are given in units of saturation density, $n_0 = 0.160 \ \txt{fm}^{-3}$. The coexistence and spinodal regions of the nuclear phase transition, depicted with black lines, are common to all shown EOSs. The phase diagrams of sets V and VI coincide, which is expected.}
	\label{phase_diagram_nB} 
\end{figure}

\begin{figure}[t]
	\centering\mbox{
	\includegraphics[width=0.65\textwidth]{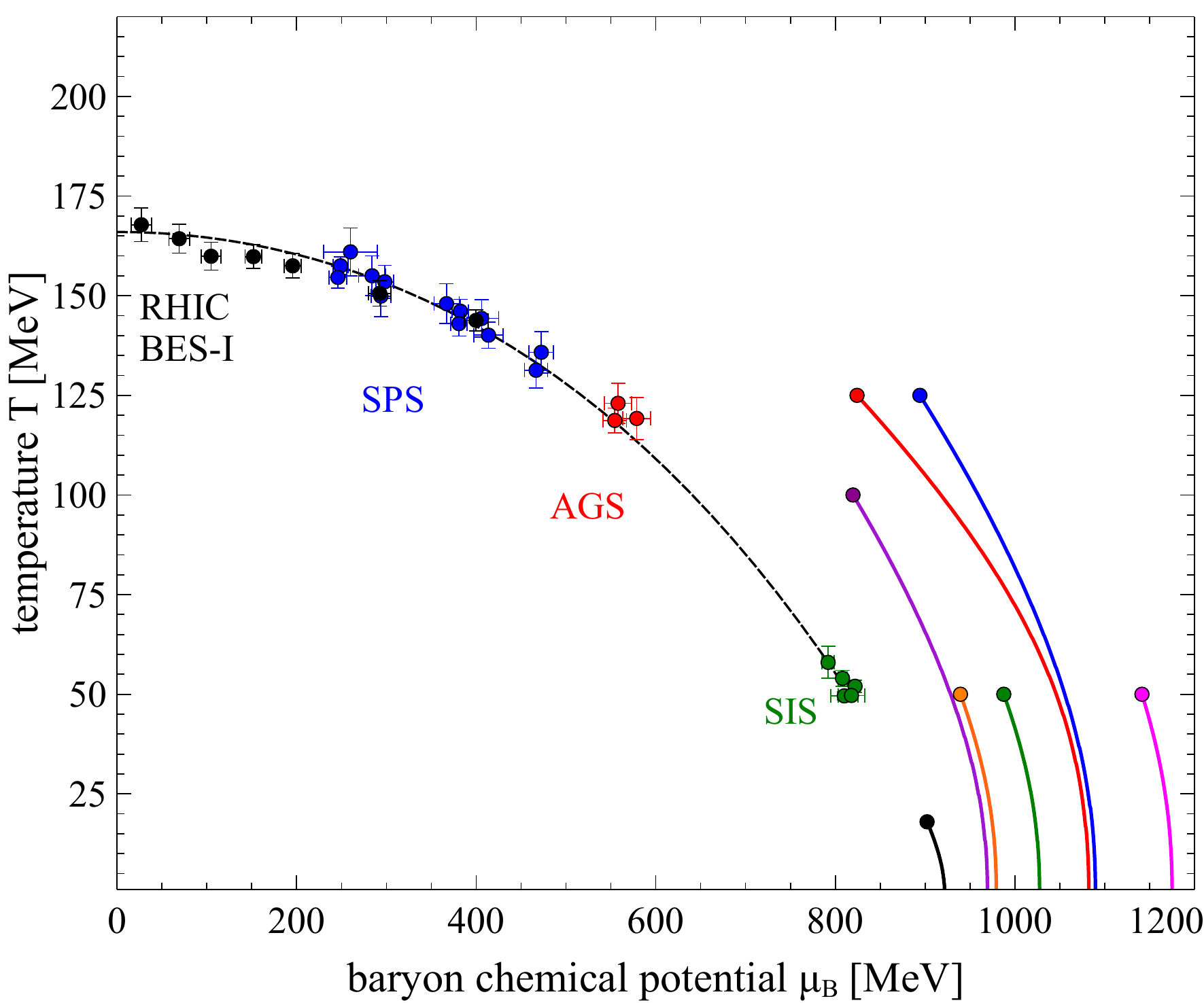} }
	\caption[Phase diagrams in the $(T,\mu_B)$ plane for several VDF EOSs]{Phase diagrams in the $(T,\mu_B)$ plane for sets of characteristics listed in Table \ref{example_characteristics}. The legend is the same as in Fig.\ \ref{phase_diagram_nB}. Solid lines represent the coexistence lines. The coexistence line of the nuclear phase transition, depicted with a solid black line, is common to all sets of characteristics. Also shown are chemical freeze-out points obtained in experiment and a parametrization of the freeze-out line from \cite{Cleymans:2005xv}. The noticeably different curvature of the coexistence line at high temperatures for set IV (solid red line) is due to employing both nucleons and thermally induced Delta resonances.}
	\label{phase_diagram_muB} 
\end{figure}

The phase diagrams for the EOSs corresponding to the characteristics listed in Table \ref{example_characteristics} are shown in Figs.\ \ref{phase_diagram_nB} and \ref{phase_diagram_muB}. Solid and dashed lines represent the boundaries of the coexistence and spinodal regions, respectively. The coexistence and spinodal regions of the nuclear phase transition, depicted with black lines, are common for all used EOSs by construction. For sets V and VI, which differ only by the number of species contributing  to the baryon number density $n_B$, the phase diagrams in Fig.\ \ref{phase_diagram_nB} coincide, which is expected.

It is immediately apparent that the QGP-like coexistence curves in the phase diagrams all look alike. This is a consequence of our choice to employ only interactions depending on vector baryon number density, as in this case the dependence of the thermal part of the pressure on temperature $T$ and effective chemical potential $\mu^*$ is just like that of an ideal Fermi gas, as can be seen from Eq.\ (\ref{summary_pressure}). Consequently, all VDF EOSs display similar behavior with increasing temperature $T$. This can be especially easily seen in the $(T,\mu_B)$ phase diagram (Fig.\ \ref{phase_diagram_muB}), where the coexistence lines exhibit the exact same curvature. An exception from this behavior as shown on this phase diagram is the curve calculated for a system with both nucleons and thermally produced $\Delta$ resonances (denoted with a red line), which bends more forcefully towards the $\mu_B = 0$ axis as the temperature increases. This is to be expected, as including an additional baryon species lowers the value of the baryon chemical potential for a given baryon number density. Including even more baryon species would strengthen this effect. 

Another feature, easily discerned in the $(T,n_B)$ phase diagram (Fig.\ \ref{phase_diagram_nB}), is that the spinodal regions $[\eta_L, \eta_R]$ (and likewise the coexistence regions $[n_L, n_R]$) are always approximately centered around the critical baryon number density $n_c^{(Q)}$. This is again a consequence of the fact that vector-like interactions do not have a temperature dependence, so that the thermal contribution to the pressure is just like that of an ideal gas of fermions with mass $m_N$ (for a detailed explanation see Appendix \ref{symmetric_spinodal_regions}). As a result, the critical baryon number density $n_c^{(Q)}$ and the boundaries of the spinodal region, $\eta_L$ and $\eta_R$, are not independent. In consequence, we have effectively one less free parameter. For example, once we set the ordinary nuclear matter properties, the critical point of the quark-hadron phase transition, and the lower spinodal boundary at $T=0$, $\eta_L$, the right spinodal boundary at $T=0$, $\eta_R$, is practically fixed. This could be already observed in Fig.\ \ref{all_pressures_favorite_plot_2}, where the minimal allowed changes in $\eta_R$, given $(n_c, T_c, \eta_L)$, result in easily discernible ``bundles'' of EOSs.

We expect that all these regularities in the behavior of the spinodal and coexistence lines would not be as prominent if scalar-type interactions were included, rendering the thermal part of the pressure non-trivial. In particular, we expect that this would allow us to obtain coexistence regions bending towards the $n_B = 0$ axis in the $(T,n_B)$ plane, which would correspond to an even stronger tendency to bend towards the $\mu_B=0$ axis in the $(T,\mu_B)$ plane. This expectation is based on the fact that, typically, scalar interactions result in a small effective mass, which in addition decreases with temperature, and that in turn produces a relatively larger thermal contribution to the pressure for a given $n_B$ and $T$. As a result, such phase transitions would more significantly affect the region of the phase diagram covered by the Beam Energy Scan program at RHIC. Parametrizations of the VSDF model leading to such effects are planned for the near future.

\section{VDF model results: Cumulants of baryon number}
\label{Cumulants_of_baryon_number}

In analyses of heavy-ion collision experiments, considerable attention has been paid to cumulants of the baryon number distribution. In the grand canonical ensemble, the $j$th cumulant of the baryon number $\kappa_j$ can be calculated from
\begin{eqnarray}
\kappa_j = VT^{j-1} \frac{d^j P}{d \mu_B^j} ~,
\label{cumulants_definition_pressure}
\end{eqnarray}
and the first four cumulants in terms of derivatives with respect to the baryon number density were already given in Eqs.\ (\ref{cumulant_1}-\ref{cumulant_4}), repeated here for convenience,
\begin{eqnarray}
&& \hspace{-15mm} \kappa_1 = Vn_B = N_B ~, 
\label{cumulant_10} \\
&& \hspace{-15mm} \kappa_2 = \frac{VTn_B}{\left( \frac{dP}{dn_B} \right)_T} ~, 
\label{cumulant_20}  \\
&& \hspace{-15mm} \kappa_3 = \frac{VT^2 n_B}{\left( \frac{dP}{dn_B} \right)_T^2} \left[ 1 - \frac{n_B}{\left( \frac{dP}{dn_B} \right)_T}  \left( \frac{d^2P}{dn_B^2} \right)_T  \right] ~, 
\label{cumulant_30} \\
&& \hspace{-15mm} \kappa_4 = \frac{VT^3 n_B}{\left(\frac{dP}{dn_B}  \right)_T^3}  \left[  1    -   \frac{4n_B}{\left(\frac{dP}{dn_B}  \right)_T} \left(\frac{d^2P}{dn_B^2}\right)_T +       \frac{3n_B^2}{\left(\frac{dP}{dn_B}  \right)_T^2}   \left(\frac{d^2P}{dn_B^2}\right)_T^2    -   \frac{n_B^2}{\left(\frac{dP}{dn_B}  \right)_T} \left(\frac{d^3P}{dn_B^3}  \right)_T \right] ~.
\label{cumulant_40} 
\end{eqnarray}
In the VDF model, analytic expressions for derivatives of the pressure $\inderr{^jP}{n_B^j}$ are available, and in particular we provide these formulas up to $j=3$ in Appendix \ref{pressure_derivatives_in_the_VDF_model}. Cumulants may also be expressed in terms of derivatives of the baryon chemical potential $\inderr{^j \mu_B}{n_B^j}$; we provide these formulas up to the sixth-order cumulant in Appendix \ref{cumulants_in_terms_of_derivatives_of_the_chemical_potential}. Which expressions are used is purely a question of preference.

\begin{figure}[t]
	\includegraphics[width=0.99\textwidth]{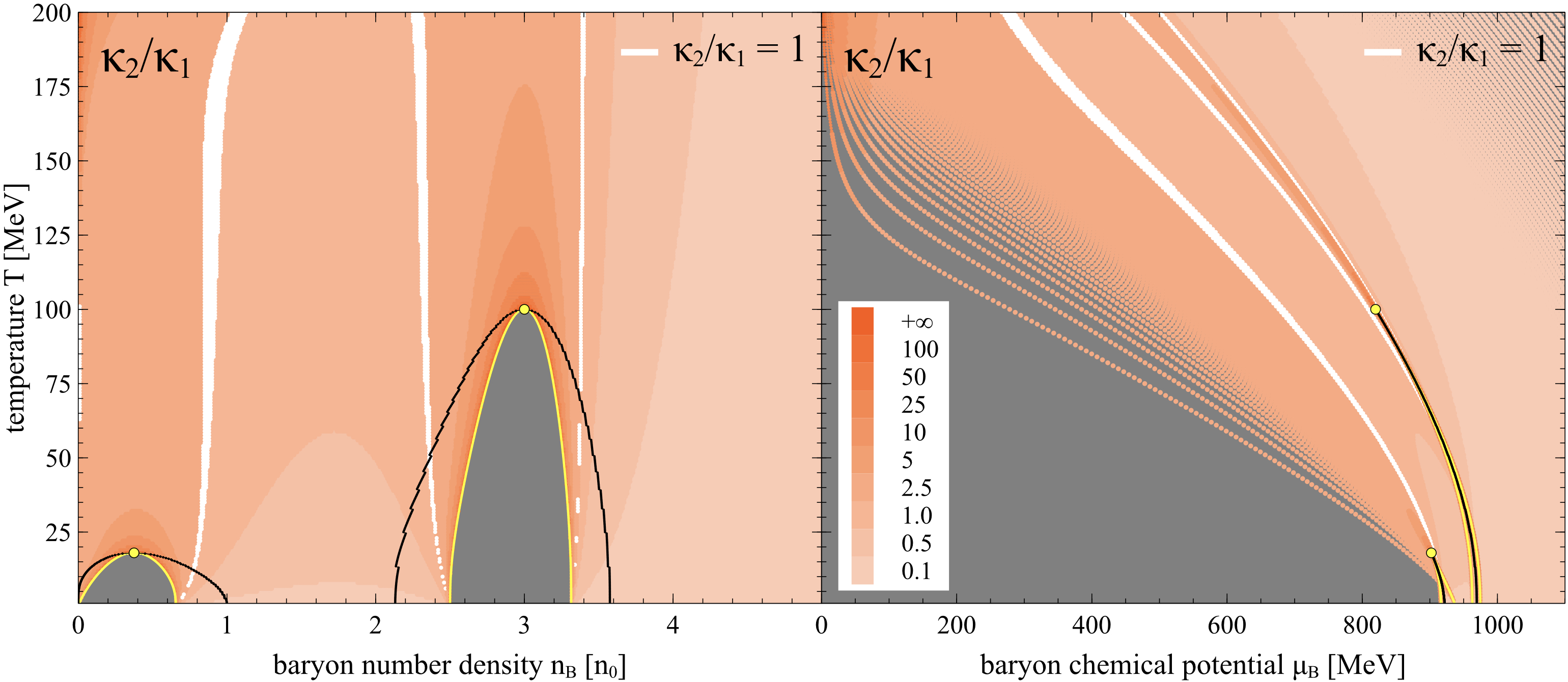} 
	\caption[Contour plots of the cumulant ratio $\kappa_2/\kappa_1$ for a VDF EOS]{Contour plots of the cumulant ratio $\kappa_2/\kappa_1$ for the fourth (IV) EOS listed in Table \ref{example_characteristics}. Coexistence (spinodal) regions are denoted with black (yellow) lines, and critical points are marked with yellow dots. White regions correspond to $\kappa_2/\kappa_1= 1 \pm 0.03$. Legend entries denote upper limits of values of the cumulant ratios.
	}
	\label{Cumulants_diagrams_k2k1}
\end{figure}

While cumulants measure the derivatives of the EOS, Eq.\ (\ref{cumulants_definition_pressure}), at the same time they can be related to the moments of the baryon number distribution, see Eqs.\ (\ref{cumulant_1_in_terms_of_moments}-\ref{cumulant_4_in_terms_of_moments}). Because cumulants of the proton number distribution (serving as a proxy for the cumulants of the baryon number distribution) can be measured in experiment, cumulants provide one of the strongest links between theoretical predictions and experimental data. The experimental values of cumulants are expected to be influenced by enhanced fluctuations of conserved charges in the vicinity of the critical point, rendering them a signal for the existence of the critical point and a first-order phase transition in QCD \cite{Stephanov:1998dy, Stephanov:1999zu, Koch:2008ia}. In particular, it is argued that, for systems crossing the phase diagram close to and above the critical point, the sign of the third-order cumulant $\kappa_3$ will change once \cite{Asakawa:2009aj}, while the sign of the fourth-order cumulant $\kappa_4$ will change twice \cite{Stephanov:2011pb}. This is explained at length in Section \ref{non-statistical_event-by-event_fluctuations_of_conserved_charges}.

\begin{figure}[t]
	\includegraphics[width=0.99\textwidth]{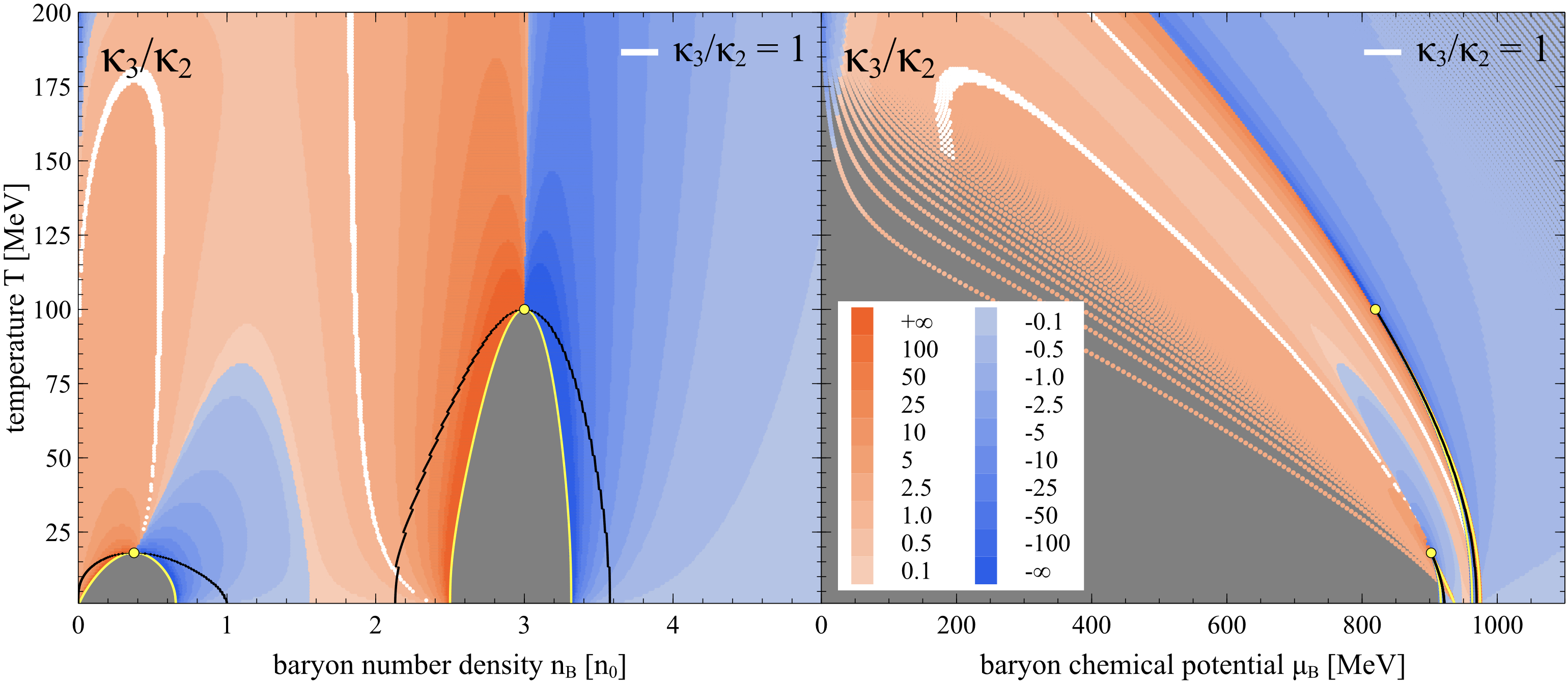} 
	\caption[Contour plots of the cumulant ratio $\kappa_3/\kappa_2$ for a VDF EOS]{Contour plots of the cumulant ratio $\kappa_3/\kappa_2$ for the fourth (IV) EOS listed in Table \ref{example_characteristics}. Coexistence (spinodal) regions are denoted with black (yellow) lines, and critical points are marked with yellow dots. White regions correspond to $\kappa_3/\kappa_2= 1 \pm 0.03$. Legend entries denote upper (lower) limits of positive (negative) values of the cumulant ratios.
	}
	\label{Cumulants_diagrams_k3k2}
\end{figure}

The explicit volume dependence of the cumulants, which is typically divided out in theoretical calculations, is difficult to control in experiment. Therefore, it is customary to consider ratios of cumulants, most commonly
\begin{eqnarray}
\frac{\sigma^2}{\mu} = \frac{\kappa_2}{\kappa_1}~ , \hspace{10mm}  S \sigma= \frac{\kappa_3}{\kappa_2}~, \hspace{10mm}  \kappa \sigma^2 = \frac{\kappa_4}{\kappa_2}~,
\label{cumulant_ratios}
\end{eqnarray}
where $\mu$ denotes the mean, $\sigma^2$ denotes variance, $S$ denotes skewness, and $\kappa$ denotes excess kurtosis. Preliminary results from the Beam Energy Scan indeed suggest that the fourth-order cumulant ratio $\infrac{\kappa_4}{\kappa_2}$ exhibits non-monotonic behavior with the collision energy \cite{STAR:2021iop}.

\begin{figure}[t]
	\includegraphics[width=0.99\textwidth]{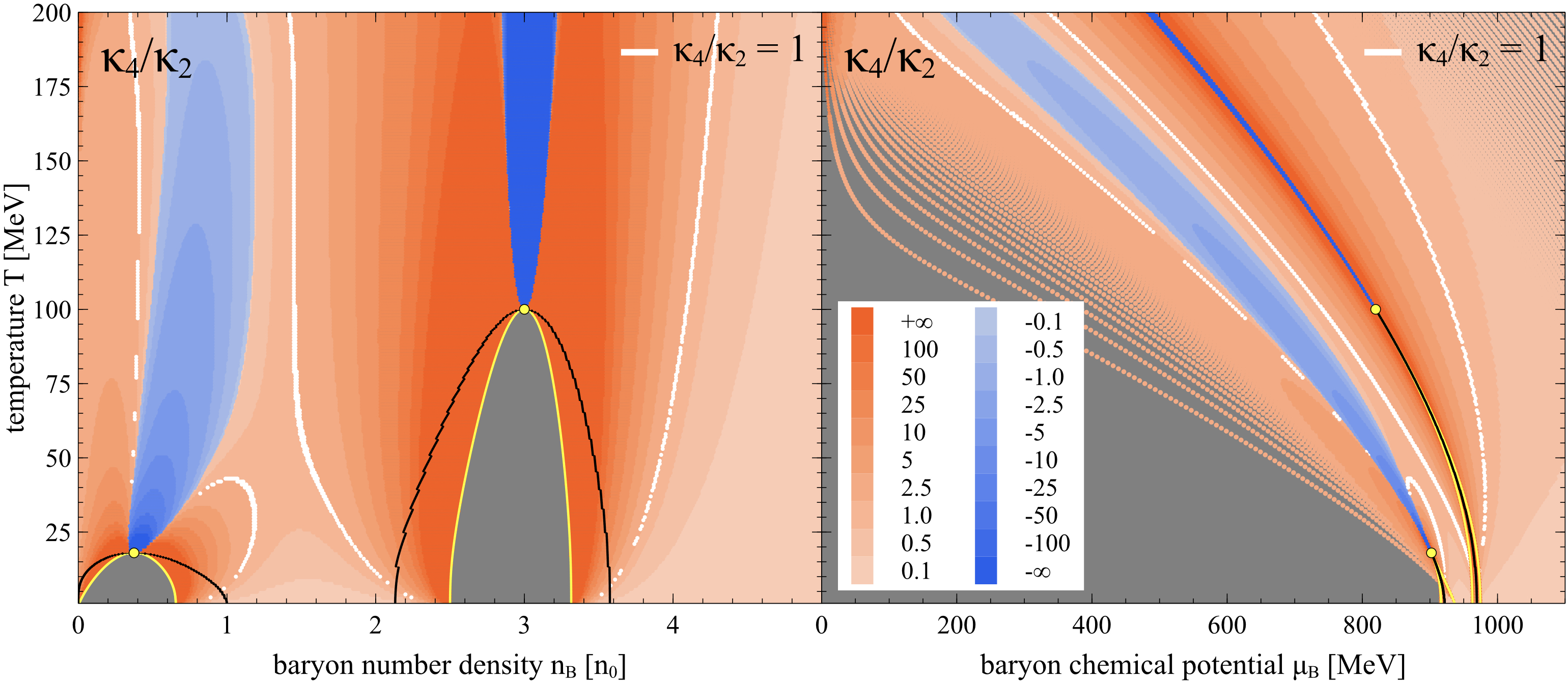} 
	\caption[Contour plots of the cumulant ratio $\kappa_4/\kappa_2$ for a VDF EOS]{Contour plots of the cumulant ratio $\kappa_4/\kappa_2$ for the fourth (IV) EOS listed in Table \ref{example_characteristics}. Coexistence (spinodal) regions are denoted with black (yellow) lines, and critical points are marked with yellow dots. White regions correspond to $\kappa_4/\kappa_2= 1 \pm 0.03$. Legend entries denote upper (lower) limits of positive (negative) values of the cumulant ratios.
	}
	\label{Cumulants_diagrams_k4k2}
\end{figure}

In this section, we will focus on results for the fourth (IV) set of characteristics listed in Table \ref{example_characteristics}. The choice of this set is arbitrary and does not reflect any preference for the location of the QCD critical point, but simply serves as an illustration of the properties of the VDF model which are qualitatively comparable for all obtained EOSs. In Figs.\ \ref{Cumulants_diagrams_k2k1}, \ref{Cumulants_diagrams_k3k2}, and \ref{Cumulants_diagrams_k4k2} we plot the cumulant ratios (\ref{cumulant_ratios}) in the $(T,n_B)$ and $(T,\mu_B)$ planes; we note here that gray areas on these diagrams signify regions in which either the cumulant calculation is invalid (inside the spinodal regions, visible in the $(T,n_B)$ diagrams) or where data has not been produced (at very small values of $n_B$, which affects the $(T,\mu_B)$ diagrams). Dramatic increase in magnitudes of cumulant ratios as well as sudden changes in sign, observed in regions close to and above the critical point, agree with the expectations for the behavior of the cumulants described above and in Section \ref{non-statistical_event-by-event_fluctuations_of_conserved_charges}. Interestingly, the effects of the nuclear phase transition are clearly present even at very high temperatures (this has been also observed in Ref.\ \cite{Vovchenko:2016rkn}). This raises the question to what extent the presence of the nuclear phase transition affects the interpretation of experimental data, either by damping the signal originating at the QGP phase transition, or by acting as an imposter. Such questions could be answered by comparing outcomes of simulations utilizing a VDF EOS with either nuclear phase transition only, or both nuclear and quark-hadron phase transitions. Studies of this type are planned for future research and are beyond the scope of this thesis.

\section{Summary}

In this chapter, we parametrized the vector density functional (VDF) model to describe two first-order phase transitions: the experimentally observed nuclear liquid-gas phase transition, and a postulated high-temperature, high-density phase transition intended to model the QGP phase transition. We then described gross properties of the obtained family of EOSs based on several EOSs which are representative of the range of EOSs available in the parametrized family. In particular, we focused on the behavior of the pressure, the speed of sound, and the energy per particle as functions of the baryon density, as well as on the features of the phase diagrams in the $(T,n_B)$ and $(T,\mu_B)$ planes and the cumulants of baryon number.

\newpage
\chapter{Implementation in a hadronic transport code}
\label{implementation}

A significant part of work underlying this thesis was devoted to implementing and testing the VDF equations of motion, Eqs.\ (\ref{EOM_covariant_formulation_x}) and (\ref{EOM_covariant_formulation_p}) with $M=0$, in the hadronic transport code \texttt{SMASH} \cite{Weil:2016zrk}. This implementation is available in \texttt{SMASH} starting with the \texttt{SMASH}-2.1 release. As a hadronic transport code, \texttt{SMASH} simulates hadronic non-equilibrium dynamics through numerically solving the Boltzmann equation. Below, we briefly sketch the theoretical foundations of hadronic transport simulations and discuss some of the technical details of the implementation. A brief introduction to the kinetic theory is provided in Appendix \ref{basics_of_the_kinetic_theory}.

\section{Transport simulations}
\label{transport_simulations}

The contents of this section are largely based on Ref.\ \cite{Sorensen:2020ygf}. 

In statistical physics, a transport equation describes the time evolution of a distribution function of a system of $N$ particles $f_N(\bm{x},\bm{p})$, where we use a short-hand notation $\bm{x} = (\bm{x}_1, \dots, \bm{x}_N)$ and $\bm{p} = (\bm{p}_1, \dots, \bm{p}_N)$. The distribution $f_N(\bm{x},\bm{p})$ can be understood as the density of particles in the phase space, or equivalently as the probability that the $i$-th particle has a given position $\bm{x}_i$ and a given momentum $\bm{p}_i$, $i = 1, \dots, N$. From the Liouville theorem (see Appendix \ref{the_Liouville_theorem}) we know that the time evolution of $f_N$ satisfies 
\begin{eqnarray}
\frac{d f_N}{dt} = \left( \parr{f_N}{t} \right)_{\txt{drift}} + \frac{d\bm{x}}{dt} \parr{f_N}{\bm{x}} + \frac{d\bm{p}}{dt} \parr{f_N}{\bm{p}} = 0 ~.
\label{the_Liouville_theorem_implementation}
\end{eqnarray}

In practice, many of the most relevant features of the system are described by the 1-body distribution function $f_1(\bm{x}_1,\bm{p}_1)$, which gives the probability that any of the $N$ particles in the system has a given position $\bm{x}_1$ and a given momentum $\bm{p}_1$. The time evolution of $f_1(\bm{x}_1,\bm{p}_1)$ is given by
\begin{eqnarray}
\parr{f_1}{t} +  \derr{\bm{x}_1}{t} \parr{f_1}{\bm{x}_1} + \derr{\bm{p}_1}{t} \parr{f_1}{\bm{p}_1} = I_{\txt{coll}}~, 
\label{transport_equation_general}
\end{eqnarray}
where on the right-hand side we introduced the collision integral $I_{\txt{coll}}$, which is an often-used common notation for all terms contributing to changes in $f_1$ due to particle-particle collisions (particles colliding into or out of the phase space volume element) and particle transformations such as decays or resonance formation. If the collision term is constructed according to the assumption of molecular chaos (meaning that particles are not correlated neither before nor after a collision takes place), also known as the \textit{Stosszahlansatz}, then Eq.\ \eqref{transport_equation_general} is known as the Boltzmann equation. A formal derivation of the Boltzmann equation from the Liouville theorem, Eq.\ \eqref{the_Liouville_theorem_implementation}, can be obtained using the BBGKY hierarchy of equations (see Appendix \ref{the_Boltzmann_equation} for a simplified derivation of the Boltzmann equation from the BBGKY hierarchy, which is itself derived in Appendix \ref{the_BBGKY_hierarchy}). In the following, we will suppress the subscript ``1'' and assume that $f(\bm{x}, \bm{p})$ always refers to a 1-body distribution function.

In the context of heavy-ion collision simulations, the Boltzmann equation is often also called the Boltzmann-Uehling-Uhlenbeck (BUU) equation, the Vlasov-Uehling-Uhlenbeck (VUU) equation, or the Boltzmann-Nordheim-Vlasov (BNV) equation. The specification comes from considering the phase space distribution in the presence of a self-consistent mean field $U(\bm{x},\bm{p})$ entering the dynamics of the system through the single-particle Hamiltonian $H_{(1)} = \sqrt{ \bm{p}^2 + m^2} + U(\bm{x},\bm{p})$,
\begin{eqnarray}
\bigg[ \parr{}{t} + \parr{H_{(1)}}{\bm{p}} \bnabla_{\bm{x}} - \parr{H_{(1)}}{\bm{x}} \bnabla_{\bm{p}} \bigg]  f(t, \bm{x},\bm{p}) =   I_{\txt{coll}}~,
\label{BUU_equation}
\end{eqnarray} 
where we inserted Hamilton's equations of motion into the Boltzmann equation \eqref{transport_equation_general}. In cases where there are no collisions, $I_{\txt{coll}} = 0$, the term ``Vlasov equation'' is also often used.

The function $f(t, \bm{x},\bm{p})$ is a continuous distribution function for a given total number of nucleons, denoted by $A$, and solving the Boltzmann equation is equivalent to obtaining the time evolution of $f(t, \bm{x},\bm{p})$. (We note here that in general, heavy-ion collision simulations evolve hundreds of hadron species, described by a system of hundreds of coupled Boltzmann equations; the following discussion, treating explicitly the Boltzmann equation in the presence of nucleons only, can be straightforwardly generalized to include all necessary particle species.) Numerically, given the initial condition in form of the distribution function at some time $t_0$, $f(t_0, \bm{x}_0,\bm{p}_0)$, we solve for the distribution at a slightly later time $t = t_0 + \delta t$, $f(t_0 + \delta t, \bm{x},\bm{p})$, and repeat the process until a final time $t = t_{\txt{end}}$ is reached. In more detail, the numerical solution of the VUU equation is achieved within a numerical approach known as the method of test particles \cite{Wong:1982zzb}, which is based on the assumption that the continuous phase space distribution $f(t,\bm{x},\bm{p})$ of a system of $A$ nucleons can be approximated by the distribution of a large number $N$ of discrete test particles with phase space coordinates $\big(\bm{x}_i (t), \bm{p}_i(t)\big)$,
\begin{eqnarray}
f(t, \bm{x}, \bm{t}) \approx \frac{1}{N_T} \sum_{i =1}^{N} \delta \Big( \bm{x} - \bm{x}_i (t)  \Big) \delta \Big( \bm{p} - \bm{p}_i (t)  \Big)~.
\label{test_particle_approximation}
\end{eqnarray}
Here, $N_T$ is the number of test particles per nucleon and $N = N_T A$. Each test particle contributes to the total charge in the system with a charge of the corresponding real particle divided by $N_T$ (for example, a ``nucleon test particle'' contributes a baryon charge of $\infrac{1}{N_T}$), so that the total charge in the simulation equals that of a system of $A$ particles. The cornerstone of the method of test particles is the realization that if we demand that these test particles are propagated according to 
\begin{eqnarray}
\frac{d \bm{x}}{dt} = \parr{H_{(1)}}{\bm{p}}~, \hspace{15mm}\frac{d \bm{p}}{dt} = -\parr{H_{(1)}}{\bm{x}} ~,
\label{Hamiltons_equations_test_particles}
\end{eqnarray}
then the Vlasov equation, which is the left-hand side of Eq.\ (\ref{BUU_equation}), immediately follows from the Liouville theorem. In other words, given some configuration of the positions and momenta of the test particles at a time $t_0$, which approximates the phase space distribution $f(t_0, \bm{x},\bm{p})$ according to Eq.\ \eqref{test_particle_approximation}, we can propagate the test particles over a time $dt$ using the Hamilton equations, Eq.\ \eqref{Hamiltons_equations_test_particles}, and use the configuration of positions and momenta of the propagated test particles to approximate $f(t_0 + dt, \bm{x}, \bm{p})$. We note here that in \texttt{SMASH}, the equations of motion propagate the kinetic momentum of particles, see Eq.\ (\ref{kinetic_momentum}), which is numerically more straightforward. An alternative approach, in which the canonical momenta are propagated, is possible \cite{Ko:1988zz}.

In practice, there exist two ways of realizing the method of test particles in hadronic transport. Within the first approach, one initializes a system with $N_TA$ test particles, which are then propagated according to the equations of motion. Scatterings are performed according to cross sections that are scaled as $\sigma/N_T$, where $\sigma$ is the physical cross section, which ensures that the average number of scatterings per test particle is the same as in a system of $A$ particles. Each test particle carries the charge of the corresponding real particle, but contributes to the density with a scaling factor of $1/N_T$, which preserves the total baryon number evolved in the simulation and, for $N_T \gg 1$, results in a mean field that is a smoothed out version of the mean field corresponding to $A$ particles. This approach is sometimes referred to as the ``full ensemble''.

An alternative approach is known as ``parallel ensembles'' \cite{Bertsch:1988ik}. In this method, $N_T$ instances of a system of $A$ particles are created. Particles in each instance are propagated according to the equations of motion, and scatterings are performed using the physical cross section $\sigma$. Each test particle carries the charge of the corresponding real particle, but contributes to the density and the corresponding mean field, which are calculated by summing contributions from all $N_T$ instances of the system, with a scaling factor of $1/N_T$. Evolving the $N_T$ systems with mean fields calculated in this fashion means that the systems are not in fact independent, and their evolution due to the mean fields is shared. At the same time, this approach is computationally much more efficient, as collision searches are performed only within individual instances of the system, thus reducing the numerical cost of performing the collisions within one event by a factor on the order of $\mathcal{O}(N_T^2)$, and the overall cost of the simulation by a factor on the order of $\mathcal{O}(N_T)$ (this is because the parallel ensembles mode evolves $N_T$ events at the same time, while the full ensemble mode evolves only one event). 

It can be checked that these two simulation paradigms lead to the same results in typical cases \cite{Xu:2016lue}. In this thesis we utilize the full ensemble approach to the test particle method, as this is the primary method supported in \texttt{SMASH}. Recently, the option to run the simulations in the parallel ensembles mode has been added to \texttt{SMASH} \cite{smash_version_2.0}, and it is currently being tested.

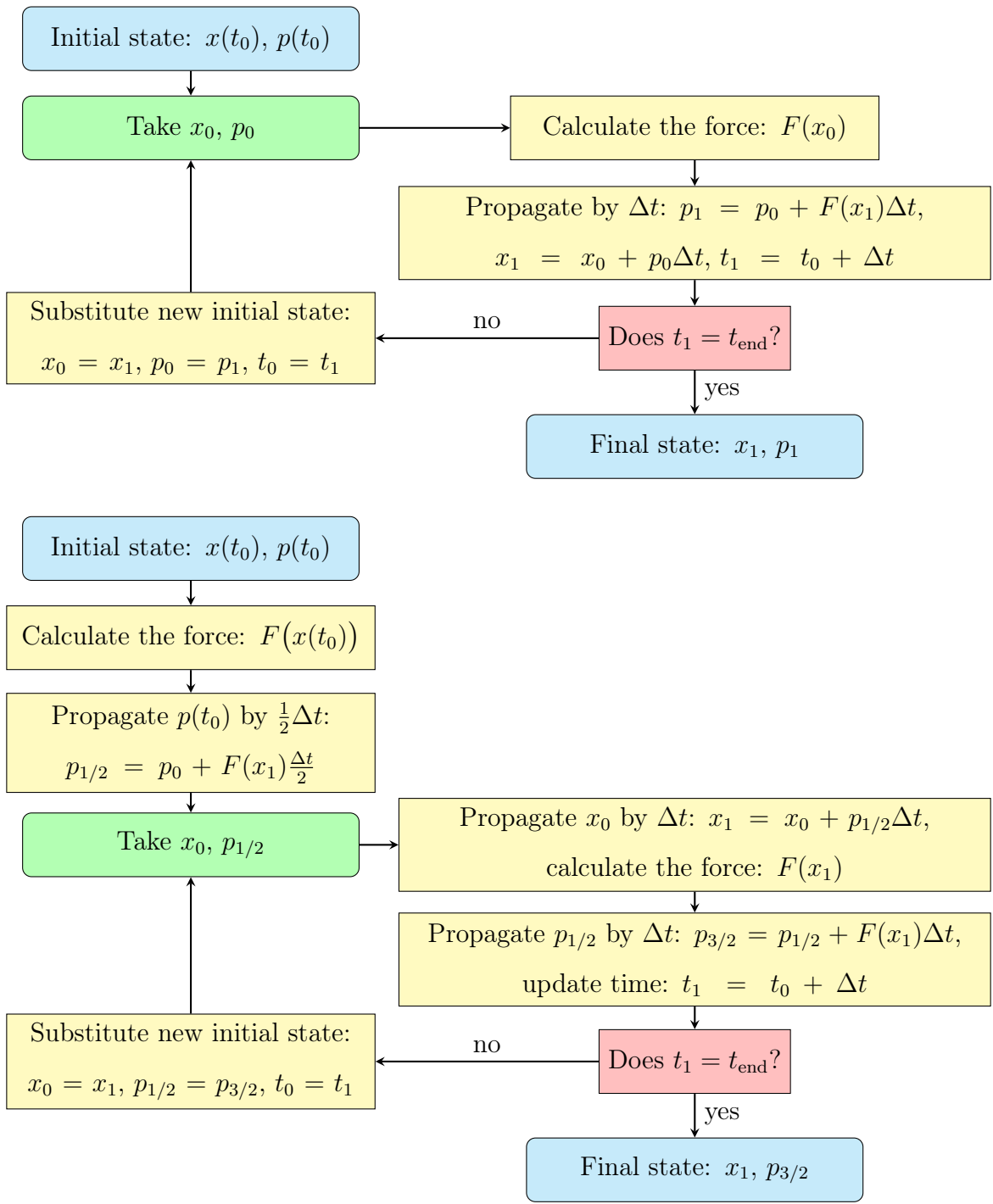
\begin{figure}
	\begin{center}
	\begin{tikzpicture}[node distance=1.9cm]
	
	%%%%% EULER
	
	\node (initial) [io, xshift = 0.0cm] {Initial state: $x(t_0)$, $p(t_0)$};
	
	\node (t0) [startingpoint, below of = initial, yshift = 0.5cm] {Take $x_0$, $p_0$};
	\draw [arrow] (initial) -- (t0);
	
	\node (in1) [operationsmall, right of=t0, xshift=6.0cm] {Calculate the force: $F(x_0)$};
	\draw [arrow] (t0) -- (in1);
	
	\node (timestep1) [operation, below of=in1, yshift = 0.25cm] {Propagate by $\Delta t$: $p_1 =  p_0 + F(x_1) \Delta t$, $x_1 =  x_0 + p_0 \Delta t$, $t_1 = t_0 + \Delta t$};
	\draw [arrow] (in1) -- (timestep1);
	
	\node (continue) [question, below of = timestep1, yshift = 0.25cm] {Does $t_1 = t_{\txt{end}}$?};
	\draw [arrow] (timestep1) -- (continue);
	
	\node (substitute) [operationsmall, left of = continue, xshift = -6.0cm] {Substitute new initial state: $x_0 = x_1$, $p_0 = p_1$, $t_0 = t_1$};
	\draw [arrow] (continue) -- node[anchor=south] {no} (substitute);
	\draw [arrow] (substitute) -- (t0);

	\node (final) [io, below of = continue, yshift = 0.2cm] {Final state: $x_1$, $p_1$};
	\draw [arrow] (continue) -- node[anchor=west] {yes} (final);

	%%%%% LEAPFROG
	
	\node (initialLF) [io, yshift = -8.0cm] {Initial state: $x(t_0)$, $p(t_0)$};
	
	\node (halftimestep1LF) [operationsmall, below of = initialLF, yshift = 0.5cm] {Calculate the force: $F\big(x(t_0)\big)$};
	\draw [arrow] (initialLF) -- (halftimestep1LF);
	
	\node (halftimestep2LF) [operationsmall, below of=halftimestep1LF, yshift = 0.25cm] {Propagate $p(t_0)$ by $\frac{1}{2}\Delta t$: $p_{1/2} = p_0 + F(x_1) \frac{\Delta t}{2}$};
	\draw [arrow] (halftimestep1LF) -- (halftimestep2LF);
	
	\node (t0LF) [startingpoint, below of = halftimestep2LF, yshift = 0.3cm] {Take $x_0$, $p_{1/2}$};
	\draw [arrow] (halftimestep2LF) -- (t0LF);
	
	\node (in1LF) [operation, right of=t0LF, xshift=6.0cm, align=center] {Propagate $x_0$ by $\Delta t$: $x_1 = x_0 + p_{1/2} \Delta t$,\\calculate the force: $F(x_1)$};
	\draw [arrow] (t0LF) -- (in1LF);
	
	\node (in3LF) [operation, below of=in1LF, yshift = 0.1cm, align = center] {Propagate $p_{1/2}$ by $\Delta t$: $p_{3/2} = p_{1/2} + F(x_1) \Delta t$,\\update time: $t_1 = t_0 +  \Delta t$};
	\draw [arrow] (in1LF) -- (in3LF);
	
	\node (continueLF) [question, below of = in3LF, yshift = 0.3cm] {Does $t_1 = t_{\txt{end}}$?};
	\draw [arrow] (in3LF) -- (continueLF);
	
	\node (substituteLF) [operationsmall, left of = continueLF, xshift = -6.0cm] {Substitute new initial state: $x_0 = x_1$, $p_{1/2} = p_{3/2}$, $t_0 = t_1$};
	\draw [arrow] (continueLF) -- node[anchor=south] {no} (substituteLF);
	\draw [arrow] (substituteLF) -- (t0LF);
	
	\node (finalLF) [io, below of = continueLF, yshift = 0.2cm] {Final state: $x_1$, $p_{3/2}$};
	\draw [arrow] (continueLF) -- node[anchor=west] {yes} (finalLF);
	\end{tikzpicture}
	\end{center}
	\caption[A schematic representation of the Euler and leapfrog integration algorithms]{A schematic representation of the Euler (top diagram) and leapfrog (bottom diagram) integration algorithms.}
	\label{leapfrog_vs_1st_order}
\end{figure}

Let us discuss in some detail the numerical method used to evolve the test particles in time. In \texttt{SMASH}, the positions and momenta of particles are advanced using the leapfrog algorithm. In this method, the evolutions of positions and momenta are ``interlaced'' in the following way: We start from some initial positions $\bm{x}_0$ and momenta $\bm{p}_0$. The momenta at time $t_0$ are advanced by half a time step to time $t_0 + \frac{1}{2}dt$ using forces calculated based on positions at the initial time $F(\bm{x}_0)$, $\bm{p}_0 \to \bm{p}_{1/2}$. Then, the positions of particles are advanced through the time step $\Delta t$ using the values of the momenta $\bm{p}_{1/2}$, $\bm{x}_0 \to \bm{x}_1$. These positions are used to calculate the forces on test particles $F(\bm{x}_1)$, which are then used to update the values of momenta from $\bm{p}_{1/2}$ to $\bm{p}_{3/2}$. Then positions $\bm{x}_1$ are advanced through the second time step $\Delta t$ using the momenta $\bm{p}_{3/2}$, and so on. Thus within the leapfrog method, the positions and momenta are updated at interleaved time points, so that the evolutions of $\bm{x}$ and $\bm{p}$ ``leapfrog'' over each other. This is markedly different from Euler integration, in which we would advance the positions from $\bm{x}_0$ to $\bm{x}_1$ using momenta $\bm{p}_0$, and advance the momenta from $\bm{p}_0$ to $\bm{p}_1$ using forces calculated based on the particle positions $\bm{x}_0$ (see Fig.\ \ref{leapfrog_vs_1st_order}).

In fact, it can be shown that the leapfrog algorithm, although it requires the same number of steps as the Euler integration (which is a first-order method), is a second-order method, and consequently it leads to more stable solutions. Indeed, we have tested the quality of solving the equations of motion using the \texttt{SMASH} leapfrog algorithm against the 4th-order Runge-Kutta integrator by tracking the energy conservation throughout the simulations (although energy should be conserved exactly by the Boltzmann equation, various effects connected with the method of integration, density calculation, and other elements of the simulation result in deviations from this theoretical expectation; studies of this nature are further discussed in Section \ref{comparison_of_different_numerical_procedures}). We confirmed that the two algorithms behave comparably for the standard simulation time step of $\Delta t = 0.1\ \txt{fm}/c$. At the same time, the leapfrog algorithm is considerably more numerically efficient. We note here for completeness that the leapfrog algorithm as implemented in \texttt{SMASH} misses the first step of advancing the initial momenta by half a time step. Such an omission could be disastrous for applications of the leapfrog integrator to problems with precise initial conditions, however, in \texttt{SMASH} the initial conditions are always created by random sampling of the desired phase space distribution with a large number of test particles. Advancing the initial momenta by half a time step is in this case equivalent to another random sample of the phase space distribution, and in result has no tangible consequences; this has been confirmed in tests.

Finally, we note that while the momentum changes due to the presence of potentials are addressed in \texttt{SMASH} at evolution times separated by a constant time interval $\Delta t$, particle-particle collisions and decays can take place at any time during advancing the particle positions. As a result, in \texttt{SMASH} the propagation within one time step is performed in the following way:
\begin{enumerate}
	\item At $t = t_0$: Take particles' momenta $\bm{p}$; find all collisions (and decays) that will occur based on particle trajectories given by $\bm{x}(t_0) + \bm{p}\Delta t$, and order them according to the time at which they occur; identify the time of the first occurring collision (or decay) $t'$.
	\item Propagate the particles in straight lines until $t = t'$. 
	\item At $t = t'$: perform the collision (decay), which in particular will produce new momenta for the interacting particles; find collisions (decays) for the particles that have just interacted, include them in the list of all collisions (decays); identify the time of the next occurring collision (decay) $t''$.
	\item Propagate the particles until $t = t''$, and repeat the above. Repeat steps 2-3 until $t = t_0 + \Delta t$ is reached. 
	\item At $t = t_0 + \Delta t$: Update particles' momenta according to the equations of motion, with forces calculated at the propagated positions, $F\big(\bm{x}(t+\Delta t)\big)$.
\end{enumerate}
Steps 1-5 are then repeated for each time step until the end of the simulation.

In the next section, we discuss some details of the density and mean field calculation in \texttt{SMASH}. There are, however, many other features and properties of \texttt{SMASH} not presented in this brief overview. A complete \texttt{SMASH} User Guide can be found online \cite{SMASH_user_guide}, and similarly for the \texttt{SMASH} development documentation \cite{smash_development_documentation}.

\section{Calculation of mean fields}
\label{calculation_of_mean-fields}

Parts of this section are based on Ref.\ \cite{Sorensen:2020ygf}. 

For an evolution without mean fields (that is for an evolution which does not require density calculations) it is most natural to take the number of test particles exactly corresponding to the actual number of nucleons present in the system, $N = A$, $N_T = 1$. However, employing mean-fields dependent on local density and its gradients requires adopting an approach in which the statistical noise due to a finite number of test particles is suppressed. This is especially important in the case of models with competing repulsive and attractive potentials of large magnitudes, where relatively small numerical fluctuations can produce significant errors in the mean field calculation. Thus, for example, for studies of nuclear matter with average density around the saturation density $n_0$, a number of test particles per nucleon $N_T = 100$ is often used.

The local baryon current can be defined on a discrete lattice, and the current at a given lattice node can be calculated as a sum of contributions from all test particles which are in the volume element $V_i$ corresponding to that lattice node,
\begin{eqnarray}
j^{\mu} (\bm{r}_i) = \frac{1}{N_T}\frac{1}{V_i} \sum_{k \in V_i} B_k \frac{\Pi^\mu_k }{\Pi^0_k} ~,
\label{baryon_current_zeroth_approx}
\end{eqnarray} 
where $B_k$ is the baryon charge of the $k$-th test particle. This prescription, known as the particle-in-cell method, naturally reproduces the baryon number in a given volume element,
\begin{eqnarray}
B(i) = j^0 (i) V_i = \frac{ N(i)}{N_T}~,
\end{eqnarray}
where $N(i)$ is the (net) number of test particles in $V_i$. In practice, in order for the local densities and currents to be smooth enough, a prescription is used in which currents at a given lattice node $i$ are weighted sums of contributions from all test particles in some chosen volume $V_s$ around that lattice node $i$, with $V_s$ taken to be larger than the volume element $V_i$, $V_s > V_i$. In this case the baryon current is given by
\begin{eqnarray}
j^{\mu} (\bm{r}_i)  = \sum_{k \in V_s} B_k\frac{\Pi^\mu_k }{\Pi^0_k}  S(\bm{r}_i - \bm{r}_k)~,
\end{eqnarray}
where  the weight $S(\bm{r}_i - \bm{r}_k)$ is known as the smearing function (often also referred to as the smearing factor or the smearing kernel), normalized such that
\begin{eqnarray}
\sum_{i} V_i  ~S (\bm{r}_i - \bm{r}_k) = \frac{1}{N_T}~.
\end{eqnarray}
Note that the volume $V_s$ from which test particles contribute to the density at a given node effectively establishes the range of the interaction and therefore it is a parameter of the algorithm that has a physical relevance; while this smearing range can be adjusted to reproduce certain physical behaviors of studied systems, in most applications it is often set at around $r \approx 2\ \txt{fm}$ \cite{Colonna:2021xuh}. Having calculated the current $j^{\mu}$, current gradients are usually calculated by taking a finite difference of baryon current values at adjacent lattice nodes, for example
\begin{eqnarray}
\frac{d j_0 ( x_i, y_j, z_k)}{dx}  = \frac{j_0 ( x_i + 1, y_j, z_k) - j_0 ( x_i - 1, y_j, z_k)}{2 L_x}~,
\end{eqnarray}
where $L_x$ is the lattice spacing in the $x$-direction. Consequently, derivatives calculated in hadronic transport can only resolve structures of size bigger than the lattice spacing. 

To achieve a higher resolution, one can in principle use a smaller lattice spacing, however, the computational cost of using finer lattices is often prohibitive. This is because finer lattices not only mean that there are more lattice nodes at which the baryon current is computed, but more importantly, the number of test particles per particle must be increased so that the number of test particles per lattice cell is high enough to ensure a smooth density calculation. This significantly affects the time cost of the simulation, and the effect is especially pronounced for simulations run in the full ensemble mode, where the number of performed particle-particle collisions scales as $N_T^2$. As a result, one needs to find a compromise between the resolution and a viable computation time. In practice, the lattice spacing in \texttt{SMASH} is often chosen at a typical value of $1\ \txt{fm}$ (which is also the case in this work), and so the constraint on the resolution coming from the range of the smearing factor is typically more important.

The time derivatives of densities and currents are most often also calculated using the finite difference method, based on the difference between the values of the baryon current at times $t_0 + \delta t$ and $t_0$ at each node. The resolution here naturally depends on the time step $\Delta t$ used to advance the evolution.

Finally, because the positions of the test particles take continuous values, while the lattice nodes are located at discrete positions $\bm{x}_i$, a question arises what density should be used for the mean-field evolution of a test particle located at a position $\bm{x} \neq \bm{x}_i ~\forall i$. Currently in \texttt{SMASH} this is treated within a zeroth-order approximation, where the density at $\bm{x}$ is assumed to be equal to the density at the nearest lattice node. Naturally, this approximation is good only as long as the density doesn't change appreciably over distances on the order of $L_i/\sqrt{2}$.

The existing transport codes employ various numerical approaches for calculating the effects of the mean-field potentials (for an exhaustive comparison of mean-field calculations in nuclear matter simulated in several different transport codes, see Ref.\ \cite{Colonna:2021xuh}). In the case of results presented in Chapter \ref{results}, we employ a triangular smearing function, originating from the lattice Hamiltonian method of solving nuclear dynamics \cite{Lenk:1989zz}, and derivatives based on the finite difference method. Below, we briefly discuss and compare three smearing schemes available in \texttt{SMASH} together with some additional details on the calculation of derivatives. The Gaussian smearing scheme is the original smearing available in \texttt{SMASH}, while the discrete and triangular smearing have been added as part of this work. Likewise, we introduced the option to calculate the gradients of currents using the chain rule or direct finite difference derivatives scheme. All of the options discussed below are available starting with the \texttt{SMASH}-2.1 release. We note here for clarity that although we always refer to calculating the baryon current, in \texttt{SMASH} it is also possible to calculate currents of different particle charges such as, e.g., isospin.

\subsection{Discrete smearing}

Within a smearing scheme which in \texttt{SMASH} is referred to as ``discrete smearing'', each test particle contributes a fraction $F_{\txt{c}} = 1/a$ of its charge to the lattice cell in which it is located (center cell) and a fraction $F_{\txt{n}} = (1 - F_{\txt{c}})/6 =  (a-1)/6a$ of its charge to each of its six nearest neighbor lattice cells, that is nearest lattice cells in the $-x$, $+x$, $-y$, $+y$, $-z$, and $+z$ directions. In order for $F_{\txt{c}} > F_{\txt{n}}$ to hold, the discrete smearing parameter must satisfy $F_{\txt{c}} \geq 1/7$; an often used choice is $F_{\txt{c}} = 1/3$. Such weighted contributions are then further scaled by the number of test particles per particle $N_T$ (if it hasn't been already taken into the account in the charge of the test particle) and the volume of the cells $V_c$ to obtain the baryon current at each node. This smearing scheme has the advantage that the integration of the baryon density calculated on the lattice equals the number of baryon charges present within the bounds of the lattice exactly, meaning that the smearing conserves the baryon number. Additionally, it is very computationally efficient. However, at the same time it often requires using a very large number of test particles per particle $N_T$, on the order of $N_T \approx 1000$, to provide reliable results. This considerably lowers the efficiency of this scheme in the full ensemble mode.

\subsection{Triangular smearing}

In the smearing scheme referred to in \texttt{SMASH} as ``triangular smearing'', originating from the lattice Hamiltonian method of solving nuclear dynamics \cite{Lenk:1989zz}, each test particle contributes to the baryon current at a given node with a smearing factor
\begin{eqnarray}
S(\bm{r}_i - \bm{r}) = \frac{B}{N_T \left( nL_x  \right)^2\left( nL_y  \right)^2\left( nL_z\right)^2} g( \Delta x ) g(\Delta y) g(\Delta z)~,
\end{eqnarray}
where 
\begin{eqnarray}
g(\Delta q) = \Big( nL_q - |\Delta q| \Big) \theta \Big( nL_q - |\Delta q| \Big)~.
\label{triangular_smearing}
\end{eqnarray}
Here, $\bm{r}_i$ is the position of the node, $\bm{r}$ is the position of the test particle, $B$ is the baryon charge of the contributing test-particle (not scaled by $N_T$), $L_x$, $L_y$, $L_z$ are the lattice spacings in the $x$-, $y$-, and $z$-direction, the smearing parameter $n$ is an integer which determines the range of $S$ in terms of the lattice spacings, and $\Delta x$, $\Delta y$, $\Delta z$ are the $x$, $y$, and $z$ components of $\bm{r}_i - \bm{r}$. Because the triangular smearing is linear, the summation of the resulting baryon density over volume integrates to the number of baryons present within the bounds of the lattice exactly (this is because Euler integration is exact for a linear function). Additionally, the triangular smearing is very efficient numerically, and a smooth density calculation is obtained for a standard number of test particles per particle on the order of $N_T \approx 100$.

\subsection{Gaussian smearing}

Finally, the type of smearing referred to in \texttt{SMASH} as ``Gaussian smearing'' and present in the original release of \texttt{SMASH} uses a Lorentz covariant Gaussian kernel as its basis \cite{Oliinychenko:2015lva}. Starting from a smearing kernel in the rest frame of a test particle defined as
\begin{eqnarray}
K (\bm{r} - \bm{r}_i) = \frac{1} { \big(2 \pi \sigma^2\big)^{3/2}} \exp \left(- \frac{\big(\bm{r} - \bm{r_i}\big)^2}{2\sigma^2} \right)~,
\label{Gaussian_kernel}
\end{eqnarray}
one can perform a boost to the computational frame in the usual manner. Additionally adopting a normalization ensuring that volume integrations over $K (\bm{r} - \bm{r}_i) $ are also Lorentz covariant leads to the expression for the baryon current in the computation frame of the form
\begin{eqnarray}
j^{\mu}_{\txt{comp}}(\bm{r}) = \frac{\gamma} { N_T\big(2 \pi \sigma^2\big)^{3/2}}  \sum_{i=1}^N B_i  \frac{\Pi^{\mu}_i}{\Pi_{0,i}}  \exp \left( - \frac{\big(\Lambda_i(\bm{r} - \bm{r}_i)\big)^2}{2\sigma^2} \right)~,
\label{SMASH_baryon_current}
\end{eqnarray}
where $\gamma = \big( 1 - v^2 \big)^{-1/2}$ is the Lorentz factor, the sum is performed over all test particles, $B_i$ is the baryon charge of the $i$-th test particle (not scaled by $N_T$), and $\Lambda_i$ is the Lorentz transformation matrix from the test particle's rest frame to the computational frame. The Gaussian smearing is characterized by two parameters: the first of them is the Gaussian width, usually taken to be $\sigma = 1$ fm, and the second is a finite cut-off radius expressed in units of the Gaussian width, often taken to be $r_{\txt{c}} = 4\sigma$, which is the radius beyond which the test particle's contribution to the density is considered negligible. This cut-off is introduced to save the computation time and not evaluate near-zero contributions to density at distances far from a given test particle; for a given chosen $r_{\txt{c}}$, the Gaussian smearing kernel is again normalized such that it integrates to $1/N_T$, as it should. 

While the use of exponentials is numerically very costly, the Gaussian smearing has the advantage in that it is Lorentz covariant and therefore properly treats relativistic effects through the contraction of the smearing kernel in the direction of a test particle's motion (in other smearing schemes, these effects are often dealt with by using \textit{ad-hoc} methods such as, for example, increasing the fineness of the lattice in a direction in which the system is expected to be contracted). Additionally, using the fact that the Gaussian smearing kernel is everywhere differentiable, in \texttt{SMASH} the derivatives of the baryon current can be calculated analytically (naturally, only in the case of the Gaussian smearing). For example, the gradient of the baryon current, Eq.\ \eqref{SMASH_baryon_current}, is given by 
\begin{eqnarray}
\frac{dj^{\mu}}{d\bm{r}} =  \frac{\gamma} { N_T\big(2 \pi \sigma^2\big)^{3/2}} \sum_{i=1}^N B_i \frac{\Pi^{\mu}_i}{\Pi_{0,i}} \exp \left( - \frac{(\Lambda_i(\bm{r} - \bm{r}_i))^2}{2\sigma^2} \right) \frac{\Lambda_i^2(\bm{r} - \bm{r}_i)}{\sigma^2} ~,
\label{SMASH_current_derivative}
\end{eqnarray}
where we can see that the derivative has been applied directly to the smearing function. Consequently, these ``Gaussian gradients'' do not depend on the size of the lattice spacing, but rather on the Gaussian smearing parameters $\sigma$ and $r_{\txt{c}}$. In certain situations this may prove to be problematic: tests have shown that because of this feature, rapidly varying gradients are not reproduced well due to the fact that the analytic expressions for derivatives are ``smoothed out'' over large distances. Additionally, this derivative scheme misses a possibly important term when mean fields are present, as in this case the kinetic momentum has itself a spacetime-dependence through $\Pi_i^{\mu} = p_i^{\mu} - \sum_k A_k^{\mu}(x)$, see Eq.\ \eqref{kinetic_momentum}; we show this explicitly below.

Finally, the complex functional form of the Gaussian smearing factor means that the sum of its contributions to discrete lattice nodes does not add up to $1/N_T$ (this is because the Euler integration is not exact in the case of a smearing kernel of the form Eq.\ \eqref{Gaussian_kernel}), which means that the baryon number is not completely preserved on the lattice.

\subsection{Gaussian smearing and the time derivative of vector baryon current}
\label{Gaussian_smearing_and_the_time_derivative_of_the_vector_baryon_current}

Let us consider the time derivative of the vector baryon current calculated within the Gaussian smearing scheme, Eq.\ \eqref{SMASH_baryon_current}, which we rewrite here with explicit time dependence, 
\begin{eqnarray}
\bm{j} \big(\bm{x}, t \big) = \sum_{i} \frac{\bm{\Pi}_i(t)}{\Pi_{0,i}(t)} K \Big( \bm{x} - \bm{x}_i(t) \Big)~,
\end{eqnarray}
where $\bm{\Pi}_i$ and $\Pi_{0,i}$ denote the spatial and temporal components of the kinetic momentum $\Pi^{\mu}$, respectively, and $K$ is the smearing factor which includes the baryon number coefficient $B_i$ as well as the normalization for clarity of calculations. The current at the time $t + \Delta t$ is given by 
\begin{eqnarray}
\bm{j} \big(\bm{x}, t + \Delta t \big) = \sum_{i} \frac{\bm{\Pi}_i(t + \Delta t)}{\Pi_{0,i}(t + \Delta t)} K \Big( \bm{x} - \bm{x}_i(t + \Delta t) \Big)~,
\end{eqnarray}
which we can expand around $t$,
\begin{eqnarray}
\hspace{-10mm}\bm{j} \big(\bm{x}, t + \Delta t \big)  &=&  \sum_{i} \frac{\bm{\Pi}_i(t) + \Delta t \parr{\bm{\Pi}_i}{ t}}{\Pi_{0,i}(t) + \Delta t \parr{\Pi_{0,i} (t)}{t}} \left[ K \Big( \bm{x} - \bm{x}_i(t) \Big) + \Delta t \parr{\bm{x}_i}{t} \parr{K}{\bm{x}_i}\right] + \mathcal{O} \Big( (\Delta t)^2\Big) \non \\
&& \hspace{-20mm} \approx \sum_{i} \frac{\bm{\Pi}_i(t) + \Delta t \parr{\bm{\Pi}_i}{ t}}{\Pi_{0,i}(t) } \left( 1 - \frac{\Delta t}{\Pi_{0,i}} \frac{\bm{\Pi}_i}{\Pi_{0,i}}\parr{\bm{\Pi}_i (t)}{t} \right) \left[ K \Big( \bm{x} - \bm{x}_i(t) \Big) + \Delta t \parr{\bm{x}_i}{t} \parr{K}{\bm{x}_i}\right] ~.
\end{eqnarray}
Keeping only terms up to linear order in $\Delta t$ results in
\begin{eqnarray}
\hspace{-5mm}\bm{j} \big(\bm{x}, t + \Delta t \big) &\approx& \sum_{i} \frac{\bm{\Pi}_i(t)}{\Pi_{0,i}(t) }   K \Big( \bm{x} - \bm{x}_i(t) \Big)   \non \\
&& \hspace{-28mm}+~ \Delta t  \sum_{i} \frac{1}{  \Pi_{0,i}(t) } \left[   \frac{m_i^2}{\Pi_{0,i}^2} \right] \parr{\bm{\Pi}_i (t)}{t}  K \Big( \bm{x} - \bm{x}_i(t) \Big)   + \Delta t \sum_{i} \frac{\bm{\Pi}_i(t)}{\Pi_{0,i}(t) }\left[  \parr{\bm{x}_i}{t} \parr{K}{\bm{x}_i}\right] ~,
\end{eqnarray}
where we have used the fact that $\Pi_{0,i}^2 = \bm{\Pi}_i^2 + m^2$. We can then calculate
\begin{eqnarray}
\hspace{-5mm} \parr{}{t} \bm{j} \big(\bm{x}, t \big) &\equiv& \lim_{\Delta t \to 0} \frac{  \bm{j} \big(\bm{x}, t + \Delta t \big) - \bm{j} \big(\bm{x}, t \big) }{\Delta t} \non \\
&=& \sum_{i} \frac{1}{\Pi_{0,i}(t) } \left[   \frac{m_i^2}{\Pi_{0,i}^2} \right] \parr{\bm{\Pi}_i (t)}{t}  K \Big( \bm{x} - \bm{x}_i(t) \Big)  +  \sum_{i} \frac{\bm{\Pi}_i(t)}{\Pi_{0,i}(t) }\left[  \parr{\bm{x}_i}{t} \parr{K}{\bm{x}_i}\right] ~.
\label{full_derivative}
\end{eqnarray}
By comparing with Eq.\ \eqref{SMASH_current_derivative}, adapted to the case of the derivative with respect to time, one can see that the analytical derivative used in the Gaussian smearing scheme includes only the second term in the above expression. This, in particular, misses contributions coming directly from changes in the momentum due to the mean-field potential.

\subsection{Chain rule and direct finite difference derivatives}

A convenient form of the equation of motion for the kinetic momentum in the VSDF model, Eq.\ \eqref{EOM_covariant_formulation_p}, has been derived in Appendix \ref{Lorentz_force}, Eq.\ \eqref{EOM_VSDF_explicit}, which we rewrite here for the VDF case,
\begin{eqnarray}
\frac{d \bm{\Pi}}{dt} =  \sum_{k=1}^K \left\{ - \Big(  \bm{\nabla} A_k^{0}  + \partial^{0} \bm{A}_k   \Big) + \frac{\bm{\Pi}}{\Pi_0} \times \big( \bnabla \times \bm{A}_k  \big)  \right\}   ~.
\label{VDF_EOM}
\end{eqnarray}
The finite difference derivatives of the vector field $\sum_{k=1}^K A_k^{\mu}$, where $A_k^{\mu}$ is given by Eq.\ \eqref{summary_vector_field}, repeated here for convenience
\begin{eqnarray}
A_k^{\mu}(x; C_k, b_k) = C_k n_B^{b_k - 2} j^{\mu} ~,
\end{eqnarray}
can be performed in two ways. We can apply the chain rule to the derivative expressions in Eq.\ \eqref{VDF_EOM}, obtaining
\begin{eqnarray}
&& \sum_{k=1}^K\bm{\nabla} A_k^{~0}   =  \sum_{k=1}^K C_k n_B^{b_k - 3}  \bigg((b_k - 2) \Big( \bm{\nabla} n_B  \Big) j^0 +  n_B \Big( \bm{\nabla} j^0\Big) \bigg)~, 
\label{finite_difference_derivative_chain_rule_1}\\
&& \sum_{k=1}^K\parr{}{t}\bm{A}_k  = \sum_{k=1}^K C_k n_B^{b_k - 3} \bigg((b_k - 2)  \left( \parr{n_B }{t} \right) \bm{j} + n_B \left( \parr{\bm{j}}{t} \right) \bigg)  ~, 
\label{finite_difference_derivative_chain_rule_2}\\
&& \sum_{k=1}^K\bm{\nabla} \times   \bm{A}_k =  \sum_{k=1}^K C_k n_B^{b_k - 3} \bigg( (b_k - 2) \Big( \bm{\nabla} n_B  \Big) \times \bm{j}  +  n_B  \Big( \bm{\nabla} \times \bm{j} \Big) \bigg)
\label{finite_difference_derivative_chain_rule_3}~,
\end{eqnarray}
and then use the finite difference derivatives of the rest frame density $n_B$ and current $j^{\mu}$ at a given node to obtain $\sum_{k=1}^K\bm{\nabla} A_k^{~0} $, $\sum_{k=1}^K\inparr{\bm{A}_k}{t}$, and $\sum_{k=1}^K\bm{\nabla} \times   \bm{A}_k$ at that node. Alternatively, we can use the values of $n_B$ and $j^{\mu}$ at a given node to compute $\sum_{k=1}^K A_k^{\mu}$ at that node, and then compute the finite difference derivatives of the relevant terms. As an example, let us consider the gradient of the zeroth component of the field $\sum_{k=1}^K \bm{\nabla} A_k^{0} $. The $z$-direction component of the gradient is then simply given by
\begin{eqnarray}
\sum_{k=1}^K \nabla_z A_k^0 (z_0) = \frac{ \sum_{k=1}^K A_k^0 (z_0 + \Delta z)  - \sum_{k=1}^K A_k^0 (z_0 - \Delta z)  }{2 \Delta z} ~,
\end{eqnarray}
where $\Delta z$ is the lattice spacing the $z$-direction. 

We refer to the former method of calculating derivatives as the chain rule finite difference derivative and to the latter as the direct finite difference derivative. In our studies, the direct finite difference derivative has proven to be numerically more stable, as evaluated by the degree to which energy is conserved in the simulation.

\section{Potentials in \texttt{SMASH}}
\label{potentials_in_SMASH}

\texttt{SMASH} was introduced with a density-based mean-field potential of the Skyrme parametrization type, which we will refer to simply as the ``Skyrme'' potential. Based on a semi-relativistic generalization of the rest-frame Skyrme potential, the equation of motion for the kinetic momentum as used in the \texttt{SMASH} Skyrme mean-field mode is given by 
\begin{equation}
\frac{d\bm{\Pi}}{dt} = \frac{\partial U}{\partial n_B} \bigg[-\bigg(\bnabla j^0 + \parr{ \bm{j}}{t} \bigg)+  \frac{d\bm{x}}{dt} \times \big(\bnabla \times \bm{j}\big)\bigg],
\label{Feng_Li's_EOMs_p}
\end{equation}
where $n_B \equiv \sqrt{j_{\mu}j^{\mu}}$ is the rest frame baryon density and $U$ is the rest frame single-particle potential given by
\begin{eqnarray}
U(n_B) = a \left( \frac{n_B}{n_0}\right) + b \left(\frac{n_B}{n_0}\right)^{\tau} ~,
\label{SMASH_Skyrme}
\end{eqnarray}
where $n_0$ is the saturation density, so that
\begin{eqnarray}
\frac{\partial U}{\partial n_B} = \frac{a}{n_0} + b\tau \frac{n_B^{\tau - 1}}{n_0^{\tau}}~. 
\end{eqnarray}
(We note here that \texttt{SMASH} allows the user to vary the values of the Skyrme parameters $a$, $b$, and $\tau$, but uses a fixed value of saturation density $n_0 = 0.168\ \txt{fm}^{-3}$.) Importantly, while Eq.\ \eqref{Feng_Li's_EOMs_p} has a form similar to that of the equations of motion in relativistic electrodynamics, it is not fully relativistic, as we show below.

Within the work presented in this thesis, we implemented the VDF potential and the corresponding fully relativistic equations of motion in \texttt{SMASH}. Using the VDF model, it is straightforward to reproduce the Skyrme potential of the form given in Eq.\ \eqref{SMASH_Skyrme} by using the vector field $\sum_{k=1}^K \bm{\nabla} A_k^{\mu} = \sum_{k=1}^K \tilde{C}_k (n_B/n_0)^{b_k - 2}( j^{\mu}/n_0) $ (see Appendix \ref{parameter_sets} for more details on the conversion of the VDF interaction coefficients to units of energy) with
\begin{eqnarray}
K = 2~, \hspace{5mm}   \tilde{C}_1 = a ~, \hspace{5mm} b_1 = 2 ~,  \hspace{5mm} \tilde{C}_2 = b ~, \hspace{5mm} b_2 = \tau + 1 ~, \hspace{5mm} n_0 = 0.168\ \txt{fm}^{-3}  ~. 
\label{VDF_reproduces_Skyrme}
\end{eqnarray}
This possibility is very useful for comparisons of simulation results obtained using the two potentials. On the theoretical level, comparing Eq.\ \eqref{Feng_Li's_EOMs_p} with Eq.\ \eqref{VDF_EOM}, supported by Eqs.\ (\ref{finite_difference_derivative_chain_rule_1}-\ref{finite_difference_derivative_chain_rule_3}), immediately reveals that the \texttt{SMASH} Skyrme equation of motion, Eq.\ \eqref{Feng_Li's_EOMs_p}, misses contributions proportional to $(\bnabla n_B)j^0$, $(\inparr{n_B}{t}) \bm{j}$, and $\bnabla n_B \times  \bm{j}$. Notably, the VDF equation of motion, Eq.\ \eqref{VDF_EOM}, becomes the \texttt{SMASH} equation of motion, Eq.\ \eqref{Feng_Li's_EOMs_p}, for $j^0 \approx n_B$, which is the nonrelativistic limit.

In addition to the Skyrme potential, \texttt{SMASH} also includes the option to use a symmetry potential, which is of relevance for systems characterized by a non-zero isospin asymmetry. This possibility is at this time only available for simulations using the Skyrme potential and therefore will not be discussed here.

\section{Comparison of different numerical procedures}
\label{comparison_of_different_numerical_procedures}

As part of the implementation of the VDF equations of motion and different baryon current calculation schemes in \texttt{SMASH}, we thoroughly compared their behavior against the original \texttt{SMASH} Skyrme potentials and the Gaussian smearing scheme. In particular, we extensively studied deviations from perfect energy conservation in systems initialized at different points of the phase diagram, with different EOSs, characterized by various geometries, using different smearing and derivative calculation schemes, and employing different mean fields.

We note that problems related to energy conservation are present in many hadronic transport codes and can be directly linked not only to the integration algorithm for the equations of motion, but also to the particular details of density and density gradient calculations (including the influence of statistical noise fluctuations on the magnitude of local density gradients) as well as to the readout of these quantities from the discrete lattice. Although methods ensuring energy conservation are known \cite{Wang:2019ghr} for transport simulations dealing with nonrelativistic systems, an application of such approaches in relativistic transport codes has not been attempted.

One of the best ways to test the robustness of the integration of equations of motion is to study systems evolving under extreme conditions. An example of such a situation is uniform nuclear matter initialized at a low temperature inside the spinodal region of the nuclear phase transition. This system is mechanically unstable, and because the initialized matter is not perfectly uniform due to the finite number of test particles used to sample the uniform density profile, there are initial density fluctuations which over time will be magnified by the mean field, leading to a spontaneous separation into two coexisting phases: a nuclear liquid and a gas of nucleons (more details on this and other simulations mentioned in this section will be provided in Chapter \ref{results}). This separation causes the system to exhibit steep density gradients, and the degree to which the density calculation and the integration of the equations of motion deals well with these gradients is reflected in the deviation of the energy conservation from the ideal case, where the energy of the whole system is conserved over time, $dE_{\txt{system}}/dt = 0$. This deviation is best measured as the energy difference per particle, denoted here as $\Delta E = \infrac{(E_{\txt{final}} - E_{\txt{initial}})}{N}$.

While the total energy of the system should stay constant, the kinetic energy and the mean-field energy are allowed to change their values, and in fact it is expected that they do so, since the non-optimal initial value of the mean-field energy is what drives the system to undergo a phase separation in the first place. Here, the evolution of the system can be, to a zeroth degree, captured within the ratio of the final mean-field energy of the system to the initial mean-field energy, $R_{\txt{MF}} = E_{\txt{MF,final}}/E_{\txt{MF,initial}}$. 

Finally, an important part of testing a given scheme for integrating the equations of motion is keeping track of the amount of time it takes to simulate a single event $t_{\txt{sim}}$, as typical simulations of heavy-ion collisions require on the order of 1,000 or even 10,000 events. For the ballpark calculations presented here we used a personal computer with a 2.6 GHz processor and 16 GB of memory.

In Tables \ref{density_and_derivatives_box_low_density_Skyrme}, \ref{density_and_derivatives_box_low_density_VDF}, and \ref{density_and_derivatives_box_low_density_VDF_2} we show the values of $\Delta E$ (in $\txt{MeV}/\txt{nucleon}$), $R_{\txt{MF}}$, and $t_{\txt{sim}}$ (in minutes), obtained from simulations initializing uniform nuclear matter in a box of side $10\ \txt{fm}$ with periodic boundary conditions (\texttt{SMASH} Box modus) at $T=1\ \txt{MeV}$ and $n_B = 0.25\ n_0$, and averaged over 10 events. The lattice used for the density calculation had 10 nodes per side, so that $L_x = L_y = L_z = 1\ \txt{fm}$, and we turned off all collisions and decays between the test particles to focus on the effects of the mean-field calculation. We tested different smearing schemes: Gaussian, triangular, and discrete, at different values of their respective parameters (with the Gaussian width always equal $\sigma = 1\ \txt{fm}$), as well as different derivative calculation schemes: Gaussian, chain rule finite difference, and direct finite difference, for both the original \texttt{SMASH} Skyrme potential, Eq.\ \eqref{SMASH_Skyrme}, and the implemented VDF potential reproducing the \texttt{SMASH} Skyrme potential, Eq.\ \eqref{VDF_reproduces_Skyrme}, with Skyrme parameters taken at their usual values of $a= -209.2\ \txt{MeV}$, $b = 156.4\ \txt{MeV}$, and $\tau = 1.35$. The difference between the two models, as discussed in Section \ref{potentials_in_SMASH}, is that the Skyrme potential is semi-relativistic, while the VDF potential is fully relativistic.

\begin{table}[t]
	\caption[Effects of different smearing and derivative calculation modes on energy conservation and simulation time for the Skyrme potential]{Values of $\Delta E$ (in MeV/nucleon), $R_{\txt{MF}}$, and $t_{\txt{sim}}$ (in minutes) for uniform nuclear matter simulated in a box with periodic boundary conditions, initialized in the spinodal region of the nuclear phase transition, as calculated using the Skyrme potential for different smearing schemes (``Sm.'', with abbreviations for Gaussian (``G.''), triangular (``tr.''), and discrete (``dis.'') smearing), values of smearing parameters (``par.''), numbers of test particles (``$N_T$''), and derivative modes (``$\bnabla$''). Gaussian smearing always used a width of $\sigma = 1\ \txt{fm}/c$.}
	\vspace{3mm}
	\label{density_and_derivatives_box_low_density_Skyrme}
	\def\arraystretch{1.12}
	\begin{tabular}{|r|r|r|r|r|r|r|r|r|}
		\hline
	\multicolumn{9}{|c|}{Uniform matter in a box, Skyrme potential: $T = 1\ \txt{MeV}$, $n_B = 0.25\ n_0$, $t_{\txt{end}} = 200\ \txt{fm}$.}  \\
	\hline
	\multirow{2}{*}{Sm.} &\multirow{2}{*}{par.} & \multirow{2}{*}{$N_T$} &  \multicolumn{3}{r|}{Gaussian $\bnabla$} & \multicolumn{3}{r|}{finite $\bnabla$ (chain rule)}\\
	\cline{4-9}
	& & &  $\Delta E$  & $R_{\txt{MF}}$ & $t_{\txt{sim}}$& $\Delta E$  & $R_{\txt{MF}}$ & $t_{\txt{sim}}$ \\
	\hline
	\hline
	G. &  $r_{\txt{c}}= 4$  & 20 & \co{$2.40\pm 0.21$}& $1.36\pm0.04$& 0.74 & $1.96\pm0.21$ & $1.34\pm0.02$ & 0.53\\
	\hline
	G. & $r_{\txt{c}} = 2$ &  20 & \cred{$1.30\pm 0.24$} & \cred{$1.15\pm0.02$} & 0.10& \co{$2.39\pm0.23$}& $1.39\pm0.03$ & 0.08 \\
	\hline
	G. & $r_{\txt{c}}= 4$ & 200 & $1.85 \pm 0.11$ & $1.39 \pm 0.01$ & 7.22 & \cg{$1.25\pm0.09$}   & $1.37\pm0.02$ & 4.68 \\
	\hline
	G. &  $r_{\txt{c}} = 2$ & 200 & \cred{$-0.04\pm0.07$} & \cred{$1.08\pm0.02$} & 1.11 & \cg{$1.41\pm0.12$} & $1.45\pm0.02$  & 0.78 \\
	\hline
	tr. &  $n=2$ & 200 & ---&  ---&--- & \cg{$1.40 \pm 0.08$}& $1.43\pm0.02$ & 0.54\\
	\hline
	tr. &  $n=4$ & 200 & ---&  ---&--- & \cred{$0.07 \pm 0.04$}& \cred{$1.01\pm0.00$} & 2.60  \\
	\hline
	dis. & $F_{\txt{c}} = \frac{1}{3}$ & 200 & ---&  ---&--- & \co{$3.66 \pm 0.20$} & $1.43\pm0.02$  & 0.21 \\
	\hline
	dis. &  $F_{\txt{c}} = \frac{1}{3}$ & 1000 & ---&  ---&---& \co{$2.02 \pm 0.19$} &  $1.51\pm0.02$  & 2.96\\
	\hline
	\end{tabular}
\end{table}

\begin{table}[t]
	\caption[Effects of different smearing and derivative calculation modes on energy conservation and simulation time for the VDF potential (1)]{Values of $\Delta E$ (in MeV/nucleon), $R_{\txt{MF}}$, and $t_{\txt{sim}}$ (in minutes) for uniform nuclear matter simulated in a box with periodic boundary conditions, initialized in the spinodal region of the nuclear phase transition, as calculated using the VDF potential for different smearing schemes (``Sm.'', with abbreviations for Gaussian (``G.''), triangular (``tr.''), and discrete (``dis.'') smearing), values of smearing parameters (``par.''), numbers of test particles (``$N_T$''), and derivative modes (``$\bnabla$''). Gaussian smearing always used a width of $\sigma = 1\ \txt{fm}/c$.}
	\vspace{3mm}
	\label{density_and_derivatives_box_low_density_VDF}
	\def\arraystretch{1.12}
	%\small\addtolength{\tabcolsep}{1pt}
	\begin{tabular}{|r|r|r|r|r|r|r|r|r|}
		\hline
		\multicolumn{9}{|c|}{Uniform matter in a box, VDF potential: $T = 1\ \txt{MeV}$, $n_B = 0.25\ n_0$, $t_{\txt{end}} = 200\ \txt{fm}$.}  \\
		\hline
		 \multirow{2}{*}{Sm.} &\multirow{2}{*}{par.} & \multirow{2}{*}{$N_T$} &  \multicolumn{3}{r|}{Gaussian $\bnabla$} & \multicolumn{3}{r|}{finite $\bnabla$ (chain rule)}\\
		\cline{4-9}
		& & & $\Delta E$ & $R_{\txt{MF}}$ & $t_{\txt{sim}}$ & $\Delta E$ & $R_{\txt{MF}}$ & $t_{\txt{sim}}$\\
		\hline
		\hline
		G. &$r_{\txt{c}} = 4$& 20 & \co{$2.26 \pm 0.17$} & $1.35\pm0.02$ & 0.72 &  $1.88\pm0.22$  & $1.35\pm0.03$& 0.46  \\
		\hline
		 G. &$r_{\txt{c}}= 2$&  20 & \cred{$1.29 \pm 0.34$}&\cred{$1.14\pm0.03$}& 0.11&  \co{$2.47\pm0.21$} & $1.39\pm0.03$  & 0.08 \\
		\hline
		G. &$r_{\txt{c}}= 4$& 200 &  $1.73 \pm 0.12$ & $1.39\pm0.01$ & 7.14&   \cg{$1.24\pm0.09$} & $1.36\pm0.02$& 4.48  \\
		\hline
		 G. &$r_{\txt{c}} = 2$& 200 &  \cred{$-0.02 \pm 0.09$} &\cred{$1.08\pm0.02$}& 1.11&  \cg{$1.38 \pm 0.13$} & $1.45\pm0.02$ & 0.79 \\
		\hline
		tr. &$n=2$& 200 & ---&---&--- &  \cg{$1.38 \pm 0.09$} & $1.43\pm0.02$  &0.50 \\
		\hline
		tr. &$n=4$& 200 & ---&---&--- &  \cred{$0.07\pm0.04$} & \cred{ $1.01\pm0.00$} & 2.46 \\
		\hline
		dis. &$F_{\txt{c}} = \frac{1}{3}$& 200 & ---&---&--- &  \co{$4.25\pm0.18$}  & $1.40\pm0.01$ &  0.23\\
		\hline
		dis. &$F_{\txt{c}} = \frac{1}{3}$& 1000 & ---& ---&--- &  \co{$2.11 \pm 0.18$} & $1.51\pm0.02$ &  2.92 \\
		\hline
	\end{tabular}
\end{table}

\begin{table}[t]
	\caption[Effects of different smearing and derivative calculation modes on energy conservation and simulation time for the VDF potential (2)]{Values of $\Delta E$ (in MeV/nucleon), $R_{\txt{MF}}$, and $t_{\txt{sim}}$ (in minutes) for uniform nuclear matter simulated in a box with periodic boundary conditions, initialized in the spinodal region of the nuclear phase transition, as calculated using the VDF potential for different smearing schemes (``Sm.'', with abbreviations for Gaussian (``G.''), triangular (``tr.''), and discrete (``dis.'') smearing), values of smearing parameters (``par.''), numbers of test particles (``$N_T$''), and derivative modes (``$\bnabla$''). Gaussian smearing always used a width of $\sigma = 1\ \txt{fm}/c$.}
	\vspace{3mm}
	\label{density_and_derivatives_box_low_density_VDF_2}
	\def\arraystretch{1.12}
	\begin{tabular}{|r|r|r|r|r|r|r|r|r|}
		\hline
		\multicolumn{9}{|c|}{Uniform matter in a box, VDF potential: $T = 1\ \txt{MeV}$, $n_B = 0.25\ n_0$, $t_{\txt{end}} = 200\ \txt{fm}$.}  \\
		\hline
		\multirow{2}{*}{Sm.} &\multirow{2}{*}{par.} & \multirow{2}{*}{$N_T$} & \multicolumn{3}{r|}{finite $\bnabla$ (chain rule)}& \multicolumn{3}{r|}{finite $\bnabla$ (direct)}\\
		\cline{4-9}
		& & &$\Delta E$&$R_{\txt{MF}}$&$t_{\txt{sim}}$&$\Delta E$&$R_{\txt{MF}}$&$t_{\txt{sim}}$\\
		\hline
		\hline
		G. &$r_{\txt{c}} = 4$& 20 &  $1.88 \pm 0.22$  & $1.35\pm0.03$& 0.46  & $1.85\pm0.25$ & $1.36\pm0.02$ &  0.46 \\
		\hline
		G. &$r_{\txt{c}}= 2$&  20 & \co{$2.47 \pm 0.21$} & $1.39\pm0.03$  & 0.08 & \co{$2.24\pm0.21$} & $1.41\pm0.04$ & 0.09\\
		\hline
		G. &$r_{\txt{c}}= 4$& 200 &   \cg{$1.24 \pm 0.09$} & $1.36\pm0.02$& 4.48 & \cg{$1.19\pm0.07$} & $1.38\pm0.02$ & 4.59\\
		\hline
		G. &$r_{\txt{c}} = 2$& 200 &  \cg{$1.38 \pm 0.13$} & $1.45\pm0.02$ & 0.79 & \cg{$1.28\pm0.08$} & $1.47\pm0.02$ & 0.78\\
		\hline
		tr. &$n=2$& 200 &  \cg{$1.38 \pm 0.09$} & $1.43\pm0.02$  &0.50& \cg{$1.29\pm0.09$} & $1.45\pm0.01$ & 0.48\\
		\hline
		tr. &$n=4$& 200 & \cred{$0.07 \pm 0.04$} & \cred{ $1.01\pm0.00$} & 2.46 & \cred{$0.07\pm0.03$} & \cred{$1.01\pm0.00$} & 2.55\\
		\hline
		dis. &$F_{\txt{c}} = \frac{1}{3}$& 200 & \co{$4.25 \pm 0.18$}  & $1.40\pm0.01$ &  0.23& \co{$4.13\pm0.24$} & $1.41\pm0.02$ & 0.23 \\
		\hline
		dis. &$F_{\txt{c}} = \frac{1}{3}$& 1000 &  \co{$2.11 \pm 0.18$} & $1.51\pm0.02$ &  2.92 & \co{$2.04\pm0.15$} & $1.52\pm0.02$ & 2.90 \\
		\hline
	\end{tabular}
\end{table}

The direct finite difference derivatives are only available for the VDF model, and Gaussian derivatives only make sense when used together with the Gaussian smearing. Additionally, finite difference derivatives properly account for the time derivative of the baryon current, as already discussed in Section \ref{potentials_in_SMASH}. Table \ref{density_and_derivatives_box_low_density_Skyrme} focuses on results using the Skyrme potential, while Tables \ref{density_and_derivatives_box_low_density_VDF} and \ref{density_and_derivatives_box_low_density_VDF_2} show results using the VDF potential (the first of these tables compares the performance of Gaussian \textit{versus} finite difference chain rule derivatives, and the second compares the performance of chain rule \textit{versus} direct finite difference derivatives).

For an easier analysis of the simulations results shown in Tables \ref{density_and_derivatives_box_low_density_Skyrme}, \ref{density_and_derivatives_box_low_density_VDF}, and \ref{density_and_derivatives_box_low_density_VDF_2}, we color coded the most important cases. A red font marks simulation schemes which result in a completely incorrect evolution. For all of these cases, in each event the system fails to separate at all or the degree of separation is very low, which can be seen from the low values of $R_{\txt{MF}}$. This failure to undergo the spinodal decomposition is connected to a faulty combination of density and gradient calculation schemes; for example, in the case of the triangular smearing with the range parameter $n=4$ (corresponding to the smearing range equal $4L_i = 4\ \txt{fm}$), the intrinsic density variations due to the finite number of test particles are spread over a considerable volume, so that fluctuations at any given part of the box never become large enough to sustain a spinodal breakup. With orange font we denoted simulation schemes which result in the deviation from energy conservation larger than $\Delta E = 2\ \txt{MeV}/\txt{nucleon}$, while green font denotes simulation schemes where $\Delta E \leq 1.5 \txt{MeV}/\txt{nucleon}$. The best performing simulations are those utilizing finite difference derivatives and either the Gaussian smearing scheme at $N_T=200$ and $r_{\txt{cut}} = \{2\sigma, 4\sigma\}$ or the triangular smearing with $N_T=200$ and the triangular smearing range parameter $n=2$. We see that using Gaussian derivatives is strongly disfavored in most cases, which is understood to be a consequence of not properly resolving the density gradients in this derivative scheme; indeed, by construction, Gaussian derivatives are affected by smearing in the same manner as the density calculation, making the effects of density fluctuations highly non-local. We also found that among the best performing simulation schemes mentioned above, characterized by $\Delta E \leq 1.5 \txt{MeV}/\txt{nucleon}$, the simulation time for the Gaussian smearing with $r_{\txt{cut}} = 4\sigma$ is significantly longer than in the other two cases. Additionally, simulations utilizing the VDF model show that using direct finite difference derivatives leads to a further improvement in energy conservation.

To conclude our study of violations of the energy conservation in hadronic transport, we simulated systems initialized at different points of the phase diagram, where we also varied the simulation time $t_{\txt{end}}$ and the time step $\Delta t$, both measured in $\txt{fm}/c$. In these simulations, we used a parametrization of the VDF model reproducing the fourth (IV) set of characteristics listed in Table \ref{example_characteristics} (the parameters of this potential can be found in Appendix \ref{parameter_sets}), which describes nuclear matter with two phase transitions: one corresponding to the ordinary nuclear matter liquid-gas phase transition, and one corresponding to a possible QGP-like phase transition with the critical temperature $T_c^{(Q)} = 100\ \txt{MeV}$, the critical density $n_c^{(Q)} = 3.0n_0$, and the boundaries of the spinodal region at $T=0$ given by $\eta_L = 2.50n_0$ and $\eta_R = 3.32n_0$, where $n_0$ is the saturation density of nuclear matter, $n_0 = 0.160\ \txt{fm}^{-3}$ (this and similar EOSs are discussed in Chapter \ref{parametrization_and_model_results}, while a detailed discussion of the results of simulations using the chosen EOS is given in Chapter \ref{results}). For the density and gradients calculation, we used triangular smearing with the triangular smearing parameter $n=2$ and finite difference direct derivatives. As previously, the systems were initialized in a box of size $10\ \txt{fm}$ with periodic boundary conditions, the lattice used had 10 nodes per side, and we averaged over 10 events.

\begin{table}[t]
	\caption[Effects of different initialization points and using varying simulation parameters on energy conservation]{Values of $\Delta E$ (in MeV/nucleon), $R_{\txt{MF}}$, and $t_{\txt{sim}}$ (in minutes) for uniform nuclear matter simulated in a box with periodic boundary conditions, initialized for a series of baryon densities (``$n_B$'', in units of the saturation density $n_0 = 0.160\ \txt{fm}^{-3}$), temperatures (``$T$'', in MeV), and evolution times (``$t_{\txt{end}}$'', in $[\txt{fm}/c]$) while varying the number of test particles (``$N_T$'') and the time step (``$\Delta t$'', in $[\txt{fm}/c]$), as calculated using the VDF potential with two phase transitions, triangular smearing with $n=2$, and direct finite difference derivatives.}
	\vspace{3mm}
	\label{different_densities_and_temperatures_box_VDF}
	\def\arraystretch{1.12}
	\begin{tabular}{|r|r|r|r|r|r|r|r|}
		\hline
		\hspace{6mm}$n_B$&\hspace{6mm}$T$&\hspace{6mm} $t_{\txt{end}}$& \hspace{6mm}$N_T$ & \hspace{6mm}$\Delta t$ &\hspace{24mm}$\Delta E$ & \hspace{25mm}$R_{\txt{MF}}$ & \hspace{2mm} $t_{\txt{sim}}$\\
		\hline
		\hline
		0.25 & 1 & 200 & 200 & 1.0  & $ 1.496\pm 0.121$  & $ 1.455 \pm 0.012$ & 0.09 \\
		\hline
		0.25 & 1 & 200 & 200 & 0.1  & $ 1.435 \pm 0.121$  & $1.463 \pm 0.012$  & 0.53 \\
		\hline
		0.25 & 1 & 200 & 200 & 0.01  & $1.424 \pm 0.117$  & $1.458 \pm 0.011$ & 4.71 \\
		\hline
		0.25 & 1 & 400 & 200 & 0.1  & $1.784 \pm 0.114$  & $ 1.465\pm 0.014$ & 1.04 \\
		\hline
		0.25 & 1 & 200 & 500 & 0.1  & $ 1.295 \pm 0.104$  & $ 1.478\pm 0.014$  & 1.64 \\
		\hline
		0.25 & 25 & 200 & 200 & 0.1  & $0.138 \pm 0.023$  & $1.002 \pm 0.001$  & 0.54 \\
		\hline
		1.00 & 1 & 200 & 200 & 0.1  & $ 0.012 \pm 0.001$  & $ 1.0 \pm 1.2 \times 10^{-5}$  &  3.27\\
		\hline
		3.00 & 1 & 50 & 50 & 0.1  & $ 1.903\pm 0.087$  & $1.054 \pm 0.004$  & 0.59 \\
		\hline
		3.00 & 50 & 50 & 50 & 0.1  & $ 0.714 \pm 0.040$  & $ 1.031\pm 0.004$  & 1.06 \\
		\hline
		3.00 & 125 & 50 & 50 & 0.1  & $0.324 \pm 0.042$  & $1.004 \pm 0.001$  & 0.82\\
		\hline
	\end{tabular}
\end{table}

A representative collection of results of these simulations is gathered in Table \ref{different_densities_and_temperatures_box_VDF}. First, in the case of nuclear matter initialized in the spinodal region of the nuclear phase transition, at $n_B = 0.25n_0$ and $T=1 \ \txt{MeV}$, we see that the degree to which the energy is not conserved depends on the time step $\Delta t$, with smaller unphysical contributions to the energy for smaller values of $\Delta t$. Consequently, the violation of energy conservation is seen to occur partially due to errors stemming from the numerical integration of the equations of motion. However, very small time steps $\Delta t$ lead to significant increases in the simulation time, therefore obtaining greater accuracy must be balanced with practical considerations. 

We also find that doubling the duration of a simulation $t_{\txt{end}}$ does not lead to doubling of the unphysical energy gain $\Delta E$. This underscores the fact that the rapid non-equilibrium evolution of the system undergoing a spontaneous spinodal decomposition leads to the most significant contributions to the violation of energy conservation. In other words, once the spinodal decomposition has taken place (which for the case of the nuclear spinodal decomposition at $T = 1 \ \txt{MeV}$, considered here, occurs around $t \approx 100 \ \txt{fm}/c$), the evolution of the system is shown to be more smooth.

Next, we consider the case of nuclear matter initialized deep inside the spinodal region of the QGP-like phase transition, at $n_B = 3.0n_0$ and $T=1 \ \txt{MeV}$. Similarly to the spontaneous spinodal decomposition in ordinary nuclear matter, we can see that the unphysical energy gain $\Delta E$ is substantial. (We note that the relatively small values of $R_{\txt{MF}}$ obtained in this case do not indicate an incorrect evolution, unlike in the example of the spinodal decomposition in ordinary nuclear matter. This can be understood as follows: After the nuclear spinodal fragmentation occurs, nearly all test particles can be found in the region where the nuclear liquid drop forms, and consequently the change in the mean-field energy of the system roughly corresponds to the difference between the mean-field energy at $n_B= 0.25n_0$ and the mean-field energy at $n_B = n_0$. In contrast, during the spinodal decomposition due to the QGP-like phase transition, the system separates into a ``less dense'' ($n_B \approx 2.13n_0$) and a ``more dense'' ($n_B \approx 3.57n_0$) phase, and the corresponding changes in the mean-field energies of these parts of the system are opposite in sign. Consequently, the total change in the mean-field energy is relatively smaller in the latter case.) At the same time, $\Delta E$ decreases appreciably for an initialization at $n_B = 3.0n_0$ and $T=50 \ \txt{MeV}$; based on the phase diagram of the chosen VDF EOS (Fig.\ \ref{phase_diagram_nB}), we can see that at this higher temperature the widths of the spinodal and coexistence regions are smaller than at $T = 1 \ \txt{MeV}$, which likely leads to less steep density gradients developing in the system and through that prevents large unphysical energy gains. Furthermore, the results obtained for systems initialized above the critical point of either of the transitions ($T = 25 \ \txt{MeV}$ and $T = 125\ \txt{MeV}$ for the nuclear and QGP-like phase transition, respectively) show values of $\Delta E$ that are fractions of the values obtained at $T = 1 \ \txt{MeV}$. This is similarly understood to be a consequence of the fact that at high temperatures, the thermal noise prevents the system from forming any sustained density gradients. Note that by construction, the VDF potential does not depend on the temperature; therefore, the differences observed in the deviations from energy conservation at different temperatures can be entirely assigned to the presence or lack of non-uniform density regions in the simulated systems.

Finally, we can easily see that $\Delta E$ is negligible for the initialization at the equilibrium point of nuclear matter, that is at $n_B = n_0$ and $T = 1 \ \txt{MeV}$. This is, naturally, connected with the fact that at the equilibrium point of nuclear matter the forces acting on the system are vanishing and, at the same time, any fluctuations away from equilibrium decay fast, leading to only minuscule violation of the conservation of energy. 

In general, we conclude that contributions to $\Delta E$ are large for systems initialized in regions of the phase diagram where the density gradients, and therefore forces acting on the test particles, are also large, e.g., inside the spinodal region of a phase transition. Conversely, energy conservation is very satisfactory when forces acting on the test particles are small, e.g., in regions of the phase diagram where nuclear matter is thermodynamically stable.

\section{Summary}

In this chapter, we have given a brief overview of hadronic transport simulations, including the description of the method of test particles, and we have also described some of the computational details specific to simulations performed with the hadronic transport code \texttt{SMASH}. In particular, we focused on the aspects of the simulation connected with the calculation of the baryon density, and we described the density calculation schemes available in \texttt{SMASH}, many of which were introduced as part of this work. Next, we described the types of mean-field potentials available in \texttt{SMASH}, including the implementation of the fully relativistic VDF model potentials, which was part of the work underlying this thesis. Finally, we provided a comparison of different numerical schemes for calculating mean fields and mean-field evolution in \texttt{SMASH}, based on the degree to which the integration of the equations of motion satisfies energy conservation. We demonstrated that both the implemented VDF potentials as well as the implemented density and mean-field calculation schemes perform better than the original options available in \texttt{SMASH}; in addition, these schemes are shown to be computationally more efficient. We concluded by discussing the numerical performance of simulations initialized at different points of the phase diagram.

\newpage
\chapter{Results from hadronic transport simulations}
\label{results}

The contents of this chapter are largely based on Ref.\ \cite{Sorensen:2020ygf}.

We investigate simulations of nuclear matter in \texttt{SMASH} \cite{Weil:2016zrk} (version 1.8 \cite{smash_version_1.8}) realized in a box with periodic boundary conditions. Such studies are best suited for testing the thermodynamic behavior following from equations of motion with mean-field interactions, as well as for exploring observables sensitive to critical phenomena in a scenario in which matter is allowed to equilibrate. While admittedly systems considered here cannot be reproduced in the laboratory, insights gained in this study will provide a useful stepping stone to understanding results of simulations of heavy-ion collisions utilizing the VDF EOS, planned for future work.

\section{Analysis}

In contrast to heavy-ion collision experiments, semi-classical hadronic transport simulations have an access to the positions of individual particles. Consequently, observables that can be used as a measure of the collective behavior of the system include the spatial pair correlation function and the distribution of particles in coordinate space. Below, we describe the details of extracting these observables.

\subsection{Pair distribution function}
\label{section_pair_distribution_function}

The radial distribution function $g(r)$ gives the probability of finding a particle at a distance $r$ from a reference particle. While in select simple cases it can be calculated analytically, in practice, given a distribution of particles, $g(r)$ is obtained by determining the distance between the reference particle and all other particles and constructing a corresponding histogram. Thus for finding the radial distribution about the $i$th (reference) particle at a given distance $r$, we count all particles within an interval $\Delta r$ around $r$, which can be written as
\begin{eqnarray}
\hspace{-5mm}g_i (r, \Delta r) = \sum_{\substack{j=1 \\ j\neq i}}^N \theta\Big(  r + \Delta r - \mathcal{R}_{ij}  \Big) \theta\Big( \mathcal{R}_{ij} - ( r - \Delta r )   \Big) ~.
\label{radial_distribution_function}
\end{eqnarray}
Here, the sum is performed over all particles (with the exception of the $i$th particle) which we index by $j$, $N$ is the total number of particles, $\theta$ is the Heaviside theta function, and $\mathcal{R}_{ij} = | \bm{r}_i - \bm{r}_j|$, where $\bm{r}_i$ is the position of the reference particle and $\bm{r}_j$ is the position of the $j$th particle. The role of the Heaviside theta functions is to only allow contributions from particles whose positions are within a distance $\mathcal{R}_{ij} \in (r - \Delta r, r + \Delta r)$ from the reference particle. The obtained histogram is then normalized with respect to an ideal gas, whose radial distribution histogram is that of completely uncorrelated particles, $g_{0}(r) \propto n ~ 4\pi r^2~ dr$, where $n$ denotes density.

\begin{figure}
	\centering\mbox{
	\includegraphics[width=0.4\textwidth]{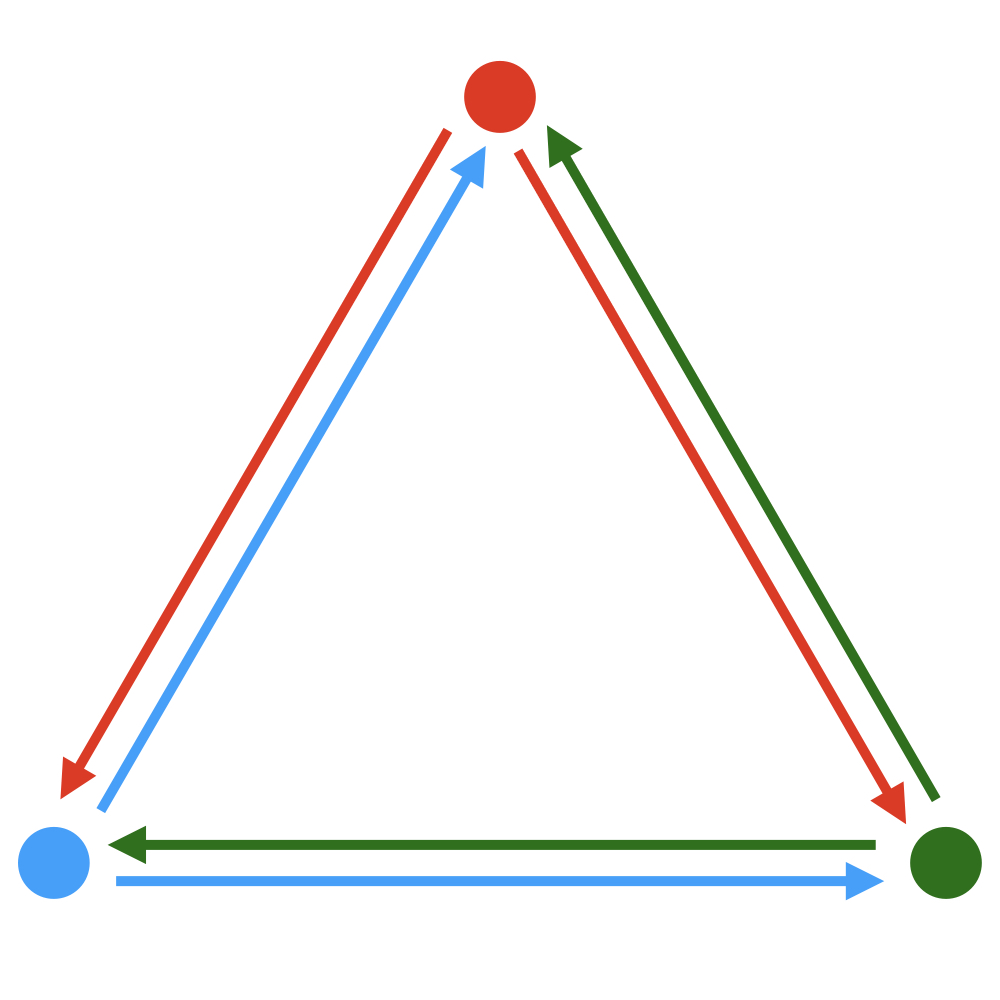} 
	}
	\caption[A diagram illustrating the meaning of the pair distribution function]{Three particles positioned at the vertices of a unilateral triangle of side $r_0$. The arrows point to particles at a distance $r_0$ from a given reference particle (color coded). The figure helps to interpret the pair distribution function $\widetilde{g}(r_0, \Delta r)$ as the number of distinct pairs of particles separated by a distance $r_0 \pm \Delta r$ (see text for details).}
	\label{example_triangle}
\end{figure} 
We can further define the radial distribution function of all distinct pairs in the system, or the pair distribution function, 
\begin{eqnarray}
\hspace{-5mm}\widetilde{g} (r, \Delta r) &=& \mathcal{N}~ \sum_{i=1}^N g_i (r, \Delta r)  = \frac{\mathcal{N}}{2} \sum_{i=1}^N \sum_{ \substack{j=1\\ j \neq  i}}^N \theta\Big(  r + \Delta r - \mathcal{R}_{ij}   \Big) \theta\Big( \mathcal{R}_{ij} - ( r - \Delta r )   \Big) ~,
\label{pair_distribution_function}
\end{eqnarray}
where the factor of $1/2$ appears to avoid counting any of the particle pairs twice, and where $\mathcal{N}$ is a normalization factor, so far unspecified (as already mentioned above, in practice the radial distribution function is compared to that of an ideal gas, in which case the normalization factors cancel out). It can be easily seen that $\widetilde{g}(r, \Delta r)$ is equal to the number of distinct (hence the factor of $1/2$) particle pairs separated by a distance $r \pm \Delta r$. As a simple example, consider a system composed of 3 particles whose relative distances are equal, $\mathcal{R}_{ij} = r_0 ~ \forall~ i,j$; naturally, these particles must sit at the vertices of a unilateral triangle of side $r_0$, see Fig.\ \ref{example_triangle}. The pair distribution function $\widetilde{g}(r_0, \Delta r)$ is equal to half the sum over the number of particles that each reference particle sees at a distance $r_0 \pm \Delta r$ away. Clearly, the first reference particle sees 2 particles at a distance $r_0$, and likewise for the second and third reference particles, so that in this case $\widetilde{g}(r_0, \Delta r) = (3 \times 2)/2$, that is the number of distinct particle pairs separated by a distance $r_0 \pm \Delta r$.

The pair distribution function of an ideal gas, denoted by $\widetilde{g}_0(r)$, is related to $g_0(r)$ through $\widetilde{g}_0(r) \simeq (N/2) g_0(r)$, where $N$ is the total number of particles in the system, which stems directly from the fact that the total number of distinct pairs in the system is equal $N(N-1)/2$. For simulations in a box with periodic boundary conditions, however, this relationship becomes more complicated for distances $r > L/2$, where $L$ is the side length of the box, due to geometry effects (see the discussion below). For this reason and because in simulations presented in this work we initialize the systems uniformly, in our analysis we use the $t=0$ histogram as the reference pair distribution function, $\widetilde{g}_0 = \widetilde{g}(t=0)$.

We stress that taking the pair distribution function of a uniform system as the reference ensures that the normalized pair distribution function $\widetilde{g}/\widetilde{g}_0$ is sensitive to density fluctuations in the system. A prominent example here is the spinodal breakup, where a spontaneous separation into two coexistent phases with different densities occurs. If the system is confined to some constant volume $V$, then the average density of the system is the same before and after the spinodal decomposition takes place. However, local fluctuations in the number of particles will be visible in the pair distribution function, as more particle pairs reside inside a high density region as compared to a low density region.

While the spinodal decomposition is the most obvious example of a situation where $\widetilde{g}/\widetilde{g}_0 \neq 1$, the normalized pair distribution function deviates from unity for any system in which the interactions between the particles affect their collective behavior. In particular, at small $r$, the normalized pair distribution function satisfies $\widetilde{g}/\widetilde{g}_0 > 1$ for correlated particles and $\widetilde{g}/\widetilde{g}_0 < 1$ for anti-correlated particles (see Appendix \ref{pair_distribution_function_and_the_second-order_cumulant} for a detailed derivation), which corresponds to attractive and repulsive interactions between the particles, respectively. Since the number of particles and thus the number of pairs is conserved, one sees an opposite trend at intermediate to large distances. 

We note that in our simulations the range of $r$ over which $\widetilde{g}(r)/\widetilde{g}_0(r)$ significantly deviates from 1 is related to the range of the interaction, which is determined by the smearing range in the density calculation (for more details see Section \ref{calculation_of_mean-fields}).

\subsubsection{Pair distribution function in a system with periodic boundary conditions}

Importantly, for a system with periodic boundary conditions the radial distance between two particles $\mathcal{R}$ is not uniquely defined. This is because for any reference particle the distance to any other particle can be calculated using the position of that other particle in the original box or in any of its 26 equivalent images. We adopt a prescription in which the smallest distance between particles is used in calculating the pair distribution function $\widetilde{g}$ (known as the minimum image criterion). This smallest distance can range from $\mathcal{R}_{\txt{min}} = 0$ to $\mathcal{R}_{\txt{max}} = \infrac{\sqrt{3}L}{2}$, where $L$ is the side length of the box.

\begin{figure}[t]
	\centering\mbox{
	\includegraphics[width=0.54\textwidth]{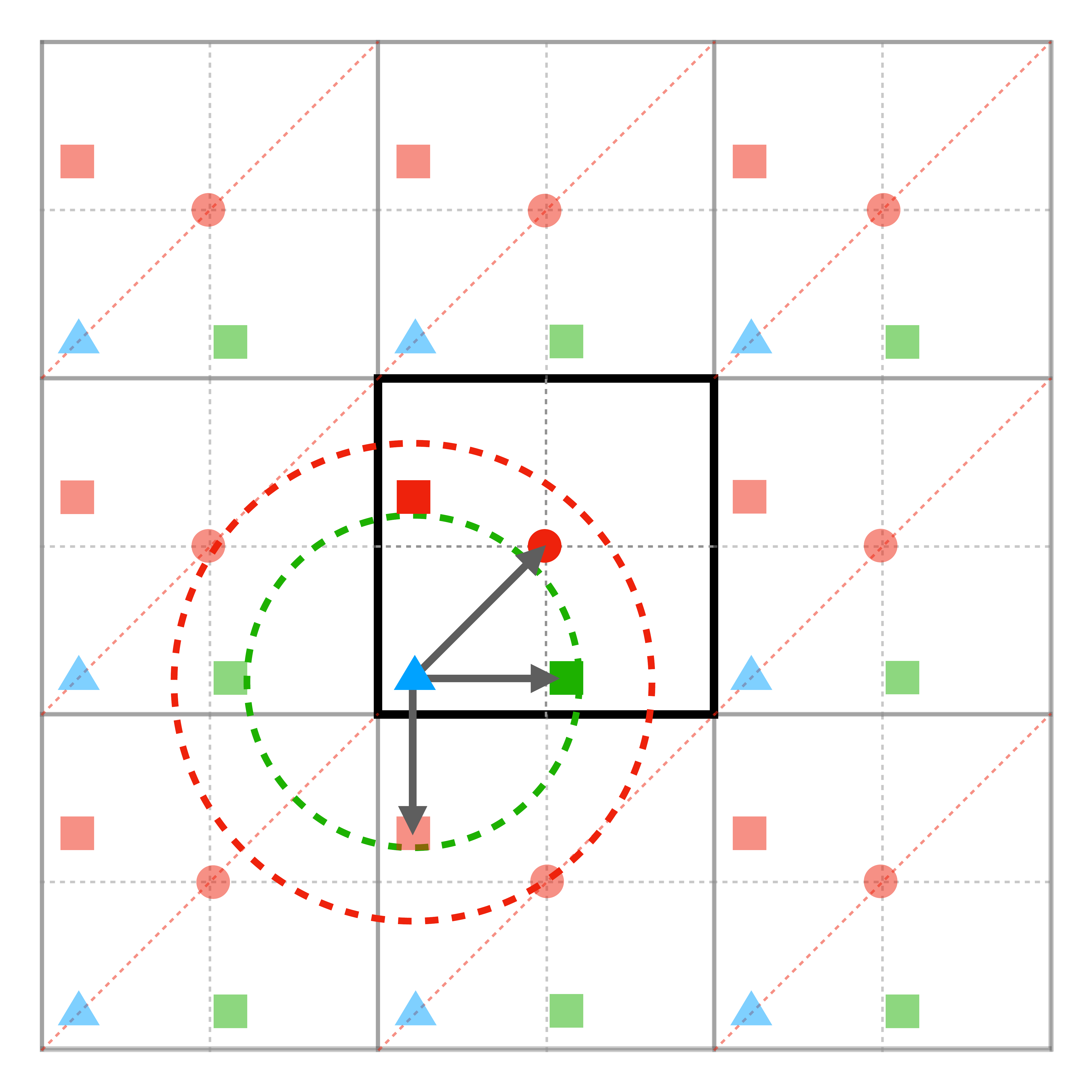} 
	}
	\caption[A diagram illustrating the effects of periodic boundary conditions on the pair distribution function at large interparticle distances]{A 2D diagram of geometry effects affecting the pair distribution function of a uniform system in a box with periodic boundary conditions. We draw the simulated box, with the reference particle denoted with a blue triangle and three other particles denoted with a green square, a red dot, and a red square, as well as its 8 equivalent images (faded colors). A green dashed circle of radius $L/2$ and a red dashed circle of radius $L/\sqrt{2}$ are drawn around the reference particle. Distances from the reference particle to other particles which contribute to the pair distribution function are implied with gray arrows.}
	\label{R_max_explained}
\end{figure}

That said, even for a uniform and uncorrelated system the geometry of the problem affects the number of particles that can be encountered at the maximal distance $\mathcal{R}_{\txt{max}}$. Let us show this based a 2D illustration of this issue, see Fig.\ \ref{R_max_explained}. To calculate the minimal distance between any two particles, we consider the original system (in the middle) and its 8 equivalent images (the latter are indicated by faded colors). We consider a particle at a position denoted with a blue triangle, and to aid the analysis we draw a green dashed circle of radius $L/2$ and a red dashed circle of radius $L/\sqrt{2}$ around it. Consider two other particles, whose positions are denoted with a red square and a red dot, separated from the blue triangle by an equal distance $\mathcal{R}$ such that $ L/2 < \mathcal{R} < L/\sqrt{2}$. For the particle pair composed of the blue triangle and the red dot, the line separating the particles lies along the diagonal of the box; since $\mathcal{R}$ is already the smallest distance between the blue triangle and any of the equivalent images of the red dot, this particle pair contributes to the pair distribution function at $\mathcal{R}$. For the particle pair composed of the blue triangle and the red square, the line separating the particles lies along one of the principle directions of the box; in contrast to the previous case, there is an equivalent image of the red square characterized by a smaller distance to the blue triangle $\mathcal{R}' < \mathcal{R}$; consequently, this pair contributes to the pair distribution function at $\mathcal{R}'$. In fact, particles pairs lying along the diagonal and along one of the principal box directions are special cases. Notice, in particular, that the only points for which it is possible to have $\mathcal{R} = \mathcal{R}_{\txt{max}}$ are points on the diagonal of the box; for any points separated by $\mathcal{R}_{\txt{max}}$ that are not on the diagonal, there exists a smaller $\mathcal{R}$ obtained by using the position of the second particle from one of the equivalent box images.  This problem also affects, to a proportionally lesser extent, interparticle distances $\mathcal{R}$ in the range  $\infrac{L}{2} < \mathcal{R} < \mathcal{R}_{\txt{max}}$. Only in the case of particles which are $\infrac{L}{2}$ or less apart the geometry of the box never affects the pair distribution function, as can be seen from the example of the particle depicted with a green square.

\begin{figure}[t]
	\centering\mbox{
	\includegraphics[width=0.5\textwidth]{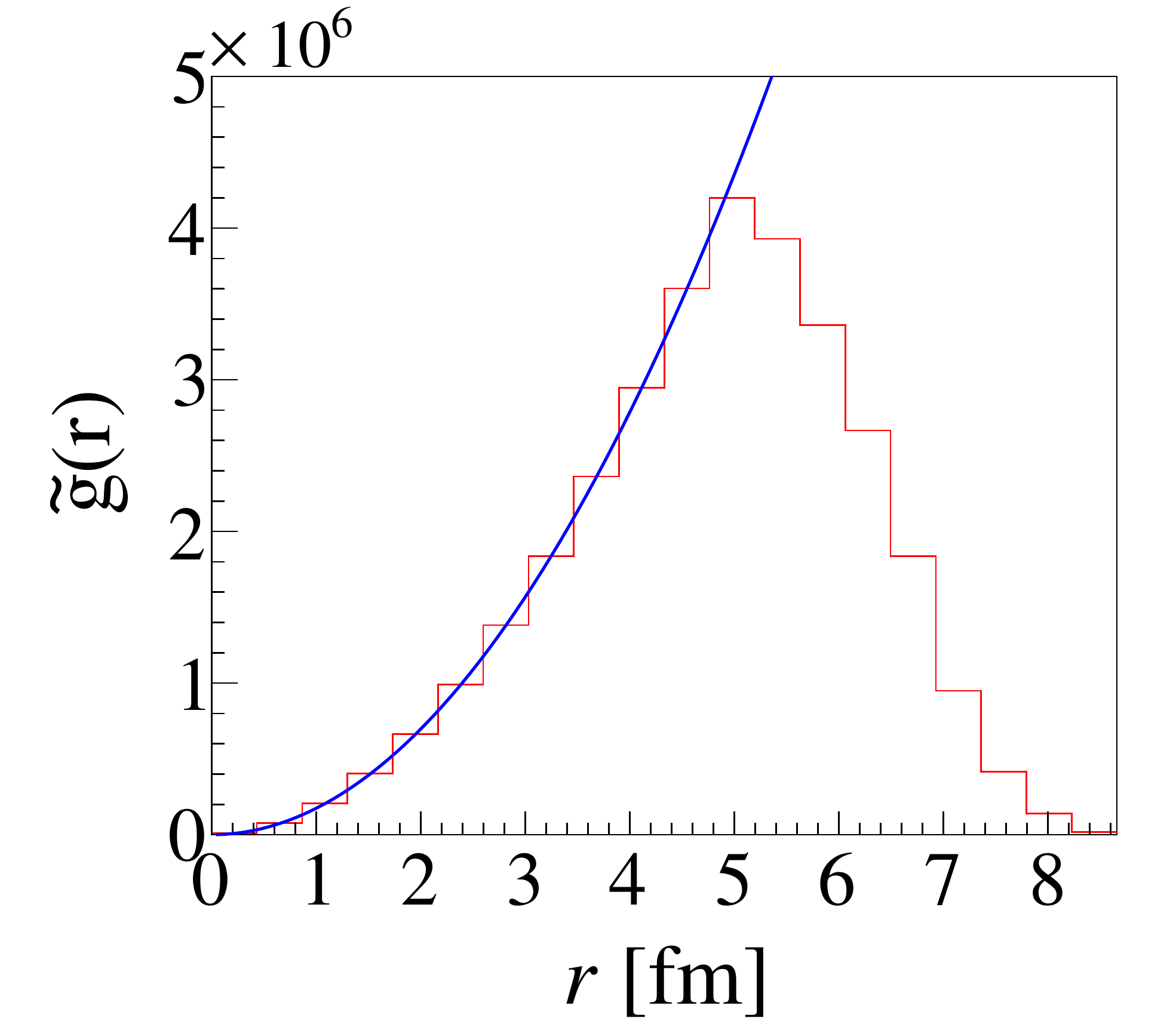} }
	\caption[Pair distribution function of spatially uniform matter in a cubic box with periodic boundary conditions]{Pair distribution function of spatially uniform matter in a cubic box with periodic boundary conditions (red histogram). To obtain the histogram, a cubic box of side length $L =10\ \txt{fm}$ was initialized with proton and neutron numbers $N_p = N_n = 20$ (corresponding to a baryon density of $0.25n_0$), and the number of test particles per particle was set at $N_T=200$, resulting in the test particle density of $n = 8\ \txt{fm}^{-3}$. For $r < L/2$, the pair distribution function is given by the pair distribution function of a uniform ideal gas, $\widetilde{g} (r) \simeq (N/2) ~ n ~ 4\pi ~ r^2 ~ \Delta r$ (solid blue line), where $\Delta r = 0.433$ is the histogram bin width. For $r > L/2$, geometric effects related to the periodic boundary conditions start playing a role; see text for more details.}
	\label{geometric_effects_with_analytic_fit}
\end{figure}

This influence of finite size effects on the pair distribution function can be clearly seen in Fig.\ \ref{geometric_effects_with_analytic_fit}, which shows the pair distribution function for a uniform matter in a cubic box of side length $L = 10 \  \txt{fm}$ and subject to periodic boundary conditions. In infinite matter, the pair distribution function of uncorrelated particles grows like $r^2$. However, finite geometry effects described above introduce an effective cut on the distribution starting at $\infrac{L}{2} = 5 \ \txt{fm}$, explaining the shape of the presented histogram.

\begin{figure}
	\centering\mbox{
	\includegraphics[width=0.65\textwidth]{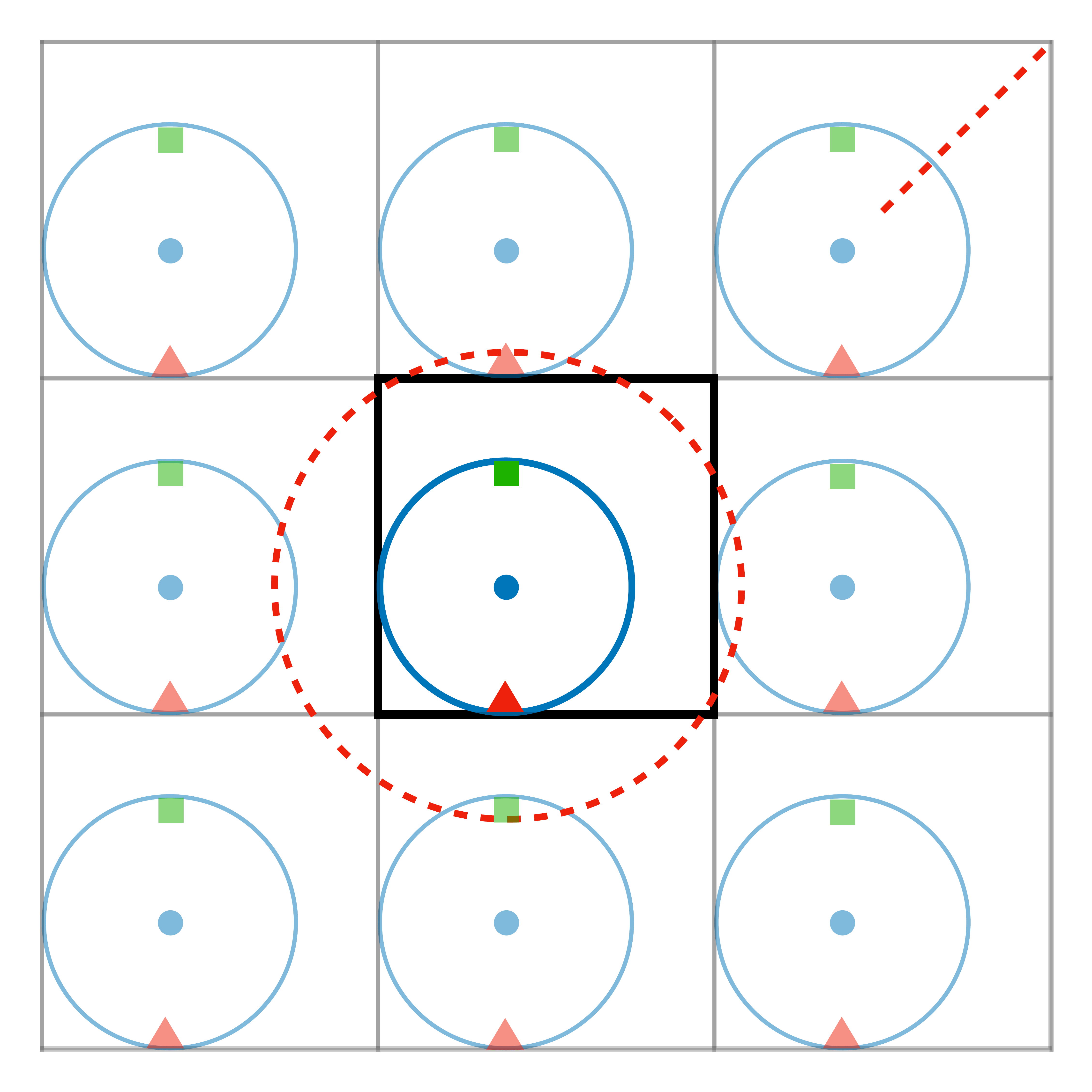} 
	}
	\caption[A diagram illustrating the effects of periodic boundary conditions on the normalized pair distribution function]{A 2D diagram of the geometry effects affecting the normalized pair distribution function of a non-uniform system. A clump of particles (blue circle) surrounded by vacuum is shown in a box of side $L$ (middle box), and positions of three hypothetical particles are indicated with a green square, a blue circle, and a red triangle. We indicate a range over which the pair distribution function is calculated for the particle in the middle of the clump (blue dot) with a circle of radius $L/\sqrt{2}$ centered around that particle, and we also show the 8 equivalent images of the box (faded colors). The smallest equivalent distance between the particles represented by the red triangle and the green box is realized ``across the vacuum'', either from the red triangle to the faded green square, or from the green square to the faded red triangle. Such a situation will arise for many particles on the edge of the clump, leading to a peculiar behavior of the pair distribution function at distances comparable to $L/\sqrt{2}$. See text for more details.}
	\label{equivalent_images_and_geometry_effects}
\end{figure}

Naturally, geometry and periodic boundary conditions also play a role in the shape of the normalized pair distribution function in non-uniform systems. As an easy illustration, consider a 2D system composed of particles forming a spherical ``clump'' of matter surrounded by vacuum, see Fig.\ \ref{equivalent_images_and_geometry_effects}. It is easy to see that if the diameter of the clump satisfies $D > \infrac{L}{2}$, then for some of the particles forming the clump the smallest distance to some of the other particles in that same clump will be ``across the vacuum'', to one of the equivalent mirror images of the latter particles; in the figure this is showcased by the example of particle positions represented by the green square and the red triangle. In result, on average, the normalized pair distribution function for this system will display a strong correlation for small values of $r$ (reflecting the existence of the clump), then an anticorrelation for intermediate values of $r$ (reflecting the vacuum), and finally a significant correlation for values of $r$ approaching $\mathcal{R}_{\txt{max}}$ (reflecting the smallest distances ``across the vacuum'').

In our simulations, which we will discuss at length in the following Sections, we clearly observe the above described effect in the case of nuclear spinodal decomposition, which yields a nuclear drop surrounded by a nearly perfect vacuum. The diameter of the nuclear drop turns out to satisfy $D > \infrac{L}{2}$, which means that for some of the test particles belonging to that drop, the smallest distance to some of the other test particles in that same drop is obtained ``across the vacuum''. This explains the rise in the normalized distribution function for $r > \infrac{L}{2}$, which can be seen, e.g., on the right-hand panel in Fig.\ \ref{spinodal_decomposition_nuclear}. The magnitude of this effect depends on the drop diameter $D$.

Finally, we note that the artifacts produced by the geometry of the simulation and periodic boundary conditions do not present a significant complication in analyzing critical behavior if we resolve to only probe the system at length scales on the order of $\infrac{L}{2}$ or smaller.

\subsubsection{Pair distribution function and Boltzmann transport}

One may ask whether calculating a pair distribution function in hadronic transport is justified in view of the fact that the BUU equation explicitly evolves a one-body distribution function which does not carry any information about the two-body distribution, usually employed in the description of two-particle correlations (see Appendix \ref{the_Boltzmann_equation} for a derivation of the Boltzmann equation from the BBGKY hierarchy, where the relevant steps leading to neglecting the two-body contributions are shown). While this may appear to be problematic, a closer look reveals that our analysis is correct. First, one needs to note that hadronic transport simulations only solve the Boltzmann equation exactly in the limit of an infinite number of test particles per particle $N_T$. The finite number of test particles employed in simulations leads to intrinsic numerical fluctuations. These numerical fluctuations are of statistical nature, similarly to variances of microscopic observables, and likewise, through both scattering and mean fields, they can become a seed for collective behavior such as spontaneous spinodal decomposition. Such effects have been described, e.g., in Ref.\ \cite{Bonasera:1994iam} (see also Refs.\ \cite{Bonasera:1990ikj,Bonasera:1992zz}), where fluctuation observables calculated using hadronic transport with the method of test particles agree with both theoretical predictions and experimental results. Additionally, it was established that for large enough $N_T$ (which the authors of that particular study found to be $N_T \gtrsim 40$) the numerical noise intrinsic to the method of test particles is negligible, while the correct statistical fluctuations are preserved. 

It is possible to construct a Boltzmann-Langevin extension of the standard BUU equation, which ensures that the simulated fluctuations are physically correct (see, e.g., \cite{Burgio:1991ej}). However, it has been found that, for example, in the case of the nuclear spinodal fragmentation the source of the noise seeding the spinodal decomposition is not essential, and it is possible to develop good approximations to the Boltzmann-Langevin equation that are numerically favorable, including the method of test particles \cite{Chomaz:2003dz}. 

We note here that a particular problem that arises in the method of test particles is that the fluctuations in the events, simulating the evolution of $N_T N_B$ test particles, are suppressed by a factor of $N_T$. The authors of \cite{Bonasera:1994iam} dealt with this issue by employing the method of parallel ensembles at final simulation times, that is \textit{a posteriori}, which allows one to obtain events with the number of test particles corresponding to the physical baryon number $N_B$ (we briefly describe this method in Section \ref{transport_simulations}, while Appendix \ref{parallel_ensembles_a_posteriori} explains the \textit{a posteriori} application of the method).

Based on the above it is apparent that the distribution function obtained through hadronic transport simulations, and in particular through the method of test particles, contains information not only about the mean of the distribution function $\langle f_{\bm{p}} \rangle$, but also about its fluctuations. Consequently, calculating fluctuation observables such as the pair distribution function is well-defined in hadronic transport. Some questions regarding the quantitative behavior of fluctuation observables obtained in simulations using the number of test particles $N_T > 1$ remain, in particular regarding the specific methods used to connect fluctuations in systems evolving $N_T N_B$ particles as compared to systems evolving $N_B$ particles. For that reason we refrain from making quantitative statements at this time, and focus on the qualitative behavior of the pair distribution functions. Future work will be devoted to a quantitative analysis of this problem, and in Section \ref{effects_of_finite_number_statistics} we give a short overview of the effects due to this issue.

\subsection{Number distribution functions}
\label{number_distribution_functions}

A complementary method of analyzing the collective behavior in a simulation utilizes coordinate space number distribution functions. To calculate number distribution functions, we divide the simulation box into $C$ cells of side length $\Delta l$ (also referred to as cell width), and construct a histogram of the number of cells in which the number of particles lies in a given interval $N_i \pm \Delta N$, where $N_i$ is the central value of the $i$th bin. We note that we scale the entries by the total number of cells $C$ so that the resulting histogram is a properly normalized representation of the corresponding probability distribution. We also note that in the subsequent parts of this work we scale the histogram entries by the volume of the cells $(\Delta l)^3$ in order to obtain the histogram as a function of number density.

The test-particle evolution in \texttt{SMASH} is governed by the mean field, which depends on the underlying continuous baryon number density for a given baryon number $N_B$,
\begin{eqnarray}
n_B(\bm{x}; N_B) = g  \int \frac{d^3p}{(2\pi)^3} ~  f(\bm{x}, \bm{p})~.
\label{continuous_baryon_number_density_distribution_function}
\end{eqnarray}
Formally, hadronic transport can give access to $n_B(\bm{x}; N_B)$ through solving the Boltzmann equation, Eq.\ (\ref{BUU_equation}), in the limit of infinitely many test particles per particle, and substituting the obtained distribution function $f (\bm{x}, \bm{p})$ in Eq.\ \eqref{continuous_baryon_number_density_distribution_function}. Below, we present three number distribution functions accessible in practice given the finite number of test particles used.

\subsubsection{Test particle number distribution function}

Hadronic transport simulations of nuclear matter are realized through evolving $N = N_B N_T$ test particles in space and time (where $N_B$ is the baryon number in the simulation and $N_T$ is the number of test particles per particle), giving a direct access to a discrete test particle number distribution function. This distribution can be written as a probability of obtaining a cell contributing to the $i$th bin of the histogram with a center value $N_i$ (that is, a cell with $N \in (N_i - \Delta N, N_i + \Delta N)$ test particles),
\begin{eqnarray}
P_N (N_i) &=& P \Big(N_i, N (N_B, N_T), \Delta l\Big) = 
\label{test_particle_number_distribution_function_def}
\\
&=& \frac{N^{(i)}_c \Big(N( N_B, N_T), \Delta l\Big)}{C} ~.
\label{test_particle_number_distribution_function}
\end{eqnarray}
Here, $C$ is the total number of cells used and $N^{(i)}_c$ is the number of cells containing a number of test particles $N$ within the range $N_i \pm \Delta N$. We note that the number of test particles in any given cell depends both on the baryon number evolved in the simulation $N_B$ and the number of test particles per particle $N_T$. We also stress that the distribution $P_N$ depends on the scale (chosen cell width $\Delta l$) at which the system is analyzed.

\subsubsection{Continuous baryon number distribution function}
\label{Continuous_baryon_number_distribution_function}

The discrete test particle distribution function, Eq.\ \eqref{test_particle_number_distribution_function_def}, can be thought of as having been obtained through sampling from the underlying continuous baryon number distribution function with a finite number $N_T N_B$ of test particles. Given access to the underlying baryon number distribution, one could use it directly to create a corresponding histogram. Indeed, the number of baryons at a cell at position $\bm{x}_k$ is given by the integral of the continuous baryon number density, Eq.\ (\ref{continuous_baryon_number_density_distribution_function}), 
\begin{eqnarray}
B ( \bm{x}_k) = \int_{V_k = (\Delta l)^3} dV~  n_B (\bm{x}, N_B) ~,
\end{eqnarray}
where $k$ indexes the histogram cells. Adding contributions from all cells yields the total baryon number in the system, $B$. We can then construct a probability distribution function for encountering a cell with a given number of baryons $N_i$,
\begin{eqnarray}
P_B (N_i) = \frac{N_{c}^{(i)}  \Big(N_i, B, \Delta l\Big)  }{C}~,
\end{eqnarray}
where $N_{c}^{(i)}$ is the number of cells containing a number of baryons $N$ within the range $N_i \pm \Delta N$. 

For a large number of test particles per particle $N_T$, statistical observables calculated using the test particle number distribution, with the number of test particles in a given sample scaled by $\infrac{1}{N_T}$, are a very good approximation to the underlying continuous baryon number distribution \cite{Steinheimer:2017dpb}. That is, it can be shown that
\begin{eqnarray}
P_B (N_i) = \lim_{N_T \to \infty} P \bigg(N_i , \frac{N(N_B, N_T)}{N_T}, \Delta l\bigg) ~.
\label{baryon_number_distribution_function}
\end{eqnarray}
Given that in our simulations we use sufficiently large numbers of test particles per particle $N_T$, we will refer to histograms constructed through the prescription on the right-hand side of Eq.\ (\ref{baryon_number_distribution_function}) as the continuous baryon number distribution function (or just baryon number distribution function) $P_B(N_i)$, with the understanding that it is only exact in the limit $N_T \to \infty$.

\subsubsection{Physical baryon number distribution function}
\label{Physical_baryon_number_distribution_function}

Both the test particle and the continuous baryon number distribution functions, Eqs.\ (\ref{test_particle_number_distribution_function_def}) and (\ref{baryon_number_distribution_function}), are markedly different from the physical baryon number distribution function corresponding to a discrete baryon number $N_B$. Here we can intuitively think of the physical baryon number distribution function as obtained through sampling from the underlying continuous baryon number distribution with $N_B$ test particles, 
\begin{eqnarray}
P_{N_B}(N_i) = P \Big(N_i, N ( N_B, N_T = 1), \Delta l\Big) ~.
\label{physical_baryon_number_distribution_function}
\end{eqnarray}

Strictly speaking, the physical baryon number distribution function could be obtained in transport by solving the Boltzmann equation in the limit of infinitely many test particles per particle, thus obtaining the underlying continuous baryon number distribution function, Eq.\ \eqref{continuous_baryon_number_density_distribution_function}, and sampling $n_B(\bm{x}, N_B)$ with $N_B$ particles. While such an approach is in principle valid, in practice it is impractical due to the enormous numerical coast necessary to approximate the limit of infinitely many test particles well. Alternatively, one can turn to the concept of parallel ensembles (introduced in Section \ref{transport_simulations}). It can be shown that the test particle distribution obtained within a parallel ensembles mode can serve as a proxy for the physical baryon number distribution. To reiterate, within the concept of parallel ensembles, a simulation corresponding to $N_B$ baryons with $N_T$ test particles per baryon is divided into $N_T$ events with $N_B$ test particles each. These $N_T$ events are not independent, as they share a common mean field. Nevertheless, at the end of the simulation we have access to $N_T$ events with the test particle number exactly corresponding to the baryon number in the ``real'' system. That is, each of the $N_T$ events is described by the probability distribution function $P_{N_B}(N_i) = P\Big(N_i; N (N_B, N_T = 1); \Delta l\Big)$. Observables calculated using $P_{N_B}(N_i)$ are probably the closest to those one would find in an experiment if one could measure positions of the  particles.

\section{Infinite matter simulation results}
\label{simulation_results}

To simulate isospin-symmetric infinite nuclear matter, we initialize equal numbers of proton and neutron test particles in a box with periodic boundary conditions. The side length of the box is taken to be $L = 10 \ \txt{fm}$; this is informed by the fact that with periodic boundary conditions, the box can be kept relatively small with no significant finite-size effects. The time step used in the simulation needs to be small enough to resolve all gradients occurring during the evolution (intuitively speaking, a test particle should not ``jump over'' a potential gradient within a single time step). We found that a time step of $\Delta t = 0.1 \ \txt{fm}/c$ is small enough to satisfy this condition, and it correctly solves the equations of motion, Eqs.\ (\ref{EOM_covariant_formulation_x}) and (\ref{EOM_covariant_formulation_p}), using the leapfrog algorithm. The mean field is calculated using the triangular smearing with triangular smearing parameter $n=2$ and finite difference direct derivatives (for more details see Chapter \ref{implementation}) on a lattice with lattice spacing $a = 1 \ \txt{fm}$, which has been tested to be sufficiently fine for accurately resolving mean-field gradients. To ensure smooth density and mean-field gradient calculations, we utilize a large number of test particles per particle, specifically, we use $N_T = 200$ for ordinary nuclear matter (Section \ref{nuclear_phase_transition}) and $N_T = 50$ for dense nuclear matter (Sections \ref{quark-hadron_phase_transition} and \ref{effects_of_finite_number_statistics}). Using different numbers of test particles in these two cases is justified by the fact that smooth density and density gradient calculations are ensured when the average number of test particles encountered in a cell of the lattice $N_{\txt{avg}}$ is large enough. As an example, within the described setup, this number will be equal to $N_{\txt{avg}} = 8$ for ordinary nuclear matter at $n_B = 0.25 n_0$, and equal to $N_{\txt{avg}} = 24$ for dense nuclear matter at $n_B = 3 n_0$. We choose $N_{\txt{avg}}$ to be bigger in the case of dense nuclear matter as mean fields encountered in that region of the phase diagram are significantly larger and require an even more smooth gradient computation.

For studying the thermodynamic behavior of nuclear matter, we are simulating systems in which all collision and decay channels are turned off. We have checked that the thermodynamic effects described here persist when collisions are allowed, and in this work we choose to omit them because our goal is to study mean-field dynamics. As in Section \ref{Cumulants_of_baryon_number}, we are considering only one of the many EOSs accessible within the VDF model, namely, the one corresponding to the fourth (IV) set of characteristics listed in Table \ref{example_characteristics}. The choice of this set is arbitrary and serves as an illustration of the properties of the VDF model which are qualitatively comparable for all obtained EOSs.

\subsection{Ordinary nuclear matter}
\label{nuclear_phase_transition}

We investigate the behavior of systems initialized at temperatures and baryon number densities specific to ordinary nuclear matter to validate the implementation of the VDF model in \texttt{SMASH} \cite{Weil:2016zrk}. For illustrative purposes, we discuss results for a single simulation run, that is one event. Remarkably, the thermodynamic behavior of the system is apparent already for this minimal statistics. This is a consequence of the large number of test particles per particle used ($N_T = 200$), as well as the fact that the investigated effects are characterized by large fluctuations, which result in clear signals.

\begin{figure}[t]
	\centering\mbox{
	\includegraphics[width=0.47\textwidth]{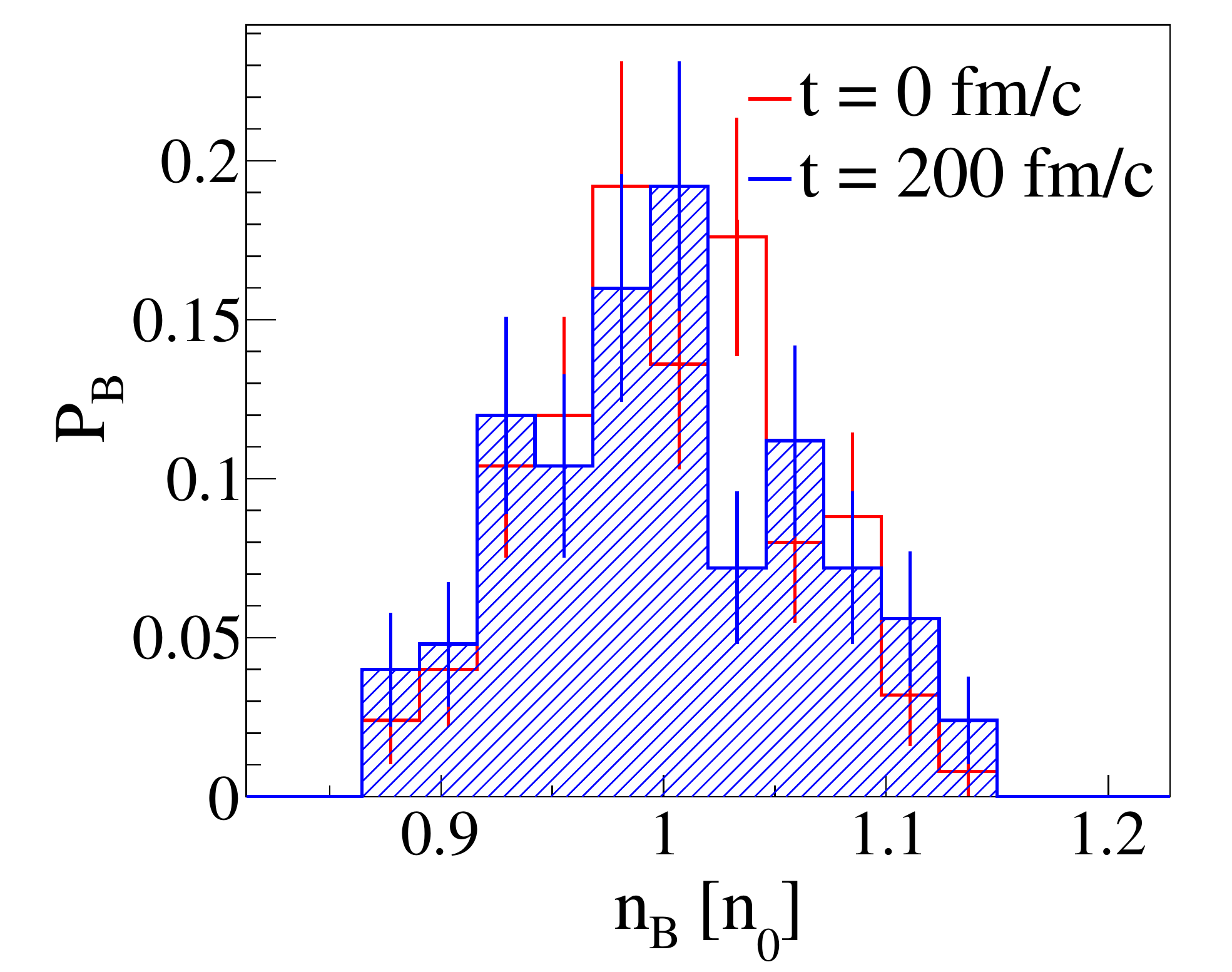} \hspace{5mm}
	\includegraphics[width=0.47\textwidth]{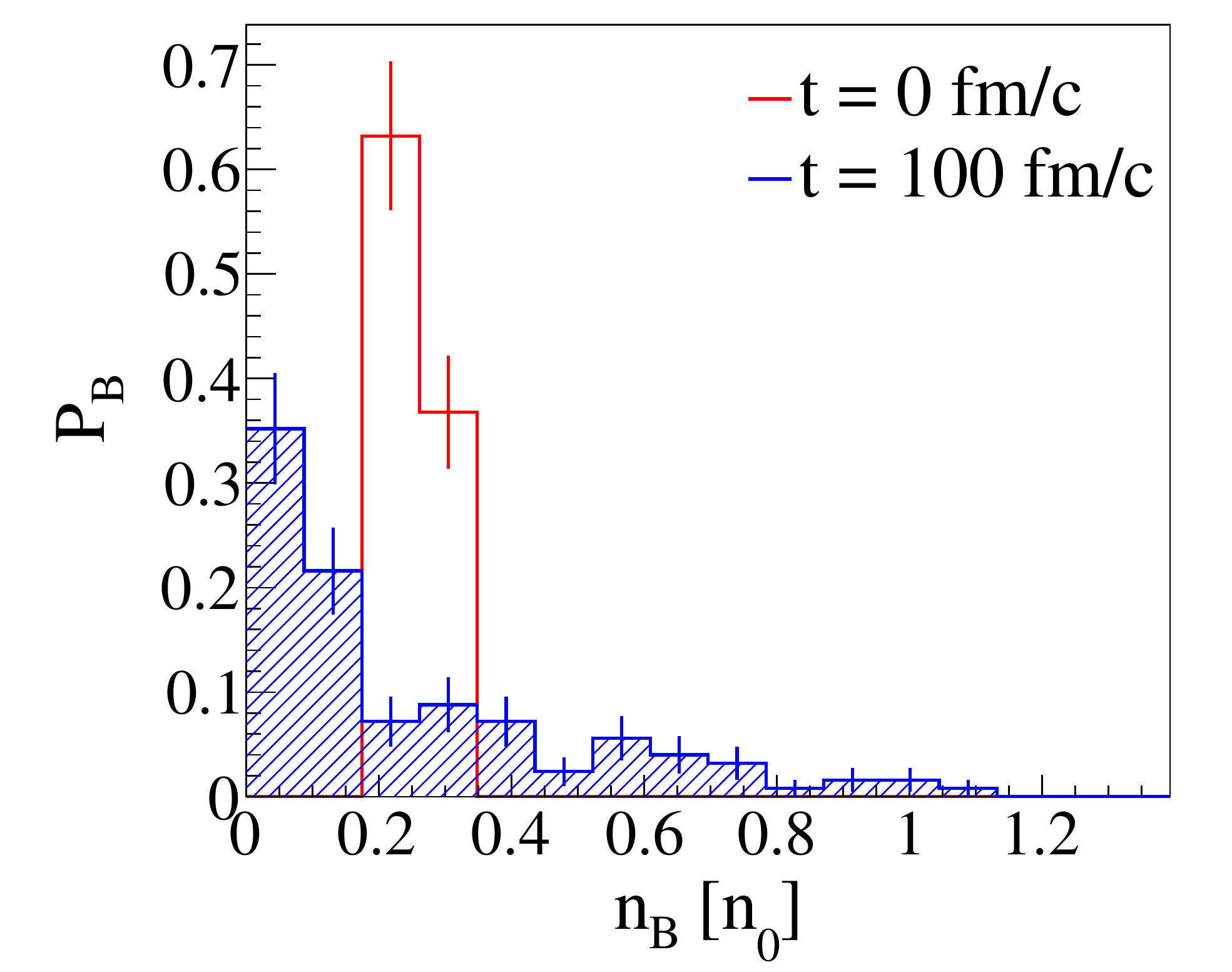} 
	}
	\caption[Baryon number distribution of nuclear matter initialized at the saturation point and inside the nuclear spinodal region]{Baryon number distribution, scaled by the volume of the cell and shown in units of the saturation density, $n_0 = 0.160\  \txt{fm}^{-3}$. The cell width is $\Delta l = 2\ \txt{fm}$. Red histograms show distributions at initialization ($t = 0$), while blue shaded histograms show distributions at the end of the evolution, $t = t_{\txt{end}}$. Left panel: Nuclear matter initialized at the saturation density $n_0$ and temperature $T = 1 \ \txt{MeV}$, evolved until $t_{\txt{end}} = 200\ \txt{fm}/c$. The system, initialized in a stable configuration, remains in the same state at $t_{\txt{end}}$. Right panel: Nuclear matter initialized in the spinodal region of the nuclear phase transition, at baryon density $n_B = 0.25 n_0$ and temperature $T = 1\ \txt{MeV}$, evolved until $t_{\txt{end}} = 100\ \txt{fm}/c$. The system, initialized in a mechanically unstable region of the phase diagram, spontaneously separates into a nucleon gas and a nuclear liquid drop with central density $n_B \approx n_0$. Figure from \cite{Sorensen:2020ygf}.}
	\label{cell_particle_number_distribution}
\end{figure}

To start, we initialize symmetric nuclear matter at saturation density $n_B = n_0$, which for the box setup described above corresponds to the number of protons and neutrons $N_p = N_n = 80$, and at temperature $T = 1 \ \txt{MeV}$. Except for a slight increase in temperature from the degenerate limit, which is not significant enough to introduce any relevant changes, this is the saturation point of nuclear matter, where the configuration is stable. We let the simulation evolve until $t_{\txt{end}} = 200 \  \txt{fm}/c$ and investigate whether the equilibrium is preserved by hadronic transport.

\begin{figure}[t]
	\centering\mbox{
	\includegraphics[width=0.99\textwidth]{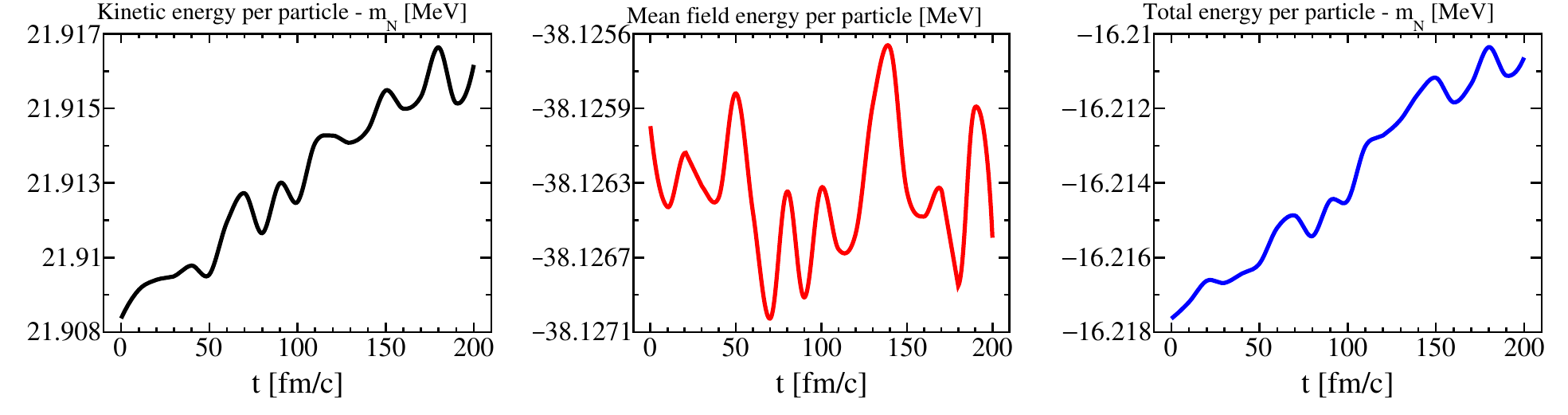} 
	}
	\caption[Time evolution of kinetic, mean-field, and total energy per particle for nuclear matter initialized at the saturation point]{Time evolution of kinetic energy per particle (left panel), mean-field energy per particle (middle panel), and total (binding) energy per particle (right panel) for a system initialized at nuclear saturation density $n_0 = 0.160 \ \txt{fm}^{-3}$ and temperature $T = 1 \ \txt{MeV}$. The binding energy per particle at initialization, $E_B (t=0) \approx - 16.218 \ \txt{MeV}$, is within 0.1\% from the value expected from model calculations. The mean-field energy oscillates slightly throughout the evolution, reflecting local fluctuations in density, but its average value remains constant. The increase in the kinetic energy per particle in time, which also causes the increase in the total energy per particle in time, is an unwanted feature of the simulation. A slight violation of the conservation of energy is a common feature of many hadronic transport codes, and is connected to the choice of the integration method for the equations of motion as well as to details of density and density gradient calculations (see Section \ref{comparison_of_different_numerical_procedures} for more details). Figure from \cite{Sorensen:2020ygf}.}
	\label{energy_graphs}
\end{figure}

To address this question, we examine the continuous baryon number distribution function (for details, see Section \ref{number_distribution_functions}), which we calculate using the cell width $\Delta l = 2 \ \txt{fm}$; we scale the histogram entries by the volume of the cell to obtain the distribution in units of the baryon number density, and further scale the results to express them in units of the saturation density, $n_0 = 0.160 \ \txt{fm}^{-3}$. As expected for matter in equilibrium, the baryon number distribution remains unchanged throughout the evolution, as can be seen in the left panel of Fig.\ \ref{cell_particle_number_distribution}. Note that the finite width of the shown distribution is a direct consequence of using a finite number of test particles; we will address this issue and related effects in Section \ref{effects_of_finite_number_statistics}. We find that throughout the simulation, the binding energy per particle agrees with the theoretically obtained value within $0.1\%$, see Fig.\ \ref{energy_graphs} (see Section \ref{comparison_of_different_numerical_procedures} for more details on energy conservation in hadronic transport). An in-depth discussion of the mean-field response to fluctuations around nuclear saturation density, comparing the results from several transport codes including \texttt{SMASH} utilizing the VDF model, can be found in Ref.\ \cite{Colonna:2021xuh}.

\begin{figure}
	\centering\mbox{
	\includegraphics[width=0.99\textwidth]{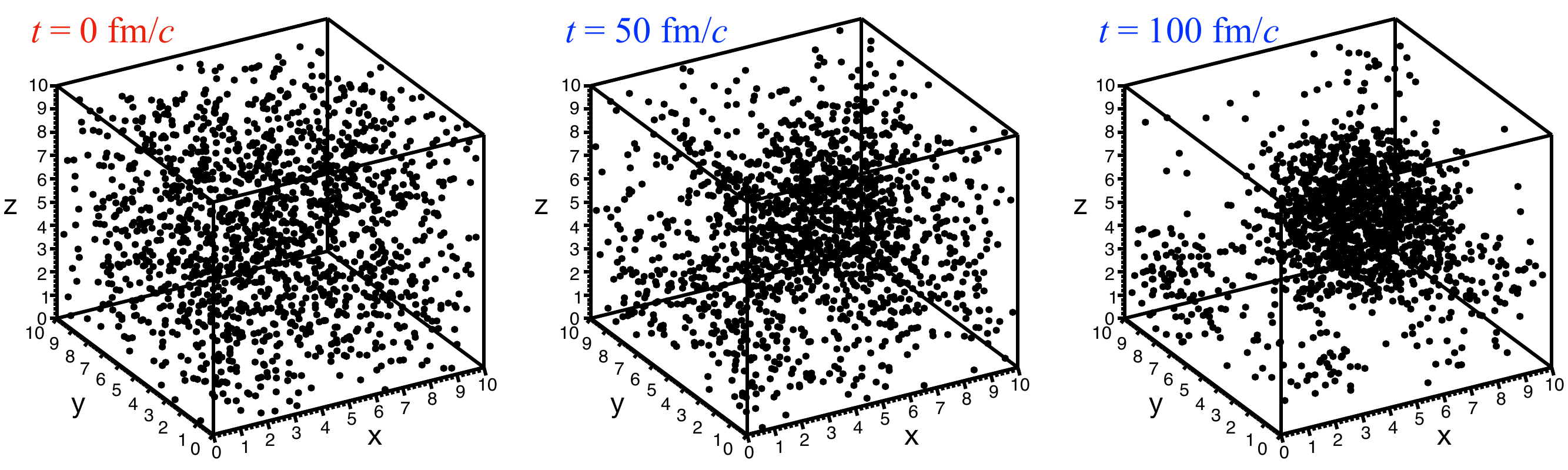} 
	}
	\caption[Visualization of spinodal decomposition in a system initialized inside the nuclear spinodal region]{Visualization of spinodal decomposition in a system initialized inside the nuclear spinodal region (at temperature $T = 1 \ \txt{MeV}$ and baryon number density $n_B = 0.25 n_0$, where $n_0$ is the saturation density, $n_0 = 0.160\  \txt{fm}^{-3}$). In the left panel one sees uniform nuclear matter at initialization, in the middle panel the mechanically unstable system is ``collapsing'' around a local random density fluctuation, and in the right panel the system has reached equilibrium where matter consists of a dense ``nuclear drop'' surrounded by a very dilute gas of nucleons. The effects due to periodic boundary conditions are clearly visible. We note that the black dots do not represent individual test particles, but rather are indicative of relative test particle density in a given region of the box.}
	\label{3D_box_nunclear_spinodal}
\end{figure}

Next, we study nuclear matter inside the spinodal region of the nuclear phase transition. Specifically, we initialize the system with the number of protons and neutrons $N_p = N_n = 20$, corresponding to a baryon number density $n_B = 0.25 n_0$, at temperature $T = 1 \ \txt{MeV}$. We let the system evolve until $t_{\txt{end}} = 100 \  \txt{fm}/c$. The spinodal region is both thermodynamically and mechanically unstable, and so we expect that local density fluctuations will drive the matter to separate into two coexisting phases: a dense phase, also known as a nuclear drop, and a dilute phase which is a nucleon gas. That this indeed happens can be seen on the right panel in Fig.\ \ref{cell_particle_number_distribution}, which shows the change in the baryon number distribution function due to the system's separation into two coexisting phases. The distribution, initially centered at $n_B = 0.25  n_0$, at the end of the evolution has a large contribution at $n_B \approx 0$, corresponding to the majority of box cells that after the separation are empty or nearly empty, and a long tail reaching out to $n_B \approx n_0$, which corresponds to the center of the nuclear drop. Even more directly, this separation can be seen in a visualization of the decomposition process, shown at three chosen evolution times, $t = 0, 50, 100\ \txt{fm}/c$, in Fig.\ \ref{3D_box_nunclear_spinodal}.

\begin{figure}[t]
	\centering\mbox{
	\includegraphics[width=0.99\textwidth]{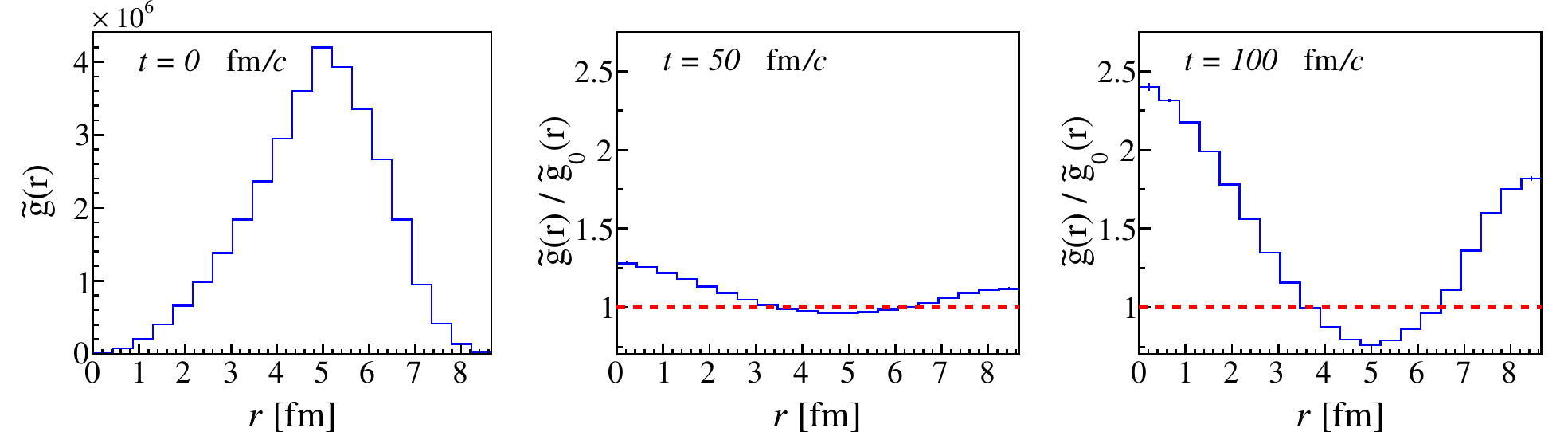} 
	}
	\caption[Time evolution of the pair distribution function for a system initialized inside the nuclear spinodal region]{Time evolution of the pair distribution function for a system initialized inside the nuclear spinodal region (at temperature $T = 1 \ \txt{MeV}$ and baryon number density $n_B = 0.25 n_0$, where $n_0$ is the saturation density, $n_0 = 0.160\  \txt{fm}^{-3}$). The $t =0$ plot (left) shows the distribution at initialization, while plots at $t = 50$ and $100 \ \txt{fm}/c$ (middle and right, respectively) show normalized distributions. Spontaneous spinodal decomposition occurs at $t > 0$ and leads to a formation of a nuclear drop surrounded by a near-perfect vacuum, resulting in a strong correlation between particles clustered within the drop. See Section \ref{section_pair_distribution_function} for a discussion of the influence of finite-size effects on the shape and large-distance behavior of the pair distribution function. Figure from \cite{Sorensen:2020ygf}.}
	\label{spinodal_decomposition_nuclear}
\end{figure}

We then proceed to calculate the pair distribution function (for details, see Section \ref{section_pair_distribution_function}) for the system initialized in the spinodal region of nuclear matter. The results are shown in Fig.\ \ref{spinodal_decomposition_nuclear}. Here, the three panels correspond to three time slices of the evolution: $t = 0, 50, 100 \  \txt{fm}/c$. The $t =0$ plot (left) shows the pair distribution function, Eq.\ (\ref{pair_distribution_function}), at initialization $\widetilde{g}_0(r, \Delta r)$, while plots at $t = 50$ and $100 \ \txt{fm}/c$ (middle and right, respectively) show normalized pair distribution functions $\widetilde{g}(r, \Delta r)/\widetilde{g}_0(r, \Delta r)$. The time evolution of the pair distribution function shows that during the spinodal decomposition the test particles cluster into the nuclear drop. The half width at half maximum of the pair distribution function is about $2\ \txt{fm}$, which corresponds to the density smearing range used, see Section \ref{section_pair_distribution_function} as well as Section \ref{calculation_of_mean-fields} for more details. The influence of the periodic boundary conditions on the shape and behavior of the pair distribution function at large inter-particle distances is also discussed in Section \ref{section_pair_distribution_function}. Additionally, the number of drops that form during spinodal decomposition depends both on the size of the box and the size of a drop, where the latter again depends on the smearing range used in density calculation.

Altogether, the results presented above demonstrate that the VDF equations of motion implemented in \texttt{SMASH} reproduce the expected bulk behavior of ordinary nuclear matter.

\subsection{Dense nuclear matter and the QGP-like phase transition}
\label{quark-hadron_phase_transition}

For simulations of critical behavior in dense symmetric nuclear matter, we run $N_{\txt{ev}} = 500$ events and average the results, calculated event-by-event. We first initialize the system at $n_B = 3 n_0$, which corresponds to the number of protons and neutrons $N_p = N_n = 240$, and at temperature $T = 1 \ \txt{MeV}$. It can be seen in Figs.\ \ref{phase_diagram_nB} and \ref{Cumulants_diagrams_k2k1}-\ref{Cumulants_diagrams_k4k2} that this corresponds to initializing dense nuclear matter inside the spinodal region of the QGP-like phase transition described by the EOS employed (the fourth (IV) set of characteristics listed in Table \ref{example_characteristics}). We evolve the system until $t_{\txt{end}} = 50 \ \txt{fm}/c$, which is sufficient for reaching equilibrium after a spinodal decomposition at high baryon number densities, since due to considerably larger values of the mean-field forces on test particles the density instabilities develop more rapidly.

\begin{figure}[t]
	\centering\mbox{
	\includegraphics[width=0.99\textwidth]{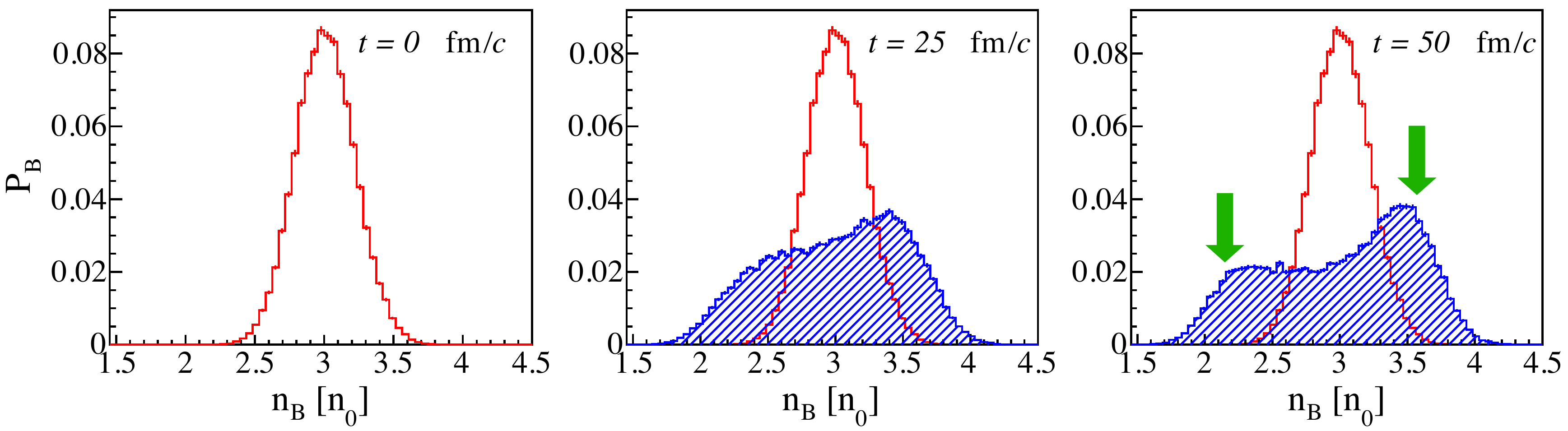} 
	}
	\caption[Time evolution of the baryon number distribution for a system initialized inside the quark-hadron spinodal region]{Time evolution of the baryon number distribution, scaled by the volume of the cell and shown in units of the saturation density of nuclear matter, $n_0 = 0.160\  \txt{fm}^{-3}$, for a system initialized inside the quark-hadron spinodal region (at baryon number density $n_B = 3 n_0$ and temperature $T = 1 \  \txt{MeV}$), averaged over $N_{\txt{ev}} = 500$ events. The cell width is chosen at $\Delta l = 2\ \txt{fm}$. Histograms delineated with red curves correspond to the baryon distribution at initialization ($t = 0$), while histograms delineated and shaded with blue curves correspond to baryon distributions at a chosen time during the evolution ($t = \{25, 50\}\ \txt{fm}/c$). The system, initialized in a mechanically unstable region of the phase diagram, undergoes a spontaneous separation into a less dense and a more dense nuclear liquid (see Section \ref{parametrization} for more discussion), resulting in a double-peaked baryon number distribution. The green arrows point to values of baryon number densities corresponding to the boundaries of the coexistence region at $T=1\ \txt{MeV}$, $n_L = 2.13 n_0$ and $n_R = 3.57 n_0$. Figure from \cite{Sorensen:2020ygf}.}
	\label{spinodal_decomposition_hadron_histograms_full}
\end{figure}

\begin{figure}[t]
	\centering\mbox{
	\includegraphics[width=0.99\textwidth]{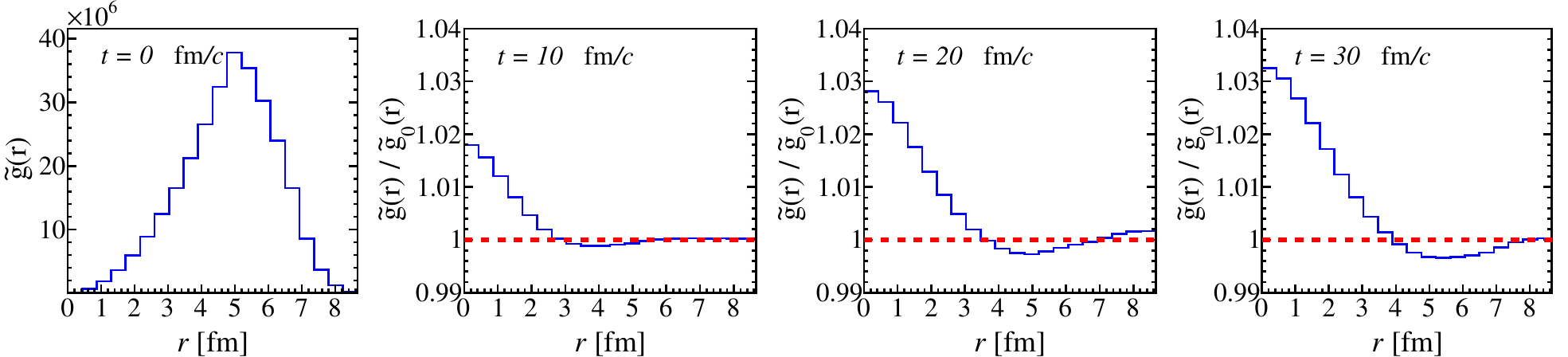} 
	}
	\caption[Time evolution of the pair distribution function for a system initialized inside the quark-hadron spinodal region]{Time evolution of the pair distribution function for a system initialized inside the QGP-like spinodal region (at baryon number density $n_B = 3.0  n_0$ and temperature $T = 1 \ \txt{MeV}$), averaged over $N_{\txt{ev}} = 500$ events. The $t = 0$ plot (first panel) shows the pair distribution at initialization, while plots at $t = 10, \ 20, \ 30\ \txt{fm}/c$ (second, third, and fourth panels) show normalized pair distributions. Spontaneous spinodal decomposition occurs at $t>0$ and leads to a formation of two coexisting phases: a less dense and a more dense nuclear liquid. The increased relative concentration of particles in the more dense phase results in an elevated normalized pair distribution at small distances. Figure from \cite{Sorensen:2020ygf}.
	}
	\label{spinodal_decomposition_hadron_correlations_full}
\end{figure}

In Fig.\ \ref{spinodal_decomposition_hadron_histograms_full}, we show the evolution of the baryon number distribution (see Section \ref{Continuous_baryon_number_distribution_function}). The cell width is chosen at $\Delta l = 2\ \txt{fm}$, and the histogram entries are scaled by the volume of the cell in order to be given in units of the baryon number density; we then further scale the results to express them in units of the saturation density, $n_0 = 0.160 \ \txt{fm}^{-3}$. In the figure, the red curve corresponds to the distribution at time $t = 0$, while the blue curves delineate the distribution at times $t>0$. At $t=0$, the distribution is peaked at the initialization density $n_B = 3 n_0$, with its width reflecting the finite number statistics. In the course of the evolution the system separates into two coexisting phases, a ``less dense'' and a ``more dense'' nuclear liquid (see Section \ref{parametrization} for more discussion). As a result, the baryon distribution displays two peaks largely coinciding with the theoretical values of the coexistence region boundaries, $n_L = 2.13 n_0$ and $n_R = 3.57 n_0$, indicated by the green arrows. We find that the prominence of the peaks depends slightly on the choice of the EOS. For example, an equation of state with the same value of critical density $n_c^{(Q)}$ and the same spinodal region $(\eta_L, \eta_R)$, but a higher critical temperature $T_c^{(Q)}$ will correspond to a more negative slope of the pressure in the spinodal region and, correspondingly, to stronger mean-field forces inside the spinodal region, leading to more prominent peaks.

Next, in Fig.\ \ref{spinodal_decomposition_hadron_correlations_full} we show the evolution of the pair distribution function. Similarly as in the case of nuclear spinodal decomposition, the ``hadron-quark'' spinodal decomposition leads to a pair distribution function indicating the formation of two phases of different densities. Unlike in nuclear spinodal decomposition, where drops of a ``nuclear liquid'' form in vacuum, in this case we have drops of a ``more dense liquid'' submerged in a ``less dense liquid'' (for a detailed discussion, see Section \ref{parametrization}). Consequently, the absolute values of the normalized pair distribution function $\widetilde{g}(r)/\widetilde{g}_0(r)$ are much smaller for the case of the ``hadron-quark'' spinodal decomposition, as the difference between the number of test particle pairs occupying the dense and dilute regions is less pronounced in this case. Nevertheless, the effect, although small, is clearly distinguishable and statistically significant.

We note here that a phase separation is such a distinct behavior of the system that the baryon distribution function and the pair distribution function as shown in Figs.\ \ref{spinodal_decomposition_hadron_histograms_full} and \ref{spinodal_decomposition_hadron_correlations_full}, respectively, can be largely recovered even in the case of minimal statistics, that is for one event. However, effects at and around the critical point, as discussed below, are much more subtle and require a relatively large number of events.

\begin{figure}[t]
	\includegraphics[width=0.93\textwidth]{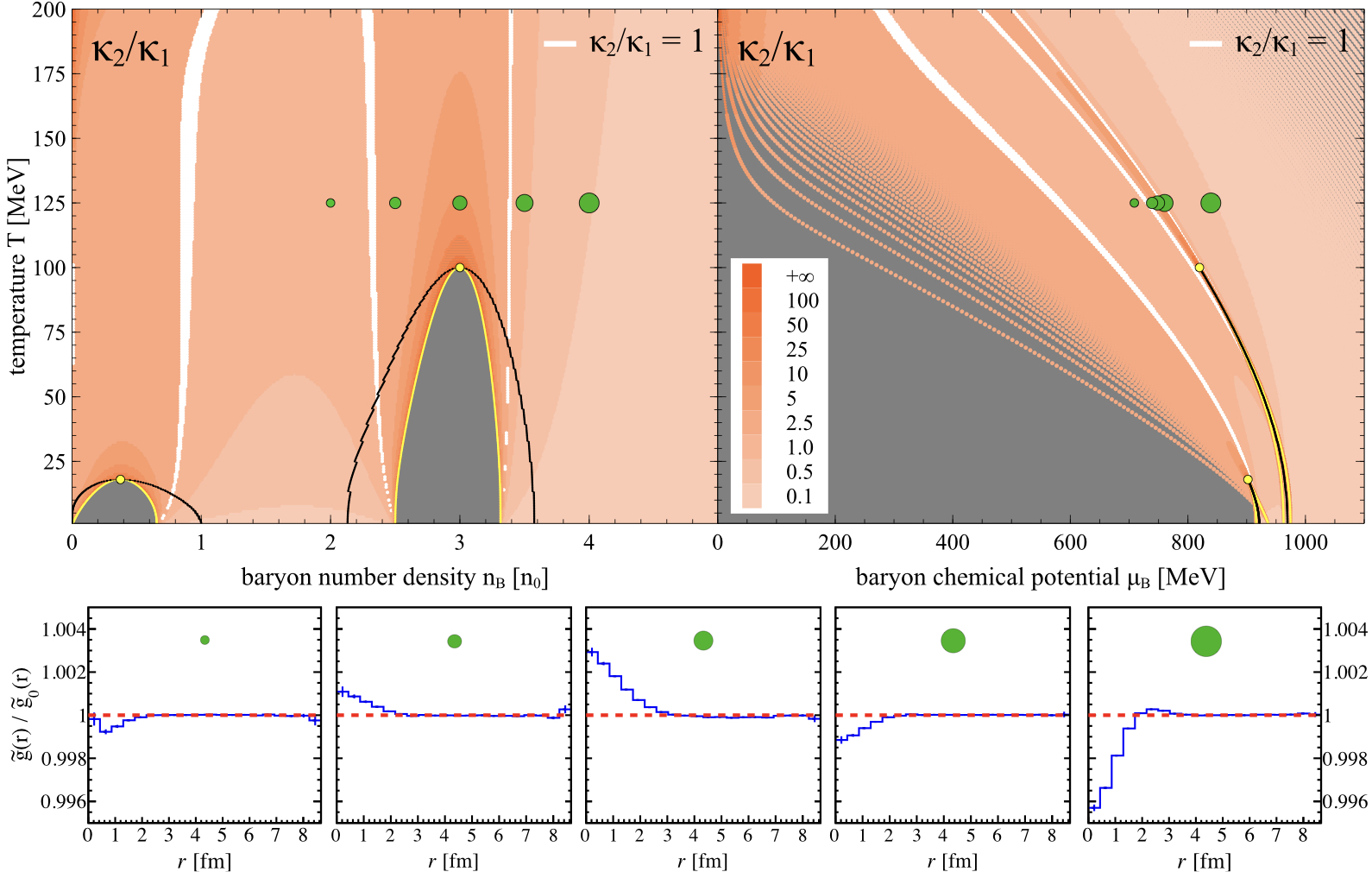}	
	\caption[Comparison of the cumulant ratio $\kappa_2/\kappa_1$ and the normalized pair distribution function for a series of initialization points]{Comparison of the cumulant ratio $\kappa_2/\kappa_1$ (upper panel), calculated in the VDF model, and the normalized pair distribution function $\widetilde{g}/\widetilde{g}_0$ at $t = 30\ \txt{fm}/c$ (lower panel), for a series of initialization points, Eq.\ (\ref{chosen_densities}), marked with green dots on the cumulant diagrams. The description of the cumulant diagrams is the same as in Fig.\ \ref{Cumulants_diagrams_k2k1}. The deviation of $\widetilde{g}/\widetilde{g}_0$ from the behavior of an uncorrelated system (red line) directly follows the deviation of $\kappa_2/\kappa_1$ from the Poissonian limit of 1. 
	Figure from \cite{Sorensen:2020ygf}.
	}
	\label{correlations_vs_cumulants}
\end{figure}

To conclude our study of dense nuclear matter in \texttt{SMASH}, we want to investigate the behavior of systems initialized at various points of the phase diagram above the critical point, inspired by possible phase diagram trajectories of heavy-ion collisions at different beam energies. Specifically, we initialize the system at one chosen temperature and a series of baryon number densities 
\begin{eqnarray}
\hspace{-5pt}T = 125~\txt{MeV}, \hspace{5pt} n_B \in \{ 2.0, 2.5, 3.0, 3.5, 4.0 \}n_0~.
\label{chosen_densities}
\end{eqnarray}
In contrast with most of the previous examples, systems initialized in this region of the phase diagram are thermodynamically stable, and there are specific predictions for the behavior of thermodynamic observables such as ratios of cumulants of baryon number (see Figs.\ \ref{Cumulants_diagrams_k2k1}-\ref{Cumulants_diagrams_k4k2}). In the upper panel of Fig.\ \ref{correlations_vs_cumulants}, we show values of the second-order cumulant ratio, $\infrac{\kappa_2}{\kappa_1}$, as calculated from the VDF model, both in the $(T,n_B)$ and the $(T,\mu_B)$ plane. The dots on the cumulant diagrams mark the points at which we initialize the system, specified in Eq.\ (\ref{chosen_densities}), and are intended to guide the eye toward the corresponding normalized pair distribution plots at the end of the evolution, $t = t_{\txt{end}}$, displayed in the lower panel of the same figure. The deviation of values of the normalized pair distributions at small distances from 1 (where 1 corresponds to a system of non-interacting particles) directly follows the deviation of values of the second-order cumulant ratio $\infrac{\kappa_2}{\kappa_1}$ from the Poissonian limit of 1,
\begin{eqnarray}
&& \frac{\widetilde{g}~(0, \Delta r)}{\widetilde{g}_0(0, \Delta r)} > 1  ~ \Leftrightarrow ~  \frac{\kappa_2}{\kappa_1} > 1  ~, 
\label{pair_distribution_kappa_2_1} \\
&& \frac{\widetilde{g}~(0, \Delta r)}{\widetilde{g}_0(0, \Delta r)} < 1    ~\Leftrightarrow ~ \frac{\kappa_2}{\kappa_1} < 1 ~.
\label{pair_distribution_kappa_2_2}
\end{eqnarray}
We show a detailed derivation of this fact in Appendix \ref{pair_distribution_function_and_the_second-order_cumulant}. It is clear that a two-particle correlation corresponds to a value of the cumulant ratio $\infrac{\kappa_2}{\kappa_1} > 1$, and a two-particle anticorrelation corresponds to a value of the cumulant ratio $\infrac{\kappa_2}{\kappa_1} < 1$. This behavior is exactly reflected in Fig.\ \ref{correlations_vs_cumulants}.

We want to stress that the pair distributions shown in Fig.\ \ref{correlations_vs_cumulants} develop relatively fast. In Fig.\ \ref{spinodal_decomposition_hadron_correlations_full}, where we explored the behavior of a system initialized at a temperature $T = 1 \ \txt{MeV}$, one can see by comparing the second and the fourth panels that already at $t = 10 \ \txt{fm}/c$ a significant part of the pair distribution has developed. This effect is further magnified at higher temperatures, where relatively larger momenta of the test particles result in a faster propagation of effects related to mean fields, and consequently the system settles faster as well. For systems shown in Fig.\ \ref{correlations_vs_cumulants}, we have verified that the majority of the pair distribution function development occurs within $\Delta t = 3 \ \txt{fm}/c$.

These results show not only that hadronic transport is sensitive to critical behavior of systems evolving above the critical point, but also that this behavior is exactly what is expected based on the underlying model. Moreover, we note that the behavior of both the second-order cumulant and the pair distribution function across the region of the phase diagram affected by the critical point is remarkably distinct. It is evident that an equilibrated system traversing the phase diagram through the series of chosen points, Eq.\ (\ref{chosen_densities}), follows a clear pattern: first displaying anticorrelation, then correlation, and then again anticorrelation. Thus already the second-order cumulant ratio presents sufficient information to explore the phase diagram, and, provided that correlations in the coordinate space are transformed into correlations in the momentum space during the expansion of the fireball, this pattern may be utilized to help locate the QCD critical point, in addition to signals carried by the third- \cite{Asakawa:2009aj} and fourth-order \cite{Stephanov:2011pb} cumulant ratios. This may prove to be especially important given that the quantity observed in heavy-ion collision experiments is not the net baryon number, but the net proton number. In calculations of the net baryon number cumulants based on the net proton number cumulants, the higher order observables are increasingly more affected by Poisson noise \cite{Kitazawa:2011wh}. In view of this, the second-order cumulant ratio (or equivalently the two-particle correlation) could be considered among the key observables utilized in the search for the QCD critical point, and it remains to be seen if this somewhat smaller signal (as compared to higher order cumulant ratios) is nevertheless noteworthy due to the much higher precision with which it can be measured in experiments.

\subsection{Effects of finite number statistics}
\label{effects_of_finite_number_statistics}

Qualitative and quantitative features of observables are influenced by the finite number of particles in the analyzed samples. When analyzing observables such as the baryon distribution, one has to keep in mind that fluctuations due to finite number statistics may wash out the expected signals. This is not only a numerical problem but, as we shall discuss below, it is also an issue relevant for experiments.

\begin{figure}[t]
	\centering\mbox{
		\includegraphics[width=0.5\textwidth]{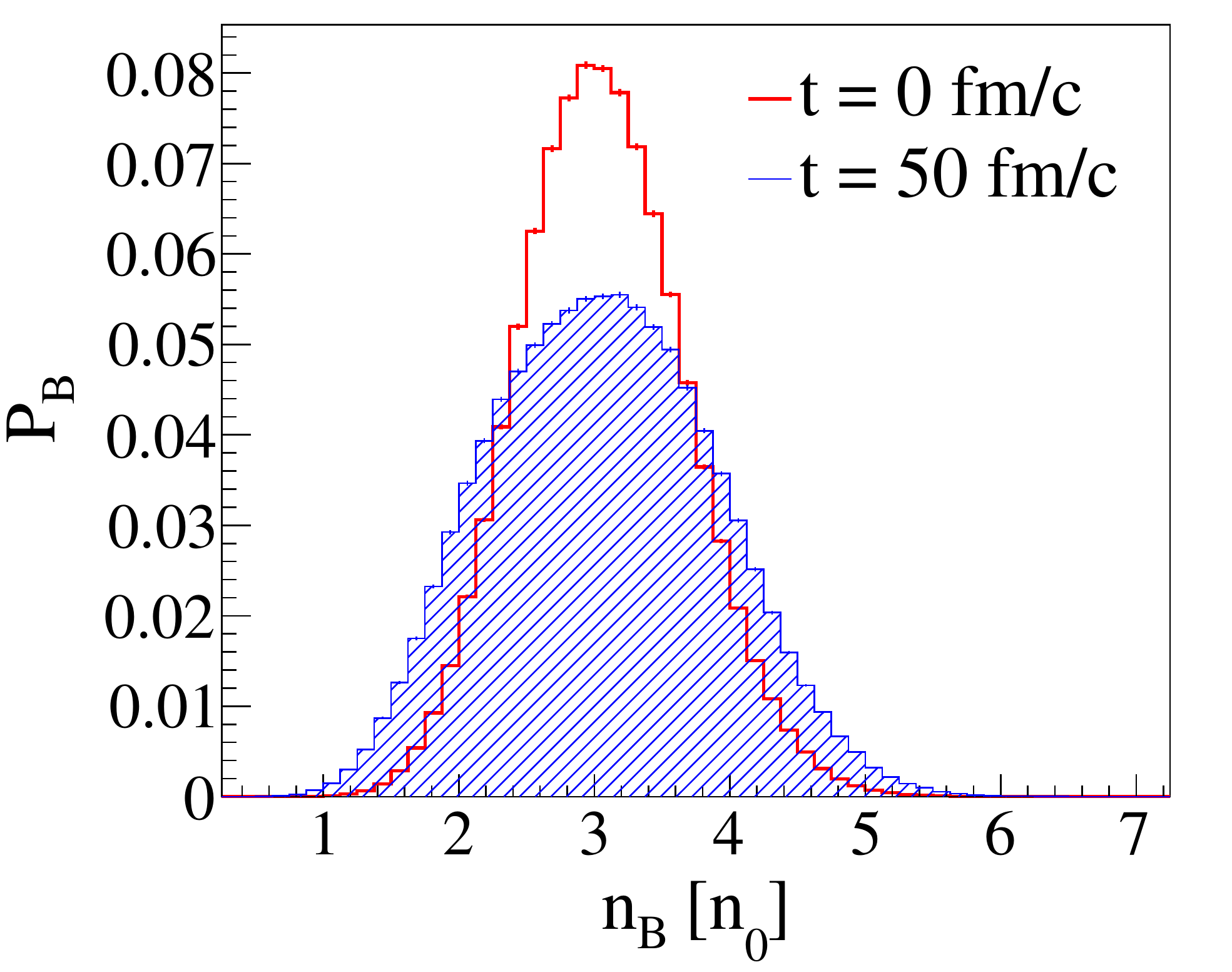} 
	}
	\caption[Demonstration of the effects on the baryon number distribution due to histogram bin width]{Time evolution of the continuous baryon number distribution, scaled by the volume of the cell and shown in units of the saturation density of nuclear matter, $n_0 = 0.160\  \txt{fm}^{-3}$, for the same system as described in Fig.\ \ref{spinodal_decomposition_hadron_histograms_full}. Here, the cell width is $\Delta l = 1\ \txt{fm}$. The red histogram shows the baryon distribution at initialization ($t = 0$), while the  blue shaded histogram shows the distribution at the end of the evolution ($t_{\txt{end}} = 50 \ \txt{fm}/c)$. Nuclear matter, initialized in a mechanically unstable region of the phase diagram, spontaneous separates into a less dense and a more dense nuclear liquid, and the distribution function becomes wider; however, due to the size of the binning cell, the average number of test particles in a cell is small and consequently the double-peaked structure, clearly seen on the right panel in Fig.\ \ref{spinodal_decomposition_hadron_histograms_full}, is washed out by Poissonian fluctuations. Figure from \cite{Sorensen:2020ygf}.
	}
	\label{spinodal_decomposition_hadron_histograms_full_smaller_cell}
\end{figure}

\begin{figure}[t]
	\centering\mbox{
	\includegraphics[width=0.5\textwidth]{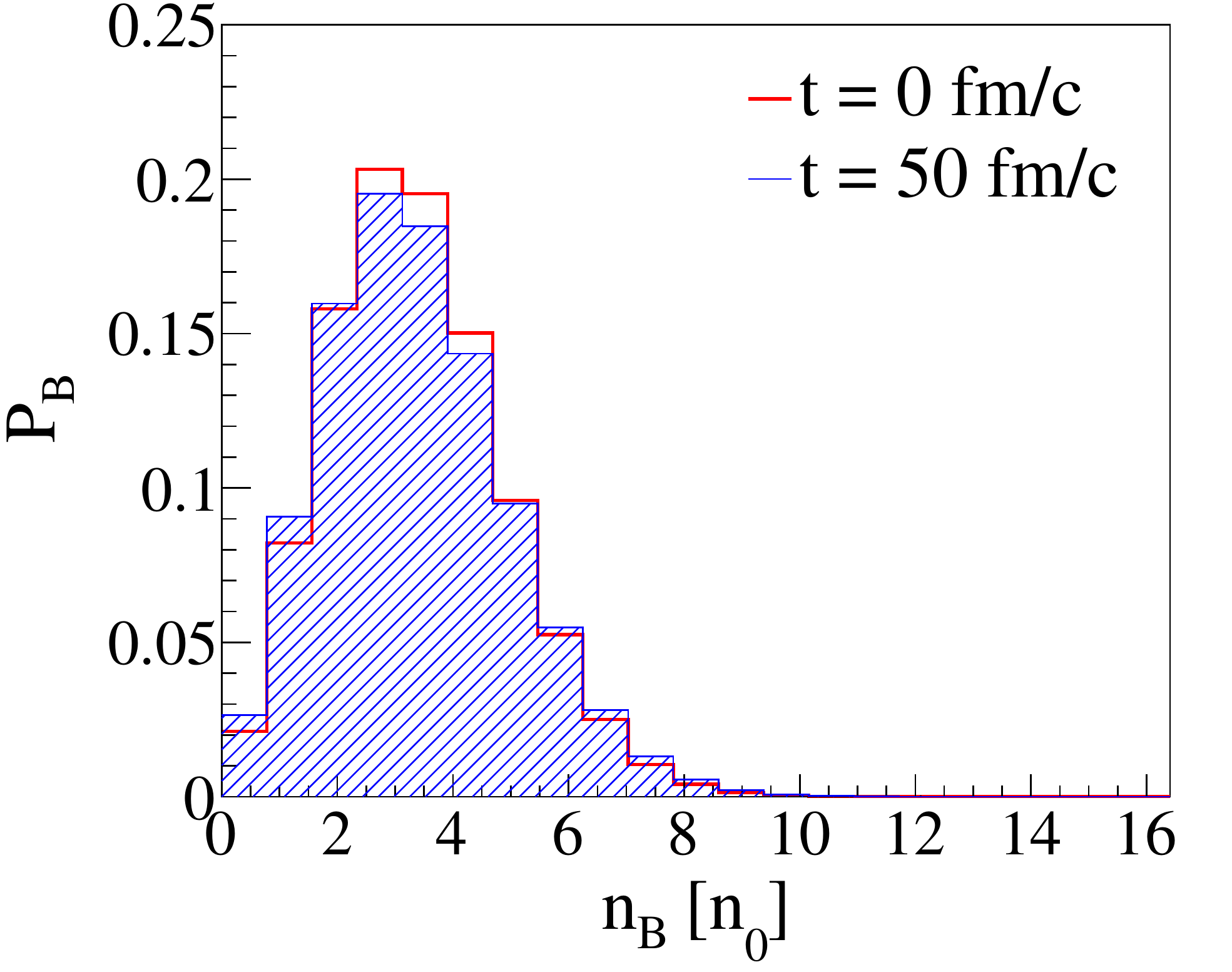} 
	}
	\caption[Time evolution of the baryon number distribution for a system initialized inside the quark-hadron spinodal region (parallel ensembles method)]{Time evolution of the continuous baryon distribution, scaled by the volume of the cell and shown in units of the saturation density of nuclear matter, $n_0 = 0.160\  \txt{fm}^{-3}$, for the same system as described in Figs.\ \ref{spinodal_decomposition_hadron_histograms_full} and \ref{spinodal_decomposition_hadron_histograms_full_smaller_cell}, but calculated using the parallel ensembles method; the results are averaged over $N_{\txt{ev}}^{(\txt{parallel})} = N_T \times N_{\txt{ev}} = 25,000$ events. Cell width is $\Delta l = 2\ \txt{fm}$. The red histogram shows the distribution at initialization ($t = 0$), while the blue shaded histogram shows the distribution at the end of the evolution ($t_{\txt{end}} = 50 \ \txt{fm}/c)$. Nuclear matter, initialized in a mechanically unstable state, spontaneously separates into a less dense and a more dense nuclear liquid, and the distribution widens; however, small numbers of particles in cells used to construct the histogram and related finite number effects wash out the structure clearly seen in the right panel of Fig.\ \ref{spinodal_decomposition_hadron_histograms_full}. Figure from \cite{Sorensen:2020ygf}.
	}
	\label{spinodal_decomposition_hadron_histograms_parallel}
\end{figure}

First, we discuss this subject in the context of the choice of binning width. In particular, the double-peak structure in the baryon number distribution shown in the right panel of Fig.\ \ref{spinodal_decomposition_hadron_histograms_full} depends on the size of the cell used to construct the histogram, chosen to be $\Delta l = 2 \ \txt{fm}$. In this case, the Poissonian finite number statistics superimposed on the underlying baryon distribution is characterized by a certain width $\sigma_{(\txt{2~fm})}$. If we reduce the cell width $\Delta l$ by a factor of 2, the average number of particles in a cell is reduced by a factor of 8. Consequently, the width of the Poissonian fluctuations will be $\sigma_{(\txt{1~fm})} =2\sqrt{2}  \sigma_{(\txt{2~fm})}$, which is considerably larger than previously and which in fact washes out the double-peak structure. This can be seen in Fig.\ \ref{spinodal_decomposition_hadron_histograms_full_smaller_cell}, where we show the baryon number distribution for a sampling cell width of $\Delta l = 1 \ \txt{fm}$ for the same events as used to create Fig.\ \ref{spinodal_decomposition_hadron_histograms_full}; the red and blue lines correspond to the distribution at time $t=0$ and $t_{\txt{end}} = 50 \ \txt{fm}/c$, respectively. For the system at hand, the Poissonian widths in the two cases, in terms of baryon density, were $\sigma_{(\txt{2~fm})} = 0.22  n_0$ and $\sigma_{(\txt{1~fm})} = 0.62  n_0$. If we then estimate the full width at half maximum as approximately given by $2.355\sigma$ (the full width at half-maximum of a normal distribution), it is clear that in the case of the cell width $\Delta l = 1 \ \txt{fm}$, the full width is comparable with the separation of the peaks given by the width of the coexistence region, $n_R - n_L = 1.44 n_0$. As a result, the two-peak structure cannot be resolved for this sampling statistics. Let us note here that decreasing the volume of the cells $(\Delta l)^3$ can be done without penalty if one proportionally increases the number of test particles per particle, $N_T$. Conversely, decreasing the number of test particles per particle $N_T$ exacerbates the effects of finite number statistics.

\begin{figure}[t]
	\centering\mbox{
	\includegraphics[width=0.5\textwidth]{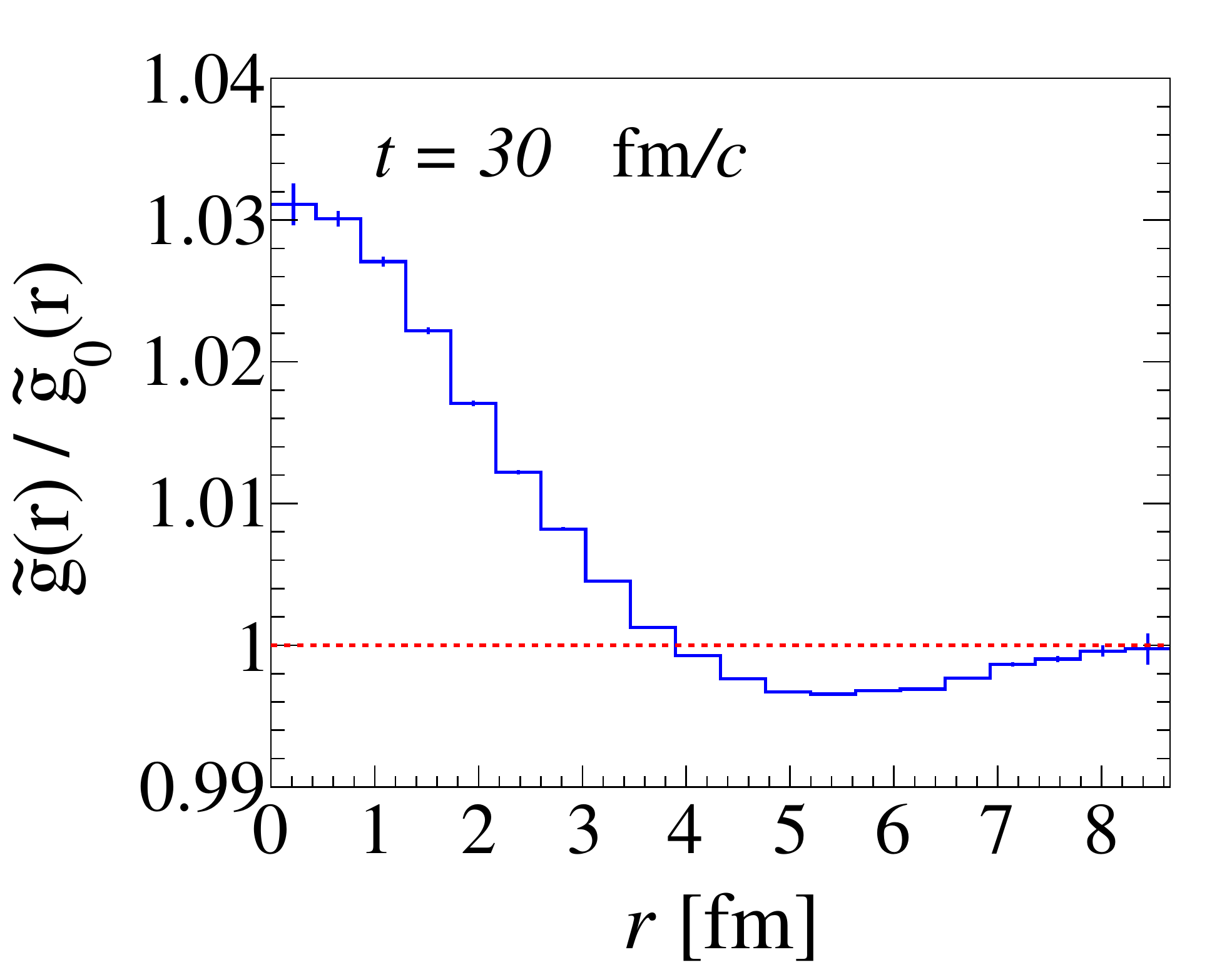} 
	}
	\caption[Pair distribution function for a system initialized inside the quark-hadron spinodal region (parallel ensembles method)]{Pair distribution function at $t = 30 \ \txt{fm}/c$ for a system initialized inside the quark-hadron spinodal region (at baryon number density $n_B = 3.0  n_0$ and temperature $T = 1 \ \txt{MeV}$), calculated with the parallel ensembles method; the results are averaged over $N_{\txt{ev}}^{(\txt{parallel})} = N_T \times N_{\txt{ev}} = 25,000$ events. Spontaneous spinodal decomposition leads to a formation of two coexisting phases: a ``less dense'' and a ``more dense'' nuclear liquid. The increased relative concentration of particles in the ``more dense'' phase results in an elevated normalized pair correlation at small distances. The correlation is exactly the same as shown on the rightmost panel in Fig.\ \ref{spinodal_decomposition_hadron_correlations_full}. Figure from \cite{Sorensen:2020ygf}.
	}
	\label{spinodal_decomposition_hadron_correlations_parallel}
\end{figure}

While this discussion may appear to be of purely numerical nature, experimental data are similarly affected by finite number statistics. In experiments, one always deals with exactly $N_B$ particles per event, which in our simulations corresponds to $N_T = 1$. Naturally, it must lead to a distribution in which any possible peaks are even more washed out. This can be seen in Fig.\ \ref{spinodal_decomposition_hadron_histograms_parallel}, where we show results for the case of $N_T = 1$ and $\Delta l = 2\ \txt{fm}$; the red and blue lines correspond to the distribution at time $t=0$ and $t_{\txt{end}} = 50 \ \txt{fm}/c$, respectively. Here, in order to ensure that we are comparing systems with identical dynamics, we used the same simulation data as in Figs.\ \ref{spinodal_decomposition_hadron_histograms_full} and \ref{spinodal_decomposition_hadron_histograms_full_smaller_cell}, but this time we accessed the baryon number distribution corresponding to $N_T = 1$ using the parallel ensembles method (for details, see Section \ref{implementation} and Appendix \ref{parallel_ensembles_a_posteriori}). Not surprisingly, the signal is almost entirely washed out and only a slight broadening of the distribution is discernible. We note that increasing the number of events does not resolve this issue, as the resolution is determined by Poissonian fluctuations in individual events. Consequently, one needs to devise other methods to extract the information about the underlying baryon distribution; this task is beyond the scope of this work.

Finally, we note that the pair distribution function is less affected by finite number statistics. In Fig.\ \ref{spinodal_decomposition_hadron_correlations_parallel}, we show the pair distribution function calculated within the parallel ensembles method, which is nearly identical to the pair distribution function calculated in the full ensemble, Fig.\ \ref{spinodal_decomposition_hadron_correlations_full}. Indeed, the normalized pair distribution function is not determined by the total number of test particles in an event or in a given subvolume of the system, but by relations between any two test particles. The only difference between the pair distribution functions obtained within the two methods is in the error bars, which are larger in the parallel ensembles case due to smaller statistics: the number of pairs in the full ensemble is given by $N_{\txt{ev}} ( N_B N_T )^2$, while in the parallel ensembles it's equal $N_{T} N_{ \txt{ev}} ( N_B)^2$. Obtaining the same pair distribution function demonstrates that the physics accessible in the full ensemble and the parallel ensembles approach is the same.

\section{Summary}

Results from simulations in \texttt{SMASH}, presented above, demonstrate that critical behavior in dense nuclear matter can be studied within a hadronic transport approach equipped with interactions corresponding to a chosen EOS. In particular, we have shown that systems initialized in unstable regions of the phase diagram undergo spontaneous spinodal decomposition, followed by an evolution towards an equilibrated mixture of two coexisting phases with compositions matching the predictions from the underlying EOS. The correct description of both the thermodynamics and the non-equilibrium phenomena implies that hadronic transport can be used as a tool with unique capabilities to investigate the dynamic evolution of matter created in heavy-ion collisions.

We have also shown that for systems initialized at various points of the phase diagram, including in the vicinity of a critical point, the pair distribution functions calculated from hadronic transport simulation data follow theoretical expectations based on the second-order cumulant ratio $\infrac{\kappa_2}{\kappa_1}$. In particular, as the baryon number density (and, consequently, baryon chemical potential) is increased in the region of the phase diagram affected by the critical point, the pair distribution function follows a clear pattern: displaying first anticorrelation, then correlation, and then again anticorrelation. The behavior of two-particle correlations (and, on the theoretical side, of the second-order cumulant ratio $\infrac{\kappa_2}{\kappa_1}$) is therefore a clear signature of crossing the phase diagram above the critical point. This is especially important in view of the experimental search for the QCD critical point, as lower order statistical observables, such as $\infrac{\kappa_2}{\kappa_1}$, are more likely to be measured with accuracy sufficient for discerning signals of critical behavior. 

Further, we have shown that finite number statistics affects both the qualitative and quantitative features of statistical observables. We have discussed two complementary simulation paradigms (described in Section \ref{transport_simulations} and in Appendix \ref{parallel_ensembles_a_posteriori}), within which hadronic transport gives access to both the continuous baryon number distribution, employed in theoretical calculations, and the physical baryon number distribution relevant to experimental results. Though driven by the same physics, these distributions lead to starkly different values for integrated statistical observables, which has consequences for comparisons with experimental data.

\newpage
\chapter{Beyond hadronic transport: the speed of sound in heavy-ion collisions}
\label{the_speed_of_sound_in_heavy-ion_collisions}

The contents of this chapter are largely based on Ref.\ \cite{Sorensen:2021zme}.

One of the ultimate goals of heavy-ion collision studies is connecting experimental observables to the underlying EOS of dense nuclear matter. Here, we present a method that may allow an estimate of the value of the speed of sound as well as its logarithmic derivative with respect to the baryon number density in matter created in heavy-ion collisions. To this end, we utilize well-known observables: cumulants of the baryon number distribution. In analyses aimed at uncovering the phase diagram of strongly interacting matter, cumulants gather considerable attention as their qualitative behavior along the explored range of collision energies is expected to aid in detecting the QCD critical point (see Sections \ref{non-statistical_event-by-event_fluctuations_of_conserved_charges} and \ref{Cumulants_of_baryon_number} for more details). We show that the cumulants may also reveal the behavior of the speed of sound in the temperature and baryon chemical potential plane. We demonstrate the applicability of such estimates within two models of nuclear matter: the Walecka model and the VDF model presented in this thesis (see Section \ref{flexible_equation_of_state_for_nuclear_matter} and Chapter \ref{parametrization_and_model_results}), and we explore what might be understood from known experimental data.

\section{Background}
\label{background}

The speed of sound $c_s$ is one of the fundamental properties of any substance. In fluids, it is the velocity with which a longitudinal compression wave propagating through the medium, and its square can be computed as the ratio of a change in the pressure $P$ corresponding to a change in the energy density $\mathcal{E}$. As such, it is directly related to a number of thermodynamic properties of the system, including its equation of state (EOS).

In dense nuclear matter, $c_s$ is of particular interest to neutron star research: its behavior as a function of baryon number density $n_B$ has a direct influence on the mass-radius relationship and, consequently, on the maximum possible mass of a neutron star \cite{Ozel:2016oaf}. Current neutron star data suggest that $c_s$ rises significantly for an $n_B$ larger than the nuclear saturation density $n_0$, and that it approaches and perhaps exceeds $c_s\sim1/\sqrt{3}$ at densities as low as a few times that of normal nuclear matter. This possibility was first suggested in Ref.\ \cite{Bedaque:2014sqa} and followed by multiple studies on the subject, see, e.g., Refs.\ \cite{Tews:2018kmu,McLerran:2018hbz,Fujimoto:2019hxv}.

Presently, heavy-ion collisions are the only means of studying dense nuclear matter in a
laboratory. Many experiments probing nuclear matter at high $n_B$, such as the Beam
Energy Scan (BES) program at the Relativistic Heavy Ion Collider (RHIC), put special significance on the search for the QCD critical point (CP). Also in this case the behavior of $c_s$ conveys relevant information: a crossover transition is characterized by a local minimum of the speed of sound, whereas at the CP and on the associated spinodal lines of the first-order phase transition, the speed of sound vanishes. Indeed, lattice QCD (LQCD) results show that at vanishing baryon chemical potential, $\mu_B=0$, a local minimum in $c_s$ occurs at temperature $T_0=156.5\pm1.5$ MeV \cite{HotQCD:2018pds} (see also Ref.\ \cite{Borsanyi:2020fev}), corresponding to a crossover transition between hadron gas and quark-gluon plasma (QGP). 

To date, a few attempts have been made to evaluate $c_s$ from heavy-ion collision data. In Ref.\ \cite{Gardim:2019xjs}, $c_s$ is estimated in ultrarelativistic collisions based on the proportionality of entropy density $s$ and temperature $T$ to charged particle multiplicity and mean transverse momentum, respectively, and the obtained value agrees with LQCD results. Unfortunately, the method used relies on the fact that at highest collision energies $\mu_B\approx n_B\approx0$, making this approach inapplicable for most of the energy range covered at RHIC. At values of the baryon chemical potential relevant to the BES program, $c_s$ was estimated in Ref.\ \cite{Steinheimer:2012bp}, where both the Landau model as well as hybrid hydrodynamics and hadronic transport simulations were used to reproduce the widths of the negatively charged pion rapidity distribution. That study purports to locate a robust minimum in $c_s$ within the collision energy range $\sqrt{s_{NN}}=4\txt{-}9\ \txt{GeV}$.

Below, we suggest a novel approach to exploring the behavior of $c_s$ by using cumulants of the baryon number distribution. The sensitivity of the cumulants to the EOS near the CP \cite{Asakawa:2009aj, Stephanov:2011pb}, which makes them central observables pursued in the Beam Energy Scan, follows directly from their sensitivity to derivatives of the pressure with respect to $\mu_B$. The key observation we make here is that, besides the vicinity of the CP, cumulants of the baryon number distribution provide rich information about the EOS at all points of the phase diagram, and in particular they allow a measurement of $c_s$ in matter created in heavy-ion collisions.

\section{Relation between cumulants and the speed of sound}
\label{relationship_cumulants_cT2}

Cumulants of net baryon number $\kappa_j$ can be obtained from $\kappa_j=VT^{j-1}\left(d^jP/d\mu_B^j\right)_T$, where $V$ is the volume. Expressed in terms of derivatives with respect to $n_B$, the first three cumulants are given by Eqs.\ (\ref{cumulant_1}-\ref{cumulant_3}), repeated here for convenience,
\begin{eqnarray}
&& \kappa_1  = V n_B ~, \label{cumulant_001} \\
&& \kappa_2 = \frac{VTn_B}{\left( \frac{dP}{dn_B} \right)_T}~, \label{cumulant_002}   \\
&& \kappa_3= \frac{VT^2n_B}{\left( \frac{dP}{dn_B} \right)_T^2} \left[  1 - \frac{n_B}{\left( \frac{dP}{dn_B} \right)_T} \left(\frac{d^2P}{dn_B^2}\right)_T \right] \label{cumulant_003} ~.
\end{eqnarray}
Importantly, cumulants can be also given in terms of moments of the net baryon distribution, which can be directly measured in experiment; in particular, for $j\leq3$, $\kappa_j\equiv\big\langle\big(N_B-\big\langle N_B\big\rangle\big)^j\big\rangle$ (for more details, see Sections \ref{non-statistical_event-by-event_fluctuations_of_conserved_charges} and \ref{Cumulants_of_baryon_number}).

The exact definition of $c_s$ depends on specifying which properties of the system can be
considered constant during the propagation of the compression wave. One often uses the speed of
sound at constant entropy $S$ per net baryon number $N_B$, $c_{\sigma}^2\equiv\left(
dP/d\mathcal E\right)_{\sigma}$, where $\sigma=S/N_B$. Similarly, the speed of sound at constant temperature is $c_T^2\equiv\left(dP/d\mathcal E\right)_{T}$. These two variants have specific regions of applicability. For example, the propagation of sound in air is governed by adiabatic
compression, so that using $c_{\sigma}^2$ is appropriate. On the other hand, when
there is a temperature reservoir (e.g., in porous media) or when the cooling timescale is very fast compared with the sound wave period (as is the case, e.g., for an interstellar medium subject to radiative cooling), $c_T^2$ is applicable.

Explicitly, $c_{\sigma}^2$ and $c_T^2$ can be written as (see Appendix \ref{the_speed_of_sound} for a detailed derivation)
\begin{eqnarray}
c_{\sigma}^2 =  \frac{  \Big( \frac{d P}{dn_B} \Big)_{T} \Big(\frac{d s}{d T} \Big)_{n_B}  +   \Big( \frac{ d P}{dT} \Big)_{n_B} \bigg[  \frac{s}{n_B} - \Big(\frac{d s}{d n_B}\Big)_T \bigg]  }{  \Big(\frac{sT}{n_B} + \mu_B  \Big) \Big( \frac{d s}{d T} \Big)_{n_B}   }  
\label{speed_isentropic} 
\end{eqnarray}
and
\begin{eqnarray}
c_T^2 = \frac{\Big( \frac{dP}{dn_B} \Big)_T}{ T  \Big(\frac{d s}{d n_B} \Big)_T   +  \mu_B  }~,
\label{speed_isothermal}
\end{eqnarray}
respectively. In the limit $T\to0$, the above expressions both lead to 
\begin{eqnarray}
c^2\Big|_{T=0} = \frac{1}{\mu_B} \bigg( \frac{dP}{dn_B} \bigg)_T ~. 
\label{cT2_approx}
\end{eqnarray}
Consequently, for $(\mu_B/T)\gg1$, the values of $c_{\sigma}^2$ and $c_T^2$ should largely coincide. Moreover, Eq.\ (\ref{speed_isothermal}) can be transformed to express $c_T^2$ as a function of the cumulants, Eqs.\ (\ref{cumulant_001}-\ref{cumulant_003}) ,
\begin{eqnarray}
c_T^{2} = \left[\bigg(\parr{\log \kappa_1}{\log T}\bigg)_{\mu_B} + \frac{\mu_B}{T} \frac{\kappa_2}{\kappa_1} \right]^{-1}~.
\label{cT2_as_function_of_cumulants}
\end{eqnarray}
Therefore, cumulants can be used to compute $c_T^2$ if they are known as functions of $T$ and $\mu_B$. The first term in Eq.\ (\ref{cT2_as_function_of_cumulants}) is challenging to estimate from experimental data, however, it can be shown to be negligible for a degenerate Fermi gas, $(\mu_B/T)\gg1$, where it constitutes an order $\left(T/\mu_B\right)^2$ correction; in that case we can write
\begin{eqnarray}
c_T^2 \approx \frac{T \kappa_1}{\mu_B \kappa_2}~.
\label{magic_equation_1}
\end{eqnarray}
We note that Eq.\ (\ref{magic_equation_1}) provides an upper limit to the value of $c_T^2$ as long as $(\partial \log \kappa_1/\partial \log T)_{\mu_B} > 0$.

Using Eq.\ (\ref{speed_isothermal}), one can also calculate the logarithmic derivative of $c_T^2$,
\begin{eqnarray}
\hspace{-1mm}\bigg(\frac{d \ln c_T^2}{d \ln n_B} \bigg)_T 
=  \frac{n_B \Big( \frac{d^2P}{dn_B^2} \Big)_T}{ \Big( \frac{dP}{dn_B} \Big)_T} - \frac{  \Big( \frac{dP}{dn_B} \Big)_T + Tn_B  \Big( \frac{d^2s}{dn_B^2} \Big)_T }{\mu_B + T  \Big( \frac{ds}{dn_B} \big)_T} ~.
\end{eqnarray}
It is again possible to represent the above equation in terms of the cumulants,
\begin{eqnarray}
\bigg(\frac{d \ln c_T^2}{d \ln n_B} \bigg)_T + c_T^2
= 1 - \frac{\kappa_3 \kappa_1}{\kappa_2^2}   - c_T^2 \bigg(\frac{d \ln (\kappa_2/T)}{d \ln T}\bigg)_{n_B} ~,
\end{eqnarray}
and neglecting the last term on the right-hand side yields
\begin{eqnarray}
\left(\frac{d \ln c_T^2}{d \ln n_B} \right)_T + c_T^2  \approx 1 - \frac{\kappa_3 \kappa_1}{\kappa_2^2} ~.
\label{magic_equation_2}
\end{eqnarray}
This approximation is again valid for $(\mu_B/T) \gg 1$, and the correction due to the neglected term is likewise of order $\left(T/\mu_B\right)^2$.

It is worth noting that in the opposite limit, $\mu_B \to 0$, Eq.\ (\ref{cT2_as_function_of_cumulants}) reveals a similarly simple form, $c_T^{2} = \left( d \ln \kappa_2/d \ln T \right)^{-1}_{\mu_B=0}$ (see Appendix \ref{the_isothermal_speed_of_sound_at_mu=0} for the derivation), suggesting that $c_T^2$ can be estimated in ultrarelativistic heavy-ion collisions, provided measurements of $\kappa_2$ are available at different temperatures with sufficient precision. It might be possible to achieve this with data from a combination of centralities, energies, collision species, or rapidity ranges. At this time, however, we are interested in utilizing Eqs.\ (\ref{magic_equation_1}) and (\ref{magic_equation_2}) applied to collisions at medium and low energies.

\begin{figure}[thp]
	\centering\mbox{
		\includegraphics[width = 0.99\textwidth]{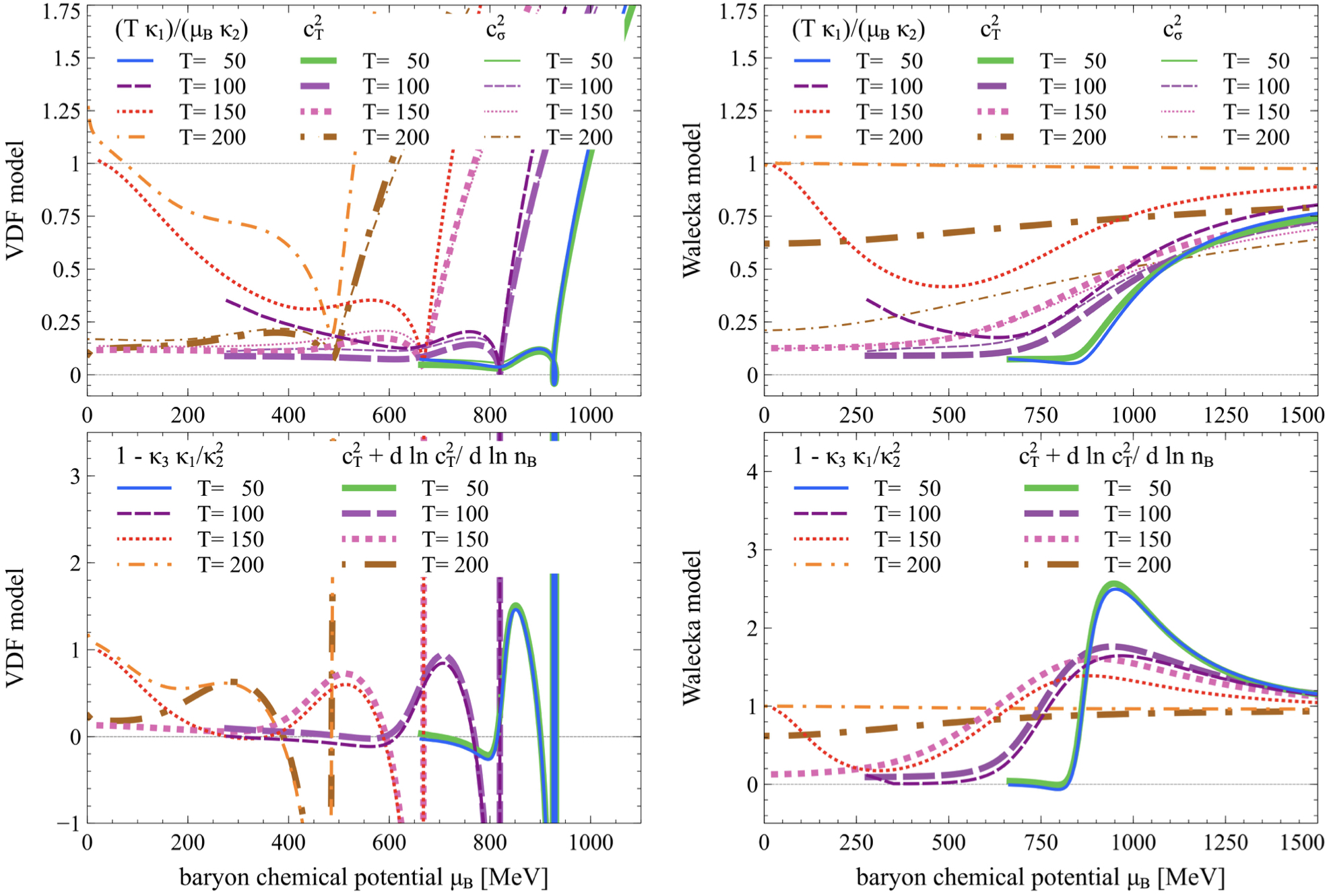} 
	}
	\caption[Model study of regions of applicability for equations connecting cumulants and the isothermal speed of sound]{Model study of regions of applicability of Eqs.\ (\ref{magic_equation_1}) and (\ref{magic_equation_2}). The left (right) panels show results obtained in the VDF (Walecka) model. The upper and lower panels show quantities entering Eq.\ (\ref{magic_equation_1}) and Eq.\ (\ref{magic_equation_2}), respectively. Results at $T=50,100,150,200\ \txt{MeV}$ are given by blue and green solid lines, dark and light purple long-dashed lines, red and pink short-dashed lines, and orange and brown dash-dotted lines. For each $T$, the thickest lines show the exact results and the medium-thick lines show the approximations, given by the right-hand sides of Eqs.\ (\ref{magic_equation_1}) and (\ref{magic_equation_2}). Additionally, in upper panels the thinnest lines show $c_{\sigma}^2$. For both models, Eq.\ (\ref{magic_equation_1}) is valid for $T\lesssim100\ \txt{MeV}$ and $\mu_B\gtrsim600\ \txt{MeV}$, while Eq.\ (\ref{magic_equation_2}) is valid for all values of $T$ and $\mu_B\gtrsim200\ \txt{MeV}$, with the exception of the Walecka model at $T=200\ \txt{MeV}$, where a phase transition to an almost massless gas of nucleons decreases the applicability of both Eqs.\ (\ref{magic_equation_1}) and (\ref{magic_equation_2}). Figure from \cite{Sorensen:2021zme}.
	}
	\label{tests_of_formulas}
\end{figure}

\section{Validation}
\label{validation_of_the_approximation}

We are interested in finding the limitations of the low-temperature approximation used to derive Eqs.\ (\ref{magic_equation_1}) and (\ref{magic_equation_2}), and for this we turn to effective models. Because the regions of the phase diagram where we expect to apply our formulas are described by hadronic degrees of freedom, we choose two models of dense nuclear matter: the VDF model with two phase transitions (see Chapter \ref{parametrization_and_model_results}) and the Walecka model \cite{Walecka:1974qa, Chin:1974sa}. The VDF model utilizes only interactions of the vector type, while the Walecka model employs both vector- and scalar-type interactions. Both models describe the nuclear liquid-gas phase transition, while the VDF model additionally describes a conjectured high-density, high-temperature phase transition modeling the QGP phase transition. In this study, the QGP-like phase transition is chosen to exhibit a CP at $T_c=100\ \txt{MeV}$ and  $n_c=3n_0$, with the $T=0$ boundaries of the spinodal region in the $(T,n_B)$ plane given by $n_{B, \txt{left spinodal} } (T=0) \equiv \eta_L=2.5 n_0$ and $n_{B, \txt{right spinodal}} (T=0) \equiv\eta_R=3.32 n_0$, where $n_0=0.160\ \txt{fm}^{-3}$ (which altogether correspond to the fourth (IV) set of characteristics listed in Table \ref{example_characteristics}); this choice is arbitrary and serves as a plausible example.

To test the derived expressions, we plot both sides of Eqs.\ (\ref{magic_equation_1}) and (\ref{magic_equation_2}) as functions of $\mu_B$ at a series of
temperatures in Fig.\ \ref{tests_of_formulas}. We note that the explored temperature range reaches beyond the region of the phase diagram where hadronic models plausibly describe matter created in heavy-ion collisions; nevertheless, it is still instructive to test the validity of our approximations at these extreme conditions.

We use natural units in which the speed of light in vacuum is $c=1$. We note that in the VDF model, $c_s$ quickly becomes acausal for $\mu_B$ exceeding the values defining the coexistence line of the QGP-like phase transition. As we already detailed in Section \ref{pressure,_speed_of_sound,_energy_per_particle}, it is an expected behavior in models using interactions dependent on high powers of $n_B$, and while generally not ideal, it does not affect the current analysis.

In all panels in Fig.\ \ref{tests_of_formulas}, the exact model calculations show expected
features as functions of $\mu_B$. In the upper left panel, showing both $c_{T}^2$ and $c_{\sigma}^2$ in the VDF model, at small $\mu_B$ we see a softening of the EOS due to the influence of the nuclear CP, followed by an increase at densities of the order of $n_0$, then a deep dive in $c_s^2$ caused by the QGP-like phase transition, and finally a steep rise for high values of $\mu_B$. In the upper right panel, showing $c_T^2$ and $c_{\sigma}^2$ in the Walecka model, we similarly observe a soft EOS at small $\mu_B$, while the value of $c_s^2$ goes asymptotically to 1 for large $\mu_B$. Additionally, for $T=200\ \txt{MeV}$, the Walecka model shows effects due to a phase transition in the nucleon-antinucleon plasma, occurring around $T\approx190\ \txt{MeV}$ and $n_B=0$; above this transition, the model describes an almost noninteracting gas of nearly massless nucleons \cite{Theis:1984qc}. The behavior of the curves in the lower panels, showing $c_T^2+\big(\infrac{d\ln c_T^2}{d\ln n_B}\big)_T$, can be directly traced to the behavior of the curves in the upper panels. In particular, for the VDF model we observe strong divergences due to the softening of the EOS in the QGP-like phase transition region.

Comparing the exact results to the approximations, we see that while Eq.\ (\ref{magic_equation_1}) is valid for small to moderate temperatures, $T\lesssim100\ \txt{MeV}$, and moderate to high baryon chemical potentials, $\mu_B\gtrsim600\ \txt{MeV}$, it behaves poorly, both qualitatively and quantitatively, for $T$ and $\mu_B$ corresponding to regions of the phase diagram probed by moderately to highly energetic heavy-ion collisions (upper panels). On the other hand, the approximation introduced in Eq.\ (\ref{magic_equation_2}) is qualitatively valid for most of the probed $T$ and $\mu_B$, with the exception of regions characterized by low chemical potentials, $\mu_B\lesssim200\ \txt{MeV}$ (lower panels).

\section{Experimental data and interpretation}

\begin{figure}[th]
	\centering\mbox{
		\includegraphics[width = 0.99\textwidth]{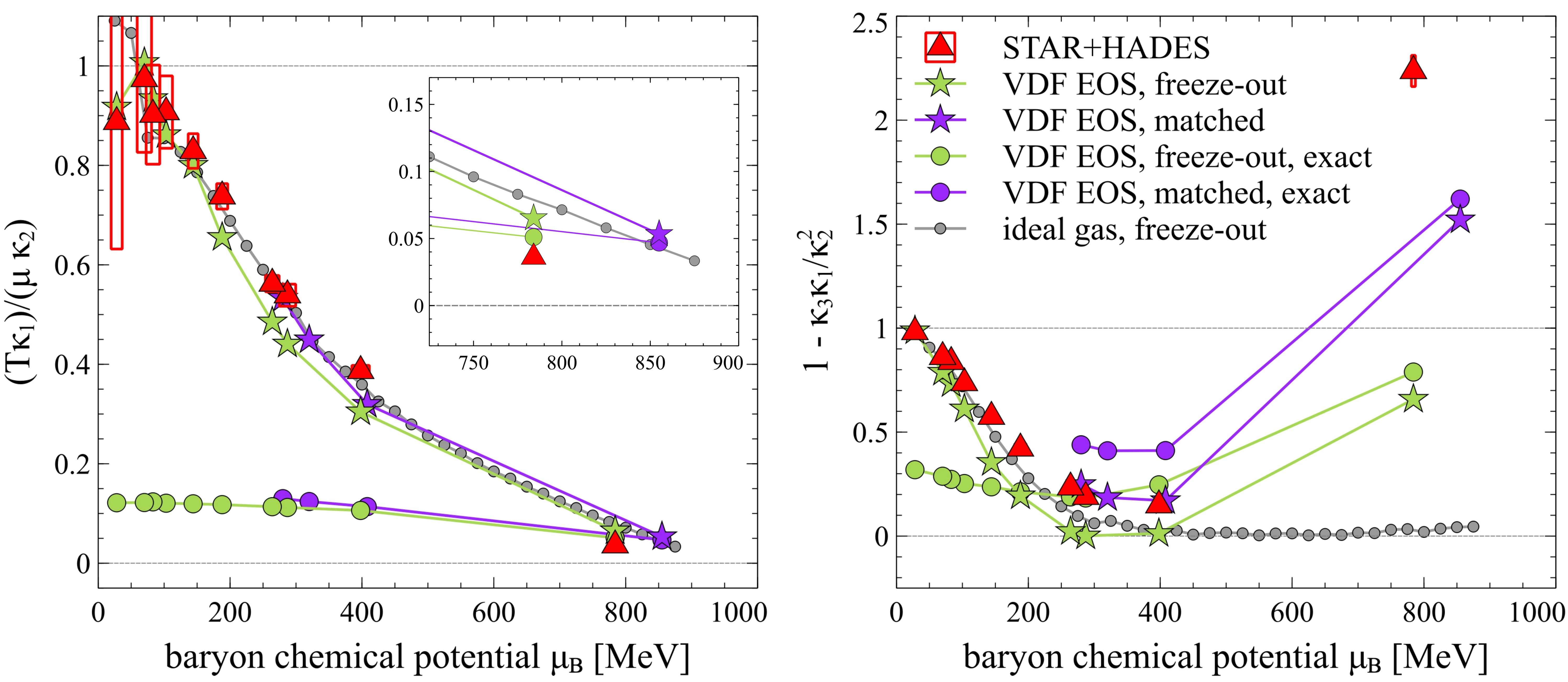} 
	}
	\caption[Plots of $\infrac{(T \kappa_1)}{(\mu_B \kappa_2)}$ and $1 - \infrac{\kappa_3 \kappa_1}{\kappa_2^2}$ for experimental data, ideal gas, and the VDF model]{Comparison of the right-hand sides of Eq.\ (\ref{magic_equation_1}) (left panel) and Eq.\ (\ref{magic_equation_2}) (right panel) for experimental data (red triangles), ideal gas at the freeze-out (small gray circles), the VDF model at the freeze-out (light green stars), and the VDF model at a set of points chosen to reproduce the data (dark purple stars); exact results, that is, the left-hand sides of Eqs.\ (\ref{magic_equation_1}) and (\ref{magic_equation_2}), are shown for the two cases considered in the VDF model (green and purple circles). The data points for the matched VDF results are chosen to reproduce experimental values of $1-\kappa_3\kappa_1/\kappa_2^2$ (see Fig.\ \ref{diagram}). We note that at $\sqrt{s}=2.4\ \txt{GeV}$, matching the value of $1-\kappa_3 \kappa_1/\kappa_2^2$ exactly is possible, but would place the matched point close to the nuclear liquid-gas CP, which we find unlikely. Figure from \cite{Sorensen:2021zme}.
	}
	\label{STAR_HADES_plots}
\end{figure}

We proceed to apply Eqs.\ \eqref{magic_equation_1} and \eqref{magic_equation_2} to heavy-ion collision data. We consider cumulants of the net proton number and chemical freeze-out parameters, $(T_{\txt{fo}},\mu_{\txt{fo}})$, in collisions at 0--5\% centrality, determined by the solenoidal tracker at RHIC (STAR) \cite{STAR:2021iop} and high acceptance dielectron spectrometer (HADES) \cite{HADES:2020wpc, HADES_MLorentz_talk} experiments, and we use them to plot Eqs.\ (\ref{magic_equation_1}) and (\ref{magic_equation_2}) (red triangles, left and right panel in Fig.\ \ref{STAR_HADES_plots}, respectively) against $\mu_B$.

Based on the model validation study presented in Section \ref{validation_of_the_approximation}, we trust the results presented in the left panel of Fig.\ \ref{STAR_HADES_plots}, approximating $c_T^2$, only for the lowest collision energy, $\sqrt{s}=2.4\ \txt{GeV}$ from the HADES experiment. At this collision energy, the value of $c_T^2$ as obtained from Eq.\ (\ref{magic_equation_1}) is small: less than half of the ideal gas value (see insert). At the same time, the value of $1-\kappa_3\kappa_1/\kappa_2^2$ (shown in the right panel in Fig.\ \ref{STAR_HADES_plots}), which we assume is dominated by $\inbfrac{d\ln c_T^2}{d\ln n_B}_T$, drops with decreasing collision energy to reach a minimum at the lowest STAR point, $\sqrt{s}=7.7\ \txt{GeV}$, and then steeply rises for the HADES point. This could mean that in matter created in $\sqrt{s} = 7.7\ \txt{GeV}$ collisions, $c_T^2$ is approximately constant as a function of $n_B$, while in $\sqrt{s} = 2.4\ \txt{GeV}$ collisions the matter is characterized by a small $c^2_T$ which nevertheless has a large positive slope as a function of $n_B$.

\begin{figure}[th]
	\centering\mbox{
		\includegraphics[width =0.85\textwidth]{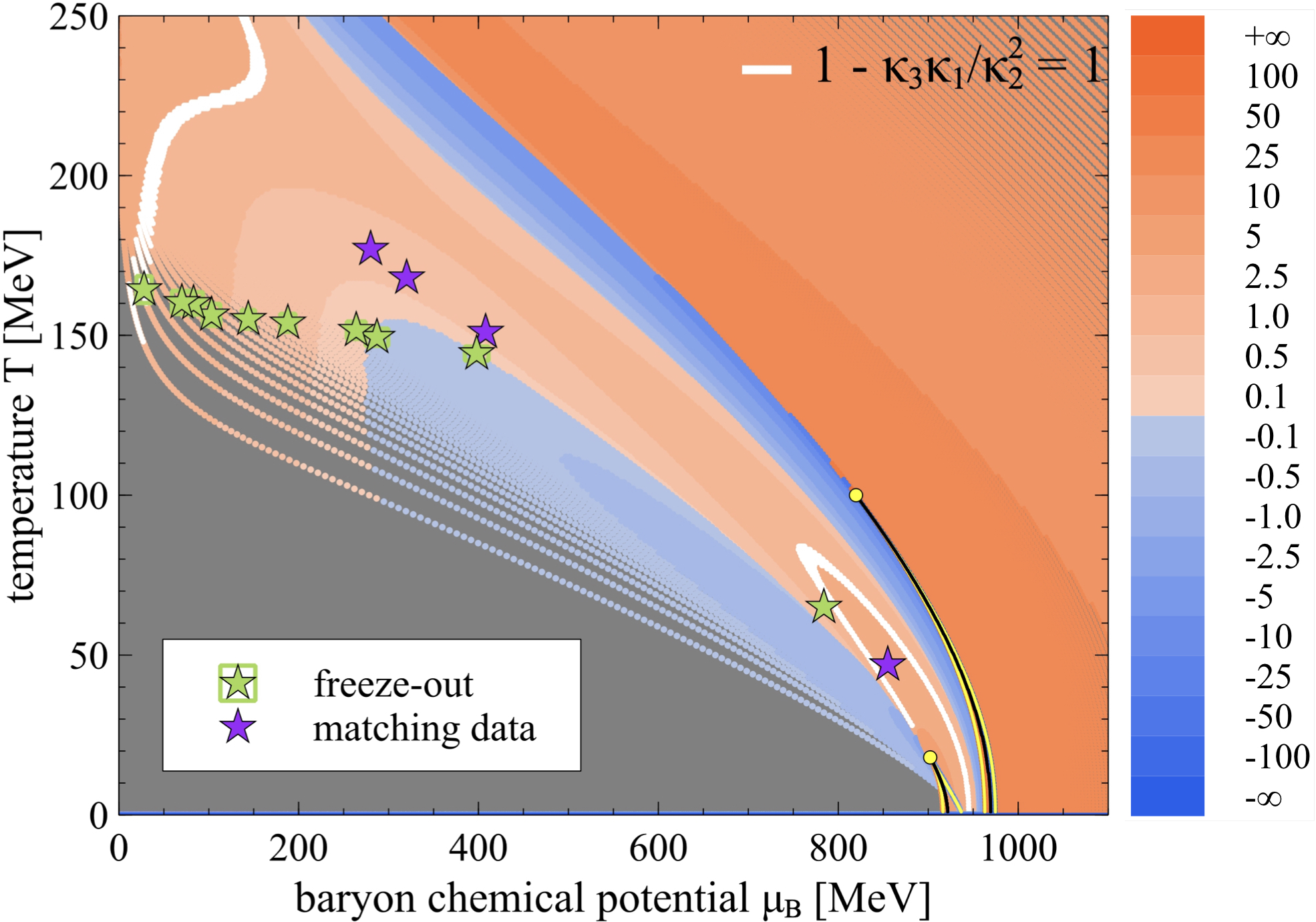} 
	}
	\caption[Contour plot of $1-\kappa_3\kappa_1/\kappa_2^2$ in the VDF model]{Contour plot of $1-\kappa_3\kappa_1/\kappa_2^2$ in the VDF model. Yellow (black) lines correspond to the spinodal (coexistence) lines, while white contours mark regions where $1-\kappa_3\kappa_1/\kappa_2^2=1\pm0.03$. Light green stars denote experimentally measured freeze-out parameters $(T_{\txt{fo}},\mu_{\txt{fo}})$, while dark purple stars denote points where $1-\kappa_3\kappa_1/\kappa_2^2$, taken along lines informed by average phase diagram trajectories for STAR collision energies \cite{Shen:2020jwv}, matches the experimentally measured values for a given collision energy. Figure from \cite{Sorensen:2021zme}.
	}
	\label{diagram}
\end{figure}

To further understand this behavior, we study the dependence of $1-\kappa_3\kappa_1/\kappa_2^2$ on $\mu_{B}$ and $T$ within the VDF model, shown as a contour plot in Fig.\ \ref{diagram}. Remarkably, the softening of the EOS, leading to negative values of $\inbfrac{d \ln c_T^2}{d\ln n_B}_T$, occurs in two regions of the phase diagram, corresponding to the ordinary nuclear matter phase transition and to the conjectured QGP-like phase transition. In order to compare model results to experiment, we need to choose at which $T$ and $\mu_B$ to take values of $1-\kappa_3 \kappa_1/\kappa_2^2$. A natural choice is to use values at the freeze-out points $(T_{\txt{fo}},\mu_{\txt{fo}})$ (marked in Fig.\ \ref{diagram} with light green stars), however, as can be seen in the right panel of Fig.\ \ref{STAR_HADES_plots}, the VDF model values of $1-\kappa_3\kappa_1/\kappa_2^2$ taken at these points (also denoted with light green stars) do not lead to an agreement with experimental data; in particular, the biggest discrepancy occurs for the HADES point. (We note here that the values of the freeze-out parameters $(T_{\txt{fo}},\mu_{\txt{fo}})$ are established with hadron interactions neglected, and the degree to which this affects our results may vary significantly across the phase diagram.) However, critical fluctuations are shown to exhibit a large relaxation time \cite{Berdnikov:1999ph,Stephanov:2017ghc,Du:2020bxp}, and as such their measured values could be affected by stages of the collision preceding the chemical freeze-out. Guided by this insight and aided by the average phase diagram trajectories of hybrid simulations of heavy-ion collisions \cite{Shen:2020jwv}, we consider values of $1-\kappa_3\kappa_1/\kappa_2^2$ at points in the phase diagram corresponding to slightly earlier stages of the collisions. Additionally demanding that the values of $1-\kappa_3\kappa_1/\kappa_2^2$ reproduce the experimentally obtained values for a given collision energy results in points denoted with dark purple stars in both Fig.\ \ref{STAR_HADES_plots} and in Fig.\ \ref{diagram} (where we choose to show points only for collisions at low energies, for which using the model is justified); here the exception from the exact matching occurs for the HADES point, for which we prioritized choosing a point in a reasonable vicinity of the measured freeze-out over obtaining a value equal to the experimental data. Comparing to the exact model results for $c_T^2$ and $c_T^2 + \big(\infrac{d\ln c_T^2}{d\ln n_B}\big)_T$, also displayed in Fig.\ \ref{STAR_HADES_plots}, as well as to the upper left panel of Fig.\ \ref{tests_of_formulas}, we can confirm that at the point reproducing the experimental value of $1-\kappa_3\kappa_1/\kappa_2^2$ for the lowest STAR energy, $c_T^2$ is nearly constant as a function of $\mu_B$ (thick short-dashed line, $T=150\ \txt{MeV}$, at $\mu_B\approx410\ \txt{MeV}$ in Fig.\ \ref{tests_of_formulas}), while at the point reproducing the result for the HADES energy, $c_T^2$ increases sharply with $\mu_B$ (thick solid line, $T=50\ \txt{MeV}$, at $\mu_B\approx850\ \txt{MeV}$ in Fig.\ \ref{tests_of_formulas}).

Naturally, the choice of $T$ and $\mu_B$ at which we compare model calculations with STAR and HADES cumulant data is driven by the wish to match the experimental results, and as such it serves mainly to show that baryon number cumulants measured in heavy-ion collisions can be connected to the speed of sound in hot and dense nuclear matter. Whether values of higher order cumulants are indeed significantly affected by stages of the evolution preceding the freeze-out needs to be further investigated (for recent developments, see Ref.\ \cite{Nahrgang:2018afz,Jiang:2017sni}). Moreover, while experiments measure proton number cumulants, the VDF model provides baryon number cumulants, putting more strain on a direct interpretation of our results. Effects due to baryon number conservation should likewise be important \cite{Bzdak:2012an,Vovchenko:2020gne}. Finally, it is understood that our model results may not be applicable in regions of the phase diagram where quark and gluon degrees of freedom become increasingly relevant.

Nonetheless, hadronic models are well-justified for describing low-energy collisions whose evolution is dominated by the hadronic stage, and it is the results in this region of the phase diagram that we want to stress. The comparison between the experimental data and the VDF model suggests that collisions at the lowest STAR and HADES energies may be probing regions of the phase diagram where the cumulants of the baryon number tell us more about hadronic physics than the QCD CP. In particular, the change in the sign of $\kappa_3$, predicted to take place in the vicinity of a critical point \cite{Asakawa:2009aj} and apparent in the HADES data (see the right panel of Fig.\ \ref{STAR_HADES_plots}), may mark the region of the phase diagram affected by the nuclear liquid-gas phase transition. If this is true, it may be worthwhile to study the cumulants at even lower collision energies, starting from $0.1\ \txt{GeV}$ projectile kinetic energy, and obtain the speed of sound around the nuclear liquid-gas CP. Conversely, at higher energies it could be possible to use collisions at different centralities and different rapidity windows to estimate the neglected terms in Eqs.\ (\ref{magic_equation_1}) and (\ref{magic_equation_2}), and obtain a stronger estimate for the speed of sound in the respective regions of the phase diagram.

\section{Summary}

In this chapter, we have used cumulants of the baryon number distribution to estimate the isothermal speed of sound squared and its logarithmic derivative with respect to the baryon number density. The approximations and the model comparisons we considered here apply to experiments at low energies, however, the presented method can be used at any collision energy provided that measurements of cumulants of the baryon number distribution as well as their temperature dependence are available. Further studies of effects due to dynamics, in particular using state-of-the-art simulations, will be essential in determining the extent to which the proposed method provides a reliable extraction of sound velocities and their derivatives. Nevertheless, this result provides a new approach to obtaining information about fundamental properties of nuclear matter studied in heavy-ion collisions, with consequences for both the search for the QCD critical point and neutron star studies.

\newpage
\chapter{Thesis summary}
\label{thesis_summary}

The work described in this thesis is devoted to developing, testing, and applying a parametrizable equation of state (EOS) of dense nuclear matter to studies centered on the phase diagram of quantum chromodynamics (QCD). The presented formalism has the potential to help interpret the measurements obtained from heavy-ion collisions experiments and, in particular, to reveal the thermodynamics of strongly-interacting QCD matter. Because the constructed family of EOSs shows a substantial flexibility in postulating the properties of the possible QCD phase transition at high temperatures and high baryon densities, it is perfectly suited for use in systematic studies of effects of different dense nuclear matter EOS on final state observables (e.g., using Bayesian analysis) and can facilitate meaningful comparisons of simulation results with experimental data. 

Below, we provide a summary of the developments presented in this work, followed by a brief outlook on future research directions.

In Chapter \ref{introduction}, we have given an overview of the search for the conjectured first-order phase transition in dense QCD matter and the most relevant results obtained within the Phase I of the Beam Energy Scan program at the Relativistic Heavy-Ion Collider. We have also discussed the state-of-the-art approaches to simulations of heavy-ion collisions and the indispensable role they will play in the interpretation of experimental data from the Phase II of the Beam Energy Scan. 

In Chapter \ref{a_flexible_model_of_dense_nuclear_matter}, after an introductory overview of the mean-field and Landau Fermi-liquid theory approaches to describing strongly-interacting many-body systems, we have developed an easily parametrizable EOS with an arbitrary number of scalar- and vector-density-functional--based interactions (the VSDF model). We have also shown that this model leads to Lorentz-covariant equations of motion, preserves conservation laws, and is thermodynamically consistent.

In Chapter \ref{parametrization_and_model_results}, we first discussed the parametrization procedure used to obtain a family of EOSs based on the version of the model introduced in Chapter \ref{a_flexible_model_of_dense_nuclear_matter} that utilizes only interactions of the vector type (the VDF model). With application to studies of the phase diagram of QCD in mind, we chose parametrizations of the VDF model with two phase transitions: one corresponding to the well-known low-temperature, low-density phase transition in ordinary nuclear matter, and one corresponding to a conjectured high-temperature, high-density QGP-like phase transition, meant to model the transition from matter described by hadronic degrees of freedom to matter described by quark and gluon degrees of freedom. Then, based on several EOSs chosen to represent the flexibility of the model, we discussed at length the thermodynamic properties of described systems.

In Chapter \ref{implementation}, we have discussed the implementation of the equations of motion stemming from the VDF model in the hadronic transport code \texttt{SMASH}, which we started with an overview of modeling many-body mean-field dynamics within hadronic transport approaches. In particular, we provided a detailed description of several different density and mean-field calculation algorithms available within \texttt{SMASH}, a few of which were introduced as part of the work underlying this thesis, and we included a comparison of the numerical performance of these options.

In Chapter \ref{results}, we have presented results of \texttt{SMASH} simulations of infinite nuclear matter, realized as matter in a cubic box with periodic boundary conditions. While the development of the V(S)DF model and the introduction of its supporting framework in \texttt{SMASH} have been done to enable large-scale comparisons between experimental data and simulations spanning a broad family of EOSs, the investigation of the qualitative behavior of dense nuclear matter presented in this thesis has been facilitated by using one specific parametrization of the dense nuclear matter EOS. After detailing the analysis methods used to study the simulation data, we discussed results pertaining to two distinct regions of the QCD phase diagram. First, we discussed simulations of nuclear matter in the density range corresponding to the ordinary nuclear matter, and we confirmed that the implemented VDF mean-field potentials lead to the correct thermodynamic behavior of nuclear matter as simulated in \texttt{SMASH}. Then we discussed results of simulations run in the high baryon density regime, affected by a conjectured QCD phase transition, stressing the excellent agreement of the behavior of the simulated systems with the thermodynamic predictions stemming from the underlying EOS. Furthermore, we have shown that the calculated fluctuation observables reproduce the thermodynamic behavior expected in the vicinity of the critical point based on the model predictions, proving that hadronic transport can be sensitive to effects due to phase transitions. In addition, we have discussed effects due to finite number statistics, whose thorough understanding is crucial to connecting theoretical calculations as well as simulation results to values of fluctuations measured in experiment.

In Chapter \ref{the_speed_of_sound_in_heavy-ion_collisions}, we have discussed a novel method of estimating the speed of sound as well as its logarithmic derivative with respect to the baryon number density in matter created in heavy-ion collisions. The method, based on a connection between the isothermal speed of sound and the susceptibilities of pressure characterizing the system, utilizes one of the most prominent observables used in the search for the QCD critical point: the cumulants of the baryon number distribution. The results of our analysis, based on available data on the cumulants measured by the STAR and HADES experiments, suggest that the logarithmic derivative of the speed of sound exhibits a non-monotonic behavior at low beam energies, on the order of $\sqrt{s_{NN}} \approx 2$--$10\ \txt{GeV}$, which signals robust changes in the thermodynamic behavior of matter created in these collisions. In particular, our results imply that matter at the lowest collision energy explored in these experiments is characterized by a very small value of the speed of sound which nevertheless grows very fast with increasing density, suggesting a stiff equation of state at high baryon densities, relevant to neutron star studies. Furthermore, comparing the experimental results with a parametrization of the VDF model including a QGP-like phase transition at high baryon density suggests that at low collision energies, the experimentally observed cumulants of baryon number may reflect the properties of the ordinary nuclear matter rather than the features of the conjectured QCD phase transition. Altogether, the presented method offers a new approach to studying the thermodynamic properties of dense nuclear matter through heavy-ion collisions, and has the potential to influence the exploration of the QCD phase diagram in regions relevant both to the Beam Energy Scan and the neutron star research.

Multiple future directions, carrying this research further, are possible, and we highlight a few of them below.

First, a natural next step in our studies is to use the VSDF model, which includes both vector- and scalar-density--dependent interactions, to parametrize a family of possible QCD EOSs. While interactions of the scalar type are significantly more computationally demanding, their addition will facilitate more robust comparisons with experiment. This is particularly important in view of the fact that scalar interactions lead to a correct momentum-dependence of the mean-field potential, which is found to play a considerable role in the description of the known behavior of nuclear matter at low energies. Additionally, extensions of the VSDF model using interaction terms with explicit momentum dependence can also be considered.

Next, hadronic transport simulations provide an excellent tool to investigate the effects of finite number statistics on both qualitative and quantitative features of statistical observables. In particular, these simulations allow one to isolate and compare effects connected to utilizing a finite number of test particles, effects related to the finite baryon number as measured in experiment, and effects driven by the physics described by the underlying theory. Studies of this kind will be crucial to establishing firm expectations for the influence of particular features of the conjectured QCD phase diagram on the behavior of heavy-ion collision observables.

Finally, our study of the connection between the speed of sound and the cumulants of baryon number can be naturally followed by an investigation of the relationship between these quantities as obtained in hadronic transport simulations, with a particular emphasis on the region of the QCD phase diagram probed by low-energy collisions. First, studies must be done to fully understand effects due to non-equilibrium evolution, conservation laws, and differences between proton and baryon number cumulants. Next, the considerable flexibility of the V(S)DF EOS can play an important role in modeling the expected thermodynamic behavior of simulated systems: while so far the VDF model has been parametrized to describe dense nuclear matter undergoing two phase transitions, one can also pursue a parametrization that reflects the properties of the speed of sound as suggested by our study, and compare results from hadronic transport simulations utilizing such a parametrization with experimental data. Additionally, generalizations of the VSDF model utilizing custom (instead of polynomial) forms of the interaction terms, allowing  for an even greater freedom in postulating the properties of nuclear matter in different regions of the phase diagram, are currently being developed.

\newpage

\appendix

\chapter{Units and notation}
\label{units_and_notation}

Throughout this thesis, we adopt natural units, that is we take $\hbar = c = k_B = 1$. 

We use the metric tensor
\begin{eqnarray}
g_{\mu\nu} = g^{\mu\nu} = \left( \begin{array}{rrrr}
1 & 0 & 0 & 0 \\
0 & -1 & 0 & 0 \\
0 & 0 & -1 & 0 \\
0 & 0 & 0 & -1
\end{array}  \right)~.
\end{eqnarray}
Greek indices run over $0$, $1$, $2$, $3$, or equivalently over $t$, $x$, $y$, $z$. Roman indices denote only the three spatial components. Repeated indices are summed over in all cases. For clarity and consistent notation of co- and contravariant vectors, we use a notation in which repeated indices occur only at different levels (superscript and subscript).

In view of this we introduce a general 4-vector as
\begin{eqnarray}
V^{\mu} = \big(V^0, \bm{V} \big)~.
\end{eqnarray}
Consequently, 
\begin{eqnarray}
V_{\mu} = g_{\mu\nu} V^{\nu} = \big(V^0, -\bm{V} \big)~. 
\end{eqnarray}
We thus establish
\begin{eqnarray}
V_i = -V^i~.
\end{eqnarray}
We note here that we may write a contravariant vector as $\bm{V} = \big(V_x, V_y, V_z\big)$, even though strictly speaking the location of the indeces in this case could be take to mean that $\bm{V}$ is covariant. The cases where such somewhat misleading notation is used are few and far in between, and we never use it in contexts where it could lead to any confusion or misconception.

The scalar product of two 4-vectors is given by
\begin{eqnarray}
V \cdot U = g_{\mu\nu} V^{\mu} U^{\nu} = V_{\nu} U^{\nu} = V_0 U^0 + V_i U^i = V^0 U^0 - \bm{V} \cdot \bm{U}~.
\end{eqnarray}
In consequence
\begin{eqnarray}
\bm{V} \cdot \bm{U} = - V_k U^k~. 
\end{eqnarray}

The four-gradient is given by
\begin{eqnarray}
\partial_{\mu} = \parr{ }{x^{\mu}} = \left(\parr{ }{x^0}, \bm{\nabla} \right)~. 
\end{eqnarray}
where $x^{\mu}$ is the position four-vector. In particular,
\begin{eqnarray}
(\bm{\nabla})_{i} = \parr{ }{x^i} = - \parr{}{x_i} ~. 
\end{eqnarray}

A specific example of the consequences of the adopted notation is that the expression for the kinetic energy becomes
\begin{eqnarray}
\epsilon_{\txt{kinetic}} = \sqrt{ \bm{p}^2 + m^2} = \sqrt{ - p_k p^k + m^2 } ~. 
\end{eqnarray}
As another example, note that the conservation of current is given by 
\begin{eqnarray}
\parr{j^0}{t} + \bm{\nabla} \cdot \bm{j} =  \parr{j^0}{x^0} + \sum_i \big(\bm{\nabla})_i j^i = \parr{j^0}{x^0} + \parr{ }{x^i} j^i =  \partial_{\mu} j^{\mu} = 0 ~.
\end{eqnarray}

\newpage
\chapter{Phase transitions}
\label{phase_transitions}

This appendix is largely based on Ref.\ \cite{Landau_Stat}.

Phase transitions are processes involving a change between two states (phases) of a given medium characterized by different properties, and they can be largely divided into two types. 

A first-order phase transition is characterized by a discontinuity of the first derivative of the thermodynamic potential (the Ehrenfest classification) and involves a latent heat (modern classification), that is, a finite amount of energy has to be released or absorbed for the system to undergo the phase transition. A canonical example of a first-order phase transition is the melting of ice: In order for this transition to occur, heat must be absorbed by the ice, and during this process of absorption there exist parts of the system that have already melted (parts that have completed the transition) and parts which are still frozen; in other words, there is a coexistence of phases. Note that the coexistence of the frozen and the melted phase means that different parts of the system are characterized by different densities corresponding to the densities of ice and water. This can be reformulated as the discontinuity of the first derivative of the pressure (the thermodynamic potential) with respect to the chemical potential, $\inparr{P}{\mu} = n$. 

A second-order phase transition is characterized by a discontinuity in the second derivative of the thermodynamic potential (the Ehrenfest classification) and involves a divergent susceptibility and an infinite correlation length (modern classification). A prominent example here is the ferromagnetic phase transition, within which the entire volume of a ferromagnet, when cooled below the transition temperature (known as the Curie temperature), develops a spontaneous magnetization, that is the magnetic moments in the system align with each other. Here the magnetization, which is the first derivative of the thermodynamic potential with respect to the external magnetic field, is zero above the Curie temperature and increases continuously from zero below the Curie temperature; consequently, the magnetic susceptibility, which is the second derivative of the thermodynamic potential with respect to the external magnetic field, is discontinuous.

It is known that the nuclear liquid-gas phase transition in the ordinary nuclear matter is of the first order, and effective models suggest that the phase transition from the hadronic to quark and gluon degrees of freedom may also be of first order in the region of the QCD phase diagram characterized by high values of the baryon chemical potential. Let us therefore discuss further certain features of this type of phase transitions.

\begin{figure}[t]
	\centering\mbox{
	\includegraphics[width=0.99\textwidth]{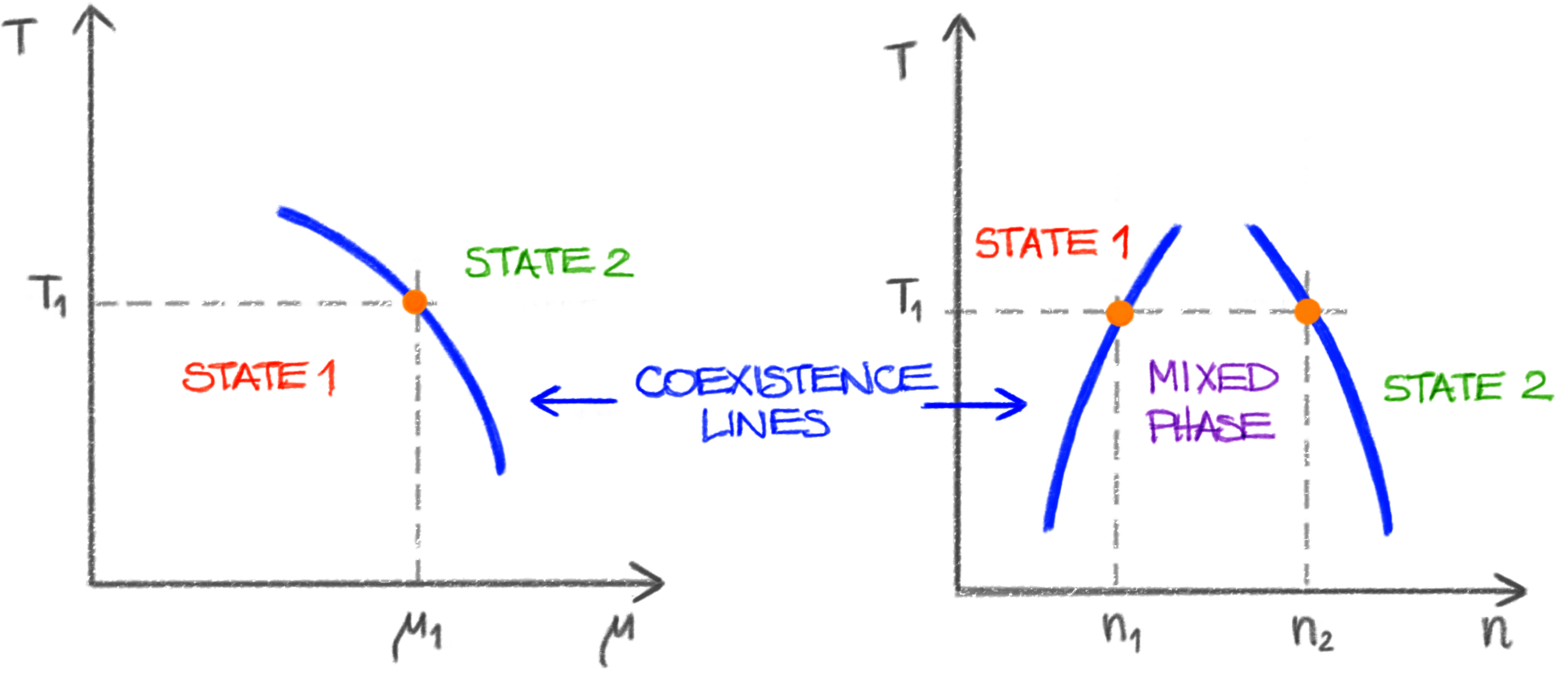} 
	}
	\caption[Phase diagrams of a hypothetical system with a first-order phase transition]{Phase diagrams in the temperature and chemical potential $(T,\mu)$, plane (left panel) and the temperature and density $(T,n)$ plane (right panel), drawn for a hypothetical system with a first-order phase transition. Orange dots mark points at which the system is characterized by a given temperature $T_1$, chemical potential $\mu_1$, and pressure $P_1$ (not shown); note that there are two such points in the $(T,n)$ plane, indicating a coexistence of two phases with different densities, $n_1$ and $n_2$. A continuum of coexistence points forms the coexistence curves, shown by the solid blue lines; note that in the $(T,n)$ plane, the coexistence curves bound a region where a mixed phase occurs. The coexistence lines terminate either on another phase transition line, or on a critical point; these possibilities are not depicted here.}
	\label{phase_coexistence_1}
\end{figure}

As already mentioned, first-order phase transitions are characterized by the occurrence of states with mixed phases, like melting ice floating in water or a liquid and its vapor. These coexisting phases are in mutual equilibrium, which means that they are characterized by the same temperature, pressure, and chemical potential,
\begin{eqnarray}
T_1 = T_2 ~, \hspace{10mm} P_1 = P_2 ~, \hspace{10mm} \mu_1 = \mu_2 ~. 
\end{eqnarray}
If we plot the points at which two phases of a given system coexist, we obtain a phase diagram of a given substance. Phase diagrams can be drawn using different primary variables. For example in the case of the liquid-gas phase transition, it can be convenient to draw a phase diagram in the temperature and chemical potential $(T,\mu)$ or temperature and density $(T,n)$ planes, see Fig.\ \ref{phase_coexistence_1}. Note that the coexistence of phases is given by a single line in the $(T,\mu)$ plane, while the phase diagram of the same system in the $(T,n)$ plane is characterized by a coexistence region bounded by two coexistence lines; this is because for every set of values $(T, P, \mu)$ at which the two phases coexist, there are two corresponding coexistence densities, $n_1$ and $n_2$. As can be clearly seen from the diagrams in Fig.\ \ref{phase_coexistence_1}, the coexistence chemical potential or, alternatively, the coexistence densities are functions of temperature. Importantly, for a given temperature $T =T_1$ at which the coexistence densities are given by $n_1$ and $n_2$, an equilibrated system of $N$ particles in volume $V$ such that the density $n = N/V$ satisfies $n_1 < n < n_2$ cannot exist. This is because the coexistence region of the phase diagram is thermodynamically and, in a specific subregion, also mechanically unstable (we will elaborate on this below). If such a system is obtained as a result of a non-equilibrium evolution (for example as a result of an adiabatic quench), the system will spontaneously separate into two phases occupying volumes $V_1 = \alpha V$ and $V_2 = (1 -\alpha)V$, where $\alpha = \infrac{(n - n_1)}{(n_2 - n_1)}$ (the latter is a simple consequence of the fact that the numbers of particles in each of the phases must satisfy $N_1 + N_2 = N$).

\begin{figure}[t]
	\centering\mbox{
	\includegraphics[width=0.99\textwidth]{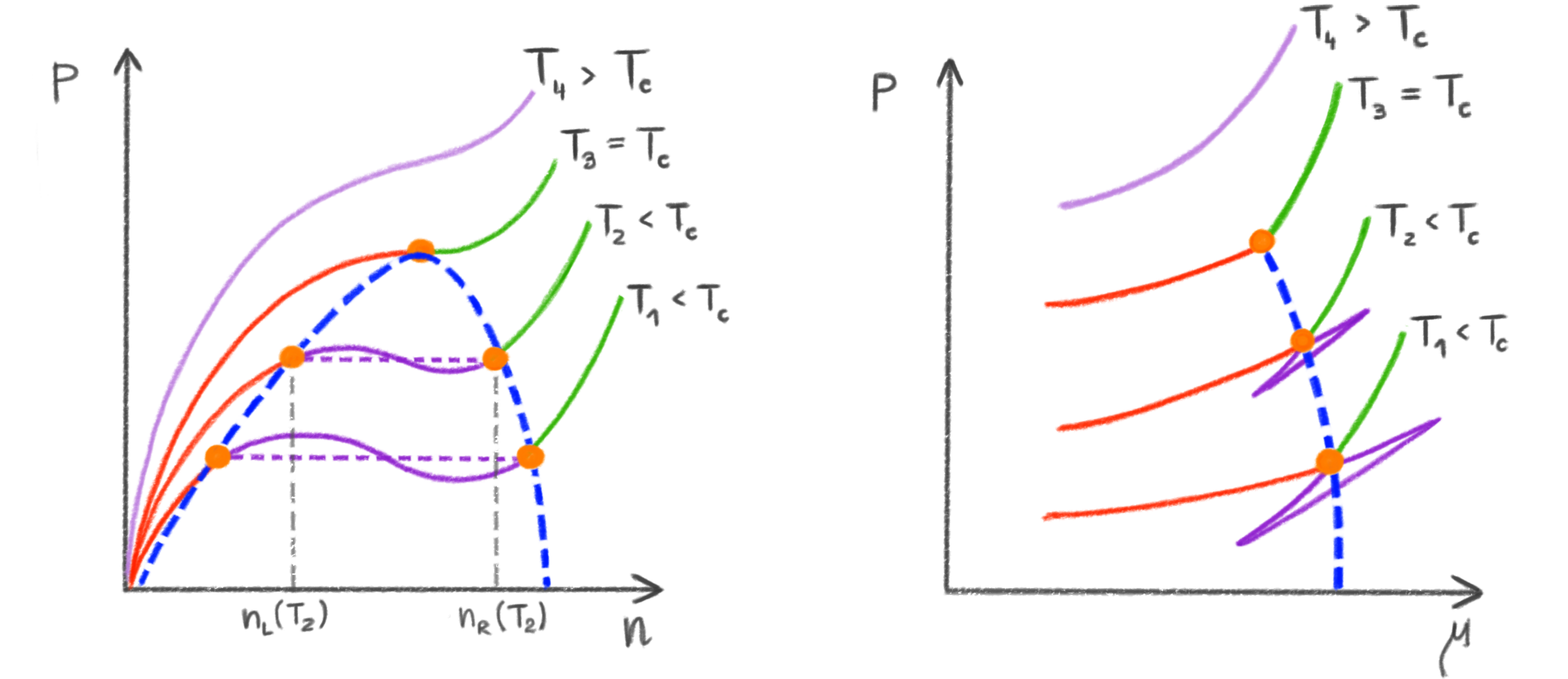}
	}
	\caption[Isotherms of pressure as functions of density and chemical potential for a hypothetical system with a first-order phase transition]{Isotherms of pressure as functions of the density $n$ (left panel) and the chemical potential $\mu$ (right panel), drawn for a hypothetical system with a first-order phase transition. Points on the pressure isotherms where two phases coexist are marked with orange dots, the coexistence regions are denoted with dashed blue lines, and the thermodynamically unstable parts of the pressure curves are denoted with purple lines; in particular, in the left panel, the solid purple lines show the functional form of the pressure, while the dashed purple lines show the Maxwell construction.}
	\label{phase_coexistence_3}
\end{figure}

Let us now discuss isotherms of pressure as a function of density, see left panel in Fig.\ \ref{phase_coexistence_3}. At a given temperature, for densities smaller than the lower (or ``left'') coexistence density $n_L = n_L(T)$, the pressure is a monotonously rising function of the density $n$. Likewise, for densities larger than the upper (or ``right'') coexistence density $n_R = n_R(T)$, the pressure is also monotonic and rising. That the pressure satisfies $dP/dn > 0$  is in fact a condition for the system to be mechanically stable (this can be easily seen by noticing that if the pressure decreased with density, the system would have a tendency to collapse in on itself).

However, in order to have $P(n_L) = P(n_R)$ the pressure must be a non-monotonic function of $n$ for $n_L(T) < n < n_R(T)$, as depicted. In particular, this means that there exist densities for which $dP(T)/dn < 0$. This shows that the matter in this region, called the spinodal region, is mechanically unstable: any system brought into the spinodal region (e.g., through a sudden quench) will undergo a spontaneous and violent separation into two coexisting phases with densities given by $n_L(T)$ and $n_R(T)$. The boundaries of the spinodal region at a given temperature $T$, here denoted by $\eta_L$ and $\eta_R$, are given by the condition $dP(\eta_{L,R}, T)/dn = 0$. The ranges of densities satisfying $n_L < n < \eta_L$ and $\eta_R < n < n_R$, are thermodynamically, but not mechanically unstable; this means that the states of the system in these regions are metastable, as is the case for example for overheated liquid or supercooled vapor. For the isotherms of pressure shown as functions of the chemical potential, see right panel of Fig.\ \ref{phase_coexistence_3}, the metastable and unstable portions of the isotherms are the ones that form a closed curve beginning and ending on the crossing point.

\begin{figure}[t]
	\centering\mbox{
	\includegraphics[width=0.9\textwidth]{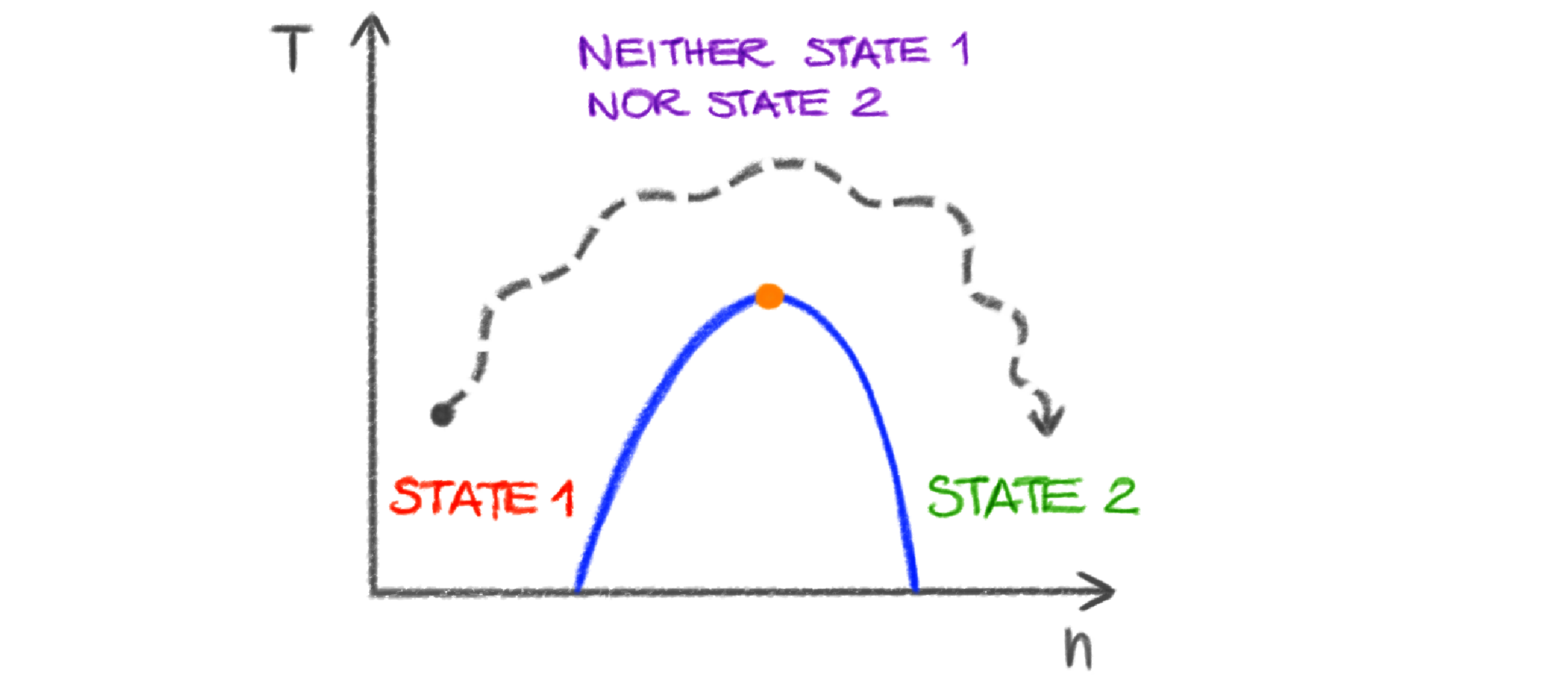}
	}
	\caption[Phase diagram for a hypothetical system with a first-order phase transition ending in a critical point]{Phase diagram in the temperature and density $(T,n)$ plane, drawn for a hypothetical system with a first-order phase transition ending in a critical point. It is possible to devise a path in the phase diagram such that the system will go from phase 1 to phase 2 without experiencing the coexistence of phases.}
	\label{phase_coexistence_5}
\end{figure}

We can further see in the left panel of Fig.\ \ref{phase_coexistence_3} that as the temperature is increased, the width of the coexistence region (and likewise the width of the spinodal region) becomes smaller, until finally at a temperature $T = T_c$, known as the critical temperature, the coexistence lines merge at $n = n_c$, known as the critical density. The point $(T_c, n_c)$ is known as the critical point. Importantly, for temperatures higher than the critical temperature $T_c$, the notion of the two phases of the studied substance does not have a well-defined sense: the system is neither in phase 1, nor in phase 2 (see Fig.\ \ref{phase_coexistence_5}). Therefore, it is possible to devise a path in the phase diagram following which brings the system from phase 1 into phase 2 without matter ever separating into two phases. This underscores the notion that the concept of phases is somewhat arbitrary and is only really well-defined when the two phases coexist and are in contact. In the case of the liquid-gas phase transition, while below $T_c$ the system may be described as being either in the liquid or the gas phase, above $T_c$, to stress the fact that these two descriptors do not apply anymore, it is usually referred to as a ``fluid''.

An in-depth review of the thermodynamics related to the first-order phase transitions in general, and to the spinodal region in particular, can be found in Ref.\ \cite{Chomaz:2003dz}.

\newpage
\chapter{Kinematic variables, collision geometry, and baryon transport in heavy-ion collisions}
\label{kinematic_variables_and_geometry_in_heavy-ion_collisions}

\section{Kinematic variables}
\label{kinematic_variables}

This section is based on Ref.\ \cite{Vogt:2007zz}.

The kinematics of relativistic particles in heavy-ion collisions is often described using two notable variables. Customarily, the direction of the heavy-ion beam defines the $z$-axis of the collision systems. If we then consider a relativistic particle with a 4-momentum $p^{\mu} = (E, \bm{p})$, where $\bm{p} = (p_x, p_y, p_z)$, then the transverse mass, $m_T$, can be defined as
\begin{eqnarray}
m_T^2 \equiv E^2 - p_z^2 = m^2 + p_T^2~,
\end{eqnarray}
where $p_z = |\bhat{p}_z|$ is the component of the particle's momentum along the beam axis, often called the longitudinal momentum, $m$ is the particle's rest mass, and $p_T = |\bm{p}_T| = |p_x \bhat{x} + p_y \bhat{y}|$ is the component of the particle's momentum transverse to the beam direction, called the transverse momentum. It is easy to see that the transverse mass is invariant under Lorentz transformations along the beam direction.

The rapidity of the particle, $y_r$, is defined as
\begin{eqnarray}
y_r \equiv \frac{1}{2} \ln \frac{E + p_z}{E - p_z} = \frac{1}{2} \ln \frac{1 + \beta}{1 - \beta} ~,
\label{rapidity_definition}
\end{eqnarray}
where the second equality uses the longitudinal velocity of the particle $\beta = p_z/E$. (Here we note that while in this appendix we denote rapidity as $y_r$ to avoid confusion with the $y$-axis in the coordinate space, it is customarily denoted with just $y$; the distinction between the two variables is usually clear from the context.) The rapidity can be thought of as a measure of the contribution of the particle's velocity in the beam direction to the particle's total energy. Note that if $p_z = 0$, then $y_r =0$. 

The advantage of $y_r$, as compared to the velocity $\beta$, is that it transforms more easily under Lorentz boosts. Consider two observers, $A$ and $B$, where the observer $B$ is moving with a constant velocity $\beta_0$ with respect to the observer $A$ (to simplify, let us consider only one spatial dimension). If $B$ observes some object $C$ moving with velocity $\beta'$, then $A$ sees the same object moving with velocity
\begin{eqnarray}
\beta \equiv \frac{dx}{dt} = \frac{\gamma_0 \big( dx' + \beta_0 dt' \big)}{\gamma_0 \big(dt' + \beta_0 dx' \big)} =\frac{\beta' + \beta_0 }{1 + \beta_0 \beta'} ~,
\end{eqnarray}
that is the velocities in the two frames are related by a composition law. On the other hand, the rapidities reported by the two observers are simply related by
\begin{eqnarray}
y_r = \frac{1}{2} \ln \frac{\gamma_0 \big( E' + \beta_0 p_z' \big) + \gamma_0\big( p_z' + \beta_0 E'\big)}{\gamma_0 \big( E' + \beta_0 p_z' \big) - \gamma_0\big( p_z' + \beta_0 E'\big)} = \frac{1}{2} \ln \frac{(E' + p_z')(1+\beta_0)}{(E' - p_z')(1 - \beta_0)} = y_r' + y_0 ~, 
\end{eqnarray}
where $y_0$ is the rapidity of $B$ with respect to $A$, so that the rapidities in the two frames are additive. In particular, this means that a difference of rapidities, $\Delta y_r = y_r^{(1)} - y_{r}^{(2)}$, is a Lorentz-invariant quantity (which is not the case for the difference of velocities). This is especially useful in the context of heavy-ion collisions, where many observables either depend on the difference in rapidities of two observed particles (e.g., two-particle correlations), or are calculated within a certain chosen rapidity window $\Delta y$ (e.g., cumulants of baryon number). 

\begin{figure}[th]
	\centering\mbox{
	\includegraphics[width=0.55\textwidth]{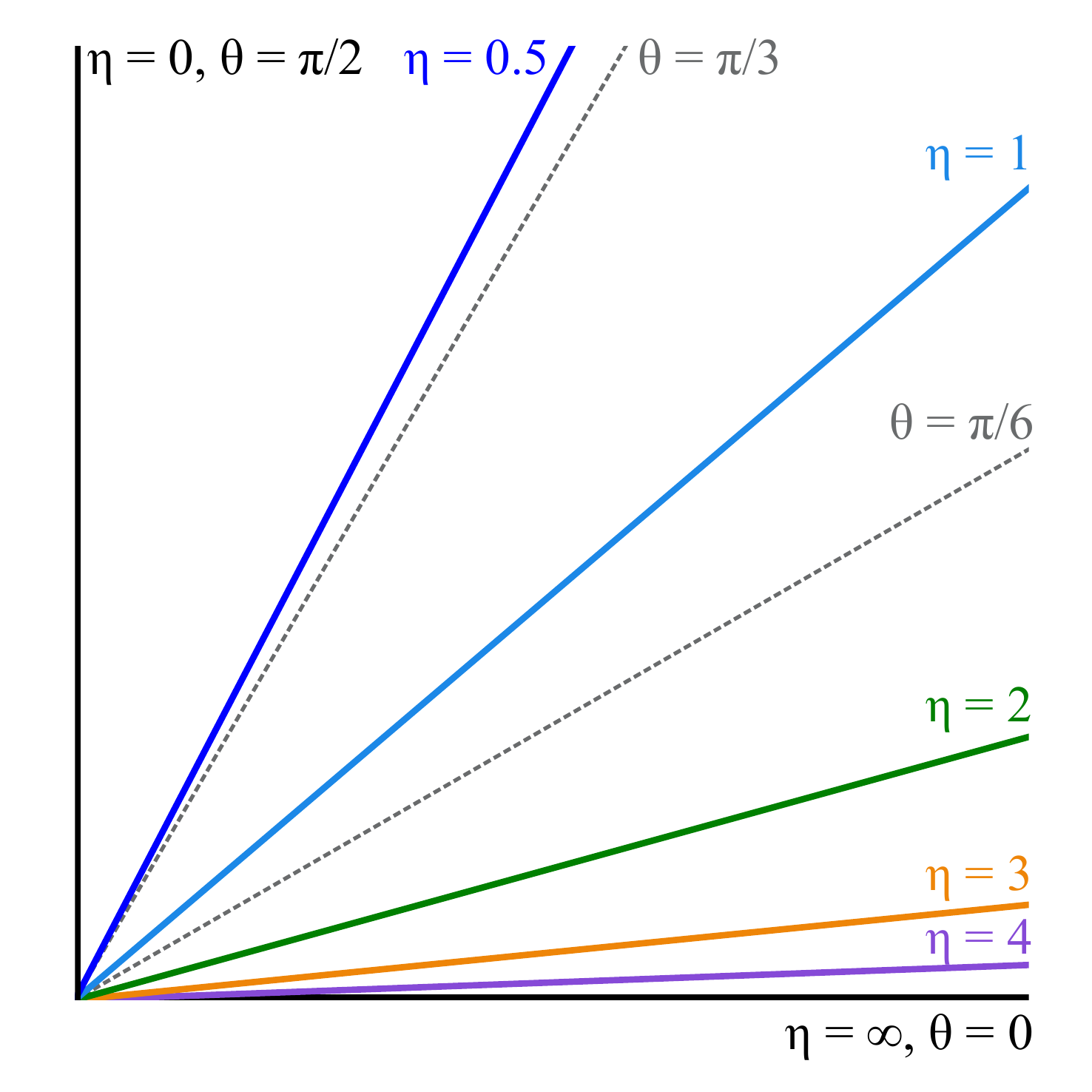} 
	}
	\caption[A plot showing the relation between values of the pseudorapidity and the polar angle]{A plot showing the relation between values of the pseudorapidity $\eta$ and the polar angle $\theta = \measuredangle \big( \bhat{p}, \bhat{z} \big)$. Note that $\eta\big(\theta = \frac{\pi}{2}\big) = 0$ and $ \eta\big(\theta = 0\big) = + \infty$.}
	\label{pseudorapidity}
\end{figure}

In experiment, it is not always possible to identify the mass of a particle, necessary to calculate the rapidity, Eq.\ \eqref{rapidity_definition}. For particles with very high momentum, however, the mass contribution to the energy can often be neglected, $E \approx p$, where $p = |\bm{p}|$. In that case the rapidity is approximately equal to the pseudorapidity $\eta$, which is defined as
\begin{eqnarray}
\eta \equiv \frac{1}{2} \ln \frac{p + p_z}{p - p_z} = \frac{1}{2} \ln \frac{1 + \frac{p_z}{p}}{1 - \frac{p_z}{p}} = \frac{1}{2} \ln \frac{1 + \cos \theta}{1 - \cos \theta} = - \ln \tan \left( \frac{\theta}{2} \right) ~, 
\end{eqnarray}
where $\theta$ is the angle between the direction of the particle's momentum and the beam axis, $\theta = \measuredangle \big( \bhat{p}, \bhat{z} \big)$. Being a function of $\theta$ only, pseudorapidity has a clear geometrical interpretation. In particular, a detector often covers a well-defined range of $\theta$ with respect to the beam axis, and it is not unusual for particles with high values of pseudorapidity to fall beyond the region where detection is possible. Note that a pseudorapidity window $\Delta \eta  = 1 $ around $\eta =0$ covers about a quarter of the entire range of the angle $\theta$, while for example the same rapidity window about $\eta = 4$ covers only a small angle, as shown in Fig.\ \ref{pseudorapidity}. This is a consequence of the non-linear dependence of $\eta$ on $\theta$. For the same reason, a particle distribution that is isotropic in the coordinate space (that is a distribution uniform in $\sin \theta  d \theta ~d\phi$) is, when expressed in terms of $\eta$, a distribution with a peak at small values of $\eta$. The same conclusions can be approximately applied to the rapidity. In consequence, experimental observables are often considered at small (relative to the beam rapidity) values of rapidity or pseudorapidity; the corresponding range of considered values of $y_r$ or $\eta$ is referred to as ``midrapity'' or ``central rapidity region''.

Notably, using the transverse mass and rapidity, it is possible to write the energy and longitudinal momentum of a particle as
\begin{eqnarray}
E = m_T \cosh y_r  \distand p_z = m_T  \sinh y_r ~,
\end{eqnarray} 
respectively.

Using the center-of-mass frame for expressing the beam energy allows one to make easy comparisons between collisions in the collider mode and collisions in the target mode. Conveniently, for collisions of identical nuclei of mass $M$ moving with velocities of equal magnitudes and opposite directions, the lab frame is the center-of-mass frame. If one denotes the 4-momentum of one of the nuclei as $p_1^{\mu} = ( E, 0, 0, p_z)$, then the 4-momentum of the other nucleus is $p_2^{\mu} = (E, 0, 0, -p_z)$, and the Lorentz invariant Mandelstam variable $s$ satisfies $\sqrt{s} = \sqrt{ \big( p_1^{\mu} + p_2^{\mu} \big)^2} = 2E$. For collisions in the target mode, in which one of the nuclei is at rest while the other nucleus is moving with energy $E$ (in the lab frame), one has $\sqrt{s} = \sqrt{2M(M + E)} = \sqrt{2M(2M + E_{\txt{kin}})}$, where $E_{\txt{kin}} \equiv  E - M$ is the kinetic energy of the projectile nucleus. To enable comparisons between collisions of nuclei with different mass numbers $A$, one further defines $\sqrt{s_{NN}}$ as the center-of-mass energy of two protons colliding with kinetic energy $E_{\txt{kin}, NN} = E_{\txt{kin}} /A$. Up to a slight discrepancy due to neglecting the binding energy of the nuclei, $\sqrt{s_{NN}} $ can be thought of as the center-of-mass energy per nucleon pair in the system, $\sqrt{s_{NN}} \approx \sqrt{s}/A$.

\section{Collision geometry}
\label{collision_geometry_appendix}

\begin{figure}[tp]
	\centering\mbox{
		\includegraphics[width=0.99\textwidth]{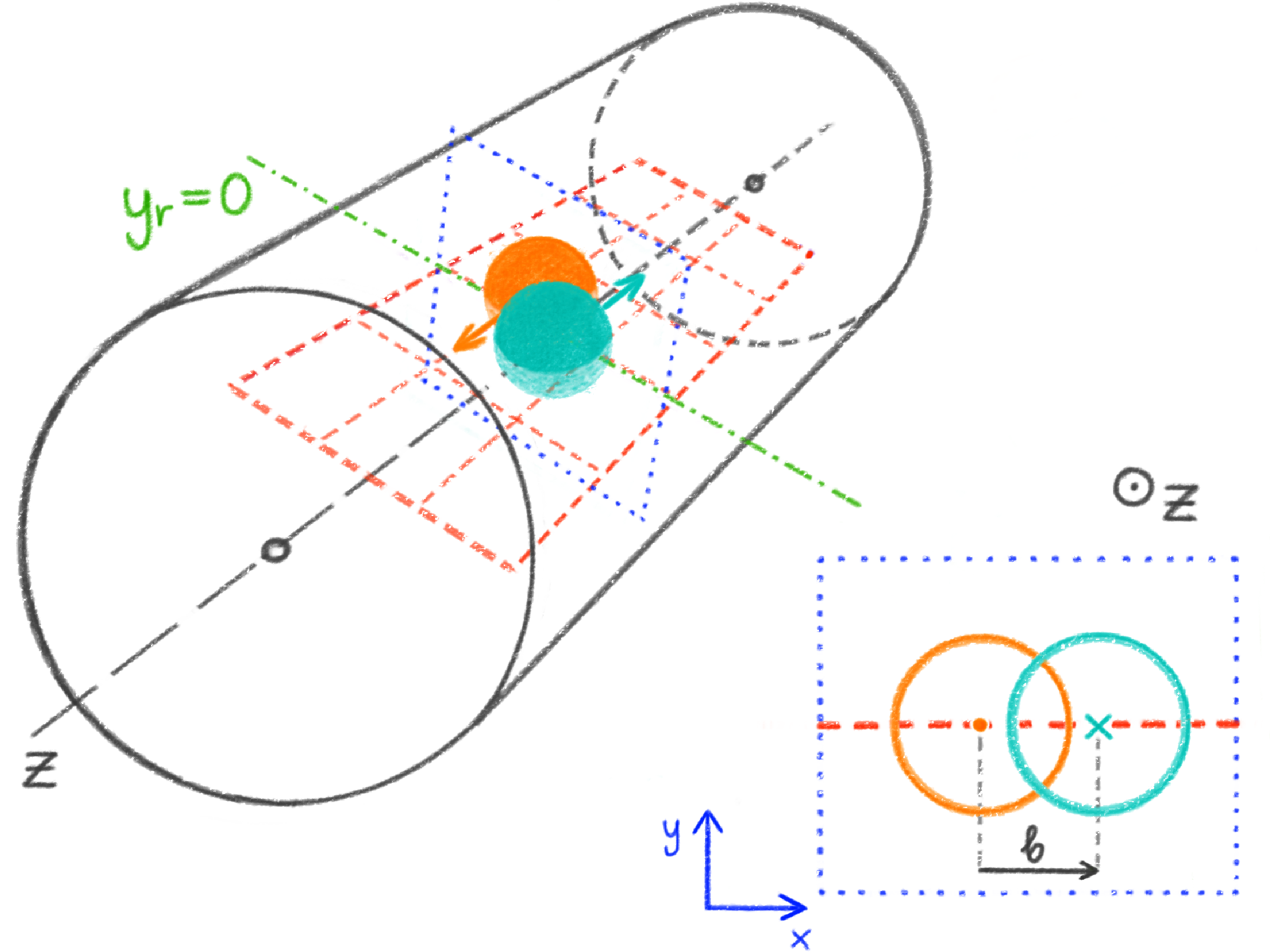} 
	}
	\caption[A sketch of a heavy-ion collision geometry in the collider mode]{A sketch of a heavy-ion collision geometry in the collider mode. Note that two nuclei moving at relativistic speeds should be Lorentz contracted, but the sketch does not include the contraction for clarity. The outer cylinder on the sketch, indicated with a solid gray line, is supposed to indicate the geometry of a typical detector such as STAR, but its relative dimensions are not to scale to make the figure more clear. The nuclei are moving along the beam pipe, coincident with the $z$-axis, indicated with a long-dashed gray line. The short-dashed red line indicates the reaction plane, the dotted blue line indicates the transverse plane, and the green dash-dotted line indicates the intersection of the reaction and transverse planes, where $y_r = 0$. The insert shows the geometry of the collision when looking along the beam pipe. See text for more details.}
	\label{collision_geometry}
\end{figure}
 
Fig.\ \ref{collision_geometry} shows a simplified sketch of a heavy-ion collision in the collider mode. Two heavy ions are moving along the beam pipe, which is coincident with the $z$-axis, with equal energies $E = \sqrt{m^2 + p_z^2}$. In the figure, the orange heavy ion moves with momentum $\bm{p}_1 = + p_z \bhat{z}$, while the turquoise heavy ion moves with momentum $\bm{p}_2 = - p_z \bhat{z}$. The impact parameter $b$ of the collision is defined as the length of the vector $\bm{b}$ connecting the centers of the colliding nuclei at the point of closest approach (see the insert in Fig.\ \ref{collision_geometry}). The $z$-axis together with $\bm{b}$ create a plane called the reaction plane, indicated in the figure by short-dashed red lines. One can also define the transverse plane, which is a plane perpendicular to the beam axis, indicated in the figure with a blue dotted line; usually, the intersection of the beam axis and the transverse plane is set at a point where the rapidity is zero, $y_r =0$; in the figure, the intersection of the reaction and the transverse planes is indicated by the green dash-dotted line. The directions in the transverse plane are given by the $x$-axis and the $y$-axis, as indicated on the insert in Fig.\ \ref{collision_geometry}.

\begin{figure}[t]
	\centering\mbox{
	\includegraphics[width=0.99\textwidth]{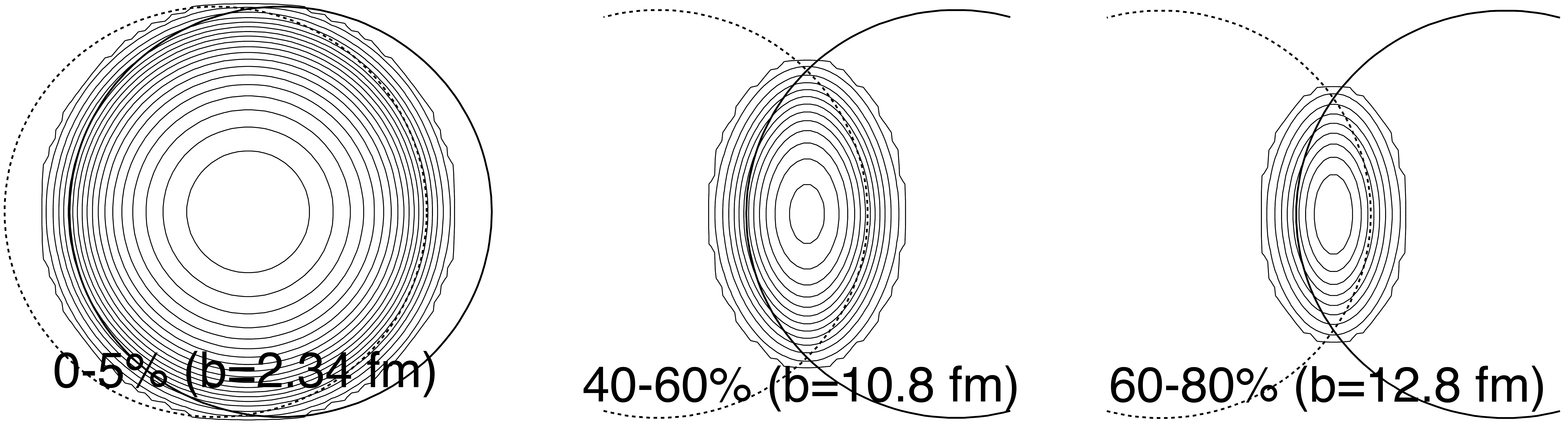}
	}
	\caption[A plot of different collision centralities]{A plot of different collision centralities, corresponding to a series of impact parameters $b$, shown in a view along the beam axis. The concentric rings indicate regions of the same density. The sketch does not include the intrinsic fluctuations in the positions of nucleons inside the nuclei. Figure from \cite{Sorensen:2003kp}.}
	\label{centrality}
\end{figure}

Collisions can occur at different values of the impact parameter $b$, see Fig.\ \ref{centrality}. Nucleons from the two nuclei which overlap in the transverse plane are called participants. The remaining nucleons, which ``detach'' from the colliding nucleus at the moment of the collision and then continue to move with the beam momentum, are called spectators. Different impact parameters will lead to different geometries of the initial collision region, numbers of participants, and initial energy depositions. Therefore, to enable meaningful comparisons between different geometries of the initial systems, collisions are sorted into classes corresponding to ranges of impact parameters. Because measuring the impact parameter itself is impossible in experiment, the collisions are instead characterized by a parameter known as centrality. There are many ways of establishing the centrality of a collision, varying between detectors, however, the intuitive picture for most of these methods is the following: Collisions with a small impact parameter will lead to a greater amount of energy deposited in the collision region and through that to a larger number of produced particles and a larger number of detected charged particles, or charged particle multiplicity, $N_{\txt{ch}}$. When the collisions are sorted by their respective values of $N_{\txt{ch}}$, one can define specific ranges of $N_{\txt{ch}}$ as corresponding to ``more central'' or ``less central'' collisions. And so, for example, one can define the most central collisions as the top 10\% of all collisions with respect to the number of detected charged particles $N_{\txt{ch}}$; such collisions are then said to be characterized by a 0--10\% centrality. Correspondingly, one can define mid-central collisions as those falling between the top 10\% and the top 40\% of all collisions, and peripheral collisions as those falling in the top 40\%--80\%. Typically, the most peripheral collisions, in this case the bottom 20\%, are discarded to avoid biases due to detector trigger inefficiencies. The entire range of used centralities, here 0--80\%, is referred to as ``minimum bias''. We note that the ranges defining the centralities are chosen to best suit a specific analysis and a particular experiment, and are by no means set to the example values provided above.

\section{Baryon transport}
\label{baryon_transport}

The initial rapidity of the beam, or equivalently the collision energy, influences the rapidity distributions of detected particles. Initially, the transverse momentum of the entire system as well as of each of the nucleons inside the colliding nuclei is zero (here, we neglect the Fermi momenta of the nucleons inside the nuclei). Because momentum is conserved, the total transverse momentum after the collision is also zero, but individual particles (both transported as well as produced) develop non-zero values of $p_T$ through scattering or collective motion of the system (e.g., collective flow). At the same time, on average, there is only so much transverse momentum that a particle can plausibly develop through these means. Indeed, head-on collisions of quarks or gluons can result in scattering products with almost zero longitudinal momentum and a large fraction of their initial energy converted into the transverse momenta; for example, at $\sqrt{s_{NN}} = 200\ \txt{GeV}$, which corresponds to $E_{\txt{kin}, NN} \approx 100 \txt{GeV}$, the individual quarks inside the proton may typically have $p_z \approx 10$--$20\ \txt{GeV}$, and so, if two such quarks collide, there is a finite chance to produce a $10$--$20\ \txt{GeV}$ transverse jet (where by a jet we understand a collection of final state particles originating from a single hard scattering at the initial stage of the collision). However, aside from these somewhat exceptional events, ordinarily the final transverse momentum of a detected particle is much smaller, on the order of $p_T \approx 0.5 \ \txt{GeV}$ (the average transverse momentum of protons tends to be $p_T \approx 1.0\ \txt{GeV}$, see below for more details).

A typical detector can detect particles with pseudorapidities $|\eta| \leq 1.5$ (and therefore, approximately, rapidities $|y_r| \leq 1.5$), while the beam rapidity $y_{\txt{beam}}$ (that is the rapidity of a single proton moving with $p_{\txt{beam}, NN}$) characterizing the collisions is often much higher. It is evident that a large change in the rapidity, $\Delta y_r = y_r' - y_r$, requires a correspondingly large change in the longitudinal momentum, $\Delta p_z$, where the latter will also grow with the magnitude of $y_r'$. Changes in the rapidity of baryons occurring during the collisions can be established by measuring the rapidity distribution of net protons $dN_{p - \bar{p}}/dy_r$ at a given collision energy and centrality. Fig.\ \ref{BRAHMS_net-proton_dNdy} in Chapter \ref{introduction} shows $dN_{p - \bar{p}}/dy_r$ as measured in collisions with 0--5\% centrality at the AGS (Au+Au collisions at $\sqrt{s_{NN}} = 5\ \txt{GeV}$, $y_{\txt{beam}} = 1.64$) \cite{E802:1999hit}, SPS (Pb+Pb collisions at $\sqrt{s_{NN}} = 17.2\ \txt{GeV}$, $y_{\txt{beam}} = 2.91$) \cite{NA49:1998gaz}, and RHIC (Au+Au collisions at $\sqrt{s_{NN}} = 200\ \txt{GeV}$, $y_{\txt{beam}} = 5.36$) \cite{BRAHMS:2003wwg}. We see that for the lowest collision energies the net proton distribution is peaked around $y_r \approx 0$, and the mean absolute value of rapidity is $\langle | y_r | \rangle \approx 0.7$ (note that mean rapidity $\langle y_r \rangle$ is always 0 due to the geometry of symmetric collisions in the center-of-mass frame), while for collisions at the SPS the distribution has two peaks around $y_r \approx \pm 1.3$, with the mean absolute rapidity of about $\langle |y_r| \rangle \approx 1.22$. For collisions at RHIC the peaks of the distribution are beyond the reach of the detector, but numerical fits suggest the peaks at about $y_r \approx 4.25$, and the corresponding mean absolute rapidity is $\langle |y_r| \rangle \approx 3.32$. The average rapidity loss, defined as $\langle \delta y_r \rangle \equiv y_{\txt{beam}} - \langle |y_r| \rangle$, for the three experiments is then approximately $1.0$, $1.7$, and $2.0$, respectively.

It is instructive to ask what is the change in the longitudinal momentum (between the initial and final state) of a typical proton measured within a given rapidity window, assuming that the proton has remained intact throughout the collision. (Here one has to bear in mind that this exercise only serves to develop a rough intuition about possible final longitudinal and transverse momenta in the system, as in a true collision any given final state particle will not only be a product of multiple scatterings and decays, but will also most probably originate in the deconfined QGP. One can hope, however, that final state protons can be used as a good proxy for the effects of the complex evolution of the baryon number, itself originating in the protons and neutrons from the colliding nuclei.) We consider protons with transverse momentum $p_T = 0.9 \ \txt{GeV}$. This value is chosen based on the BRAHMS experiment analysis of Au+Au collisions at $\sqrt{s_{NN}} = 200\ \txt{GeV}$ and 0--5\% centrality, which measured $\langle p_T \rangle = 1.01 \pm 0.01  (\txt{stat})\ \txt{GeV}$ at $y_r \approx 0$ and $\langle p_T \rangle = 0.84 \pm 0.01  (\txt{stat})\ \txt{GeV}$ at $y_r \approx 3$ \cite{BRAHMS:2003wwg}. For comparison, at lower energies, for example in Pb+Pb collisions at $\sqrt{s_{NN}} = 17.2\ \txt{GeV}$ and 0--5\% centrality, the mean transverse momentum of protons is $\langle p_T \rangle = 0.825\pm37 \ \txt{GeV}$ at $y_r = 0$ and $\langle p_T \rangle = 0.600\pm 17 \ \txt{GeV}$ at $y_r = 2.4$ \cite{NA49:1998gaz}.

\begin{table}[th]
	\caption[Change in the longitudinal momentum of a hypothetical proton traveling with $p_{\txt{beam},NN}$ and detected at rapidity $y$]{Center-of-mass energies $\sqrt{s_{NN}}$ (in GeV), beam momenta $p_{\txt{beam}, NN}$ (in GeV), beam rapidities $y_{\txt{beam}}$, and differences between the final and initial state longitudinal momenta $\Delta p_z (y_0)$ (in GeV) of a hypothetical proton undergoing an evolution from the initial state inside the colliding nucleus to the final state characterized by $p_T = 0.9 \ \txt{GeV}$ and a particular value of rapidity $y_r = y_0$. See text for more details.}
	\begin{center}
		\def\arraystretch{1.4}
		\begin{tabular}{|r|r|r|r|r|r|r|}
			\hline
			\hline
			\hspace{-0.5mm}$\sqrt{s_{NN}}$&\hspace{-0.5mm}$p_{\txt{beam}, NN}$&\hspace{-0.5mm}$y_{\txt{beam}}$&\hspace{-0.5mm}$\Delta p_z (y_0 = 1.0)$&\hspace{-0.5mm}$\Delta  p_z (y_0 = 2.0)$&\hspace{-0.5mm}$\Delta p_z (y_0 = 3.0)$&\hspace{-0.5mm}$\Delta p_z (y_0 = 4.0)$  \\ 
			\hline	
			3.0 &  1.170 & 1.046 &  0.358 &  3.545 & 11.854 & 34.310 \\  
			7.7 &  3.734 & 2.090 & -2.206 &  0.981 &  9.291 & 31.747 \\  
			19.6 &  9.755 & 3.037 & -8.227 & -5.040 &  3.270 & 25.726\\  
			62.4 & 31.186 & 4.197 & -29.658 & -26.470 & -18.161 &  4.295 \\  
			200.0 & 99.996 & 5.362 & -98.468 & -95.280 & -86.971 & -64.515  \\  
			\hline
		\end{tabular}
		\label{example_rapidities_given_pT}
	\end{center}
\end{table}

Table \ref{example_rapidities_given_pT} lists values of $p_{\txt{beam}, NN}$ and the beam rapidity $y_{\txt{beam}}$ corresponding to different values of the center-of-mass energy $\sqrt{s_{NN}}$. Also listed are values of the longitudinal momentum difference $\Delta p_z \equiv p_{\txt{beam}, NN} - p_{z, \txt{final}}$ required for the final state proton to have a transverse momentum of $p_T = 0.9\ \txt{GeV}$ and a rapidity of $y_0$, where we consider $y_0 = \{ 1, 2, 3, 4\}$. Negative values of $\Delta p_z$ signify a longitudinal momentum loss ($y_0 < y_{\txt{beam}}$), while positive values correspond to a gain in longitudinal momentum ($y_0 > y_{\txt{beam}}$). In experiment, collision participants tend to lose momentum due to the initial energy deposition in the collision region and scattering, although momentum gains are also possible through either collective effects or as excited remnants of the sheared off nuclei emit particles.

From the results of our simple calculation, displayed in Table \ref{example_rapidities_given_pT}, we see that decreasing the rapidity of a proton moving with $p_{\txt{beam}, NN}$ at $\sqrt{s_{NN}} = 200.0 \ \txt{GeV}$ by about 2 units of rapidity requires a momentum loss of $65$--$85\ \txt{GeV}$. We can compare this value with the experiment, where an average energy of measured baryons can be calculated through 
\begin{eqnarray}
\langle E \rangle = \frac{1}{N_{\txt{part}}} \int_{-y_{\txt{beam}}}^{+ y_{\txt{beam}}} dy_r ~ \Big\langle E (y_r) \Big\rangle ~ \frac{d N_{B - \bar{B}}}{dy_r} ~.
\end{eqnarray}
The difference $\delta E \equiv E_{\txt{beam}} - \langle E \rangle$ is the average energy loss, and in $\sqrt{s_{NN}} = 200.0 \ \txt{GeV}$ Au+Au collisions at 0--5\% centrality it has been established to be equal $\Delta E = 73 \pm 6 \ \txt{GeV}$ \cite{BRAHMS:2003wwg}, which agrees with our basic estimate. Importantly, $\Delta E$ can be thought of as the energy deposited in the collision region, available for, e.g., particle production.

We stress that although the values of $\Delta p_z (y_0)$ listed in Table \ref{example_rapidities_given_pT} may aid in building an intuition about the kinematics of a heavy-ion collision, one should remember that the evolution of heavy-ion collision is much more complicated than the simple picture suggested here and includes, among others, formation of the fireball, deconfinement, evolution of a strongly-interacting QGP, and hadronization.

The energy dependence of net proton rapidity distributions (see Fig.\ \ref{BRAHMS_net-proton_dNdy} in Chapter \ref{introduction}) shows that although in very high-energy collisions a significant amount of longitudinal momentum is converted into energy available for particle production, the midrapidity region is mostly baryon-free. The opposite is true at the lowest energies, where almost all baryon number participating in the collision is found within one unit of rapidity around $y_r = 0$. Models explaining this behavior exist, however, the exact mechanism of transporting baryon number to smaller absolute values of rapidity, also known as ``baryon stopping'', is unknown and continues to be a subject of ongoing research.

\newpage
\chapter{Cumulants in terms of derivatives of the pressure with respect to the net baryon number density}
\label{cumulants_of_the_net_baryon_number}

We start by noting that, based on Eq.\ \eqref{cumulants_from_susceptibilities}, the cumulants satisfy 
\begin{eqnarray}
\kappa_{j+1} \equiv T (\infrac{d \kappa_j}{d\mu_B})_T ~.
\label{cumulants_recursive}
\end{eqnarray}
The first cumulant is given by
\begin{eqnarray}
\kappa_1 \equiv V \left(\frac{d P}{d\mu_B}\right)_T = Vn_B ~.
\end{eqnarray}
Following Eq.\ \eqref{cumulants_recursive}, the second cumulant is given by
\begin{eqnarray}
\kappa_2 \equiv VT \left( \derr{n_B}{\mu_B}\right)_T~.
\label{k2_eq1}
\end{eqnarray}
From $dP = T ds + n_B d\mu_B$ we know that
\begin{eqnarray}
\left(\derr{\mu_B}{n_B} \right)_T = \frac{1}{n_B} \left(\derr{P}{n_B}\right)_T~,
\label{dmuB_dnB}
\end{eqnarray} 
so that Eq.\ \eqref{k2_eq1} can be rewritten as
\begin{eqnarray}
\kappa_2 = \frac{VT n_B}{\left( \derr{P}{n_B}\right)_T } ~.
\label{k2}
\end{eqnarray}
Similarly, the third cumulant is given by
\begin{eqnarray}
\kappa_{3} \equiv  T \left(\derr{\kappa_2}{\mu_B}\right)_T &=& VT^2 ~\left(\frac{dn_B}{d\mu_B}\right)_T \frac{d}{dn_B}\bigg|_T \left[\frac{n_B}{\left(\frac{dP}{dn_B} \right)_T} \right] \non \\
&=&  \frac{VT^2  n_B}{\left(\frac{dP}{dn_B}  \right)^2}  \left[ 1 - \frac{n_B}{\left(\frac{dP}{dn_B} \right)_T} \left(\frac{d^2P}{dn_B^2} \right)_T \right] ~.
\end{eqnarray}
where in the second equality we have used Eq.\ \eqref{k2} and in the third equality we again utilized Eq.\ \eqref{dmuB_dnB}. Finally, the fourth cumulant can be obtained in the same way, yielding
\begin{eqnarray}
\hspace{-10mm}\kappa_4 &\equiv& T \left(\frac{d \kappa_3}{d\mu_B} \right)_T =  T \left( \derr{n_B}{\mu_B}\right)_T \left(\frac{d\kappa_3}{dn_B} \right)_T\non \hspace{80mm} \\
&=&   \frac{VT^3  n_B}{\left(\frac{dP}{dn_B}  \right)_T^3}  \left[  1    -   \frac{4n_B}{\left(\frac{dP}{dn_B}  \right)_T} \left(\frac{d^2P}{dn_B^2}\right)_T     +       \frac{3n_B^2}{\left(\frac{dP}{dn_B}  \right)_T^2}   \left(\frac{d^2P}{dn_B^2}\right)_T^2    -   \frac{n_B^2}{\left(\frac{dP}{dn_B}  \right)_T} \left(\frac{d^3P}{dn_B^3}\right)_T  \right]~.
\label{4th_cumulant_in_pressure}
\end{eqnarray}

\newpage
\chapter{The static limit of the Walecka model}
\label{static_limit_of_the_Walecka_model}

This appendix is based on Ref.\ \cite{Walecka:1995mi} .

\section{A simple case of a scalar interaction field}

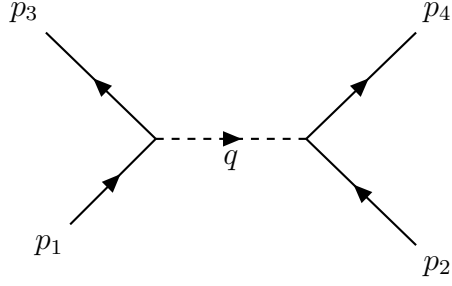
\begin{figure}[t]
	\begin{center}
		\begin{tikzpicture}
		\begin{feynman}[large]
		\vertex (a) {\(p_1\)};
		\vertex [above right=of a] (b) ;
		\vertex [above left=of b] (f1) {\(p_3\)};
		\vertex [right=of b] (c);
		\vertex [above right=of c] (f2) {\(p_4\)};
		\vertex [below right=of c] (f3) {\(p_2\)};
		
		\diagram* {
			(a) -- [fermion] (b) [dot]  -- [fermion] (f1),
			(b) -- [charged scalar, edge label'=\(q\)] (c),
			(f3) -- [fermion] (c) [dot] --  [fermion] (f2),
		};
		\end{feynman}
		\end{tikzpicture}
	\end{center}
	\caption[A diagram of two-nucleon scattering occurring through an exchange of a scalar interaction boson]{A diagram of two-nucleon scattering occurring through an exchange of a scalar interaction boson.}
	\label{2bodscattering}
\end{figure}

Let us consider a simple 2-nucleon scattering as shown in Fig.\ \ref{2bodscattering} (note that by considering only this diagram we neglect the exchange term, which is equivalent to a simplifying assumption that the nucleons taking part in the interaction are distinguishable). We can explicitly write the conditions for energy and momentum conservation (throughout the entire scattering event and at each vertex) as
\begin{eqnarray}
&& \bm{p}_1 + \bm{p}_2 = \bm{p}_3 + \bm{p}_4 ~, \\
&& E_1 + E_2 = E_3 + E_4~, \\
&& \bm{p}_3 = \bm{p}_1 - \bm{q}~, \\
&& \bm{p}_4 = \bm{p}_2 + \bm{q}~, \\
&& E_3 = E_1 - q_0~, \\
&& E_4 = E_2 + q_0~. 
\end{eqnarray}
Then we easily see that we can write, for example,
\begin{eqnarray}
q_0 = E_4 - E_2 = \sqrt{\bm{p}_4^2 + M^2 } - \sqrt{\bm{p}_2^2  + M^2 } ~.
\end{eqnarray}
If the nucleon mass $M$ is very large (or, conversely, the nucleons are not too fast), which is known as the limit of static sources, then we can approximate
\begin{eqnarray}
q_0 = M \left(\sqrt{1 + \frac{\bm{p}_4^2}{M^2}} - \sqrt{1 + \frac{\bm{p}_2^2}{M^2 }} \right) \approx  \frac{\bm{p}_4^2 - \bm{p}_2^2}{2M }~.
\label{static_sources}
\end{eqnarray}

Now, let us assume that the scattering is described by the interaction Lagrangian density
\begin{eqnarray}
\mathcal{L}_I = g \bar{\psi} \psi \phi
\end{eqnarray}
where $\phi$ denotes a scalar interaction field and $g$ is the coupling constant. The propagator of the scalar interaction field quanta is given by
\begin{eqnarray}
D(q ) = \frac{i}{q^2 - m_s^2}~,
\end{eqnarray} 
where $q$ is the momentum transfer and $m_s$ denotes the mass of the field quanta, so that the lowest order scattering matrix element in the momentum space is given by
\begin{eqnarray}
S_{fi} = ( -ig)^2~ (2\pi)^4 \delta^4 \big(k_1 + k_2 - k_3 -k_4\big) ~ \frac{1}{\Omega^2} ~ \frac{i}{q^2 + m_f^2} ~\bar{u}_a(k_3) u_a(k_1) \bar{u}_b(k_4) u_b(k_2)~,
\label{scattering_matrix_0}
\end{eqnarray}
where $\Omega$ is the quantization volume. Using the limit of static sources, Eq.\ \eqref{static_sources}, we get
\begin{eqnarray}
D(q ) = \frac{i}{q_0^2 - \vec{q}^2 - m_s^2} \approx \frac{i}{ \left(\frac{\bm{p}_4^2 - \bm{p}_2^2}{2M }\right)^2 - \bm{q}^2 - m_s^2 } \approx -  \frac{i}{ |\bm{q}|^2 + m_s^2 }~,
\end{eqnarray}
so that Eq.\ (\ref{scattering_matrix_0}) becomes
\begin{eqnarray}
S_{fi} =  (2\pi)^4 \delta^4 \big(k_1 + k_2 - k_3 -k_4\big) ~ \frac{1}{\Omega^2} ~   \frac{ ig^2}{ |\bm{q}|^2 + m_s^2 } ~\bar{u}_a(k_3) u_a(k_1) \bar{u}_b(k_4) u_b(k_2)~.
\label{scattering_matrix_1}
\end{eqnarray}

On the other hand, to the lowest order in nonrelativistic potential scattering we should have
\begin{eqnarray}
S_{fi}^{(1)} = - i\int d^4x ~ \mathcal{H}_I(x) ~.
\end{eqnarray}
In the formalism of the second quantization this can be rewritten as
\begin{eqnarray}
S_{fi}^{(1)} = - i \int d^4x \int d^4y ~ \psi^{\dagger}_a (x)  \psi^{\dagger}_b (y) ~ V_{\txt{eff}} (y- x) ~ \psi_b(y) \psi_a(x)~.
\end{eqnarray}
We can always express $\psi$ in terms of its Fourier components
\begin{eqnarray}
\psi = \frac{1}{\sqrt{\Omega}} \sum_{\lambda} \sum_{\bm{k}} a_{\lambda \bm{k}} \chi_{\lambda} e^{- k\cdot x }~,
\end{eqnarray}
which leads to
\begin{eqnarray}
S_{fi}^{(1)} = - \frac{i}{ \Omega^2} (2\pi)^4 \delta^4 (k_1 + k_2 - k_3 - k_4) ~\widetilde{V}_{\txt{eff}} (\bm{q})~,
\label{scattering_matrix_2}
\end{eqnarray}
where $\bm{q} = \bm{k}_1 - \bm{k}_3 = \bm{k}_4 - \bm{k}_2$. Upon comparing Eqs.\ (\ref{scattering_matrix_1}) and (\ref{scattering_matrix_2}) we see that we must have
\begin{eqnarray}
\widetilde{V}_{\txt{eff}} (\bm{q}) = -    \frac{ g^2}{ |\bm{q}|^2 + m_s^2 }   ~.   
\end{eqnarray}
To recover the potential in the position space we need to Fourier transform $\widetilde{V}_{\txt{eff}}(q)$,
\begin{eqnarray}
V_{\txt{eff}}(\bm{r} )  &=& \int \frac{d^3\bm{q}}{(2\pi)^3} ~\widetilde{V}_{\txt{eff}}(q)  e^{i \bm{q} \cdot \bm{r}} \non \\
&=& \frac{1}{4 \pi^2} \int_0^{\infty} dq ~ q^2 \int_{-1}^{+1} d(\cos \theta)   ~\widetilde{V}_{\txt{eff}}(q )  e^{i q r \cos\theta } \non \\
&=&   - \frac{g^2}{4 \pi^2~ i r} \int_0^{\infty} dq ~ q ~\frac{e^{i qr } - e^{- i qr }}{ q^2 + m_s^2}  \non \\
&=&   - \frac{g^2}{4 \pi^2~ i r} \int_{-\infty}^{\infty} dq ~ q ~\frac{e^{i qr } }{ q^2 + m_s^2} ~. 
\end{eqnarray}
The last integral has to be done by means of complex integration. We can always rewrite
\begin{eqnarray}
\frac{q e^{i qr } }{ q^2 + m_s^2}   =  \frac{e^{i qr } }{ 2} \left( \frac{1}{q + i m_s}+  \frac{1}{q - i m_s} \right) ~,
\end{eqnarray}
from which we immediately see that closing the contour in the upper half plane encloses the $q =  i m_s$ pole. Since now the integral is just $2\pi i$ times the residue of the pole with an overall plus sign as the contour is traversed counterclockwise, we have
\begin{eqnarray}
V_{\txt{eff}}(\bm{r} )  = - \frac{g^2}{4 \pi}   ~  \frac{e^{- m_s r}}{r}~,
\end{eqnarray}
which is an attractive Yukawa potential.

\section{The Walecka model}

In the Walecka model the interaction Lagrangian density is of the form
\begin{eqnarray}
\mathcal{L}_I = ig_\omega \hat{\psi} \gamma_{\lambda} \psi \omega^{\lambda} + g_\sigma \bar{\psi} \psi \sigma~,
\end{eqnarray}
which leads to the momentum-space potential
\begin{eqnarray}
V_W(q) = g_\omega ^2 \frac{\gamma_{\lambda}^{(1)} \cdot \gamma_{\lambda}^{(2)} }{q^2 - m_{\omega}^2 - i \epsilon} - g_\sigma^2\frac{\unity^{(1)} \cdot \unity^{(2)}}{q^2- m_{\sigma}^2 - i \epsilon}~,
\end{eqnarray}
where $q$ is the four-momentum transfer and $\gamma_{\lambda}^{(1,2)}$ refer to the first and second particle, respectively. If we assume that the baryons are nonrelativistic (which is equivalent to assuming that they are heavy), we can approximate
\begin{eqnarray}
\gamma_{\lambda}^{(1)} \cdot \gamma_{\lambda}^{(2)} \hfp \to \hfp \unity^{(1)} \cdot \unity^{(2)}
\end{eqnarray}
and, just as before,
\begin{eqnarray}
|q_0| \ll |\bm{q}|~,
\end{eqnarray}
in which case the potential becomes instantaneous and spin-independent. Following similar steps as before we can be write $V_W(q) $ in the coordinate space as
\begin{eqnarray}
V(r) = \frac{1}{4\pi} \left( g_\omega^2 \frac{e^{- mr}}{r} - g_\sigma^2 \frac{e^{- m_{\sigma} r}}{r} \right)~.
\end{eqnarray}
If $g_\omega^2 > g_\sigma^2$, this potential is repulsive at short distances. If additionally $m_{\omega} > m_{\sigma}$, the potential will be attractive at large distances. In result, such an interaction will display all main features of the nucleon-nucleon interaction, and indeed we can see that its form is just like that of a Lennard-Jones--type potential.

\newpage
\chapter{Dirac field with vector and scalar interactions}
\label{Dirac_field_with_vector_and_scalar_interactions}

\section{The Dirac equation}

In general, the Dirac equation for a fermion interacting with a classical vector field $A^{\mu}$ and a classical scalar field $S$ is given by
\begin{eqnarray}
\Big(  i \gamma^{\mu} \partial_{\mu}   -  \gamma^{\mu}A_{\mu} - m  + S \Big) \psi(x) = 0 ~;
\label{Dirac_equation_with_general}
\end{eqnarray}
the above equation can be brought to the form displayed in Eq.\ \eqref{nucleon_field_eq_2} by making the fields $A_{\mu}$ and $S$ appropriately dependent on mean fields. Multiplying Eq.\ \eqref{Dirac_equation_with_general} by $\gamma^0$ from the left and writing out the terms leads to
\begin{eqnarray}
i  \partial_{0}\psi(x)  = \Big( - i \gamma^{0}\bm{\gamma} \cdot \bm{\nabla}     -  \gamma^{0}\bm{\gamma} \cdot \bm{A} +  A_{0} + \gamma^{0}(m  - S) \Big) \psi(x)  ~.
\label{EOM_of_Dirac_field_start}
\end{eqnarray}
We know that the momentum operator is $\bhat{\mathcal{P}} \equiv - i \bm{\nabla}$, so that furthermore we have
\begin{eqnarray}
&&\hspace{-13mm} i  \partial_{0}\psi(x)  = \Big(  \gamma^{0}\bm{\gamma} \cdot \bhat{\mathcal{P}}     -   \gamma^{0}\bm{\gamma} \cdot \bm{A} +  A_{0} + \gamma^{0}(m - S)  \Big) \psi(x)   \\
&& \hspace{-10mm} = \Big( \left[ \begin{array}{cc}
0 & \bm{\sigma} \cdot \bhat{\mathcal{P}} \\
\bm{\sigma} \cdot \bhat{\mathcal{P}} & 0  
\end{array} \right]      -   \left[ \begin{array}{cc}
0 & \bm{\sigma} \cdot \bm{A} \\
\bm{\sigma} \cdot \bm{A} & 0  
\end{array} \right] + A_{0} \unity_{4} + \left[ \begin{array}{cc}
\unity_2 & 0 \\
0 & - \unity_2  
\end{array} \right](m- S)  \Big) \psi(x)  ~.
\end{eqnarray}
where in the second line we have used the explicit form of the gamma matrices in the Dirac representation,
\begin{eqnarray}
\gamma^0 =  \left[ \begin{array}{cc}
\unity & 0 \\
0& - \unity  
\end{array} \right] ~,  \hffp  
\gamma^i =  \left[ \begin{array}{cc}
0 & \sigma_i \\
-\sigma_i& 0   
\end{array} \right] ~,  
\end{eqnarray}
where $\sigma_1, \sigma_2, \sigma_3$ are Pauli matrices. Taking the wavefunction $\psi$ to be a stationary solution and expressing it in the bispinor form,
\begin{eqnarray}
\psi = e^{- i Et} \left( \begin{array}{c}
\varphi \\
\chi
\end{array} \right)~,
\end{eqnarray} 
leads to two coupled equations,
\begin{eqnarray}
&& E\varphi = \left[\big(  \bm{\sigma} \cdot \bhat{\mathcal{P}}\big)  -  \Big( \bm{\sigma}  \cdot \bm{A}(x) \Big)\right] \chi + \Big[A_0 + (m - S) \Big] \varphi~,\\
&& E\chi = \left[\big(  \bm{\sigma}  \cdot \bhat{\mathcal{P}}\big)  -  \Big( \bm{\sigma} \cdot \bm{A}(x) \Big)\right] \varphi + \Big[A_0 - (m - S) \Big] \chi~,
\end{eqnarray}
where we have made the dependence of the $\bm{A}$ field on the position explicit. The second of the above equations can be solved for $\chi$ in terms of $\varphi$, 
\begin{eqnarray}
\chi = \frac{\left[ \bm{\sigma}  \cdot \Big( \bhat{\mathcal{P}} - \bm{A}(x)\Big) \right]}{E - \Big[A_0 - (m - S) \Big] }~ \varphi ~,
\end{eqnarray}
which inserted into the first equation results in
\begin{eqnarray}
E\varphi = \left( \frac{\left[\bm{\sigma} \cdot \Big( \bhat{\mathcal{P}} - \bm{A}(x)\Big) \right]^2}{E - \Big[A_0 - (m - S) \Big] }    + \Big[A_0 + (m - S) \Big] \right) \varphi ~.
\label{plane_wave_solution_OK_eq_0}
\end{eqnarray}
Denoting $\bm{\Pi} = \bhat{\mathcal{P}} - \bm{A}(x)$ we further arrive at
\begin{eqnarray}
\bigg( \big[E - A_0\big]^2 - (m - S)^2 \bigg)\varphi = \big[  \bm{\sigma}  \cdot  \bm{\Pi} \big]^2  \varphi  ~.
\label{plane_wave_solution_OK_eq_1}
\end{eqnarray}

\section{Evaluating $\big[ \sigma  \cdot  \Pi \big]^2 $}

In order to evaluate the right-hand side of Eq.\ \eqref{plane_wave_solution_OK_eq_1} we need to remember that $\bhat{\mathcal{P}}$ does not commute with any function of the coordinate $x$, which in particular includes $\bm{A}(x)$. Specifically, we have
\begin{eqnarray}
\Big[ F(\bm{x}) , \hat{\mathcal{P}}_k \Big] = i \parr{F(\bm{x})}{x_k} 
\label{p_function_of_x_commutation}
\end{eqnarray}
(where we note that here and generally in the following part of this appendix, the subscripts do not denote covariant quantities, and in particular in Eq.\ \eqref{p_function_of_x_commutation} the subscripts $k$ both refer to components of contravariant vectors). This means that we need to perform the square in Eq.\ (\ref{plane_wave_solution_OK_eq_1}) carefully, as $\big[\Pi_i, \Pi_j \big]  \neq 0$.

We can write out the square in Eq.\ \eqref{plane_wave_solution_OK_eq_1} as
\begin{eqnarray}
\big[  \sigma \cdot \bm{\Pi} \big]^2 = \sum_i \sum_j \sigma_i \sigma_j \Pi_i \Pi_j ~.
\label{plane_wave_solution_OK_eq_11}
\end{eqnarray}
Because the Pauli matrices satisfy the anticommutation relation $\{\sigma_i , \sigma_j \} = 2\delta_{ij}$, we can write
\begin{eqnarray}
 \sigma_i \sigma_j   = 2\delta_{ij} - \sigma_j \sigma_i ~,
\end{eqnarray}
which used in Eq.\ \eqref{plane_wave_solution_OK_eq_11} along with $\Pi_i \Pi_j = \big[\Pi_i, \Pi_j \big] + \Pi_j \Pi_i$ yields
\begin{eqnarray}
\sum_i \sum_j \sigma_i \sigma_j \Pi_i \Pi_j  &=& \sum_i \sum_j 2\delta_{ij} \big[\Pi_i, \Pi_j \big] + 2   \sum_i \sum_j\delta_{ij} \Pi_j \Pi_i  \non \\
&& \hspace{10mm}  - ~  \sum_i \sum_j\sigma_j \sigma_i\big[\Pi_i, \Pi_j \big] -  \sum_i \sum_j\sigma_j \sigma_i \Pi_j \Pi_i  ~.
\end{eqnarray}
The first term on the right-hand side is identically zero, the second term is simply equal $2\bm{\Pi}^2$, and the fourth term, after a change of indeces according to $i \to j$, $j \to i$, is equal in magnitude to the left-hand side of the equation, so that altogether we get
\begin{eqnarray}
\sum_i \sum_j \sigma_i \sigma_j \Pi_i \Pi_j  &=&    \bm{\Pi}^2   -  \frac{1}{2} \sum_i \sum_j\sigma_j \sigma_i\big[\Pi_i, \Pi_j \big]  ~.
\end{eqnarray}
We can always rewrite the last term according to 
\begin{eqnarray}
&&\hspace{-10mm} -  \frac{1}{2} \sum_i \sum_j\sigma_j \sigma_i\big[\Pi_i, \Pi_j \big]  = -  \frac{1}{4} \sum_i \sum_j\sigma_j \sigma_i\big[\Pi_i, \Pi_j \big]  -  \frac{1}{4} \underset{i \to j, ~  j\to i} {\sum_i \sum_j\sigma_j \sigma_i\big[\Pi_i, \Pi_j \big] } \non \\
&& \hspace{10mm} = -  \frac{1}{4} \sum_i \sum_j\sigma_j \sigma_i\big[\Pi_i, \Pi_j \big]  -  \frac{1}{4}\sum_i \sum_j \sigma_i \sigma_j \big[\Pi_j, \Pi_i \big]  \non \\
&& \hspace{10mm} = -  \frac{1}{4} \sum_i \sum_j\sigma_j \sigma_i\big[\Pi_i, \Pi_j \big]  +  \frac{1}{4}\sum_i \sum_j \sigma_i \sigma_j \big[\Pi_i, \Pi_j \big] \non \\
&& \hspace{10mm} = \frac{1}{4}\sum_i \sum_j \big[ \sigma_i ,\sigma_j \big]\big[\Pi_i, \Pi_j \big]  ~.
\end{eqnarray}
We know that $\big[\sigma_i, \sigma_j \big] = 2i \epsilon_{ijk} \sigma_k$, and we can explicitly calculate
\begin{eqnarray}
\hspace{-3mm}\big[\Pi_i, \Pi_j \big]  &=& \Big[\hat{\mathcal{P}}_i  - A_i(x), \hat{\mathcal{P}}_j  -  A_j \Big]  = \big[\hat{\mathcal{P}}_i, \hat{\mathcal{P}} \big] - \big[\hat{\mathcal{P}}_i, A_j \big] - \big[A_i, \hat{\mathcal{P}}_j \big] + ^2 \big[A_i, A_j \big]  \\
&=&   \left(i \parr{A_j}{x_i} - i \parr{A_i}{x_j} \right)~,
\end{eqnarray}
where Eq.\ (\ref{p_function_of_x_commutation}) has been used. We thus arrive at
\begin{eqnarray}
\left[  \sigma \cdot \bm{\Pi} \right]^2 = \sum_i \sum_j \sigma_i \sigma_j \Pi_i \Pi_j  =  \bm{\Pi}^2 + \frac{1}{2}\sum_i \sum_j  \epsilon_{ijk} \sigma_k   \left(\parr{A_i}{x_j}  - \parr{A_j}{x_i}   \right)  ~.
\end{eqnarray}
Finally, we perform the summation explicitly. Writing out all non-zero terms we arrive at
\begin{eqnarray}
&& \hspace{-10mm}\sum_i \sum_j  \epsilon_{ijk} \sigma_k   \left(\parr{A_i}{x_j}  - \parr{A_j}{x_i}   \right)    \\
&& \hspace{-3mm}  = 2 \left[ \epsilon_{12k} \sigma_k  \left( \parr{A_1}{x_2} - \parr{A_2}{x_1} \right) + \epsilon_{23k} \sigma_k \left( \parr{A_2}{x_3} - \parr{A_3}{x_2} \right) + \epsilon_{13k} \left( \parr{A_1}{x_3} - \parr{A_3}{x_1} \right) \right]   ~.
\end{eqnarray}
At this point, we define a ``magnetic'' field $\bm{B}$ such that
\begin{eqnarray}
\bm{B} \equiv \bm{\nabla} \times \bm{A} = \bhat{x}_1 \left( \parr{A_{3}}{x_2} - \parr{A_2}{x_3} \right) + \bhat{x}_2 \left( \parr{A_1}{x_3} - \parr{A_3}{x_1}  \right) + \bhat{x}_3 \left( \parr{A_2}{x_1} - \parr{A_1}{x_2} \right)~. 
\end{eqnarray}
Clearly then,
\begin{eqnarray}
\sum_i \sum_j  \epsilon_{ijk} \sigma_k   \left(\parr{A_i}{x_j}  - \parr{A_j}{x_i}   \right)   = 2 \Big[- \epsilon_{12k} \sigma_k  B_3 - \epsilon_{23k} \sigma_k B_1 + \epsilon_{13k} B_2 \Big]   ~.
\end{eqnarray}
Inserting the only values of $k$ that result in a non-zero Levi-Civita tensor we obtain
\begin{eqnarray}
\sum_i \sum_j  \epsilon_{ijk} \sigma_k   \left(\parr{A_i}{x_j}  - \parr{A_j}{x_i}   \right)  = - 2 \bm{\sigma} \cdot \bm{B} ~,
\end{eqnarray}
and in this way we ultimately arrive at
\begin{eqnarray}
\left[  \sigma \cdot \bm{\Pi} \right]^2 =  \sum_i \sum_j \sigma_i \sigma_j \Pi_i \Pi_j  =  \bm{\Pi}^2 -  \bm{\sigma} \cdot \bm{B}  ~.
\label{sigma_dot_pi}
\end{eqnarray}

\section{The Pauli equation}
\label{the_Pauli_equation}

Using Eq.\ \eqref{sigma_dot_pi}, Eq.\ \eqref{plane_wave_solution_OK_eq_0} becomes
\begin{eqnarray}
E\varphi = \left( \frac{ \bm{\Pi}^2 -  \bm{\sigma} \cdot \bm{B}  }{E - A_0 + (m - S)  }    + \Big[A_0 + (m - S) \Big] \right) \varphi ~.
\label{large_componet_EOM_2}
\end{eqnarray}
Let us for a moment neglect the scalar field by putting $S=0$, and let us consider the non-relativistic limit, in which the total energy $E = m + \epsilon_{\txt{kin}}$, where $ \epsilon_{\txt{kin}}$ is the kinetic energy of the particle such that $ \epsilon_{\txt{kin}} \ll m$. Then
\begin{eqnarray}
(E - m)\varphi = \left( \frac{ \bm{\Pi}^2 -  \bm{\sigma} \cdot \bm{B}  }{2m + \epsilon_{\txt{kin}} - A_0   }    + A_0  \right) \varphi ~.
\end{eqnarray}
Since both $A_0 \ll m$ and $\epsilon_{\txt{kin}} \ll m$, we can neglect these small terms in the denominator, leading to
\begin{eqnarray}
\epsilon_{\txt{kin}}\varphi = \left( \frac{ \bm{\Pi}^2 -  \bm{\sigma} \cdot \bm{B}  }{2m   }    + A_0  \right) \varphi ~.
\end{eqnarray}
By realizing that $\epsilon_{\txt{kin}} \varphi = i \partial_t \varphi$ we arrive at the Pauli equation, which describes a spin-$\frac{1}{2}$ particle in an external electromagnetic field,
\begin{eqnarray}
i \partial_t \varphi = \left( \frac{ \big(\bm{p} - \bm{A}\big)^2  }{2m   }  - \frac{  \bm{\sigma} \cdot \bm{B}  }{2m   }   + A_0  \right) \varphi ~.
\label{Pauli_equation}
\end{eqnarray}
(Note that in our notation the coupling constant is absorbed into the field $A^{\mu}$.)

\section{Energy solutions in uniform nuclear matter}

Similarly, using Eq.\ \eqref{sigma_dot_pi} in Eq.\ \eqref{plane_wave_solution_OK_eq_1} results in
\begin{eqnarray}
\big[E - A_0\big]^2\varphi   = \bigg[ \Big(  \bm{\Pi}^2 - e \bm{\sigma} \cdot \bm{B} \Big)  +  (m - S)^2 \bigg]\varphi ~.
\label{plane_wave_solution_OK_eq_5}
\end{eqnarray}
In nuclear matter, the potential field $A^{\mu}(x)$ is spatially uniform, which in particular means that $\bm{A} = 0$, so that $\bm{B} = 0$ . With this, Eq.\ (\ref{plane_wave_solution_OK_eq_5}) becomes
\begin{eqnarray}
\big[E - A_0\big]^2\varphi   = \Big[ \bm{\Pi}^2  +  (m - S)^2 \Big]\varphi  ~,
\end{eqnarray}
and we immediately obtain
\begin{eqnarray}
E = \pm \sqrt{ \bm{\Pi}^2  + (m-S)^2} + A_0~.
\label{energy_solutions}
\end{eqnarray}
Note that because the matter in uniform, we in fact have $\bm{\Pi} = \bm{p}$.

\newpage
\chapter{Form of the quasiparticle distribution function}
\label{form_of_the_quasiparticle_distribution_function}

The functional form of the quasiparticle distribution function $f_{\bm{p}}$ can be obtained using fundamental thermodynamic relations. Any variation in the energy density is connected to a variation in entropy density $s$ and particle density $n$ through the fundamental equation of thermodynamics,
\begin{eqnarray}
\delta \mathcal{E} = T ~\delta s + \mu ~\delta n~.
\label{energy_variation}
\end{eqnarray}
It follows from Eq.\ (\ref{Landau_quasiparticle_energy_definition}) that $\delta \mathcal{E}\mathcal \ \equiv \  \eps_{\bm{p}} ~ \delta f_{\bm{p}}$. The dependence of $\delta s$ on $f_{\bm{p}}$ can be easily established by remembering that it is possible to calculate the entropy of a given state by combinatorial considerations only. In view of the one-to-one correspondence between the states of the Fermi liquid and the free Fermi gas, it is natural to assume that the entropy density must have the same form as in the case of the free Fermi gas,
\begin{eqnarray}
s = - \sum_{\bm{p}} \Big[ f_{\bm{p}} \ln f_{\bm{p}} + (1 - f_{\bm{p}}) \ln (1 - f_{\bm{p}})  \Big]~. 
\end{eqnarray}
Consequently,
\begin{eqnarray}
\delta s = -\sum_{\bm{p}} \left[\delta f_{\bm{p}}  \ln \frac{f_{\bm{p} }}{1 - f_{\bm{p}}}  \right]~.
\end{eqnarray}
Similarly, because the number of quasiparticles in the interacting system directly corresponds to the number of particles in the corresponding state of the free Fermi gas, and the interaction between the particles is assumed to conserve the particle number, the total number of particles in a state of the interacting system must be the same as in the non-interacting system. In consequence, the quasiparticle density can be expressed using the quasiparticle distribution function in the familiar way,
\begin{eqnarray}
n = \sum_{\bm{p}} f_{\bm{p}} ~,
\label{number_density_from_distribution_function}
\end{eqnarray}
from which it follows that $\delta n =  \sum_{\bm{p}} \delta f_{\bm{p}}$. With all this, Eq.\ (\ref{energy_variation}) can be rewritten as
\begin{eqnarray}
\sum_{\bm{p}} \eps_{\bm{p}} ~\delta f_{\bm{p}} &=& - T \sum_{\bm{p}} \ln \frac{f_{\bm{p} }}{1 - f_{\bm{p}}} ~\delta f_{\bm{p}} + \mu \sum_{\bm{p}} \delta f_{\bm{p}} ~,
\end{eqnarray}
which can be further rearranged as
\begin{eqnarray}
\sum_{\bm{p}} \left[ \eps_{\bm{p}} + T \ln \frac{f_{\bm{p} }}{1 - f_{\bm{p}}}  - \mu \right]~ \delta f_{\bm{p}} = 0 ~.
\end{eqnarray}
The above equality will hold for any variation $\delta f_{\bm{p}}$ if and only if the term in the square bracket vanishes for any $\bm{p}$, and this fact can be used to solve for the quasiparticle distribution function,
\begin{eqnarray}
f_{\bm{p}} = \frac{1}{\exp\left( \frac{\eps_{\bm{p}} - \mu }{T} \right) + 1} ~,
\end{eqnarray}
which turns out to be of the Fermi-Dirac form. This naturally follows one's basic expectations, as the Landau Fermi-liquid theory is based on an assumption that the states of the Fermi liquid can be obtained as a result of a continuous deformation of the states of the ideal Fermi gas, and it is natural that they would be governed by a distribution function of the same form. We stress, however, that because the quasiparticle energy $\eps_{\bm{p}}$ itself depends on the quasiparticle distribution $f_{\bm{p}}$, the above equation is in fact a rather complicated self-consistent equation for $f_{\bm{p}}$, in contrast to the free Fermi gas case.

\newpage
\chapter{Vanishing of the collision integral}
\label{vanishing_of_the_collision_integral}

Vanishing of the right-hand side of Eq.\ (\ref{Boltzmann_quasiparticle_energy}), multiplied by $X$ and integrated over all momenta (the available phase space), where $X = \{ 1, \eps_{\bm{p}}, p^j \}$ (corresponding to laws of the conservation of particle number, energy, and momentum, respectively), can be shown as follows. Let us first consider $X=1$, and let us take the discrete limit, in which the integration over all momenta becomes a sum over all particles,
\begin{eqnarray}
g \int \pdens  \left(\derr{f_{\bm{p}}}{t} \right)_{\txt{coll}}   \hspace{5mm} \to \hspace{5mm} \sum_i \left(\derr{f_i}{t} \right)_{\txt{coll}} ~;
\end{eqnarray}
here the index $i$ numbers the momenta states. In the discrete limit, the Fermi quasiparticle distribution function is just a list of occupation numbers $\{0, 1\}$ for states indexed by $i$, and so $\big(\derr{f_i}{t} \big)_{\txt{coll}}$ is the change in the occupation numbers due to collisions. Consider an event in which two quasiparticles in states $i = 1$ and $i = 2$ collide and as a result change their states to $i = 3$ and $i = 4$. We have a situation in which before the collision, the states $i = \{1, 2\}$ were occupied and the states $i' = \{3,4\}$ were unoccupied, while the opposite is true after the collision. Therefore we can see that
\begin{eqnarray}
\left(\derr{f_i}{t} \right)_{\txt{coll}} \bigg|_{i = \{1,2\}} = -1 \distand \left(\derr{f_i}{t} \right)_{\txt{coll}}\bigg|_{i = \{3,4\}}  = +1  ~.
\label{occupation_changes}
\end{eqnarray}
Thus in this event (let us assume that only the states indexed by $i=\{1,2,3,4\}$ have been changed, so that contributions from all other states are zero, $\big(\derr{f_i}{t} \big)_{\txt{coll}} \big|_{i >4} = 0$)
\begin{eqnarray}
\sum_{i=1}^4  \left(\derr{f_i}{t} \right)_{\txt{coll}} = - 1- 1+ 1+ 1 = 0~.
\label{conservation_of_number_boltzmann}
\end{eqnarray}
In the continuous limit, this corresponds to $ g \int \pdens \big(\derr{f_{\bm{p}}}{t} \big)_{\txt{coll}} = 0$.

For the case $X = \eps_{\bm{p}}$, it is again convenient to use the discrete limit. Consider an event in which two quasiparticles with energies $\eps_1$ and $\eps_2$ collide and as a result change their energies to $\eps_3$ and $\eps_4$,
\begin{eqnarray}
\eps_1  + \eps_2  \to \eps_3 + \eps_4 ~. 
\end{eqnarray}
Before the collision, the states $i = \{1, 2\}$ were occupied and the states $i' = \{3,4\}$ were unoccupied, while the opposite is true after the collision, as in Eq.\ (\ref{occupation_changes}). Thus in this event (let us again assume that only the states indexed by $i=\{1,2,3,4\}$ have been changed, $\big(\parr{f_i}{t} \big)_{\txt{coll}} \big|_{i >4} = 0$)
\begin{eqnarray}
\sum_{i=1}^4 \eps_i \left(\parr{f_i}{t} \right)_{\txt{coll}} = - \eps_1 - \eps_2 + \eps_3 + \eps_4 ~.
\label{conservation_of_energy_boltzmann}
\end{eqnarray}
By conservation of energy, the right-hand side of Eq.\ (\ref{conservation_of_energy_boltzmann}) must be zero, and in the continuous limit this corresponds to $ g \int \pdens \eps_{\bm{p}} \big(\parr{f_{\bm{p}}}{t} \big)_{\txt{coll}} = 0$.

The case for $X = p^j$ can be shown analogously.

\newpage
\chapter{VSDF model derivations}
\label{model_derivations}

\section{The quasiparticle energy}
\label{the_quasiparticle_energy}

To calculate the quasiparticle energy, we calculate a functional differential of the energy density,
\begin{eqnarray}
\eps_{\bm{p}}^* \equiv \frac{\delta \mathcal{E}}{\delta f_{\bm{p}}}~.
\end{eqnarray}
Taking the energy density to be given by $\mathcal{E}_{v1s1}$, Eq.\  (\ref{energy_density_one_scalar_one_vector_term}), and varying the first term, corresponding to the kinetic energy of the particles in the system, yields
\begin{eqnarray}
\delta \mathcal{E}_{v1s1,\txt{kin}} &=& \delta \int \ptilde  \epsilon_{\txt{kin}}^* ~ f_{\bm{p}}  = \int \ptilde  \delta \epsilon_{\txt{kin}}^* ~ f_{\bm{p}}  +  \int \ptilde  \epsilon_{\txt{kin}}^* ~ \delta f_{\bm{p}}  ~,
\end{eqnarray}
where we take into account that the kinetic energy $\epsilon_{\txt{kin}}$, Eq.\ (\ref{kinetic_energy_definition}), is also a functional of the quasiparticle distribution function through its dependence on the baryon current and effective mass. Explicitly, 
\begin{eqnarray}
\hspace{-5mm} \delta \epsilon_{\txt{kin}}^* &=& \delta \Bigg[ \sqrt{ \Big( \bm{p} - C_1 \big(j_{\mu} j^{\mu} \big)^{\alpha_1 - 1} \bm{j} \Big)^2 + {m^*}^2  } \Bigg] = \frac{\delta \left[  \Big( \bm{p} - C_1 \big(j_{\mu} j^{\mu} \big)^{\alpha_1 - 1} \bm{j} \Big)^2 + {m^*}^2  \right]}{2 \sqrt{ \Big( \bm{p} - C_1 \big(j_{\mu} j^{\mu} \big)^{\alpha_1 - 1} \bm{j} \Big)^2 + {m^*}^2  } }  \non \\
&=& \frac{  \Big( \bm{p} - C_1 \big(j_{\mu} j^{\mu} \big)^{\alpha_1 - 1} \bm{j} \Big) \cdot \left[ \delta  \Big( \bm{p} - C_1 \big(j_{\mu} j^{\mu} \big)^{\alpha_1 - 1} \bm{j} \Big) \right] + {m^*} ~\delta m^*  }{ \sqrt{ \Big( \bm{p} - C_1 \big(j_{\mu} j^{\mu} \big)^{\alpha_1 - 1} \bm{j} \Big)^2 + {m^*}^2  } } ~.
\end{eqnarray}
We have
\begin{eqnarray}
\delta  \Big( \bm{p} - C_1 \big(j_{\mu} j^{\mu} \big)^{\alpha_1 - 1} \bm{j} \Big) &=& - C_1  \delta \Big(\big(j_{\mu} j^{\mu} \big)^{\alpha_1 - 1} \bm{j} \Big)  \non \\
&=& - C_1 \left[ 2(\alpha_1 -1) \big(j_{\mu} j^{\mu} \big)^{\alpha_1 - 2} j_{\mu} ~\delta  j^{\mu} ~ \bm{j} + \big(j_{\mu} j^{\mu} \big)^{\alpha_1 - 1}  ~ \delta \bm{j}  \right] 
\end{eqnarray}
and
\begin{eqnarray}
\delta m^* =  - G_1  \delta \Big[ n_s^{\gamma_1 - 1}  \Big] = - G_1  (\gamma_1 - 1)  n_s^{\gamma_1 - 2} ~ \delta n_s  ~,
\end{eqnarray}
where in the second equality we used Eq.\ (\ref{effective_mass_definition}), so that altogether
\begin{eqnarray}
\hspace{-5mm}\int \ptilde  \delta \epsilon_{\txt{kin}}^* ~ f_{\bm{p}} && \non\\
&& \hspace{-30mm}  =~  - C_1 \left[ 2(\alpha_1 -1)  \big(j_{\mu} j^{\mu} \big)^{\alpha_1 - 2} j_{\mu} ~\delta  j^{\mu} ~ \bm{j} + \big(j_{\mu} j^{\mu} \big)^{\alpha_1 - 1} ~ \delta \bm{j}  \right] \non \\
&& \hspace{-20mm} \cdot ~ \int \ptilde  \frac{  \bm{p} - C_1 \big(j_{\mu} j^{\mu} \big)^{\alpha_1 - 1} \bm{j}   }{ \sqrt{ \Big( \bm{p} - C_1 \big(j_{\mu} j^{\mu} \big)^{\alpha_1 - 1} \bm{j} \Big)^2 + {m^*}^2  } } ~ f_{\bm{p}}  \non \\
&& \hspace{-10mm} - ~  G_1  (\gamma_1 - 1)  n_s^{\gamma_1 - 2} ~ \delta n_s   \int \ptilde  \frac{  {m^*}   }{ \sqrt{ \Big( \bm{p} - C_1 \big(j_{\mu} j^{\mu} \big)^{\alpha_1 - 1} \bm{j} \Big)^2 + {m^*}^2  } } ~ f_{\bm{p}} ~.
\end{eqnarray}
By Eqs.\ (\ref{current_definition}) and (\ref{scalar_density_definition}) this becomes
\begin{eqnarray}
\int \ptilde  \delta \epsilon_{\txt{kin}}^* ~ f_{\bm{p}} &=& - C_1 \left[ 2(\alpha_1 -1) \big(j_{\mu} j^{\mu} \big)^{\alpha_1 - 2} j_{\mu} ~\delta  j^{\mu} ~ \bm{j}  +  \big(j_{\mu} j^{\mu} \big)^{\alpha_1 - 1}  ~ \delta \bm{j}  \right]   \cdot \bm{j} \non \\
&& \hspace{10mm} - ~   G_1  (\gamma_1 - 1)  n_s^{\gamma_1 - 1} ~ \delta n_s    ~,
\end{eqnarray}
and consequently
\begin{eqnarray}
\delta \mathcal{E}_{v1s1,\txt{kin}} &=&\int \ptilde \epsilon_{\txt{kin}}^* ~ \delta f_{\bm{p}} - 2C_1  (\alpha_1 -1) \big(j_{\mu} j^{\mu} \big)^{\alpha_1 - 2}  \bm{j} \cdot \bm{j} ~ j_{\mu} ~\delta  j^{\mu}    \nonumber\\
&& \hspace{5mm} - ~  C_1  \big(j_{\mu} j^{\mu} \big)^{\alpha_1 - 1}  ~  \bm{j} \cdot \delta \bm{j}    -  G_1  (\gamma_1 - 1)  n_s^{\gamma_1 - 1} ~ \delta n_s   ~.
\label{variation_kinetic_part}
\end{eqnarray}

Varying the interaction part of the energy density $\mathcal{E}_{\txt{v1s1}}$, Eq.\ (\ref{energy_density_one_scalar_one_vector_term}), term by term yields
\begin{eqnarray}
&& \hspace{-8mm} \delta \Big[ C_1 \big(j_{\mu}j^{\mu}\big)^{\alpha_1 - 1} \big(j^0\big)^2 \Big] = 2C_1 \left[ (\alpha_1 - 1) \big( j_{\mu}j^{\mu}\big)^{\alpha_1 - 2} \big( j^0 \big)^2 j_{\mu} ~\delta j^{\mu}   +  \big( j_{\mu} j^{\mu} \big)^{\alpha_1 - 1} j_0 ~ \delta j^0 \right]~, \\
&&  \hspace{-8mm}  
\delta \left[ - C_1 \left(\frac{2 \alpha_1 - 1}{2\alpha_1}\right) \big(j_{\mu} j^{\mu} \big)^{\alpha_1}   \right] =  - C_1  (2 \alpha_1 - 1)  \big(j_{\mu} j^{\mu} \big)^{\alpha_1 - 1} j_{\mu} ~ \delta j^{\mu}~, \\
&&  \hspace{-8mm}  \delta \left[  G_1 \frac{\gamma_1 - 1}{\gamma_1} n_s^{\gamma_1} \right] =  G_1 (\gamma_1 - 1) n_s^{\gamma_1 - 1} ~ \delta n_s ~,
\end{eqnarray}
so that the variation of the entire interaction part is 
\begin{eqnarray}
\delta  \mathcal{E}_{v1s1,\txt{int}}  &=& C_1(2\alpha_1 - 2) \big( j_{\mu}j^{\mu}\big)^{\alpha_1 - 2} \big( j^0 \big)^2 j_{\mu} ~\delta j^{\mu} - C_1  (2 \alpha_1 - 1)  \big(j_{\mu} j^{\mu} \big)^{\alpha_1 - 1} j_{\mu} ~ \delta j^{\mu}   \non \\
&& \hspace{5mm} +~   2 C_1\big( j_{\mu} j^{\mu} \big)^{\alpha_1 - 1} j_0 ~ \delta j^0    + G_1 (\gamma_1 - 1) n_s^{\gamma_1 - 1} ~ \delta n_s ~.
\label{variation_interaction_part}
\end{eqnarray}

Putting Eqs.\ (\ref{variation_kinetic_part}) and (\ref{variation_interaction_part}) together, we obtain the variation of the energy density as given by Eq.\ (\ref{energy_density_one_scalar_one_vector_term}),
\begin{eqnarray}
\hspace{-5mm}\delta \mathcal{E}_{\txt{v1s1}} &=& \int \ptilde \epsilon_{\txt{kin}}^* ~ \delta f_{\bm{p}}  - C_1(2\alpha_1 - 2) \big(j_{\mu} j^{\mu} \big)^{\alpha_1 - 2} \bm{j}  \cdot \bm{j} ~j_{\mu} ~\delta  j^{\mu} \non \\
&& \hspace{3mm} - ~ C_1  \big(j_{\mu} j^{\mu} \big)^{\alpha_1 - 1}  ~   \bm{j} \cdot \delta \bm{j}    -  G_1  (\gamma_1 - 1)  n_s^{\gamma_1 - 1} ~ \delta n_s   \non \\
&&  \hspace{8mm} + ~  C_1(2\alpha_1 - 2) \big( j_{\mu}j^{\mu}\big)^{\alpha_1 - 2} \big( j^0 \big)^2 j_{\mu} ~\delta j^{\mu} - C_1  (2 \alpha_1 - 1)  \big(j_{\mu} j^{\mu} \big)^{\alpha_1 - 1} j_{\mu} ~ \delta j^{\mu}  \non \\
&& \hspace{13mm} +~   2 C_1\big( j_{\mu} j^{\mu} \big)^{\alpha_1 - 1} j_0 ~ \delta j^0    + G_1 (\gamma_1 - 1) n_s^{\gamma_1 - 1} ~ \delta n_s ~.
\end{eqnarray}
The fourth and eighth term, induced by variation of the scalar density and thus proportional to $G_1$, cancel out. Further, the second, fifth, and sixth term can be combined using the fact that $
\big( j^0 \big)^2  - \bm{j} \cdot \bm{j}  = j_{\mu} j^{\mu}$, so we can write
\begin{eqnarray}
\delta \mathcal{E}_{\txt{v1s1}} &=& \int \ptilde \epsilon_{\txt{kin}}^* ~ \delta f_{\bm{p}}  - C_1 \big(j_{\mu} j^{\mu} \big)^{\alpha_1 - 1}  ~j_{\mu} ~\delta  j^{\mu}  \non \\
&& \hspace{5mm} - ~ C_1  \big(j_{\mu} j^{\mu} \big)^{\alpha_1 - 1}  ~   \bm{j} \cdot \delta \bm{j}    +   2 C_1\big( j_{\mu} j^{\mu} \big)^{\alpha_1 - 1} j_0 ~ \delta j^0    ~.
\end{eqnarray}
Noting that $j_{\mu} \delta j^{\mu} = j_0 \delta j^0 - \bm{j} \cdot \delta \bm{j}$, we further reduces the above equation to
\begin{eqnarray}
\delta \mathcal{E}_{\txt{v1s1}} &=& \int \ptilde \epsilon_{\txt{kin}}^* ~ \delta f_{\bm{p}}  +  C_1\big( j_{\mu} j^{\mu} \big)^{\alpha_1 - 1} j_0 ~ \delta j^0    ~.
\end{eqnarray}
Finally, using the definition of baryon density $j^0$, Eq.\ (\ref{density_definition}), we arrive at
\begin{eqnarray}
\delta \mathcal{E}_{\txt{v1s1}} &=& \int \ptilde \bigg[ \epsilon_{\txt{kin}}^* ~ +  C_1\big( j_{\mu} j^{\mu} \big)^{\alpha_1 - 1} j_0  \bigg] ~  \delta f_{\bm{p}}     ~,
\end{eqnarray}
from which we immediately obtain the quasiparticle energy,
\begin{eqnarray}
\eps_{\bm{p}}^* =  \epsilon_{\txt{kin}}^* ~ +  C_1\big( j_{\mu} j^{\mu} \big)^{\alpha_1 - 1} j_0 ~. 
\end{eqnarray}

\section{Construction of the energy-momentum tensor}
\label{the_energy-momentum_tensor}

The energy-momentum tensor, $T^{\mu\nu}$, can be constructed by taking moments of the Boltzmann equation, Eq.\ (\ref{Boltzmann_eq}), and identifying the components of $T^{\mu\nu}$ which must satisfy the energy and momentum conservation, $\partial_{\nu} T^{\mu\nu} = 0$.  

First, we multiply both sides of Eq.\ (\ref{Boltzmann_eq}) by the quasiparticle energy, $\eps_{\bm{p}}^*$, and integrate over all momenta,
\begin{eqnarray}
g \int \pdens \eps_{\bm{p}}^* ~ \parr{f_{\bm{p}}}{t} - g \int \pdens \eps_{\bm{p}}^* \parr{\eps_{\bm{p}}^*}{p_i}~ \parr{f_{\bm{p}}}{x^i} + g \int \pdens \eps_{\bm{p}}^* \parr{\eps_{\bm{p}}^*}{x_i} ~\parr{f_{\bm{p}}}{p^i} = 0 ~,
\label{Boltzmann_eq_energy_conservation}
\end{eqnarray}
where get zero on the right-hand side due to the fact that quasiparticle energy is conserved in collisions (see Appendix \ref{vanishing_of_the_collision_integral} for more details). By using the product rule for derivatives, 
\begin{eqnarray}
\hspace{-5mm} g \int \pdens \eps_{\bm{p}} \parr{\eps_{\bm{p}}^*}{p_i} ~\parr{f_{\bm{p}}}{x^i}  &=& \parr{}{x^i} ~ g \int \pdens \eps_{\bm{p}} \parr{\eps_{\bm{p}}^*}{p_i} ~ f_{\bm{p}}  \non \\
&& \hspace{5mm} -  g \int \pdens \parr{\eps_{\bm{p}}}{x^i}  \parr{\eps_{\bm{p}}^*}{p_i} ~ f_{\bm{p}} - g \int \pdens \eps_{\bm{p}} \frac{\partial^2\eps_{\bm{p}}^*}{\partial x^i \partial p_i} ~ f_{\bm{p}}
\end{eqnarray}
and
\begin{eqnarray}
\hspace{-5mm} g \int \pdens \eps_{\bm{p}} \parr{\eps_{\bm{p}}^*}{x_i} ~\parr{f_{\bm{p}}}{p^i} &=&  \parr{}{p^i}  ~ g \int \pdens \eps_{\bm{p}} \parr{\eps_{\bm{p}}^*}{x_i} ~f_{\bm{p}} \non \\
&& \hspace{5mm} - ~  g \int \pdens \parr{\eps_{\bm{p}}}{p^i}  \parr{\eps_{\bm{p}}^*}{x_i} ~f_{\bm{p}} -  g \int \pdens \eps_{\bm{p}} \frac{\partial^2 \eps_{\bm{p}}^*}{\partial p^i \partial x_i} ~f_{\bm{p}} ~,
\end{eqnarray}
we can rewrite Eq.\ (\ref{Boltzmann_eq_energy_conservation}) as
\begin{eqnarray}
g \int \pdens \eps_{\bm{p}}^* ~\parr{f_{\bm{p}}}{t} - \parr{}{x^i} ~ g \int \pdens \eps_{\bm{p}}^* \parr{\eps_{\bm{p}}^*}{p_i} ~ f_{\bm{p}}   +  \parr{}{p^i}  ~ g \int \pdens \eps_{\bm{p}}^* \parr{\eps_{\bm{p}}^*}{x_i} ~f_{\bm{p}} = 0 ~.
\end{eqnarray}
The last term is a definite integral of the integrand, and it is identically zero because the distribution function must vanish for $|\bm{p}| = + \infty$, so that we have
\begin{eqnarray}
g \int \pdens \eps_{\bm{p}}^* ~\parr{f_{\bm{p}}}{t} - \partial_{i} ~ g \int \pdens \eps_{\bm{p}}^* \parr{\eps_{\bm{p}}^*}{p_i} ~ f_{\bm{p}}   = 0 ~.
\end{eqnarray}
Because from the definition $\delta \mathcal{E} \equiv g \int \pdens \eps_{\bm{p}}^* ~ \delta f_{\bm{p}}$ (see Section \ref{Landau_Fermi-liquid_theory}), where $\mathcal{E}$ is the generic energy density of the system, we can further rewrite the first term, thus arriving at
\begin{eqnarray}
\partial_0 \mathcal{E} - \partial_{i} ~ g \int \pdens \eps_{\bm{p}}^* \parr{\eps_{\bm{p}}^*}{p_i} ~ f_{\bm{p}}   = 0 ~.
\label{E_conservation_1}
\end{eqnarray}
Finally, using $- \inparr{\eps_{\bm{p}}^*}{p_i} \equiv v^i$, where $v^i$ is the velocity of a quasiparticle, the second term in the above equation can be identified as the quasiparticle energy flux, and thus it becomes apparent that Eq.\ \eqref{E_conservation_1} must be an energy density conservation equation. By comparing with $\partial_{\nu} T^{0\nu} = 0$ we see that we must have
\begin{eqnarray}
T^{00} = \mathcal{E} 
\end{eqnarray}
and
\begin{eqnarray}
T^{0i} = - g \int \pdens \eps_{\bm{p}}^* \parr{\eps_{\bm{p}}^*}{p_i} ~ f_{\bm{p}}  ~.
\end{eqnarray}

To obtain momentum conservation, we multiply both sides of Eq.\ (\ref{Boltzmann_eq}) by the $k$-th component of the quasiparticle's momentum, $p^k$, and again integrate over all momenta,
\begin{eqnarray}
g \int \pdens p^k ~\parr{f_{\bm{p}}}{t} - g \int \pdens p^k \parr{\eps_{\bm{p}}^*}{p_i} ~\parr{f_{\bm{p}}}{x^i} + g \int \pdens p^k \parr{\eps_{\bm{p}}^*}{x_i} ~ \parr{f_{\bm{p}}}{p^i} = 0 ~,
\label{Boltzmann_eq_momentum_conservation}
\end{eqnarray}
where the right-hand side vanishes because momentum, like energy, is also conserved in collisions (see Appendix \ref{vanishing_of_the_collision_integral} for more details). Using the product rule for derivatives,
\begin{eqnarray}
\hspace{-5mm} g \int \pdens p^k \parr{\eps_{\bm{p}}^*}{p_i} ~\parr{f_{\bm{p}}}{x^i} = \parr{}{x^i} ~ g \int \pdens p^k \parr{\eps_{\bm{p}}^*}{p_i}  ~ f_{\bm{p}}  -   g \int \pdens p^k   \parrtwo{\eps_{\bm{p}}^*}{x^i}{p_i}  ~ f_{\bm{p}} 
\end{eqnarray}
and
\begin{eqnarray}
\hspace{-5mm} g \int \pdens p^k \parr{\eps_{\bm{p}}^*}{x_i} ~\parr{f_{\bm{p}}}{p^i} &=& \parr{}{p^i}  ~ g \int \pdens p^k \parr{\eps_{\bm{p}}^*}{x_i} ~f_{\bm{p}}  \non \\
&& \hspace{5mm} - ~  g \int \pdens \parr{p^k }{p^i} \parr{\eps_{\bm{p}}^*}{x_i} ~f_{\bm{p}} - g \int \pdens p^k \parrtwo{\eps_{\bm{p}}^*}{p^i}{x_i} ~f_{\bm{p}} ~,
\end{eqnarray}
leads to
\begin{eqnarray}
&& g \int \pdens p^k ~\parr{f_{\bm{p}}}{t} -  \parr{}{x^i} ~ g \int \pdens p^k \parr{\eps_{\bm{p}}^*}{p_i}  ~ f_{\bm{p}} \non \\
&& \hspace{10mm}+~ \parr{}{p^i}  ~ g \int \pdens p^k \parr{\eps_{\bm{p}}^*}{x_i} ~f_{\bm{p}} - g \int \pdens g^k_{~i} \parr{\eps_{\bm{p}}^*}{x_i} ~f_{\bm{p}}  = 0 ~,
\end{eqnarray}
where we have used $\parr{p^k }{p^i} = \delta^k_i = g^{kj}g_{ji} = g^k_{~i}$, with $g^k_{~i}$ being a component of the metric tensor. The third term vanishes due to the fact that $f_{\bm{p}}\big(|\bm{p}|=+\infty\big) = 0$, while the fourth term can be further rewritten as
\begin{eqnarray}
g^k_{~i} ~ g \int \pdens  \parr{\eps_{\bm{p}}^*}{x_i} ~f_{\bm{p}} &=&  g^k_{~i} ~ \partial^i  g \int \pdens \eps_{\bm{p}}^* ~f_{\bm{p}}  - g^k_{~i} ~\partial^i \mathcal{E} ~,
\end{eqnarray}
where we again used the derivative product rule and $\delta \mathcal{E} \equiv g \int \pdens \eps_{\bm{p}}^* ~ \delta f_{\bm{p}}$. Altogether we then have
\begin{eqnarray}
\hspace{-5mm } \parr{}{t} ~ g \int \pdens p^k  ~ f_{\bm{p}}   - \partial_i ~ g \int \pdens p^k \parr{\eps_{\bm{p}}^*}{p_i}  ~ f_{\bm{p}}     + \partial_i  ~g^{ik} \left[ \mathcal{E} -   g \int \pdens \eps_{\bm{p}}^* ~f_{\bm{p}} \right]=0 ~,
\end{eqnarray}
and, using $\partial_{\nu} T^{i\nu} = 0$, we can identify
\begin{eqnarray}
T^{i0} = g \int \pdens p^i  ~ f_{\bm{p}}  
\end{eqnarray}
and
\begin{eqnarray}
T^{ij} = - g \int \pdens p^i \parr{\eps_{\bm{p}}^*}{p_j}  ~ f_{\bm{p}}  + g^{ij} \left[ \mathcal{E} -   g \int \pdens \eps_{\bm{p}}^* ~f_{\bm{p}} \right] ~.
\end{eqnarray}
In this way, we constructed all components of the energy-momentum tensor $T^{\mu\nu}$ as introduced in Eqs.\ (\ref{T00_general}-\ref{Tij_general}).

\section{Relativistic covariance of the equations of motion}
\label{EOMs_covariant}

With the definition of the kinetic momentum $\Pi^{\mu}$, Eq.\ (\ref{kinetic_momentum}), the Hamilton's equations, Eqs.\ (\ref{x_equation_of_motion_generalized}) and (\ref{p_equation_of_motion_generalized}), can be rewritten as 
\begin{eqnarray}
\frac{dx^i}{dt} &=& \frac{\Pi^i }{\Pi_0}  ~, 
\label{Hamilton_x_i} \\
\frac{dp^i}{dt} &=& \sum_{k=1}^K \frac{\sum_j  \Pi_j    }{ \Pi_0} \left( \parr{(A_k)^j}{x_i}\right) + \frac{ m^* }{\Pi_0} \parr{m^*}{x_i} + \sum_{k=1}^K \parr{(A_k)^0}{x_i}   ~.
\label{Hamilton_p_i}
\end{eqnarray}

Using the fact that $H_{(1)} = \eps_{\bm{p}} = p_0$, we can see that for the temporal component of $x^{\mu}$ we have trivially
\begin{eqnarray}
\frac{dx^0}{dt} = \parr{H_{(1)}}{p_0}  = 1 = \frac{\Pi_0}{\Pi_0}~,
\label{Hamilton_x_0}
\end{eqnarray}
which allows us to write Eqs.\ (\ref{Hamilton_x_i}) and (\ref{Hamilton_x_0}) together as
\begin{eqnarray}
\frac{dx^{\mu}}{dt}  = \frac{\Pi^{\mu}}{\Pi_0}   ~.
\label{x_relativistic}
\end{eqnarray}
For the temporal part of $p^{\mu}$ we can likewise write
\begin{eqnarray}
\frac{dp^0}{dt} = \frac{dp^0}{dx_0} = \sum_{k=1}^K \frac{\sum_{j} \Pi_j}{\Pi_0} \parr{(A_k)^j}{x_0} + \frac{m^*}{\Pi_0} \parr{m^*}{x_0} +  \sum_{k=1}^K \parr{(A_k)^0}{x_0} ~,
\label{Hamilton_p_0}
\end{eqnarray}
where on the right-hand side we have simply carried out the differentiation with respect to $x_0$, and it follows that Eqs.\ (\ref{Hamilton_p_i}) and (\ref{Hamilton_p_0}) can be jointly written as
\begin{eqnarray}
\frac{dp^{\mu}}{dt} &=& \sum_{k=1}^K \left[ \frac{\sum_{j} \Pi_j}{\Pi_0} \parr{(A_k)^j}{x_{\mu}}  + \parr{(A_k)^0}{x_{\mu}} \right]  + \frac{m^*}{\Pi_0} \parr{m^*}{x_0}   \non  \\
&=&\sum_{k=1}^K \left[ \frac{\sum_{j} \Pi_j}{\Pi_0} \parr{(A_k)^j}{x_{\mu}}  + \frac{\Pi_0}{\Pi_0}\parr{(A_k)^0}{x_{\mu}} \right]  + \frac{m^*}{\Pi_0} \parr{m^*}{x_0}  \non \\
&=&\sum_{k=1}^K \left[ \sum_{\nu} \frac{\Pi_{\nu}}{\Pi_0} \parr{(A_k)^{\nu}}{x_{\mu}}    \right] + \frac{m^*}{\Pi_0} \parr{m^*}{x_0} ~.
\label{eq345}
\end{eqnarray}
Let us note that from the definition of the kinetic momentum $\Pi$ we have
\begin{eqnarray}
\frac{d \Pi^{\mu}}{dt} = \frac{dp^{\mu}}{dt} - \sum_{k=1}^K \frac{d(A_k)^{\mu}}{dt}~.
\label{eq_EOM_3}
\end{eqnarray}
Using Eq.\ (\ref{eq345}), the above equation becomes
\begin{eqnarray}
\frac{d \Pi^{\mu}}{dt} = \sum_{k=1}^K \left[ \sum_{\nu} \frac{\Pi_{\nu}}{\Pi_0} \parr{(A_k)^{\nu}}{x_{\mu}}    \right] + \frac{m^*}{\Pi_0} \parr{m^*}{x_0}   - \sum_{k=1}^K \frac{d(A_k)^{\mu}}{dt}~.
\label{eq987}
\end{eqnarray}
We can always write
\begin{eqnarray}
\frac{d(A_k)^{\mu}}{dt} = \sum_\nu \parr{(A_k)^{\mu}}{x^{\nu}} \frac{dx^{\nu}}{dt}  = \sum_\nu \parr{(A_k)^{\mu}}{x^{\nu}} \frac{\Pi^{\nu}}{\Pi_0} ~ ,
\label{eq_EOM_4}
\end{eqnarray}
so that Eq.\ (\ref{eq987}) can be transformed into
\begin{eqnarray}
\frac{d \Pi^{\mu}}{dt} &=&  \sum_{k=1}^K  \sum_{\nu} \left[ \frac{\Pi_{\nu}}{\Pi_0} \parr{(A_k)^{\nu}}{x_{\mu}}  -  \frac{\Pi^{\nu}}{\Pi_0} \parr{(A_k)^{\mu}}{x^{\nu}}    \right] + \frac{m^*}{\Pi_0} \parr{m^*}{x_0}   \non  \\ 
&=&  \sum_{k=1}^K  \sum_{\nu}  \frac{\Pi_{\nu}}{\Pi_0} \Big[ \partial^{\mu}(A_k)^{\nu}  - \partial^{\nu} (A_k)^{\mu}   \Big] + \frac{m^*}{\Pi_0} \parr{m^*}{x_0}  \\
&=&  \sum_{\nu}  \frac{\Pi_{\nu}}{\Pi_0} \sum_{k=1}^K  (F_k)^{\mu\nu} + \frac{m^*}{\Pi_0} \parr{m^*}{x_0}  ~,
\label{p_relativistic}
\end{eqnarray}
where $(F_k)^{\mu\nu}$ is defined similarly as the field strength in relativistic electrodynamics. 

Both Eq.\ (\ref{x_relativistic}) and Eq.\ (\ref{p_relativistic}) are written in a relativistically covariant form.

\section{Lorentz force}
\label{Lorentz_force}

The relativistically covariant equation of motion for the quasiparticle momentum, Eq.\ (\ref{EOM_covariant_formulation_p}), naturally includes the Lorentz force. Explicitly, let us see that for $\mu = i$ we get
\begin{eqnarray}
\frac{d \Pi^{i}}{dt} &=&  \sum_{k=1}^K  \sum_{\nu}  \frac{\Pi_{\nu}}{\Pi_0} \Big[ \partial^{i}(A_k)^{\nu}  - \partial^{\nu} (A_k)^{i}   \Big] + \frac{m^*}{\Pi_0} \parr{m^*}{x_0}  \\
&=&  \sum_{k=1}^K \left\{\Big[ \partial^{i}(A_k)^{0}  - \partial^{0} (A_k)^{i}   \Big] + \sum_{j}  \frac{\Pi_{j}}{\Pi_0} \Big[ \partial^{i}(A_k)^{j}  - \partial^{j} (A_k)^{i}   \Big] \right\}  + \frac{m^*}{\Pi_0} \parr{m^*}{x_0}   ~.
\end{eqnarray}
By summing over $i$ we obtain (here one needs to remember that $(\bm{\nabla})_{i} = \parr{ }{x^i} = - \parr{}{x_i} $, see Appendix \ref{units_and_notation})
\begin{eqnarray}
\hspace{-5mm} \frac{d \bm{\Pi}}{dt} &=&  \sum_{k=1}^K \left\{ \Big[ - \bm{\nabla} A_k^{0}  - \partial^{0} \bm{A}_k   \Big] + \sum_i \sum_{j}  \frac{\Pi_{j}}{\Pi_0} \Big[ \partial^{i}(A_k)^{j}  - \partial^{j} (A_k)^{i}   \Big] \right\}  + \frac{m^*}{\Pi_0} \parr{m^*}{x_0}   ~.
\label{Lorentz_force_12}
\end{eqnarray}

To interpret the second term in the above equation, let us see that 
\begin{eqnarray}
\bm{\Pi} \times \big( \bm{\nabla} \times \bm{A} \big)  &=&  \left| \begin{array}{ccc}
\bhat{x} & \bhat{y} & \bhat{z} \\
\Pi_x & \Pi_y & \Pi_z \\
\big( \bm{\nabla} \times \bm{A} \big)_x & \big( \bm{\nabla} \times \bm{A} \big)_y & \big( \bm{\nabla} \times \bm{A} \big)_z
\end{array} \right| \non  \\
&=&  \bhat{x} \bigg[ \Pi_y \big( \bm{\nabla} \times \bm{A} \big)_z - \Pi_z \big( \bm{\nabla} \times \bm{A} \big)_y    \bigg]  \nonumber \\
&& \hspace{5mm} +~ \bhat{y} \bigg[\Pi_z\big( \bm{\nabla} \times \bm{A} \big)_x - \Pi_x \big( \bm{\nabla} \times \bm{A} \big)_z  \bigg]  \nonumber \\
&& \hspace{10mm} + ~  \bhat{z} \bigg[ \Pi_x \big( \bm{\nabla} \times \bm{A} \big)_y  - \Pi_y \big( \bm{\nabla} \times \bm{A} \big)_x  \bigg] \non \\
&=&  \sum_{a} \left[ \sum_b \sum_c \epsilon_{abc}~   \bhat{x}^a \Pi^b \big( \bm{\nabla} \times \bm{A} \big)^c \right] ~,
\end{eqnarray}
where $\epsilon_{abc}$ is the Levi-Civita symbol and $\bhat{x}^a$ are the unit vectors. In the same way, we can further rewrite
\begin{eqnarray}
\big( \bm{\nabla} \times \bm{A} \big)^c = \sum_{d} \sum_f \epsilon_{cdf} ~ \partial_{d} A^f ~,
\end{eqnarray}
so that altogether
\begin{eqnarray}
\bm{\Pi} \times \big( \bm{\nabla} \times \bm{A} \big)  &=& \sum_{a} \left[ \sum_b \sum_c \sum_{d} \sum_f \epsilon_{abc}~  \epsilon_{cdf}  ~  \bhat{x}^a \Pi^b ~  \partial_{d} A^f \right] ~.
\label{Lorentz_force_23}
\end{eqnarray}
It can be shown that
\begin{eqnarray}
\sum_i \epsilon_{ijk} \epsilon^{imn} = \delta_j^{~m} \delta_k^{~n} - \delta_j^{~n} \delta_k^{~m} ~,
\end{eqnarray}
which, bearing in mind that $\epsilon_{ijk} = \epsilon_{jki}$, can be used to obtain
\begin{eqnarray}
\sum_c \epsilon_{abc} \epsilon_{cdf} = \delta_a^{~d} \delta_b^{~f} - \delta_a^{~f} \delta_b^{~d} ~.  
\end{eqnarray}
Then Eq.\ (\ref{Lorentz_force_23}) becomes
\begin{eqnarray}
&& \hspace{-7mm}\bm{\Pi}\times \big( \bm{\nabla} \times \bm{A} \big) = \sum_{a} \left[ \sum_b \sum_{d} \sum_f \Big( \delta_a^{~d} \delta_b^{~f}  \Big)  ~  \bhat{x}^a \Pi^b ~  \partial_{d} A^f -  \sum_b \sum_{d} \sum_f \Big(  \delta_a^{~f} \delta_b^{~d} \Big)  ~  \bhat{x}^a \Pi^b~  \partial_{d} A^f \right] \non \\
&& \hspace{20mm} = \sum_{a} \left[ \sum_b     \bhat{x}^a \Pi^b ~  \partial_{a} A^b -  \sum_b   \bhat{x}^a \Pi^b ~  \partial_{b} A^a \right] \non \\
&& \hspace{20mm} =  -\sum_{a} \sum_b \Pi^b ~  \left[     \partial^{a} A^b~ \bhat{x}^a -  \partial^{b} A^a~ \bhat{x}^a \right] \non \\
&& \hspace{20mm} =  \sum_{a} \sum_b \Pi_b ~  \left[     \partial^{a} A^b~ \bhat{x}^a -  \partial^{b} A^a~ \bhat{x}^a \right] ~,
\end{eqnarray}
where in the two last lines we have used the fact that $\partial_i = - \partial^i$ and $\Pi_i = - \Pi_i$. Comparing Eq.\ (\ref{Lorentz_force_12}) against the above formula reveals that
\begin{eqnarray}
\frac{d \bm{\Pi}}{dt} &=&  \sum_{k=1}^K \left\{ \Big[ - \bm{\nabla} A_k^{0}  - \partial^{0} \bm{A}_k   \Big] + \frac{\bm{\Pi}}{\Pi_0} \times \big( \bnabla \times \bm{A}_k  \big)  \right\}  + \frac{m^*}{\Pi_0} \parr{m^*}{x_0}   ~.
\label{EOM_VSDF_explicit}
\end{eqnarray}
Noting that by the equation of motion for the quasiparticle position, Eq.\ (\ref{EOM_covariant_formulation_x}), we have $\frac{\bm{\Pi}}{\Pi_0} = \frac{d\bm{x}}{dt}$, we see that the middle term in the above equation is evidently the Lorentz force, $\bm{v} \times \bm{B}$, where $\bm{B} \equiv  \bnabla \times \bm{A}_k $.

\section{Minimization of the auxiliary fields}
\label{minimization_of_the_auxiliary_fields}

Thermodynamic consistency demands that the thermodynamic potential is minimized by the auxiliary fields, $n_s$ and $n$. In the grand canonical ensemble, this is equivalent to minimizing the pressure,
\begin{eqnarray}
\derr{P}{n_s} \bigg|_{T, \mu}  = 0  \distand  \derr{P}{n} \bigg|_{T, \mu} = 0 ~.
\end{eqnarray}
With the density dependence of the kinetic term explicit, the pressure, Eq.\ (\ref{VSDF_pressure}), is given by
\begin{eqnarray}
P_{(K,M)} &=& g \int \pdens T ~  \ln \bigg[1 + e^{- \beta \big( \sqrt{p^2 + {m^*}^2} + \sum_k C_k n^{b_k - 1} - \mu \big)} \bigg]  \nonumber\\
&& \hspace{10mm} + ~\sum_{k=1}^K C_k \left(\frac{b_k - 1}{b_k}\right) n^{b_k} - \sum_{m=1}^M G_m\left(\frac{d_m - 1}{d_m}\right) n_s^{d_m}  ~.
\end{eqnarray}
Then we can calculate
\begin{eqnarray}
\derr{P}{n_s} \bigg|_{T, \mu} &=& - g\int \pdens ~ \left(  \frac{dm^*}{dn_s} \frac{m^*}{\epsilon^*} + \frac{dn}{dn_s} \sum_k C_k (b_k - 1) n^{b_k - 2}  \right) ~ f_{\bm{p}}   \nonumber \\
&& \hspace{10mm} + ~  \frac{dn}{dn_s} \sum_{k=1}^K C_k (b_k - 1) n^{b_k -1}  - \sum_{m=1}^M G_m (d_m - 1 ) n_s^{d_m - 1}   \\
&=& -n_s \bigg(\frac{dm^*}{dn_s}  + \sum_{m=1}^M G_m (d_m - 1 ) n_s^{d_m - 2}\bigg)    \non \\
&& \hspace{10mm} + ~  \frac{dn}{dn_s} \sum_{k=1}^K C_k (b_k - 1) n^{b_k - 2} \bigg( n - g\int \pdens  f_{\bm{p}}  \bigg)  = 0 ~;
\end{eqnarray}
here, the the second term vanishes from the definition of the baryon current, Eq.\ (\ref{baryon_4current}), while demanding that the first term disappears as well yields, after integration, the gap equation, 
\begin{eqnarray}
m^* = m_N - \sum_{m=1}^M G_m n_s^{d_m -1} ~,
\label{the_gap_equation}
\end{eqnarray}
which is identical to the definition of the effective mass, Eq.\ (\ref{effective_mass_definition_generalized}).

Similarly, we calculate
\begin{eqnarray}
\derr{P}{n} \bigg|_{T, \mu_B} &=& \frac{dm^*}{dn} \left[n_s + \derr{n_s}{m^*} \sum_{m=1}^M G_m (d_m - 1) n_s^{d_m - 1} \right]   \non \\
&& \hspace{10mm} + ~  \sum_{k=1}^K C_k (b_k - 1) n^{b_k - 2} \bigg( n - g\int \pdens  f_{\bm{p}}  \bigg) = 0~,
\label{eq9876}
\end{eqnarray}
where the vanishing of the first term can be easily established by taking the derivative of the gap equation, Eq.\ (\ref{effective_mass_definition_generalized}) (or equivalently Eq.\ \eqref{the_gap_equation}), with respect to the effective mass $m^*$, while the second term again disappears by Eq.\ (\ref{baryon_4current}).

Thus we show that the pressure is minimized by the auxiliary fields, $n_s$ and $n$, provided that these auxiliary fields are defined self-consistently by Eqs.\ (\ref{effective_mass_definition_generalized}) and (\ref{baryon_4current}), respectively.

\chapter{Form of the energy density functional}
\label{form_of_the_energy_density_functional}

We want to write down a generic energy density in which the interaction couples to the mean-field vector and scalar currents (here we will consider only one vector and one scalar term, which corresponds to $K = 1$ and $M = 1$ in the notation used in Section \ref{multiple_terms} and onward, however, the derivation can be easily generalized). We know that in such cases the energy of the quasiparticle takes the form
\begin{equation}
\eps_{\bm{p}} = \epsilon_{\txt{kin}} + A_0~,
\label{quasiparticle_energy}
\end{equation}
where $\epsilon_{\txt{kin}}$ is kinetic energy which also depends on the interaction,
\begin{equation}
\epsilon_{\txt{kin}} = \sqrt{\big(\bm{p} - \bm{A}\big)^2 + {m^*}^2} ~,
\end{equation}
and the interaction term is in general given by 
\begin{eqnarray}
A^{\mu} = \alpha(n) j^{\mu} ~,
\end{eqnarray}
where $\alpha(n)$ an arbitrary function of the rest frame baryon density $n$ and $j^{\mu}$ is the baryon 4-current,
\begin{equation}
j^{\mu} = \int \ptilde \frac{p^{\mu} - A^{\mu}}{\epsilon_{\txt{kin}}} ~ f_{\bm{p}} ~,
\end{equation}
while the effective mass is given by
\begin{eqnarray}
m^* = m_0  - B ,
\label{general_effective_mass}
\end{eqnarray}
where $B = B (n_s)$ is some function of the scalar density $n_s$. In general, the energy density will then look like
\begin{eqnarray}
\mathcal{E} &=& g \int \pdens \eps_{\bm{p}} ~ f_{\bm{p}} - g^{00} \Gamma  A_{\lambda} j^{\lambda} + \Lambda B n_s  \non \\
&=& g \int \pdens \epsilon_{\txt{kin}} ~ f_{\bm{p}} + A_0 j^0 - g^{00} \Gamma   A_{\lambda} j^{\lambda}  + \Lambda B  n_s ~,
\label{energy_density}
\end{eqnarray}
where $\Gamma$ and $\Lambda$ are constants in front of the counterterms $A_{\lambda} j^{\lambda}$ and $B  n_s$, respectively. The counterterms are necessary to avoid overcounting potential energy contributions  from the kinetic term, which sums quasiparticle energies at all possible momenta; intuitively, the counterterms can be understood to ``adjust'' the energy density to the correct value (for an elementary example, see the discussion following Eq.\ \eqref{Landau_quasiparticle_energy_definition} in Chapter \ref{a_flexible_model_of_dense_nuclear_matter}).

The quasiparticle energy is from the definition given by
\begin{equation}
\eps_{\bm{p}} \equiv \frac{\delta \mathcal{E}}{\delta f_{\bm{p}}} ~. 
\end{equation}
Taking the functional differential of Eq.\ \eqref{energy_density},
\begin{eqnarray}
\delta \mathcal{E} &=& g\int \pdens \frac{\bm{p} - \bm{A}}{\epsilon_{\txt{kin}}} \big(- \delta \bm{A}\big) ~ f_{\bm{p}} + g\int \pdens \frac{m^*}{\epsilon_{\txt{kin}}} \delta m^* ~ f_{\bm{p}}  +  \int \pdens \epsilon_{\txt{kin}} ~ \delta f_{\bm{p}}  \non \\
&& \hspace{10mm} +  ~ \big( \delta A_0 \big) j^0 + A_0 \big( \delta j^0 \big) - \Gamma \big( \delta A_{\lambda} \big) j^{\lambda} - \Gamma A_{\lambda} \big( \delta j^{\lambda} \big) \non \\
&& \hspace{20mm} + ~  \Lambda\big(\delta B \big)n_s + \Lambda B \big(\delta n_s \big)  \\
&=&  \big(- \delta \bm{A}\big) \cdot \bm{j} + n_s \big( \delta m^* \big) + g\int \pdens \epsilon_{\txt{kin}} ~ \delta f_{\bm{p}}  \non \\
&& \hspace{10mm} + ~ \big( \delta A_0 \big) j^0 + A_0 ~g \int \pdens \delta f_{\bm{p}} - \Gamma \big( \delta A_{\lambda} \big) j^{\lambda} - \Gamma A_{\lambda} \big( \delta j^{\lambda} \big) \non \\
&& \hspace{20mm} + ~  \Lambda\big(\delta B \big)n_s +\Lambda B \big(\delta n_s \big) ~,
\end{eqnarray}
we see that the quasiparticle energy will only take the form as in Eq.\ \eqref{quasiparticle_energy} if
\begin{equation}
\begin{cases}
\big(- \delta \bm{A}\big) \cdot \bm{j} + \big( \delta A_0 \big) j^0  - \Gamma \big( \delta A_{\lambda} \big) j^{\lambda} - \Gamma A_{\lambda} \big( \delta j^{\lambda} \big) = 0 \\
n_s \big( \delta m^* \big) +  \Lambda\big(\delta B \big)n_s + \Lambda B \big(\delta n_s \big)   = 0  
\end{cases}
\label{two_equations}
\end{equation}

The first of the equations can be further rewritten as
\begin{eqnarray}
\big[1 - \Gamma \big] \big( \delta A_\mu \big) j^\mu  - \Gamma A_{\lambda} \big( \delta j^{\lambda} \big) = 0~.
\label{eq_8765}
\end{eqnarray}
We then calculate
\begin{eqnarray}
\delta A_{\mu} &=& \Big( \delta \alpha(n) \Big) j_{\mu} + \alpha(n) \big(\delta j_{\mu} \big)  \\
&=& \bigg( \frac{d \alpha(n)}{d n} \delta n \bigg) j_{\mu} + \alpha(n) \big(\delta j_{\mu} \big)  \\
&=& \bigg( \frac{d \alpha(n)}{d n} \frac{1}{n} j^{\mu} \big(\delta j_{\mu}\big) \bigg) j_{\mu} + \alpha(n) \big(\delta j_{\mu} \big) ~,
\end{eqnarray}
where we have used the fact that $n = \big(j_{\mu} j^{\mu}\big)^{1/2}$, so that Eq.\ \eqref{eq_8765} becomes
\begin{eqnarray}
\big[1 - \Gamma \big] \Bigg(  \rho\frac{d \alpha(n)}{d n} j^{\mu} \big(\delta j_{\mu}\big)   + \alpha(n) \big(\delta j_{\mu} \big) j^\mu\Bigg)   - \Gamma A^{\mu} \big( \delta j_{\mu} \big) = 0~,
\end{eqnarray}
which further reduces to 
\begin{eqnarray}
\big[1 - \Gamma \big] \Bigg(  \rho\frac{d \alpha(n)}{d n}   + \alpha(n) \Bigg)   - \Gamma \alpha(n)  = 0~.
\end{eqnarray}
This is a differential equation for $\alpha (n)$,
\begin{eqnarray}
\frac{d \alpha(n)}{d n}   = \frac{2\Gamma - 1 }{ 1 - \Gamma }  \frac{\alpha(n)}{  n}  ~,
\end{eqnarray}
and because $\Gamma$ is a constant, it follows that we must have 
\begin{eqnarray}
\alpha(n) = C_k n^{\beta}~,
\end{eqnarray}
where $C_k$ is some constant and $\Gamma$ is related to $\beta$ through
\begin{eqnarray}
\Gamma = \frac{\beta + 1}{\beta + 2} ~.
\end{eqnarray}
Identifying $\beta = b_k - 2$ leads to the same vector-current--dependent term as in Eq.\ (\ref{summary_energy_density}) in the case $K =1$.

Using the fact that the effective mass is given by Eq.\ (\ref{general_effective_mass}), from which we have $\delta m^* = - \delta B$ where $ \delta B(n_s) =  (dB(n_s)/dn_s)\delta n_s$, the second of the conditions in Eq.\ (\ref{two_equations}) becomes
\begin{equation}
-n_s \frac{dB(n_s)}{dn_s} \big( \delta n_s \big) +  \Lambda \frac{dB(n_s)}{dn_s} \big(\delta n_s \big)n_s + \Lambda B(n_s) \big(\delta n_s \big)   = 0  ~,
\end{equation}
which can be immediately rewritten as a differential equation for $B(n_s)$,
\begin{equation}
\frac{dB(n_s)}{dn_s}   = \frac{\Lambda}{1 - \Lambda} \frac{B(n_s)}{n_s}  ~.
\end{equation}
Because $\Lambda$ is also a constant, here it also follows that 
\begin{eqnarray}
B (n_s) = G_m n_s^{\gamma} ~,
\end{eqnarray}
where $G_m$ is some constant and $\Lambda$ is related to $\gamma$ through
\begin{eqnarray}
\Lambda = \frac{\gamma}{\gamma + 1} ~.
\end{eqnarray}
Identifying $\gamma = d_m - 1 $ leads to the same scalar-current--dependent term as in Eq.\ (\ref{summary_energy_density}) in the case $M =1$.

\newpage
\chapter{Parametrization equations}
\label{parametrization_equations}

Here, we derive explicit forms of the parametrization equations, Eqs.\ (\ref{minimum}-\ref{spinodal_2}), in the VSDF model with an arbitrary number of vector and scalar interaction terms, $K$ and $M$. For simplicity, we consider only matter composed of protons and neutrons with degenerate masses, and their antiparticles.

The parameters of the EOS are established based on the thermodynamic properties of the system, that is for uniform nuclear matter in the rest frame. In particular, the energy density in the VSDF model, Eq.\ (\ref{summary_energy_density}), becomes
\begin{eqnarray}
\mathcal{E}_{(K,M)} \big|_{\substack{\text{rest} \\ \text{frame}}} = g \int \pdens \epsilon^{*}_{\txt{kin}}~ \big( f_{\bm{p}} + \bar{f}_{\bm{p}} \big) + \sum_{k=1}^K  \frac{C_k}{b_k} n_B^{b_k}  + \sum_{m=1}^M G_m \frac{d_m - 1}{d_m} n_s^{d_m} ~. 
\end{eqnarray}
In the following we will drop the subscript ``rest frame'' for clarity. In the limit $T \to 0$, the Fermi-Dirac distribution function for particles approaches the behavior of a Heaviside-theta function, while the distribution for antiparticles becomes zero, so that at $T=0$ we have
\begin{eqnarray}
\mathcal{E}_{(K,M)}^{(T=0)} = \frac{g}{2 \pi^2} \int_0^{p_F} dp ~  p^2 \sqrt{p^2 + {m^*}^2}    + \sum_{k=1}^K \frac{C_k }{b_k} n_B^{b_k}  + \sum_{m=1}^M G_m \frac{d_m - 1}{d_m} n_s^{d_m}  ~.
\label{energy_density_T=0}
\end{eqnarray}
The integral in the above expression can be explicitly calculated,
\begin{eqnarray}
\frac{g}{2 \pi^2} \int_0^{p_F} dp ~  p^2 \sqrt{p^2 + {m^*}^2}  = \frac{g}{16 \pi^2} \left[ 2 {E_F^*}^3 p_F - {m^*}^2 E_F^* p_F - {m^*}^4 \ln \left(\frac{E_F^* + p_F}{m^*} \right) \right] ~;
\end{eqnarray}
we note, however, that often the integral form is more convenient for calculations such as shown below. 

Similarly, the pressure in the VSDF model, Eq.\ (\ref{summary_pressure}), is easily generalized to the case with antiparticles, and adopting the notation utilizing the effective chemical potential, Eq.\ (\ref{effective_chemical_potential}), can be written as
\begin{eqnarray}
P_{(K,M)}  &=& g \int \pdens T ~  \ln \Big[1 + e^{- \beta \big(\sqrt{p^2 + {m^*}^2} - \mu^* \big)} \Big] \non \\
&& \hspace{5mm} ~ +  g \int \pdens T ~  \ln \Big[1 + e^{- \beta \big(\sqrt{p^2 + {m^*}^2} + \mu^* \big)} \Big] \non \\
&& \hspace{10mm} ~ + \sum_{k=1}^K C_k \left(\frac{b_k - 1}{b_k}\right) n_B^{b_k} - \sum_{m=1}^M G_m\left(\frac{d_m - 1}{d_m}\right) n_s^{d_m}  ~.
\label{working_pressure_formula}
\end{eqnarray}
Here as well the expression simplifies in the $T \to 0$ limit,
\begin{eqnarray}
P_{(K,M)}^{(T=0)} =  \frac{g}{2\pi^2} \int_0^{p_F} dp  ~ \frac{p^4}{3E^*}  + \sum_{k=1}^K C_k \left(\frac{b_k - 1}{b_k} \right) n_B^{b_k } - \sum_{m=1}^M G_m \frac{d_m - 1}{d_m} n_s^{d_m}  ~,
\label{pressure_T=0}
\end{eqnarray}
and the integral for the kinetic energy term can be calculated explicitly, 
\begin{eqnarray}
\frac{g}{2\pi^2} \int_0^{p_F} dp  ~ \frac{p^4}{3E^*}  =  \frac{g}{16\pi^2} \left[ \frac{2}{3} {E_F^*} p_F^3  - {m^*}^2 E_F^* p_F + {m^*}^4 \ln \left( \frac{E_F^* + p_F}{m^*} \right) \right] ~.
\end{eqnarray}

\section{The minimum of the energy per particle}
\label{the_minimum_of_the_energy_per_particle}

The position of the minimum of the binding energy per particle, which is equivalent to the minimum of the energy per particle (where we note that $E/N_B = \mathcal{E}/n_B$), is determined by the equation
\begin{eqnarray}
\frac{d}{dn_B} \left(\frac{\mathcal{E}^{(T=0)}_{(K,M)}}{n_B} \right) = 0  ~,
\label{binding_energy_minimum_0}
\end{eqnarray}
the solution to which gives the saturation density $n_0$. At zero temperature, the energy density is a function of number density $n_B$ only. However, due to the presence of many terms with complicated $n_B$-dependence, see Eq.\ (\ref{energy_density_T=0}), we have
\begin{eqnarray}
\mathcal{E}^{(T=0)}_{(K,M)} = \mathcal{E}^{(T=0)}_{(K,M)} \Big(p_F(n_B), m^*(n_s), n_B, n_s(n_B)\Big)
\end{eqnarray} 
and it is convenient to rewrite the derivative using the chain rule,
\begin{eqnarray}
\frac{d}{dn_B} =  \parr{ }{n_B}\bigg|_{p_F, n_s} + \frac{d p_F}{d n_B} \parr{ }{p_F}\bigg|_{n_B, n_s} +  \frac{d n_s}{d n_B}\left(\parr{ }{n_s}\Big|_{m^*} + \frac{d m^*}{ dn_s} \parr{ }{m^*}\Big|_{n_s}  \right)\bigg|_{n_B, p_F}~.
\end{eqnarray}
First, we can see that 
\begin{eqnarray}
&&\hspace{-15mm} \left(\parr{ }{n_s}\Big|_{m^*} + \frac{d m^*}{ dn_s} \parr{ }{m^*}\Big|_{n_s}  \right)\bigg|_{n_B, p_F}  \mathcal{E}^{(T=0)}_{(K,M)} \non \\
&& \hspace{-3mm} =  \sum_{m=1}^M G_m (d_m - 1) n_s^{d_m - 1} - \sum_{m=1}^M G_m (d_m - 1) n_s^{d_m - 2} ~ \frac{g}{2\pi^2} \int_0^{p_F} dp ~  p^2 ~  \frac{m^*}{\epsilon_{\txt{kin}}^*}  = 0 ~. 
\end{eqnarray}
Here we note that while the vanishing of the derivative of the energy density with respect to the auxiliary field $n_s$ is reminiscent of the vanishing of the same derivative of the pressure, where the latter is a consequence of the minimization of the thermodynamic potential with respect to the auxiliary field (see Appendix \ref{minimization_of_the_auxiliary_fields}), in the case of the energy density this derivative vanishes only at $T=0$ due to the trivial form of the distribution function, and should not be thought of as the minimization condition. Next, we calculate
\begin{eqnarray}
\left(\parr{ }{n_B}\bigg|_{p_F, n_s} + \frac{d p_F}{d n_B} \parr{ }{p_F}\bigg|_{n_B, n_s}  \right) \mathcal{E}^{(T=0)}_{(K,M)}  &=& \sum_{k=1}^K C_k n_B^{b_k - 1} + \frac{d p_F}{d n_B}  ~ \frac{g}{2\pi^2} p_F^2 \sqrt{ p_F^2 + {m^*}^2 } \non \\
&=& \sqrt{ p_F^2 + {m^*}^2 } + \sum_{k=1}^K C_k n_B^{b_k - 1}    ~,
\end{eqnarray}
where we have used $p_F = \left( \frac{6 \pi^2 n_B}{g} \right)^{1/3}$ (see Eq.\ (\ref{Fermi_momentum}) and below). Thus we have
\begin{eqnarray}
\frac{d \mathcal{E}^{(T=0)}_{(K,M)} }{dn_B} = \epsilon_{F}  + \sum_{k=1}^K C_k n_B^{b_k - 1} ~,
\label{derivative_of_energy_density_over_density}
\end{eqnarray}
where $\epsilon_{F}$ is the kinetic energy at the Fermi surface, $\epsilon_{F} = \sqrt{p_F^2 + {m^*}^2}$.

Let us note that the above result can be obtained in a much easier way: at $T=0$, the first law of thermodynamics $d\mathcal{E} = T ds + \mu_B dn_B$ reduces to $ d\mathcal{E} = \mu_B dn_B$, so that 
\begin{eqnarray}
\frac{d \mathcal{E}^{(T=0)}_{(K,M)} }{dn_B}  = \mu_B = \eps_{F} ~,
\end{eqnarray}
where by $\eps_{F}$ we denote the quasiparticle energy at the Fermi surface.

Altogether, in the VSDF model Eq.\ (\ref{binding_energy_minimum_0}) can be explicitly written as
\begin{eqnarray}
\frac{1}{n_B} \left( \eps_{F}   - \frac{\mathcal{E}^{(T=0)}_{(K,M)}}{n_B} \right) = 0  ~.
\label{eq98763}
\end{eqnarray}
This agrees with the Hugenholtz-van-Hove theorem \cite{Hugenholtz:1958zz}, which states that $\eps_{F}  = (E_0/N) + (P/n)$, where $E_0$ is the system's ground state energy, and from which it follows that for systems at equilibrium (where pressure $P=0$) the Fermi energy and the energy per particle in the system, $ E/N = \mathcal{E}/n_B$, are equal. The fact that $P$ vanishes at equilibrium an be easily seen for systems at zero temperature, as at $T=0$ the pressure is given by
\begin{eqnarray}
P = n_B^2 \derr{\left(\frac{\mathcal{E}}{n_B} \right)}{n_B} ~,
\label{pressure_definition_T=0}
\end{eqnarray}
which is automatically zero if Eq.\ (\ref{binding_energy_minimum_0}) holds, that is if the system is in equilibrium.

\section{Spinodal region boundary at $T = 0$}

The boundaries of the spinodal region at $T=0$ are given by the extremal points of the pressure, given by
\begin{eqnarray}
\frac{dP_{(K,M)}^{(T=0)} }{dn_B} = 0 ~. 
\end{eqnarray}
Similarly as in the case of energy density considered in the previous section, pressure has a complicated dependence on baryon density $n_B$, see Eq.\ (\ref{pressure_T=0}), and so it is convenient to use the chain rule,
\begin{eqnarray}
\frac{d}{dn_B}\bigg|_{T}  =  \parr{}{n_B}\bigg|_{T, p_F,n_s} +  \frac{dp_F}{dn_B} \parr{}{p_F}\bigg|_{T, n_B, n_s} + \frac{dn_s}{dn_B} \parr{}{n_s} \bigg|_{T, n_B, p_F}  ~.
\end{eqnarray}
We note that while the last term in the equations above disappears in minimization of the thermodynamic potential when $T$ and $\mu_B$ are held fixed, in this case it is finite. We calculate
\begin{eqnarray}
&&\hspace{-18mm} \parr{P_{(K,M)}^{(T=0)}}{n_B}\bigg|_{p_F,n_s}  = \sum_{k=1}^K C_k (b_k - 1) n_B^{b_k -1}~, \\
&& \hspace{-18mm} \frac{dp_F}{dn_B} \parr{P_{(K,M)}^{(T=0)}}{p_F}\bigg|_{n_B, n_s}  = \frac{p_F^2  }{3E_F^*} ~, \\
&& \hspace{-18mm} \frac{dn_s}{dn_B} \parr{P_{(K,M)}^{(T=0)}}{n_s} \bigg|_{n_B, p_F}  =  \frac{dn_s}{dn_B}  \bigg[   \frac{g}{16\pi^2} \frac{dm^*}{dn_s}  \bigg( \frac{8p_F^3 m^*      }{3E_F^*}    \bigg) - n_s \frac{dm^*}{dn_s}  -  \sum_{m=1}^M G_m ( d_m - 1 ) n_s^{d_m - 1}   \bigg]~, \label{eq456}
\end{eqnarray}
where we note that the last equation requires a considerable amount of algebra, including using the expression for the scalar density at $T=0$, 
\begin{eqnarray}
n_s = \frac{g}{4\pi^2} \left[ m^* E_F^* p_F - {m^*}^3 \ln \left( \frac{E_F^* + p_F}{m^*} \right)  \right] ~.
\label{scalar_density_T=0}
\end{eqnarray}
By the gap equation, Eq.\ (\ref{summary_effective_mass_definition}), we have
\begin{eqnarray}
-n_s \frac{dm^*}{dn_s} =  \sum_{m=1}^M G_m (d_m- 1) n_s^{d_m - 1}~,
\end{eqnarray}
so that Eq.\ (\ref{eq456}) can be rewritten as
\begin{eqnarray}
\frac{dn_s}{dn_B} \parr{P_{(K,M)}^{(T=0)}}{n_s} \bigg|_{n_B, p_F}  &=&    \frac{g}{16\pi^2}  \bigg(  \frac{8p_F^3 m^*      }{3E_F^*}    \Bigg)  \frac{dm^*}{dn_B}  =      \frac{p_F  m^* }{3E_F^*}  \frac{dm^*}{dp_F}~.
\end{eqnarray}

The gap equation can be further used to calculate
\begin{eqnarray}
\parr{m^*}{p_F} = - \sum_{m=1}^M G_m (d_m - 1) n_s^{d_m - 2} \parr{n_s}{p_F} = - \mathcal{G} \parr{n_s}{p_F}~,
\label{dmstar_dpF}
\end{eqnarray}
where we have introduced a short-hand notation for the sum over interaction terms,
\begin{eqnarray}
\mathcal{G} = \sum_{m=1}^M G_m (d_m - 1) n_s^{d_m - 2} ~.
\label{sum_scalar_interaction_terms}
\end{eqnarray} 
Using Eq.\ (\ref{scalar_density_T=0}), we further have
\begin{eqnarray}
\hspace{-15mm}\frac{dn_s}{dp_F} &=& \frac{g}{4\pi^2} \Bigg\{  \frac{2m^* p_F^2  }{E_F^*}      \non\\
&& \hspace{-10mm} + ~ \left[ E_F^* p_F  + \frac{ {m^*}^2 p_F }{E_F^*}  - 3{m^*}^2 \ln \left( \frac{E_F^* + p_F }{m^*}\right)     -   \frac{  {m^*}^4  }{E_F^*(E_F^* + p_F )}   + {m^*}^2   \right] \frac{dm^*}{dp_F} \Bigg\}~.
\label{dns_dpF}
\end{eqnarray}
With
\begin{eqnarray}
{m^*}^4 = \Big(  {m^*}^2   \Big)^2 =  \Big(  {E_F^*}^2 - p_F^2   \Big)^2 =  \Big(  ({E_F^*} - p_F)  ({E_F^*} + p_F) \Big)^2 
\end{eqnarray}
the term in the square bracket can be rewritten as
\begin{eqnarray}
\Big[ \dots\Big] =   E_F^*p_F  + \frac{  2{m^*}^2p_F}{E_F^*}   - 3{m^*}^2 \ln \left( \frac{E_F^* + p_F }{m^*}\right) ~,
\end{eqnarray}
and inserting Eq.\ (\ref{dns_dpF}) back into Eq.\ (\ref{dmstar_dpF}) finally yields
\begin{eqnarray}
\frac{dm^*}{dp_F}  =  - \frac {   \frac{2m^* p_F^2  }{E_F^*}    }{   \infrac{4\pi^2} {g\mathcal{G}} +    \left[ E_F^* p_F  + \frac{  2{m^*}^2p_F}{E_F^*}   - 3{m^*}^2 \ln \left( \frac{E_F^* + p_F }{m^*}\right)   \right]    }~,
\label{d_mstar_d_pfermi}
\end{eqnarray}
where $\mathcal{G}$ is given by Eq.\ (\ref{sum_scalar_interaction_terms}). 

Altogether we then have
\begin{eqnarray}
\frac{dP_{(K,M)}^{(T=0)}}{dn_B} = \frac{p_F^2  }{3E_F^*}  + \sum_{k=1}^K C_k (b_k - 1) n_B^{b_k -1} +   \frac{p_F  m^* }{3E_F^*}  \frac{dm^*}{dp_F}   ~,
\label{dP_dnB_at_T=0}
\end{eqnarray}
where $\frac{dm^*}{dp_F}$ is given by Eq.\ (\ref{d_mstar_d_pfermi}).

\section{Incompressibility}

Classically, compressibility is defined as
\begin{eqnarray}
\beta \equiv -  \frac{1}{V} \left( \parr{P}{V} \right)^{-1} =  \frac{1}{V} \left(\frac{n_B^2}{N_B}  \parr{P}{n_B} \right)^{-1}~,
\end{eqnarray}
so that the incompressibility is
\begin{eqnarray}
\beta^{-1} = V \frac{n_B^2}{N_B}  \parr{P}{n_B} = n_B \parr{P}{n_B} ~.
\end{eqnarray}
In nuclear physics, however, it is customary to use the following expression for incompressibility,
\begin{eqnarray}
\beta^{-1} \to K = 9  \parr{P}{n_B} ~.
\label{the_real_definition_of_incompressibility}
\end{eqnarray}

Formally, the incompressibility of nuclear matter at zero temperature and saturation density, $n_B = n_0$, is defined as
\begin{eqnarray}
K_0 \equiv \left( p_F^2 \frac{d^2}{dp_F^2} \left( \frac{\mathcal{E}^{(T=0)}_{(K,M)}}{n_B}\right) \right)_{n_B=n_0} ~. 
\end{eqnarray}
By expressing the derivative with respect to the Fermi momentum $p_F$ in terms of the derivative with respect to the baryon density, $\infrac{d}{dp_F} = \infrac{3n_B}{p_F} \infrac{d}{dn_B}$, and using the fact that at $n_B = n_0$ we have $d  \infrac{\mathcal{E}^{(T=0)}_{(K,M)}}{n_B}/dn_B = 0$, we can rewrite the above equation to obtain
\begin{eqnarray}
K_0 = \left( 9n_B^2\left[\frac{d^2}{dn_B^2}\left( \frac{\mathcal{E}^{(T=0)}_{(K,M)}}{n_B}\right) \right] \right)_{n_B=n_0} ~.
\label{compressibility_10}
\end{eqnarray}
Furthermore, again using the fact that $n_0$ is  the saturation density, this can be rewritten as
\begin{eqnarray}
K_0 = \left(9n_B  \frac{d^2\mathcal{E}^{(T=0)}_{(K,M)}}{dn_B^2} \right)_{n_B=n_0} ~.
\label{compressibility_20}
\end{eqnarray}
Noting that the pressure at $T=0$ is given by Eq.\ (\ref{pressure_definition_T=0}), we can at the same time calculate
\begin{eqnarray}
\frac{dP}{dn_B} = \bigg|_{n_B = n_0} \left(n_B^2 \frac{d^2}{dn_B^2} \left(\frac{\mathcal{E}}{n_B} \right) \right)_{n_B=n_0} ~,
\end{eqnarray}
from which, by comparison with Eq.\ (\ref{compressibility_10}), it follows immediately that
\begin{eqnarray}
K_0 = 9 \left(\frac{dP}{dn_B} \right)_{n_B = n_0} ~,
\label{compressibility_30}
\end{eqnarray}
in agreement with Eq.\ (\ref{the_real_definition_of_incompressibility}).

To calculate the incompressibility, we will use Eq.\ (\ref{compressibility_20}). We know the first derivative of energy density from (\ref{derivative_of_energy_density_over_density}). Using the chain rule we can write
\begin{eqnarray}
\frac{d}{dn_B} =  \parr{}{n_B}\bigg|_{p_F,m^*} +  \frac{dp_F}{dn_B} \parr{}{p_F}\bigg|_{n_B, m^*} + \frac{dp_F}{dn_B} \frac{dm^*}{dp_F} \parr{}{m^*} \bigg|_{n_B, p_F} ~,
\end{eqnarray}
from which we get
\begin{eqnarray}
\frac{d}{dn_B} \left(\frac{d \mathcal{E}^{(T=0)}_{(K,M)}}{dn_B} \right) =  \frac{p_F}{3n_B} \left[ \frac{p_F}{E_F^*} +  \parr{m^*}{p_F} \frac{m^*}{E_F^*} \right] + \sum_{k=1}^K C_k (b_k - 1) n_B^{b_k - 2}  ~,
\label{incompressibility_9876}
\end{eqnarray}
leading to
\begin{eqnarray}
K_0\bigg|_{n_B=n_0} =  \left[ 9n_B  \left( \frac{p_F^2}{3n_BE_F^*}   + \sum_{k=1}^K C_k (b_k - 1) n_B^{b_k - 2}  +  \frac{p_F m^*}{3n_BE_F^*} \frac{dm^*}{dp_F} \right)  \right]_{n_B=n_0} ~,
\label{incompressibility_from_energy_density}
\end{eqnarray}
where $\frac{dm^*}{dp_F}$ is given by (\ref{d_mstar_d_pfermi}).

We note that using Eq.\ (\ref{compressibility_30}) leads to the same result, as is clear from Eq.\ (\ref{dP_dnB_at_T=0}).

\section{The critical point}

To apply the conditions for identifying the location of the critical point at $(n_c, T_c)$,
\begin{eqnarray}
&& \left(\frac{dP }{dn_B} \right)_T(n_c, T_c) = 0~,
\label{P_first_derivative_00} \\
&& \left(\frac{d^2P }{dn_B^2} \right)_T(n_c, T_c) = 0~.
\label{P_second_derivative_00}
\end{eqnarray}
one needs to know the first and second derivative of the pressure with respect to baryon number density. 

Here we note that the resulting formulas are relatively complicated. In view of this, it may be more convenient to utilize numerical derivatives in solving Eqs.\ (\ref{P_first_derivative_00}) and (\ref{P_second_derivative_00}). Nevertheless, obtaining analytical formulas is possible, and in particular they work very well in the case of the VDF model, where scalar type interactions are neglected ($M=0$). To illustrate the calculation, below we compute the first derivative of the pressure in the VSDF model. The formula for the VDF model is easily obtained by taking $M=0$. The second and third derivatives of the pressure in the VDF model are provided in Appendix \ref{pressure_derivatives_in_the_VDF_model}.

\subsection{First derivative of the pressure }

We want to calculate $\inbfrac{dP_{(K,M)}}{dn_B}\big|_{T}$. It's clear from Eq.\ (\ref{working_pressure_formula}) that the pressure is a function of
\begin{eqnarray}
P_{(K,M)} = P_{(K,M)} \Big(T, m^*\big( n_s(T, n_B) \big), \mu^*(T, n_B) , n_B, n_s(T, n_B) \Big).
\end{eqnarray}
We will keep $T$ constant, so that the derivative with respect to $n_B$ can be written as
\begin{eqnarray}
\frac{d}{dn_B} = \parr{}{n_B}\bigg|_{T, \mu^*, n_s} + \frac{d\mu^*}{dn_B}\parr{}{\mu^*}\bigg|_{T, n_B, n_s}  +  \frac{dn_s}{dn_B}\parr{}{n_s}\bigg|_{T, n_B, \mu^*} ~.
\label{pressure_derivatives_chain_rule_anti}
\end{eqnarray}
Note here that the effective mass is an explicit function of $n_s$, $m^* = m^*(n_s)$, and that we take $\mu^*$, as opposed to $\mu_B$, as our chosen variable; an important thing to bear in mind is that one needs to be consistent about this decision.

We have
\begin{eqnarray}
\parr{P}{n_B}\bigg|_{T, \mu^*, n_s} = \sum_{k=1}^{K} C_k (b_k - 1) n_B^{b_k -1} 
\end{eqnarray}
and 
\begin{eqnarray}
\frac{d\mu^*}{dn_B}\parr{P}{\mu^*}\bigg|_{T, n_B, n_s} = \left(\frac{dn_B}{d\mu^*}\right)^{-1} n_B~,
\end{eqnarray}
where the latter equation follows directly from Eq.\ (\ref{working_pressure_formula}). The last derivative term stemming from Eq.\ (\ref{pressure_derivatives_chain_rule_anti}) disappears because the scalar field minimizes the thermodynamic potential (see Appendix \ref{minimization_of_the_auxiliary_fields}, which can be easily generalized to the case with antiparticles, considered here), so that altogether
\begin{eqnarray}
\frac{dP}{dn_B}\bigg|_{T} =  n_B\left(\frac{dn_B}{d\mu^*}\right)^{-1}  + \sum_{k=1}^{K} C_k (b_k - 1) n_B^{b_k -1} ~.
\end{eqnarray}

We introduce the following vector-density-- and scalar-density--like integrals over powers of the distribution function,
\begin{eqnarray}
&& \mathcal{B}_a = g \int \pdens f_{\bm{p}}^a~,  \\
&& \barr{\mathcal{B}}_a = g \int \pdens \bar{f}_{\bm{p}}^a~,  \\
&& \mathcal{S}_a = g \int \pdens \frac{m^*}{E^*} ~ f_{\bm{p}}^a~, \\ 
&& \barr{\mathcal{S}}_a = g \int \pdens \frac{m^*}{E^*} ~ \bar{f}_{\bm{p}}^a ~.
\end{eqnarray}
Using these, we calculate the following derivative,
\begin{eqnarray}
\frac{d\big(\mathcal{B}_k \pm \barr{\mathcal{B}}_k \big)}{d\mu^*}  &=& \beta k \Bigg[ \big( \mathcal{B}_k - \mathcal{B}_{k+1}  \big)  \mp  \big( \barr{\mathcal{B}}_k - \barr{\mathcal{B}}_{k+1}  \big) \non \\
&& \hspace{10mm}  - ~  \derr{m^*}{\mu^*} \bigg( \big( \mathcal{S}_k  - \mathcal{S}_{k+1} \big)\pm \big( \barr{\mathcal{S}}_k  - \barr{\mathcal{S}}_{k+1} \big) \bigg) \Bigg] ~.
\end{eqnarray}
In particular, we have $n_B = \mathcal{B}_1 - \barr{\mathcal{B}}_1$, so that 
\begin{eqnarray}
\frac{dn_B}{d\mu^*} &=&\beta \Bigg[ \Big( \big(\mathcal{B}_1 - \mathcal{B}_2\big) + \big(\barr{\mathcal{B}}_1 - \barr{\mathcal{B}}_2\big)   \Big)    - \frac{dm^*}{d\mu^*} \bigg( \big(  \mathcal{S}_1 - \mathcal{S}_2 \big) - \big(  \barr{\mathcal{S}}_1 - \barr{ \mathcal{S}}_2 \big) \bigg) \Bigg] ~.
\end{eqnarray}

Next, we can always rewrite
\begin{eqnarray}
\frac{dm^*}{d\mu^*} = \frac{dn_s}{d\mu^*}\frac{dm^*}{dn_s} = - \frac{dn_s}{d\mu^*} \sum_{m=1}^M G_m (d_m -1)n_s^{d_m - 2} = - \mathcal{G}\frac{dn_s}{d\mu^*} ~,
\label{eq876_anti}
\end{eqnarray}
where $\mathcal{G}$ is given by Eq.\ (\ref{sum_scalar_interaction_terms}). In general, we can calculate
\begin{eqnarray}
\frac{d\big( \mathcal{S}_k \pm \barr{\mathcal{S}}_k   \big)}{d\mu^*} &=& \frac{dm^*}{d\mu^*} \Bigg[ g \int \pdens \frac{p^2}{{\epsilon_{\txt{kin}}^*}^3} ~ \big( f_{\bm{p}}^k \pm  \bar{f}_{\bm{p}}^k \big) \\
&& \hspace{10mm} - ~   \beta k ~  g \int \pdens   \frac{{m^*}^2}{{\epsilon_{\txt{kin}}^*}^2} ~ \bigg( \big(f_{\bm{p}}^k \mp  \bar{f}_{\bm{p}}^k\big) - \big( f_{\bm{p}}^{k+1} \mp \bar{f}_{\bm{p}}^{k+1} \big) \bigg)   \Bigg] \\
&& \hspace{20mm} + ~ \beta k \Bigg[ \big( \mathcal{S}_k - \mathcal{S}_{k+1} \big) \pm    \big( \barr{\mathcal{S}}_k - \barr{\mathcal{S}}_{k+1} \big)   \Bigg]~.
\end{eqnarray}
In particular, the scalar density $n_s = \mathcal{S}_1 + \barr{\mathcal{S}}_1$, so that
\begin{eqnarray}
\frac{dn_s}{d\mu^*} &=& \frac{dm^*}{d\mu^*} \Bigg[ g \int \pdens \frac{p^2}{{\epsilon_{\txt{kin}}^*}^3} ~ \big( f_{\bm{p}} +  \bar{f}_{\bm{p}} \big) \\
&& \hspace{10mm} - ~   \beta ~  g \int \pdens   \frac{{m^*}^2}{{\epsilon_{\txt{kin}}^*}^2} ~ \bigg( \big(f_{\bm{p}} -  \bar{f}_{\bm{p}}\big) - \big( f_{\bm{p}}^{2} - \bar{f}_{\bm{p}}^{2} \big) \bigg)   \Bigg] \\
&& \hspace{20mm} + ~ \beta k \Bigg[ \big( \mathcal{S}_1- \mathcal{S}_{2} \big) +   \big( \barr{\mathcal{S}}_1 - \barr{\mathcal{S}}_{2} \big)   \Bigg] ~. 
\label{eq987654_anti}
\end{eqnarray}
Inserting Eq.\ (\ref{eq987654_anti}) into Eq.\ (\ref{eq876_anti}) leads to
\begin{eqnarray}
\frac{dm^*}{d\mu^*} = \frac{- \mathcal{G} \beta k \Bigg[ \big( \mathcal{S}_1- \mathcal{S}_{2} \big) +   \big( \barr{\mathcal{S}}_1 - \barr{\mathcal{S}}_{2} \big)   \Bigg] }{1 +  \mathcal{G}  g \int \pdens \Bigg[   \frac{p^2}{{\epsilon_{\txt{kin}}^*}^3} ~ \big( f_{\bm{p}} +  \bar{f}_{\bm{p}} \big) -   \beta   \frac{{m^*}^2}{{\epsilon_{\txt{kin}}^*}^2} ~ \bigg( \big(f_{\bm{p}} -  \bar{f}_{\bm{p}}\big) - \big( f_{\bm{p}}^{2} - \bar{f}_{\bm{p}}^{2} \big) \bigg)   \Bigg] } ~.
\label{d_mstar_d_mustar_anti}
\end{eqnarray}
Putting all of these terms together, the first derivative of the pressure with respect to $n_B$ is
\begin{eqnarray}
\frac{dP}{dn_B}\bigg|_{T} &=&  \frac{Tn_B}{ \Big( \big(\mathcal{B}_1 - \mathcal{B}_2\big) + \big(\barr{\mathcal{B}}_1 - \barr{\mathcal{B}}_2\big)   \Big)    - \frac{dm^*}{d\mu^*} \bigg( \big(  \mathcal{S}_1 - \mathcal{S}_2 \big) - \big(  \barr{\mathcal{S}}_1 - \barr{ \mathcal{S}}_2 \big) \bigg) } \non \\
&& \hspace{20mm} + ~ \sum_{k=1}^{K} C_k (b_k - 1) n_B^{b_k -1} ~,
\end{eqnarray}
where $\infrac{dm^*}{d\mu^*}$ is given by Eq.\ (\ref{d_mstar_d_mustar_anti}).

\newpage
\chapter{VDF model parameter sets}
\label{parameter_sets}

\begin{table*}[b]
	\caption[Parameter sets for several VDF EOSs]{Parameter sets corresponding to the VDF EOSs reproducing sets of the QGP-like phase transition characteristics $\big( T_c^{(Q)}, n_c^{(Q)}, \eta_L, \eta_R\big)$, listed in Table \ref{example_characteristics}.}
	\label{parameters}
	\begin{center}
		\bgroup
		\def\arraystretch{1.4}
		\begin{tabular}{c r r r r }
			\hline
			\hline
			\hspace{3mm}set\hspace{3mm} & \hspace{25mm} $b_1$ \hspace{0mm} & \hspace{25mm} $b_2$ \hspace{0mm} &  \hspace{25mm} $b_3$ \hspace{0mm} & \hspace{25mm} $b_4$ \hspace{0mm} \\ 
			& \hspace{8mm} $\tilde{C}_1$ [MeV] \hspace{0mm} & \hspace{8mm} $\tilde{C}_2$ [MeV] \hspace{0mm} & \hspace{8mm} $\tilde{C}_3$ [MeV] \hspace{0mm} & \hspace{8mm} $\tilde{C}_4$ [MeV] \hspace{0mm} \\
			\hline
			\hline
			I & 1.7614679  & 3.8453863  & 4.4772660  & 6.7707861  \\
			& -8.315987e+01  &  6.144706e+01   &  -3.108395e+01   & 3.127069e-01   \\  
			\hline
			II & 1.8033077  & 3.0693813  & 7.9232548  & 10.7986978  \\
			&   -9.204350e+01  & 3.968766e+01  & -1.306487e-01  & 2.434034e-03 \\  
			\hline
			III & 1.8042024  & 3.0631798  & 6.6860893  & 20.7276154  \\
			& -9.224000e+01  & 3.986263e+01  & -1.066766e-01  & 2.160279e-11 \\  
			\hline
			IV & 1.7681391  & 3.5293515  & 5.4352787  & 6.3809823   \\
			& -8.450948e+01  & 3.843139e+01  & -7.958557e+00  & 1.552593e+00\\  
			\hline
			V & 1.8007135    &  3.0931706   &   6.6396492   &   8.1076981  \\
			&   -9.142377e+01   &  3.925851e+01   &  -3.439980e-01  &   2.520940e-02 \\  
			\hline
			VI & 1.7989835  & 3.1098389 & 6.3017683  & 8.0937872 \\
			& -9.101665e+01 & 3.899891e+01 &-4.856681e-01 & 1.935808e-02 \\
			\hline
			\hline
		\end{tabular}
		\egroup
	\end{center}
\end{table*}

Here we provide parameters corresponding to the VDF EOSs reproducing sets of the QGP-like phase transition characteristics $\big( T_c^{(Q)}, n_c^{(Q)}, \eta_L, \eta_R\big)$, listed in Table \ref{example_characteristics}. It is important to note that the values of the coefficients of the interaction terms, $\{C_1, C_2, C_3, C_4 \}$, depend on a chosen system of units. Here, we adopt a convention used in many Skyrme-like parametrizations, in which the single-particle potential is written in the form
\begin{eqnarray}
U = \sum_{i=1}^N  \tilde{C}_i \left( \frac{n_B}{n_0} \right)^{b_i - 1} ~,
\end{eqnarray}
where $n_0$ is the saturation density, so that $\tilde{C}_i$ must have a dimension of energy.  Naturally, $\tilde{C}_i$ and $C_i$ are related by
\begin{eqnarray}
C_i = \frac{\tilde{C}_i }{n_0^{b_i - 1}} ~. 
\end{eqnarray}

In Table \ref{parameters}, we list coefficients $\{\tilde{C}_1,  \tilde{C}_2, \tilde{C}_3, \tilde{C}_4\}$ in units of MeV. Note that in particular, the sum of all coefficients yields the (rest frame) value of the single-particle potential at $n_B = n_0$, $\sum_{i=1}^N \tilde{C}_i = -52.484 \ \txt{MeV}$.

\newpage
\chapter{Symmetric spinodal regions}
\label{symmetric_spinodal_regions}

The spinodal region is the range of baryon number densities between two local extrema of pressure, a maximum at $\eta_L$ and a minimum at $\eta_R$, with $\eta_L < \eta_R$. A curve exhibiting two extrema will most naturally have an inflection point approximately in between them. We can see this by considering the following polynomial:
\begin{eqnarray}
f(x) = ax^3 + bx^2 + cx + d ~,
\end{eqnarray}
which is a ``minimal'' polynomial needed to produce two local extrema. The condition for an extremum at some point $x_0$ is
\begin{eqnarray}
\frac{df}{dx}\bigg|_{x = x_0} = 3ax^2 + 2bx + c \bigg|_{x = x_0}  = 0 ~.  
\end{eqnarray}
We can solve this equation to yield the positions of the extrema $x_L$ and $x_R$,
\begin{eqnarray}
&& x_L = \frac{-b - \sqrt{b^2 - 3ac}}{3a} ~, \\
&& x_R = \frac{-b + \sqrt{b^2 - 3ac}}{3a}  ~. 
\end{eqnarray}
The position of the inflection point is established through the condition
\begin{eqnarray}
\frac{d^2f}{dx^2}\bigg|_{x = x_{\txt{infl}}} = 6ax + 2b \bigg|_{x = x_{\txt{infl}}} = 0 ~,
\end{eqnarray}
from which we get 
\begin{eqnarray}
x_{\txt{infl}} = - \frac{b}{3a}~.
\end{eqnarray}
It is immediately apparent that 
\begin{eqnarray}
x_{\txt{infl}} = \frac{x_L + x_R}{2}~,
\end{eqnarray}
placing the inflection point exactly in the middle between the two extrema. This result is only exact for a third-order polynomial, and will be changed if the polynomial includes additional terms with which one is able to manipulate the behavior of the curve between the extrema.

We will now argue that in a model with vector-type interactions only, the inflection point of the pressure curve at zero temperature, 
\begin{eqnarray}
\frac{d^2P(T=0)}{dn_B^2} \bigg|_{n_B = n_{\txt{infl}}} = 0~,
\end{eqnarray}
will coincide with the location of the critical point on the $n_B$ axis. Let us first write the pressure as a sum of an ideal gas term and an interaction term,
\begin{eqnarray}
P = P_{\txt{ideal}} + P_{\txt{int}} ~.
\end{eqnarray}
In particular, at $T=0$ the ideal part of the pressure is given by the ideal Fermi gas, $P_{\txt{ideal}}(T=0) = P^{\txt{FG}}_0$. Because the Fermi gas at zero temperature depends on the baryon density as $P^{\txt{FG}}_0 \propto n_B^{4/3}$, for large densities we can safely assume that
\begin{eqnarray}
\frac{d^2 P^{\txt{FG}}_0 }{dn_B^2} = \frac{4}{9} n_B^{-2/3} \approx 0 ~.
\label{eq457}
\end{eqnarray}
It then follows that at the inflection point we must have
\begin{eqnarray}
\frac{d^2P_{\txt{int}}}{dn_B^2} \bigg|_{n_B = n_{\txt{infl}}} \approx 0 ~.
\end{eqnarray}
At the same time, the condition for the position of the critical point at some location $(T_c, n_c)$ leads to
\begin{eqnarray}
\frac{d^2 P_{\txt{int}}}{dn_B^2} \bigg|_{n_B = n_c} = - \frac{d^2 P_{\txt{ideal}}}{dn_B^2} \bigg|_{\substack{n_B = n_c\\ T = T_c }} ~.
\label{critical_point_two_pressure_parts}
\end{eqnarray}
For large enough temperatures, the ideal Fermi gas is well approximated by the ideal Boltzmann gas, and we can write the ideal part of the pressure as 
\begin{eqnarray}
P_{\txt{ideal}} \approx T n_B ~.
\end{eqnarray}
As a result, Eq.\ (\ref{critical_point_two_pressure_parts}) becomes
\begin{eqnarray}
\frac{d^2 P_{\txt{int}}}{dn_B^2} \bigg|_{n_B = n_c} = 0~,
\end{eqnarray}
which immediately confirms that in this case, the location of the critical density $n_c$ coincides with the location of the inflection point $n_{\txt{infl}}$ of the pressure at zero temperature. Moreover, going beyond the approximation used in Eq.\ \eqref{eq457}, we see that at zero temperature the pressure at $n_B= n_c$ will have a very small and positive curvature, which means that the critical density is somewhat larger than the inflection point density, $n_c \gtrsim n_{\txt{infl}} $.

The VDF model largely reproduces the behavior described above. First, due to the fact that the pressure fits in the VDF model are ``minima'' fits reproducing (among other constraints) two local extrema, a maximum at $\eta_L$  and a minimum at $\eta_R$, the inflection point of the pressure lies roughly in the middle between $\eta_L$ and $\eta_R$. Second, due to the thermal part of the pressure being just like that of an ideal gas, the location of the critical point $n_c$ and the location of the inflection point of the pressure at zero temperature $n_{\txt{infl}}$ are related by $n_c = n_{\txt{infl}} + \delta n$, where $\delta n$ is a small positive correction. This explains why in the VDF model the critical baryon number density $n_c$ lies roughly in the middle of the spinodal region $(\eta_L, \eta_R)$.

\newpage
\chapter{The speed of sound}
\label{the_speed_of_sound}

The isentropic speed of sound is defined as the following total derivative
\begin{eqnarray}
c_{\sigma}^2 \equiv \left( \derr{P}{\mathcal{E}} \right)_{S/N} ~,
\label{the_isentropic_speed_of_sound_definition}
\end{eqnarray} 
where $\sigma \equiv S/N$, $S$ is the entropy, and $N$ is the number of particles, so that $\sigma$ is the entropy per particle. The isothermal speed of sound is defined as
\begin{eqnarray}
c_{T}^2 \equiv \left( \derr{P}{\mathcal{E}} \right)_{T} ~,
\label{the_isothermal_speed_of_sound_definition}
\end{eqnarray} 
In practice, carrying out either of these derivatives depends strongly on the primary variables in terms of which the variations in pressure, energy density, and entropy per particle can be obtained.

In general, the total differential of the pressure can be given in terms of its two primary variables $A$ and $B$, 
\begin{eqnarray}
dP(A,B) = \left(\derr{P}{A} \right)_{B} dA + \left( \derr{P}{B} \right)_{A}  dB ~.
\label{pressure_general}
\end{eqnarray}
In order to calculate the speed of sound as defined in Eq.\ (\ref{the_isentropic_speed_of_sound_definition}) or (\ref{the_isothermal_speed_of_sound_definition}), we instead need to have 
\begin{eqnarray}
dP\big(\mathcal{E}, Y\big) = \left(\derr{P}{\mathcal{E}} \right)_{Y} d\mathcal{E}  + \left(\derr{P}{Y}\right)_{\mathcal{E}} dY ~,
\end{eqnarray}
where $Y = \{ (S/N), T \}$. The above equation has the obvious advantage that we can explicitly set $dY$ to zero. 

The following is based on the assumption that since $A$ and $B$ are the primary variables in which the pressure is given (either as an analytic formula or as a table in $A$ and $B$), we know what the derivatives $\left( \inderr{P}{A} \right)_{B}$ and $\left( \inderr{P}{B} \right)_{A}$ are. Therefore, we only need to find expressions for $dA$ and $dB$ in terms of $d\mathcal{E}$ and $dY$, as this will allow us to compute $\left(\derr{P}{\mathcal{E}} \right)_{Y}$ and $\left(\derr{P}{Y}\right)_{\mathcal{E}}$. To this end we express the total differentials of $\mathcal{E}$ and $C$ in terms of $A$ and $B$,
\begin{eqnarray}
&& d\mathcal{E} = \left(\derr{\mathcal{E}}{A} \right)_{B} dA + \left(\derr{\mathcal{E}}{B} \right)_{A} dB~, \\
&& dY = \left( \derr{Y}{A} \right)_{B} dA + \left(\derr{Y}{B} \right)_{A} dB~. 
\end{eqnarray}
We can solve the above equations for the differentials $dA$ and $dB$ in terms of $d\mathcal{E}$ and $dY$,
\begin{eqnarray}
&& dA = \left[ d\mathcal{E} - \frac{\left(\derr{\mathcal{E}}{B} \right)_{A}}{\left(\derr{Y}{B} \right)_{A} } dY \right] \left[ \left(\derr{\mathcal{E}}{A} \right)_{B}  - \frac{\left( \derr{Y}{A} \right)_{B} }{\left(\derr{Y}{B} \right)_{A} }  \left(\derr{\mathcal{E}}{B} \right)_{A} \right]^{-1 }  ~,
\label{eq_derivatives_1} \\
&& dB = \left[ \frac{ \left(\derr{\mathcal{E}}{A} \right)_{B}  }{\left(\derr{Y}{B} \right)_{A} } dY  - \frac{\left( \derr{Y}{A} \right)_{B} }{\left(\derr{Y}{B} \right)_{A} } d\mathcal{E}  \right] \left[ \left(\derr{\mathcal{E}}{A} \right)_{B}  - \frac{\left( \derr{Y}{A} \right)_{B} }{\left(\derr{Y}{B} \right)_{A} }  \left(\derr{\mathcal{E}}{B} \right)_{A} \right]^{-1}    ~,
\label{eq_derivatives_2}
\end{eqnarray}
Because the derivatives in Eqs.\ (\ref{the_isentropic_speed_of_sound_definition}) and (\ref{the_isothermal_speed_of_sound_definition}) are taken at constant $Y$, we can set $dY = 0 $, which allows us to write
\begin{eqnarray}
&& dA\Big|_{dY = 0} = \frac{\left(\derr{Y}{B} \right)_{A} }{  \left[ \left(\derr{\mathcal{E}}{A} \right)_{B}\left(\derr{Y}{B} \right)_{A}   - \left( \derr{Y}{A} \right)_{B}   \left(\derr{\mathcal{E}}{B} \right)_{A} \right]} d\mathcal{E} ~, \\
&& dB\Big|_{dY = 0} = -  \frac{ \left( \derr{Y}{A} \right)_{B}  }{ \left[ \left(\derr{\mathcal{E}}{A} \right)_{B}\left(\derr{Y}{B} \right)_{A}   - \left( \derr{Y}{A} \right)_{B}   \left(\derr{\mathcal{E}}{B} \right)_{A} \right] } d\mathcal{E}    ~.
\end{eqnarray}
Inserting the above expressions into Eq.\ (\ref{pressure_general}) immediately allows us to take the derivative of the pressure with respect to the energy density with $Y$ kept constant, that is
\begin{eqnarray}
c_{Y}^2 =  \frac{\left(\derr{P}{A} \right)_{B} \left(\derr{Y}{B} \right)_{A}   -  \left( \derr{P}{B} \right)_{A} \left( \derr{Y}{A} \right)_{B}  }{  \left(\derr{\mathcal{E}}{A} \right)_{B} \left(\derr{Y}{B} \right)_{A}  -  \left(\derr{\mathcal{E}}{B} \right)_{A} \left( \derr{Y}{A} \right)_{B}  }   ~.
\label{general_speed_of_sound}
\end{eqnarray}
By identifying $Y = \{(S/N), T \}$, explicit expressions for Eqs.\ (\ref{the_isentropic_speed_of_sound_definition}) and (\ref{the_isothermal_speed_of_sound_definition}) can be obtained, where it is useful to remember that we have
\begin{eqnarray}
&& dP = T ~ ds + n ~ d\mu~, \\
&& d\mathcal{E} = T ~ ds + \mu ~ dn~, \\
&& d(S/N) = d(s/n) = \frac{1}{n}~ds - \frac{s}{n^2} ~ dn~. 
\end{eqnarray}

\section{The isentropic speed of sound $c_{\sigma}^2$ given pressure as a function of $T$ and $\mu$}

In this case we set $A = T$, $B = \mu$, and $Y = (S/N)$, and we can immediately calculate
\begin{eqnarray}
\left(\derr{P}{T} \right)_{\mu} = s \distand\left(\derr{P}{\mu} \right)_{T} = n ~, 
\label{cs2_eq_1}
\end{eqnarray}
as well as
\begin{eqnarray}
&& \left(\derr{\mathcal{E}}{T} \right)_{\mu}  = T  \left(\derr{s}{T}\right)_\mu + \mu \left(\derr{n}{T} \right)_\mu   ~, 
\label{cs2_eq_2} \\
&& \left(\derr{\mathcal{E}}{\mu} \right)_{T}  = T  \left(\derr{s}{\mu}\right)_T + \mu \left(\derr{n}{\mu} \right)_T ~, 
\label{cs2_eq_3} \\
&& \left( \derr{Y}{T} \right)_{\mu} = \left( \derr{(S/N)}{T} \right)_{\mu} = \frac{1}{n} \left(  \derr{s}{T} \right)_\mu - \frac{s}{n^2} \left(\derr{n}{T}\right)_\mu     ~, \\
&& \left(\derr{Y}{\mu} \right)_{T} = \left(\derr{(S/N)}{\mu} \right)_{T}  = \frac{1}{n} \left(  \derr{s}{\mu} \right)_T - \frac{s}{n^2} \left(\derr{n}{\mu}\right)_T ~. 
\end{eqnarray}
Inserting all od these into Eq.\ (\ref{general_speed_of_sound}) yields
\begin{eqnarray}
c_{\sigma}^2 = \frac{ sn \left(  \derr{s}{\mu} \right)_T - s^2 \left(\derr{n}{\mu}\right)_T    - n^2 \left(  \derr{s}{T} \right)_\mu +  sn \left(\derr{n}{T}\right)_\mu }{ \left( sT   +   \mu n \right) \left[ \left(\derr{s}{\mu}\right)_T \left(\derr{n}{T}\right)_\mu   -  \left(\derr{s}{T}\right)_\mu \left(\derr{n}{\mu}\right)_T  \right]  }  ~.
\label{the_isentropic_speed_of_sound_given_T_and_mu}
\end{eqnarray}
We note that from Gibbs relation we have
\begin{eqnarray}
sT + \mu n = \mathcal{E} + P ~,
\end{eqnarray}
while from Maxwell's relations
\begin{eqnarray}
\left( \derr{s}{\mu}  \right)_T = \left( \derr{n}{\mu} \right)_T ~. 
\end{eqnarray}

\section{The isentropic speed of sound $c_{\sigma}^2$ given pressure as a function of $T$ and $n$}

In this case we set $A = T$, $B = n$, and $Y = (S/N)$. We assume that both
\begin{eqnarray}
\left(\derr{P}{T} \right)_{n}  \distand\left(\derr{P}{n} \right)_{T} 
\end{eqnarray}
are known; for example they may be obtained by means of taking numerical derivatives based on a table of the values of $P$ at different points $(T,n)$. We also have
\begin{eqnarray}
&& \left(\derr{\mathcal{E}}{T} \right)_{n}  = T  \left(\derr{s}{T}\right)_n   ~, 
\label{cs2_eq_4} \\
&& \left(\derr{\mathcal{E}}{n} \right)_{T}  = T  \left(\derr{s}{n}\right)_T + \mu ~,
\label{cs2_eq_5} \\
&& \left( \derr{Y}{T} \right)_{n}  = \left( \derr{(S/N)}{T} \right)_{n} = \frac{1}{n} \left(  \derr{s}{T} \right)_n     ~, \\
&&  \left(\derr{Y}{n} \right)_{T} = \left(\derr{(S/N)}{n} \right)_{T}  = \frac{1}{n} \left(  \derr{s}{n} \right)_T - \frac{s}{n^2}  ~. 
\end{eqnarray}
Inserting all od these into Eq.\ (\ref{general_speed_of_sound}) yields
\begin{eqnarray}
c_{\sigma}^2  =  \frac{s \left(\derr{P}{T} \right)_{n}  + n\left( \derr{P}{n} \right)_{T} \left(  \derr{s}{T} \right)_n  -  n \left(\derr{P}{T} \right)_{n} \left(  \derr{s}{n} \right)_T    }{  \big( sT  +  \mu n \big) \left(  \derr{s}{T} \right)_n  }   ~.
\label{the_isentropic_speed_of_sound_given_T_and_n}
\end{eqnarray}

\section{The isothermal speed of sound $c_T^2$ given pressure as a function of $T$ and $\mu$}

In this case we set $A = T$, $B = \mu$, and $Y = T$. We reuse Eqs.\ (\ref{cs2_eq_1}-\ref{cs2_eq_3}), while for the other two derivatives we get trivial results
\begin{eqnarray}
&& \left( \derr{Y}{T} \right)_{\mu} = \left( \derr{T}{T} \right)_{\mu} = 1 ~, \\
&& \left(\derr{Y}{\mu} \right)_{T} = \left(\derr{T}{\mu} \right)_{T}  = 0~. 
\end{eqnarray}
Thus
\begin{eqnarray}
c_{T}^2 =  \frac{   n }{ T \left( \derr{s}{\mu} \right)_T + \mu \left( \derr{n}{\mu}  \right)_T   }   ~.
\label{the_isothermal_speed_of_sound_given_T_and_mu}
\end{eqnarray}

\section{The isothermal speed of sound $c_T^2$ given pressure as a function of $T$ and $n$}

In this case we set $A = T$, $B = n$, and $Y = T$. We reuse Eqs.\ (\ref{cs2_eq_4}-\ref{cs2_eq_5}) , while for the other two derivatives we again get trivial results
\begin{eqnarray}
&& \left( \derr{Y}{T} \right)_{n} = \left( \derr{T}{T} \right)_{n} = 1 ~, \\
&& \left(\derr{Y}{n} \right)_{T} = \left(\derr{T}{n} \right)_{T}  = 0~. 
\end{eqnarray}
Therefore
\begin{eqnarray}
c_{T}^2 =  \frac{ \left( \derr{P}{n} \right)_{T}  }{  T  \left(\derr{s}{n} \right)_{T} + \mu }   ~.
\label{the_isothermal_speed_of_sound_given_T_and_n}
\end{eqnarray}

\section{Showing the equivalency of formulas}

The equivalence of the two formulas for the isentropic speed of sound squared, Eqs.\ (\ref{the_isentropic_speed_of_sound_given_T_and_mu}) and (\ref{the_isentropic_speed_of_sound_given_T_and_n}), as well as of the two formulas for the isothermal speed of sound squared, Eqs.\ (\ref{the_isothermal_speed_of_sound_given_T_and_mu}) and (\ref{the_isothermal_speed_of_sound_given_T_and_n}), can be easily shown. It is convenient to start from Eq.\ (\ref{the_isentropic_speed_of_sound_given_T_and_n}). Using the same strategy as that leading to Eqs.\ (\ref{eq_derivatives_1}) and (\ref{eq_derivatives_2}), but adapted a change of variables according to $(T,n) \to (T, \mu)$, we arrive at  
\begin{eqnarray}
&& \left( \derr{P }{n} \right)_{T} = \frac{n}{ \left( \derr{n}{\mu} \right)_T }  ~ , \\
&& \left( \derr{P}{T} \right)_{n}  = s - n \frac{\left( \derr{n}{T} \right)_{\mu}}{\left( \parr{n}{\mu} \right)_T} ~, 
\label{cs2_eq_6}  \\
&& \left( \derr{s}{n}  \right)_T  = \frac{ \left(  \derr{s}{\mu}  \right)_T}{  \left( \parr{n}{\mu}   \right)_T} ~, \\
&& \left( \derr{s}{T} \right)_{n} = \left( \derr{s}{T} \right)_{\mu}  - \left( \derr{s}{\mu} \right)_T  \frac{\left( \derr{n}{T} \right)_{\mu}}{\left( \derr{n}{\mu} \right)_T} ~.
\label{cs2_eq_7} 
\end{eqnarray}
The above formulas, inserted into Eq.\ (\ref{the_isentropic_speed_of_sound_given_T_and_n}), after some algebra yield Eq.\ (\ref{the_isentropic_speed_of_sound_given_T_and_mu}). The same allows to transform Eq.\ (\ref{the_isothermal_speed_of_sound_given_T_and_n}) into Eq.\ (\ref{the_isothermal_speed_of_sound_given_T_and_mu}).

\section{Behavior in the $\mu \to 0$ and $T \to 0$ limits}

\subsection{The limit of $\mu \to 0$}

Let us consider the $\mu \to 0$ limit of the isentropic speed of sound, which also implies $n \to 0$. Using Eq.\ (\ref{the_isentropic_speed_of_sound_given_T_and_n}) as well as Eqs.\ (\ref{cs2_eq_6}) and (\ref{cs2_eq_7}) together with the fact that $\lim_{\mu\to 0} ~\left( \inderr{n}{T} \right)_\mu =0$, we immediately get
\begin{eqnarray}
\lim_{\mu \to 0} c_{\sigma}^2  =  \frac{s   }{  T  \left(  \derr{s}{T} \right)_\mu  }   ~.
\end{eqnarray}
This can be rewritten by realizing that the specific heat at constant volume is defined as
\begin{eqnarray}
C_V = \left(\frac{dE}{dT} \right)_{V,N} = T \left( \derr{S}{T}\right)_{n}~,
\label{heat_capacity_partial_eq}
\end{eqnarray}
where $E$ is the total energy of the system. Because at the same time
\begin{eqnarray}
S = V \left(\derr{P}{T} \right)_{\mu}~,
\end{eqnarray}
we can rewrite Eq.\ (\ref{heat_capacity_partial_eq}) as
\begin{eqnarray}
C_V &=& TV \derr{ }{T}\bigg|_{n} \left(\derr{P}{T}\right)_{\mu} = TV \derr{}{T}\bigg|_{\mu}  \left(\derr{P}{T}\right)_{n} = TV \derr{}{T}\bigg|_{\mu} \left[ s + n  \left(\derr{\mu}{T}\right)_{n} \right] ~.
\end{eqnarray}
In the limit $n \to 0$ the second term in the square bracket drops out, so that we get
\begin{eqnarray}
c_v = \frac{C_V}{V} = T \left( \derr{s}{T} \right)_{\mu} ~.
\end{eqnarray}
We will obtain 
\begin{eqnarray}
\lim_{\mu \to 0} c_{\sigma}^2  =  \frac{s   }{ c_V }   
\label{limit_isentropic_speed_of_sound_mu=0}
\end{eqnarray}
as long as $\left( \inderr{}{T} \right)_{\mu} = \left( \inderr{}{T} \right)_{n}$. This indeed is the case in the limit $\mu \to 0$, which can be seen as follows.

Let us consider quite generally the derivative
\begin{eqnarray}
\left( \frac{d A(T,\mu_B) }{d C(T,\mu_B)} \right)_{n_B} ~,
\end{eqnarray}
where $A$ and $C$ are any functions of $T$ and $\mu_B$. We can always write it out in the following way,
\begin{eqnarray}
\left( \frac{d A(T,\mu_B) }{d C(T,\mu_B)} \right)_{n_B}  = \left( \frac{ \left( \frac{dA}{dT} \right)_{\mu_B} dT + \left( \frac{dA}{d\mu_B} \right)_T d \mu_B }{ \left( \frac{dC}{dT} \right)_{\mu_B} dT + \left( \frac{dC}{d\mu_B} \right)_T d \mu_B  } \right)_{n_B}  ~.
\label{eqD1}
\end{eqnarray}
We can also write
\begin{eqnarray}
dn_B = \left( \frac{dn_B}{dT} \right)_{\mu_B} dT + \left( \frac{dn_B}{d\mu_B} \right)_T d \mu_B ~.
\end{eqnarray}
If we demand that $dn_B = 0$, then we can solve for $dT$,
\begin{eqnarray}
d T = - \frac{\left( \frac{dn_B}{d\mu_B} \right)_T}{\left( \frac{dn_B}{dT} \right)_{\mu_B}} d \mu_B  ~. 
\end{eqnarray}
Inserting the above into Eq.\ (\ref{eqD1}) results in
\begin{eqnarray}
\left( \frac{d A(T,\mu_B) }{d C(T,\mu_B)} \right)_{n_B}  &=& \frac{ \left( \frac{dA}{d\mu_B} \right)_T\left( \frac{dn_B}{dT} \right)_{\mu_B}  - \left( \frac{dA}{dT} \right)_{\mu_B}  \left( \frac{dn_B}{d\mu_B} \right)_T }{\left( \frac{dC}{d\mu_B} \right)_T\left( \frac{dn_B}{dT} \right)_{\mu_B} - \left( \frac{dC}{dT} \right)_{\mu_B}  \left( \frac{dn_B}{d\mu_B} \right)_T   }    ~.
\end{eqnarray}
We can now take the limit $\mu_B \to 0$. Remembering that  for a gas with antiparticles $\left(  \infrac{d n_B}{dT}\right)_{\mu =0} = 0 $, we arrive at
\begin{eqnarray}
\lim_{\mu_B \to 0} \left( \frac{d A(T,\mu_B) }{d C(T,\mu_B)} \right)_{n_B}   = \lim_{\mu_B \to 0}  \frac{  \left( \frac{dA}{dT} \right)_{\mu_B}   }{  \left( \frac{dC}{dT} \right)_{\mu_B}   }  = \lim_{\mu_B \to 0}   \left( \frac{dA}{dC} \right)_{\mu_B}   =  \left( \frac{dA}{dC} \right)_{\mu_B = 0}   ~.
\end{eqnarray}
Thus the $\mu \to 0$ limit of the isentropic speed of sound $c_{\sigma}^2$ is indeed given by Eq.\ \eqref{limit_isentropic_speed_of_sound_mu=0}.

\subsection{The limit of $T \to 0$}

In the limit $T \to 0$ we also have $s \to 0$, and using Eq.\ (\ref{the_isentropic_speed_of_sound_given_T_and_mu}) we immediately get
\begin{eqnarray}
\lim_{T \to 0} c_{\sigma}^2 = \frac{   n }{ \mu \left(\derr{n}{\mu}\right)_T   }  ~.
\end{eqnarray}
Similarly, using Eq.\ (\ref{the_isothermal_speed_of_sound_given_T_and_n}) it is immediately clear that
\begin{eqnarray}
\lim_{T \to 0} c_{T}^2 = \frac{\left( \derr{P}{n} \right)_T}{\mu} = \lim_{T \to 0} c_{\sigma}^2 ~.
\end{eqnarray}
That is, the isothermal and isetropic speed of sound are identical at $T=0$.

Additionally, let us note that in the limit $T \to 0$, for a non-interacting gas we have $\mu_B = \mu_F =\epsilon_{F}=  \sqrt{p_F^2 + m^2}$, as well as $n = \infrac{g p_F^3}{6 \pi^2}$, so that
\begin{eqnarray}
\derr{n}{\mu_B} = \frac{dp_F}{d \epsilon_{F}} \derr{n}{p_F} =  \frac{g\epsilon_F p_F}{2\pi^2}
\end{eqnarray}
and
\begin{eqnarray}
{c_\sigma^2} \Big|_{T = 0}  =  \frac{1}{3}\frac{ p_F^2}{\epsilon_F^2 } ~.
\end{eqnarray}
In the nonrelativistic limit we have $\epsilon_F \approx m$ and thus
\begin{eqnarray}
{c_{\sigma}^2} \Big|_{ \substack{T = 0 \\ \textrm{non-rel}}}  =  \frac{1}{3}\frac{ p_F^2}{m^2 } ~.
\end{eqnarray}
Conversely, in the ultrarelativistic limit the speed of sound of a non-interacting Fermi gas becomes
\begin{eqnarray}
{c_{\sigma}^2} \Big|_{ \substack{T = 0 \\ \textrm{ultra-rel}}}  =  \frac{1}{3} ~,
\end{eqnarray}
which is consistent with the conformal limit.

\newpage
\chapter{Pressure derivatives in the VDF model}
\label{pressure_derivatives_in_the_VDF_model}

The pressure in the VDF model, Eq.\ (\ref{summary_pressure_antiparticles}), can be divided into a Fermi-gas--like term and the interaction term,
\begin{eqnarray}
P_{(K)} = P_{\txt{FG}} + P_{\txt{int}}~.
\label{pressure_divided}
\end{eqnarray}
In the presence of antiparticles, the first term in the above equation is given by 
\begin{eqnarray}
\hspace{-3mm}P_{\txt{FG}} &=& P_{\txt{FG}}^{\txt{(particles)}} + P_{\txt{FG}}^{\txt{(antiparticles)}} \non \\
&=&  g\int \pdens T ~ \ln \Big[ 1 + e^{- \beta \big( \epsilon_{\txt{kin}} - \mu^* \big)}  \Big] + g \int \pdens T ~ \ln \Big[ 1 +  e^{- \beta \big( \epsilon_{\txt{kin}} + \mu^* \big)} \Big] ~,
\end{eqnarray}
while the second term is simply
\begin{eqnarray}
P_{\txt{int}} =  \sum_{k=1}^K C_k \left(\frac{b_k - 1}{b_k} \right) n_B^{b_k }  ~,
\end{eqnarray}
where $n_B$ is the net baryon number density. With the distribution functions for baryons and antibaryons given by
\begin{eqnarray}
f_{\bm{p}} = \frac{1}{e^{\beta \big( \epsilon_{\txt{kin}} - \mu^*  \big)} + 1 }    \distand     \bar{f}_{\bm{p}} = \frac{1}{e^{\beta \big( \epsilon_{\txt{kin}} + \mu^*  \big)} + 1 } ~,
\end{eqnarray}
respectively, we can define the following density-like integrals over powers of the distribution functions,
\begin{eqnarray}
\mathcal{B}_a = g \int \pdens f_{\bm{p}}^a   \distand  \barr{\mathcal{B}}_a = g \int \pdens \bar{f}_{\bm{p}}^a  ~.
\end{eqnarray}
In particular, the net baryon density, entering explicitly in the second term in Eq.\ (\ref{pressure_divided}), is given by
\begin{eqnarray}
n_B \equiv \mathcal{B}_1 - \barr{\mathcal{B}}_1 ~. 
\end{eqnarray}

In calculating $\inderr{^jP_{(K)}}{n_B^j}$, the derivatives of the interaction part of the pressure $P_{\txt{int}}$ are straightforward, leading to
\begin{eqnarray}
&& \frac{dP_{(K)}}{dn_B} = \frac{d P_{\txt{FG}}}{dn_B}  +  \sum_{k=1}^K C_k \big(b_k - 1\big) n_B^{b_k - 1} ~, \\
&& \frac{d^2P_{(K)}}{dn_B^2} = \frac{d^2 P_{\txt{FG}}}{dn_B^2}  +  \sum_{k=1}^K C_k \big(b_k - 1\big)^2 n_B^{b_k - 2} ~, \\
&& \frac{d^3P_{(K)}}{dn_B^3} = \frac{d^3 P_{\txt{FG}}}{dn_B^3}  +  \sum_{k=1}^K C_k \big(b_k - 1\big)^2\big(b_k - 2\big) n_B^{b_k - 3} ~.
\end{eqnarray}
Below, we derive expressions for the derivatives of the Fermi-gas--like part of the pressure $\inderr{^jP_{\txt{FG}}}{n_B^j}$.

We take $P_{\txt{FG}}$ to be a function of temperature $T$ and effective chemical potential $\mu^*$, $P_{\txt{FG}} = P_{\txt{FG}} (T, \mu^*)$ (we note that this choice is arbitrary and one can likewise use $P_{\txt{FG}} = P_{\txt{FG}} (T, \mu_B)$ or $P_{\txt{FG}} = P_{\txt{FG}} (T, n_B)$, although we do find that dealing with $P_{\txt{FG}} = P_{\txt{FG}} (T, \mu^*)$ is the most straightforward). In this case we can write
\begin{eqnarray}
\frac{d P_{\txt{FG}}}{dn_B} = \left(  \parr{}{n_B} + \frac{d\mu^*}{dn_B} \parr{}{\mu^*} \right) P_{FG}~. 
\end{eqnarray}
There is no explicit dependence on $n_B$ in $P_{FG} = P_{\txt{FG}} (T, \mu^*)$, so that the first term is zero. First, we calculate
\begin{eqnarray}
\parr{P_{FG}}{\mu^*} &=&  g \int \pdens T ~   \frac{\beta e^{- \beta \big( \epsilon_{\txt{kinetic}} - \mu^* \big)} }{ 1 + e^{- \beta \big( \epsilon_{\txt{kinetic}} - \mu^* \big)} }    - g \int \pdens T ~    \frac{ \beta e^{- \beta \big( \epsilon_{\txt{kinetic}} + \mu^* \big)}}{ 1 + e^{- \beta \big( \epsilon_{\txt{kinetic}} + \mu^* \big)}}   \non \\
&=& g\int \pdens   f_{\bm{p}}    - g\int \pdens  \bar{f}_{\bm{p}} = \mathcal{B}_1 - \barr{\mathcal{B}}_1  = n_{B} ~,
\end{eqnarray}
which is what we should expect. Then we want to calculate
\begin{eqnarray}
\frac{d \mu^*}{dn_B}  = \left( \frac{dn_B}{d \mu^*} \right)^{-1} = \left( \frac{d}{d\mu^*} \big( \mathcal{B}_1 - \barr{\mathcal{B}}_1  \big) \right)^{-1} ~.
\label{eq45676545}
\end{eqnarray}
Let us see that in general 
\begin{eqnarray}
&&\hspace{-15mm}\frac{d}{d\mu^*} \Big( \mathcal{B}_k \pm \barr{\mathcal{B}}_k \Big) = \frac{d}{d\mu^*} \left[ g \int \pdens \frac{1}{\Big[ e^{\beta \big(\epsilon_{\txt{kin}} - \mu^* \big)} + 1 \Big]^k}  \pm  g\int \pdens \frac{1}{\Big[ e^{\beta \big(\epsilon_{\txt{kin}} + \mu^* \big)} + 1 \Big]^k} \right] \non \\
&& \hspace{8mm} =  \beta k ~ \left[ g \int \pdens \frac{e^{\beta \big(\epsilon_{\txt{kin}} - \mu^* \big)} }{\Big[ e^{\beta \big(\epsilon_{\txt{kin}} - \mu^* \big)} + 1 \Big]^{k+1}}  \mp g\int \pdens \frac{e^{\beta \big(\epsilon_{\txt{kin}} + \mu^* \big)}}{\Big[ e^{\beta \big(\epsilon_{\txt{kin}} + \mu^* \big)} + 1 \Big]^{k+1}} \right] \non \\
&& \hspace{8mm} =  \beta k ~ \left[ g \int \pdens \Big( f_{\bm{p}}^k - f_{\bm{p}}^{k+1}     \Big)  \mp g\int \pdens \Big( \bar{f}_{\bm{p}}^{k} - \bar{f}_{\bm{p}}^{k+1}   \Big) \right] \non \\
&& \hspace{8mm} =   \beta k ~ \Big[  \big( \mathcal{B}_k - \mathcal{B}_{k+1}     \big)  \mp  \big( \barr{\mathcal{B}}_{k} - \barr{\mathcal{B}}_{k+1}   \big)  \Big] ~,
\label{general_mu_star_derivative_of_B_k}
\end{eqnarray}
where in going from the second to the third line we used the fact that the numerators can be rewritten according to $x = x+1 - 1$. In particular, we can use Eq.\ (\ref{general_mu_star_derivative_of_B_k}) to evaluate Eq.\ (\ref{eq45676545}),
\begin{eqnarray}
\derr{\mu^*}{n_B} =  \frac{T}{\Big[  \big( \mathcal{B}_1 + \barr{\mathcal{B}}_{1}   \big)  -  \big(  \mathcal{B}_{2}  + \barr{\mathcal{B}}_{2}   \big)  \Big]  } ~, 
\label{dmustar_dnB}
\end{eqnarray}
and thus we obtain 
\begin{eqnarray}
\derr{P_{FG}}{n_B}  = \frac{T n_B}{\Big[  \big( \mathcal{B}_{1}  + \barr{\mathcal{B}}_{1}   \big)  -  \big( \mathcal{B}_{2} + \barr{\mathcal{B}}_{2}   \big)  \Big] } ~. 
\label{pressure_1st_derivative}
\end{eqnarray}

Using Eq.\ (\ref{pressure_1st_derivative}), we further calculate
\begin{eqnarray}
\hspace{-3mm}\frac{d^2 P_{\txt{FG}}}{dn_B^2} &=& \frac{T }{ \big( \mathcal{B}_1  + \barr{\mathcal{B}}_1  \big) - \big( \mathcal{B}_2  + \barr{\mathcal{B}}_2 \big) } \left( 1 -  n_B \frac{ \frac{d}{dn_B} \Big[ \big( \mathcal{B}_1  + \barr{\mathcal{B}}_1  \big) - \big( \mathcal{B}_2  + \barr{\mathcal{B}}_2 \big)\Big] }{ \Big[\big( \mathcal{B}_1  + \barr{\mathcal{B}}_1  \big) - \big( \mathcal{B}_2  + \barr{\mathcal{B}}_2 \big) \Big]} \right) ~.
\label{pressure_der_eq987}
\end{eqnarray}
Eq.\ (\ref{general_mu_star_derivative_of_B_k}) allows us to compute that in general
\begin{eqnarray}
\hspace{-10mm}\frac{d \big(\mathcal{B}_k \pm \barr{\mathcal{B}}_k \big)}{dn_B} = \left(  \frac{d \big(\mathcal{B}_1 - \barr{\mathcal{B}}_1 \big)}{d\mu^*} \right)^{-1}   \frac{d \big(\mathcal{B}_k \pm \barr{\mathcal{B}}_k \big)}{d\mu^*}   = k ~  \frac{\Big[  \big( \mathcal{B}_k - \mathcal{B}_{k+1}     \big)  \mp  \big( \barr{\mathcal{B}}_{k} - \barr{\mathcal{B}}_{k+1}   \big)  \Big]}{  \Big[  \big( \mathcal{B}_1  + \barr{\mathcal{B}}_{1}    \big)  -  \big( \mathcal{B}_{2}  + \barr{\mathcal{B}}_{2}   \big)  \Big]}  
\label{general_nB_derivative_of_B_k}
\end{eqnarray}
so that we have
\begin{eqnarray}
\frac{d}{dn_B} \Big[ \big( \mathcal{B}_1  + \barr{\mathcal{B}}_1  \big) - \big( \mathcal{B}_2  + \barr{\mathcal{B}}_2 \big)\Big]  =    \frac{  \big( \mathcal{B}_1   -   \barr{\mathcal{B}}_{1} \big)   - 3 \big(  \mathcal{B}_2 -   \barr{\mathcal{B}}_{2} \big)  +  2 \big( \mathcal{B}_{3}        - \barr{\mathcal{B}}_{3} \big)  }{  \Big[  \big( \mathcal{B}_1  + \barr{\mathcal{B}}_{1}    \big)  -  \big( \mathcal{B}_{2}  + \barr{\mathcal{B}}_{2}   \big)  \Big]}      ~,
\end{eqnarray}
with which Eq.\ (\ref{pressure_der_eq987}) becomes
\begin{eqnarray}
\hspace{-4mm} \frac{d^2 P_{\txt{FG}}}{dn_B^2} = \frac{T }{ \big( \mathcal{B}_1  + \barr{\mathcal{B}}_1  \big) - \big( \mathcal{B}_2  + \barr{\mathcal{B}}_2 \big) } \Bigg( 1 -  n_B ~  \frac{  \big( \mathcal{B}_1   -   \barr{\mathcal{B}}_{1} \big)   - 3 \big(  \mathcal{B}_2 -   \barr{\mathcal{B}}_{2} \big)  +  2 \big( \mathcal{B}_{3}        - \barr{\mathcal{B}}_{3} \big)  }{  \Big[  \big( \mathcal{B}_1  + \barr{\mathcal{B}}_{1}    \big)  -  \big( \mathcal{B}_{2}  + \barr{\mathcal{B}}_{2}   \big)  \Big]^2 } \Bigg) ~.
\label{pressure_2nd_derivative}
\end{eqnarray}

Using the same approach, we can likewise calculate
\begin{eqnarray}
\hspace{-5mm}\frac{d^3 P_{\txt{FG}}}{dn_B^3} &=&  - 2T ~\frac{  \big( \mathcal{B}_1   -   \barr{\mathcal{B}}_{1} \big)   - 3 \big(  \mathcal{B}_2 -   \barr{\mathcal{B}}_{2} \big)  +  2 \big( \mathcal{B}_{3}        - \barr{\mathcal{B}}_{3} \big)  }{  \Big[  \big( \mathcal{B}_1  + \barr{\mathcal{B}}_{1}    \big)  -  \big( \mathcal{B}_{2}  + \barr{\mathcal{B}}_{2}   \big)  \Big]^3}   \non \\
&& \hspace{5mm}+~  3Tn_B ~  \frac{ \Big[ \big( \mathcal{B}_1   -   \barr{\mathcal{B}}_{1} \big)   - 3 \big(  \mathcal{B}_2 -   \barr{\mathcal{B}}_{2} \big)  +  2 \big( \mathcal{B}_{3}        - \barr{\mathcal{B}}_{3} \big) \Big]^2 }{  \Big[  \big( \mathcal{B}_1  + \barr{\mathcal{B}}_{1}    \big)  -  \big( \mathcal{B}_{2}  + \barr{\mathcal{B}}_{2}   \big)  \Big]^5}   \non \\
&& \hspace{10mm} - ~ Tn_B  ~\frac{\big( \mathcal{B}_1 + \barr{\mathcal{B}}_{1}    \big)  - 7\big( \mathcal{B}_2 + \barr{\mathcal{B}}_{2}    \big)  +  12 \big( \mathcal{B}_3 + \barr{\mathcal{B}}_{3}     \big)  -  6\big( \mathcal{B}_{4}  + \barr{\mathcal{B}}_{4}   \big)   }{\Big[ \big( \mathcal{B}_1  + \barr{\mathcal{B}}_1  \big) - \big( \mathcal{B}_2  + \barr{\mathcal{B}}_2 \big) \Big]^4} ~.
\label{pressure_3rd_derivative}
\end{eqnarray}

\newpage
\chapter{Cumulants in terms of derivatives of the chemical potential}
\label{cumulants_in_terms_of_derivatives_of_the_chemical_potential}

In calculation of the cumulants, it may be more convenient to consider the derivatives of the chemical potential instead of the derivatives of the pressure. In particular, while the latter are easier to interpret based on the expected behavior of the EOS, the former lead to simpler expressions. Here we provide formulas for the first six cumulants, 
\begin{eqnarray}
&& \kappa_1 = Vn_B  ~, \\ 
&& \kappa_2 = VT \Bigg[\frac{1}{\Big(\frac{d\mu_B}{dn_B} \Big)_T} \Bigg]~, \\
&& \kappa_3 =  VT^2  \Bigg[-  \frac{\left(\derr{^2\mu_B}{n_B^2}\right)_T}{\Big( \derr{\mu_B}{n_B}\Big)^3 } \Bigg]  ~, \\
&& \kappa_4 = VT^3   \Bigg[  3  \frac{ \left(\derr{^2\mu_B}{n_B^2}\right)_T^2}{\Big( \derr{\mu_B}{n_B}\Big)_T^5 }   -    \frac{1}{\Big( \derr{\mu_B}{n_B}\Big)_T^4 } \left(\derr{^3\mu_B}{n_B^3} \right)_T\Bigg] ~, \\
&& \kappa_5 =  VT^4 \Bigg[ - 15 \frac{\left(\derr{^2\mu_B}{n_B^2}\right)_T^3 }{\Big( \derr{\mu_B}{n_B}\Big)_T^7 } + 10 \frac{\left(\derr{^2\mu_B}{n_B^2}\right)_T \left(\derr{^3\mu_B}{n_B^3}\right)_T }{\Big( \derr{\mu_B}{n_B}\Big)_T^6 }   -    \frac{\left(\derr{^4\mu_B}{n_B^4}\right)_T }{\Big( \derr{\mu_B}{n_B}\Big)_T^5 } \Bigg] ~, \\ 
&& \kappa_6 = VT^5  \Bigg[ 105 \frac{\left(\derr{^2\mu_B}{n_B^2}\right)_T^4 }{\Big( \derr{\mu_B}{n_B}\Big)_T^9 } - 105 \frac{\left(\derr{^2\mu_B}{n_B^2}\right)_T^2 \left(\derr{^3\mu_B}{n_B^3}\right)_T}{\Big( \derr{\mu_B}{n_B}\Big)_T^8 }  + 10 \frac{\left(\derr{^3\mu_B}{n_B^3}\right)_T^2 }{\Big( \derr{\mu_B}{n_B}\Big)_T^7 }   \non \\
&& \hspace{30mm} + ~  15  \frac{\left(\derr{^2\mu_B}{n_B^2}\right)_T \left(\derr{^4\mu_B}{n_B^4}\right)_T}{\Big( \derr{\mu_B}{n_B}\Big)_T^7 }   -    \frac{\left( \derr{^5\mu_B}{n_B^5}\right)_T}{\Big( \parr{\mu_B}{n_B}\Big)^6 }  \Bigg]~.
\end{eqnarray}

In the VDF model the derivatives of the chemical potential are given by
\begin{eqnarray}
\derr{^j \mu_B}{ n_B^j} = \derr{^j \mu^*}{ n_B^j}  + \sum_i C_i \prod_{k=1}^j \big(b_i  - j \big) n_B^{b_i - 1 - j} ~,
\end{eqnarray}
and the consecutive derivatives of the effective chemical potential $\mu^*$ can be calculated using Eqs.\ (\ref{dmustar_dnB}) and (\ref{general_nB_derivative_of_B_k}) from Appendix \ref{pressure_derivatives_in_the_VDF_model}.

\newpage
\chapter{Basics of the kinetic theory}
\label{basics_of_the_kinetic_theory}

In this appendix, Sections \ref{phase_space} and \ref{ensembles} are based on Ref.\ \cite{Kittel_Elementary_statistical_physics}, Section \ref{the_Liouville_theorem} is based on Refs.\ \cite{P_Eastman_lecture_notes, Kardar_Statistical_physics_of_particles}, Section \ref{the_BBGKY_hierarchy} is based on Ref.\ \cite{Kardar_Statistical_physics_of_particles}, and Section \ref{the_Boltzmann_equation} is based on Refs.\ \cite{Kardar_Statistical_physics_of_particles,H_van_Hees_lecture_notes}.

\section{Phase space}
\label{phase_space}

We consider a system of a large number $N$ of particles, described by positions $\bm{q}_i$ and momenta $\bm{p}_i$, $i=1, ..., N$, which are in a given configuration $(\bm{q}_1, \dots, \bm{q}_N  ,\bm{p}_1, \dots, \bm{p}_N )$; for simplicity of notation we denote $\bm{q} = (\bm{q}_1, \dots, \bm{q}_N)$ and $\bm{p} = (\bm{p}_1, \dots, \bm{p}_N)$, so that a particular configuration of the system can be written as $(\bm{q}, \bm{p})$. The considered system is characterized by a given thermodynamic property $X$; in particular, $X$ may be a collection of properties, e.g., for a system characterized by given values of temperature and density we have $X = (T, n)$. We say that the system is in a macrostate characterized by $X$, and the corresponding configuration $(\bm{q}, \bm{p})$ is called a microstate of the system. Even though the macrostate of the system is a product of a specific microstate, there is a large number of distinct microstates that also lead to the same macrostate characterized by $X$. Moreover, while the macrostate of the system may remain unchanged, the system in consideration is dynamic and over time it will explore the available microstates (here ``available'' may for example mean ``microstates with the same energy'', etc.). This can be conveniently visualized using the concept of the phase space $\Gamma$, which is a $6N$-dimensional space of all possible positions and momenta of particles in the system. Within $\Gamma$, there is a subset space $\Gamma_X$ corresponding to the macrostate $X$. The thermodynamic postulate of equal \textit{a priori} probabilities states that any microstate $(\bm{q}, \bm{p})$ consistent with a given macrostate $X$ has an equal probability. If $X$ is an equilibrium state, the system will over time move along a trajectory in phase space contained within $\Gamma_X$, and given a sufficiently long time, it will explore all possible microstates associated with $X$.

\section{Statistical ensemble}
\label{ensembles}

For any thermodynamic quantity $X$, taking $S$ samples (measurements) of $X$ allows one to compute the average of $X$, 
\begin{eqnarray}
\langle X \rangle = \frac{1}{S} \sum_{s=1}^{S} X_s~.
\end{eqnarray}
If the measurements are consecutive in time, as is often the case for macroscopic thermodynamic properties, the above equation may be also written as
\begin{eqnarray}
\langle X \rangle = \frac{1}{S} \sum_{s=1}^{S} X (t_s)~,
\end{eqnarray}
where $t_s$ are moments in time when the measurement is made; in particular, in the limit of infinitesimal time intervals between the measurements we have
\begin{eqnarray}
\langle X \rangle = \frac{1}{T} \int_{t_0}^{t_0 + T} dt ~ X(t) ~.
\end{eqnarray}
At different instances of the measurement, the system is found in different microstates in the phase space, leading to particular values of $X(t_s)$ at the moments of measurement $t_1, t_2, ..., t_S$. Note that this measurement of the states of the system over time is equivalent to considering $S$ snapshots of the system with particular values of $\big(\bm{q}(t_s), \bm{p}(t_s)\big)$, corresponding to a particular value of $X(t_s) = X\big(\bm{q}(t_s), \bm{p}(t_s)\big)$. This in turn is equivalent to considering $S$ copies of the system characterized by particular configurations in phase space $\big(\bm{q}_s, \bm{p}_s\big)$ and the corresponding values of $X\big(\bm{q}_s, \bm{p}_s\big)$.

In fact, rather than calculating averages of observables over time, in general it is more convenient to consider a hypothetical collection of copies of the system corresponding to different phase space configurations, called a statistical ensemble. The members of the ensemble (that is, hypothetical copies of the system) reflect the possible phase space configurations that the system may explore. One can then define the probability density $f(\bm{q}, \bm{p})$ for a member of the ensemble to occupy the volume element $d\bm{q} ~d\bm{p}$ in the phase space, normalized such that the number of ensemble systems found within $d\bm{q} ~d\bm{p}$ is given by 
\begin{eqnarray}
f (\bm{q}, \bm{p}) ~ d\bm{q} ~d\bm{p} ~. 
\label{ensembles_prob}
\end{eqnarray}
(We note that the particular normalization used isn't very important as it ultimately drops out of any expressions of interest, and therefore one can choose a normalization which is most convenient for a given discussion; an alternative normalization to the one used here could prioritize that $f(\bm{q}, \bm{p})$ is a properly defined probability, i.e.\ that $\int d\bm{q} ~d\bm{p} ~ f(\bm{q}, \bm{p}) = 1$ holds.) The ensemble average of an observable $X(\bm{q}, \bm{p})$ is then naturally given by
\begin{eqnarray}
\langle X \rangle = \frac{\int d\bm{q} ~d\bm{p} ~ X(\bm{q}, \bm{p}) ~ f(\bm{q}, \bm{p})}{\int d\bm{q} ~d\bm{p} ~ f(\bm{q}, \bm{p})} ~. 
\end{eqnarray}

\section{The Liouville theorem}
\label{the_Liouville_theorem}

Let us now consider a small region of phase space around the phase space point $(\bm{q},\bm{p})$, whose boundaries extend from $\bm{q}$ to $\bm{q} + \Delta \bm{q}$ and from $\bm{p}$ to $\bm{p} + \Delta \bm{p}$ and whose volume is given by $\Delta V = \Delta \bm{q} \Delta\bm{p}$. This region contains a number of ensemble members $N = N(\Delta V)$, which is given by Eq.\ \eqref{ensembles_prob}. Consequently, by definition, the probability density associated with this subset of the ensemble members can be written as
\begin{eqnarray}
f(\Delta V) = \frac{N}{\Delta V} ~.
\label{probability_ensemble_2}
\end{eqnarray}
Each of the $N$ ensemble members considered here is characterized by a particular set of coordinates and momenta $(\bm{q}, \bm{p})$, and is in fact a snapshot, taken at some time $t$, of a dynamical system whose positions and momenta are evolving in time. The question that can arise here is: if we were to consider the same $N$ ensemble members at some later time $t +  d t$, what would be the associated probability?  

From the setup of the problem the number of considered ensemble members $N$ is constant, therefore any change in the associated probability can only come from the change in the phase space volume $\Delta V$ occupied by these ensemble members, see Eq.\ \eqref{probability_ensemble_2}. We then consider
\begin{eqnarray}
\frac{d\Delta V}{dt} = \frac{d \big( \Delta \bm{q} \Delta \bm{p} \big) }{dt} &=& \frac{d \Delta \bm{q}}{dt}   \Delta \bm{p} +  \Delta \bm{q} \frac{d  \Delta \bm{p} }{dt}  \non \\
&=& \frac{d \Big[ \big(\bm{q} + \Delta \bm{q}\big) - \bm{q}\Big]}{dt}   \Delta \bm{p} +  \Delta \bm{q} \frac{d  \Big[ \big( \bm{p} + \Delta \bm{p}\big) - \bm{p} \Big] }{dt}\non \\
&=& \bigg[ \frac{d \big(\bm{q} + \Delta \bm{q}\big) }{dt} - \frac{ d\bm{q}}{dt}  \bigg]  \Delta \bm{p} +  \Delta \bm{q} \bigg[ \frac{d \big(\bm{p} + \Delta \bm{p}\big) }{dt} - \frac{ d\bm{p}}{dt}  \bigg] ~.
\label{Liouville_derivation_1}
\end{eqnarray}
Because $\Delta q$ is small, we can approximate
\begin{eqnarray}
\frac{d \big(\bm{q} + \Delta \bm{q}\big) }{dt}  \approx \frac{d \bm{q} }{dt}  +  \Delta \bm{q} \frac{d}{d\bm{q}}\frac{d \bm{q}  }{dt} 
\label{approximation_Liouville}
\end{eqnarray}
and similarly for the second term in Eq.\ \eqref{Liouville_derivation_1} involving $\bm{p}$ (here we note that we're using a somewhat abstract notation, as the derivatives with respect to the variables $\bm{q} = (\bm{q}_1, \dots, \bm{q}_N)$ and $\bm{p} = (\bm{p}_1, \dots, \bm{p}_N)$  are really sums over derivatives with respect to individual particle positions $\bm{q}_i$ and momenta $\bm{p}_i$; introduction of such more explicit notation is trivial and we omit it here for clarity). These approximations will become exact in the limit $\Delta\bm{q}\to 0, \Delta\bm{p} \to 0$, leading to 
\begin{eqnarray}
\frac{d\Delta V}{dt} =    \Delta \bm{q}\Delta \bm{p} \bigg( \frac{d}{d\bm{q}} \frac{d \bm{q}}{dt} + \frac{d}{d\bm{p}} \frac{d\bm{p}}{dt}    \bigg) ~.
\end{eqnarray}
The equations of motion, $\infrac{d \bm{q}}{dt} $ and $\infrac{d \bm{p}}{dt}$, are given by the Hamilton equations, yielding
\begin{eqnarray}
\frac{d\Delta V}{dt} =    \Delta \bm{q}\Delta \bm{p} \bigg( \frac{d}{d\bm{q}} \frac{d H}{d\bm{p}} - \frac{d}{d\bm{p}} \frac{dH}{d\bm{q}}    \bigg) = 0 ~,
\label{Liouville_derivation_2}
\end{eqnarray}
where in the second equality we used the fact that the Hamiltonian is a well-behaved function satisfying $ \inbderr{^2 H}{\bm{q} d\bm{p}} = \inbderr{^2 H}{\bm{p} d\bm{q}} $. Thus we have shown that the volume of the phase space associated with the chosen subset of the statistical ensemble does not change in time, and from this it immediately follows that the associated probability density doesn't change as well,
\begin{eqnarray}
\frac{df(\Delta V) }{dt} = 0 ~. 
\end{eqnarray}
This, in fact, is the Liouville theorem, usually written more explicitly as
\begin{eqnarray}
\frac{df(t, \bm{q}, \bm{p})}{dt} = \parr{f}{t} + \derr{\bm{q}}{t} \parr{f}{\bm{q}} + \derr{\bm{p}}{t} \parr{f}{\bm{p}} = 0 ~, 
\label{Liouville_long}
\end{eqnarray}
or, using the Poisson bracket, as
\begin{eqnarray}
\frac{df(t, \bm{q}, \bm{p})}{dt} = \parr{f}{t} + \big\{ f, H \big\} = 0 ~, 
\label{Liouville_Poisson_bracket}
\end{eqnarray}
and it simply states that the probability density associated with a given volume element of the phase space does not change in time. 

That the Liouville theorem is satisfied is extremely important to some of the most basic thermodynamic assumptions. Let us take for example the assumption of equal \textit{a priori} probabilities, which states that all microstates $(\bm{q},\bm{p})$ corresponding to a chosen macrostate $X$ have an equal probability density. The Liouville theorem ensures that if this statement is initially true, then it will also remain true with time evolution. More generally, any state characterized by a given initial probability density will over time evolve through states characterized by the same probability density.

Let us see what the Liouville theorem implies for the time evolution of observable averages,
\begin{eqnarray}
\derr{\langle \mathcal{O} \rangle}{t} &=& \mathcal{N}~ \derr{}{t} \int d\bm{q} d\bm{p} ~ \mathcal{O} (\bm{q}, \bm{p}) ~ f(t, \bm{q}, \bm{p})\non  \\
&=&\mathcal{N}~  \int d\bm{q} d\bm{p} ~ \mathcal{O} (\bm{q}, \bm{p}) ~ \parr{ f(t, \bm{q}, \bm{p}) }{t}~,
\end{eqnarray}
where $\mathcal{N}$ is a normalization factor. Eq.\ \eqref{Liouville_long} together with Hamilton's equations allows one to write this as
\begin{eqnarray}
\derr{\langle \mathcal{O} \rangle}{t} &=& - \mathcal{N}~  \int d\bm{q} d\bm{p} ~ \mathcal{O} (\bm{q}, \bm{p}) \left( \parr{H}{\bm{p}} \parr{f(t, \bm{q}, \bm{p})}{\bm{q}} - \parr{H}{\bm{q}} \parr{f(t, \bm{q}, \bm{p})}{\bm{p}}  \right)~,
\end{eqnarray}
which can be further rewritten using integration by parts,
\begin{eqnarray}
\derr{\langle \mathcal{O} \rangle}{t} &=&  \mathcal{N}~  \int d\bm{q} d\bm{p} ~ \left(\parr{\mathcal{O} (\bm{q}, \bm{p}) }{\bm{q}}  \parr{H}{\bm{p}} -   \parr{\mathcal{O}(\bm{q}, \bm{p})}{\bm{p}}  \parr{H}{\bm{q}}  \right) ~ f(t, \bm{q}, \bm{p})~,
\end{eqnarray}
where some of the terms disappear by the same arguments as used in Eq.\ \eqref{Liouville_derivation_2}. Using the definition of the Poisson bracket we arrive at
\begin{eqnarray}
\derr{\langle \mathcal{O} \rangle}{t} =  \mathcal{N}~  \int d\bm{q} d\bm{p} ~ \big\{\mathcal{O} (\bm{q}, \bm{p}) , H \big\} ~ f(t, \bm{q}, \bm{p}) = \langle  \big\{\mathcal{O} (\bm{q}, \bm{p}) , H \big\}  \rangle     ~.
\label{time_evolution_observable}
\end{eqnarray}

Interestingly, the Liouville theorem implies that it is impossible for any closed system, initialized away from equilibrium, to come to an equilibrium. This follows from the discussion above: If initially the system occupies a region of phase space represented by a given microstate $(\bm{q},\bm{p})$ which is characterized by a particular probability density $f(t,\bm{x},\bm{p})$ and corresponds to some non-equilibrium macrostate $X$, then even after arbitrarily long evolution in the phase space the system will be still described by the same probability density $f$ and therefore it will correspond to the same macrostate $X$; that is, the system will only explore the subspace $\Gamma_X$ of the full phase space.

Naturally, everyday experience tells us that in general systems do equilibrate. First, this is because one can never prepare a truly isolated system. Second, the Liouville theorem rests on the assumption that the considered region of the phase space is infinitesimally small, see Eq.\ \eqref{approximation_Liouville}. Any small but finite region of the phase space can be considered to consist of many even smaller regions. While these regions will initially follow very similar evolutions (and so the associated occupied volumes of the phase space will be initially adjacent), over time their phase space trajectories will spread out. This is true for any region of the phase space, so that after some time any small but finite region of phase space will contain states that originate from many different points in the phase space. Each of these microstates maintains the same probability density as at the initial time, but the probability density associated with a region spanned by any finite values of $\Delta x$ and $\Delta p$ will be an average of the many probability densities from different contributing microstates. Ultimately, the system comes to an equilibrium when the probability densities become sufficiently mixed, and the probability density associated with any finite phase space volume element becomes a uniform average. Most importantly, the Liouville theorem only considers the effects due to the drift of particles in the phase space (note that the left-hand side of Eq.\ \eqref{Liouville_long} is in fact a hydrodynamic derivative, and the Liouville theorem states that the phase space density behaves like an incompressible fluid), and in particular it does not take into account the effects of particle collisions and particle transformations such as decays or resonance formation. An isolated system in which particle collisions and particle-number--changing processes take place will, over time, come to an equilibrium.

\section{The BBGKY hierarchy}
\label{the_BBGKY_hierarchy}

The phase space distribution function $f(t, \bm{q}, \bm{p})$ contains detailed information about the system. Note that using an unabbreviated notation, the phase space distribution function for $N$ particles can be explicitly written as
\begin{eqnarray}
f (t, \bm{q}, \bm{p}) = f_N \big(t,  \bm{q}_1, \bm{p}_1, \bm{q}_2, \bm{p}_2, \dots, \bm{q}_N, \bm{p}_N     \big) ~. 
\end{eqnarray}
This means that $f_N$ gives the probability that at time $t$ particle 1 has position $\bm{q}_1$ and momentum $\bm{p}_1$, particle 2 has position $\bm{q}_2$ and momentum $\bm{p}_2$, and so on.  For most applications, however, we are not interested in such detailed knowledge. For example, calculating the pressure of the gas only requires knowing whether any particle can be found at a given position $\bm{q}'$ with a momentum of $\bm{p}'$. This information is contained in a one-particle distribution function $f_1 (t, \bm{q}', \bm{p}')$, which can be obtained from the $N$-particle distribution function through
\begin{eqnarray}
\hspace{-10mm} f_1(t, \bm{q}', \bm{p}') &=& \int d\bm{q}_1 d\bm{p}_1 \int d\bm{q}_2 d\bm{p}_2 ~\dots \int d\bm{q}_N d\bm{p}_N  ~ f_N \big(t,  \bm{q}_1, \bm{p}_1,\bm{q}_2,\bm{p}_2,  \dots, \bm{q}_N , \bm{p}_N     \big) \non\\
&& \hspace{5mm} \times ~  \left( \sum_{i=1}^N \delta^3 \big( \bm{q}' - \bm{q}_i \big)  \delta^3 \big( \bm{p}' - \bm{p}_i \big)   \right) ~.
\end{eqnarray}
By assuming that the distribution function $f_N$ is symmetric with respect to permuting the particles, the above equation becomes
\begin{eqnarray}
\hspace{-10mm} f_1(t, \bm{q}', \bm{p}') &=& N ~ \int  \prod_{i=2}^N d\bm{q}_i d\bm{p}_i  ~ f_N \big(t,  \bm{q}_1 = \bm{q}', \bm{p}_1 = \bm{p}', \bm{q}_2,\bm{p}_2, \dots, \bm{q}_N, \bm{p}_N     \big)  ~.
\end{eqnarray}
Similarly, a two-particle distribution is given by
\begin{eqnarray}
\hspace{-10mm} f_1(t, \bm{q}', \bm{p}', \bm{q}'', \bm{p}'') &=& N(N-1) ~ \int  \prod_{i=3}^N d\bm{q}_i d\bm{p}_i  \non \\
&& \hspace{5mm} \times ~ f_N \big(t,  \bm{q}_1 = \bm{q}', \bm{p}_1 = \bm{p}',\bm{q}_2 = \bm{q}'', \bm{p}_2 = \bm{p}'', \dots, \bm{q}_N ,  \bm{p}_N     \big)  ~,
\end{eqnarray}
and in general the $s$-particle distribution is defined by
\begin{eqnarray}
f_s (t, \bm{q}_1, \bm{p}_1, \dots, \bm{q}_s, \bm{p}_s) = \frac{N!}{(N-s)!} \int \prod_{i=s+1}^N d\bm{q}_i d\bm{p}_i  ~ f_N (t, \bm{q}, \bm{p}) ~.
\label{s-body_distribution_function}
\end{eqnarray}

The evolution equation for the $s$-particle distribution $f_s$ can be derived within an approach named after the works of Nikolay Bogolyubov, Max Born, Herbert Green, John Kirkwood, and Jacques Yvon, and often called the Bogoliubov-Born-Green-Kirkwood-Yvon (BBGKY) hierarchy for reasons that will become apparent. Consider a system of $N$ particles whose Hamiltonian contains terms related to the kinetic energy $T$, the external potential $U$, and the two-body interaction $V$,
\begin{eqnarray}
H(\bm{q},\bm{p}) = \sum_{i=1}^N T(\bm{p}_i) +  \sum_{i=1}^N U(\bm{q}_i)  + \frac{1}{2} \sum_{i=1}^N \sum_{j=1}^N V (\bm{q}_i - \bm{q}_j) ~. 
\label{N-body_Hamiltonian}
\end{eqnarray}
(Note that in principle one should further include terms corresponding to three-body interactions, four-body interactions, and so on; in practice, these terms can be often neglected, which is an excellent approximation, e.g., in the case of dilute gases.) The strategy to compute the $s$-particle distribution function is to divide the Hamiltonian $H(\bm{q},\bm{p})$, Eq.\ \eqref{N-body_Hamiltonian}, into a part related to the $s$ particles in question, a part related to the remaining $N-s$ particles, and an interaction term between the two groups of particles,
\begin{eqnarray}
H = H_s + H_{N-s} + H'~,
\end{eqnarray}
with
\begin{eqnarray}
&& H_s = \sum_{i=1}^s T(\bm{p}_i) + \sum_{i=1}^s U(\bm{q}_i) + \frac{1}{2} \sum_{i=1}^s \sum_{j=1}^{s} V (\bm{q}_i - \bm{q}_j) ~,
\label{Hamiltonian_s} \\
&& H_{N-s} = \sum_{i=s+1}^N T(\bm{p}_i) + \sum_{i=s+1}^N U(\bm{q}_i) + \frac{1}{2} \sum_{i=s+1}^N \sum_{j=s+ 1}^{N} V (\bm{q}_i - \bm{q}_j) ~,
\label{Hamiltonian_N-s} \\
&& H' = \sum_{i=1}^s \sum_{j=s+ 1}^{N} V (\bm{q}_i - \bm{q}_j)  ~.
\end{eqnarray}
Using Eqs.\ \eqref{Liouville_Poisson_bracket} and \eqref{s-body_distribution_function} we can then write the time evolution of the $s$-body distribution function as
\begin{eqnarray}
\parr{f_s}{t } &=& \frac{N!}{(N-s)!} \int \prod_{i=s+1}^N d\bm{q}_i d\bm{p}_i  ~ \parr{f_N}{t} \non\\
&=& - \frac{N!}{(N-s)!} \int \prod_{i=s+1}^N d\bm{q}_i d\bm{p}_i  ~ \bigg[\big\{ f_N, H_s  \big\} + \big\{ f_N,H_{N-s}  \big\} + \big\{ f_N,  H'  \big\} \bigg]~.
\label{BBGYK_derivation_1}
\end{eqnarray}
Note that the first of the Poisson brackets is explicitly given by
\begin{eqnarray}
\hspace{-5mm}\big\{ f_N, H_s  \big\}  = \sum_{a=1}^N \bigg( \parr{f_N}{\bm{q}_a} \cdot \parr{H_s}{\bm{p}_a} - \parr{f_N}{\bm{p}_a} \cdot  \parr{H_s}{\bm{q}_a}  \bigg) = \sum_{a=1}^s \bigg( \parr{f_N}{\bm{q}_a} \cdot \parr{H_s}{\bm{p}_a} - \parr{f_N}{\bm{p}_a} \cdot \parr{H_s}{\bm{q}_a}  \bigg) ~,
\end{eqnarray}
so that the integrations and differentiations present in the first term in Eq.\ \eqref{BBGYK_derivation_1} are performed over different sets of variables, allowing us to exchange their order and arrive at
\begin{eqnarray}
- \bigg\{\frac{N!}{(N-s)!}  \int \prod_{i=s+1}^N d\bm{q}_i d\bm{p}_i  f_N, H_s  \bigg\} = - \big\{ f_s , H_s\big\} ~.
\end{eqnarray}
The second term in Eq.\ \eqref{BBGYK_derivation_1} vanishes, which can be seen by writing the integrand explicitly using Eq.\ \eqref{Hamiltonian_N-s},
\begin{eqnarray}
\hspace{-8mm}\int \prod_{i=s+1}^N d\bm{q}_i d\bm{p}_i  ~  \sum_{a=s+1}^N\bigg[ \parr{f_N}{\bm{q}_a} \cdot  \parr{T(\bm{p}_a)}{\bm{p}_a}  - \parr{f_N}{\bm{p}_a} \cdot  \bigg(\parr{U(\bm{q}_a)}{\bm{q}_a} +    \frac{1}{2}  \sum_{j=s+ 1}^{N} \parr{ V (\bm{q}_s - \bm{q}_j)}{\bm{q}_a}   \bigg) \bigg] ~,
\end{eqnarray}
and performing integration by parts with respect to $\bm{q}_a$ in the first term and with respect to $\bm{p}_a$ in the second term in the square bracket. Finally, because $H' = H'(\bm{q})$, the third term in Eq.\ \eqref{BBGYK_derivation_1} is explicitly
\begin{eqnarray}
- \frac{N!}{(N-s)!} \int \prod_{i=s+1}^N d\bm{q}_i d\bm{p}_i  ~ \bigg[ \bigg(\sum_{a=1}^s  \parr{f_N}{\bm{p}_a}     + \sum_{a=s+1}^N  \parr{f_N}{\bm{p}_a}    \bigg) \cdot  \sum_{j=s+ 1}^{N}  \parr{V (\bm{q}_a - \bm{q}_j)}{\bm{q}_a} \bigg] ~,
\label{BBGYK_derivation_2}
\end{eqnarray}
where the sum over all particles has been divided into two parts. The second term disappears through integration by parts, and because the derivatives over $\bm{q}_a$ are the same for each term in the sum over $j$, Eq.\ \eqref{BBGYK_derivation_2} is equal to
\begin{eqnarray}
&&\hspace{-10mm} - (N-s) \frac{N!}{(N-s)!} \int \prod_{i=s+1}^N d\bm{q}_i d\bm{p}_i  ~ \sum_{a=1}^s  \parr{f_N}{\bm{p}_a}    \cdot    \parr{V (\bm{q}_a - \bm{q}_{s+1})}{\bm{q}_a} \non \\
&& \hspace{0mm} = -  \frac{N!}{\big(N-(s+1)\big)!} \sum_{a=1}^s  \int  d\bm{q}_{s+1} d\bm{p}_{s+1} ~\parr{V (\bm{q}_a - \bm{q}_{s+1})}{\bm{q}_a} \cdot \parr{}{\bm{p}_a}   \int \prod_{i=s+2}^N d\bm{q}_i d\bm{p}_i  ~  f_N \non \\
&& \hspace{0mm} = -  \sum_{a=1}^s  \int  d\bm{q}_{s+1} d\bm{p}_{s+1} ~\parr{V (\bm{q}_a - \bm{q}_{s+1})}{\bm{q}_a} \cdot \parr{f_{s+1}}{\bm{p}_a}  ~,         
\end{eqnarray}
where in the last equality we have used Eq.\ \eqref{s-body_distribution_function}. Altogether, Eq.\ \eqref{BBGYK_derivation_1} becomes
\begin{eqnarray}
\parr{f_s}{t } + \big\{ f_s , H_s\big\} =  \sum_{a=1}^s  \int  d\bm{q}_{s+1} d\bm{p}_{s+1} ~\parr{V (\bm{q}_a - \bm{q}_{s+1})}{\bm{q}_a} \cdot \parr{f_{s+1}}{\bm{p}_a} ~.
\label{BBGKY_hierarchy}
\end{eqnarray}

It is easy to see that if there are no interactions between the particles described by the $s$-particle distribution function, then we arrive at the Liouville theorem, Eq.\ \eqref{Liouville_Poisson_bracket}, for $f_s$. If, on the other hand, the interactions with the other $N-s$ particles are non-zero, then the change of $f_s$ in time is equal to the right-hand side of Eq.\ \eqref{BBGKY_hierarchy}, often called the collision term or collision integral because of the fact that it sums over possible interactions (collisions) of particles from the group of $s$ particles with particles from the group of $N-s$ particles, weighted by the probability $f_{s+1}$ of finding a particle from the $N-s$ group. This results in a sequence, or hierarchy, of equations, where $df_1/dt$ depends on $f_2$, $df_2/dt$ depends on $f_3$, and so on, which is the origin of the name ``BBGKY hierarchy of equations''.

As it is, the BBGKY hierarchy, Eq.\ \eqref{BBGKY_hierarchy}, retains all the information originally contained within the $N$-body distribution function $f_N$, and solving for the consecutive $s$-body distribution functions $f_s$ using Eq.\ \eqref{BBGKY_hierarchy} is as complicated as solving for the original $N$-body distribution function $f_N$. However, in contrast to the Liouville equation specifying $f_N$, the BBGKY hierarchy can be systematically truncated, leading to solvable problems.

\section{The Boltzmann equation}
\label{the_Boltzmann_equation}

A detailed discussion of the derivation of the Boltzmann equation from the BBGKY hierarchy in the dilute regime can be found in Ref.\ \cite{Kardar_Statistical_physics_of_particles}; here we briefly present the main steps. We focus on the first two equations in the BBGKY hierarchy, obtained by setting $s=1$ and $s=2$ in Eq.\ \eqref{BBGKY_hierarchy}. Using Eq.\ \eqref{Hamiltonian_s} and assuming that the potential is symmetric, $V(\bm{q}_1 - \bm{q}_2) = V(\bm{q}_2 - \bm{q}_1)$ (from which it follows that $\inparr{V(\bm{q}_1 - \bm{q}_2) }{\bm{q}_1} = - \inparr{V(\bm{q}_2 - \bm{q}_1) }{\bm{q}_2} $), we arrive at
\begin{eqnarray}
\parr{f_1}{t } +  \parr{T(\bm{p}_1)}{\bm{p}_1} \cdot  \parr{f_1}{\bm{q}_1}- \parr{U(\bm{q}_1)}{\bm{q}_1} \cdot  \parr{f_1}{\bm{p}_1}=    \int  d^3\bm{q}_{2} d^3\bm{p}_{2}~ \parr{V (\bm{q}_1 - \bm{q}_{2})}{\bm{q}_1} \cdot\parr{f_{2}}{\bm{p}_1} 
\label{BBGKY_hierarchy_s=1_explicit_2}
\end{eqnarray}
and
\begin{eqnarray}
&&\hspace{-10mm} \parr{f_2}{t } +\left[ \parr{T(\bm{p}_1)}{\bm{p}_1} \cdot \parr{f_2}{\bm{q}_1}  +  \parr{T(\bm{p}_2)}{\bm{p}_2} \cdot \parr{f_2}{\bm{q}_2}  \right]  - \left[  \parr{U(\bm{q}_1)}{\bm{q}_1} \cdot   \parr{f_2}{\bm{p}_1}+   \parr{U(\bm{q}_2)}{\bm{q}_2}\cdot  \parr{f_2}{\bm{p}_2} \right] \non \\
&& \hspace{15mm} -~ \parr{V(\bm{q}_1 - \bm{q}_2)}{\bm{q}_1} \cdot   \left[  \parr{f_2}{\bm{p}_1} -  \parr{f_2}{\bm{p}_2}   \right] \non \\
&& \hspace{5mm} =   \int  d^3\bm{q}_{3} d^3\bm{p}_{3} ~   \left[\parr{V (\bm{q}_1 - \bm{q}_{3})}{\bm{q}_1} \cdot\parr{}{\bm{p}_1} + \parr{V (\bm{q}_2 - \bm{q}_{3})}{\bm{q}_2}\cdot \parr{}{\bm{p}_2}  \right] f_{3}~.
\label{BBGKY_hierarchy_s=2_explicit_2}
\end{eqnarray}
All terms acting on $f_s$ in above equations have dimensions of inverse time, and their magnitude can be estimated by dimensional analysis. We can define the timescale $\tau_U$ associated with terms of the form
\begin{eqnarray}
\parr{U(\bm{q})}{\bm{q}} \cdot   \parr{}{\bm{p}} \propto \frac{1}{\tau_U}   \propto \frac{v}{L}~,
\end{eqnarray}
where $v$ is the typical particle velocity and $L$ is the length scale over which the external potential $U$ varies. Likewise, we can define the timescale $\tau_V$ associated with terms involving the particle-particle interactions (collisions),
\begin{eqnarray}
\parr{V }{\bm{q}} \cdot\parr{}{\bm{p}} \propto \frac{1}{\tau_V} \propto \frac{v}{d},
\end{eqnarray}
where $d$ is the effective range of the scattering potential. Finally, we note that while a scattering term only occurs on the right-hand side of Eq.\ \eqref{BBGKY_hierarchy_s=1_explicit_2}, in case of Eq.\ \eqref{BBGKY_hierarchy_s=2_explicit_2} there are scattering terms both on the left- and right-hand side of the equation. The relative magnitude of these terms will be proportional to the ratio $f_3/f_2$. Over small volumes $\Delta V \approx d^3$ corresponding to the effective range of the potential $d$, $f_2$ is proportional to the number of particle pairs in $\Delta V$, while $f_3$ is proportional to the number of particle triplets, therefore $f_3/f_2 \propto n$, where $n$ is the number density of particles. Consequently, the time scale $\tau_I$ associated with the collision integral on the right-hand side of Eq.\ \eqref{BBGKY_hierarchy_s=1_explicit_2} is given by
\begin{eqnarray}
\int  d^3\bm{q}_{3} d^3\bm{p}_{3} ~   \parr{V }{\bm{q}} \cdot\parr{}{\bm{p}}  ~\frac{f_{3}}{f_{2}} \propto \frac{1}{\tau_I} \propto \frac{nd^3}{\tau_V} \propto nvd^2~. 
\end{eqnarray}
From the above equation one can see that $\tau_I$ is in fact the mean free time, that is a typical period between subsequent collisions of the same particle; in agreement with the basic intuition, the mean free time is inversely proportional to particle density, the particle's velocity, and the effective ``cross section area'' of the particle given by the square of interaction range $d$. 

For dilute systems, the average number of particles in the volume associated with the scattering range is very small, $nd^3 \ll 1$. Therefore the scattering term on the right-hand side of Eq.\ \eqref{BBGKY_hierarchy_s=1_explicit_2} will be negligible as compared to the scattering term on the left-hand side of the same equation, and it is reasonable to truncate the BBGKY hierarchy at $s=2$ by setting the right-hand side of Eq.\ \eqref{BBGKY_hierarchy_s=1_explicit_2}  to zero. Furthermore, typically the time scale associated with scattering $\tau_V$ is much shorter than the time scale associated with the variation of the external potential $\tau_U$, which allows one to neglect terms proportional to $1/\tau_U$ in Eq.\ \eqref{BBGKY_hierarchy_s=1_explicit_2}. Finally, additionally assuming that $\inparr{f}{t} \ll 1$ yields
\begin{eqnarray}
\left[ \parr{T(\bm{p}_1)}{\bm{p}_1} \cdot \parr{f_2}{\bm{q}_1}  +  \parr{T(\bm{p}_2)}{\bm{p}_2} \cdot \parr{f_2}{\bm{q}_2}  \right]  - \parr{V(\bm{q}_1 - \bm{q}_2)}{\bm{q}_1} \cdot   \left[  \parr{f_2}{\bm{p}_1} -  \parr{f_2}{\bm{p}_2}   \right]  =   0 ~.
\label{BBGKY_hierarchy_s=2_explicit_3}
\end{eqnarray}
The derivatives with respect to $ \inparr{}{\bm{q}_1} $ and $ \inparr{}{\bm{q}_2} $ can be expressed in terms of the center-of-mass and relative coordinate, $\bm{Q} = (\bm{q}_1 + \bm{q}_2)/2$ and $\bm{q} = (\bm{q}_2 - \bm{q}_1)/2$, as we have $ d\bm{q}_1 =  d\bm{Q} + d\bm{q}$ and  $ d\bm{q}_2 = d\bm{Q} - d\bm{q}$. The two-body distribution function $f_2$ is expected to vary slowly along the $\bm{Q}$ direction and appreciably over the $\bm{q}$ direction (this can be understood based on the fact that the average position of two particles $\bm{Q}$ is governed by the long-range part of the interaction, while their relative position $\bm{q}$ is governed by the short range part of the interaction), so that $ \inparr{}{\bm{q}_1} \approx -  \inparr{}{\bm{q}_2} \approx \inparr{}{\bm{q}} $, and we can rewrite Eq.\ \eqref{BBGKY_hierarchy_s=2_explicit_3} as
\begin{eqnarray}
\left[ \parr{T(\bm{p}_1)}{\bm{p}_1}  -  \parr{T(\bm{p}_2)}{\bm{p}_2} \right] \cdot \parr{f_2}{\bm{q}}  = \parr{V(\bm{q}_1 - \bm{q}_2)}{\bm{q}_1} \cdot   \left[  \parr{f_2}{\bm{p}_1} -  \parr{f_2}{\bm{p}_2}   \right]  ~.
\label{BBGKY_hierarchy_s=2_explicit_4}
\end{eqnarray}
We can use the above relation on the right-hand side of Eq.\ \eqref{BBGKY_hierarchy_s=1_explicit_2}, where we can perform the substitution $\inparr{f_{2}}{\bm{p}_1}  \to \inparr{f_{2}}{\bm{p}_1}  - \inparr{f_{2}}{\bm{p}_2} $ because $\inparr{f_{2}}{\bm{p}_1} $ is a complete derivative which integrates to zero. Additionally, we use the fact that for a relativistic system $\inparr{T(\bm{p})}{\bm{p}} = \bm{p}/E = \bm{v}$, and in this way we arrive at
\begin{eqnarray}
\hspace{-3mm}\parr{f_1}{t } +  \frac{\bm{p}_1}{E_1} \cdot  \parr{f_1}{\bm{q}_1}- \parr{U(\bm{q}_1)}{\bm{q}_1} \cdot  \parr{f_1}{\bm{p}_1}=    \int  d^3\bm{q} d^3\bm{p}_{2}~ \big( \bm{v}_1  -  \bm{v}_2 \big) \cdot \parr{f_2(t, \bm{q}_1,\bm{p}_1, \bm{q},  \bm{p}_2 )}{\bm{q}} ~.
\label{BBGKY_hierarchy_s=2_explicit_5}
\end{eqnarray}
On the right-hand side of the above equation we have a derivative of $f_2$ with respect to $\bm{q}$ along the direction of th relative motion $\bm{p} = \bm{p}_2 - \bm{p}_1$, and it is convenient to express $\bm{q}$ in a new basis in which one axis, denoted with $a$, is parallel to $\bm{p} = \bm{p}_2 - \bm{p}_1$, while the remaining two coordinates form a plane perpendicular to $\bm{p}$,
\begin{eqnarray}
\bm{q} \to \bm{a} + \bm{b}  \distand d^3\bm{q} \to d\bm{a} ~d^2\bm{b} ~;
\end{eqnarray}
we note that $\bm{b}$ is an impact parameter vector, which in particular is zero for a head-on collision where $\bm{q}_1 - \bm{q}_2$ is parallel to $\bm{p}_1 - \bm{p}_2$. We can then attempt to integrate Eq.\ \eqref{BBGKY_hierarchy_s=2_explicit_5} over the $a$ coordinate. As this integral is performed, we go from negative relative distances, starting from some lower boundary value $a_-$, through a zero relative distance which is where the collision takes place, and then through positive relative distances up to some upper boundary value $a_+$ (on a side note, $|a_+ - a_-|$ can be taken to be small because $f_2$ changes appreciably only over distances comparable with the collision range $d$). In this way, we ``observe'' the system from some time before the collision to some time after the collision. Importantly, the momenta of the individual particles are changed throughout the collision even though the total momentum is conserved,
\begin{eqnarray}
\bm{p}_1 + \bm{p}_2 = \bm{p}_1' + \bm{p}_2 ' ~. 
\end{eqnarray}
Therefore the integral of Eq.\ \eqref{BBGKY_hierarchy_s=2_explicit_5} over the $a$ coordinate can be written as
\begin{eqnarray}
\derr{f_1}{t } =    \int  d^2\bm{b} d^3\bm{p}_{2}~ \big| \bm{v}_1  -  \bm{v}_2 \big| \Big[ f_2(t, \bm{q}_1, \bm{p}_1', \bm{b}, a_+ ,  \bm{p}_2' ) - f_2(t, \bm{q}_1,\bm{p}_1, \bm{b}, a_-,   \bm{p}_2 )\Big] ~,
\end{eqnarray}
where $|\bm{v}_1 - \bm{v}_2|$ is the relative speed. As the next step, we change the integration variable from the impact parameter vector to the solid angle $\Omega$,
\begin{eqnarray}
\hspace{-5mm}\derr{f_1}{t } =    \int  d\Omega d^3\bm{p}_{2}~ \left| \frac{d\sigma}{d \Omega} \right| \big| \bm{v}_1  -  \bm{v}_2 \big| \Big[ f_2(t, \bm{q}_1, \bm{p}_1', \bm{b}, a_+ ,  \bm{p}_2' ) - f_2(t, \bm{q}_1, \bm{p}_1, \bm{b}, a_-,  \bm{p}_2 )\Big] ~,
\end{eqnarray}
where the Jacobian of the transformation $\left| \infrac{d\sigma}{d \Omega} \right| $ has a dimension of area and is known as the differential cross section. Finally, we use the assumption of molecular chaos, according to which the two-body distribution function is given by a product of one-body distribution functions,
\begin{eqnarray}
f_2(t, \bm{q}_1,\bm{p}_1,  \bm{b}, a_{\pm},  \bm{p}_2 ) = f_1 (t, \bm{q}_1, \bm{p}_1)  f_1 (t, \bm{q}_1, \bm{p}_2)  ~,
\end{eqnarray}
and which means that, except for the moment in which the collision take place, the particles are assumed to be uncorrelated. This allows us to arrive at the Boltzmann equation, 
\begin{eqnarray}
&& \hspace{-15mm} \parr{f_1}{t } +  \frac{\bm{p}_1}{E_1} \cdot  \parr{f_1}{\bm{q}_1}- \parr{U(\bm{q}_1)}{\bm{q}_1} \cdot  \parr{f_1}{\bm{p}_1} \non \\
&& \hspace{-10mm}  =    \int  d\Omega d^3\bm{p}_{2}~ \left| \frac{d\sigma}{d \Omega} \right| \big| \bm{v}_1  -  \bm{v}_2 \big| \Big[ f_1 (t, \bm{q}_1, \bm{p}_1')  f_1 (t, \bm{q}_1, \bm{p}_2')  - f_1 (t, \bm{q}_1, \bm{p}_1)  f_1 (t, \bm{q}_1, \bm{p}_2) \Big] ~,
\label{Boltzmann_equation_derived}
\end{eqnarray}
which is an integrodifferential equation for the evolution of the one-body distribution $f_1$.

Although various parts of the above derivation sketch have characteristics of a \textit{deux ex machina}, the form of Eq.\ \eqref {Boltzmann_equation_derived} can be also argued on phenomenological grounds. The terms on the left-hand side are connected to a motion of a single particle in the external potential $U$. The collision term on the right-hand side describes the change in the probability to find a particle of momentum $\bm{p}_1$ at $\bm{q}_1$ that can happen due to a collision with another particle of momentum $\bm{p}_2$, in result of which both particles acquire new momenta $\bm{p}_1'$ and $\bm{p}_2'$. This probability is proportional to the cross section for the scattering $\left| \infrac{d\sigma}{d \Omega} \right| \big| $, the flux of incident particles $\big| \bm{v}_1  -  \bm{v}_2 \big| $, the probability of finding the particles at a position $\bm{q}$ with momenta $\bm{p}_1$ and $\bm{p}_2$, and the probability of the final states with momenta $\bm{p}_1'$ and $\bm{p}_2'$.

For describing relativistic heavy-ion collisions, it is useful to write Eq.\ \eqref{Boltzmann_equation_derived} in a manifestly covariant way,
\begin{eqnarray}
p^{\mu} \parr{f}{x^{\mu}} + m \parr{\big(K^{\mu} f\big)}{p^{\mu}} =  \int \frac{d^3 \bm{p}_2}{E_2}  \frac{d^3 \bm{p}_1'}{E_1'} \frac{d^3 \bm{p}_2'}{E_2'} ~     W \big( \bm{p}_1', \bm{p}_2' \leftarrow \bm{p}, \bm{p}_2  \big) \Big( f_1' f_2' - f f_2   \Big)
\end{eqnarray}
where $K^{\mu}$ is a four-force vector and $W\big( \bm{p}_1', \bm{p}_2' \leftarrow \bm{p}, \bm{p}_2  \big) $ is a Lorentz invariant cross section for scattering of particles with momenta $\bm{p}$, $\bm{p}_2$ into particles of momenta $\bm{p}_1'$, $\bm{p}_2'$. Here, it is useful to adopt an interpretation of the expression $\Big( f_1' f_2' - f f_2   \Big)$ in which the first term corresponds to particles scattering into the considered volume element (gain term), while the second term corresponds to particles scattering out of the considered volume element (loss term). Importantly, Boltzmann equation can also include quantum effects, which was first studied by Edwin Uehling and George Uhlenbeck \cite{Uehling_Uhlenbeck_on_Boltzmann}, leading to 
\begin{eqnarray}
\hspace{-10mm} p^{\mu} \parr{f}{x^{\mu}} + m \parr{\big(K^{\mu} f\big)}{p^{\mu}} &=&  \int \frac{d^3 \bm{p}_2}{E_2}  \frac{d^3 \bm{p}_1'}{E_1'} \frac{d^3 \bm{p}_2'}{E_2'} ~  \non  \\
&& \hspace{-35mm} \times ~  W \big( \bm{p}_1', \bm{p}_2' \leftarrow \bm{p}, \bm{p}_2  \big) \Bigg( f_1' f_2' \big(1 -a f\big) \big(1 - af_2\big)  - f f_2\big(1  - a f_1'\big)\big(1 - a f_2'\big)   \Bigg)~,
\label{Boltzmann_quantum}
\end{eqnarray}
where $a=1$ for fermions and $a =-1$ for bosons. The additional terms in the big round bracket can be easily understood: for example, in the case of fermions, a particle in a state 2' can only scatter into a state 2 if that state is unoccupied, the probability of which is given by $1 - f_2$. The above equation is known as the Boltzmann-Uehling-Uhlenbeck (BUU) equation.

So far we have only dealt with the Boltzmann equation for one particle species. To describe heavy-ion collisions one needs to include hundreds of hadron species which can collide with each other, decay, and form resonances. This means that the system is described by hundreds (as many as particle species) of coupled Boltzmann equations. This extremely complex system can be solved by means of Monte-Carlo simulations, where the particles are propagated using the equations of motion entering the left-hand side of Eq.\ \eqref{Boltzmann_quantum} and the collision integral on the right-hand side is realized by explicitly simulating collisions, decays, and resonance formation. For more details see Chapter \ref{implementation}.

\newpage
\chapter{Pair distribution function and the second-order cumulant}
\label{pair_distribution_function_and_the_second-order_cumulant}

The procedure to compute the radial distribution function $g_i(r)$, given by Eq.\ \eqref{radial_distribution_function}, can be generalized to the case of a continuous system described by a particle density distribution $n(\bm{r}')$, 
\begin{eqnarray}
g_i (r, \Delta r) &=&  \int d\bm{r}'  \Bigg(n(\bm{r}') - 1 ~\delta(\bm{r}_i - \bm{r}')     \Bigg) \non \\
&&\hspace{15mm} \times ~\theta\Big(  r + \Delta r - |\bm{r}_i - \bm{r}' |   \Big) \theta\Big( |\bm{r}_i - \bm{r}' | - ( r - \Delta r )   \Big) ~,
\end{eqnarray}
where care must be taken to subtract the self-contribution from the reference particle. Similarly, the pair distribution function $\widetilde{g}(r)$, Eq.\ \eqref{pair_distribution_function}, can be rewritten as
\begin{eqnarray}
\hspace{-5mm}\widetilde{g} (r, \Delta r) &=&\frac{\mathcal{N}}{2}   \int d \bm{r}'  \int d\bm{r}''  ~   n(\bm{r}') \Big(n(\bm{r}'')  - \delta( \bm{r}' - \bm{r}'') \Big)  \non  \\
&& \hspace{10mm} \times ~\theta\Big(  r + \Delta r - |\bm{r}' - \bm{r}'' |   \Big) \theta\Big( |\bm{r}' - \bm{r}'' | - ( r - \Delta r )   \Big)\non  \\
&=& \frac{\mathcal{N}}{2}   \int d \bm{r}'  \int d\bm{r}''  ~   n(\bm{r}') n(\bm{r}'')   ~ \theta\Big(  r + \Delta r - |\bm{r}' - \bm{r}'' |   \Big) \theta\Big( |\bm{r}' - \bm{r}'' | - ( r - \Delta r )   \Big) \non  \\
&& \hspace{5mm} - ~ \frac{\mathcal{N}}{2}   \int d \bm{r}' ~  n(\bm{r}')    ~ \theta\Big( \Delta r  - r    \Big) ~.
\label{pair_distr_kappa_2_1}
\end{eqnarray}
We note that the second term is only non-zero when $r < \Delta r$, which is correct given that the self-contribution only needs to be subtracted if we consider the pair distribution function within a distance $\Delta r$ around the reference particles. 

It is possible to establish a connection between the pair distribution function and the second-order cumulant $\kappa_2$. For this, we consider the pair distribution function $\widetilde{g}(r)$ at distances close to the reference particle, that is we put $r =0$, by means of which Eq.\ \eqref{pair_distr_kappa_2_1} becomes
\begin{eqnarray}
\hspace{-0mm} \widetilde{g} (0, \Delta r) &=& \frac{\mathcal{N}}{2} \bigg[  \int d \bm{r}'  \int d\bm{r}''  ~   n(\bm{r}') n(\bm{r}'')   ~\theta\Big(  \Delta r - |\bm{r}' - \bm{r}'' |   \Big)  -    \int d \bm{r}' ~  n(\bm{r}')    \bigg] ~.
\label{pair_distr_kappa_2_2}
\end{eqnarray}
Let us assume that $\Delta r$ is small and that within the distance $\Delta r$ from $\bm{r}'$ the density is smooth enough for $n(\bm{r}'')  \approx n(\bm{r'})$ to hold, in which case
\begin{eqnarray}
\hspace{-7mm} \widetilde{g} (0, \Delta r) &=& \frac{\mathcal{N}}{2} \bigg[  \int d \bm{r}' ~ \big[n(\bm{r}') \big]^2 ~  \int d\bm{r}''    ~\theta\Big(  \Delta r - |\bm{r}' - \bm{r}'' |   \Big)  -    \int d \bm{r}' ~  n(\bm{r}')    \bigg] \non \\
&=&  \frac{\mathcal{N}}{2}  \left[ V_{\Delta} \int d \bm{r}' ~ \big[n(\bm{r}') \big]^2  -    \int d \bm{r}' ~  n(\bm{r}')    \right] ~,
\label{pair_distr_kappa_2_3}
\end{eqnarray}
where $V_{\Delta} = (4/3)\pi r_{\Delta}^3$. Furthermore, let us divide the volume of the system $V$ into cubes of volume $V_{\Delta}$, $N_{\txt{cubes}} = V / V_{\Delta}$, and assume that we can safely discretize the remaining integrals in Eq.\ \eqref{pair_distr_kappa_2_3} according to $\int d \bm{r}' ~ f(\bm{r'})  \to  \sum_{i=1}^{N_{\txt{cubes}}} V_{\Delta}  f(\bm{r}_i)$, where $\bm{r}_i$ points to the center of each cube. With this and taking the number of particles in the $i$-th cube to be $N_i(\bm{r}_i) \equiv V_{\Delta} n(\bm{r}_i)$, Eq.\ (\ref{pair_distr_kappa_2_3}) becomes
\begin{eqnarray}
\hspace{-5mm}\widetilde{g} (0, \Delta r) &\approx& \mathcal{N}~ \frac{1}{2} \left[  \sum_{i=1}^{N_{\txt{cubes}}}   \big[ N(\bm{r}_i)\big]^2  -   \sum_{i=1}^{N_{\txt{cubes}}}  N(\bm{r}_i)    \right] ~.
\end{eqnarray}
Since the normalization can be freely chosen given that $\widetilde{\rho} (0, \Delta r) $ should be compared to a reference distribution for an ideal gas $\widetilde{\rho}_0(0, \Delta r)$, in particular we can take $\mathcal{N} = 2/N_{\txt{cubes}}$, so that finally 
\begin{eqnarray}
\hspace{-7mm}\widetilde{g} (0, \Delta r) &=& \frac{1}{N_{\txt{cubes}}} \left[  \sum_{i=1}^{N_{\txt{cubes}}}   \big[ N(\bm{r}_i)\big]^2  -   \sum_{i=1}^{N_{\txt{cubes}}}  N(\bm{r}_i)    \right] ~,
\label{pair_distr_kappa_2_4}
\end{eqnarray}
where $N_{\txt{cubes}}$ is determined by $\Delta r$.

It is clear from Eq.\ \eqref{pair_distr_kappa_2_4} that the radial distribution function of all distinct particle pairs at distances close to the reference particles is
\begin{eqnarray}
\widetilde{g}  (0, \Delta r) = M_2 - M_1 = F_2 = \langle N (N-1) \rangle ~,
\label{pair_distr_kappa_2_5}
\end{eqnarray}
where $M_i$ and $F_i$ are moments and factorial moments of the distribution, respectively. Moreover, assuming that the pair distribution function for uncorrelated pairs $\widetilde{g}_0 (0, \Delta r)$ is described by the Poisson distribution, for which $\langle N \rangle = \lambda$ and $\langle N^2 \rangle = \lambda^2 + \lambda$ (where $\lambda$ is the mean), we have
\begin{eqnarray}
\widetilde{g}_0(0, \Delta r) = \langle N \rangle^2 ~. 
\label{pair_distr_kappa_2_6}
\end{eqnarray}

Let us consider the deviation of the behavior of the pair distribution function $\widetilde{g}(0, \Delta r)$ from the ideal case of $\widetilde{g}_0(0, \Delta r)$, which can be conveniently done by considering the measure
\begin{eqnarray}
R = \frac{\widetilde{g}~(0, \Delta r)}{\widetilde{g}_0(0, \Delta r)} - 1 ~.
\end{eqnarray}
Using Eqs.\ \eqref{pair_distr_kappa_2_5} and \eqref{pair_distr_kappa_2_6} we can immediately rewrite this as
\begin{eqnarray}
R = \frac{\langle N^2 \rangle - \langle N \rangle  - \langle N \rangle^2}{\langle N \rangle^2} = \frac{\kappa_2 - \kappa_1  }{\kappa_1^2}  ~.
\end{eqnarray}
In particular, provided that $\kappa_1 > 0$, we immediately obtain that $R$ is bigger (smaller) than 0 if and only if the second-order cumulant ratio $\kappa_2/\kappa_1$ is bigger (smaller) than 1, which can be alternatively expressed as in Eqs.\ (\ref{pair_distribution_kappa_2_1}) and (\ref{pair_distribution_kappa_2_2}).

We would like to stress that the above relations hold for an arbitrary distribution of particles, without any assumptions on the underlying physics, provided that the corresponding uncorrelated system can be described by the Poisson distribution. In any such system the sign of $[\widetilde{g}(r, \Delta r)/\widetilde{g}_0(r, \Delta r)] - 1$ at $r \to 0$ is the same as the sign of $(\kappa_2/\kappa) - 1$. In particular, it follows that $\widetilde{g}(r, \Delta r)/\widetilde{g}_0(r, \Delta r) < 1$ for systems where a repulsive interaction dominates at short distances (leading to a distribution more uniform than that of an ideal gas), while $\widetilde{g}(r, \Delta r)/\widetilde{g}_0(r, \Delta r) > 1$ for systems where an attractive interaction dominates at short distances (which leads to a distribution that is less uniform than that of an ideal gas).

\newpage

\chapter{\textit{A posteriori} application of the parallel ensembles method}
\label{parallel_ensembles_a_posteriori}

\begin{figure}[bh]
	\centering\mbox{
	\includegraphics[width=0.99\textwidth]{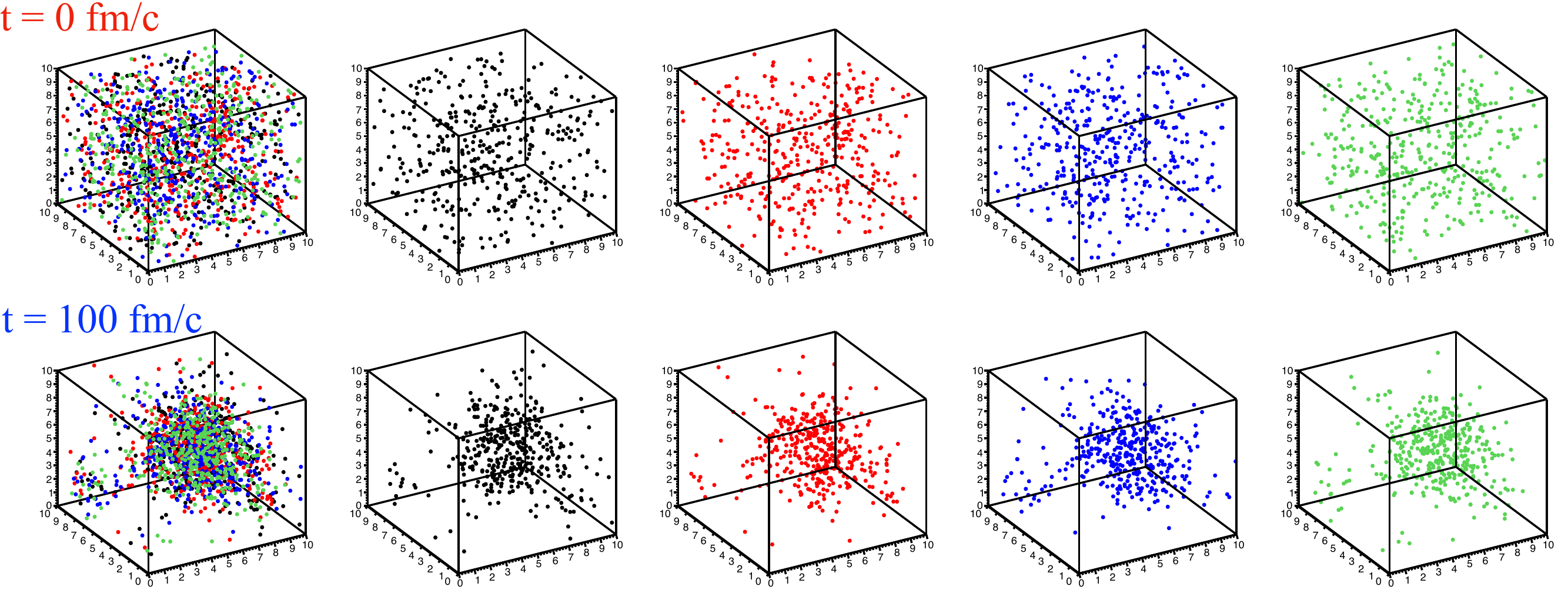} 
	}
	\caption[Spinodal decomposition in a hadronic transport simulation using the parallel ensembles mode]{\textit{A posteriori} construction of the parallel ensembles mode for a system undergoing nuclear spinodal fragmentation (for details see Section \ref{nuclear_phase_transition}). The upper row corresponds to the initialization time, while the lower row corresponds to $t = 100\ \txt{fm}/c$. The left-most boxes in both rows show the full ensemble event, in which test particles have been divided into $N_T = 4$ groups, marked with different colors. The remaining boxes in both rows show the state of each of the parallel ensembles obtained by dividing the original full ensemble event into $N_T = 4$ separate events. It is evident that the parallel ensemble systems mirror the evolution of the full ensemble.}
	\label{parallel_ensembles_3D_box}
\end{figure}

The version of \texttt{SMASH} that we used did not have the option to run in a parallel ensembles mode (this option has been recently added to \texttt{SMASH} \cite{smash_version_2.0} and is currently being tested). However, for simulations with all collision and decay channels turned off (such as we study in Chapter \ref{results}), the concept of parallel ensembles can be safely employed \textit{a posteriori}, that is at the analysis stage. Specifically, in each event we divide the $N_T N_B$ test particles obtained from a full ensemble \texttt{SMASH} simulation (where $N_B$ is the baryon number evolved in the simulation and $N_T$ is the number of test particles per particle) into $N_T$ separate groups; this process is shown in Fig.\ \ref{parallel_ensembles_3D_box} for $N_T=4$. We then treat these groups as separate events. Each of these \textit{a posteriori} constructed events is governed by $P_{N_B}(N_i)$ (see Section \ref{Physical_baryon_number_distribution_function}). We stress that for a \texttt{SMASH} simulation run in the full ensemble mode with $N_{\txt{ev}}$ events and $N_T$ test particles per particle, the corresponding calculation in the parallel ensembles mode will be characterized by $N_T N_{\txt{ev}}$ events with $N_T = 1$ test particles per particle.

\newpage

\chapter{The isothermal speed of sound at $\mu = 0$}
\label{the_isothermal_speed_of_sound_at_mu=0}

We want to compute the $\mu =0$ limit of Eq.\ (\ref{magic_equation_1}), repeated here for convenience,
\begin{eqnarray}
c_{T}^2 =  \left(\frac{\mu }{T}\frac{\kappa_2}{\kappa_1} +   \left( \frac{ d \ln \kappa_1 }{d \ln T}\right)_{\mu} \right)^{-1}  ~. 
\label{magic_equation_1_aux_1}
\end{eqnarray}
We will use the fact that at $\mu =0$ we have by symmetry, for all values of the temperature $T$, $\kappa_1 = 0$ and $\kappa_3 = 0$, from which it follows that  $\left(  \infrac{d \kappa_1}{dT}\right)_{\mu =0} = 0$. We will also use the l'Hospital rule, according to which the $x \to x_0$ limit of the ratio of two functions $f(x)$ and $g(x)$ such that either $\lim_{x \to x_0} f(x) = \lim_{x \to x_0} g(x) = 0 $ or $\lim_{x \to x_0} f(x) = \lim_{x \to x_0} g(x) = \infty$ is given by
\begin{eqnarray}
\lim_{x \to x_0} \frac{f(x)}{g(x)} = \lim_{x \to x_0} \frac{f'(x)}{ g'(x)}  ~,
\end{eqnarray}
provided that $g'(x) \neq 0 ~\forall ~ x \neq x_0$ and the limit $\lim_{x \to x_0} \frac{f'(x)}{ g'(x)} $ exists. We denote the evoking of the l'Hospital rule with a subscript ``LH''. 

For the first term in Eq.\ \eqref{magic_equation_1_aux_1}, assuming that $\kappa_2 \neq 0$ (which is equivalent to assuming that there are finite fluctuations in the system), we have
\begin{eqnarray}
\frac{1}{T} ~\lim_{\mu \to 0}  \frac{\mu \kappa_2}{\kappa_1} =\bigg|_{\txt{LH}} \frac{1}{T} ~\lim_{\mu \to 0} \frac{\left( \derr{(\mu \kappa_2)}{\mu} \right)_T}{ \left(\frac{d \kappa_1}{d\mu}\right)_T } = \frac{ \kappa_2}{T} \lim_{\mu \to 0}~ \frac{\kappa_2 + \mu \frac{\kappa_3}{T}}{ \frac{\kappa_2}{T} } = 1 ~.
\end{eqnarray}
Then we deal with the second term,
\begin{eqnarray}
\lim_{\mu \to 0} ~\frac{T}{\kappa_1} \left( \frac{ d \kappa_1 }{d  T}\right)_{\mu} &=& \bigg|_{\txt{LH}}  \lim_{\mu \to 0} ~ \frac{T}{ \left( \frac{d \kappa_1}{d \mu}\right)_{T} } \frac{d}{d\mu} \bigg|_{T} \left( \frac{ d \kappa_1 }{d  T}\right)_{\mu}   = \lim_{\mu \to 0}  ~\frac{T}{ \left( \frac{d \kappa_1}{d \mu}\right)_{T} } \frac{d}{dT} \bigg|_{\mu} \left( \frac{ d \kappa_1 }{d  \mu}\right)_{T} \non \\
&=&    \lim_{\mu \to 0} ~ \frac{T}{  \frac{\kappa_2}{T}} \frac{d}{dT} \bigg|_{\mu} \left( \frac{  \kappa_2 }{T}\right) = \lim_{\mu \to 0} ~ \left( \frac{d \ln \left( \frac{\kappa_2}{T} \right)}{d \ln T} \right)_{\mu} ~.
\end{eqnarray}
Here the limit can be now taken simply by taking the value at $\mu =0$, 
\begin{eqnarray}
\left( \frac{d \ln \left( \frac{\kappa_2}{T} \right)}{d \ln T} \right)_{\mu = 0} = \frac{T^2}{ \kappa_2 \big|_{\mu=0}} \left[ \frac{1}{T} \left(\frac{d \kappa_2}{dT}\right)_{\mu = 0}   - \frac{\kappa_2 \big|_{\mu=0}}{T^2}\right] = \frac{T}{\kappa_2\big|_{\mu=0}} \left(\frac{d\kappa_2}{dT}\right)_{\mu=0} - 1 ~. 
\end{eqnarray}
Altogether,
\begin{eqnarray}
\lim_{\mu \to 0} c_T^2 = \left( \frac{T}{\kappa_2\big|_{\mu=0}} \left(\frac{d\kappa_2}{dT}\right)_{\mu=0}    \right)^{-1} =  \left(\frac{d \ln \kappa_2}{d \ln T}\right)_{\mu=0}^{-1} ~.
\label{cT2_limit_mu=0}
\end{eqnarray}

%%%%%%%%%%%%%%%%%%%%%%%%%%%%%%%%%%%%%%%%%%

\bibliography{Thesis_my_bibliography}

\begin{thebibliography}{100}
\providecommand{\url}[1]{#1}
\csname url@samestyle\endcsname
\providecommand{\newblock}{\relax}
\providecommand{\bibinfo}[2]{#2}
\providecommand{\BIBentrySTDinterwordspacing}{\spaceskip=0pt\relax}
\providecommand{\BIBentryALTinterwordstretchfactor}{4}
\providecommand{\BIBentryALTinterwordspacing}{\spaceskip=\fontdimen2\font plus
\BIBentryALTinterwordstretchfactor\fontdimen3\font minus
  \fontdimen4\font\relax}
\providecommand{\BIBforeignlanguage}[2]{{%
\expandafter\ifx\csname l@#1\endcsname\relax
\typeout{** WARNING: IEEEtran.bst: No hyphenation pattern has been}%
\typeout{** loaded for the language `#1'. Using the pattern for}%
\typeout{** the default language instead.}%
\else
\language=\csname l@#1\endcsname
\fi
#2}}
\providecommand{\BIBdecl}{\relax}
\BIBdecl

\bibitem{Fritzsch:1973pi}
H.~Fritzsch, M.~Gell-Mann, and H.~Leutwyler,
  ``{\href{https://doi.org/10.1016/0370-2693(73)90625-4}{Advantages of the
  Color Octet Gluon Picture}},'' \emph{Phys. Lett. B}, vol.~47, pp. 365--368,
  1973.

\bibitem{Gross:1973id}
D.~J. Gross and F.~Wilczek,
  ``{\href{https://doi.org/10.1103/PhysRevLett.30.1343}{Ultraviolet Behavior of
  Nonabelian Gauge Theories}},'' \emph{Phys. Rev. Lett.}, vol.~30, pp.
  1343--1346, 1973.

\bibitem{Politzer:1973fx}
H.~D. Politzer, ``{\href{https://doi.org/10.1103/PhysRevLett.30.1346}{Reliable
  Perturbative Results for Strong Interactions?}}'' \emph{Phys. Rev. Lett.},
  vol.~30, pp. 1346--1349, 1973.

\bibitem{Heinz:2000bk}
U.~W. Heinz and M.~Jacob,
  ``{\href{https://arxiv.org/pdf/nucl-th/0002042}{Evidence for a new state of
  matter: An Assessment of the results from the CERN lead beam program}},'' 1
  2000.

\bibitem{BRAHMS:2004adc}
I.~Arsene \emph{et~al.}, ``{\href{https://arxiv.org/pdf/nucl-ex/0410020}{Quark
  gluon plasma and color glass condensate at RHIC? The Perspective from the
  BRAHMS experiment}},'' \emph{Nucl. Phys. A}, vol. 757, pp. 1--27, 2005.

\bibitem{PHENIX:2004vcz}
K.~Adcox \emph{et~al.},
  ``{\href{https://arxiv.org/pdf/nucl-ex/0410003}{Formation of dense partonic
  matter in relativistic nucleus-nucleus collisions at RHIC: Experimental
  evaluation by the PHENIX collaboration}},'' \emph{Nucl. Phys. A}, vol. 757,
  pp. 184--283, 2005.

\bibitem{PHOBOS:2004zne}
B.~B. Back \emph{et~al.}, ``{\href{https://arxiv.org/pdf/nucl-ex/0410022}{The
  PHOBOS perspective on discoveries at RHIC}},'' \emph{Nucl. Phys. A}, vol.
  757, pp. 28--101, 2005.

\bibitem{STAR:2005gfr}
J.~Adams \emph{et~al.},
  ``{\href{https://arxiv.org/pdf/nucl-ex/0501009}{Experimental and theoretical
  challenges in the search for the quark gluon plasma: The STAR Collaboration's
  critical assessment of the evidence from RHIC collisions}},'' \emph{Nucl.
  Phys. A}, vol. 757, pp. 102--183, 2005.

\bibitem{Muller:2012zq}
B.~Muller, J.~Schukraft, and B.~Wyslouch,
  ``{\href{https://arxiv.org/pdf/1202.3233}{First Results from Pb+Pb collisions
  at the LHC}},'' \emph{Ann. Rev. Nucl. Part. Sci.}, vol.~62, pp. 361--386,
  2012.

\bibitem{LongRangePlan1983}
``{\href{https://science.osti.gov/-/media/np/nsac/pdf/docs/lrp_1983.pdf}{A Long
  Range Plan for Nuclear Science}},'' 1983.

\bibitem{Akiba:2015jwa}
Y.~Akiba \emph{et~al.}, ``{\href{https://arxiv.org/pdf/1502.02730}{The Hot QCD
  White Paper: Exploring the Phases of QCD at RHIC and the LHC}},'' 2 2015.

\bibitem{Elliott:2013pna}
J.~B. Elliott, P.~T. Lake, L.~G. Moretto, and L.~Phair,
  ``{\href{https://doi.org/10.1103/PhysRevC.87.054622}{Determination of the
  coexistence curve, critical temperature, density, and pressure of bulk
  nuclear matter from fragment emission data}},'' \emph{Phys. Rev. C}, vol.~87,
  no.~5, p. 054622, 2013.

\bibitem{Ozel:2016oaf}
F.~\"Ozel and P.~Freire, ``{\href{https://arxiv.org/pdf/1603.02698}{Masses,
  Radii, and the Equation of State of Neutron Stars}},'' \emph{Ann. Rev.
  Astron. Astrophys.}, vol.~54, pp. 401--440, 2016.

\bibitem{Silbar:2003wm}
R.~R. Silbar and S.~Reddy,
  ``{\href{https://arxiv.org/pdf/nucl-th/0309041}{Neutron stars for
  undergraduates}},'' \emph{Am. J. Phys.}, vol.~72, pp. 892--905, 2004,
  [Erratum: Am.J.Phys. 73, 286 (2005)].

\bibitem{Fujimoto:2019hxv}
Y.~Fujimoto, K.~Fukushima, and K.~Murase,
  ``{\href{https://arxiv.org/pdf/1903.03400}{Mapping neutron star data to the
  equation of state using the deep neural network}},'' \emph{Phys. Rev. D},
  vol. 101, no.~5, p. 054016, 2020.

\bibitem{Raaijmakers:2021uju}
G.~Raaijmakers, S.~K. Greif, K.~Hebeler, T.~Hinderer, S.~Nissanke, A.~Schwenk,
  T.~E. Riley, A.~L. Watts, J.~M. Lattimer, and W.~C.~G. Ho,
  ``{\href{https://arxiv.org/pdf/2105.06981}{Constraints on the dense matter
  equation of state and neutron star properties from NICER's mass-radius
  estimate of PSR J0740+6620 and multimessenger observations}},'' 5 2021.

\bibitem{Pethick:2015jma}
C.~J. Pethick, T.~Schaefer, and A.~Schwenk,
  ``{\href{https://arxiv.org/pdf/1507.05839}{Bose-Einstein condensates in
  neutron stars}},'' 7 2015.

\bibitem{Alford:2007xm}
M.~G. Alford, A.~Schmitt, K.~Rajagopal, and T.~Sch\"afer,
  ``{\href{https://arxiv.org/pdf/0709.4635}{Color superconductivity in dense
  quark matter}},'' \emph{Rev. Mod. Phys.}, vol.~80, pp. 1455--1515, 2008.

\bibitem{Kapusta:2006pm}
J.~I. Kapusta and C.~Gale,
  \emph{{\href{https://doi.org/10.1017/CBO9780511535130}{Finite-Temperature
  Field Theory: Principles and Applications}}}, {Cambridge University Press,
  2011}.

\bibitem{Gasser:1983yg}
J.~Gasser and H.~Leutwyler,
  ``{\href{https://doi.org/10.1016/0003-4916(84)90242-2}{Chiral Perturbation
  Theory to One Loop}},'' \emph{Annals Phys.}, vol. 158, p. 142, 1984.

\bibitem{Hagedorn:1965st}
R.~Hagedorn,
  ``{\href{https://inspirehep.net/files/0f47965fc72ecec80d0e4f8f71f7d9e5}{Statistical
  thermodynamics of strong interactions at high-energies}},'' \emph{Nuovo Cim.
  Suppl.}, vol.~3, pp. 147--186, 1965.

\bibitem{Andronic:2005yp}
A.~Andronic, P.~Braun-Munzinger, and J.~Stachel,
  ``{\href{https://arxiv.org/pdf/nucl-th/0511071}{Hadron production in central
  nucleus-nucleus collisions at chemical freeze-out}},'' \emph{Nucl. Phys. A},
  vol. 772, pp. 167--199, 2006.

\bibitem{Aoki:2006we}
Y.~Aoki, G.~Endrodi, Z.~Fodor, S.~D. Katz, and K.~K. Szabo,
  ``{\href{https://arxiv.org/pdf/hep-lat/0611014}{The order of the quantum
  chromodynamics transition predicted by the standard model of particle
  physics}},'' \emph{Nature}, vol. 443, pp. 675--678, 2006.

\bibitem{Pratt:2015zsa}
S.~Pratt, E.~Sangaline, P.~Sorensen, and H.~Wang,
  ``{\href{https://arxiv.org/pdf/1501.04042}{Constraining the Eq. of State of
  Super-Hadronic Matter from Heavy-Ion Collisions}},'' \emph{Phys. Rev. Lett.},
  vol. 114, p. 202301, 2015.

\bibitem{HotQCD:2018pds}
A.~Bazavov \emph{et~al.}, ``{\href{https://arxiv.org/pdf/1812.08235}{Chiral
  crossover in QCD at zero and non-zero chemical potentials}},'' \emph{Phys.
  Lett. B}, vol. 795, pp. 15--21, 2019.

\bibitem{Aoki:2006br}
Y.~Aoki, Z.~Fodor, S.~D. Katz, and K.~K. Szabo,
  ``{\href{https://arxiv.org/pdf/hep-lat/0609068}{The QCD transition
  temperature: Results with physical masses in the continuum limit}},''
  \emph{Phys. Lett. B}, vol. 643, pp. 46--54, 2006.

\bibitem{Borsanyi:2020fev}
S.~Borsanyi, Z.~Fodor, J.~N. Guenther, R.~Kara, S.~D. Katz, P.~Parotto,
  A.~Pasztor, C.~Ratti, and K.~K. Szabo,
  ``{\href{https://arxiv.org/pdf/2002.02821}{QCD Crossover at Finite Chemical
  Potential from Lattice Simulations}},'' \emph{Phys. Rev. Lett.}, vol. 125,
  no.~5, p. 052001, 2020.

\bibitem{Karsch:2001cy}
F.~Karsch, ``{\href{https://arxiv.org/pdf/hep-lat/0106019}{Lattice QCD at high
  temperature and density}},'' \emph{Lect. Notes Phys.}, vol. 583, pp.
  209--249, 2002.

\bibitem{Pisarski:1983ms}
R.~D. Pisarski and F.~Wilczek,
  ``{\href{https://doi.org/10.1103/PhysRevD.29.338}{Remarks on the Chiral Phase
  Transition in Chromodynamics}},'' \emph{Phys. Rev. D}, vol.~29, pp. 338--341,
  1984.

\bibitem{Stephanov:2004wx}
M.~A. Stephanov, ``{\href{https://arxiv.org/pdf/hep-ph/0402115}{QCD phase
  diagram and the critical point}},'' \emph{Prog. Theor. Phys. Suppl.}, vol.
  153, pp. 139--156, 2004.

\bibitem{Cuteri:2021ikv}
F.~Cuteri, O.~Philipsen, and A.~Sciarra,
  ``{\href{https://arxiv.org/pdf/2107.12739}{On the order of the QCD chiral
  phase transition for different numbers of quark flavours}},'' 7 2021.

\bibitem{Petersen:2008dd}
H.~Petersen, J.~Steinheimer, G.~Burau, M.~Bleicher, and H.~St\"ocker,
  ``{\href{https://arxiv.org/pdf/0806.1695}{A Fully Integrated Transport
  Approach to Heavy Ion Reactions with an Intermediate Hydrodynamic Stage}},''
  \emph{Phys. Rev. C}, vol.~78, p. 044901, 2008.

\bibitem{Bass:1998ca}
S.~A. Bass \emph{et~al.},
  ``{\href{https://arxiv.org/pdf/nucl-th/9803035}{Microscopic models for
  ultrarelativistic heavy ion collisions}},'' \emph{Prog. Part. Nucl. Phys.},
  vol.~41, pp. 255--369, 1998.

\bibitem{Bleicher:1999xi}
M.~Bleicher \emph{et~al.},
  ``{\href{https://arxiv.org/pdf/hep-ph/9909407}{Relativistic hadron hadron
  collisions in the ultrarelativistic quantum molecular dynamics model}},''
  \emph{J. Phys. G}, vol.~25, pp. 1859--1896, 1999.

\bibitem{Rischke:1995ir}
D.~H. Rischke, S.~Bernard, and J.~A. Maruhn,
  ``{\href{https://arxiv.org/pdf/nucl-th/9504018}{Relativistic hydrodynamics
  for heavy ion collisions. 1. General aspects and expansion into vacuum}},''
  \emph{Nucl. Phys. A}, vol. 595, pp. 346--382, 1995.

\bibitem{Rischke:1995mt}
D.~H. Rischke, Y.~Pursun, and J.~A. Maruhn,
  ``{\href{https://arxiv.org/pdf/nucl-th/9504021}{Relativistic hydrodynamics
  for heavy ion collisions. 2. Compression of nuclear matter and the phase
  transition to the quark - gluon plasma}},'' \emph{Nucl. Phys. A}, vol. 595,
  pp. 383--408, 1995, [Erratum: Nucl.Phys.A 596, 717--717 (1996)].

\bibitem{MADAI_collaboration}
{MADAI collaboration}, {\url{http://madai.us}.}

\bibitem{Weil:2016zrk}
J.~Weil \emph{et~al.}, ``{\href{https://arxiv.org/pdf/1606.06642}{Particle
  production and equilibrium properties within a new hadron transport approach
  for heavy-ion collisions}},'' \emph{Phys. Rev. C}, vol.~94, no.~5, p. 054905,
  2016.

\bibitem{SMASH_website}
{SMASH hadronic transport, available at
  \href{https://smash-transport.github.io}{https://smash-transport.github.io}.}

\bibitem{H1:2009pze}
F.~D. Aaron \emph{et~al.}, ``{\href{https://arxiv.org/pdf/0911.0884}{Combined
  Measurement and QCD Analysis of the Inclusive e+- p Scattering Cross Sections
  at HERA}},'' \emph{JHEP}, vol.~01, p. 109, 2010.

\bibitem{BRAHMS:2003wwg}
I.~G. Bearden \emph{et~al.},
  ``{\href{https://arxiv.org/pdf/nucl-ex/0312023}{Nuclear stopping in Au + Au
  collisions at s(NN)**(1/2) = 200-GeV}},'' \emph{Phys. Rev. Lett.}, vol.~93,
  p. 102301, 2004.

\bibitem{E802:1999hit}
L.~Ahle \emph{et~al.},
  ``{\href{https://doi.org/10.1103/PhysRevC.60.064901}{Proton and deuteron
  production in Au + Au reactions at 11.6/A-GeV/c}},'' \emph{Phys. Rev. C},
  vol.~60, p. 064901, 1999.

\bibitem{NA49:1998gaz}
H.~Appelshauser \emph{et~al.},
  ``{\href{https://arxiv.org/pdf/nucl-ex/9810014}{Baryon stopping and charged
  particle distributions in central Pb + Pb collisions at 158-GeV per
  nucleon}},'' \emph{Phys. Rev. Lett.}, vol.~82, pp. 2471--2475, 1999.

\bibitem{BRAHMS:2004dwr}
I.~G. Bearden \emph{et~al.},
  ``{\href{https://arxiv.org/pdf/nucl-ex/0403050}{Charged meson rapidity
  distributions in central Au+Au collisions at s(NN)**(1/2) = 200-GeV}},''
  \emph{Phys. Rev. Lett.}, vol.~94, p. 162301, 2005.

\bibitem{PHOBOS:2004hlv}
B.~B. Back \emph{et~al.},
  ``{\href{https://arxiv.org/pdf/nucl-ex/0405027}{Collision geometry scaling of
  Au+Au pseudorapidity density from s(NN)**(1/2) = 19.6-GeV to 200-GeV}},''
  \emph{Phys. Rev. C}, vol.~70, p. 021902, 2004.

\bibitem{Capella:1999cz}
A.~Capella and C.~A. Salgado,
  ``{\href{https://arxiv.org/pdf/hep-ph/9903414}{Baryon stopping and hyperon
  enhancement in the improved dual parton model}},'' \emph{Phys. Rev. C},
  vol.~60, p. 054906, 1999.

\bibitem{Mehtar-Tani:2009wji}
Y.~Mehtar-Tani and G.~Wolschin,
  ``{\href{https://arxiv.org/pdf/0907.5444}{Baryon stopping and saturation
  physics in relativistic collisions}},'' \emph{Phys. Rev. C}, vol.~80, p.
  054905, 2009.

\bibitem{McLerran:2018avb}
L.~D. McLerran, S.~Schlichting, and S.~Sen,
  ``{\href{https://arxiv.org/pdf/1811.04089}{Spacetime picture of baryon
  stopping in the color-glass condensate}},'' \emph{Phys. Rev. D}, vol.~99,
  no.~7, p. 074009, 2019.

\bibitem{Denicol:2018wdp}
G.~S. Denicol, C.~Gale, S.~Jeon, A.~Monnai, B.~Schenke, and C.~Shen,
  ``{\href{https://arxiv.org/pdf/1804.10557}{Net baryon diffusion in fluid
  dynamic simulations of relativistic heavy-ion collisions}},'' \emph{Phys.
  Rev. C}, vol.~98, no.~3, p. 034916, 2018.

\bibitem{Li:2018fow}
M.~Li and C.~Shen, ``{\href{https://arxiv.org/pdf/1809.04034}{Longitudinal
  Dynamics of High Baryon Density Matter in High Energy Heavy-Ion
  Collisions}},'' \emph{Phys. Rev. C}, vol.~98, no.~6, p. 064908, 2018.

\bibitem{Cleymans:2005xv}
J.~Cleymans, H.~Oeschler, K.~Redlich, and S.~Wheaton,
  ``{\href{https://arxiv.org/pdf/hep-ph/0511094}{Comparison of chemical
  freeze-out criteria in heavy-ion collisions}},'' \emph{Phys. Rev. C},
  vol.~73, p. 034905, 2006.

\bibitem{STAR:2017sal}
L.~Adamczyk \emph{et~al.}, ``{\href{https://arxiv.org/pdf/1701.07065}{Bulk
  Properties of the Medium Produced in Relativistic Heavy-Ion Collisions from
  the Beam Energy Scan Program}},'' \emph{Phys. Rev. C}, vol.~96, no.~4, p.
  044904, 2017.

\bibitem{Odyniec:2019kfh}
G.~Odyniec,
  ``{\href{https://inspirehep.net/files/4106a0a25a6830993f9f4f1faaf3eb37}{Beam
  Energy Scan Program at RHIC (BES I and BES II) \textendash{} Probing QCD
  Phase Diagram with Heavy-Ion Collisions}},'' \emph{PoS}, vol. CORFU2018, p.
  151, 2019.

\bibitem{STAR:2010vob}
M.~M. Aggarwal \emph{et~al.}, ``{\href{https://arxiv.org/pdf/1007.2613}{An
  Experimental Exploration of the QCD Phase Diagram: The Search for the
  Critical Point and the Onset of De-confinement}},'' 7 2010.

\bibitem{STAR:2003wqp}
J.~Adams \emph{et~al.},
  ``{\href{https://arxiv.org/pdf/nucl-ex/0306007}{Particle type dependence of
  azimuthal anisotropy and nuclear modification of particle production in Au +
  Au collisions at s(NN)**(1/2) = 200-GeV}},'' \emph{Phys. Rev. Lett.},
  vol.~92, p. 052302, 2004.

\bibitem{Galatyuk:2014vha}
T.~Galatyuk, ``{\href{https://doi.org/10.1016/j.nuclphysa.2014.10.044}{HADES
  overview}},'' \emph{Nucl. Phys. A}, vol. 931, pp. 41--51, 2014.

\bibitem{NA49-future:2006qne}
N.~Antoniou \emph{et~al.},
  ``{\href{http://weblib.cern.ch/abstract?CERN-SPSC-2006-034}{Study of hadron
  production in hadron nucleus and nucleus nucleus collisions at the CERN
  SPS}},'' 11 2006.

\bibitem{HanburyBrown:1954amm}
R.~Hanbury~Brown and R.~Q. Twiss,
  ``{\href{https://doi.org/10.1080/14786440708520475}{A new type of
  interferometer for use in radio astronomy}},'' \emph{Phil. Mag. Ser. 7},
  vol.~45, pp. 663--682, 1954.

\bibitem{HanburyBrown:1956bqd}
------, ``{\href{https://doi.org/10.1038/1781046a0}{A Test of a New Type of
  Stellar Interferometer on Sirius}},'' \emph{Nature}, vol. 178, pp.
  1046--1048, 1956.

\bibitem{Zajc:1984vb}
W.~A. Zajc \emph{et~al.},
  ``{\href{https://doi.org/10.1103/PhysRevC.29.2173}{Two pion correlations in
  heavy-ion collisions}},'' \emph{Phys. Rev. C}, vol.~29, pp. 2173--2187, 1984.

\bibitem{Fung:1978eq}
S.~Y. Fung, W.~Gorn, G.~P. Kiernan, J.~J. Lu, Y.~T. Oh, and R.~T. Poe,
  ``{\href{https://doi.org/10.1103/PhysRevLett.41.1592}{Observation of Pion
  Interferometry in Relativistic Nuclear Collisions}},'' \emph{Phys. Rev.
  Lett.}, vol.~41, pp. 1592--1594, 1978.

\bibitem{Heinz:1999rw}
U.~W. Heinz and B.~V. Jacak,
  ``{\href{https://arxiv.org/pdf/nucl-th/9902020}{Two particle correlations in
  relativistic heavy ion collisions}},'' \emph{Ann. Rev. Nucl. Part. Sci.},
  vol.~49, pp. 529--579, 1999.

\bibitem{Lisa:2005dd}
M.~A. Lisa, S.~Pratt, R.~Soltz, and U.~Wiedemann,
  ``{\href{https://arxiv.org/pdf/nucl-ex/0505014}{Femtoscopy in relativistic
  heavy ion collisions}},'' \emph{Ann. Rev. Nucl. Part. Sci.}, vol.~55, pp.
  357--402, 2005.

\bibitem{Gutierrez:2004rm}
T.~D. Gutierrez, ``{\href{https://arxiv.org/pdf/nucl-ex/0403012}{Pion
  interferometry from pp and dAu collisions at RHIC}},'' in \emph{{17th
  International Conference on Ultra Relativistic Nucleus-Nucleus Collisions
  (Quark Matter 2004)}}, 3 2004.

\bibitem{Bertsch:1989vn}
G.~F. Bertsch, ``{\href{https://doi.org/10.1016/0375-9474(89)90597-6}{Pion
  Interferometry as a Probe of the Plasma}},'' \emph{Nucl. Phys. A}, vol. 498,
  pp. 173C--180C, 1989.

\bibitem{Pratt:1990zq}
S.~Pratt, T.~Csorgo, and J.~Zimanyi,
  ``{\href{https://doi.org/10.1103/PhysRevC.42.2646}{Detailed predictions for
  two pion correlations in ultrarelativistic heavy ion collisions}},''
  \emph{Phys. Rev. C}, vol.~42, pp. 2646--2652, 1990.

\bibitem{Lacey:2014wqa}
R.~A. Lacey, ``{\href{https://arxiv.org/pdf/1411.7931}{Indications for a
  Critical End Point in the Phase Diagram for Hot and Dense Nuclear Matter}},''
  \emph{Phys. Rev. Lett.}, vol. 114, no.~14, p. 142301, 2015.

\bibitem{STAR:2014shf}
L.~Adamczyk \emph{et~al.},
  ``{\href{https://arxiv.org/pdf/1403.4972}{Beam-energy-dependent two-pion
  interferometry and the freeze-out eccentricity of pions measured in heavy ion
  collisions at the STAR detector}},'' \emph{Phys. Rev. C}, vol.~92, no.~1, p.
  014904, 2015.

\bibitem{ALICE:2011dyt}
K.~Aamodt \emph{et~al.}, ``{\href{https://arxiv.org/pdf/1012.4035}{Two-pion
  Bose-Einstein correlations in central Pb-Pb collisions at $\sqrt{{s}_{NN}} =$
  2.76 TeV}},'' \emph{Phys. Lett. B}, vol. 696, pp. 328--337, 2011.

\bibitem{Pratt:1988vn}
S.~Pratt, ``{\href{https://doi.org/10.1063/1.37709}{Two-pion interferometry as
  a signal for the quark-gluon plasma}},'' \emph{AIP Conf. Proc.}, vol. 176,
  pp. 1003--1008, 2008.

\bibitem{Bertsch:1988db}
G.~Bertsch, M.~Gong, and M.~Tohyama,
  ``{\href{https://doi.org/10.1103/PhysRevC.37.1896}{Pion Interferometry in
  Ultrarelativistic Heavy Ion Collisions}},'' \emph{Phys. Rev. C}, vol.~37, pp.
  1896--1900, 1988.

\bibitem{Hung:1994eq}
C.~M. Hung and E.~V. Shuryak,
  ``{\href{https://arxiv.org/pdf/hep-ph/9412360}{Hydrodynamics near the QCD
  phase transition: Looking for the longest lived fireball}},'' \emph{Phys.
  Rev. Lett.}, vol.~75, pp. 4003--4006, 1995.

\bibitem{Rischke:1996em}
D.~H. Rischke and M.~Gyulassy,
  ``{\href{https://arxiv.org/pdf/nucl-th/9606039}{The time-delay signature of
  quark-gluon plasma formation in relativistic nuclear collisions}},''
  \emph{Nucl. Phys. A}, vol. 608, pp. 479--512, 1996.

\bibitem{STAR:2020dav}
J.~Adam \emph{et~al.}, ``{\href{https://arxiv.org/pdf/2007.14005}{Flow and
  interferometry results from Au+Au collisions at $\sqrt{s_{NN}} = 4.5$
  GeV}},'' \emph{Phys. Rev. C}, vol. 103, no.~3, p. 034908, 2021.

\bibitem{Voloshin:2008dg}
S.~A. Voloshin, A.~M. Poskanzer, and R.~Snellings,
  ``{\href{https://arxiv.org/pdf/0809.2949}{Collective phenomena in non-central
  nuclear collisions}},'' \emph{Landolt-Bornstein}, vol.~23, pp. 293--333,
  2010.

\bibitem{Rischke:1995pe}
D.~H. Rischke, Y.~Pursun, J.~A. Maruhn, H.~Stoecker, and W.~Greiner,
  ``{\href{https://arxiv.org/pdf/nucl-th/9505014}{The Phase Transition to the
  Quark-Qluon Plasma and Its Effect on Hydrodynamic Flow}},'' \emph{Acta Phys.
  Hung. A}, vol.~1, pp. 309--322, 1995.

\bibitem{Stoecker:2004qu}
H.~Stoecker, ``{\href{https://arxiv.org/pdf/nucl-th/0406018}{Collective flow
  signals the quark gluon plasma}},'' \emph{Nucl. Phys. A}, vol. 750, pp.
  121--147, 2005.

\bibitem{STAR:2014clz}
L.~Adamczyk \emph{et~al.},
  ``{\href{https://arxiv.org/pdf/1401.3043}{Beam-Energy Dependence of the
  Directed Flow of Protons, Antiprotons, and Pions in Au+Au Collisions}},''
  \emph{Phys. Rev. Lett.}, vol. 112, no.~16, p. 162301, 2014.

\bibitem{Steinheimer:2014pfa}
J.~Steinheimer, J.~Auvinen, H.~Petersen, M.~Bleicher, and H.~St\"ocker,
  ``{\href{https://arxiv.org/pdf/1402.7236}{Examination of directed flow as a
  signal for a phase transition in relativistic nuclear collisions}},''
  \emph{Phys. Rev. C}, vol.~89, no.~5, p. 054913, 2014.

\bibitem{Nara:2016hbg}
Y.~Nara, H.~Niemi, J.~Steinheimer, and H.~St\"ocker,
  ``{\href{https://arxiv.org/pdf/1611.08023}{Equation of state dependence of
  directed flow in a microscopic transport model}},'' \emph{Phys. Lett. B},
  vol. 769, pp. 543--548, 2017.

\bibitem{Hillmann:2018nmd}
P.~Hillmann, J.~Steinheimer, and M.~Bleicher,
  ``{\href{https://arxiv.org/pdf/1802.01951}{Directed, elliptic and triangular
  flow of protons in Au+Au reactions at 1.23 A GeV: a theoretical analysis of
  the recent HADES data}},'' \emph{J. Phys. G}, vol.~45, no.~8, p. 085101,
  2018.

\bibitem{Danielewicz:2002pu}
P.~Danielewicz, R.~Lacey, and W.~G. Lynch,
  ``{\href{https://arxiv.org/pdf/nucl-th/0208016}{Determination of the equation
  of state of dense matter}},'' \emph{Science}, vol. 298, pp. 1592--1596, 2002.

\bibitem{Molnar:2003ff}
D.~Molnar and S.~A. Voloshin,
  ``{\href{https://arxiv.org/pdf/nucl-th/0302014}{Elliptic flow at large
  transverse momenta from quark coalescence}},'' \emph{Phys. Rev. Lett.},
  vol.~91, p. 092301, 2003.

\bibitem{Schwarzschild:1963zz}
A.~Schwarzschild and C.~Zupancic,
  ``{\href{https://doi.org/10.1103/PhysRev.129.854}{Production of Tritons,
  Deuterons, Nucleons, and Mesons by 30-GeV Protons on A-1, Be, and Fe
  Targets}},'' \emph{Phys. Rev.}, vol. 129, pp. 854--862, 1963.

\bibitem{Sato:1981ez}
H.~Sato and K.~Yazaki,
  ``{\href{https://doi.org/10.1016/0370-2693(81)90976-X}{On the coalescence
  model for high-energy nuclear reactions}},'' \emph{Phys. Lett. B}, vol.~98,
  pp. 153--157, 1981.

\bibitem{STAR:2013cow}
L.~Adamczyk \emph{et~al.},
  ``{\href{https://arxiv.org/pdf/1301.2347}{Observation of an Energy-Dependent
  Difference in Elliptic Flow between Particles and Antiparticles in
  Relativistic Heavy Ion Collisions}},'' \emph{Phys. Rev. Lett.}, vol. 110,
  no.~14, p. 142301, 2013.

\bibitem{Mishra:2007tw}
A.~P. Mishra, R.~K. Mohapatra, P.~S. Saumia, and A.~M. Srivastava,
  ``{\href{https://arxiv.org/pdf/0711.1323}{Super-horizon fluctuations and
  acoustic oscillations in relativistic heavy-ion collisions}},'' \emph{Phys.
  Rev. C}, vol.~77, p. 064902, 2008.

\bibitem{Sorensen:2010zq}
P.~Sorensen, ``{\href{https://arxiv.org/pdf/1002.4878}{Implications of
  space-momentum correlations and geometric fluctuations in heavy-ion
  collisions}},'' \emph{J. Phys. G}, vol.~37, p. 094011, 2010.

\bibitem{Petersen:2012qc}
H.~Petersen, R.~La~Placa, and S.~A. Bass,
  ``{\href{https://arxiv.org/pdf/1201.1881}{A systematic study of the
  sensitivity of triangular flow to the initial state fluctuations in
  relativistic heavy-ion collisions}},'' \emph{J. Phys. G}, vol.~39, p. 055102,
  2012.

\bibitem{Landau_Fluid_Mechanics}
L.~D. Landau and E.~M. Lifshitz, \emph{Fluid Mechanics, Course of Theoretical
  Physics, Vol.\ 6}, {Pergamon Press, 1987}.

\bibitem{STAR:2016vqt}
L.~Adamczyk \emph{et~al.}, ``{\href{https://arxiv.org/pdf/1601.01999}{Beam
  Energy Dependence of the Third Harmonic of Azimuthal Correlations in Au+Au
  Collisions at RHIC}},'' \emph{Phys. Rev. Lett.}, vol. 116, no.~11, p. 112302,
  2016.

\bibitem{STAR:2013qio}
------, ``{\href{https://arxiv.org/pdf/1301.2187}{Third Harmonic Flow of
  Charged Particles in Au+Au Collisions at sqrtsNN = 200 GeV}},'' \emph{Phys.
  Rev. C}, vol.~88, no.~1, p. 014904, 2013.

\bibitem{Cueto-Felgueroso_VdW_EOS}
L.~Cueto-Felgueroso, X.~Fu, and R.~Juanes,
  ``\href{https://link.aps.org/doi/10.1103/PhysRevFluids.3.084302}{Pore-scale
  modeling of phase change in porous media},'' \emph{Phys. Rev. Fluids},
  vol.~3, p. 084302, Aug 2018.

\bibitem{Stephanov:1998dy}
M.~A. Stephanov, K.~Rajagopal, and E.~V. Shuryak,
  ``{\href{https://arxiv.org/pdf/hep-ph/9806219}{Signatures of the tricritical
  point in QCD}},'' \emph{Phys. Rev. Lett.}, vol.~81, pp. 4816--4819, 1998.

\bibitem{Stephanov:1999zu}
------, ``{\href{https://arxiv.org/pdf/hep-ph/9903292}{Event-by-event
  fluctuations in heavy ion collisions and the QCD critical point}},''
  \emph{Phys. Rev. D}, vol.~60, p. 114028, 1999.

\bibitem{Koch:2008ia}
V.~Koch, ``{\href{https://arxiv.org/pdf/0810.2520}{Hadronic Fluctuations and
  Correlations}},'' {\textit{Relativistic Heavy Ion Physics}, pp. 626--652,
  Springer-Verlag Berlin Heidelberg, 2010}.

\bibitem{Asakawa:2009aj}
M.~Asakawa, S.~Ejiri, and M.~Kitazawa,
  ``{\href{https://arxiv.org/pdf/0904.2089}{Third moments of conserved charges
  as probes of QCD phase structure}},'' \emph{Phys. Rev. Lett.}, vol. 103, p.
  262301, 2009.

\bibitem{Stephanov:2011pb}
M.~A. Stephanov, ``{\href{https://arxiv.org/pdf/1104.1627}{On the sign of
  kurtosis near the QCD critical point}},'' \emph{Phys. Rev. Lett.}, vol. 107,
  p. 052301, 2011.

\bibitem{Stephanov:2008qz}
------, ``{\href{https://arxiv.org/pdf/0809.3450}{Non-Gaussian fluctuations
  near the QCD critical point}},'' \emph{Phys. Rev. Lett.}, vol. 102, p.
  032301, 2009.

\bibitem{Zinn-Justin:2002ecy}
J.~Zinn-Justin, \emph{{Quantum Field Theory and Critical Phenomena}},
  {Clarendon Press, 2002}.

\bibitem{Bzdak:2019pkr}
A.~Bzdak, S.~Esumi, V.~Koch, J.~Liao, M.~Stephanov, and N.~Xu,
  ``{\href{https://arxiv.org/pdf/1906.00936}{Mapping the Phases of Quantum
  Chromodynamics with Beam Energy Scan}},'' \emph{Phys. Rept.}, vol. 853, pp.
  1--87, 2020.

\bibitem{Kitazawa:2011wh}
M.~Kitazawa and M.~Asakawa, ``{\href{https://arxiv.org/pdf/1107.2755}{Revealing
  baryon number fluctuations from proton number fluctuations in relativistic
  heavy ion collisions}},'' \emph{Phys. Rev. C}, vol.~85, p. 021901, 2012.

\bibitem{Vovchenko:2020gne}
V.~Vovchenko, R.~V. Poberezhnyuk, and V.~Koch,
  ``{\href{https://arxiv.org/pdf/2007.03850}{Cumulants of multiple conserved
  charges and global conservation laws}},'' \emph{JHEP}, vol.~10, p. 089, 2020.

\bibitem{STAR:2021iop}
M.~Abdallah \emph{et~al.}, ``{\href{https://arxiv.org/pdf/2101.12413}{Cumulants
  and correlation functions of net-proton, proton, and antiproton multiplicity
  distributions in Au+Au collisions at energies available at the BNL
  Relativistic Heavy Ion Collider}},'' \emph{Phys. Rev. C}, vol. 104, no.~2, p.
  024902, 2021.

\bibitem{Berdnikov:1999ph}
B.~Berdnikov and K.~Rajagopal,
  ``{\href{https://arxiv.org/pdf/hep-ph/9912274}{Slowing out-of-equilibrium
  near the QCD critical point}},'' \emph{Phys. Rev. D}, vol.~61, p. 105017,
  2000.

\bibitem{Nahrgang:2018afz}
M.~Nahrgang, M.~Bluhm, T.~Schaefer, and S.~A. Bass,
  ``{\href{https://arxiv.org/pdf/1804.05728}{Diffusive dynamics of critical
  fluctuations near the QCD critical point}},'' \emph{Phys. Rev. D}, vol.~99,
  no.~11, p. 116015, 2019.

\bibitem{Shen:2020gef}
C.~Shen, ``{\href{https://arxiv.org/pdf/2001.11858}{Studying QGP with flow: A
  theory overview}},'' \emph{Nucl. Phys. A}, vol. 1005, p. 121788, 2021.

\bibitem{Batyuk:2016qmb}
P.~Batyuk, D.~Blaschke, M.~Bleicher, Y.~B. Ivanov, I.~Karpenko, S.~Merts,
  M.~Nahrgang, H.~Petersen, and O.~Rogachevsky,
  ``{\href{https://arxiv.org/pdf/1608.00965}{Event simulation based on
  three-fluid hydrodynamics for collisions at energies available at the Dubna
  Nuclotron-based Ion Collider Facility and at the Facility for Antiproton and
  Ion Research in Darmstadt}},'' \emph{Phys. Rev. C}, vol.~94, p. 044917, 2016.

\bibitem{Hohenberg:1977ym}
P.~C. Hohenberg and B.~I. Halperin,
  ``{\href{https://doi.org/10.1103/RevModPhys.49.435}{Theory of Dynamic
  Critical Phenomena}},'' \emph{Rev. Mod. Phys.}, vol.~49, pp. 435--479, 1977.

\bibitem{Akamatsu:2018vjr}
Y.~Akamatsu, D.~Teaney, F.~Yan, and Y.~Yin,
  ``{\href{https://arxiv.org/pdf/1811.05081}{Transits of the QCD critical
  point}},'' \emph{Phys. Rev. C}, vol. 100, no.~4, p. 044901, 2019.

\bibitem{Asakawa:2019kek}
M.~Asakawa, M.~Kitazawa, and B.~M\"uller,
  ``{\href{https://arxiv.org/pdf/1912.05840}{Issues with the search for
  critical point in QCD with relativistic heavy ion collisions}},'' \emph{Phys.
  Rev. C}, vol. 101, no.~3, p. 034913, 2020.

\bibitem{Bjorken:1982qr}
J.~D. Bjorken, ``{\href{https://doi.org/10.1103/PhysRevD.27.140}{Highly
  Relativistic Nucleus-Nucleus Collisions: The Central Rapidity Region}},''
  \emph{Phys. Rev. D}, vol.~27, pp. 140--151, 1983.

\bibitem{Teaney:2000cw}
D.~Teaney, J.~Lauret, and E.~V. Shuryak,
  ``{\href{https://arxiv.org/pdf/nucl-th/0011058}{Flow at the SPS and RHIC as a
  quark gluon plasma signature}},'' \emph{Phys. Rev. Lett.}, vol.~86, pp.
  4783--4786, 2001.

\bibitem{Teaney:2001av}
------, ``{\href{https://arxiv.org/pdf/nucl-th/0110037}{A Hydrodynamic
  Description of Heavy Ion Collisions at the SPS and RHIC}},'' 10 2001.

\bibitem{Steinheimer:2016cir}
J.~Steinheimer, V.~Vovchenko, J.~Aichelin, M.~Bleicher, and H.~St\"ocker,
  ``{\href{https://arxiv.org/pdf/1608.03737}{Conserved charge fluctuations are
  not conserved during the hadronic phase}},'' \emph{Phys. Lett. B}, vol. 776,
  pp. 32--37, 2018.

\bibitem{Miller:2007ri}
M.~L. Miller, K.~Reygers, S.~J. Sanders, and P.~Steinberg,
  ``{\href{https://arxiv.org/pdf/nucl-ex/0701025}{Glauber modeling in high
  energy nuclear collisions}},'' \emph{Ann. Rev. Nucl. Part. Sci.}, vol.~57,
  pp. 205--243, 2007.

\bibitem{Schenke:2012wb}
B.~Schenke, P.~Tribedy, and R.~Venugopalan,
  ``{\href{https://arxiv.org/pdf/1202.6646}{Fluctuating Glasma initial
  conditions and flow in heavy ion collisions}},'' \emph{Phys. Rev. Lett.},
  vol. 108, p. 252301, 2012.

\bibitem{Schenke:2012hg}
------, ``{\href{https://arxiv.org/pdf/1206.6805}{Event-by-event gluon
  multiplicity, energy density, and eccentricities in ultrarelativistic
  heavy-ion collisions}},'' \emph{Phys. Rev. C}, vol.~86, p. 034908, 2012.

\bibitem{Moreland:2014oya}
J.~S. Moreland, J.~E. Bernhard, and S.~A. Bass,
  ``{\href{https://arxiv.org/pdf/1412.4708}{Alternative ansatz to wounded
  nucleon and binary collision scaling in high-energy nuclear collisions}},''
  \emph{Phys. Rev. C}, vol.~92, no.~1, p. 011901, 2015.

\bibitem{Karpenko:2015xea}
I.~A. Karpenko, P.~Huovinen, H.~Petersen, and M.~Bleicher,
  ``{\href{https://arxiv.org/pdf/1502.01978}{Estimation of the shear viscosity
  at finite net-baryon density from $A+A$ collision data at
  $\sqrt{s_\mathrm{NN}} = 7.7-200$ GeV}},'' \emph{Phys. Rev. C}, vol.~91,
  no.~6, p. 064901, 2015.

\bibitem{Du:2018mpf}
L.~Du, U.~Heinz, and G.~Vujanovic,
  ``{\href{https://arxiv.org/pdf/1807.04721}{Hybrid model with dynamical
  sources for heavy-ion collisions at BES energies}},'' \emph{Nucl. Phys. A},
  vol. 982, pp. 407--410, 2019.

\bibitem{Shen:2017bsr}
C.~Shen and B.~Schenke, ``{\href{https://arxiv.org/pdf/1710.00881}{Dynamical
  initial state model for relativistic heavy-ion collisions}},'' \emph{Phys.
  Rev. C}, vol.~97, no.~2, p. 024907, 2018.

\bibitem{Kurkela:2018vqr}
A.~Kurkela, A.~Mazeliauskas, J.-F. Paquet, S.~Schlichting, and D.~Teaney,
  ``{\href{https://arxiv.org/pdf/1805.00961}{Effective kinetic description of
  event-by-event pre-equilibrium dynamics in high-energy heavy-ion
  collisions}},'' \emph{Phys. Rev. C}, vol.~99, no.~3, p. 034910, 2019.

\bibitem{STAR:2000ekf}
K.~H. Ackermann \emph{et~al.},
  ``{\href{https://arxiv.org/pdf/nucl-ex/0009011}{Elliptic flow in Au + Au
  collisions at (S(NN))**(1/2) = 130 GeV}},'' \emph{Phys. Rev. Lett.}, vol.~86,
  pp. 402--407, 2001.

\bibitem{Schenke:2010nt}
B.~Schenke, S.~Jeon, and C.~Gale,
  ``{\href{https://arxiv.org/pdf/1004.1408}{(3+1)D hydrodynamic simulation of
  relativistic heavy-ion collisions}},'' \emph{Phys. Rev. C}, vol.~82, p.
  014903, 2010.

\bibitem{Noronha-Hostler:2013gga}
J.~Noronha-Hostler, G.~S. Denicol, J.~Noronha, R.~P.~G. Andrade, and F.~Grassi,
  ``{\href{https://arxiv.org/pdf/1305.1981}{Bulk Viscosity Effects in
  Event-by-Event Relativistic Hydrodynamics}},'' \emph{Phys. Rev. C}, vol.~88,
  no.~4, p. 044916, 2013.

\bibitem{Noronha-Hostler:2014dqa}
J.~Noronha-Hostler, J.~Noronha, and F.~Grassi,
  ``{\href{https://arxiv.org/pdf/1406.3333}{Bulk viscosity-driven suppression
  of shear viscosity effects on the flow harmonics at energies available at the
  BNL Relativistic Heavy Ion Collider}},'' \emph{Phys. Rev. C}, vol.~90, no.~3,
  p. 034907, 2014.

\bibitem{Du:2019obx}
L.~Du and U.~Heinz,
  ``{\href{https://arxiv.org/pdf/1906.11181}{(3+1)-dimensional dissipative
  relativistic fluid dynamics at non-zero net baryon density}},'' \emph{Comput.
  Phys. Commun.}, vol. 251, p. 107090, 2020.

\bibitem{Wu:2021fjf}
X.-Y. Wu, G.-Y. Qin, L.-G. Pang, and X.-N. Wang,
  ``{\href{https://arxiv.org/pdf/2107.04949}{(3+1)-D viscous hydrodynamics
  CLVisc at finite net baryon density: identified particle spectra, anisotropic
  flows and flow fluctuations across BES energies}},'' 7 2021.

\bibitem{Ahmad:2016ods}
S.~Ahmad, H.~Holopainen, and P.~Huovinen,
  ``{\href{https://arxiv.org/pdf/1608.03444}{Dynamical freeze-out criterion in
  a hydrodynamical description of Au + Au collisions at
  $\sqrt{s_\mathrm{NN}}=200$ GeV and Pb + Pb collisions at
  $\sqrt{s_\mathrm{NN}}=2760$ GeV}},'' \emph{Phys. Rev. C}, vol.~95, no.~5, p.
  054911, 2017.

\bibitem{McNelis:2019auj}
M.~McNelis, D.~Everett, and U.~Heinz,
  ``{\href{https://arxiv.org/pdf/1912.08271}{Particlization in fluid dynamical
  simulations of heavy-ion collisions: The iS3D module}},'' \emph{Comput. Phys.
  Commun.}, vol. 258, p. 107604, 2021.

\bibitem{Ryu:2019atv}
S.~Ryu, J.~Staudenmaier, and H.~Elfner,
  ``{\href{https://inspirehep.net/files/4f51ace18aadd6e03c9709dd36b54a75}{Bulk
  Observables within a Hybrid Approach for Heavy Ion Collisions with SMASH
  Afterburner}},'' \emph{MDPI Proc.}, vol.~10, no.~1, p.~44, 2019.

\bibitem{Oliinychenko:2019zfk}
D.~Oliinychenko and V.~Koch,
  ``{\href{https://arxiv.org/pdf/1902.09775}{Microcanonical Particlization with
  Local Conservation Laws}},'' \emph{Phys. Rev. Lett.}, vol. 123, no.~18, p.
  182302, 2019.

\bibitem{Oliinychenko:2020cmr}
D.~Oliinychenko, S.~Shi, and V.~Koch,
  ``{\href{https://arxiv.org/pdf/2001.08176}{Effects of local event-by-event
  conservation laws in ultrarelativistic heavy-ion collisions at
  particlization}},'' \emph{Phys. Rev. C}, vol. 102, no.~3, p. 034904, 2020.

\bibitem{Martinez:2019rlp}
P.~Carzon, M.~Martinez, M.~D. Sievert, D.~E. Wertepny, and J.~Noronha-Hostler,
  ``{\href{https://arxiv.org/pdf/1911.12454}{Toward Initial Conditions of
  Conserved Charges Part II: The ICCING Monte Carlo Algorithm}},'' 11 2019.

\bibitem{Rougemont:2017tlu}
R.~Rougemont, R.~Critelli, J.~Noronha-Hostler, J.~Noronha, and C.~Ratti,
  ``{\href{https://arxiv.org/pdf/1704.05558}{Dynamical versus equilibrium
  properties of the QCD phase transition: A holographic perspective}},''
  \emph{Phys. Rev. D}, vol.~96, no.~1, p. 014032, 2017.

\bibitem{Fotakis:2021diq}
J.~A. Fotakis, O.~Soloveva, C.~Greiner, O.~Kaczmarek, and E.~Bratkovskaya,
  ``{\href{https://arxiv.org/pdf/2102.08140}{Diffusion coefficient matrix of
  the strongly interacting quark-gluon plasma}},'' \emph{Phys. Rev. D}, vol.
  104, no.~3, p. 034014, 2021.

\bibitem{Fotakis:2019nbq}
J.~A. Fotakis, M.~Greif, C.~Greiner, G.~S. Denicol, and H.~Niemi,
  ``{\href{https://arxiv.org/pdf/1912.09103}{Diffusion processes involving
  multiple conserved charges: A study from kinetic theory and implications to
  the fluid-dynamical modeling of heavy ion collisions}},'' \emph{Phys. Rev.
  D}, vol. 101, no.~7, p. 076007, 2020.

\bibitem{Stephanov:2017ghc}
M.~Stephanov and Y.~Yin,
  ``{\href{https://arxiv.org/pdf/1712.10305}{Hydrodynamics with parametric
  slowing down and fluctuations near the critical point}},'' \emph{Phys. Rev.
  D}, vol.~98, no.~3, p. 036006, 2018.

\bibitem{Rajagopal:2019xwg}
K.~Rajagopal, G.~Ridgway, R.~Weller, and Y.~Yin,
  ``{\href{https://arxiv.org/pdf/1908.08539}{Understanding the
  out-of-equilibrium dynamics near a critical point in the QCD phase
  diagram}},'' \emph{Phys. Rev. D}, vol. 102, no.~9, p. 094025, 2020.

\bibitem{Pradeep2021}
M.~S. Pradeep,
  ``{\href{https://indico.cern.ch/event/985460/contributions/4264616/}{Freezing
  out critical fluctuations}},'' {contributed talk at the CPOD 2021
  conference}.

\bibitem{Parotto:2020fwu}
P.~Parotto, M.~Bluhm, D.~Mroczek, M.~Nahrgang, J.~Noronha-Hostler,
  K.~Rajagopal, C.~Ratti, T.~Sch\"afer, and M.~Stephanov,
  ``{\href{https://arxiv.org/pdf/1805.05249}{QCD equation of state matched to
  lattice data and exhibiting a critical point singularity}},'' \emph{Phys.
  Rev. C}, vol. 101, no.~3, p. 034901, 2020.

\bibitem{Stafford:2021wik}
J.~M. Karthein, D.~Mroczek, A.~R. Nava~Acuna, J.~Noronha-Hostler, P.~Parotto,
  D.~R.~P. Price, and C.~Ratti,
  ``{\href{https://arxiv.org/pdf/2103.08146}{Strangeness-neutral equation of
  state for QCD with a critical point}},'' \emph{Eur. Phys. J. Plus}, vol. 136,
  no.~6, p. 621, 2021.

\bibitem{Day:1967zza}
B.~D. Day, ``{\href{https://doi.org/10.1103/RevModPhys.39.719}{Elements of the
  Brueckner-Goldstone Theory of Nuclear Matter}},'' \emph{Rev. Mod. Phys.},
  vol.~39, pp. 719--744, 1967.

\bibitem{Walecka1986_lectures}
J.~Walecka, ``{\href{https://doi.org/10.1007/978-1-4684-5179-5_8}{The
  Relativistic Nuclear Many-Body Problem}},'' {\textit{New Vistas in Nuclear
  Dynamics}, pp. 229--271, Springer US, 1986}.

\bibitem{Blaettel:1993uz}
B.~Blaettel, V.~Koch, and U.~Mosel,
  ``{\href{https://doi.org/10.1088/0034-4885/56/1/001}{Transport theoretical
  analysis of relativistic heavy ion collisions}},'' \emph{Rept. Prog. Phys.},
  vol.~56, pp. 1--62, 1993.

\bibitem{Abrikosov_Methods_of_QFT_in_SP}
A.~A. Abrikosov, L.~P. Gorkov, and I.~E. Dzyaloshinski, \emph{Methods of
  Quantum Field Theory in Statistical Physics}, {Dover Publications, 1975}.

\bibitem{BaymPethickLandauFermiLiquidTheory}
G.~Baym and C.~Pethick, \emph{Landau Fermi-Liquid Theory}, {John Wiley \& Sons,
  Ltd, 2004}.

\bibitem{Brown:1971zza}
G.~E. Brown, ``{\href{https://doi.org/10.1103/RevModPhys.43.1}{Landau,
  Brueckner-Bethe, and Migdal Theories of Fermi Systems}},'' \emph{Rev. Mod.
  Phys.}, vol.~43, pp. 1--14, 1971.

\bibitem{Mayer:1949pd}
M.~G. Mayer, ``{\href{https://doi.org/10.1103/PhysRev.75.1969}{On closed shells
  in nuclei. 2}},'' \emph{Phys. Rev.}, vol.~75, pp. 1969--1970, 1949.

\bibitem{Haxel:1949fjd}
O.~Haxel, J.~H.~D. Jensen, and H.~E. Suess,
  ``{\href{https://doi.org/10.1103/PhysRev.75.1766.2}{On the ''Magic Numbers''
  in Nuclear Structure}},'' \emph{Phys. Rev.}, vol.~75, no.~11, pp. 1766--1766,
  1949.

\bibitem{WongIntroductoryNuclearPhysics}
S.~M. Wong, \emph{Introductory Nuclear Physics}, {John Wiley \& Sons, Ltd,
  1998}.

\bibitem{Bethe:1971xm}
H.~A. Bethe,
  ``{\href{https://doi.org/10.1146/annurev.ns.21.120171.000521}{Theory of
  nuclear matter}},'' \emph{Ann. Rev. Nucl. Part. Sci.}, vol.~21, pp. 93--244,
  1971.

\bibitem{Walecka:1974qa}
J.~D. Walecka, ``{\href{https://doi.org/10.1016/0003-4916(74)90208-5}{A Theory
  of highly condensed matter}},'' \emph{Annals Phys.}, vol.~83, pp. 491--529,
  1974.

\bibitem{Chin:1974sa}
S.~A. Chin and J.~D. Walecka,
  ``{\href{https://doi.org/10.1016/0370-2693(74)90708-4}{An Equation of State
  for Nuclear and Higher-Density Matter Based on a Relativistic Mean-Field
  Theory}},'' \emph{Phys. Lett. B}, vol.~52, pp. 24--28, 1974.

\bibitem{Yukawa:1935xg}
H.~Yukawa, ``{\href{https://doi.org/10.1143/PTPS.1.1}{On the Interaction of
  Elementary Particles I}},'' \emph{Proc. Phys. Math. Soc. Jap.}, vol.~17, pp.
  48--57, 1935.

\bibitem{Machleidt:1987hj}
R.~Machleidt, K.~Holinde, and C.~Elster,
  ``{\href{https://doi.org/10.1016/S0370-1573(87)80002-9}{The Bonn Meson
  Exchange Model for the Nucleon Nucleon Interaction}},'' \emph{Phys. Rept.},
  vol. 149, pp. 1--89, 1987.

\bibitem{Wiringa:1994wb}
R.~B. Wiringa, V.~G.~J. Stoks, and R.~Schiavilla,
  ``{\href{https://arxiv.org/pdf/nucl-th/9408016}{An Accurate nucleon-nucleon
  potential with charge independence breaking}},'' \emph{Phys. Rev. C},
  vol.~51, pp. 38--51, 1995.

\bibitem{Walecka:1985my}
J.~D. Walecka, ``{The relativistic nuclear many-body problem},''
  {ITP-783-STANFORD, 1985}.

\bibitem{Clark:1982sa}
B.~C. Clark, S.~Hama, and R.~L. Mercer,
  ``{\href{https://doi.org/10.1063/1.33993}{Dirac phenomenology and the nuclear
  optical model}},'' \emph{AIP Conf. Proc.}, vol.~97, pp. 260--287, 1983.

\bibitem{McNeil:1983yh}
J.~A. McNeil, J.~R. Shepard, and S.~J. Wallace,
  ``{\href{https://doi.org/10.1103/PhysRevLett.50.1439}{Impulse-Approximation
  Dirac Optical Potential}},'' \emph{Phys. Rev. Lett.}, vol.~50, p. 1439, 1983.

\bibitem{Clark:1983wa}
B.~C. Clark, S.~Hama, R.~l. Mercer, L.~Ray, and B.~d. Serot,
  ``{\href{https://doi.org/10.1103/PhysRevLett.50.1644}{Dirac-Equation Impulse
  Approximation for Intermediate-Energy Nucleon-Nucleus Scattering}},''
  \emph{Phys. Rev. Lett.}, vol.~50, pp. 1644--1647, 1983.

\bibitem{Durso:1980vn}
J.~W. Durso, A.~D. Jackson, and B.~J. Verwest,
  ``{\href{https://doi.org/10.1016/0375-9474(80)90351-6}{Models of
  pseudophysical $N\bar{N} \to \pi\pi$ amplitudes}},'' \emph{Nucl. Phys. A},
  vol. 345, pp. 471--492, 1980.

\bibitem{Noether:1918zz}
E.~Noether, ``{\href{https://doi.org/10.1080/00411457108231446}{Invariant
  Variation Problems}},'' \emph{Gott. Nachr.}, vol. 1918, pp. 235--257, 1918.

\bibitem{Landau:1956_1}
L.~D. Landau, ``{The Theory of a Fermi Liquid},'' \emph{Sov. Phys. JETP},
  vol.~3, no.~6, p. 920, 1956, [Zh. Eksp. Teor. Fiz.30,no.6,1058(1956)].

\bibitem{Landau:1957_2}
------, ``{Oscillations in a Fermi Liquid},'' \emph{Sov. Phys. JETP}, vol.~5,
  no.~1, p. 101, 1957, [Zh. Eksp. Teor. Fiz.32,no.1,59(1957)].

\bibitem{Landau:1959_3}
------, ``{On the Theory of the Fermi Liquid},'' \emph{Sov. Phys. JETP},
  vol.~8, no.~1, p.~97, 1959, [Zh. Eksp. Teor. Fiz.35,no.1,97(1959)].

\bibitem{Verresen_2019}
R.~Verresen, R.~Moessner, and F.~Pollmann,
  ``\href{https://doi.org/10.1038/s41567-019-0535-3}{Avoided quasiparticle
  decay from strong quantum interactions},'' \emph{Nature Physics}, vol.~15,
  no.~8, pp. 750--753, 2005.

\bibitem{Wheatley1970}
J.~C. Wheatley,
  ``{\href{https://www.sciencedirect.com/science/article/abs/pii/S0079641708600621}{Experimental
  Properties of Pure He$^3$ and Dilute Solutions of He$^3$ in Superfluid He$^4$
  at Very Low Temperatures. Application to Dilution Refrigeration}},''
  \emph{Progr. Low Temp. Phys.}, vol.~6, pp. 77--161, 1970.

\bibitem{Platzman1973}
P.~M. Platzman and P.~A. Wolff, \emph{Waves and Interactions in Solid State
  Plasmas}, {Academic Press, New York, 1973}.

\bibitem{Migdal_theory_of_finite_fermi_systems}
A.~B. Migdal, \emph{Theory of Finite Fermi Systems and Applications to Finite
  Nuclei}, {Interscience, London, 1967}.

\bibitem{Baym:1975va}
G.~Baym and S.~A. Chin,
  ``{\href{https://doi.org/10.1016/0375-9474(76)90513-3}{Landau Theory of
  Relativistic Fermi Liquids}},'' \emph{Nucl. Phys. A}, vol. 262, pp. 527--538,
  1976.

\bibitem{Matsui:1981ag}
T.~Matsui, ``{\href{https://doi.org/10.1016/0375-9474(81)90103-2}{Fermi-liquid
  properties of nuclear matter in a relativistic mean-field theory}},''
  \emph{Nucl. Phys. A}, vol. 370, pp. 365--388, 1981.

\bibitem{Hohenberg:1964zz}
P.~Hohenberg and W.~Kohn,
  ``{\href{https://doi.org/10.1103/PhysRev.136.B864}{Inhomogeneous Electron
  Gas}},'' \emph{Phys. Rev.}, vol. 136, pp. B864--B871, 1964.

\bibitem{Kohn:1965zzb}
W.~Kohn and L.~J. Sham,
  ``{\href{https://doi.org/10.1103/PhysRev.140.A1133}{Self-Consistent Equations
  Including Exchange and Correlation Effects}},'' \emph{Phys. Rev.}, vol. 140,
  pp. A1133--A1138, 1965.

\bibitem{Bender:2003jk}
M.~Bender, P.-H. Heenen, and P.-G. Reinhard,
  ``{\href{https://doi.org/10.1103/RevModPhys.75.121}{Self-consistent
  mean-field models for nuclear structure}},'' \emph{Rev. Mod. Phys.}, vol.~75,
  pp. 121--180, 2003.

\bibitem{Sorensen:2020ygf}
A.~Sorensen and V.~Koch, ``{\href{https://arxiv.org/pdf/2011.06635}{Phase
  transitions and critical behavior in hadronic transport with a relativistic
  density functional equation of state}},'' \emph{Phys. Rev. C}, vol. 104,
  no.~3, p. 034904, 2021.

\bibitem{Kruse:1985hy}
H.~Kruse, B.~V. Jacak, and H.~Stoecker,
  ``{\href{https://doi.org/10.1103/PhysRevLett.54.289}{Microscopic theory of
  pion production and sidewards flow in heavy ion collisions}},'' \emph{Phys.
  Rev. Lett.}, vol.~54, pp. 289--292, 1985.

\bibitem{Landau_Classical_Theory_of_Fields}
L.~D. Landau and E.~M. Lifshitz, \emph{The Classical Theory of Fields, Course
  of Theoretical Physics, Vol.\ 2}, {Butterworth-Heinemann, 1980}.

\bibitem{Novak:2013bqa}
J.~Novak, K.~Novak, S.~Pratt, J.~Vredevoogd, C.~Coleman-Smith, and R.~Wolpert,
  ``{\href{https://arxiv.org/pdf/1303.5769}{Determining Fundamental Properties
  of Matter Created in Ultrarelativistic Heavy-Ion Collisions}},'' \emph{Phys.
  Rev. C}, vol.~89, no.~3, p. 034917, 2014.

\bibitem{Bernhard:2016tnd}
J.~E. Bernhard, J.~S. Moreland, S.~A. Bass, J.~Liu, and U.~Heinz,
  ``{\href{https://arxiv.org/pdf/1605.03954}{Applying Bayesian parameter
  estimation to relativistic heavy-ion collisions: simultaneous
  characterization of the initial state and quark-gluon plasma medium}},''
  \emph{Phys. Rev. C}, vol.~94, no.~2, p. 024907, 2016.

\bibitem{Ko:1987gp}
C.~M. Ko, Q.~Li, and R.-C. Wang,
  ``{\href{https://doi.org/10.1103/PhysRevLett.59.1084}{Relativistic Vlasov
  Equation for Heavy Ion Collisions}},'' \emph{Phys. Rev. Lett.}, vol.~59, pp.
  1084--1087, 1987.

\bibitem{Buss:2011mx}
O.~Buss, T.~Gaitanos, K.~Gallmeister, H.~van Hees, M.~Kaskulov, O.~Lalakulich,
  A.~B. Larionov, T.~Leitner, J.~Weil, and U.~Mosel,
  ``{\href{https://arxiv.org/pdf/1106.1344}{Transport-theoretical Description
  of Nuclear Reactions}},'' \emph{Phys. Rept.}, vol. 512, pp. 1--124, 2012.

\bibitem{McLerran:2018hbz}
L.~McLerran and S.~Reddy, ``{\href{https://arxiv.org/pdf/1811.12503}{Quarkyonic
  Matter and Neutron Stars}},'' \emph{Phys. Rev. Lett.}, vol. 122, no.~12, p.
  122701, 2019.

\bibitem{Hempel:2013tfa}
M.~Hempel, V.~Dexheimer, S.~Schramm, and I.~Iosilevskiy,
  ``{\href{https://arxiv.org/pdf/1302.2835}{Noncongruence of the nuclear
  liquid-gas and deconfinement phase transitions}},'' \emph{Phys. Rev. C},
  vol.~88, no.~1, p. 014906, 2013.

\bibitem{Landau_Stat}
L.~D. Landau and E.~M. Lifshitz, \emph{Statistical Physics, Course of
  Theoretical Physics, Vol.\ 5}, {Pergamon Press, New York, 1980}.

\bibitem{Chomaz:2003dz}
P.~Chomaz, M.~Colonna, and J.~Randrup,
  ``{\href{http://hal.in2p3.fr/in2p3-00020100}{Nuclear spinodal
  fragmentation}},'' \emph{Phys. Rept.}, vol. 389, pp. 263--440, 2004.

\bibitem{Dutra:2012mb}
M.~Dutra, O.~Lourenco, J.~S. Sa~Martins, A.~Delfino, J.~R. Stone, and P.~D.
  Stevenson, ``{\href{Dutra:2012mb}{Skyrme Interaction and Nuclear Matter
  Constraints}},'' \emph{Phys. Rev. C}, vol.~85, p. 035201, 2012.

\bibitem{Zeldovich:1962emp}
Y.~B. Zel'dovich, ``{The equation of state at ultrahigh densities and its
  relativistic limitations},'' \emph{Zh. Eksp. Teor. Fiz.}, vol.~41, pp.
  1609--1615, 1961.

\bibitem{Bedaque:2014sqa}
P.~Bedaque and A.~W. Steiner, ``{\href{https://arxiv.org/pdf/1408.5116}{Sound
  velocity bound and neutron stars}},'' \emph{Phys. Rev. Lett.}, vol. 114,
  no.~3, p. 031103, 2015.

\bibitem{Annala:2019puf}
E.~Annala, T.~Gorda, A.~Kurkela, J.~N\"attil\"a, and A.~Vuorinen,
  ``{\href{https://arxiv.org/pdf/1903.09121}{Evidence for quark-matter cores in
  massive neutron stars}},'' \emph{Nature Phys.}, vol.~16, no.~9, pp. 907--910,
  2020.

\bibitem{Kapusta:1984ij}
J.~I. Kapusta, ``{\href{https://doi.org/10.1103/PhysRevC.29.1735}{Deuteron and
  entropy production and the nuclear liquid-gas phase transition}},''
  \emph{Phys. Rev. C}, vol.~29, pp. 1735--1743, 1984.

\bibitem{Poberezhnyuk:2017yhx}
R.~V. Poberezhnyuk, V.~Vovchenko, D.~V. Anchishkin, and M.~I. Gorenstein,
  ``{\href{https://arxiv.org/pdf/1708.05605}{Quantum van der Waals and Walecka
  models of nuclear matter}},'' \emph{Int. J. Mod. Phys. E}, vol.~26, no.~10,
  p. 1750061, 2017.

\bibitem{Vovchenko:2016rkn}
V.~Vovchenko, M.~I. Gorenstein, and H.~Stoecker,
  ``{\href{https://arxiv.org/pdf/1609.03975}{van der Waals Interactions in
  Hadron Resonance Gas: From Nuclear Matter to Lattice QCD}},'' \emph{Phys.
  Rev. Lett.}, vol. 118, no.~18, p. 182301, 2017.

\bibitem{Wong:1982zzb}
C.-Y. Wong, ``{\href{https://doi.org/10.1103/PhysRevC.25.1460}{Dynamics of
  nuclear fluid. VIII. Time-dependent Hartree-Fock approximation from a
  classical point of view}},'' \emph{Phys. Rev. C}, vol.~25, pp. 1460--1475,
  1982.

\bibitem{Ko:1988zz}
C.-M. Ko and Q.~Li,
  ``{\href{https://doi.org/10.1103/PhysRevC.37.2270}{Relativistic
  Vlasov-Uehling-Uhlenbeck model for heavy-ion collisions}},'' \emph{Phys. Rev.
  C}, vol.~37, pp. 2270--2273, 1988.

\bibitem{Bertsch:1988ik}
G.~F. Bertsch and S.~Das~Gupta,
  ``{\href{https://doi.org/10.1016/0370-1573(88)90170-6}{A Guide to microscopic
  models for intermediate-energy heavy ion collisions}},'' \emph{Phys. Rept.},
  vol. 160, pp. 189--233, 1988.

\bibitem{Xu:2016lue}
J.~Xu \emph{et~al.}, ``{\href{https://arxiv.org/pdf/1603.08149}{Understanding
  transport simulations of heavy-ion collisions at 100A and 400A MeV:
  Comparison of heavy-ion transport codes under controlled conditions}},''
  \emph{Phys. Rev. C}, vol.~93, no.~4, p. 044609, 2016.

\bibitem{smash_version_2.0}
{SMASH version 2.0, available at
  \href{https://doi.org/10.5281/zenodo.4336358}{https://doi.org/10.5281/zenodo.4336358}.}

\bibitem{SMASH_user_guide}
{SMASH User Guide, available at
  \href{http://theory.gsi.de/~smash/userguide/current/}{http://theory.gsi.de/$_{\widetilde{~}}$smash/userguide/current/}.}

\bibitem{smash_development_documentation}
{SMASH documentation, available at
  \href{http://theory.gsi.de/~smash/doc/current}{http://theory.gsi.de/$_{\widetilde{~}}$smash/doc/current}.}

\bibitem{Colonna:2021xuh}
M.~Colonna \emph{et~al.}, ``{\href{https://arxiv.org/pdf/2106.12287}{Comparison
  of heavy-ion transport simulations: Mean-field dynamics in a box}},''
  \emph{Phys. Rev. C}, vol. 104, no.~2, p. 024603, 2021.

\bibitem{Lenk:1989zz}
R.~J. Lenk and V.~R. Pandharipande,
  ``{\href{https://doi.org/10.1103/PhysRevC.39.2242}{Nuclear mean field
  dynamics in the lattice Hamiltonian Vlasov method}},'' \emph{Phys. Rev. C},
  vol.~39, pp. 2242--2249, 1989.

\bibitem{Oliinychenko:2015lva}
D.~Oliinychenko and H.~Petersen,
  ``{\href{https://arxiv.org/pdf/1508.04378}{Deviations of the Energy-Momentum
  Tensor from Equilibrium in the Initial State for Hydrodynamics from Transport
  Approaches}},'' \emph{Phys. Rev. C}, vol.~93, no.~3, p. 034905, 2016.

\bibitem{Wang:2019ghr}
R.~Wang, L.-W. Chen, and Z.~Zhang,
  ``{\href{https://arxiv.org/pdf/1902.01256}{Nuclear collective dynamics in the
  lattice Hamiltonian Vlasov method}},'' \emph{Phys. Rev. C}, vol.~99, no.~4,
  p. 044609, 2019.

\bibitem{smash_version_1.8}
{SMASH version 1.8, available at
  \href{https://doi.org/10.5281/zenodo.3742965}{https://doi.org/10.5281/zenodo.3742965}.}

\bibitem{Bonasera:1994iam}
A.~Bonasera, M.~Colonna, M.~Di~Toro, F.~Gulminelli, and A.~Smerzi,
  ``{\href{https://doi.org/10.1016/0375-9474(94)90428-6}{Equilibrium and
  nonequilibrium fluctuations of the one-body distribution function}},''
  \emph{Nucl. Phys. A}, vol. 572, pp. 171--180, 1994.

\bibitem{Bonasera:1990ikj}
A.~Bonasera, M.~Colonna, M.~Di~Toro, F.~Gulminelli, and H.~H. Wolter,
  ``{\href{https://doi.org/10.1016/0370-2693(90)90049-C}{Microscopic
  description of dissipative fragmentation}},'' \emph{Phys. Lett. B}, vol. 244,
  pp. 169--172, 1990.

\bibitem{Bonasera:1992zz}
A.~Bonasera, F.~Gulminelli, and P.~Schuck,
  ``{\href{https://doi.org/10.1103/PhysRevC.46.1431}{Fluctuations of the
  one-body distribution function}},'' \emph{Phys. Rev. C}, vol.~46, pp.
  1431--1436, 1992.

\bibitem{Burgio:1991ej}
G.~F. Burgio, P.~Chomaz, and J.~Randrup,
  ``{\href{https://lib-extopc.kek.jp/preprints/PDF/2000/0037/0037491.pdf}{Dynamical
  clusterization in presence of instabilities}},'' \emph{Phys. Rev. Lett.},
  vol.~69, pp. 885--888, 1992.

\bibitem{Steinheimer:2017dpb}
J.~Steinheimer and V.~Koch, ``{\href{https://arxiv.org/pdf/1705.08538}{Effect
  of finite particle number sampling on baryon number fluctuations}},''
  \emph{Phys. Rev. C}, vol.~96, no.~3, p. 034907, 2017.

\bibitem{Sorensen:2021zme}
A.~Sorensen, D.~Oliinychenko, V.~Koch, and L.~McLerran,
  ``{\href{https://arxiv.org/pdf/2103.07365}{Speed of Sound and Baryon
  Cumulants in Heavy-Ion Collisions}},'' \emph{Phys. Rev. Lett.}, vol. 127,
  no.~4, p. 042303, 2021.

\bibitem{Tews:2018kmu}
I.~Tews, J.~Carlson, S.~Gandolfi, and S.~Reddy,
  ``{\href{https://arxiv.org/pdf/1801.01923}{Constraining the speed of sound
  inside neutron stars with chiral effective field theory interactions and
  observations}},'' \emph{Astrophys. J.}, vol. 860, no.~2, p. 149, 2018.

\bibitem{Gardim:2019xjs}
F.~G. Gardim, G.~Giacalone, M.~Luzum, and J.-Y. Ollitrault,
  ``{\href{Gardim:2019xjs}{Thermodynamics of hot strong-interaction matter from
  ultrarelativistic nuclear collisions}},'' \emph{Nature Phys.}, vol.~16,
  no.~6, pp. 615--619, 2020.

\bibitem{Steinheimer:2012bp}
J.~Steinheimer and M.~Bleicher,
  ``{\href{https://arxiv.org/pdf/1207.2792}{Extraction of the sound velocity
  from rapidity spectra: Evidence for QGP formation at FAIR/RHIC-BES
  energies}},'' \emph{Eur. Phys. J. A}, vol.~48, p. 100, 2012.

\bibitem{Theis:1984qc}
J.~Theis, G.~Graebner, G.~Buchwald, J.~A. Maruhn, W.~Greiner, H.~Stoecker, and
  J.~Polonyi, ``{\href{https://doi.org/10.1103/PhysRevD.28.2286}{Phase
  transition of the nucleon-antinucleon plasma in a relativistic mean-field
  theory}},'' \emph{Phys. Rev. D}, vol.~28, pp. 2286--2290, 1983.

\bibitem{HADES:2020wpc}
J.~Adamczewski-Musch \emph{et~al.},
  ``{\href{https://arxiv.org/pdf/2002.08701}{Proton-number fluctuations in
  $\sqrt {s_{NN}}$ =2.4 GeV Au + Au collisions studied with the High-Acceptance
  DiElectron Spectrometer (HADES)}},'' \emph{Phys. Rev. C}, vol. 102, no.~2, p.
  024914, 2020.

\bibitem{HADES_MLorentz_talk}
M.~Lorenz,
  ``{\href{https://indico.gsi.de/event/9423/contributions/40798/}{Strangeness
  production at HADES}},'' {contribution to the 3rd EMMI Workshop, 2019}.

\bibitem{Shen:2020jwv}
C.~Shen and S.~Alzhrani,
  ``{\href{https://arxiv.org/pdf/2003.05852}{Collision-geometry-based 3D
  initial condition for relativistic heavy-ion collisions}},'' \emph{Phys. Rev.
  C}, vol. 102, no.~1, p. 014909, 2020.

\bibitem{Du:2020bxp}
L.~Du, U.~Heinz, K.~Rajagopal, and Y.~Yin,
  ``{\href{https://arxiv.org/pdf/2004.02719}{Fluctuation dynamics near the QCD
  critical point}},'' \emph{Phys. Rev. C}, vol. 102, no.~5, p. 054911, 2020.

\bibitem{Jiang:2017sni}
L.~Jiang, J.-H. Zheng, and H.~Stoecker,
  ``{\href{https://arxiv.org/pdf/1711.05339}{Ising parameterization of QCD
  Landau free energy and its dynamics}},'' 11 2017.

\bibitem{Bzdak:2012an}
A.~Bzdak, V.~Koch, and V.~Skokov,
  ``{\href{https://arxiv.org/pdf/1203.4529}{Baryon number conservation and the
  cumulants of the net proton distribution}},'' \emph{Phys. Rev. C}, vol.~87,
  no.~1, p. 014901, 2013.

\bibitem{Vogt:2007zz}
R.~Vogt, \emph{{Ultrarelativistic Heavy-Ion Collisions}}, {Elsevier, Amsterdam,
  2007}.

\bibitem{Sorensen:2003kp}
P.~R. Sorensen, ``{\href{https://arxiv.org/pdf/nucl-ex/0309003}{Kaon and Lambda
  production at intermediate p(T): Insights into the hadronization of the bulk
  partonic matter created in Au+Au collisions at RHIC}},'' 9 2003.

\bibitem{Walecka:1995mi}
J.~D. Walecka, \emph{{Theoretical Nuclear and Subnuclear Physics}}, {Oxford
  University Press, 1995}.

\bibitem{Hugenholtz:1958zz}
N.~M. Hugenholtz and L.~van Hove,
  ``{\href{https://doi.org/10.1016/S0031-8914(58)95281-9}{A theorem on the
  single particle energy in a Fermi gas with interaction}},'' \emph{Physica},
  vol.~24, pp. 363--376, 1958.

\bibitem{Kittel_Elementary_statistical_physics}
C.~Kittel, \emph{Elementary Statistical Physics}, {John Wiley \& Sons, New
  York, 1958}.

\bibitem{P_Eastman_lecture_notes}
{Lecture notes on Statistical Mechanics by P.~Eastman, Chapter 8, available at
  \href{https://web.stanford.edu/~peastman/statmech/}{https://web.stanford.edu/$_{\widetilde{~}}$peastman/statmech/}}.

\bibitem{Kardar_Statistical_physics_of_particles}
M.~Kardar, \emph{Statistical Physics of Particles}, {Cambridge University
  Press, New York, 2007}.

\bibitem{H_van_Hees_lecture_notes}
{Lecture notes on Relativistic Transport Theory by H.~van Hees, available at
  \href{https://itp.uni-frankfurt.de/~hees/publ/roorkee.pdf}{https://itp.uni-frankfurt.de/$_{\widetilde{~}}$hees/publ/roorkee.pdf}}.

\bibitem{Uehling_Uhlenbeck_on_Boltzmann}
E.~A. Uehling and G.~E. Uhlenbeck,
  ``\href{https://link.aps.org/doi/10.1103/PhysRev.43.552}{Transport Phenomena
  in Einstein-Bose and Fermi-Dirac Gases. I},'' \emph{Phys. Rev.}, vol.~43, pp.
  552--561, Apr 1933.

\end{thebibliography}
\bibliographystyle{IEEEtran}

\end {document}